\pdfoutput=1
\documentclass{aa}
\usepackage{natbib}
\usepackage{graphicx}
\usepackage{pdflscape}
\usepackage{xcolor}
\usepackage{subfig}
\usepackage{threeparttable}
\usepackage{amsmath}
\usepackage{txfonts}
\newcommand{\methanol}{\mbox{CH$_3$OH}}

\newcommand{\hii}{H{\sc ii}}

\newcommand{\uchii}{UC\,H{\sc ii}}

\newcommand{\hchii}{HC\,H{\sc ii}}

\newcommand{\KS}{Kolmogorov-Smirnov}

\newcommand{\lsun}{L$_\odot$}

\newcommand{\meth}{\mbox{CH$_{3}$OH}\,}
\newcommand{\msol}{\mbox{M$_{\odot}$}\,}

\newcommand{\asec}{$^{\prime\prime}$\,}

\newcommand{\fasec}{.\!\!$^{\prime\prime}$}
\newcommand{\dirty}{``dirty''\,}
\newcommand{\clean}{\texttt{CLEAN}}

\DeclareRobustCommand{\kms}{km\,${\rm s}^{-1}$}
\DeclareRobustCommand{\jyb}{Jy\,beam${}^{-1}$\,}

\newcommand{\dconf}{D-configuration\,}
\newcommand{\bconf}{B-configuration\,}

\newif\iflong
\longtrue
\newcommand{\C}{\color{black}}

\DeclareGraphicsExtensions{.pdf,.png}
\usepackage[colorlinks=true, allcolors=blue]{hyperref}
\begin{document} 

\title{A Global View on Star Formation: The GLOSTAR Galactic Plane Survey}
\subtitle{V. 6.7\,GHz Methanol Maser Catalogue}
\titlerunning{GLOSTAR: 6.7\,GHz Methanol Maser Catalogue}

   \author{H.\,Nguyen
          \inst{1}\fnmsep\thanks{Member of the International Max Planck Research School (IMPRS) for Astronomy and Astrophysics at the Universities of Bonn and Cologne.}
          \and
          M.\,R.\,Rugel\inst{1}
          \and
          C.\,Murugeshan \inst{2,1}
          \and
          K.\,M.\,Menten\inst{1}
          \and
          A.\,Brunthaler\inst{1}
          \and
          J.\,S.\,Urquhart \inst{3}
          \and
          R.\,Dokara \inst{1,\star}
          \and
          S.\,A.\,Dzib\inst{4,1}
          \and
          Y.\,Gong \inst{1}
          \and
          S.\,Khan \inst{1,\star}
          \and
          S-N.\,X.\,Medina \inst{1}
          \and
          G.\,N.\,Ortiz-Le{\'o}n \inst{5,1}
          \and
          W.\,Reich \inst{1}
          \and
          F. Wyrowski \inst{1}
          \and
          A.\,Y.\,Yang\inst{1}          
          \and
          H.\,Beuther \inst{6}
          \and
          W.\,D.\,Cotton \inst{7,8}
          %\and
          %T.\,Csengeri \inst{8}
          \and
          J.\,D.\,Pandian \inst{9}
          %\and
          %N.\,Roy \inst{10}
          }

   \institute{Max-Planck-Institut f\"ur Radioastronomie, Auf dem H\"ugel 69, 53121 Bonn, Germany\\
            \email{hnguyen@mpifr-bonn.mpg.de}
            \and
             CSIRO Space and Astronomy, PO Box 1130, Bentley WA 6102, Australia
            \and
            Centre for Astrophysics and Planetary Science, University of Kent, Ingram Building, Canterbury, Kent CT2 7NH, UK
            \and
            IRAM, 300 rue de la piscine, 38406 Saint Martin d'H\`eres, France
            \and
            Instituto de Astronom\'ia, Universidad Nacional Aut\'onoma de M\'exico (UNAM), Apdo Postal 70-264, Ciudad de M\'exico, M\'exico.
            \and
            Max Planck Institute for Astronomy, Königstuhl 17, D-69117 Heidelberg, Germany
            \and
            National Radio Astronomy Observatory, 520 Edgemont Road, Charlottesville, VA 22903, USA
            \and
            South African Radio Astronomy Observatory, 2 Fir St, Black River Park, Observatory 7925, South Africa
            %\and
            %Laboratoire d'astrophysique de Bordeaux, Univ. Bordeaux, CNRS, B18N, all\'ee Geoffroy Saint-Hilaire, 33615 Pessac, France
            \and
            Department of Earth \& Space Sciences, Indian Institute of Space Science and Technology, Trivandrum 695547, India
            %\and
            %Department of Physics, Indian Institute of Science, Bengaluru 560012, India
            }

   \date{Received May 25, 2022 / Accepted July 18, 2022}
   %\date{Draft \today}

% \abstract{}{}{}{}{} 
% 5 {} token are mandatory
 
  \abstract
   {Class II methanol (CH$_{3}$OH) masers are amongst the clearest signposts of recent high-mass star formation (HMSF). A complete catalogue outlines the distribution of star formation in the Galaxy, the number of young star-forming cores, and the physical conditions of their environment. The Global View on Star Formation (GLOSTAR) survey, which is a blind survey in the radio regime of 4--8\,GHz, maps the Galactic mid-plane in the radio continuum, 6.7\,GHz methanol line, the 4.8\,GHz formaldehyde line, and several radio recombination lines. We present the analysis of the observations of the 6.7\,GHz CH$_{3}$OH maser transition using data from the D-configuration of the Very Large Array (VLA). We analyse the data covering Galactic longitudes from $-2^{\circ}< l <60^{\circ}$ and Galactic latitudes of $|\textit{b}|<1^{\circ}$. We detect a total of 554 methanol masers, out of which 84 are new, and catalogue their positions, velocity components, and integrated fluxes. With a typical noise level of $\sim$18\,mJy\,beam$^{-1}$, this is the most sensitive unbiased methanol survey for methanol masers to date. We search for dust continuum and radio continuum associations, and find that 97\% of the sources are associated with dust, and 12\% are associated with radio continuum emission.} 
   %{b}
   %{c}
   %{d}
   %{e}

   \maketitle

\section{Introduction}\label{sect:intro}

The Global View On Star Formation (GLOSTAR) survey \citep{BrunthalerPilot} is an unbiased survey that observed the Galactic plane with the Karl G. Jansky Very Large Array (VLA) in D- and B-configurations and the Effelsberg 100\,m radio telescope in order to find and characterise star-forming regions in the Milky Way. Being both sensitive and having high angular resolution, the data contain a wealth of information that has already been used to catalogue new radio sources \citep{sac2019}, to identify supernova remnants (SNR; \citealt{dokara2021}) and to find new methanol (\meth) masers in the Cygnus~X region \citep{giselacygnus}. 

In studying high-mass star formation (HMSF), methanol masers have proven to be indispensable. Interstellar methanol maser emission was first discovered by \cite{barret1971} at 25\,GHz towards Orion-KL. Since then, 
many other methanol maser lines have since been discovered, such as those at 6.7 and 12.2\,GHz \citep{batrla1987,menten1991b}. These are divided into two types based on their pumping mechanism, collisional \citep[Class I:][]{batrla1987, cragg1992, voronkov2010, voronkov2014, leurini2016} or radiative \citep[Class II:][]{menten1991a, menten1991b, mmb1C}. In particular, Class II \meth masers have already proven to be one of the clearest signposts of HMSF with the 6.7\,GHz transition being the brightest and most widespread in the Galaxy \citep{menten1991b, menten1993, walsh1997, walsh1998}. Second only to the 22.2\,GHz H$_{2}$O maser in its intensity and abundance, methanol maser emission at 6.7 GHz is unique in that it exclusively traces high-mass star forming regions \citep{minier2003, ellingsen2006, xu2008}. The 6.7~GHz line from the $5_{1}-6_{0}$A$^{+}$ transition of the methanol molecule requires specific conditions in order to begin masing. These are met in the surrounding dust and gas of massive young stellar objects (MYSOs) with densities greater than $10^{6}$\,cm$^{-3}$ and temperatures $>150$\,K \citep[e.g.,][]{sobolev1994,cragg2005} due to the intense radiation of the MYSOs. While some of the earliest detections of Class II methanol masers were made toward ultra-compact \hii~regions, namely the archetypical W3(OH) \citep{batrla1987, menten1992}, it was found that in fact
very few of these methanol masers have radio continuum counterparts \citep{walsh1998, beuther2002, urquhart2013, urquhart2015, hu2016, billington2019}. It is thus clear, that most 6.7\,GHz methanol masers probe MYSOs located in regions of recent high-mass star formation, and finding and studying them can provide insight on the distribution of these regions in the Galaxy and to characterise them.

Due to their usefulness in the study of high-mass star formation, many targeted surveys \citep[e.g.,][]{menten1991b,macleod1992,caswell1995,caswell1996GC,ellingsen1996,vanderwalt1996,walsh1997,ellingsen2007,yang2019} and unbiased surveys \citep{rickert2019,pesta2005,Pandian2007, mmb1C,mmb2G, mmb3C, mmb4G, mmb5B, giselacygnus} have been performed, culminating in over 1000 Class II 6.7\,GHz methanol masers being discovered in our Milky Way Galaxy. However, with the technological upgrades to the VLA \citep{perley2011}, the methanol data from the GLOSTAR survey provides the most sensitive and unbiased catalogue to date. % with which to study HMSF regions.  

\begin{figure*}[]
\centering 
   \includegraphics[trim={0.0cm 0.2cm 0.0cm 0.2cm}, clip, width=0.71\textwidth]{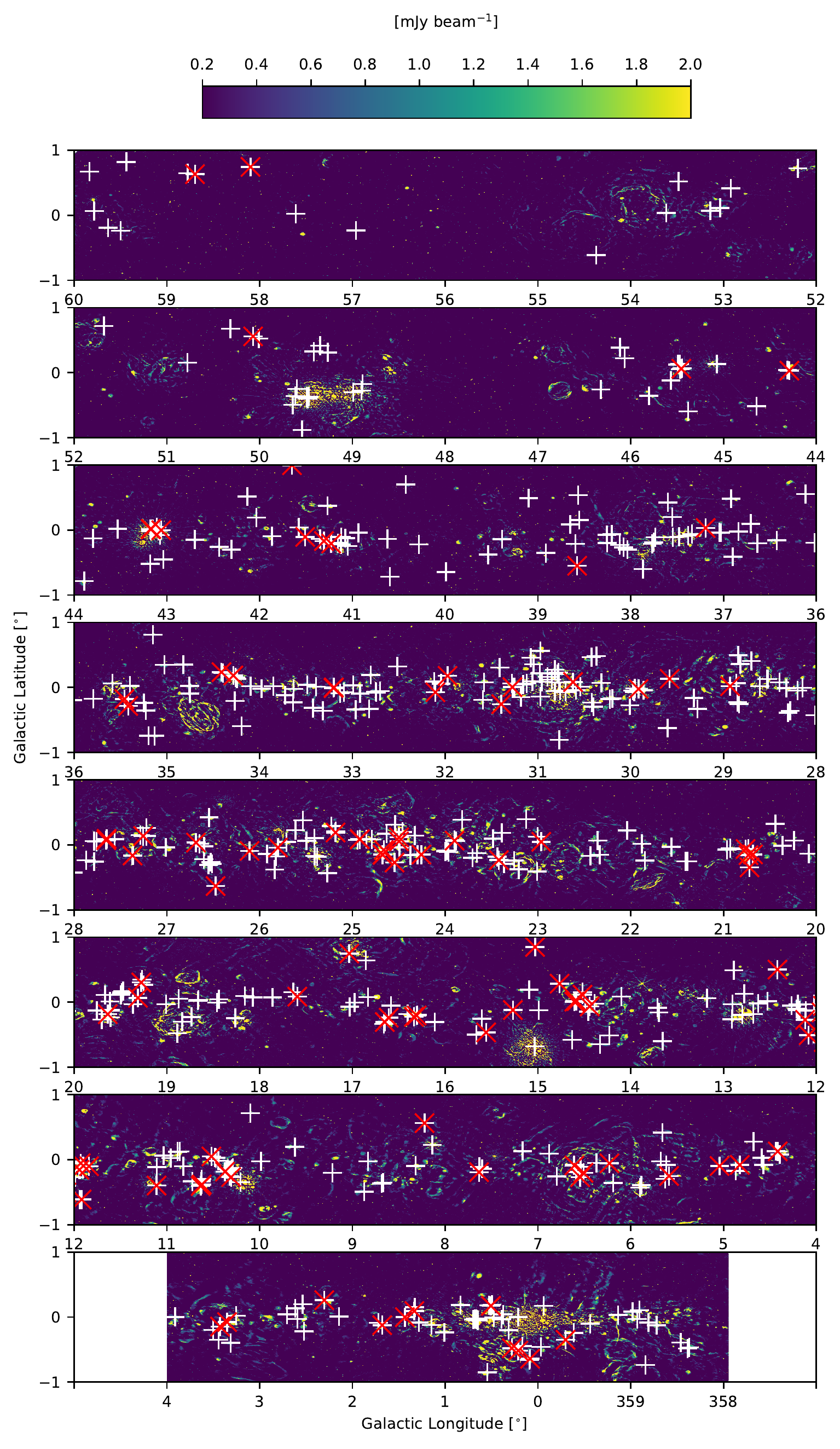}
    \caption{Methanol maser detections plotted as white crosses on top of the \dconf continuum images from GLOSTAR where the flux has been limited to be between 0.2 and 2\,m\jyb for visibility. The red `x's show the positions of all new methanol masers as discussed in Section~\ref{sect:otherSurveys}.}
    \label{fig:maser_overlay_dconf}
\end{figure*}

Here, we report on the detection of 554 6.7\,GHz methanol masers in the region of $l=-2^{\circ}$~to~$60^{\circ}$ and $|b|=<1^{\circ}$ (see Fig.~\ref{fig:maser_overlay_dconf} for survey coverage); 84 of these represent new detections. We used an automated search algorithm to search and verify all detections manually. We look for associations at other wavelengths to identify the physical properties of the population of sources with detected maser emission and the difference in these properties for newly detected sources with respect to the total population. 

In addition to the part of the Galactic plane listed above, the Cygnus~X star formation complex was also covered by the GLOSTAR survey. The 6.7 GHz methanol maser content of Cygnus~X has been discussed in a recent article by \citet{giselacygnus}.

We structure this paper as follows: In Section~\ref{sect:obs} we give a summary of the data used in this paper along with its calibration and imaging. Section~\ref{sect:sec} describes the algorithm used to detect masers in the VLA data. Section~\ref{sect:results} details the production of our methanol maser catalogue and their general properties.
Section~\ref{sect:discussion} discusses our comparison with other surveys and associations with other wavelengths. We present
the conclusions and summary in Section~\ref{sect:summary}.
%%%%%%%%%%%%%%%%%%%%%%%%%%%%%%%%%%%%%%%%%%%%%%%%%%%%%%%%%%%%%%%%%%%%%%%%%

%%%%%%%%%%%%%%%%%%%%%%%%%%%%%%%%%%%%%%%%%%%%%%%%%%%%%%%%%%%%%%%%%%%%%%%%%

\section{Observations}\label{sect:obs}

\begin{table}[]
\tiny
    \centering
    \caption{Summary of the VLA observations}
    \label{tab:dconf_obs}
    \begin{threeparttable}
    \begin{tabular}{l r r r}
         \hline
         \hline
         Observing Date & Galactic coverage & Program & Calibrator\\
         D-conf.&  & ID\\
         \hline
         
 2014-10-05$^{a}$& $-2^{\circ}<l<-1^{\circ}$ ; $|b|<1.0^{\circ}$& 14B-254&J1820-2528\\ 
 2014-09-26$^{a}$& $-1^{\circ}<l<0^{\circ}$ ; $|b|<1.0^{\circ}$& 14B-254&J1820-2528\\ 
 2014-09-28$^{a}$& $0^{\circ}<l<1^{\circ}$ ; $|b|<1.0^{\circ}$& 14B-254&J1820-2528\\ 
 2016-01-09$^{a}$& $1^{\circ}<l<2^{\circ}$ ; $|b|<1.0^{\circ}$& 15B-175&J1820-2528\\ 
 2016-01-17$^{a}$& $2^{\circ}<l<3^{\circ}$ ; $|b|<1.0^{\circ}$& 15B-175&J1820-2528\\ 
 2016-01-21$^{a}$& $3^{\circ}<l<4^{\circ}$ ; $|b|<1.0^{\circ}$& 15B-175&J1820-2528\\ 
 2016-01-22$^{a}$& $4^{\circ}<l<5^{\circ}$ ; $|b|<1.0^{\circ}$& 15B-175&J1820-2528\\ 
 2016-01-16$^{a}$& $5^{\circ}<l<6^{\circ}$ ; $|b|<1.0^{\circ}$& 15B-175&J1820-2528\\ 
 2017-04-03& $6^{\circ}<l<7^{\circ}$ ; $|b|<1.0^{\circ}$& 17A-197&J1820-2528\\ 
 2017-03-31& $7^{\circ}<l<8^{\circ}$ ; $|b|<1.0^{\circ}$& 17A-197&J1820-2528\\ 
 2017-02-20& $8^{\circ}<l<9^{\circ}$ ; $|b|<1.0^{\circ}$& 17A-197&J1820-2528\\ 
 2016-01-24$^{a}$& $9^{\circ}<l<10^{\circ}$ ; $|b|<1.0^{\circ}$& 15B-175&J1820-2528\\ 
 2013-05-16$^{a}$& $10^{\circ}<l<11^{\circ}$ ; $|b|<1.0^{\circ}$& 13A-334&J1811-2055\\ 
 2013-05-17& $11^{\circ}<l<12^{\circ}$ ; $|b|<1.0^{\circ}$& 13A-334&J1825-0737\\ 
 2017-03-06& $12^{\circ}<l<13^{\circ}$ ; $|b|<1.0^{\circ}$& 17A-197&J1825-0737\\ 
 2017-04-04& $13^{\circ}<l<14^{\circ}$ ; $|b|<1.0^{\circ}$& 17A-197&J1825-0737\\ 
 2017-02-19& $14^{\circ}<l<15^{\circ}$ ; $|b|<1.0^{\circ}$& 17A-197&J1825-0737\\ 
 2014-07-14& $15^{\circ}<l<16^{\circ}$ ; $|b|<1.0^{\circ}$& 14A-420&J1825-0737\\ 
 2014-07-24& $16^{\circ}<l<17^{\circ}$ ; $|b|<1.0^{\circ}$& 14A-420&J1825-0737\\ 
 2014-08-05& $17^{\circ}<l<18^{\circ}$ ; $|b|<1.0^{\circ}$& 14A-420&J1825-0737\\ 
 2014-08-14& $18^{\circ}<l<19^{\circ}$ ; $|b|<1.0^{\circ}$& 14A-420&J1825-0737\\ 
 2014-07-12& $19^{\circ}<l<20^{\circ}$ ; $|b|<1.0^{\circ}$& 14A-420&J1825-0737\\ 
 2014-07-23& $20^{\circ}<l<21^{\circ}$ ; $|b|<1.0^{\circ}$& 14A-420&J1825-0737\\ 
 2014-07-28& $21^{\circ}<l<22^{\circ}$ ; $|b|<1.0^{\circ}$& 14A-420&J1825-0737\\ 
 2014-07-27& $22^{\circ}<l<23^{\circ}$ ; $|b|<1.0^{\circ}$& 14A-420&J1825-0737\\ 
 2014-08-26& $23^{\circ}<l<24^{\circ}$ ; $|b|<1.0^{\circ}$& 14A-420&J1825-0737\\ 
 2014-07-16& $24^{\circ}<l<25^{\circ}$ ; $|b|<1.0^{\circ}$& 14A-420&J1825-0737\\ 
 2014-07-29& $25^{\circ}<l<26^{\circ}$ ; $|b|<1.0^{\circ}$& 14A-420&J1825-0737\\ 
 2014-08-13& $26^{\circ}<l<27^{\circ}$ ; $|b|<1.0^{\circ}$& 14A-420&J1825-0737\\ 
 2014-08-28& $27^{\circ}<l<28^{\circ}$ ; $|b|<1.0^{\circ}$& 14A-420&J1825-0737\\ 
 2013-04-09& $28^{\circ}<l<29^{\circ}$ ; $|b|<1.0^{\circ}$& 13A-334&J1804+0101\\ 
 2013-04-06& $29^{\circ}<l<30^{\circ}$ ; $|b|<1.0^{\circ}$& 13A-334&J1804+0101\\ 
 2013-04-11& $30^{\circ}<l<31^{\circ}$ ; $|b|<1.0^{\circ}$& 13A-334&J1804+0101\\ 
 2013-04-15& $31^{\circ}<l<32^{\circ}$ ; $|b|<1.0^{\circ}$& 13A-334&J1804+0101\\ 
 2013-04-16& $32^{\circ}<l<33^{\circ}$ ; $|b|<1.0^{\circ}$& 13A-334&J1804+0101\\ 
 2013-04-20& $33^{\circ}<l<34^{\circ}$ ; $|b|<1.0^{\circ}$& 13A-334&J1804+0101\\ 
 2013-04-29& $34^{\circ}<l<35^{\circ}$ ; $|b|<1.0^{\circ}$& 13A-334&J1804+0101\\ 
 2013-05-02& $35^{\circ}<l<36^{\circ}$ ; $|b|<1.0^{\circ}$& 13A-334&J1804+0101\\ 
 2014-07-07& $36^{\circ}<l<37^{\circ}$ ; $|b|<1.0^{\circ}$& 14A-420&J1907+0127\\ 
 2014-07-04& $37^{\circ}<l<38^{\circ}$ ; $|b|<1.0^{\circ}$& 14A-420&J1907+0127\\ 
 2014-08-01& $38^{\circ}<l<39^{\circ}$ ; $|b|<1.0^{\circ}$& 14A-420&J1907+0127\\ 
 2014-08-25& $39^{\circ}<l<40^{\circ}$ ; $|b|<1.0^{\circ}$& 14A-420&J1907+0127\\ 
 2014-08-07& $40^{\circ}<l<41^{\circ}$ ; $|b|<1.0^{\circ}$& 14A-420&J1907+0127\\ 
 2014-07-21& $41^{\circ}<l<42^{\circ}$ ; $|b|<1.0^{\circ}$& 14A-420&J1907+0127\\ 
 2014-07-09& $42^{\circ}<l<43^{\circ}$ ; $|b|<1.0^{\circ}$& 14A-420&J1907+0127\\ 
 2014-07-17& $43^{\circ}<l<44^{\circ}$ ; $|b|<1.0^{\circ}$& 14A-420&J1907+0127\\ 
 2014-08-03& $44^{\circ}<l<45^{\circ}$ ; $|b|<1.0^{\circ}$& 14A-420&J1907+0127\\ 
 2014-06-29& $45^{\circ}<l<46^{\circ}$ ; $|b|<1.0^{\circ}$& 14A-420&J1907+0127\\ 
 2015-11-25& $46^{\circ}<l<47^{\circ}$ ; $|b|<1.0^{\circ}$& 15B-175&J1922+1530\\ 
 2015-11-13& $47^{\circ}<l<48^{\circ}$ ; $|b|<1.0^{\circ}$& 15B-175&J1922+1530\\ 
 2015-11-21& $48^{\circ}<l<49^{\circ}$ ; $|b|<1.0^{\circ}$& 15B-175&J1922+1530\\ 
 2015-11-14& $49^{\circ}<l<50^{\circ}$ ; $|b|<1.0^{\circ}$& 15B-175&J1922+1530\\ 
 2015-11-22& $50^{\circ}<l<51^{\circ}$ ; $|b|<1.0^{\circ}$& 15B-175&J1922+1530\\ 
 2015-11-11& $51^{\circ}<l<52^{\circ}$ ; $|b|<1.0^{\circ}$& 15B-175&J1922+1530\\ 
 2015-11-20& $52^{\circ}<l<53^{\circ}$ ; $|b|<1.0^{\circ}$& 15B-175&J1922+1530\\ 
 2015-11-10& $53^{\circ}<l<54^{\circ}$ ; $|b|<1.0^{\circ}$& 15B-175&J1922+1530\\ 
 2015-11-27& $54^{\circ}<l<55^{\circ}$ ; $|b|<1.0^{\circ}$& 15B-175&J1922+1530\\ 
 2015-12-17& $55^{\circ}<l<56^{\circ}$ ; $|b|<1.0^{\circ}$& 15B-175&J1922+1530\\ 
 2015-11-28& $56^{\circ}<l<57^{\circ}$ ; $|b|<1.0^{\circ}$& 15B-175&J1925+2106\\ 
 2015-11-08& $57^{\circ}<l<58^{\circ}$ ; $|b|<1.0^{\circ}$& 15B-175&J1925+2106\\ 
 2011-12-15& $58^{\circ}<l<59^{\circ}$ ; $|b|<1.0^{\circ}$& 11B-168&J1931+2243\\ 
 2011-12-29& $59^{\circ}<l<60^{\circ}$ ; $|b|<1.0^{\circ}$& 11B-168&J1931+2243\\ 
         \hline
    \end{tabular}
    \begin{tablenotes}
    \item[a] These observations were conducted with the VLA in DnC configuration.
    \end{tablenotes}
    
\end{threeparttable}
\end{table}

The GLOSTAR survey \citep{sac2019, BrunthalerPilot} is an on-going survey with the VLA and the Effelsberg 100\,m telescope between 4--8 GHz of the Galactic mid-plane from
$-2^{\circ}<\textit{l}<60^{\circ}$ and $|\textit{b}|<1^{\circ}$, and
the Cygnus~X star-forming complex. The VLA observations used in this work were conducted in D-configuration with a typical angular resolution of 18\asec at 6.7\,GHz. % and 1.5\arcsec~at 5.8\,GHz, respectively, 
Using methanol, formaldehyde, and radio recombination lines as well as radio continuum data, the survey aims to detect various tracers of different stages of early star formation in order to gain information on the start of the stellar evolution process of massive stars. The full details can be found in \cite{BrunthalerPilot}. The observations used in this work were carried out using $\sim$300\,hours during the time period from December 2011 until April 2017 where the program IDs and details are summarised in Table~\ref{tab:dconf_obs}. Observations that were performed in the DnC hybrid configuration are marked in the table.

\subsection{VLA data calibration}\label{sect:cali}
As detailed in \cite{BrunthalerPilot}, a modified version of the VLA scripted pipeline\footnote{\url{https://science.nrao.edu/facilities/vla/data-processing/pipeline/scripted-pipeline}} (version 1.3.8) for CASA\footnote{\url{https://casa.nrao.edu/}} (version 4.6.0) was adapted to work with spectral line data. We highlight the relevant changes here: no Hanning smoothing was performed on the first pass when producing the preliminary images to preserve the spectral resolution where possible; Hanning smoothing is performed on a select few sources after an initial inspection; the \texttt{rflag} flagging command was only applied to the calibration scans to avoid flagging spectral lines erroneously; \texttt{statwt} was not used to modify the statistical weights. 
The complex gain calibrators used for different fields include: J1804+0101, J1820-2528, J1811-2055, J1825-0737, J1907+0127, J1955+1530, J1925+2106, and J1931+2243, and the flux calibrators are 3C~286 and 3C48.
%%%%%%%%%%%%%%%%%%%%%%%%%%%%%%%%%%%%%%%%%%%%%%%%%%%%%%%%%%%%%%%%%%%%%%%%%
%%%%%%%%%%%%%%%%%%%%%%%%%%%%%%%%%%%%%%%%%%%%%%%%%%%%%%%%%%%%%%%%%%%%%%%%%
\subsection{Spectral line data imaging}\label{sect:clean}
To process the spectral line data, we image it in two steps. We first produce a so called \dirty image (or un\clean ed), which is just the direct Fourier transform of the $uv$ data
with the \texttt{tclean} task in CASA. We search these cubes for preliminary sources. We then properly \clean~smaller sub cubes centered on these sources
to search for additional sources. This approach was chosen to address the computation limitations imposed by the sheer volume of data used. With the available computing resources at the time, it would take a month to produce a $2^{\circ}\times1^{\circ}$ \clean ed data cube for the methanol data, where a dirty image would take only three days.%For more details, see \cite{BrunthalerPilot}.

\subsubsection{Imaging ``dirty'' cubes}
Using the CASA task \texttt{tclean} we produce preliminary, mosaicked, and primary beam corrected images with the following task parameters: \texttt{niter}=0, a cube size
(\texttt{imsize}) of $2500\times2500$ pixels, and a pixel size (\texttt{cell}) of 2.5\asec for the \dconf data. The number of channels (\texttt{nchan}) and the rest frequency
(\texttt{restfreq}) are set to 1800 channels and 6668.518\,MHz, respectively. We use a `natural' \texttt{weighting} parameter for better sensitivity in detecting sources. On
average, the 1$\sigma$ root-mean-square (rms) noise in the line free channels of the dirty cubes is found to be $\sim$18\,mJy\,beam$^{-1}$ for a single channel (0.18\,\kms), which is
better than the estimated noise for a single pointing of $\sim$40\,mJy\,beam$^{-1}$ based on a 15 second integration time. This is due to the mapping strategy where each field has been overlapped by six neighbouring fields resulting in sensitivities that are at least a factor of two better. The achieved sensitivity is about a factor of 2 better than previous VLA targeted surveys \citep[e.g.,][]{hu2016}. We are able to detect weak and isolated masers in the \dirty image itself which leads to savings in computational time. Weak masers that are in the vicinity of strong masers and with similar velocity may be missed but will be found in the next stage of our imaging approach.

\subsubsection{Imaging of individual masers}
\label{sect:subcubes}
In order to find weaker masers that are hidden in the side lobes of stronger masers, we use the \clean~algorithm on small subsets of the data. We select on average 16 pointings that cover the positions of sources detected from the \dirty images. They are split out for further imaging and deconvolution to make smaller cubes ($\sim$0.2$^\circ \times 0.2^\circ$, $\Delta v\sim$55\,\kms). These sub-cubes are imaged with a cell/pixel size of 2.5$\arcsec$ and the imaging is restricted to the velocity ranges over which significant signal was detected. The sub-cubes have a spatial extent of $350\times350$ pixels and spectral extents of 300 channels centered on the peak velocity. Included at the beginning and end of the 300 channels are line free channels to estimate the spectral noise. For deconvolution, a \clean ing threshold of 38\,~m\jyb\ was chosen ($\sim$2$\sigma$ of final cubes). Using the \texttt{tclean} task of CASA (version 5.4.0), the images were made with parameters: \texttt{gridder}=`mosaic', \texttt{deconvolver}=`hogbom', \texttt{weighting}=`uniform', and a variable \texttt{niter}.
The number of minor \clean~cycles was set to either 1000, 5000 or 10000 depending on the strength of the maser involved. These iteration values were found to optimize the automatic cleaning for many sources by maximizing image fidelity while avoiding over-cleaning. The weighting parameter was set to `uniform', since it gives a better angular resolution and thus more accurate positional information.

\subsection{Complementary continuum data}\label{sect:contData}

In addition to the methanol line data, we also use GLOSTAR-VLA radio 5.8\,GHz continuum data to search for associations with methanol masers. Radio sources such as ultra-compact \hii~regions (UC\hii) are a clear indicator of HMSF, however their relationship with methanol masers is not yet fully understood and thus, a study of the associations and the physical properties of these sources may give insight on the overall formation process of high-mass stars in this stage of their evolution. The full analysis of continuum maps in the \dconf and \bconf will be presented in forthcoming papers \citep{medinaFull, dzibBconf, yangBconf} while the \dconf continuum catalogue for the pilot region is already complete \citep{sac2019}. We also make use of the Co-Ordinated Radio `N' Infrared Survey for High-mass star formation \citep[CORNISH][]{hoare2012,cornish_source} which used the VLA in the B and BnA-configuration at 5\,GHz to supplement our comparisons.

%%%%%%%%%%%%%%%%%%%%%%%%%%%%%%%%%%%%%%%%%%%%%%%%%%%%%%%%%%%%%%%%%%%%%%%%%
\section{Source extraction}\label{sect:sec}
Here we give a technical description of the algorithm used for automatically selecting maser candidates from the \dirty~images. A description of the final catalogue is given in Section~\ref{sect:detections}.

\subsection{Source extraction code}

As explained in Section~\ref{sect:clean}, given the computational challenge of imaging and deconvolving large mosaics, we adopted an approach of searching for the methanol masers in the dirty images. We wrote a simple Source Extraction Code \citep[SEC;][]{murugeshan2015,nguyen2015} to deal specifically with this data in a rapid manner. This SEC was written to detect methanol masers that have high brightness values in comparison to the surrounding noise.
It was also adapted for absorption searches \citep[see][]{BrunthalerPilot}. It takes the \dirty or fully \clean ed images as input.
The code was first used on the \dirty~images to produce a preliminary catalogue of detections which were verified by visual inspection. The verified detections were \clean ed over a small spatial and spectral extent (see Section~\ref{sect:subcubes}). The code was then used again on these sub-cubes to find any other masers that were previously not detected on account of their proximity to stronger sources.

The code scans through the images, saving the coordinates where the brightness meets two criteria.
The first and foremost is that the emission surpasses a certain signal to noise ratio (S/N) threshold. 
Secondly, we require that the emission is above this threshold for at least two consecutive channels. If these two criteria are satisfied, the code considers
such a detection as real. 

\begin{figure}[ht]
\centering
\includegraphics[trim={0.0cm 3.0cm 0.0cm 0.0cm}, clip,width = 0.50\textwidth]{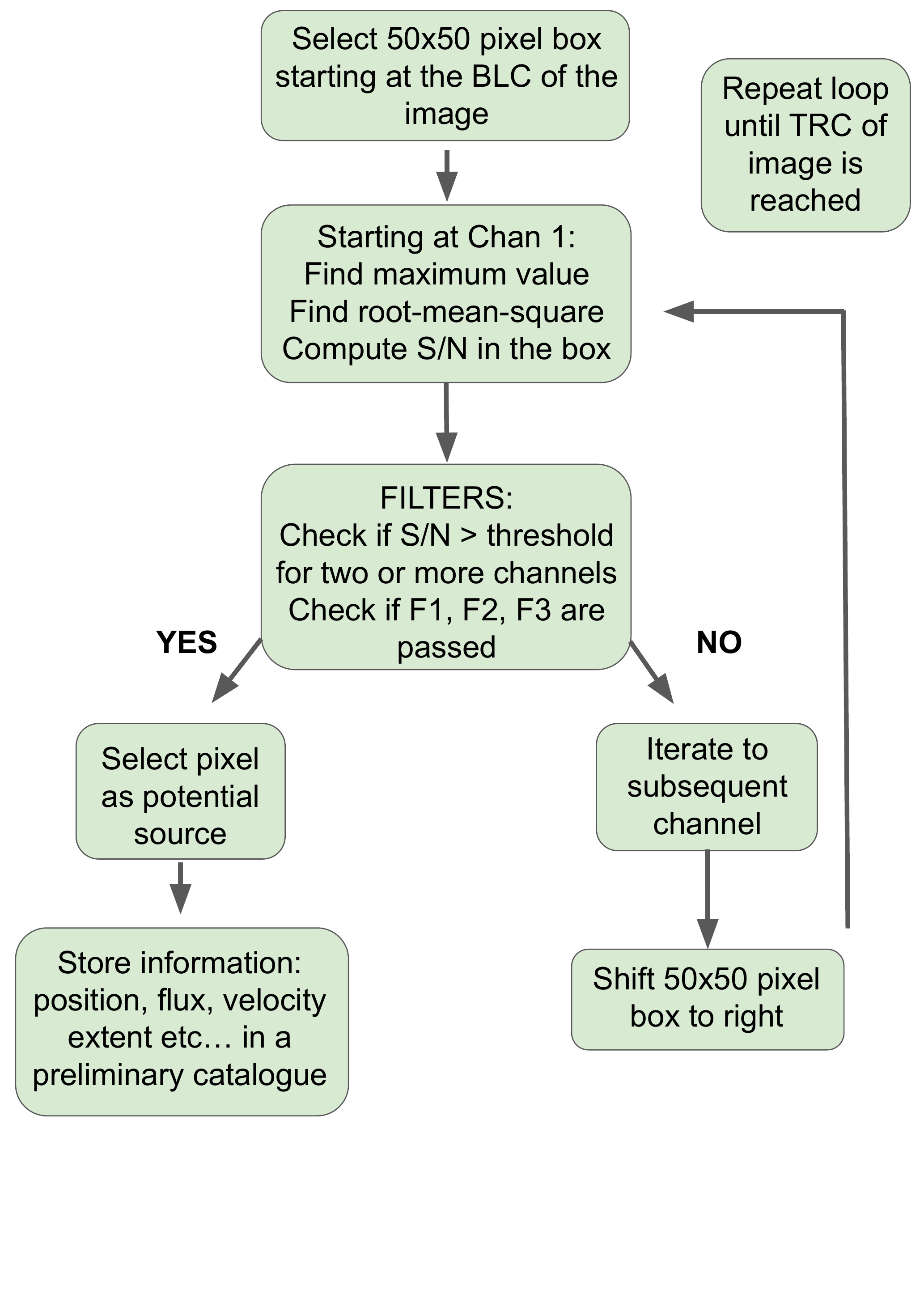}\\
\includegraphics[width = 0.40\textwidth]{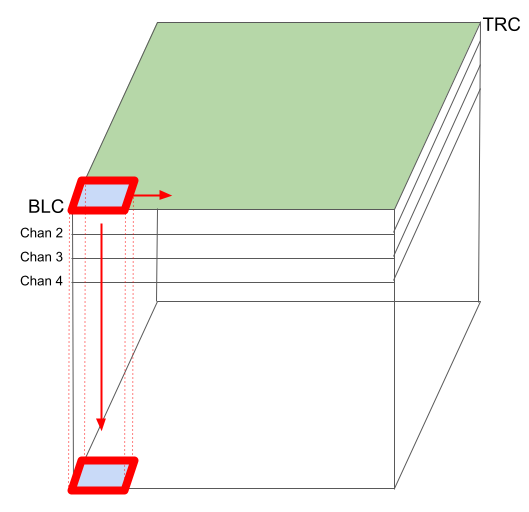}
\caption{\emph{Top}: Algorithm flow chart for the Source Extraction Code (SEC) which details the selection of the positions and channels of maser candidates \C{based on the signal-to-noise (S/N) ratio (see Section~\ref{sect:sec}).} \emph{Bottom}: Illustration of process used by the SEC. A 50$\times$50 pixel box \C{that starts from the bottom left corner (BLC) iterates first through channels to detect sources until it reaches the top right corner (TRC).}}
\label{fig:search_chart}
\end{figure}

We select a $50\times50$ pixel ``box'' to start. Beginning from the first channel, the root mean square (rms) of the box is calculated. This is chosen to be the noise within the box. Next, the code selects the pixel with the maximum flux. The ratio between
the maximum flux and the rms of the box then defines the S/N. This is done iteratively for each channel within the boundaries 
of the $50\times50$ pixel box for the entire spectral extent of the given image. For a given channel, where the S/N is above a user-defined
threshold, we check if it continues for at least another contiguous channel in order to satisfy our main criteria described above. If it also passes through three additional selection filters, which we will define
below, it qualifies as a potential source. We then record the pixel coordinates, right ascension (RA), declination (DEC), channel range, peak flux,
the associated channel of the peak flux, and the S/N. The code moves onto the next spatial box and repeats the search along the spectral axis. After its completion,
a catalogue of the potential sources is produced. As a final check, we always verify each source visually. A flow diagram illustrating the algorithm is presented
in the top of Fig.~\ref{fig:search_chart} and an illustration of the physical movements of the search algorithm is shown in the bottom.

In order to reduce false detections and repeated detections, we describe the aforementioned filters used to improve the quality of detections:
\begin{itemize}
\item {\bf First filter (F1)}:  This filter checks if the current potential detection is within 200 pixels from
the previous detection and also appears in the same channel range. If these conditions are met, it checks if
the S/N of current detection is greater than the previous detection. If this is the case, the current detection is updated
as the true detection and its coordinates kept as a reference.
\item {\bf Second filter (F2)}:  The next filter checks if the current detection is within 200 pixels of the
previous detection, but appears in a different channel range. Since side lobe artefacts appear in the same channel range as the source,
we consider the detection a potential source if the channel ranges are different. This helps to differentiate multiple sources within the same box.
\item {\bf Third filter (F3)}: The final filter checks if the current detection is more than 200 pixels away from any previous detection.
An angular separation of 200 pixels ($\sim$8.3$^{\prime}$) is chosen as typically this is the largest extent of any
side lobe feature. If this criterion is met, the code selects the pixel as a potential source.
\end{itemize}

\begin{figure*}[]
    \begin{tabular}{l r}
    \centering
    \includegraphics[width = 0.5\textwidth]{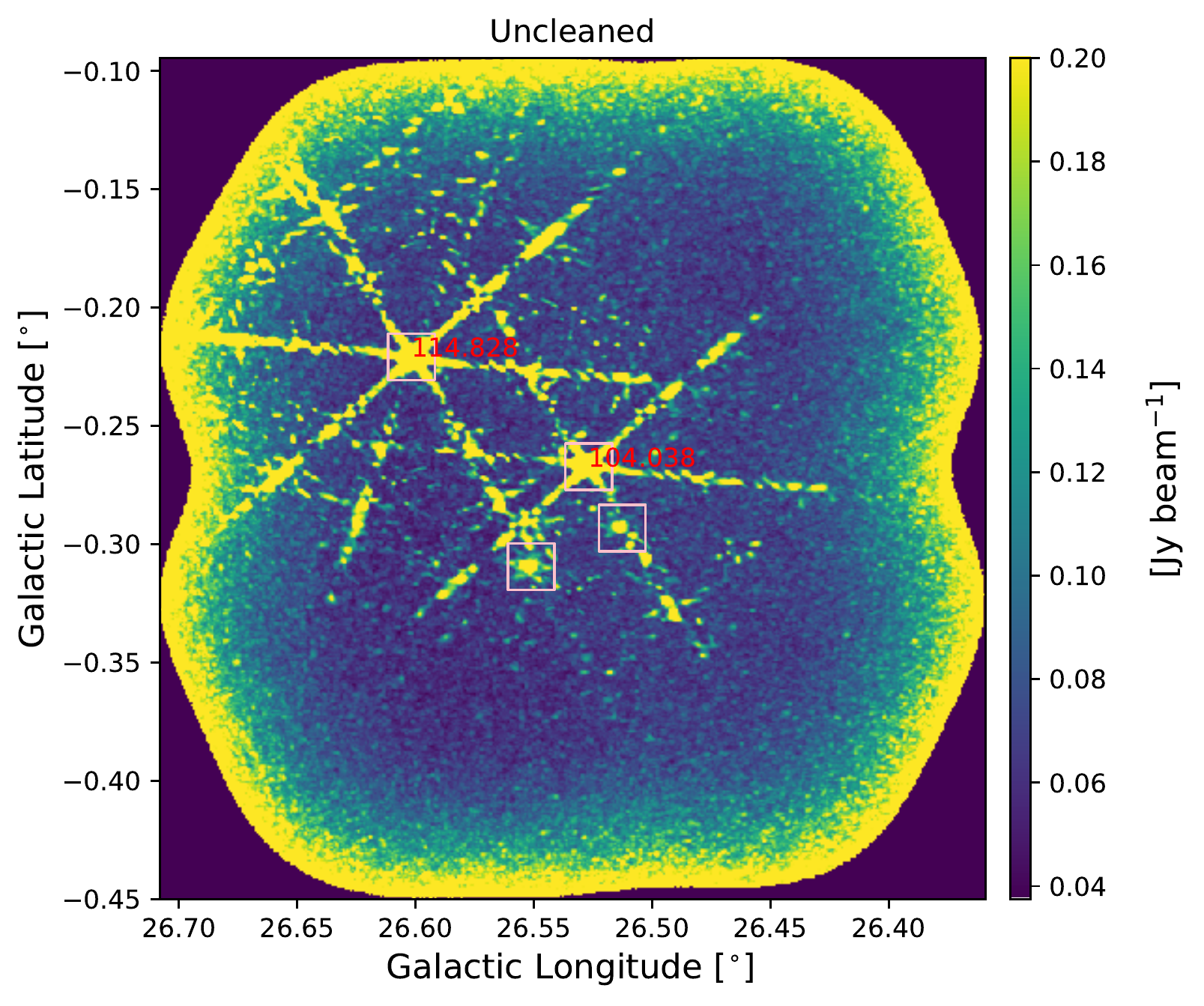}
    \includegraphics[width = 0.5\textwidth]{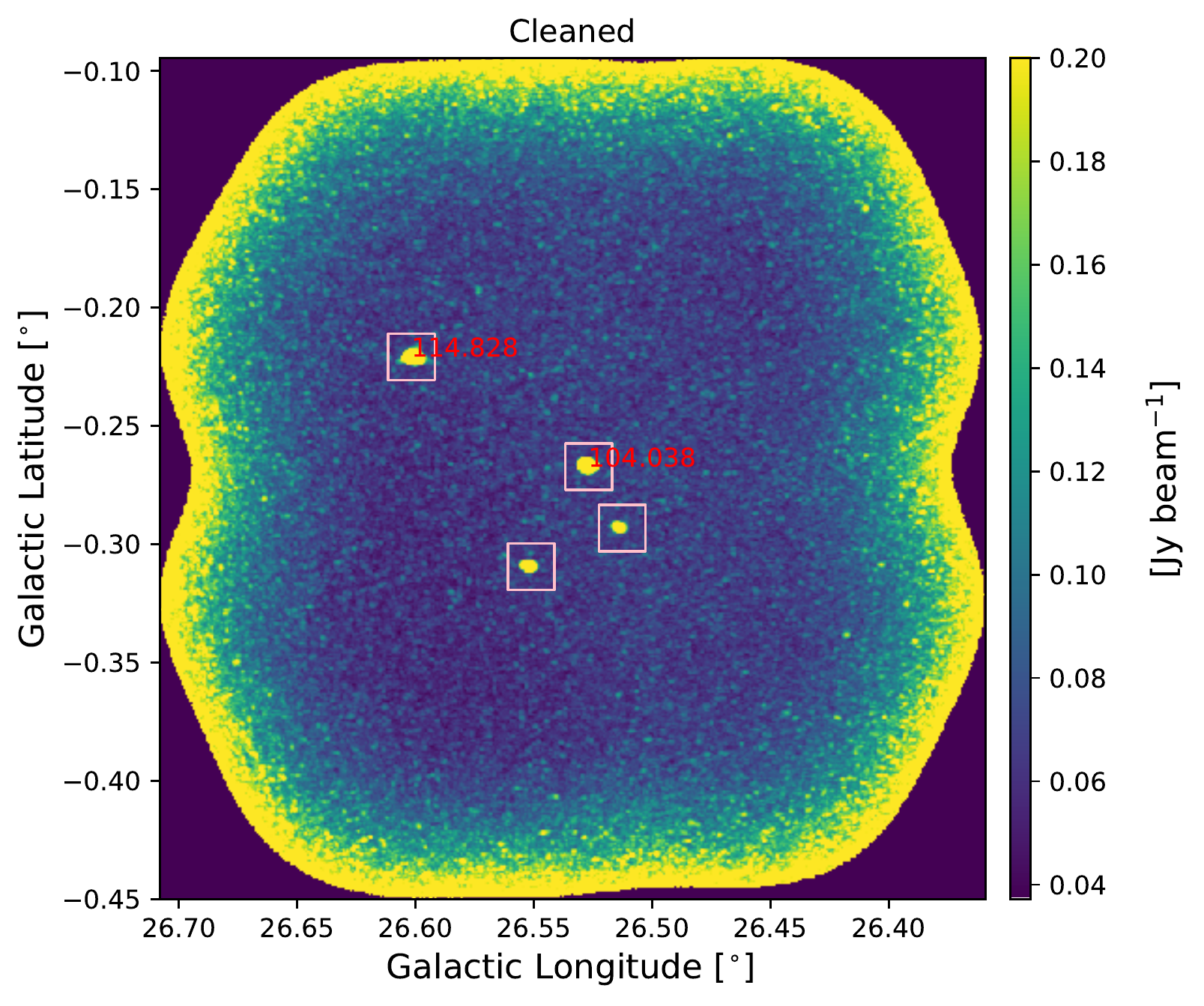}
\end{tabular}
\caption{Example of the \C{Source Extraction Code} (SEC) on un\clean ed and \clean ed data sub-cubes. \emph{Left}: Example peak intensity map of a dirty sub-cube towards $l=26.5270$\,deg, $b=-0.2674$\,deg which is an example of the data used when making preliminary detection catalogues with our SEC. Velocities of known sources are shown in red text while pink squares highlight the final source detections. In this \dirty image, the bottom two sources would not have been detected. \emph{Right}: Once we \clean~the image, we can see how the two bottom sources would indeed be detected by our SEC and as such, illustrates how running the SEC again on the cleaned sub-cubes can lead to new detections.}
\label{fig:cleaned_sub_sec}
\end{figure*}

We chose a S/N value of 6 when searching through the \dirty images. This choice is  based on the general statistics of the cube in order to balance dismissing artefacts as well as picking up weak sources. For the \clean ed sub-cubes, we use a lower S/N value of 4 as it was found to be the most efficient in detecting the generally weaker sources that were initially hidden in the side lobes of the nearby stronger sources. 

Since the algorithm picks out just one maximum in a given box, it is possible that real but weaker sources are missed if they lie in the same box.
It was found that a box size of 50 pixels (corresponding to $\sim$2$^{\prime}$\,; in comparison the primary beam of the VLA at 6.7\,GHz is 6.5$^{\prime}$) substantially minimizes the run time without losing too many additional sources. Furthermore, because we are using a \dirty~image, there are noisy structures in the neighbourhood of strong sources. 
In the direct vicinity of a potential strong source, the code will reliably pick out the pixel coordinates of the real source as side lobe features will never have a higher flux than the actual source. The spillage of side lobe features into adjacent boxes is mitigated by the utilisation of the filters defined above.

Given the simple nature of the algorithm, the code can be easily parallelized.
The parallelization is done by splitting the spatial extent of the image into smaller patches. The code was then run individually on each of these patches. Each process creates its own search catalogue which is then combined when all processes are finished. For a strong source, side lobe artefacts can spill over onto the next patch and be picked up in a different catalogue.
We resolve this by verifying visually.

\subsection{Catalogue creation}
\label{sect:create_cat}
After performing the initial pass with the SEC on the \dirty~images, we cross check with all known masers in the survey range in the case that there are weaker known sources that do not pass our initial noise threshold criterion. We further inspect all the potential SEC detections, looking at their spectra and moment maps. We examine the positions, velocities, and intensities of the detections to decide which SEC candidates are to be considered real. We visually check for any velocity features that have very large offsets from other velocity features along the line of sight to also be considered as a new maser. From this, we compile a list of candidates for further cleaning. 

For these maser candidates, we make smaller \clean~images as described in Section~\ref{sect:clean}. We run our SEC code with modified settings (box size of 5 pixels and no filters) in order to pick up weaker sources that would have been hidden by a nearby strong source. We visually inspect each cube at the end to ensure that we pick up all possible masers in the data. Shown in Fig.~\ref{fig:cleaned_sub_sec} is an example of two new methanol maser detections that were not found in the \dirty image cube due to the presence of nearby strong sources.

From this final visual inspection, we compile a final list of detections.
As we are interested in determining the methanol masers' properties, we perform a 2D Gaussian fit of the brightness distribution in each channel whose peak intensity is above 4$\sigma$. We use the CASA task \texttt{imfit} on a $14\times14$ pixel box ($\sim$2$\times$ the restoring beam) which is centered on the pixel with maximum intensity. As discussed in \cite{giselacygnus}, the error in maser position is determined by the astrometric uncertainty $\theta_{\text{res}}/(2\times$S/N), where $\theta_{\text{res}}$ is the (VLA) restoring beam, and S/N is the ratio between source intensity and rms \citep{thompson2017}. Given that the average beam size in the D-configuration methanol maps across the whole Galactic plane is 15$^{\prime\prime}$, and a maser detection with S/N=10, the precision in position is $\approx0$\fasec7. We further discuss the reliability of our position measurements in Section~\ref{sect:otherSurveys}. We visually verify the result of the source fits, as well as determine the peaks of each maser site. We refer to each velocity peak as a maser spot inside one maser site \citep{walsh2014}. These are listed in Table~{\ref{tab:maser_spots_eg}} for a few examples of new masers sites while the complete catalogue can be found online. The positions for all emission channels of a given maser site are given as an offset to the position of the maser spot with the strongest emission within the maser site.

%%%%%%%%%%%%%%%%%%%%%%%%%%%%%%%%%%%%%%%%%%%%%%%%%%%%%%%%%%%%%%%%%%%%%%%%%
\section{Results}\label{sect:results}
\subsection{Detections}
\label{sect:detections}
\begin{figure}
\centering
\includegraphics[width = 0.49\textwidth]{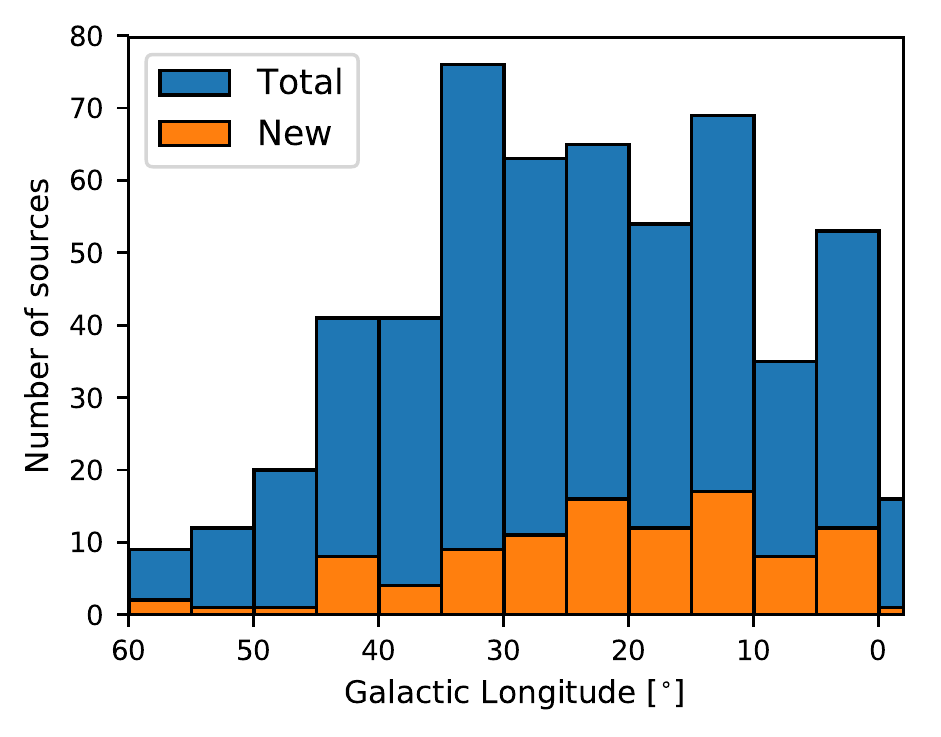}
\caption{Distribution of detected masers along Galactic longitude. The bin width used is $5^{\circ}$ from $l=60^{\circ}$ to $0^{\circ}$ and $2^{\circ}$ for the last bin.}
\label{fig:distribution_longitude}
\end{figure}

\begin{figure}
\centering
\includegraphics[width = 0.49\textwidth]{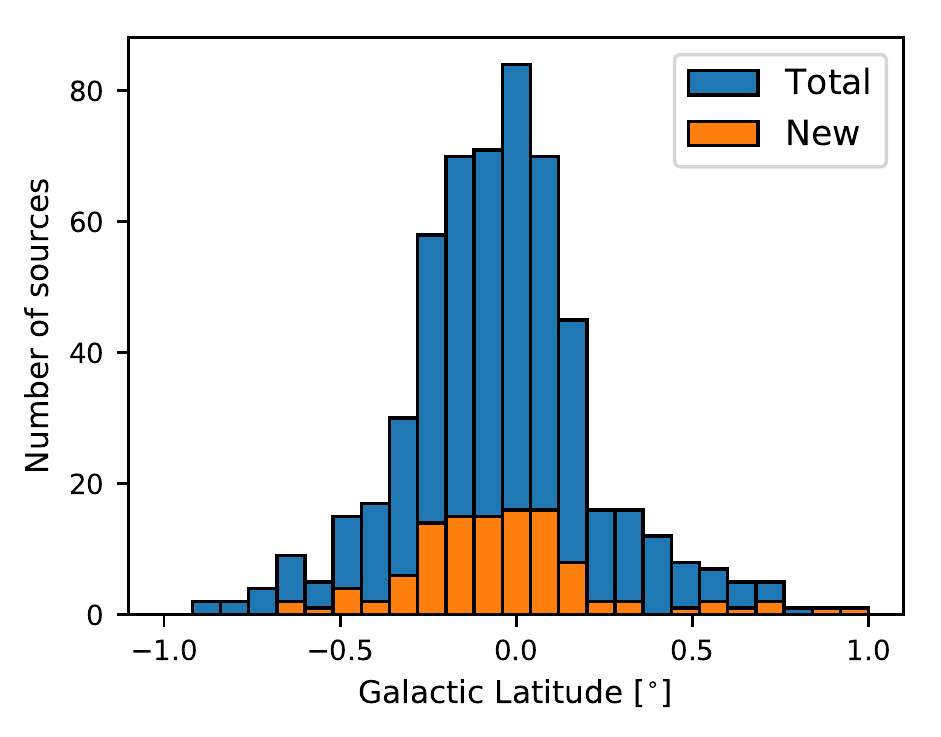}
\caption{Distribution of detected masers along Galactic latitude. The bin width used is $0.08^{\circ}$.}
\label{fig:distribution_latitude}
\end{figure}

\begin{figure*}
\centering
\includegraphics[width = 1\textwidth]{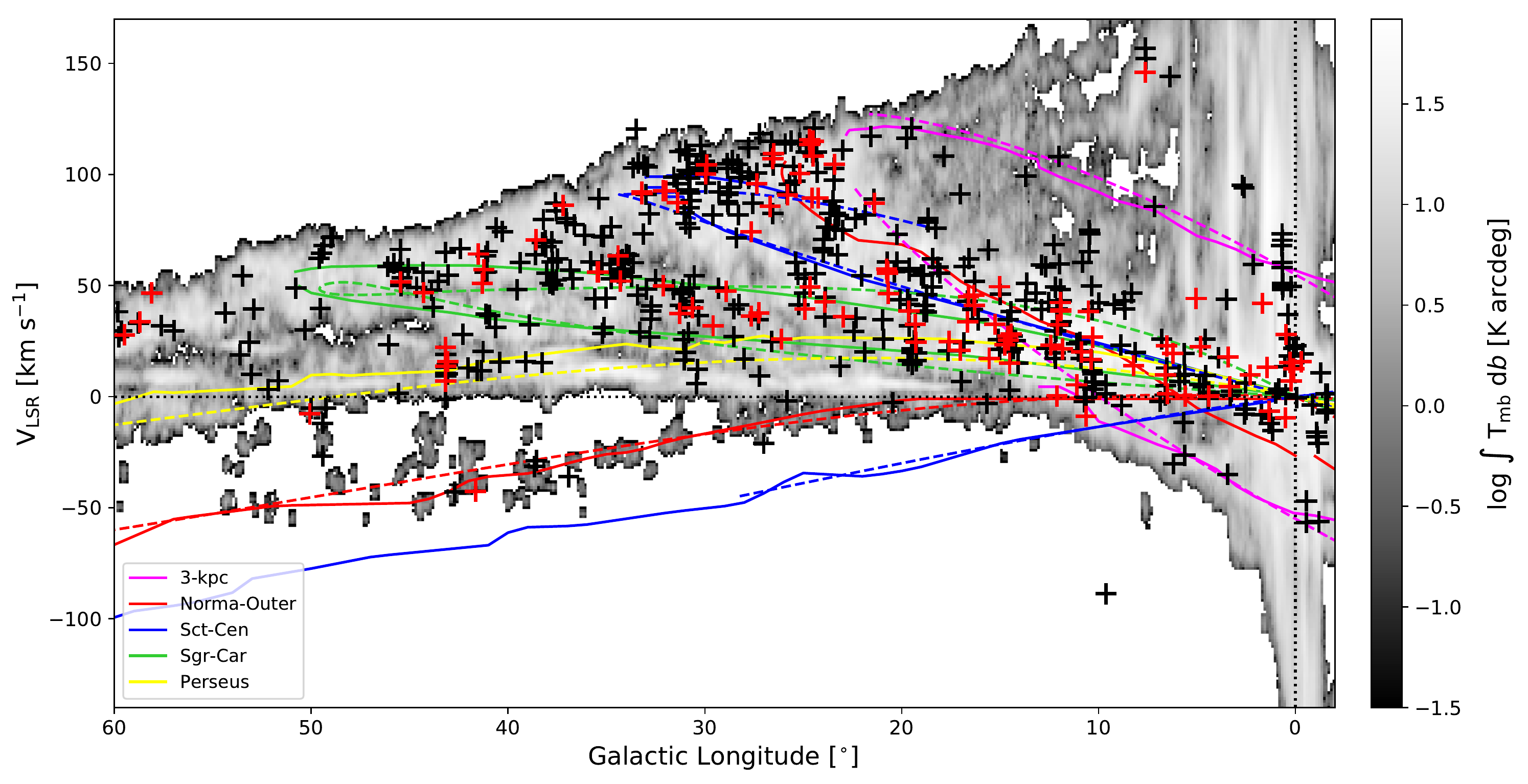}
\caption{Distribution of detected 6.7\,GHz methanol maser velocity with respect to Galactic longitude. Black crosses represent detections of known sources in the GLOSTAR survey and red crosses represent new detections. The coloured spiral arms are as follows: magenta -- 3~kpc, red -- Norma-Outer, blue -- Scutum Centaurus, green -- Sagittarius Carina, and yellow -- Perseus. The dashed lines represent the updated spiral arm models of \cite{taylor1993} as used in e.g., \cite{schuller2021} while the solid lines are the spiral arm descriptions from \cite{reid2019}. The background shows the CO emission from \cite{dame2001}. }
\label{fig:spiral_arms_PV}
\end{figure*}

We have detected a total of 554 \meth maser emission and 6 cases of methanol absorption in the range of GLOSTAR survey coverage
($-2^{\circ} < l < 60^{\circ}$, $|b| \leq 1^{\circ}$; see \cite{giselacygnus} for the 13 masers found in the Cygnus~X region, or the online table for all 567 GLOSTAR masers). Of these detections, we have determined 84 ($\sim$15$\%$) of them to be new detections. As we require that for a source to be detected, a minimum of two adjacent channels must meet our S/N threshold, our estimated completeness level may be higher than the $4\sigma$ noise level. We have overlaid the detections on top of the GLOSTAR \dconf continuum emission in Fig.~\ref{fig:maser_overlay_dconf} and displayed their spatial distributions in Figures~\ref{fig:distribution_longitude} and \ref{fig:distribution_latitude}. We note that within the GLOSTAR Galactic longitude range, the \meth 6.7\,GHz maser source distribution seems to peak towards $l=30^{\circ}$, which is not surprising given the multiple crossings of spiral arms of the Milky Way
as is evident from Fig.~\ref{fig:spiral_arms_PV}. The distribution in Galactic latitude
is presented in Fig.~\ref{fig:distribution_latitude}, which shows that the majority of sources
are in the range of $|b|\leq0.5^{\circ}$.

\begin{table*}[!tbhp]
\caption{Properties of methanol maser spots from D-configuration maps for a selection of sources.}             
\label{tab:maser_spots_eg}      
\centering          
\begin{threeparttable}
\begin{tabular}{l c c r r c c}     % 7 columns 
\hline\hline
Name   &     $\alpha$/$\Delta\alpha$      &    $\delta$/$\Delta\delta$ &   V$_{\text{LSR}}$    &  $S_{v,\,\mathrm{Peak}}$    &  $S_{v,\,\mathrm{Int}}$ &Notes \\

& h:m:s/\asec & d:m:s/\asec   &km s$^{-1}$ & Jy beam$^{-1}$ &Jy &\\
(1) &(2) &(3) &(4) &(5) &(6) &(7) \\
\hline
G35.2260--0.3544 & 18:56:53.2332 & +01:52:47.068 &59.49 & 0.38$\pm$0.02 & 0.39$\pm$0.04&YANG19\\ 
G35.2260--0.3544 & -0.17 & -0.34 &59.13 & 0.12$\pm$0.02 & 0.16$\pm$0.04&\\ 
G35.2260--0.3544 & 0.27 & 0.03 &59.31 & 0.20$\pm$0.02 & 0.19$\pm$0.03&\\ 
G35.2260--0.3544 & 1.32 & 0.84 &59.67 & 0.14$\pm$0.02 & 0.23$\pm$0.04&\\ 
G35.2476--0.2365 & 18:56:30.3917 & +01:57:08.664 &72.25 & 1.40$\pm$0.02 & 1.47$\pm$0.03&MMB\\ 
G35.2476--0.2365 & -1.30 & 1.68 &71.53 & 0.12$\pm$0.02 & 0.14$\pm$0.03&\\ 
G35.2476--0.2365 & -0.91 & -0.42 &71.89 & 0.08$\pm$0.02 & 0.10$\pm$0.03&\\ 
G35.2476--0.2365 & 0.12 & 0.16 &72.43 & 1.34$\pm$0.02 & 1.33$\pm$0.04&\\ 
G35.2476--0.2365 & 2.08 & 1.03 &72.79 & 0.10$\pm$0.02 & 0.22$\pm$0.05&\\ 
G35.3974+0.0252 & 18:55:50.7873 & +02:12:18.699 &89.07 & 0.36$\pm$0.01 & 0.36$\pm$0.02&MMB\\ 
G35.3974+0.0252 & -2.04 & -2.05 &88.89 & 0.12$\pm$0.01 & 0.15$\pm$0.03&\\ 
G35.3974+0.0252 & -0.29 & 0.39 &89.25 & 0.27$\pm$0.02 & 0.32$\pm$0.03&\\ 
G35.3974+0.0252 & 0.76 & -0.28 &89.43 & 0.19$\pm$0.02 & 0.25$\pm$0.04&\\ 
G35.3974+0.0252 & 0.76 & -0.50 &89.61 & 0.11$\pm$0.02 & 0.14$\pm$0.04&\\ 
G35.4166--0.2839 & 18:56:59.0536 & +02:04:54.463 &56.11 & 1.67$\pm$0.02 & 1.69$\pm$0.03&NEW\\ 
G35.4166--0.2839 & -0.05 & -0.49 &55.75 & 0.21$\pm$0.01 & 0.23$\pm$0.02&\\ 
G35.4166--0.2839 & 0.00 & -0.02 &55.93 & 0.73$\pm$0.01 & 0.72$\pm$0.03&\\ 
G35.4166--0.2839 & 0.17 & -0.07 &56.29 & 1.35$\pm$0.02 & 1.35$\pm$0.03&\\ 
G35.4166--0.2839 & -0.04 & -0.20 &56.47 & 0.48$\pm$0.02 & 0.55$\pm$0.04&\\ 
G35.4571--0.1782 & 18:56:41.0152 & +02:09:57.411 &56.11 & 0.26$\pm$0.02 & 0.31$\pm$0.05&NEW\\ 
G35.4571--0.1782 & -1.16 & 0.92 &54.67 & 0.12$\pm$0.02 & 0.21$\pm$0.04&\\ 
G35.4571--0.1782 & -0.08 & -0.11 &54.85 & 0.19$\pm$0.02 & 0.30$\pm$0.05&\\ 
G35.4571--0.1782 & -1.41 & -2.16 &55.21 & 0.09$\pm$0.01 & 0.17$\pm$0.04&\\ 
G35.4571--0.1782 & 1.16 & 0.48 &55.39 & 0.11$\pm$0.01 & 0.09$\pm$0.02&\\ 
G35.4571--0.1782 & -0.66 & 0.20 &55.57 & 0.24$\pm$0.02 & 0.32$\pm$0.04&\\ 
G35.4571--0.1782 & -1.01 & 0.30 &55.75 & 0.20$\pm$0.02 & 0.21$\pm$0.04&\\ 
G35.4571--0.1782 & -2.15 & -0.36 &55.93 & 0.23$\pm$0.02 & 0.27$\pm$0.04&\\ 
G35.4571--0.1782 & -0.45 & -0.46 &56.29 & 0.17$\pm$0.02 & 0.21$\pm$0.04&\\ 
G35.4571--0.1782 & -2.15 & -2.03 &56.65 & 0.12$\pm$0.02 & 0.18$\pm$0.04&\\ 

\hline      
\end{tabular}
\begin{tablenotes}
\item[] \textbf{Notes.} Column (1) gives the GLOSTAR source name based on the GLOSTAR Galactic coordinates.
Columns (2) and (3) are the GLOSTAR equatorial coordinates of the position of the maser velocity component with the highest intensity. For sources with multiple components, we list their position offsets with respect to the component with the highest intensity. The position uncertainties are $\sim$1.1$^{\prime\prime}$ (see Sect~\ref{sect:otherSurveys}). Column (4) gives the LSR radial velocity of the peak of the component. Columns (5) and (6) give the peak and integrated fluxes at the peak velocity, given by (4). Column (7) Source references.
\end{tablenotes}
\end{threeparttable}
\end{table*} %table of masers

\begin{table*}[!tbhp]
\caption{Estimated distances and maser luminosities from D-configuration maps for the maser sources listed in Table~\ref{tab:maser_spots_eg}. Refer to the online table for a complete list of sources.} 
\label{tab:maser_dist_lum_eg}      
\centering          
\begin{threeparttable}
\begin{tabular}{l r r r r r}     % 7 columns 
\hline\hline
Name & $\Delta V_{\mathrm{D}}$ & Dist. & Note & $S_{\mathrm{Int}}$  & $L_{\mathrm{maser,D}}$   \\
(glon, glat) & (\kms) &(kpc)& &(Jy \kms)&($L_{\odot}$) \\
(1) & (2)&(3)&(4)&(5)&(6) \\
\hline
G35.2260--0.3544 & 0.72 & 9.43$\pm$0.36 & B & 0.96$\pm$0.07 & $5.8\times 10^{-7}$\\
G35.2476--0.2365 & 0.90 & 4.84$\pm$0.61 & B & 3.26$\pm$0.08 & $5.2\times 10^{-7}$\\
G35.3974+0.0252 & 0.90 & 5.90$\pm$0.54 & A & 1.22$\pm$0.07 & $2.2\times 10^{-7}$\\
G35.4166--0.2839 & 0.90 & 3.20$\pm$0.37 & A & 4.53$\pm$0.07 & $2.7\times 10^{-6}$\\
G35.4571--0.1782 & 1.80 & 4.10$\pm$0.38 & A & 2.27$\pm$0.13 & $1.4\times 10^{-6}$\\
\hline      
\end{tabular}
\begin{tablenotes}
\item[] \textbf{Notes.} Column (1) is the GLOSTAR source name. Column (2) gives the total velocity extent of maser emission above the local 4$\sigma$ level. Column (3) gives the distance obtained from the Bayesian distance estimator \citep{reid2019} or from the ATLASGAL \C{compact source catalogue} (CSC) \citep{urquhart2018,urquhart2022} as marked in column (4) with a B or A respectively. Column (5) and (6) give the maser integrated flux and luminosity, respectively.
\end{tablenotes}
\end{threeparttable}
\end{table*}
%table with luminosities

The properties of a few example detections are listed in Table~\ref{tab:maser_spots_eg}, and we refer the reader to the online catalogue for the full list of detections. We identify each maser source by its galactic coordinates. The equatorial coordinates for the maser spot with the highest peak flux, that is  the main velocity component, is given while the remaining maser spots (if any) are given as an offset to this main maser spot. The velocity of each maser spot as well as its peak flux and integrated flux are listed in the table. The fluxes
were determined using 2D Gaussian fitting (see Section~\ref{sect:create_cat}). Additionally, we report in the online catalogues every velocity channel where emission was detected above the $4\sigma$ level for a given maser.

\subsection{Flux densities}
\label{sect:basic_flux}

The brightest 6.7\,GHz \meth maser we detect is the well known G9.6213+0.1961 with a
peak flux density of $\sim$5700\,Jy. This source is known not only as the methanol maser that reaches the highest flux density, but, given its distance of 5.2 kpc, also has the highest line luminosity  \citep{sanna2015}. Conversely, the weakest maser we detect is the source G25.1772+0.2111 with a peak flux density of $\sim$0.09\,Jy. For the newly detected masers, the fluxes range from 0.16\,Jy to 5.4\,Jy. The median peak flux density of the newly detected masers is 0.47\,Jy. The surface density of new detections across the survey coverage is $\sim$0.8 masers per sq.\,degree. We detect a total of 80 masers above 20\,Jy, none of which are new detections.

\subsection{Distance determination}
\label{sect:distance_determination}
\begin{figure}
\centering
\includegraphics[trim={2.0cm 0.0cm 0.5cm 0.0cm}, clip, width = 0.59\textwidth]{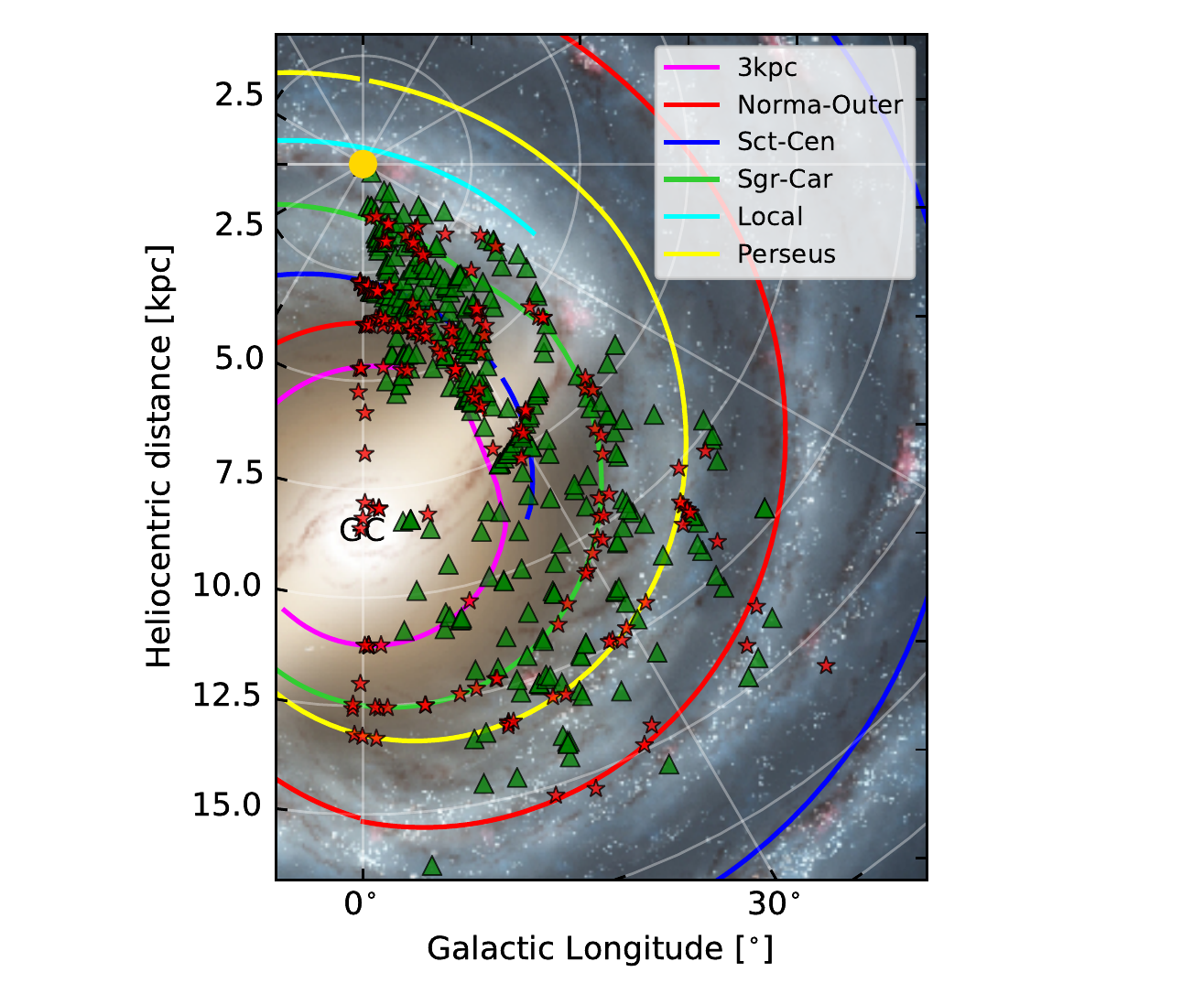}
\caption{Distances of 6.7\,GHz methanol masers from GLOSTAR \dconf VLA observations plotted on top an artist's rendition of the Milky Way. We have used the  distance estimator \citep{reid2019} to assign the near or far distance to a source (red stars) except for maser sources with an ATLASGAL dust clump association, for which we use the reported distance from \cite{urquhart2018,urquhart2022} (green triangle).}
\label{fig:mw_kda}
\end{figure}

To calculate the luminosity of a maser source, the distance information is required. This can be obtained from the peak velocity of the maser and comparing
it to a Galactic rotation curve. However, there is an inherent kinematic distance ambiguity that affects all sources within the Solar Circle (see \cite{romanDuval2009} for an overview). To resolve this, we use the distances obtained from associated ATLASGAL 870\,$\mu$m emission sources \citep{schuller2009} from the compact source catalogue \citep{urquhart2018,urquhart2022} as they have been individually checked for HI self absorption. However, not every maser source has a
dust clump association from ATLASGAL and in these cases, we use the Bayesian distance estimator from \cite{reid2019} to help resolve the kinematic distance ambiguity. We show the distribution of our maser sample overlaid on an artist's impression
of the Milky Way\footnote{\url{https://photojournal.jpl.nasa.gov/catalog/PIA19341}} in Fig.~\ref{fig:mw_kda}.
We note that this method is biased to the location of the spiral arms which can be seen by its very smooth distribution along curves.

\subsection{Luminosity}
\label{sect:lum_calc}
\begin{figure}
\centering
\includegraphics[width = 0.50\textwidth]{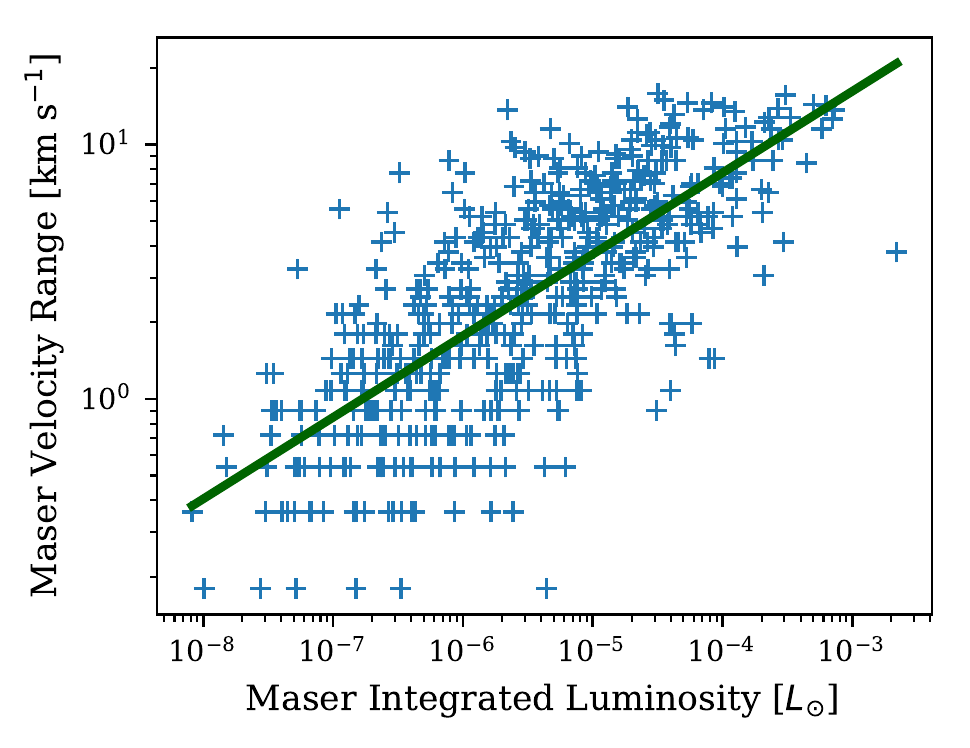}
\caption{Total velocity range of maser emission vs. integrated maser luminosity as blue crosses. We see a positive trend. The green line represents a power law fit to the data with an exponent of $0.32\pm0.01$. A Spearman correlation test yields a rank coefficient of $r=0.74$ and $p$-value\,$\ll0.0013$ which indicates a positive correlation and is consistent with previous studies using a smaller sample \cite[e.g.,][]{giselacygnus}.}
\label{fig:vel_v_lum}
\end{figure}

The luminosity is estimated across all the velocity channels in which we have emission $>4\sigma$. In this way, we use the line flux, that is the velocity integrated flux density, $S_{\mathrm{Int}}$ in units of Jy\,\kms\,to determine the luminosity:
\begin{equation}
    L_{\mathrm{maser}} = 4\pi D^{2}S_{\mathrm{Int}}f/c
    %Lmas  = 4pi d^2 \int S_{f} df = 4pi d^2 \int S_{\rm v} f/c d{\rm v} = 4pi d^2 S_{\rm int,v} f/c.
\end{equation}
where $D$ is the heliocentric distance to the source, $f$ is the rest frequency of the maser line ($5_{1}-6_{0}$A$^{+}$, 6668.5192\,MHz) and $c$ is the speed of light.
The total velocity covered, $\Delta V_{\mathrm{D}}$, the estimated distance, the velocity integrated flux density, $S_{\mathrm{Int}}$, and the final isotropic maser luminosity,
$L_{\mathrm{maser}}$, are listed for a few example sources in Table~\ref{tab:maser_dist_lum_eg} and the rest are listed in an online catalogue. As mentioned by previous studies \citep{breen2011, billington2019, giselacygnus}, a positive trend between the total velocity width, $\Delta V_{\mathrm{D}}$,  of a maser source and its maser luminosity is shown in Fig.~\ref{fig:vel_v_lum}. Given that we can take the velocity range of a maser as a proxy for line complexity, i.e., many components, it might naively be expected that stronger sources are more complex. However, it could also be that for
weaker sources, there may be other velocity components but that these components might not meet the S/N threshold and are thus not considered. We have fit a power law in the form of $\Delta V_{\mathrm{D}} \propto L_{\mathrm{maser}}^{\alpha}$ to the relation and found $\alpha=0.32\pm0.01$ which supports the positive trend we see (a Spearman correlation test yields a coefficient $r=0.74$ and $p$-value\,$\ll0.0013$ which strongly supports the correlation).

\subsection{Comparison with other maser surveys}\label{sect:otherSurveys}
The Arecibo Methanol Maser Galactic Plane Survey \citep[AMMGPS,][]{Pandian2007} which used the 305\,m Arecibo radio telescope to cover the ranges of $35.2^{\circ} \leq l \leq 53.7^{\circ}$ and $|b|\leq0.41^{\circ}$ to detect 86 masers. The survey has an rms noise level of $\sim$85\,mJy in each spectral channel after Hanning smoothing and averaging both polarizations. Of these masers, only G35.374+0.018 at 96.9\,\kms~and
G36.952--0.245 at 61.7\,\kms~were not detected by our survey. This may be due to the time variability of methanol masers, making these sources candidates for long term observations.

The Methanol Multibeam (MMB) survey \citep{mmbiG} is a comprehensive
unbiased survey that covers a large portion of the Galactic plane ranging from $186^{\circ} \leq l \leq 60^{\circ}$ and $|b|<2^{\circ}$. It uses the Parkes 64\,m
radio \C{telescope} to make preliminary detections which are followed up using the
higher resolution interfermometers: Australia Telescope Compact Array (ATCA) or Multi-Element Radio Linked Interferometer Network \citep[MERLIN,][]{MERLIN} to improve the accuracy
in their reported positions to be better than $<1$\asec. The survey sensitivity is $\sim$0.2\,Jy. Of the $\sim$1000 sources
in their complete catalogue \citep{mmb1C,mmb2G, mmb3C, mmb4G, mmb5B}, 404 lie within the GLOSTAR data presented here. Of these, 394 were detected in our survey. As GLOSTAR is an unbiased survey, it is useful to compare the
catalogues resulting from these surveys.

To supplement our comparison, we also use the \cite{yang2019} catalogue of 6.7\,GHz \meth masers that were obtained through
a targeted search towards sources from the Wide Field Infrared Survey Explorer (WISE) point source catalogue using the 65\,m Shanghai Tianama Radio Telescope (TMRT). These surveys were used to help in the identification of our SEC detections in Section~\ref{sect:sec} where weak sources that were close to the $4\sigma$ threshold were conclusively kept as a detection given matching coordinates and
velocities. This results in 113 detections that were not detected by \cite{yang2019}. There are some masers from \cite{yang2019} that were undetected in the GLOSTAR survey, with the majority having fluxes below our sensitivity limit. While the sensitivity of the observations done by \cite{yang2019} with the TMRT is 1.5\,Jy\,K$^{-1}$, they also include weaker masers that were previously detected in literature in their catalogue.

We make a final check with the webtool, Maserdb\footnote{\url{maserdb.net}} \citep{maserdb}, which is an online collection of catalogues of many maser species (e.g., OH, H$_{2}$O). We find that our catalogue has 84 new 6.7\,GHz \meth maser emission
sources (see Fig.~\ref{fig:maser_overlay_dconf}).

\begin{figure}
\centering
\includegraphics[width = 0.50\textwidth]{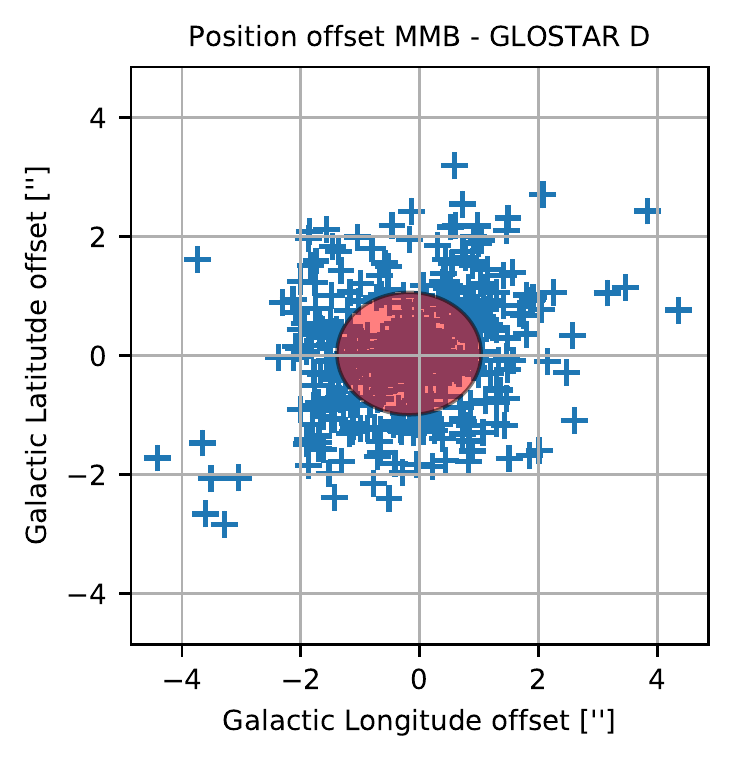}
\caption{Comparison of the positions of matching MMB and GLOSTAR methanol maser sources. The red shaded ellipse is centered on the mean offsets of $\delta l=-0.18$\,arcsec, and $\delta b=0.04$\,arcsec. It shows the half-axes standard deviations in Galactic Longitude and Latitude which are 1.21\,arcsec and 1.03\,arcsec respectively.}
\label{fig:pos_off_mmb}
\end{figure}

Next, we compare the GLOSTAR and the MMB positions. Shown in Fig.~\ref{fig:pos_off_mmb} are the position offsets to the GLOSTAR \dconf detections. The mean offsets are $\delta l=-0.18\pm1.21$\,arcsec and $\delta b=0.04\pm1.03$\,arcsec.
This indicates that there are no systematic offset in the astrometry with respect to the MMB catalogue. However, the standard deviation of the offsets suggests that the astrometric uncertainty is closer to $\sim$1.1 arcseconds, which is slightly larger than the statistical uncertainties of 0.7$^{\prime\prime}$ determined for a $10\sigma$ detection.

\begin{figure}
\centering
\includegraphics[width = 0.50\textwidth]{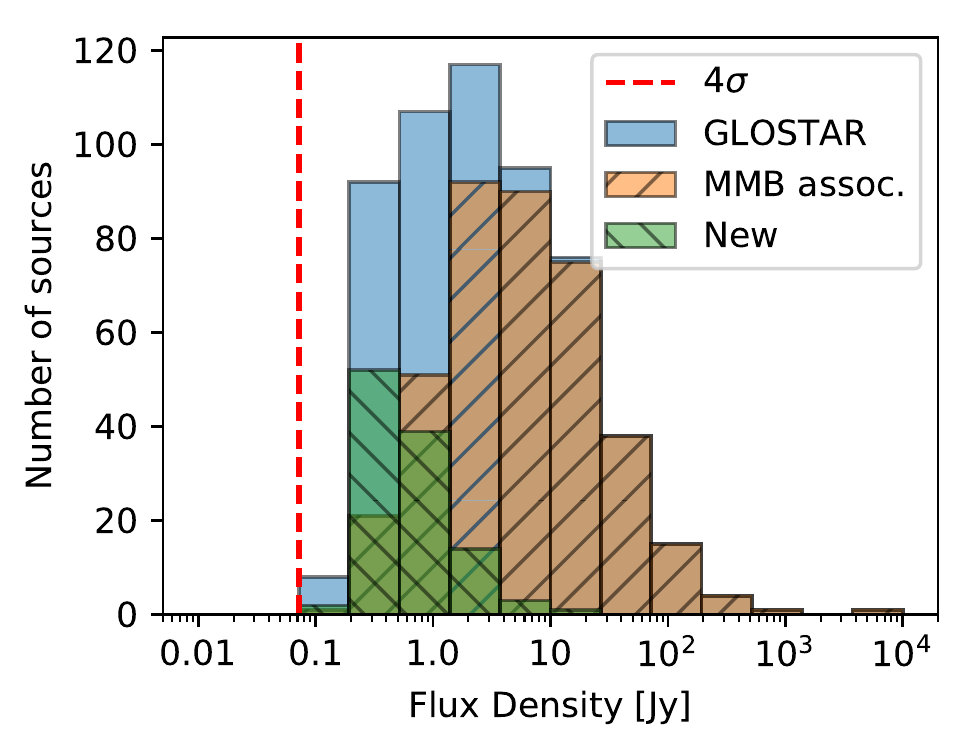}
\caption{Flux distribution of detected masers. The vertical dashed red line corresponds to the average 4$\sigma$ noise level for the methanol D-configuration data ($\sim$70\,mJy\,beam$^{-1}$). We show the full set of GLOSTAR fluxes, as well as the subset with MMB associations, and all new detections. As expected, the bulk of the  new detections peaks at lower flux densities.}
\label{fig:glo_flux_hist}
\end{figure}

We show in Fig.~\ref{fig:glo_flux_hist} the flux distributions of the GLOSTAR methanol maser detections. We highlight the subset of sources that have a MMB counterpart as well as the subset of new sources. 
In comparison with the MMB catalogue, we directly see that our increased sensitivity finds new weaker sources. However, there are 74 sources above the survey cube detection threshold of 0.7\,Jy that one
would have expected to be detected by the MMB. There are several reasons as to why they were not detected by the MMB. Given that these observations were taken years apart, it could be the result of variability, making these sources possible candidates for long term observations. Furthermore, $\sim$30 of these sources are situated near bright sources and as such, were not able to be initially resolved by the MMB in their blind survey. Examples of these are shown in Appendix~\ref{app:strong_new}.

\begin{figure}
\centering
\includegraphics[width = 0.50\textwidth]{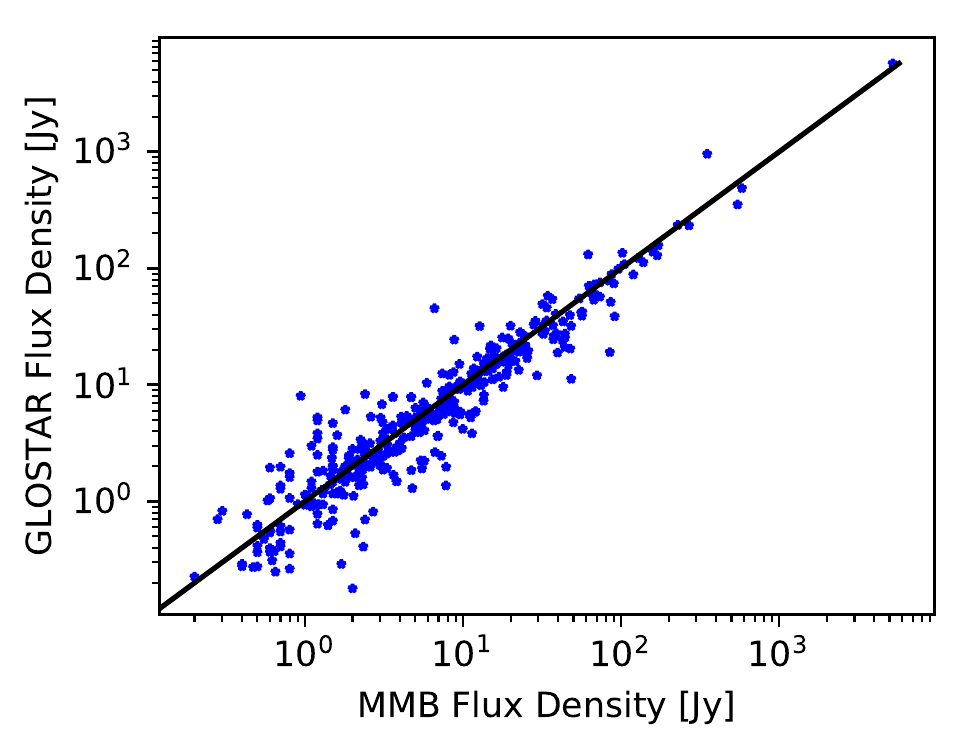}
\caption{Comparison of GLOSTAR vs. MMB peak flux densities for masers detected in both surveys. The black line indicates the 1:1 equality line. Red circles denote sources for which the percent difference was greater than 50$\%$.}
\label{fig:mmb_peakf}
\end{figure}

For the masers with counterparts in both surveys, we compare the peak fluxes in Fig.~\ref{fig:mmb_peakf}. The fluxes do not show a systematic difference. G9.6211+0.1956 is already known to be a periodic Class II \methanol\, maser with a period of $\sim$244\,days \citep[e.g.,][]{goedhart2007, vanderwalt2009}\footnote{In periodic methanol masers, certain velocity components show periodic variability with a wide range of periods, ranging from 20 to $> 500$\,days. As discussed in a future publication, the several  epochs at which the GLOSTAR data were taken and comparison with MMB spectra will lead to the detection of new candidates of these interesting sources \citep[e.g.,][and references therein]{goedhart2018}} . We detected two main velocity components for this source at the known velocity of 1.3\,\kms and at a new velocity of $-88.7$\,\kms. Given the large difference in velocity, we consider these to be distinct entries in our catalogue. Both of these spatially coincide with the known MMB maser at the same position. The component at 1.3\,\kms\,has a high peak flux density of >5700\,Jy\,beam$^{-1}$ in both the GLOSTAR and the MMB spectra with matching velocity, which corresponds to the velocity of the host source. As such we consider this the true association for the following analysis. The component at the velocity of $-88.7$\,\kms\, is much weaker with a peak flux density $\sim$1\,Jy\,beam$^{-1}$. Emission at this velocity seems to not yet have been detected towards this source and deserves further study.

\begin{figure}
\centering
\includegraphics[width = 0.50\textwidth]{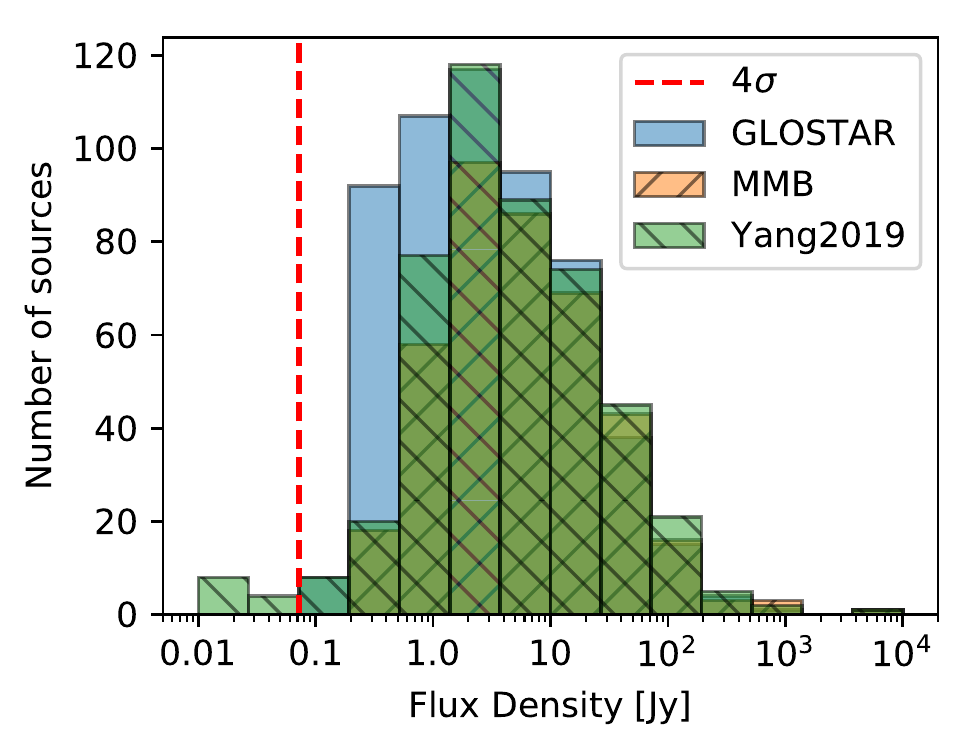}
\caption{Flux distribution of masers as reported in the GLOSTAR, the MMB \citep{mmbiG} and the \cite{yang2019} catalogues. The vertical dashed red line corresponds to the average 4$\sigma$ noise level for the GLOSTAR methanol D-configuration data ($\sim$70\,mJy\,beam$^{-1}$). The sources of the other surveys that were not detected in the GLOSTAR survey are well below the sensitivity level of our unbiased search.}
\label{fig:all_cat_flux}
\end{figure}

\begin{figure}
\centering
\includegraphics[width = 0.50\textwidth]{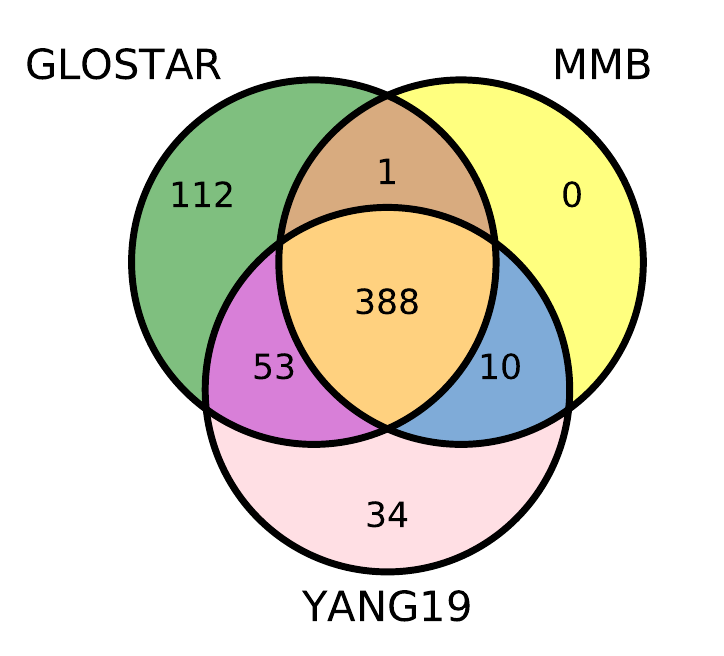}
\caption{Venn diagram presenting the overlap between the GLOSTAR methanol maser catalogue, the MMB \citep{mmbiG} and the \cite{yang2019} catalogues over the same region as GLOSTAR. We detect 112 masers that were not detected by the MMB or the \cite{yang2019} catalogues, while not detecting 44 known masers.}
\label{fig:venn_maser_cats}
\end{figure}

In Fig.~\ref{fig:all_cat_flux} we compare the flux distributions between the GLOSTAR, MMB, and \cite{yang2019} catalogues for masers within the GLOSTAR survey coverage. 
It is evident that there are masers that were below our detection threshold and were not detected. The detection comparison between the three maser surveys is shown in Fig.~\ref{fig:venn_maser_cats} and details how many masers are detected by each survey and their overlap. There are 44 sources that we did not detect but were detected by \cite{yang2019} or by both \cite{yang2019} and the MMB. Of these, 14 have reported fluxes below our sensitivity limit. Variability may also account for some of the other non-detections, however, long term observations would be needed to confirm this nature of maser activity. We also see that while there are some known masers that we do not detect, the majority of our new detections are in the lower flux bins as expected.

\subsection{Absorption detections}
\label{sect:abs}
\begin{figure}
\centering
\includegraphics[width = 0.5\textwidth]{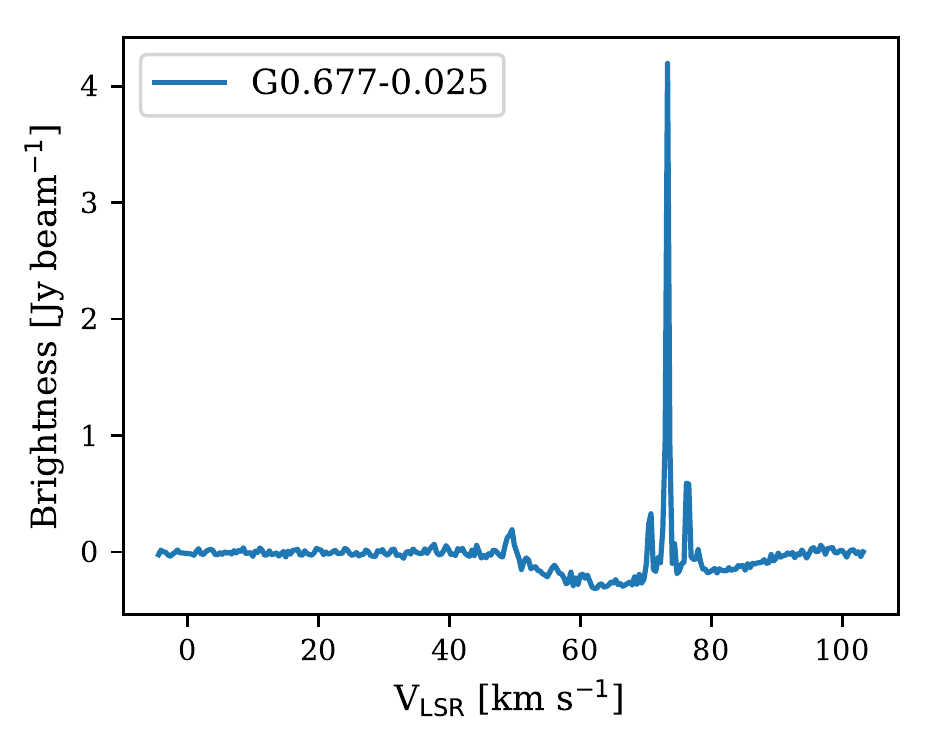}
\caption{\meth\,6.7\,GHz spectra of the maser source G0.677-0.025 near Sagittarius~B2. Broad absorption can be seen between the velocity range of 50\,\kms~and 90\,\kms.}
\label{fig:spec_abs}
\end{figure}

While detections of the 6.7\,GHz $5_{1}-6_{0}$A$^{+}$ \meth line are widespread,
the absorption detections are comparatively more sparse. Only a few studies
have confirmed the 6.7\,GHz line in absorption thus far \citep[e.g.,][]{menten1991b, pandian2008, impell2008, giselacygnus,wenjinYang2022}. Absorption in this line can occur towards radio continuum emission and the cosmic microwave background (CMB). In conjunction with maser emission detections, we are also sensitive enough to detect absorption features and indeed we find a few absorption sources (listed in Table~\ref{tab:abs_sources}), where an example is shown in Fig.~\ref{fig:spec_abs}. A systematic search was not yet performed and thus a comprehensive list of all absorption detections is not presented in this work. An in-depth analysis
of the absorption sources detected in the GLOSTAR data will be performed in a future work, for which we will use the complete GLOSTAR \dconf continuum source catalogue \citep{medinaFull} to check for methanol in absorption.

\begin{table*}[]
    \centering
    \caption{Examples of absorption features detected. The source is listed with equatorial coordinates (J2000), the peak velocity feature, and the peak flux density.}
    \begin{tabular}{l|r r r r r l}
         \hline
         \hline
         Common Name& $\alpha$ & $\delta$ & $v_{\mathrm{Peak}}$& $S_{\mathrm{Peak}}$\\
         & h:m:s & d:m:s &km\,s$^{-1}$ & Jy\,beam$^{-1}$\\
         \hline
         Sgr B2 & 17:47:18.71 & -28:22:53.54 & 70.4 &-2.0 \\
         G08.67-0.36 &18:06:19.02 &-21:37:30.29 & 35.6& -0.2\\
         G10.62-0.38 &18:10:28.62 &-19:55:48.40 &-1.3 & -0.4 \\
         G012.81-0.20 &18:14:13.95 &-17:55:38.31 & 36.0 &-1.6  \\
         G34.26+0.16 &18:53:18.03 &01:15:00.09&59.7 &-0.2 \\
         W49 &19:10:12.97 &09:06:10.98 &12.55 &-0.3 \\
         \hline
    \end{tabular}
    \label{tab:abs_sources}
\end{table*}

%%%%%%%%%%%%%%%%%%%%%%%%%%%%%%%%%%%%%%%%%%%%%%%%%%%%%%%%%%%%%%%%%%%%%%%%%
\section{Discussion}\label{sect:discussion}
\subsection{Association with ATLASGAL sources}
\label{sect:atlasgal}
Methanol masers are known tracers of star formation and the Class II 6.7\,GHz methanol maser is thought to exclusively trace the early stages of HMSF \citep{minier2003,ellingsen2006,xu2008}. \cite{billington2019} recently used the MMB and ATLASGAL surveys to do a comprehensive study on the physical environments of the regions these masers originate from. They used newly available distances and luminosities to compare with the clump properties as determined in ATLASGAL and JPS \citep[JCMT Plane Survey:][]{moore2015,eden2017} to determine correlations for maser associated sources. As seen by previous studies \citep[e.g.,][]{urquhart2013, urquhart2015,billington2019}, there is a ubiquitous association
with the MMB masers and dust continuum sources ($99\%$), strongly correlating these masers with the earlier stages of HMSF.
\cite{billington2019} determined that clump masses and radii are not indicative if a clump has a 6.7\,GHz methanol maser, whereas clump density may be able to do so. Further, they determined a lower density threshold of $n$(H$_{2})\geq 10^{4.1}$\,cm$^{-3}$ for the ``turn on" of maser emission.
As such, with our sample, especially our new weaker detections, it is interesting to see if these correlations hold.

We perform a cross-match with ATLASGAL sources using an emission map and distance threshold of 12$^{\prime\prime}$, which is three times the pointing uncertainty of ATLASGAL and was determined from analysing the surface density distribution of matches. We use the ATLASGAL CSC from $l=3^{\circ}$ to $60^{\circ}$ as the source properties for sources in the Galactic Centre region are not to the same confidence level.
We find 363 associations within 12\asec ($\sim$65$\%$) between GLOSTAR masers and the ATLASGAL compact source
catalogue \citep[CSC;][]{urquhart2018,urquhart2022} for which the dust clump properties (e.g., clump mass, clump temperature, bolometric luminosity) were calculated. The $\sim$65$\%$ association with dust emission is lower than expected from the studies mentioned above. For masers without an ATLASGAL CSC association, it is possible that they are still are associated with dust as they could be associated with more distant clumps well below the ATLASGAL threshold. They could also be situated at a closer heliocentric distance to us but are associated instead with low-mass clumps.

\begin{figure}
\centering
\includegraphics[width = 0.50\textwidth]{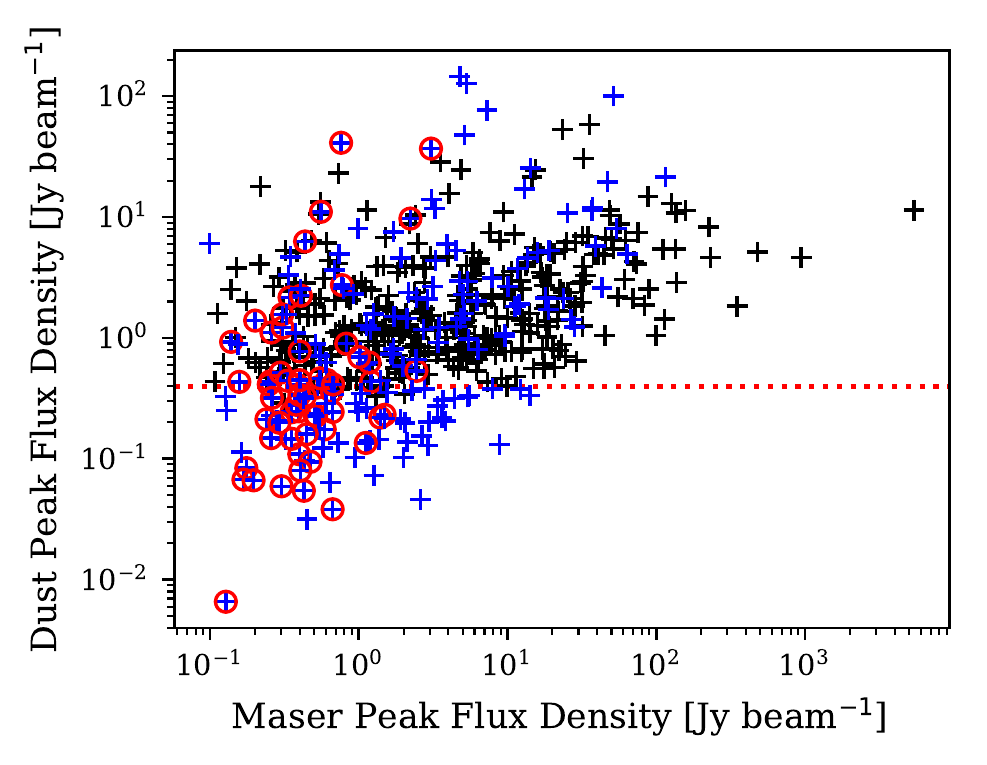}
\caption{\meth 6.7\,GHz maser peak flux density compared to 870\,$\mu$m dust emission extracted from ATLASGAL maps towards the maser position. Black crosses correspond to sources that were matched to the ATLASGAL \C{compact source catalogue} (CSC). The remaining sources are marked in blue. The red dashed line corresponds to the ATLASGAL $6\sigma$ noise level and the red circles highlight the new masers found in this work.}
\label{fig:dust_v_maser_flux}
\end{figure}

To address this discrepancy, we visually inspect ATLASGAL cutouts centered on the positions of the masers (examples can be found in Appendix~\ref{app:atlasgal_maser_examples}) using dust continuum contours from 1$\sigma$ to 5$\sigma$. We take this approach as the ATLASGAL CSC uses a threshold of at least 6$\sigma$ for the sources they report but there are still many potential dust continuum sources below this limit.
In this way, we find that there are an additional 72 maser
sources that show compact dust emission above 3$\sigma$. There are a further 93 maser sources that can be associated to an extended ATLASGAL feature above $6\sigma$ and 7 maser sources that are offset slightly further than 12$^{\prime\prime}$. As such, we find that only 18 maser sources have no dust continuum emission which corresponds to a
methanol maser and dust continuum association of $\sim$97$\%$. The result is in agreement with previous studies \citep[e.g.,][]{billington2019}. From Fig.~\ref{fig:dust_v_maser_flux} we find that there seems to be a cluster of new maser detections that are centered on the lower end of the dust emission around the 6$\sigma$ noise level of ATLASGAL.

\begin{figure*}
\centering
\begin{tabular}{cc}
\includegraphics[width = 0.49\textwidth]{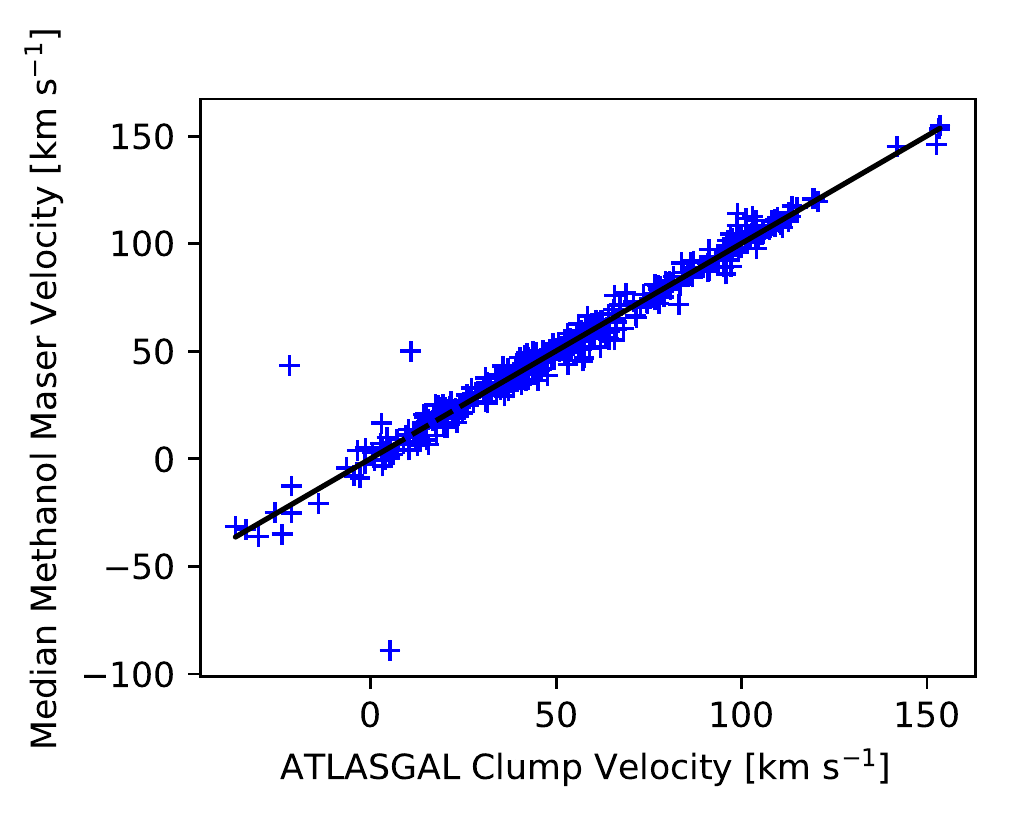} 
\includegraphics[width = 0.49\textwidth]{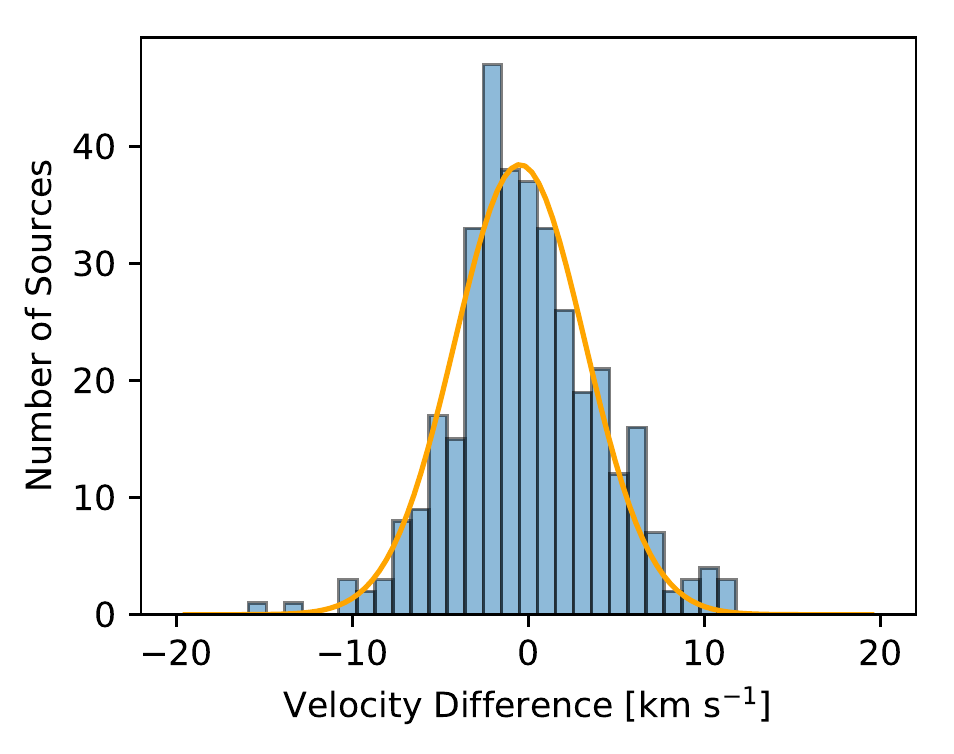}
\end{tabular}
\caption{\emph{Left}: The median methanol maser velocity plotted as a function of the ATLASGAL clump velocity for sources with an association to an ATLASGAL \C{compact source catalogue} (CSC) source. The black line is the fitted linear result, with a Spearman's rank coefficient of $r=0.98$ and $p$-value$\ll0.0013$. \emph{Right}: Distribution of the offsets between the median methanol maser velocities and
the molecular line velocities from ATLASGAL. \C{Fitting the distribution with a Gaussian yields a mean of $0.49\pm0.18$\,\kms and standard deviation of $3.69\pm0.1$\,\kms. We use 3$\sigma$ (11\,\kms) as a confidence threshold to identify outliers.}}
\label{fig:cf_maser_clump_vel}
\end{figure*}

We compare the velocities of the maser's median velocity and the velocities of the matched dust clumps in Fig.~\ref{fig:cf_maser_clump_vel}. The ATLASGAL velocities were assigned by matching clump positions with observations of molecular line transitions from multiple molecular line surveys \citep[see][Section 2.1 for details]{urquhart2018}. A linear fit
yields a slope and $y$-intercept of $1.00\pm0.13$ and $ 0.01\pm0.75$ respectively and Spearman's rank coefficient of $r=0.98$ and $p$-value\,$\ll0.0013$ supporting the positive correlation. In fitting a Gaussian to the distribution of the velocity offsets, we find a mean offset of $0.49\pm0.18$\,\kms\,and dispersion of $3.69\pm0.1$\,\kms. The result is in agreement with \cite{billington2019}, which compared the total MMB sample with corresponding ATLASGAL sources where they use sources with offsets of $<3\sigma$. There are three sources, however, with larger velocity offsets: G3.5022$-$0.2005, G9.6211+0.1956, and G10.3563$-$0.1484.
They were all previously detected in the MMB with G3.5022$-$0.2005 being associated with millimeter dust continuum \citep{rosolowsky2010} and G10.3563$-$0.1484 associated to a YSO candidate \citep{deharveng2015}. G9.6211+0.1956 is associated with the well studied \hii~region of similar name where its shock fronts have been studied \citep[e.g.,][and references therein]{liu2017}. As mentioned in Section~\ref{sect:otherSurveys}, there are two sources in our catalogue with vastly different velocities associated with this ATLASGAL source. The maser at 1.3\,\kms\,is consistent with the well studied clump velocity. This other velocity component at $-$88.7\,\kms\, would be an interesting target for future studies.

%%%%%CDF
\begin{figure*}[]
\begin{tabular}{ccc}
    \includegraphics[width = 0.33\textwidth]{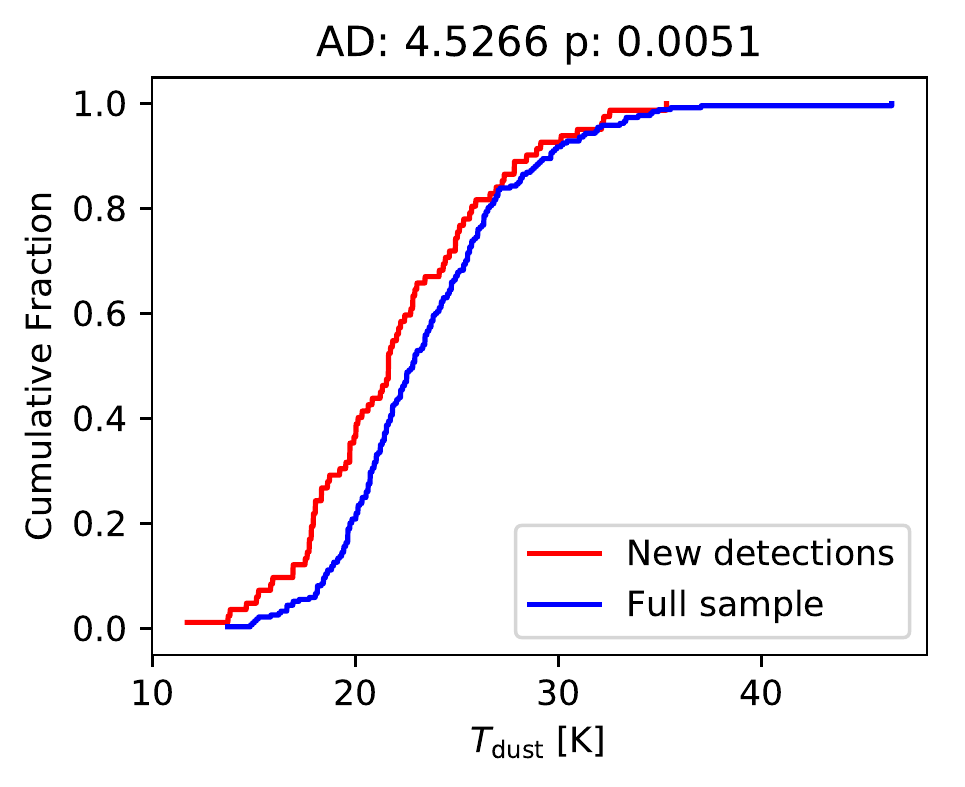}
    \includegraphics[width = 0.33\textwidth]{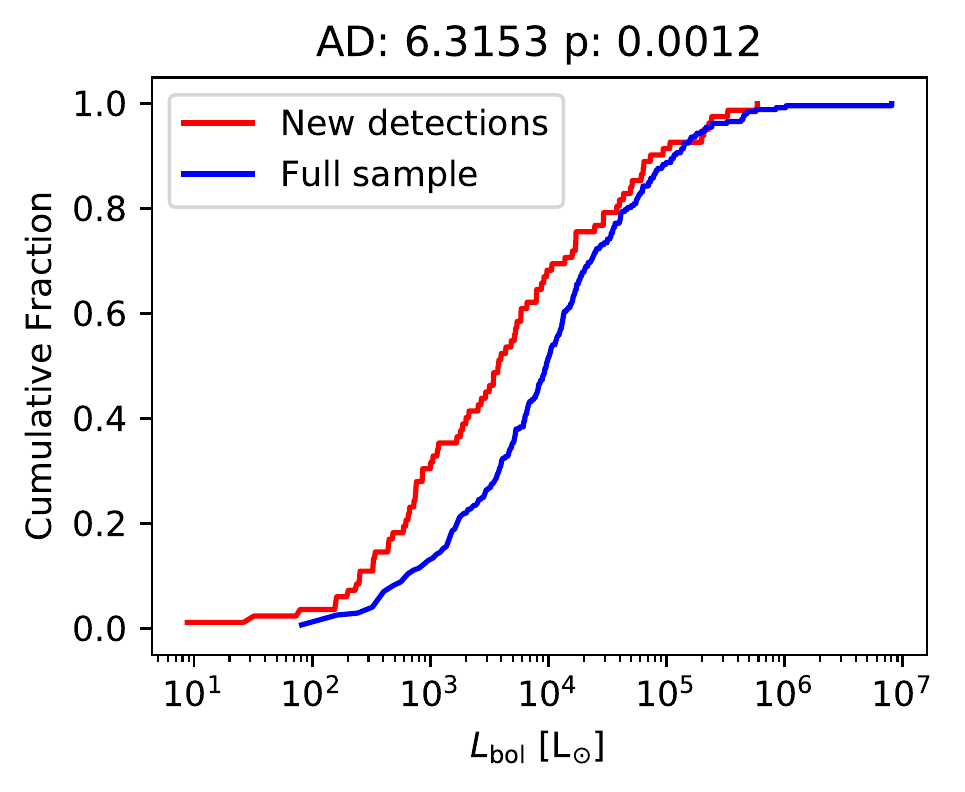}
    \includegraphics[width = 0.33\textwidth]{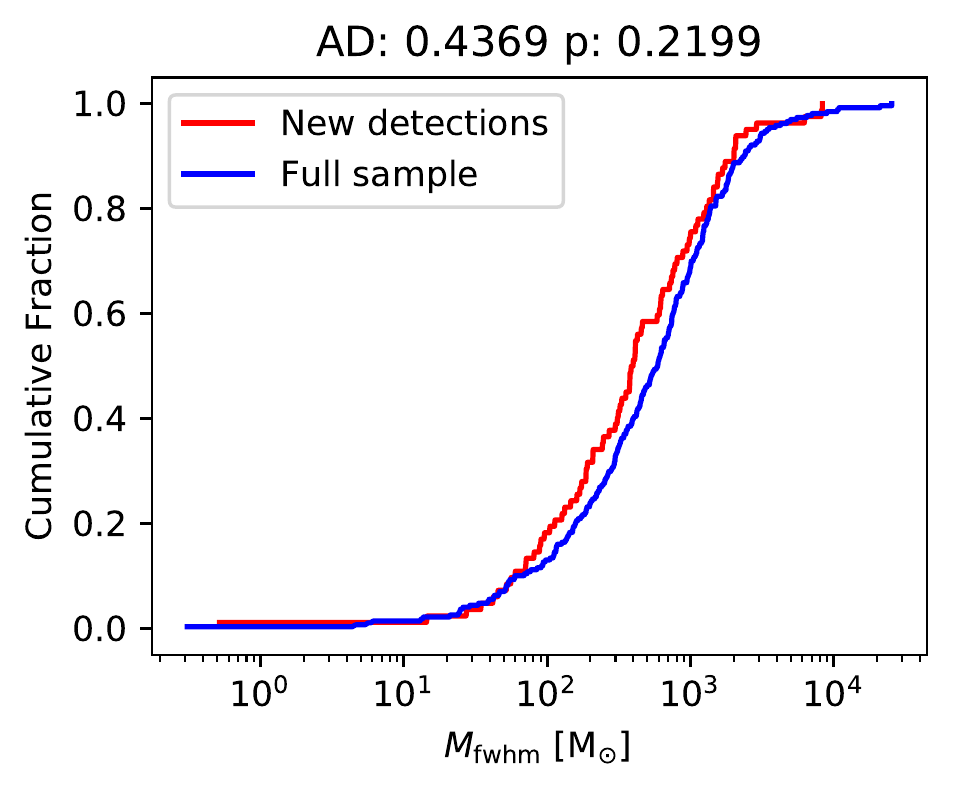}\\
    \includegraphics[width = 0.33\textwidth]{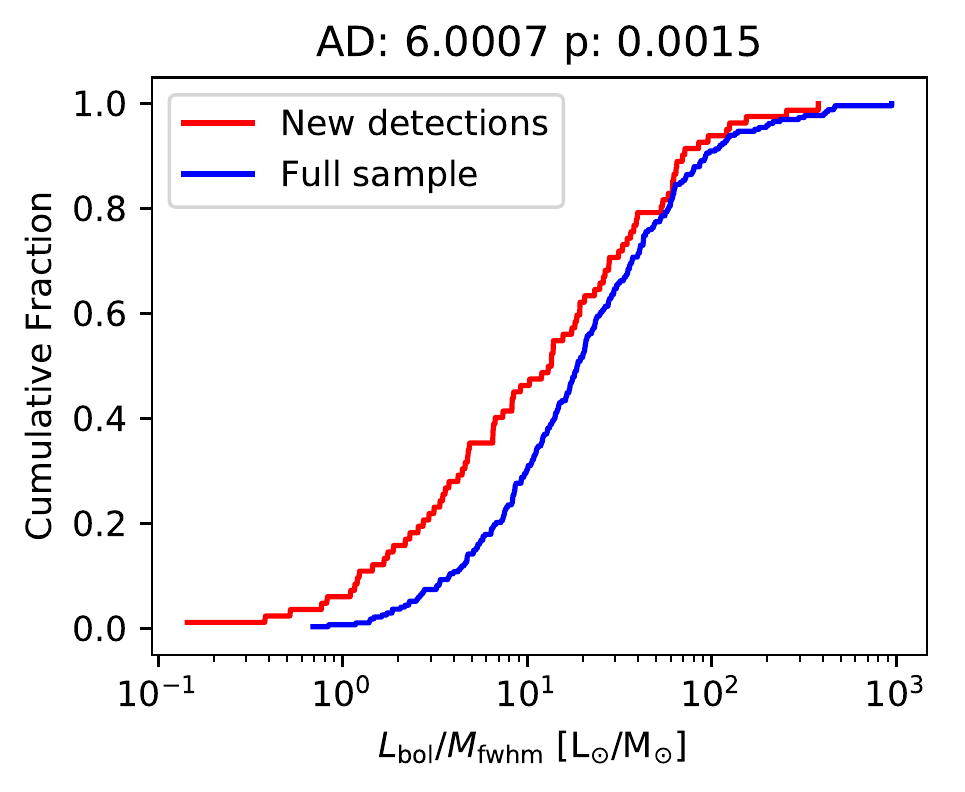}
    \includegraphics[width = 0.33\textwidth]{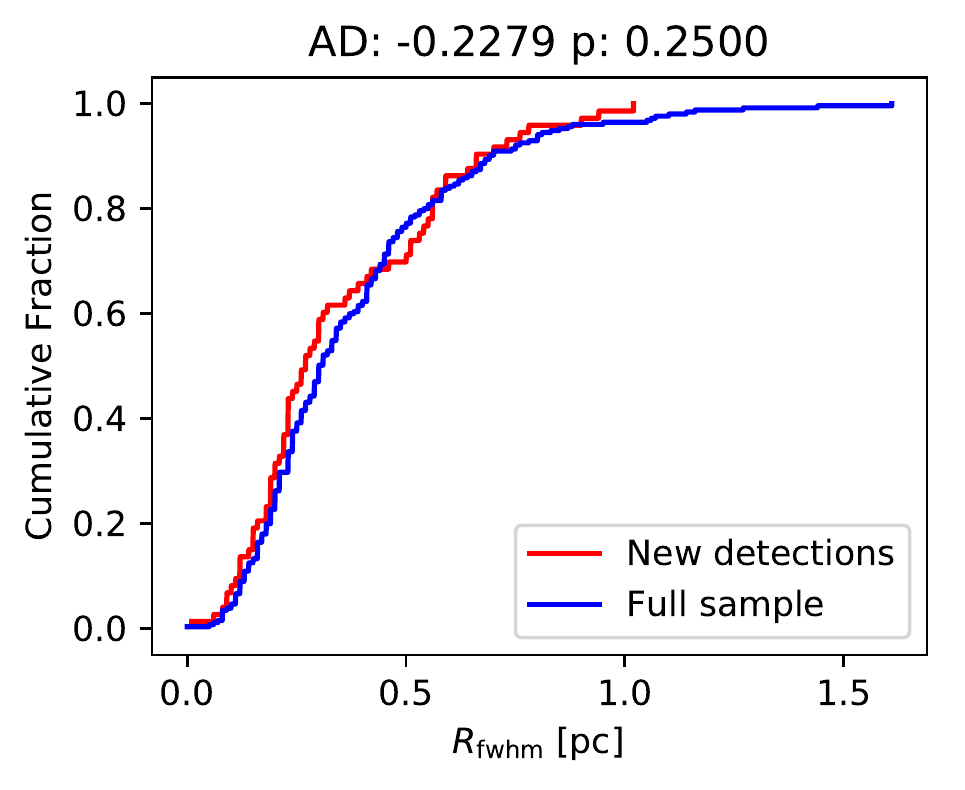}
    \includegraphics[width = 0.33\textwidth]{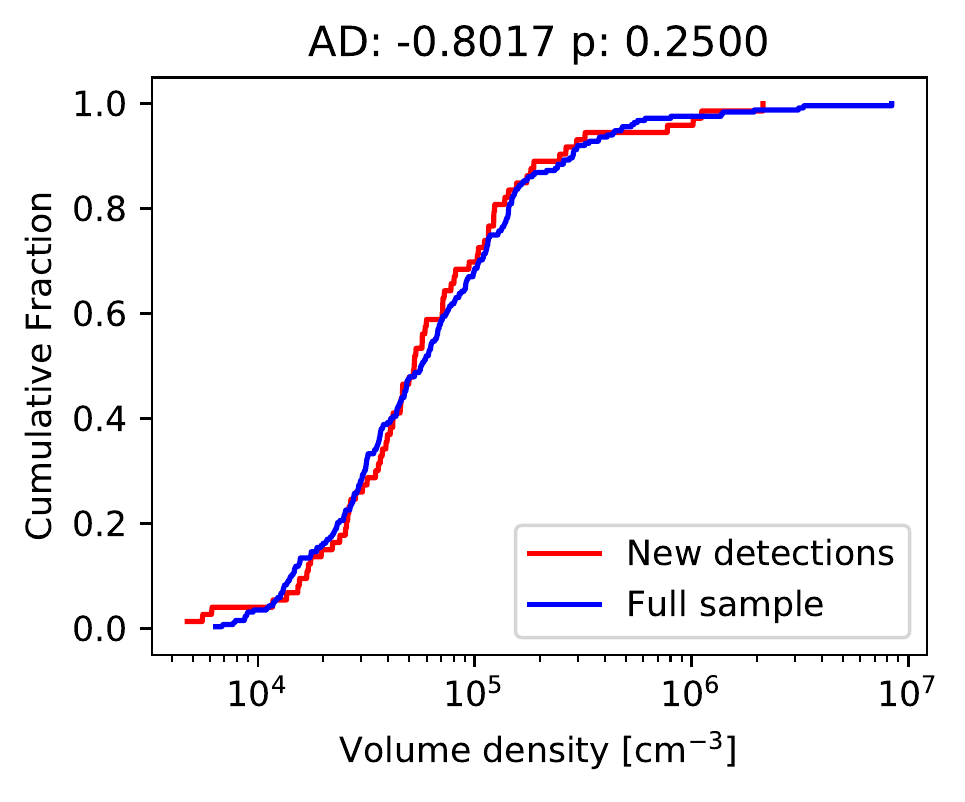}\\
\end{tabular}
\caption{Cumulative distribution functions \C{(CDFs)} for different properties of ATLASGAL clumps associated with GLOSTAR methanol masers. We compare the sample clumps associated with new methanol maser detections (red) to the sample of clumps with associations to the full methanol maser catalogue (blue). The results of the Anderson-Darling \C{(AD)} test are shown above each plot.} 
\label{fig:atlasgal_core_prop}
\end{figure*}
We show the cumulative distribution functions (CDF) of the clumps associated with GLOSTAR masers and clumps associated with just the new detections in Fig.~\ref{fig:atlasgal_core_prop} for different associated clump properties such as dust temperature, bolometric luminosity, clump mass, luminosity to mass ratio, clump size, and H$_{2}$~density. This compares a sample of 364 maser associated clumps, with that of 45 clumps associated with newly detected GLOSTAR masers. \cite{urquhart2022} showed CDFs (in their Figure~7) that compared the clump properties for different evolutionary stages of ATLASGAL sources, showing indeed that one can distinguish evolutionary stages by certain properties. Furthermore, \cite{billington2019} compared the properties of ATLASGAL clumps with the subset that have MMB methanol maser associations. They found that maser associated sources have higher dust temperatures, bolometric luminosities and luminosity-to-mass ratios, which is expected as these are regions in the process of developing high mass stars (see Fig.~10 from \citealt{billington2019}). In contrast to these works, we investigate the ATLASGAL sample of sources that have GLOSTAR methanol maser associations with the subset of those that are newly detected masers.
We perform Anderson-Darling tests for all CDFs instead of Kolmogorov–Smirnov tests as the Anderson-Darling test is more sensitive to changes at the boundaries, which is the subset of our sample that we are more interested in, given that most of our newly detected masers are weaker. The results are shown on the plots in Fig.~\ref{fig:atlasgal_core_prop}. Clump luminosity is the only statistically different sample to the $3\sigma$ level ($p$-value\,$<0.0013$), while dust temperature and luminosity-to-mass ratio are significant only to the $2\sigma$ level ($p$-value\,$<0.05$). In obtaining more clump properties for the masers without an ATLASGAL CSC counterpart, it will help to better determine the significance of the Anderson-Darling tests on these properties.
Furthermore, we see that for the dust temperature, bolometric luminosity of the clump, and luminosity-to-mass ratio, the mean properties for the new maser detections are slightly lower but they extend to similar limits on the high end as the general population of masers. One might naively expect this as lower luminosity masers may trace earlier stages of development.

\begin{figure*}
\centering
\begin{tabular}{ccc}
\includegraphics[width = 0.32\textwidth]{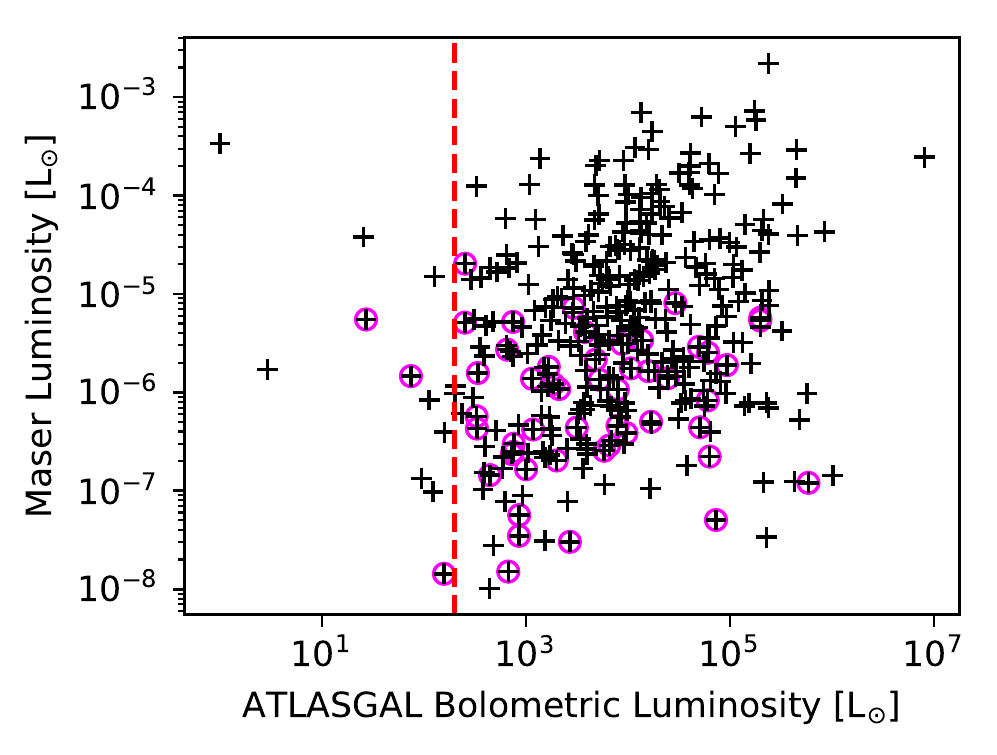} 
\includegraphics[width = 0.32\textwidth]{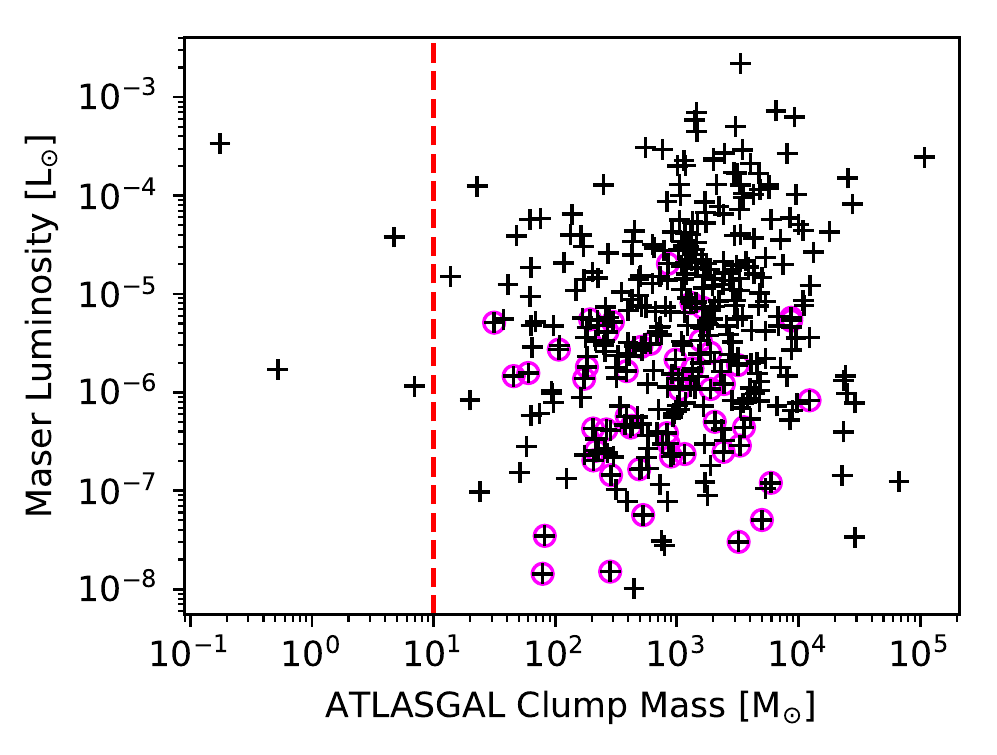}
\includegraphics[width = 0.32\textwidth]{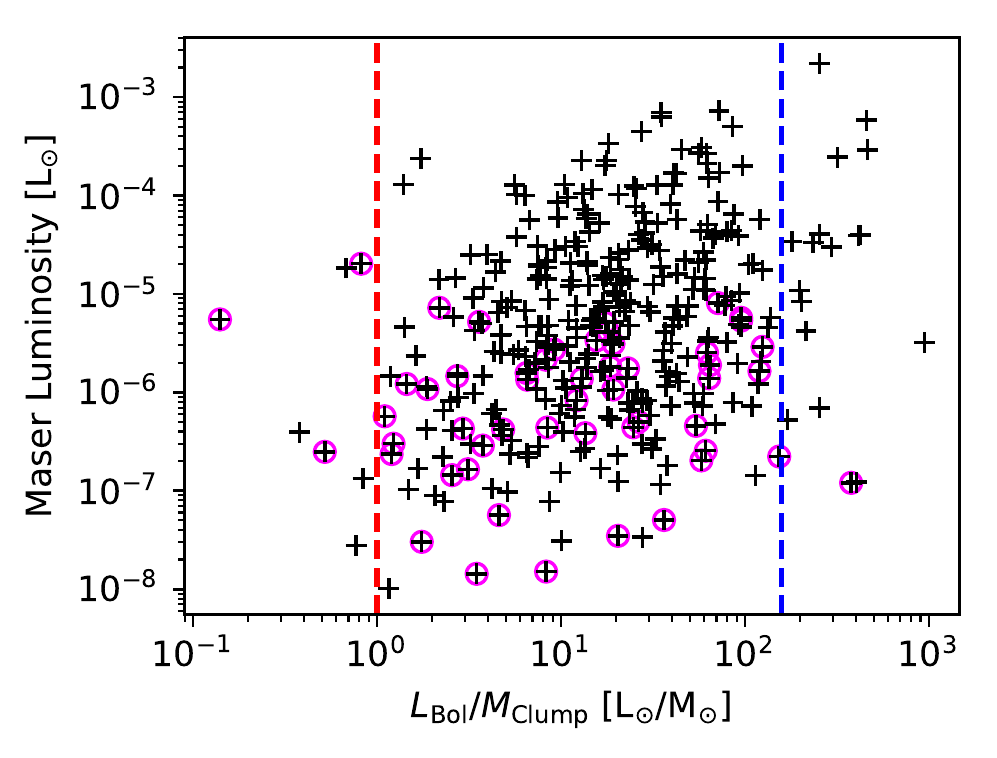}\\ 
\includegraphics[width = 0.32\textwidth]{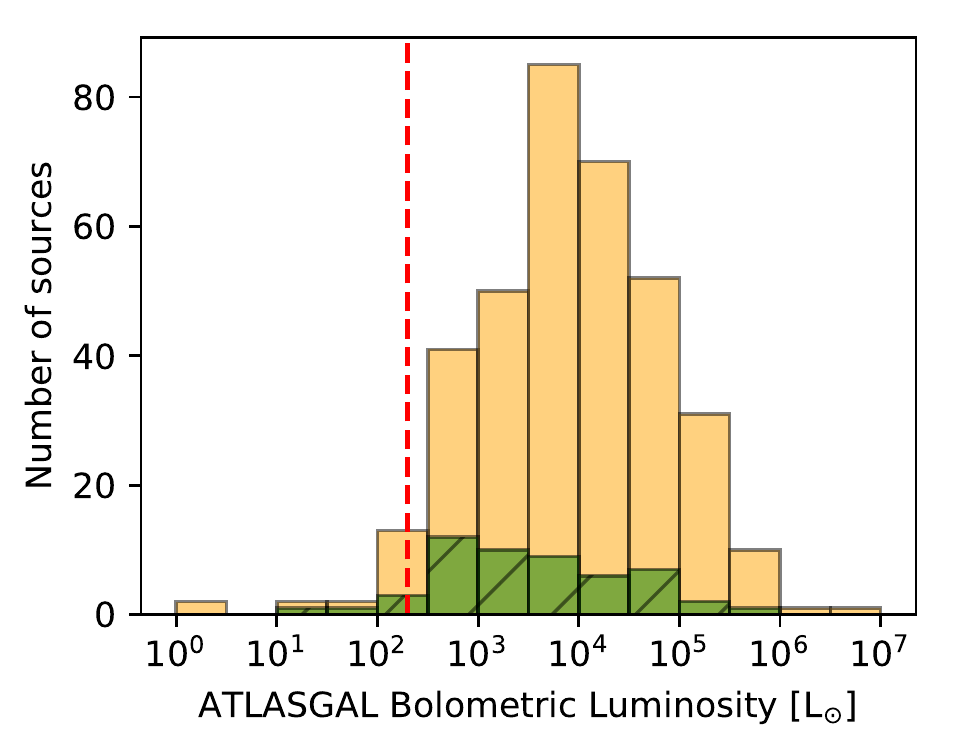} 
\includegraphics[width = 0.32\textwidth]{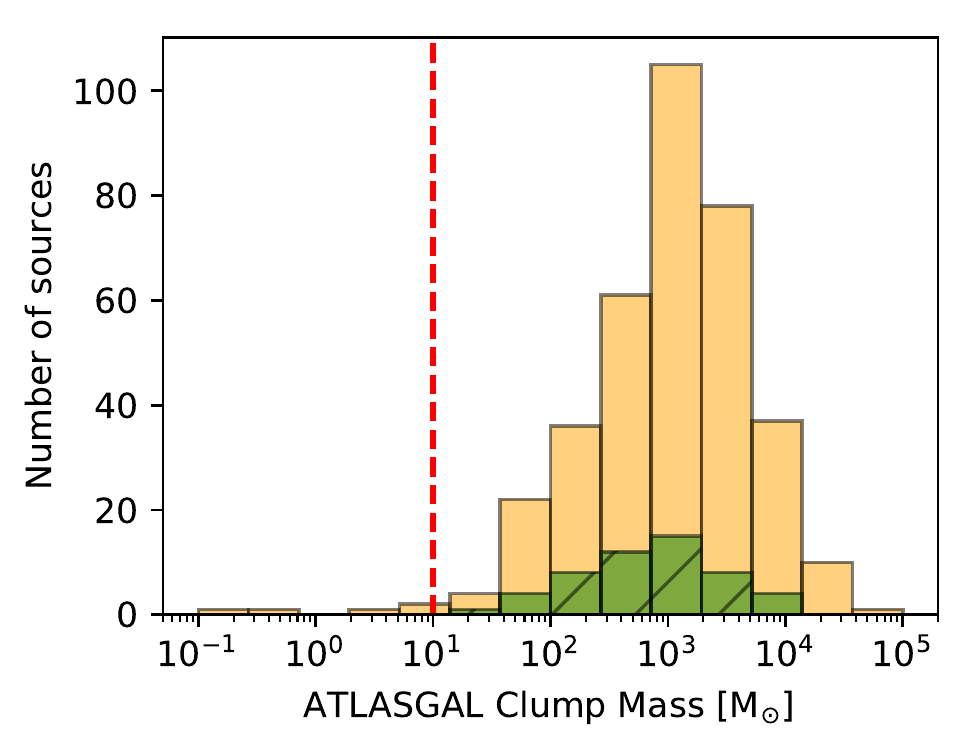}
\includegraphics[width = 0.32\textwidth]{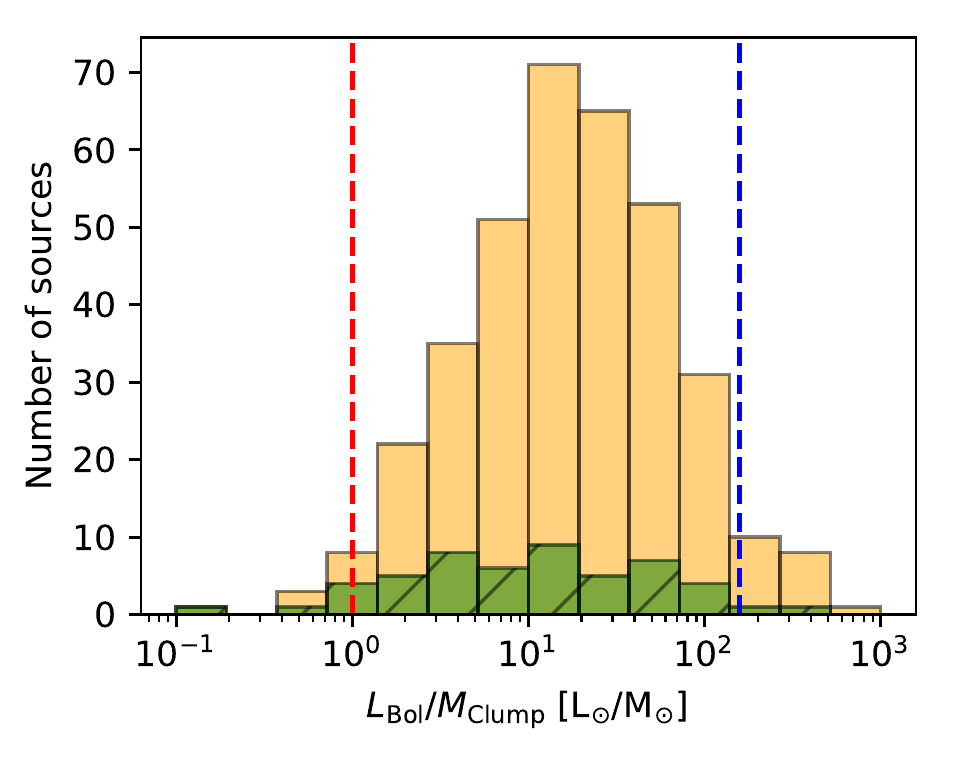}
\end{tabular}
\caption{The top row shows maser integrated luminosity measured in D-configuration GLOSTAR data as a function of ATLASGAL clump properties: bolometric luminosity (\emph{left}), FWHM mass (\emph{middle}), and luminosity-to-mass ratio (\emph{right}). The black crosses represent masers with ATLASGAL \C{compact source catalogue (CSC)} counterparts, while the crosses surrounded by magenta circles highlight masers that were newly detected as part of the GLOSTAR survey. The bottom row shows the distribution of ATLASGAL molecular clump properties for the new masers (hatched green) in comparison to the full sample (yellow) of clumps associated with masers. The red-dashed lines denote the lower limits of the respective properties determined by \cite{giselacygnus}. The blue-dashed line denotes the upper bound at which maser emission is expected to decline due to the disruption of the physical conditions required for maser emission \citep[e.g., expanding \hii~regions and dispersion of the host clump;][]{walsh1997,walsh1998,vanderwalt2003}.}
\label{fig:lum_v_agal}
\end{figure*}

We also compare the 6.7\,GHz methanol maser luminosity to the dust clump properties of luminosity and mass in the left and middle panels of Fig.~\ref{fig:lum_v_agal}. To test if these two dust clump properties are correlated to the maser luminosity, we perform a Spearman's rank correlation test that yields values of $r=0.28$ and $r=0.18$ respectively (with $p$-value\,$\ll 0.0013$) and so there is a weak but significant correlation. The left panels of Fig.~\ref{fig:lum_v_agal} shows the comparison between maser and clump luminosities, and the distribution of the clump luminosities where we have plotted for reference, the dust core luminosity value of $\sim$200\,\lsun\,as found by \cite{giselacygnus} to be the lower limit of methanol maser associated clumps in the Cygnus~X region. There are a few sources in our sample that show luminosities lower than this. The sharp cutoff does lie, however, close to this value and not around 10$^{3}\,$L$_{\odot}$ as estimated by \cite{bourke2005}. Our results are in agreement with other recent studies such as \cite{giselacygnus}, which uses similar data but for a small sample in the Cygnus~X region, \cite{paulson2020} which used a sample of 320 MMB masers, and \cite{billington2019} which used a sample of 958 methanol masers from the MMB.

The middle panel in Fig.~\ref{fig:lum_v_agal} similarly shows the investigation using the ATLASGAL FWHM masses (mass within 50$\%$ of the 870\,$\mu$m contour) for the clumps. \cite{giselacygnus} determined the minimum core mass in Cygnus~X for maser associated cores to be $\sim$10\,\msol, while \cite{billington2019} used the FWHM clump masses of the ATLASGAL sample to estimate a lower limit of $\sim$17\,\msol. \cite{paulson2020} determine a value similar to \cite{giselacygnus} of 11\,\msol for their sample. These clump values are sufficient to produce high-mass stars if one assumes a 10$\%$ star formation efficiency. We used the FWHM clump masses of ATLASGAL and find a minimum mass of 0.175\,\msol. Rather than being the true lower limit,
we see that it is likely an exception as all but four data points have clump masses above the lower limit estimated by \cite{giselacygnus}. Since we use masses from the same sample as \cite{billington2019}, we similarly see a cutoff at around $\sim$17\,\msol but the 1st percentile mass is $\sim$10\,\msol.

In combining the clump properties of luminosity and mass, the luminosity-to-mass ($L/M$) ratio has been shown to be a statistical indicator of the evolutionary stage of high-mass star forming clumps \citep{molinari2008}.
Furthermore, \cite{billington2019} found a weak correlation between the $L/M$ ratio of maser associated clumps and maser integrated luminosity. In the right most of Fig.~\ref{fig:lum_v_agal}, we find a Spearman's rank coefficient of $r=0.3$ with $p$-value $\ll0.0013$, which suggests that there is a weak correlation between the properties. \C{We find that 90$\%$ of the data points lie between the values of 1\,L$_{\odot}$\,M$_{\odot}^{-1}$ and $10^{2.2}$\,L$_{\odot}$\,M$_{\odot}^{-1}$.} These values estimate the lower and upper limits of the $L/M$ ratio, which depict the onset of maser emission and the decline of the maser due to the formation of the \hii~region having disruptive effects on the maser's environment. Our results are in agreement with previous studies \citep[e.g.,][]{breen2010,billington2019,billington2020,giselacygnus}.

We also highlight the dust clumps that are associated with newly detected masers in Fig.~\ref{fig:lum_v_agal}. Contrary to our hypothesis that the newly detected and weak masers would strongly trace the earliest stages of high-mass star forming clumps, we see that except for clump luminosity, the histograms shown in Fig.~\ref{fig:lum_v_agal} have similar shapes. Furthermore, the Anderson-Darling tests in Fig.~\ref{fig:atlasgal_core_prop}, save for the bolometric luminosity, show no significant correlation between the samples to the $3\sigma$ level.
We note that many newly detected masers have low
maser luminosity ($<10^{-6}\,$L$_{\odot}$), and the lack of a strong correlation of this sample is in agreement with \cite{paulson2020} which suggests that other properties, e.g. gas density and gas temperature are perhaps more important factors for the maser luminosity than the dust clump bolometric luminosity. However, there are some of our new masers (53) that have ATLASGAL associations, but for which we do not have the clump properties and, as stated above, are generally associated with lower 870\,$\mu$m emission dust clumps. As this is a significant portion of our new detections, the outcome of the sample comparison presented here may differ once we obtain the host clump properties for these masers in future works. The results presented here are in agreement with previous results \citep[e.g.,][]{billington2019} as the majority of the values are derived from known methanol maser and ATLASGAL clump associations.

\subsection{Association with radio continuum}
\label{sect:dconf_cont}

In general, one does not expect to see a close relationship between methanol masers and radio
emission from \hii~regions, as the latter is a more developed stage of HMSF where methanol maser emission begins to decline \citep[e.g.,][]{beuther2002}. 
However, this may not necessarily be the case with the properties and early evolution of the more compact \hii~regions, (i.e., hyper-compact (HC) and ultra-compact (UC) \hii~regions) as proposed by e.g., \cite{walsh1998,aiyuan2019,aiyuan2021}.
These more compact \hii~regions are younger sites of HMSF and may still have
maser emission in their surroundings as they evolve. This is supported by the observational results seen in \cite{aiyuan2021} from the largest sample of \hchii~regions showing a maser detection rate of 100\%, and the detection rate decreases as \hii~regions evolve from \hchii~regions to \uchii~regions. Radio continuum emission at this stage is difficult to detect, however, due to the compactness of the optically thick free-free radiation of \hchii, which results in low fluxes. 

\begin{figure}
    \centering
    \includegraphics[width =0.50\textwidth]{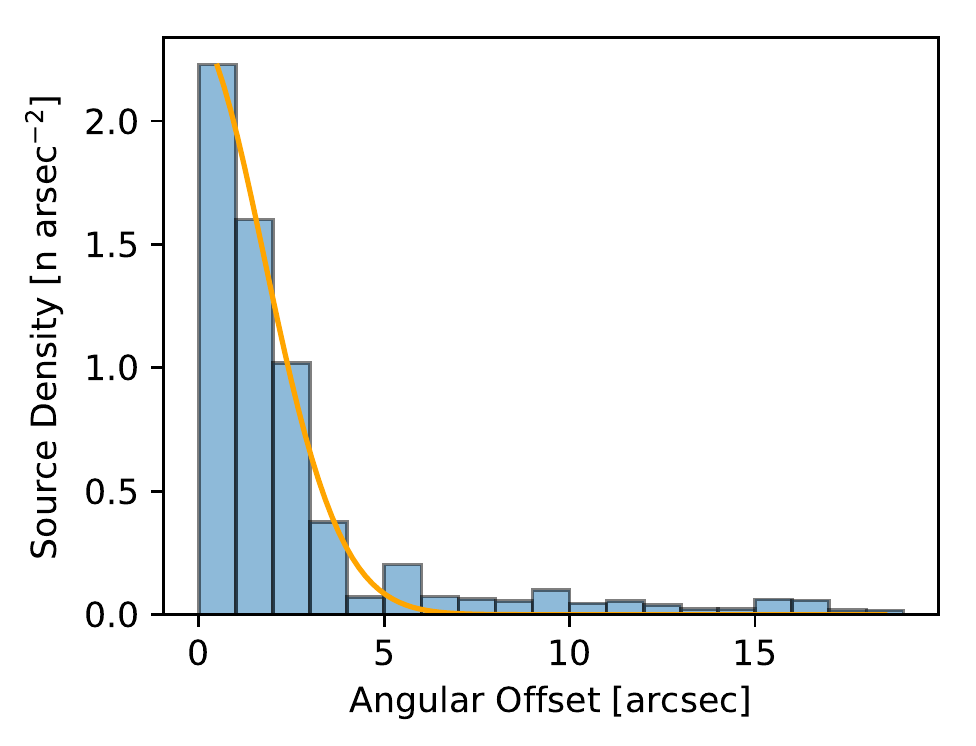}
    \caption{Surface density plot of the offset of GLOSTAR radio continuum sources associated with GLOSTAR 6.7\,GHz methanol masers.}
    \label{fig:surface_density}
\end{figure}

We attempt a search for associations of masers with the 5.8\,GHz GLOSTAR \dconf continuum (from $3^{\circ}<l<60^{\circ}$).
We find there to be 111 sources within 20\asec, which is the size of the VLA D-array beam dropping to 64 when we use 
an angular distance threshold of 6\asec. This threshold value corresponds to the $3\sigma$ level of the distribution shown in the source surface density plot in Fig~\ref{fig:surface_density}. This 12$\%$ association rate is smaller than that reported  by \cite{hu2016} \C{(with a better sensitivity of $\sim$45\,$\mu$Jy\,beam$^{-1}$ in the Galactic mid-plane for the radio continuum data)}, who found that $\sim$30$\%$ of masers were associated with \uchii~regions. This is not unexpected as the resolution of the \dconf continuum catalogue is not as well suited to sample \uchii\ and \hchii~regions. The lack of association of 6.7\,GHz methanol masers with radio continuum, however, indicates that these masers trace the earliest stages of high mass star formation.

\begin{figure}
\centering
\includegraphics[width =0.50\textwidth]{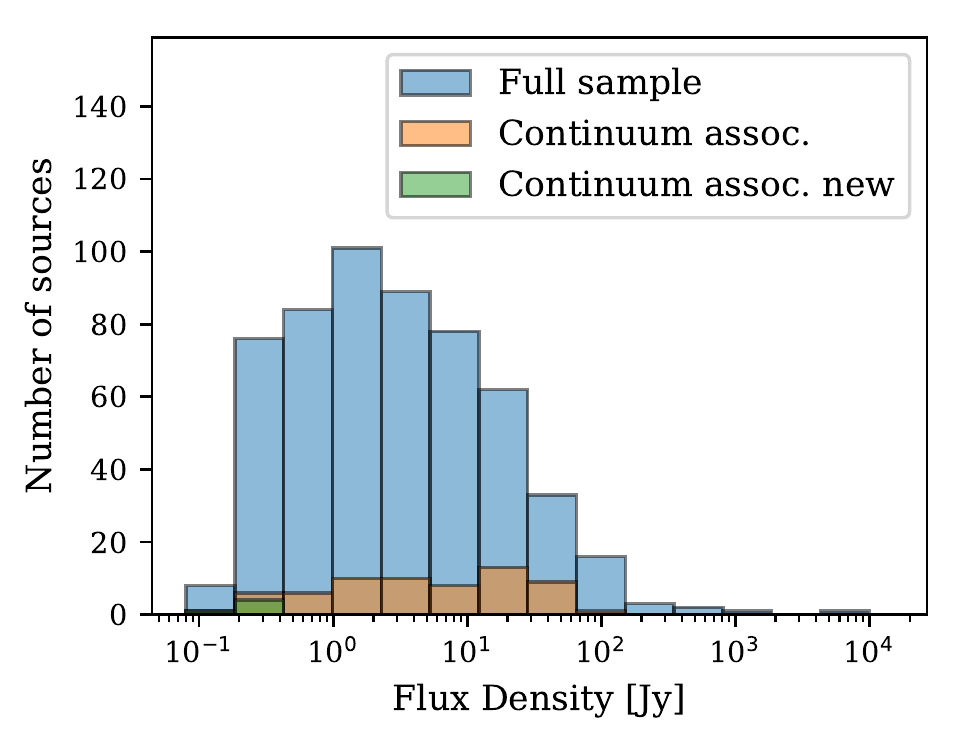}
\caption{As Fig.~\ref{fig:glo_flux_hist} except for subsets of GLOSTAR maser detections that have GLOSTAR \dconf continuum source detections.}
\label{fig:meth_cont_flux_hist}
\end{figure}

\begin{figure}
\centering
\includegraphics[width = 0.50\textwidth]{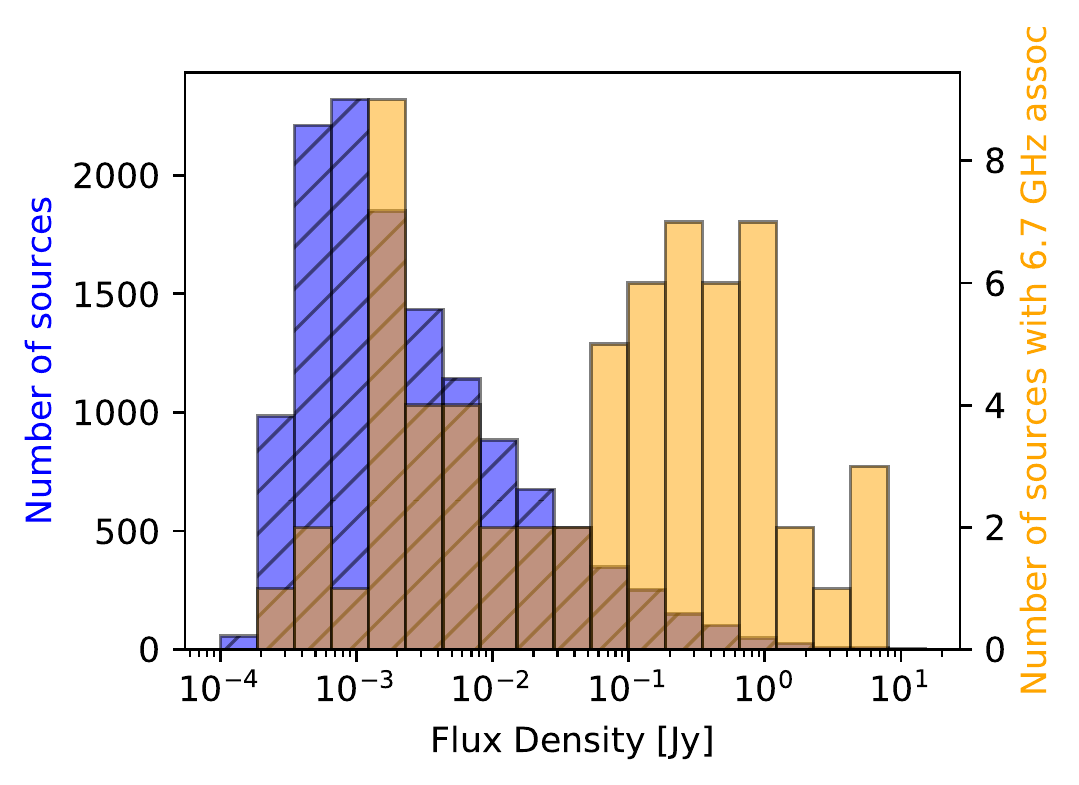}\\
\includegraphics[width = 0.45\textwidth]{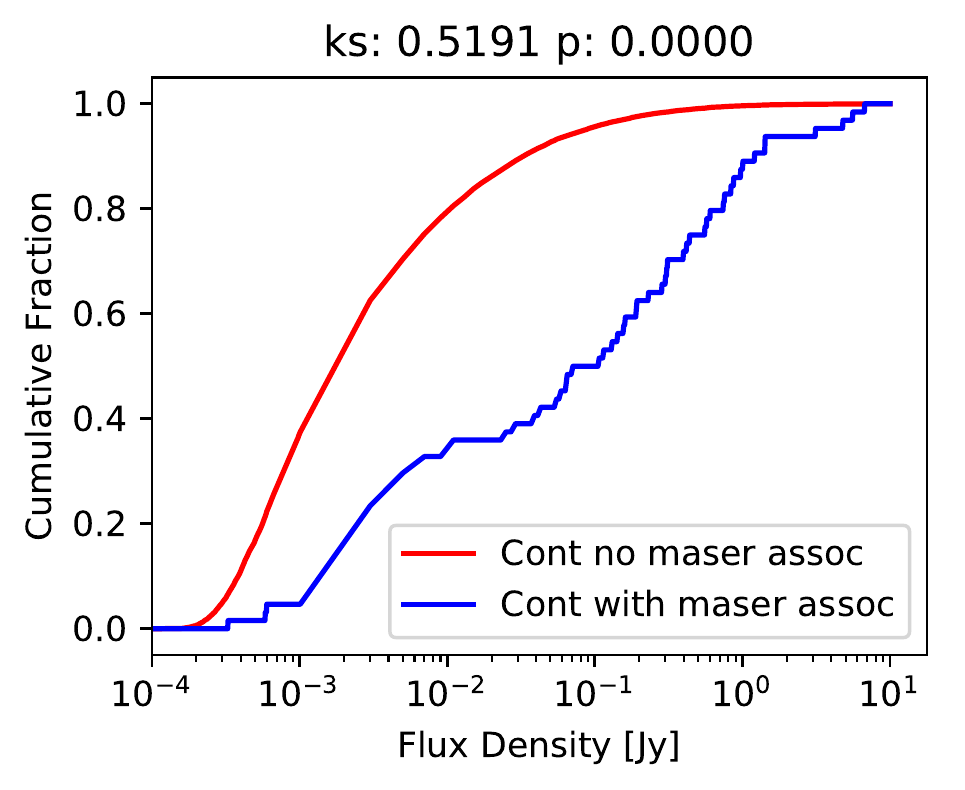}
\caption{\emph{Top}: The blue hatched histogram shows the flux distribution of the GLOSTAR \dconf continuum sources (\citealt{sac2019}, \citealt{medinaFull}). The orange histogram shows the distribution of the radio sources that are associated with GLOSTAR 6.7\,GHz methanol masers and has been rescaled for better visibility (the axis is indicated on the right).  \emph{Bottom}: The CDFs for the flux density of radio continuum sources that have 6.7\,GHz methanol maser associations (blue) and those without (red). The result of the Anderson-Darling test is reported above the figure and indicates that both distributions are distinct, with continuum sources associated with methanol masers typically being stronger than the overall distribution of continuum sources.}
\label{fig:cont_flux_hist}
\end{figure}

We show in Fig.~\ref{fig:meth_cont_flux_hist} again the flux distribution of the masers
and their associations to continuum sources. By comparing the fraction of masers with an associated continuum source for each flux bin, there seems to be a trend in that the association with continuum sources increases with maser flux density. In Fig.~\ref{fig:cont_flux_hist}, we show instead the flux distribution of the continuum sources to see if weaker radio sources (that are \hii~regions) are more correlated with methanol masers as one might attribute weaker sources to younger stages of development. Despite, the low association rate, we see that a Kolmogorov–Smirnov test of the distribution (bottom panel of Fig.~\ref{fig:cont_flux_hist}) results in $p$-value\,$\ll0.0013$ showing that there is a correlation with the continuum source flux for sources that have methanol maser associations. We see that radio sources with methanol masers are significantly brighter than the general population of radio sources.

The nature of radio continuum sources however can vary from being our desired \hii~regions, planetary nebulae, or even have extragalactic origin. Following \cite{sac2019}, where it was determined that for the GLOSTAR \dconf sensitivity, there would be a source density of 0.0172\,arcmin$^{2}$, this suggests that there would be $\sim$7600 extragalactic sources which is close to 60$\%$ of sources in the catalogue.
As such, we need to take into account the likelihood of an extragalactic background source being inside our matching radius. This is given by $N_{\mathrm{bg}}=(\mathrm{source\,density})\times(\mathrm{search\,area})$. We used a search radius of 6\,arcsec around the maser positions, which means that the estimated number of background sources is then $N_{\mathrm{bg}}\ll1$. This implies that line of sight associations between GLOSTAR 6.7\,GHz masers and GLOSTAR 5.8\,GHz radio continuum sources have a low probability of being purely coincidental.
We also used the CORNISH catalogue to supplement our comparison. 
They have classified their sources which helps us to determine the nature of the continuum sources we have associated with our methanol masers. We find that 34 masers have CORNISH counterparts, which are all labelled as \uchii~regions. Of these associations, 15 do not have GLOSTAR \dconf radio catalogue counterparts. These are likely involved in extended emission seen through the \dconf. Conversely, 45 sources have GLOSTAR radio counterparts but no CORNISH counterparts. A future analysis of the sources' spectral index can provide further insight on the astrophysical nature of the remaining continuum sources \citep{medinaFull}.

\begin{figure}
\centering

\includegraphics[width = 0.50\textwidth]{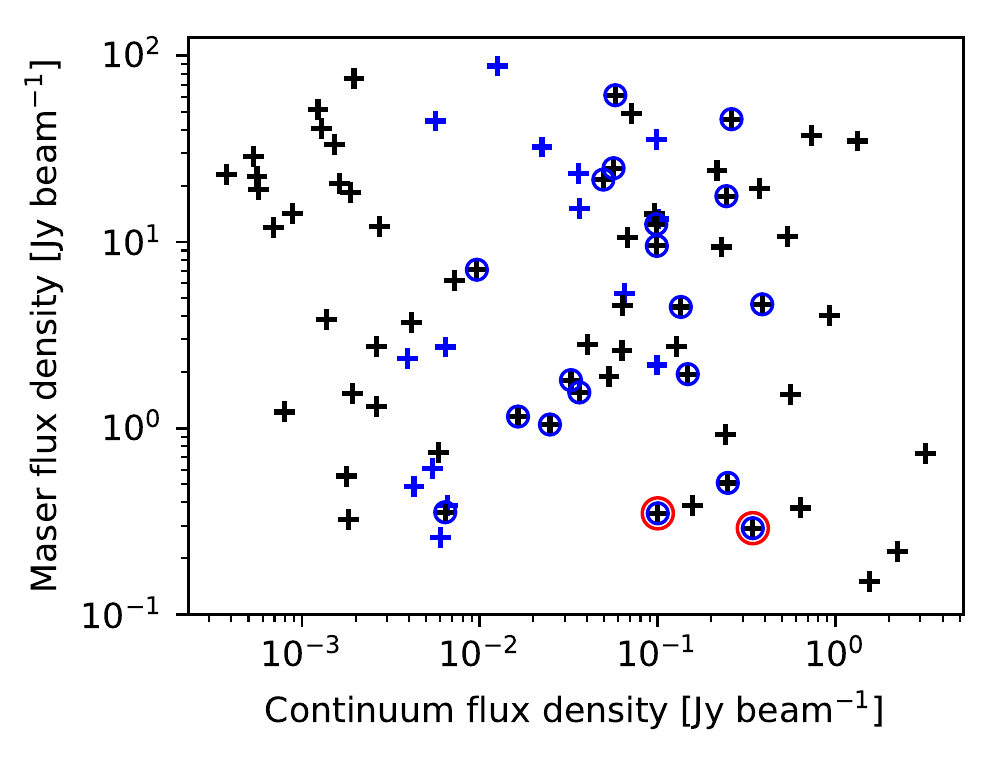}
\caption{Maser flux density against radio continuum flux density. Black crosses correspond to GLOSTAR D-configuration values for the radio continuum whereas blue crosses correspond to CORNISH B-configuration values. Blue circles show the sources that have counterparts in both continuum catalogues, but are plotted with the GLOSTAR flux density. Red circles denote sources that are new maser detections.}
\label{fig:mflux_v_cont_flux}
\end{figure}

\begin{figure}
\centering

\includegraphics[width = 0.50\textwidth]{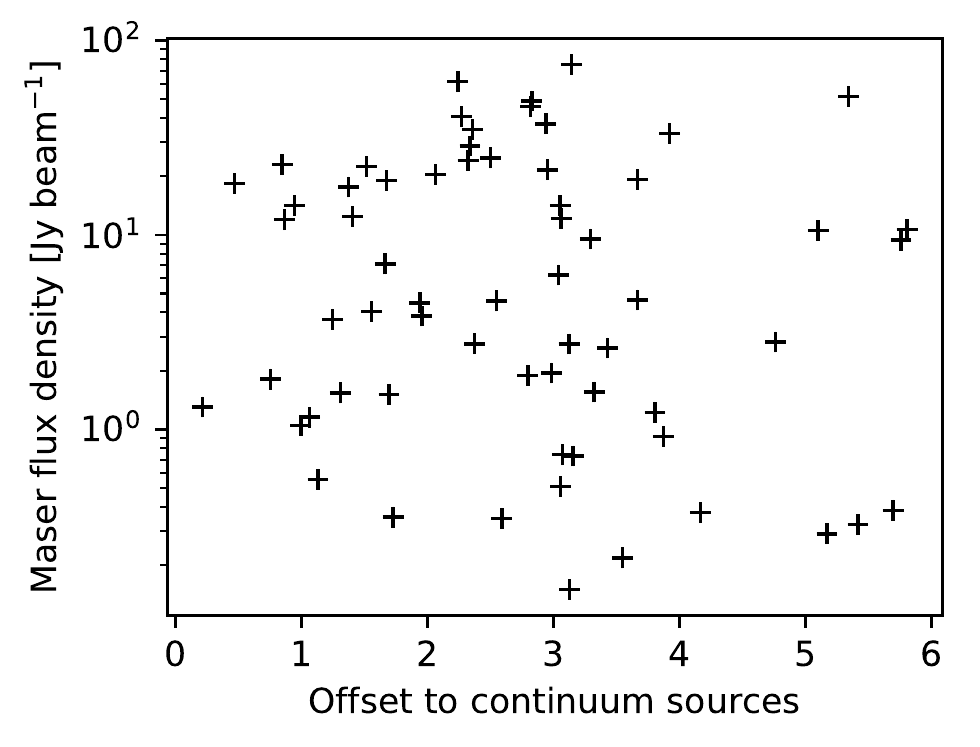}
\caption{Methanol maser flux as a function of offset to the peak flux position of its associated radio continuum source.}
\label{fig:mflux_v_cont_flux_offset}
\end{figure}

We show in Fig.~\ref{fig:mflux_v_cont_flux} the maser flux density as a function of continuum flux density, using the CORNISH catalogue to supplement the GLOSTAR radio continuum catalogue, where we use the flux from GLOSTAR where available. The Spearman's rank coefficient is $r=-0.11$ with $p$-value$=0.32$ thus showing no correlation between the properties. As discussed above, we know that there is a positional correlation between radio sources and methanol masers, however in Fig.~\ref{fig:mflux_v_cont_flux_offset}, we do not see any relation between the flux of a maser and its proximity to the radio continuum source (\KS test results of $r=-0.11$ and $p$-value\,$=0.3$). If the methanol maser was intimately connected with the radio source, one would expect to see increasing maser strength with decreasing position offset. Given that the maser flux and continuum flux show no correlation, and that there is no correlation with maser flux and position offset, this suggests that the mechanisms powering the two kinds of sources are unrelated as expected, despite there being a positional correlation.

\subsection{Luminosity function}\label{sect:luminosity}

\begin{figure*}
\centering
\begin{tabular}{ccc}
\includegraphics[width = 0.33\textwidth]{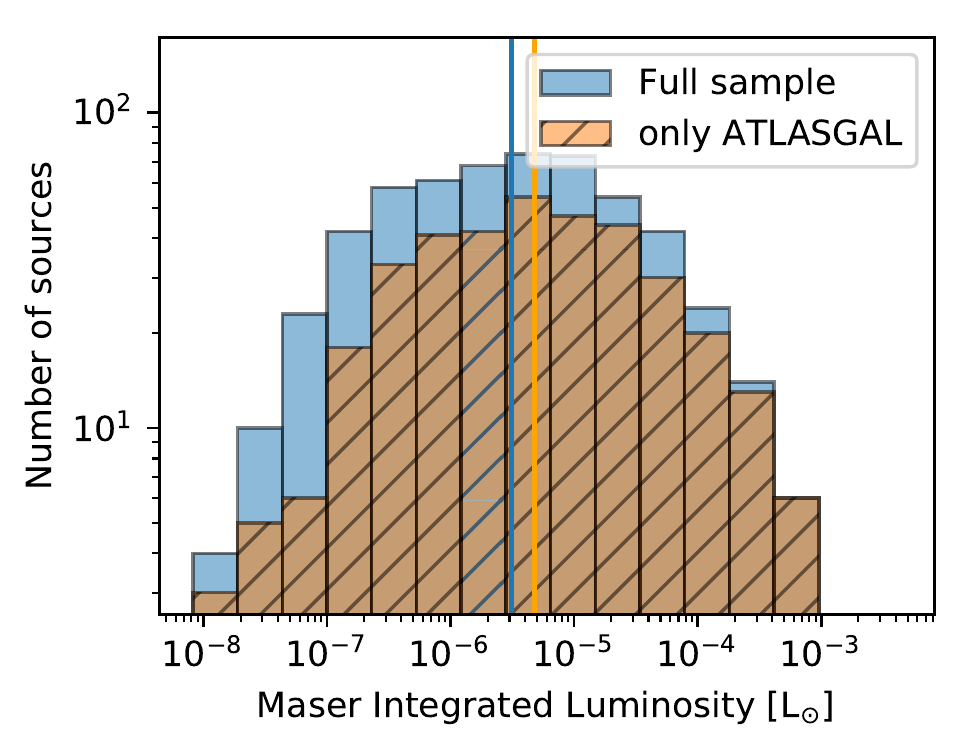}
\includegraphics[width = 0.33\textwidth]{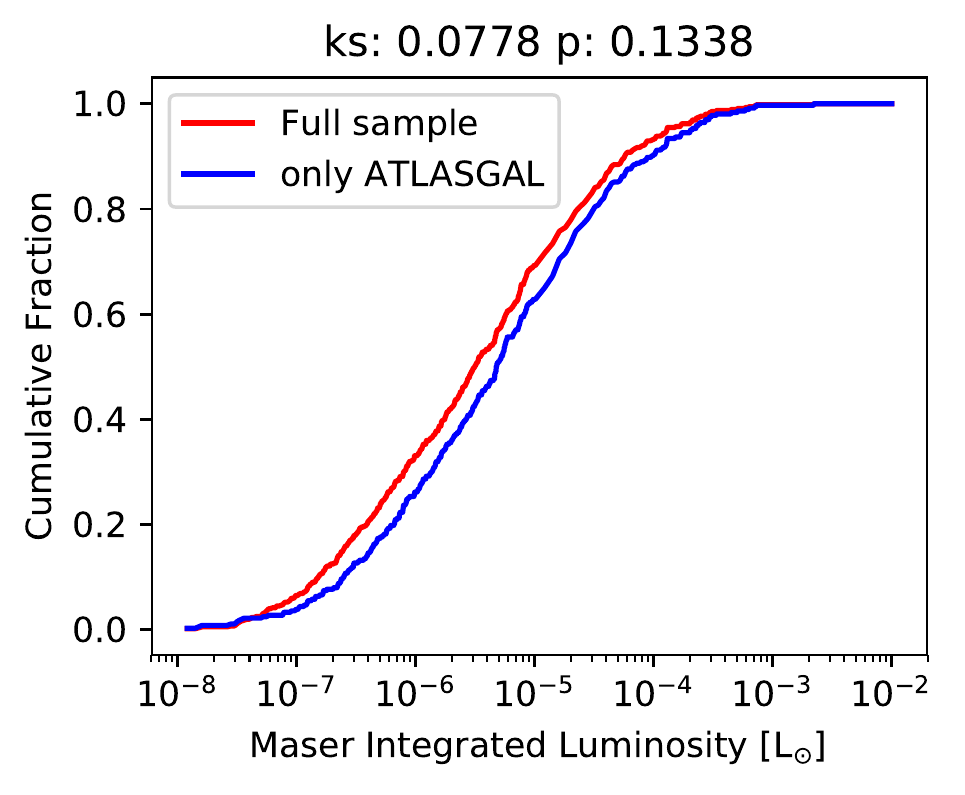}
\includegraphics[width = 0.33\textwidth]{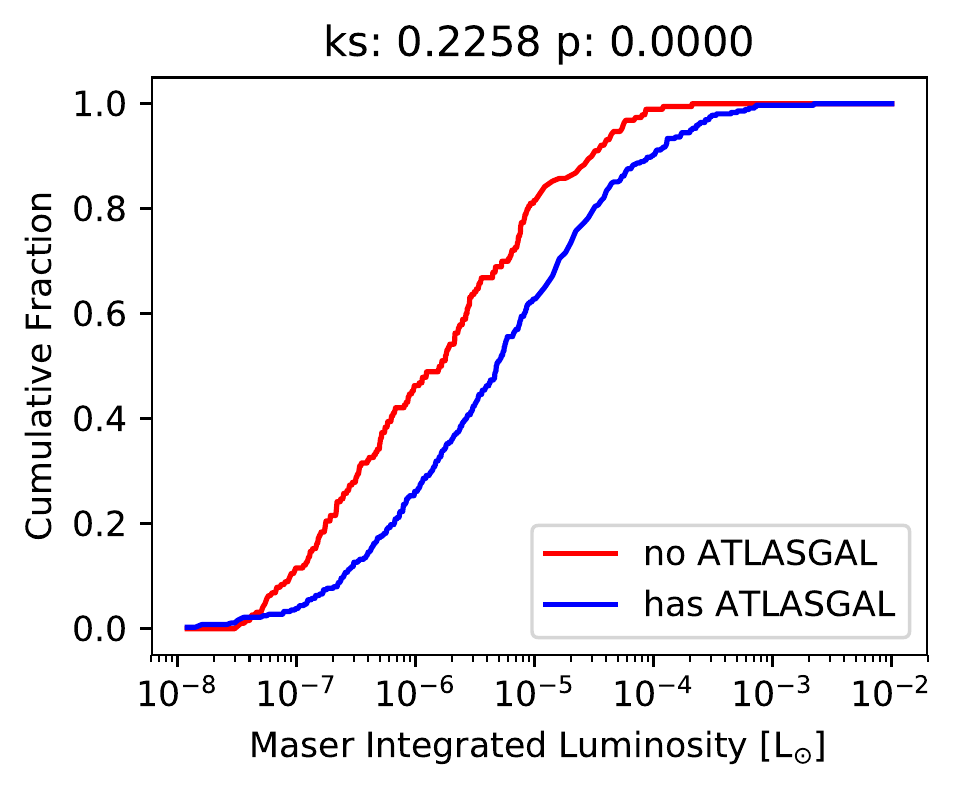}
\end{tabular}
\caption{\emph{Left}: Distribution of luminosities of sources with and without association with ATLASGAL sources.. The median value is $3.1\times10^{-6}\,$L$_{\odot}$ for the full sample and $4.7\times10^{-6}\,$L$_{\odot}$ and the subset of detections peaks that have ATLASGAL \C{compact source catalogue (CSC)} associations. \emph{Middle}: \C{Cumulative distribution function (CDF)} comparing the full sample of luminosities to subset of sources that are associated with ATLASGAL where the two samples are not distinct. \emph{Right}: CDF comparing the sample of luminosities without an ATLASGAL association to those with where they are seen to be statistically distinct.}
\label{fig:lum_func_agal_cf}
\end{figure*}

\begin{figure}
\centering
\includegraphics[width = 0.50\textwidth]{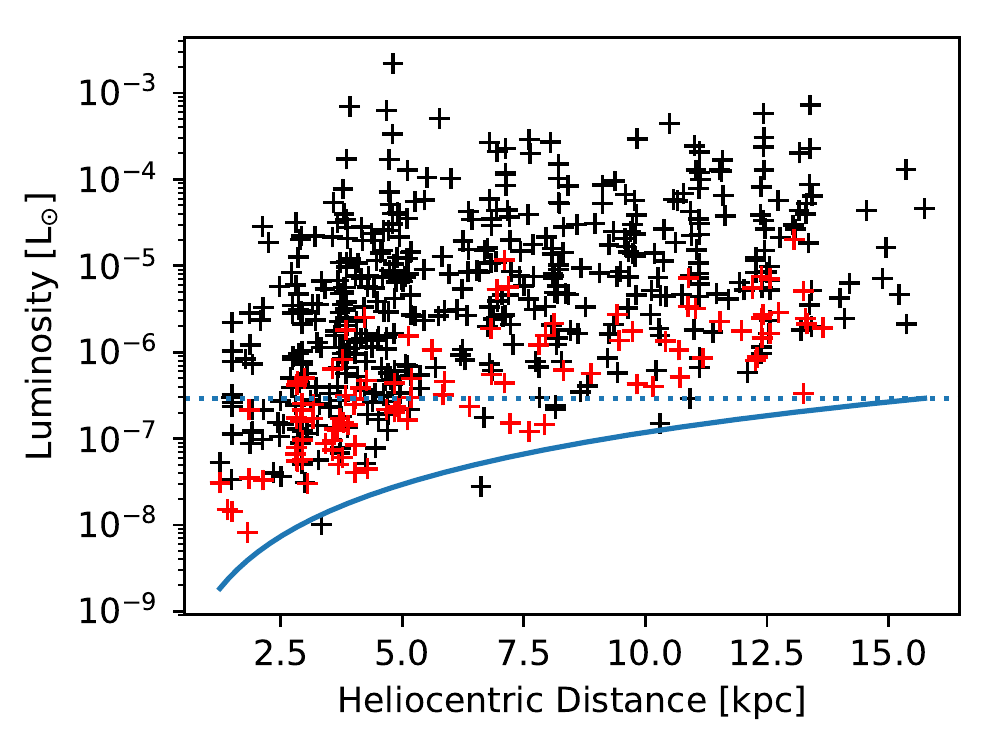}
\caption{Maser integrated luminosity as a function of heliocentric distance marked as black crosses where red crosses highlight the new maser detections from this work. The blue curve denotes the 5$\sigma$ luminosity threshold and the dotted blue line corresponds to the completeness level of $100\%$.}
\label{fig:heliodist_v_lum}
\end{figure}

\begin{figure}
\centering
\includegraphics[width = 0.50\textwidth]{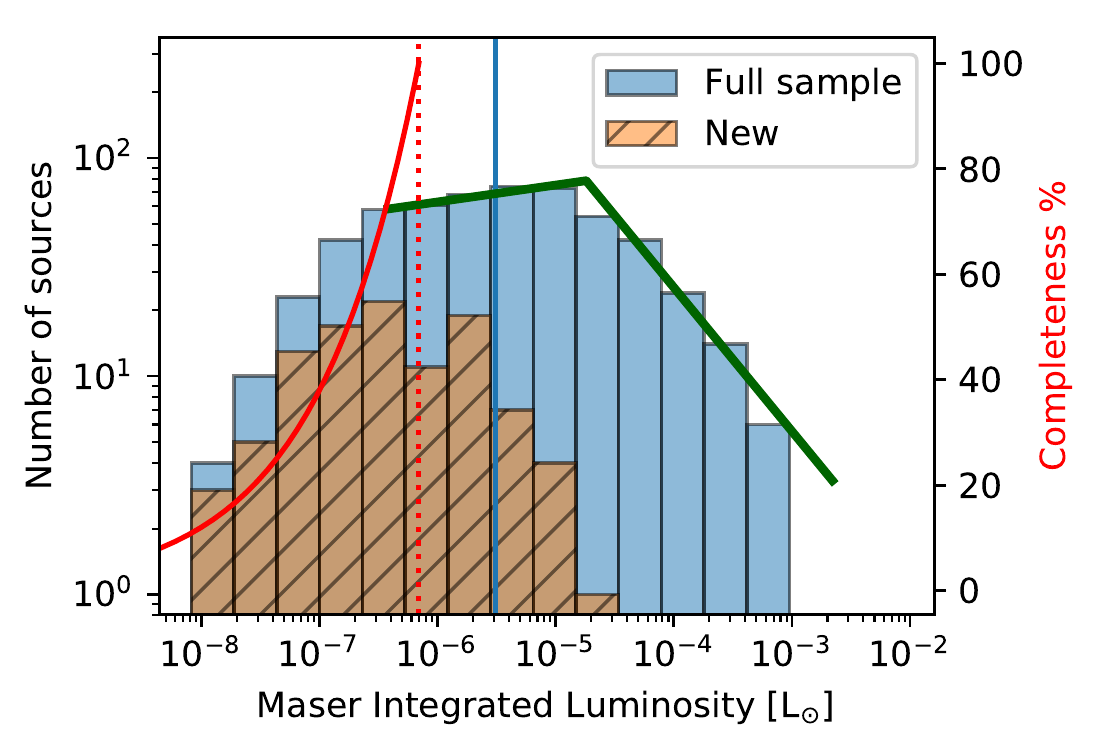}
\caption{Distribution of luminosities derived. The blue vertical line represents the median value of $3.1\times10^{-6}$\,\lsun\,. The subset of new detections peaks at a lower luminosity as expected. The red curve corresponds to the completeness level with values on the right axis, and the dotted red line indicates the level of 100$\%$ completeness. The two green lines represent the broken power law fit with powers of $0.08\pm0.05$ for the lower luminosity range and $-0.66\pm0.05$ for the upper luminosity range.}
\label{fig:lum_func}
\end{figure}
Given that the 6.7\,GHz methanol maser is a tracer of HMSF, a luminosity function of these sources would allow us to compare the amount of HMSF in the Milky Way and other nearby Galaxies. Studies have shown that the luminosity function for these masers cannot be fit by a single power law but may be fit with a broken power law \citep{pesta2007,pandian2009, jag2011}. As discussed in Section~\ref{sect:distance_determination}, we have allocated distances to the maser sources using a Bayesian distance estimator as well as ATLASGAL clump velocities. However, if we were to use only the sources with an ATLASGAL association to determine the luminosity function, we would be biased towards sources with higher luminosities. This is evident in Fig.~\ref{fig:lum_func_agal_cf}, which shows a clear fractional difference in the number of sources in the lower luminosity bins compared to the full sample to ATLASGAL CSC only sources. While a CDF of these two samples show that they are not statistically distinct, which is to be expected, we do find that in examining the right panel of Fig.~\ref{fig:lum_func_agal_cf}, we see that there is a statistical difference in maser properties between the sources with ATLASGAL counterparts and those without (\KS test result of $r=0.74$ and $p$-value\,$\ll0.0013$). As such, we choose to use the full sample. In Fig.~\ref{fig:heliodist_v_lum}, we plot the luminosity as a function of heliocentric distance. As expected, the new maser detections cover the lower luminosity ranges for a given distance. This also allows us to determine the completeness level. Given a minimum flux, we can calculate the minimum luminosity of a maser we can detect for a given heliocentric distance. This in turn can be turned around to give the maximum distance at which a maser of a given luminosity can be detected. Then, within the limits of the survey coverage, the fraction of the Milky Way disk covered at a given distance will give us the completeness that has been normalised over the survey area. This is shown in the luminosity function in Fig.~\ref{fig:lum_func}. We find that we are 100$\%$ complete at $6.9\times10^{-7}$\,\lsun. We find that the median luminosity is $3\times10^{-6}\,$L$_{\odot}$ and this agrees with previous
studies \citep[e.g.,][]{pandian2009} that the distribution peaks around $10^{-6}\,$L$_{\odot}$. Our sample size is $\sim$6 times larger than \cite{Pandian2007,pandian2009} and therefore our median luminosity is statistically more robust. To characterise the luminosity function, we use only the luminosity bins for above which we are complete. We simultaneously fit two power laws and find indexes of $0.08\pm0.05$ for the lower luminosity range and $-0.66\pm0.05$ for the higher luminosity range where the turnover has been determined to be $\sim$2$\times10^{-5}$\,L$_{\odot}$. However, we see that while it is possible to fit a broken power law to the data, we do not sample well the lower luminosities as our 100$\%$ completeness is around $6.9\times10^{-7}\,$L$_{\odot}$.

\section{Summary and Conclusions}\label{sect:summary}

We have conducted the most sensitive, unbiased survey of Class II 6.7\,GHz \meth masers
to date in the region covered by the GLOSTAR survey in the Galactic plane. A total of 554 masers were detected, with 84 of them being new detections. \C{Over 50\,$\%$ of the new detections have fluxes $<0.5$\,Jy and could be detected due to the improved sensitivity of GLOSTAR compared to other unbiased surveys. A summary of the main results of this work are listed below:}
\C{
\begin{itemize}
    \item Comparing with the ATLASGAL Compact Source Catalogue (CSC) we find that 65$\%$ of the \meth masers are associated with dense gas, with many of the newly detected masers being unassociated. However, a visual inspection reveals a much higher association rate of 97$\%$, indicating that many of the new masers are associated with weak dust emission that is below the sensitivity required for inclusion in the ATLASGAL CSC. \\

    \item The newly detected masers are weaker both in terms of their maser emission and associated dust emission. This might indicate they are either more distant than the previously detected masers or could be associated with lower mass stars or less evolved stars. Given the lower range of maser luminosities and that the $L/M$ distribution of the new masers are consistent with the previous masers, this indicates that they are more likely to be associated with lower mass stars.\\

    \item The high correlation between methanol masers and dust emission and the high bolometric luminosities are consistent with the picture of methanol masers being associated with the early stages of high-mass star formation. We have derived an $L/M$ threshold for the onset of the methanol maser emission of $\sim$1\,L$_{\odot}$\,M$_{\odot}^{-1}$, which is consistent with values determined by \cite{giselacygnus} from a study of the Cygnus~X region with GLOSTAR data and previous work on the MMB catalogue by \cite{billington2019}.\\

    \item We find that 12\% of the masers are coincident with radio continuum emission (i.e. $< 12\arcsec$) but in comparing the radio and maser flux distribution, we find no correlation as a function of angular offset. This suggests that the mechanisms powering maser and continuum emission are unrelated.\\

    \item We use our sample of masers to construct a luminosity function using a broken power-law. Our results agree with previous studies in that the distribution has a median luminosity $10^{-6}$\,\lsun. We sample well the high luminosity maser population but are limited in the lower luminosity bins. \\

\end{itemize}

This work is the first step in our study of 6.7\,GHz \meth~masers using GLOSTAR data in the Galactic plane.  Methanol absorption sources have also been detected and a systematic search is forthcoming. Further study of the properties of these masers would be best served with higher resolution data that we will present in future works.
}

\begin{acknowledgements}
    We would like to thank the anonymous referee for their useful comments.
    
    H.N. is a member of the International Max-Planck Research School at the universities of Bonn and Cologne (IMPRS).
    
    This research was partially funded by the ERC Advanced Investigator Grant GLOSTAR (247078).

    H. B. acknowledges  support from the European Research Council under the European Community's Horizon 2020 framework program (2014-2020) via the ERC Consolidator grant ‘From Cloud to Star Formation (CSF)' (project number 648505). H. B. further acknowledges support from the Deutsche Forschungsgemeinschaft (DFG) via Sonderforschungsbereich (SFB) 881 “The Milky Way System” (sub-project B1).
    
    The National Radio Astronomy Observatory is a facility of the National Science Foundation, operated under a cooperative agreement by Associated Universities, Inc.
    
    This research made use of information from the ATLASGAL database at \url{http://atlasgal.mpifr-bonn.mpg.de/cgi-bin/ATLASGAL_DATABASE.cgi} supported by the MPIfR in Bonn.
    
    This research made use of Astropy,\footnote{http://www.astropy.org} a community-developed core Python package for Astronomy \citep{Astropy-CollaborationRobitaille:2013ab,Astropy-CollaborationPrice-Whelan:2018aa}.
    
    This research has made use of the SIMBAD database, operated at CDS, Strasbourg, France.

\end{acknowledgements}

\bibliographystyle{aa}
\bibliography{hng2022}

%-------------------------------------------------------------------
%APPENDIX 

%\newpage
%\clearpage
\begin{appendix}

%%%%%%%%%%%%%%%%%%%%%%%%%%%%%%%%%%%%%%%%%%%%%%%%%%%%%%%%%%%%%%%%%%%%%%

\section{Strongest New sources}
\label{app:strong_new}
Cutouts of the strongest new 6.7\,GHz \meth maser detections at their respective peak velocities. We visually inspect these sources that should have been detected by the MMB given their high fluxes. While some sources were probably missed previously due to its proximity to a stronger maser source, it is likely maser variability that plays a role in their previous non detection.

\begin{figure*}
\centering
\begin{tabular}{cc}
\includegraphics[width = 0.50\textwidth]{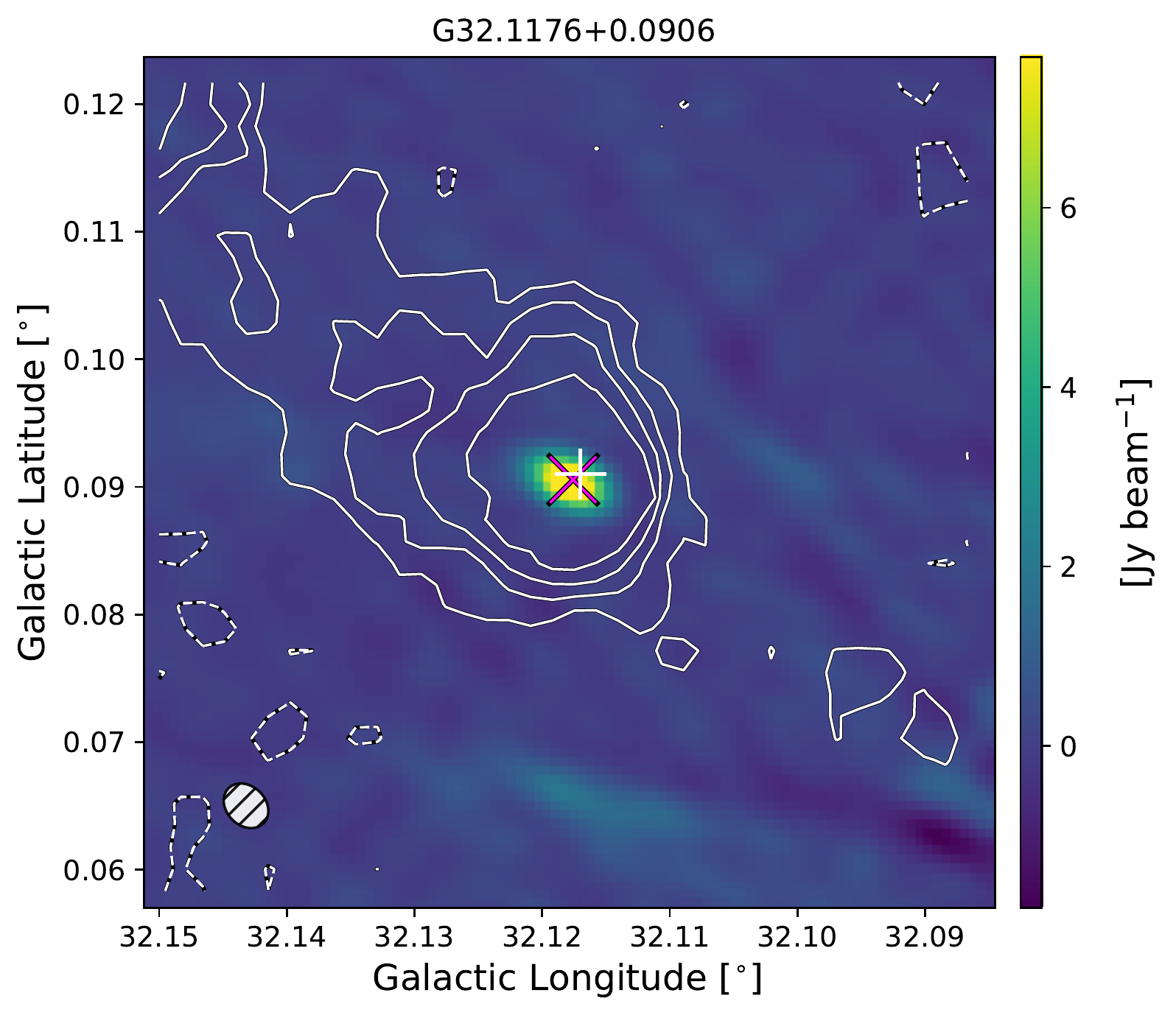} 
\includegraphics[width = 0.50\textwidth]{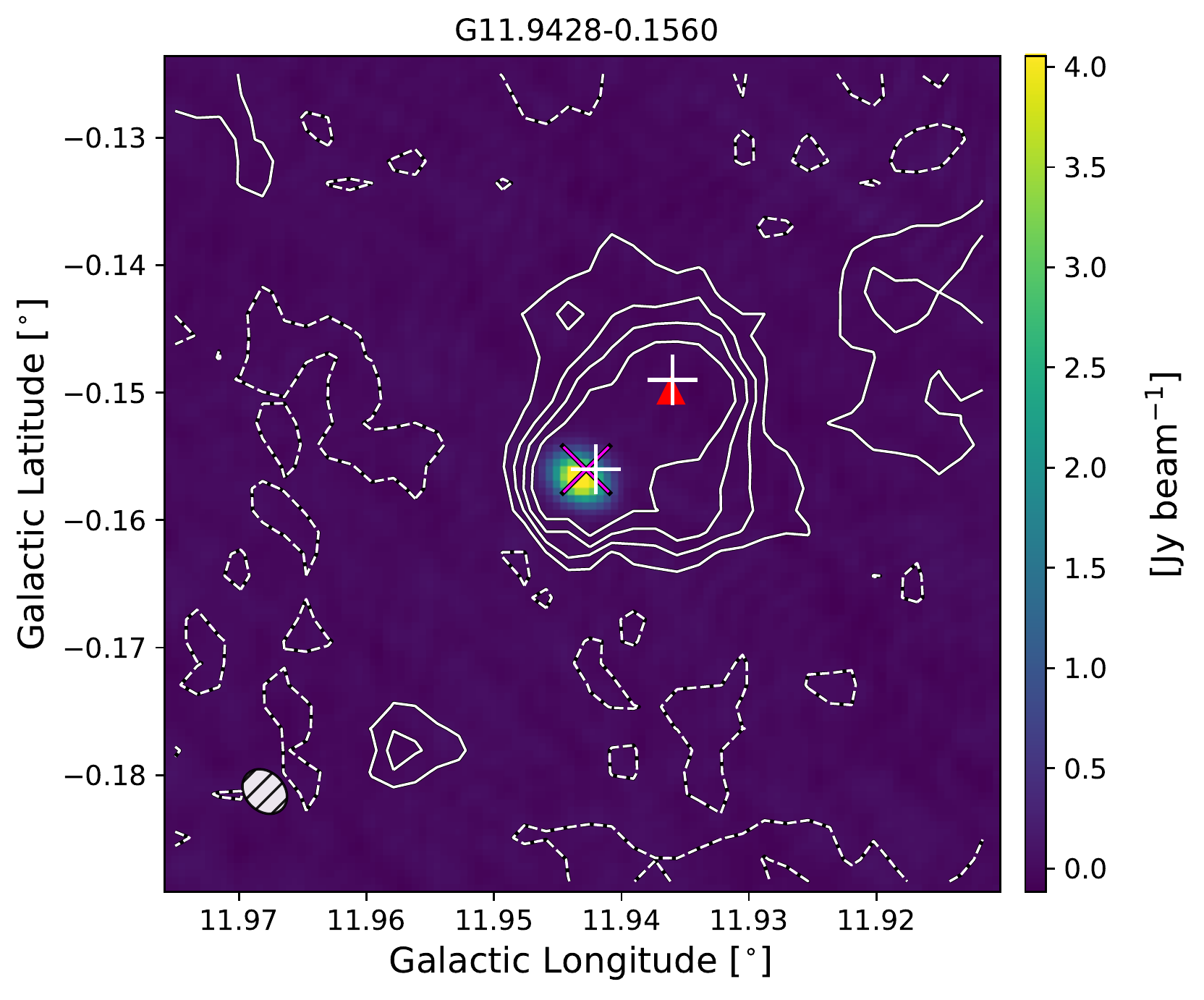} \\
\includegraphics[width = 0.50\textwidth]{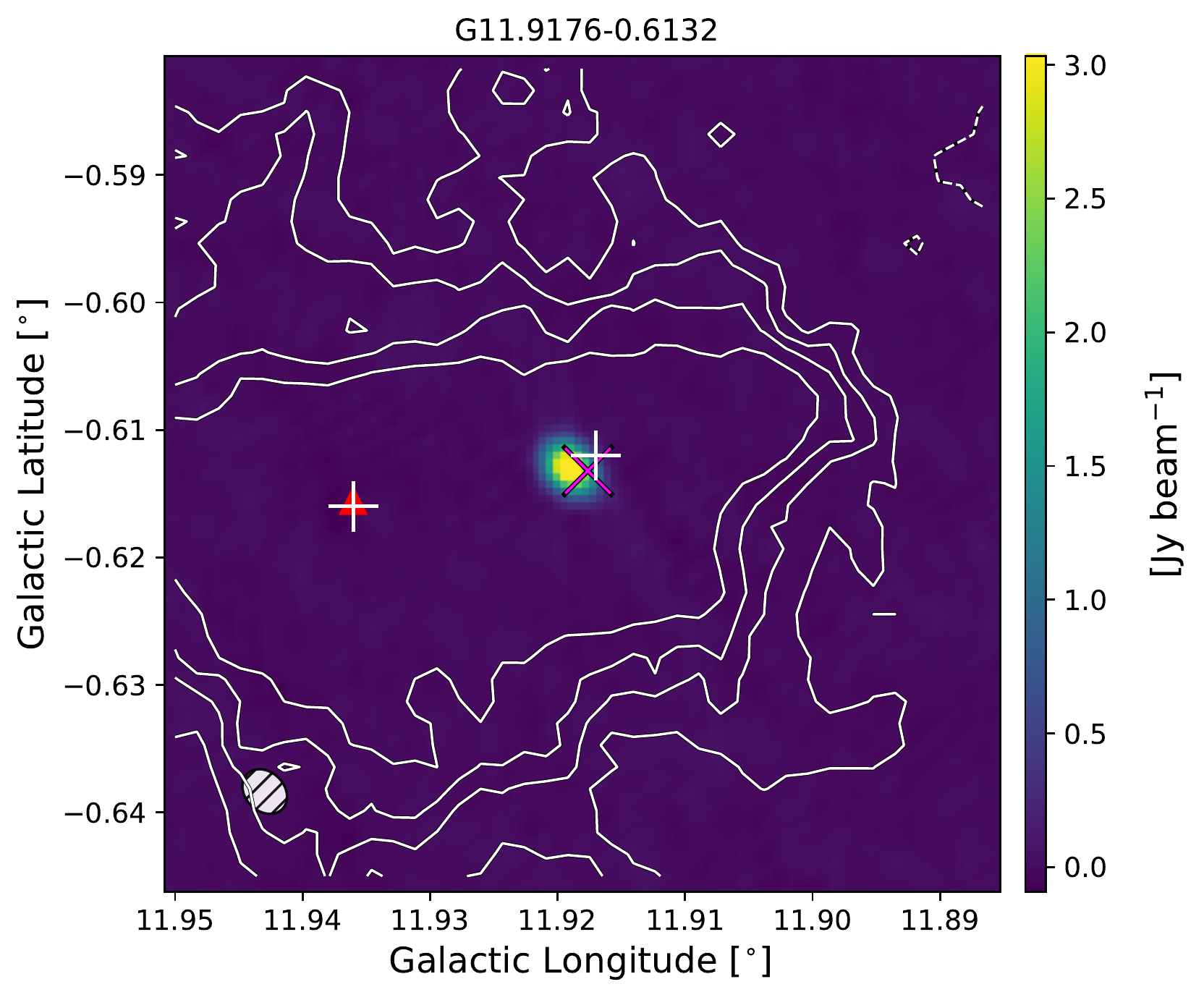} 
\includegraphics[width = 0.50\textwidth]{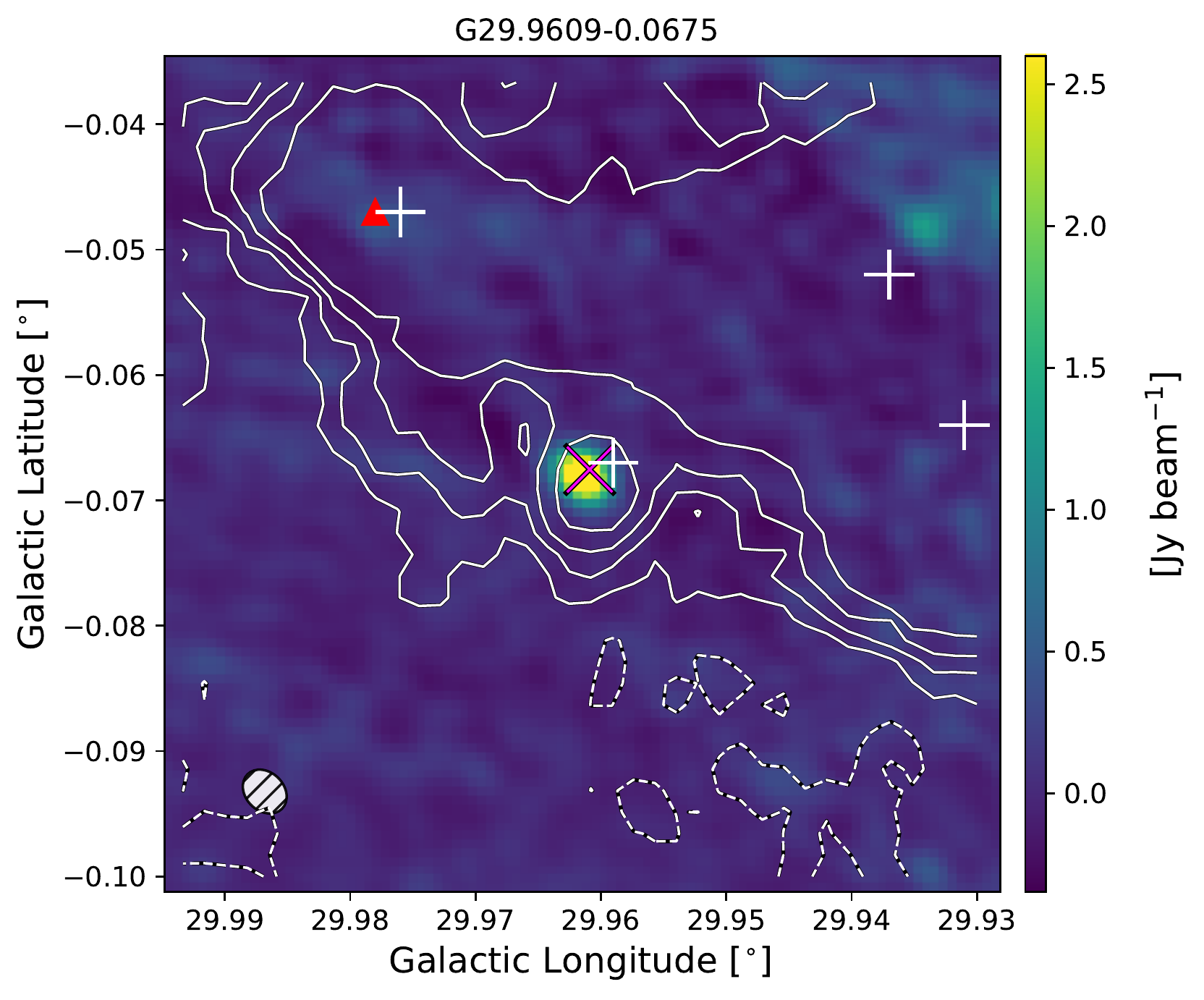} \\
\end{tabular}
\caption{GLOSTAR 6.7\,GHz emission maps at the velocity of the maser emission peak of the 10 strongest new masers which are denoted by the magenta `X'. Red triangles denote the position of known MMB masers. The flux levels were limited to 75$\%$ of the maser peak to better illustrate low intensity features. The white `+' \C{signs show} the \C{positions} of known compact ATLASGAL sources and the white contours are from the ATLASGAL 870\,$\mu$m dust emission map with contour levels at -3,3,5,7, and 10 $\sigma$ noise levels.}
\label{fig:new_maser_cutout}
\end{figure*}

\begin{figure*}
\centering
\begin{tabular}{cc}
\includegraphics[width = 0.50\textwidth]{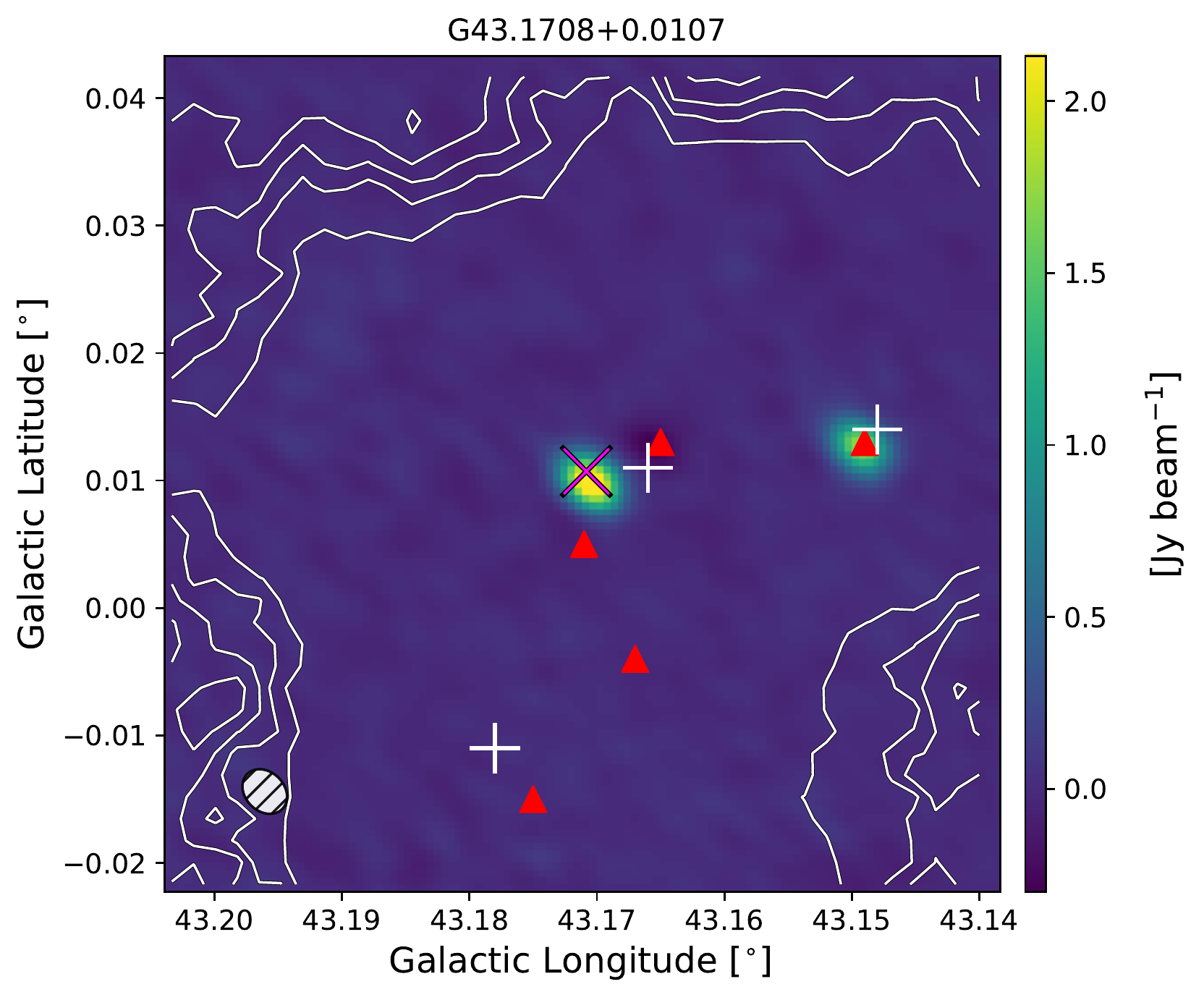} 
\includegraphics[width = 0.50\textwidth]{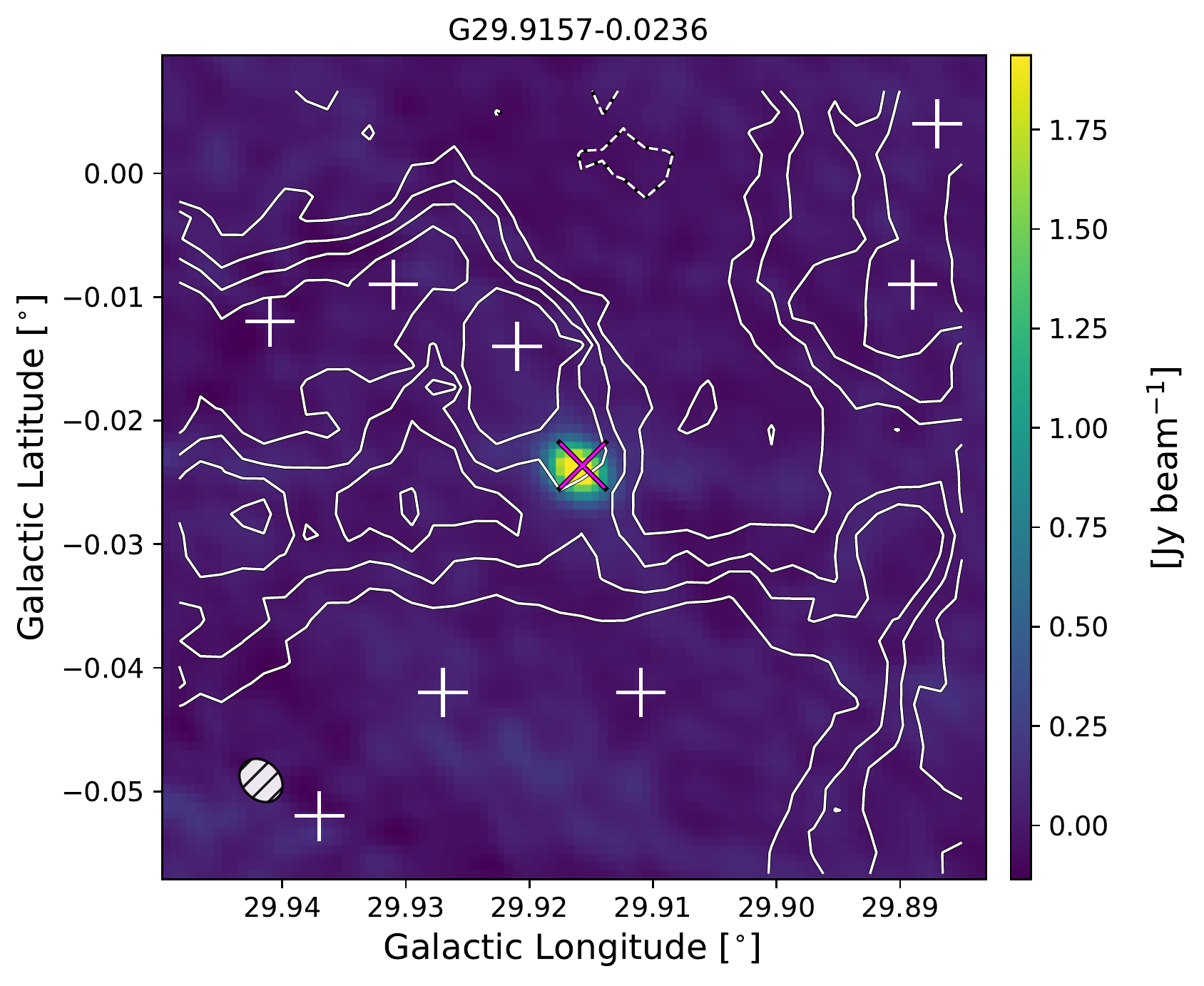} \\
\includegraphics[width = 0.50\textwidth]{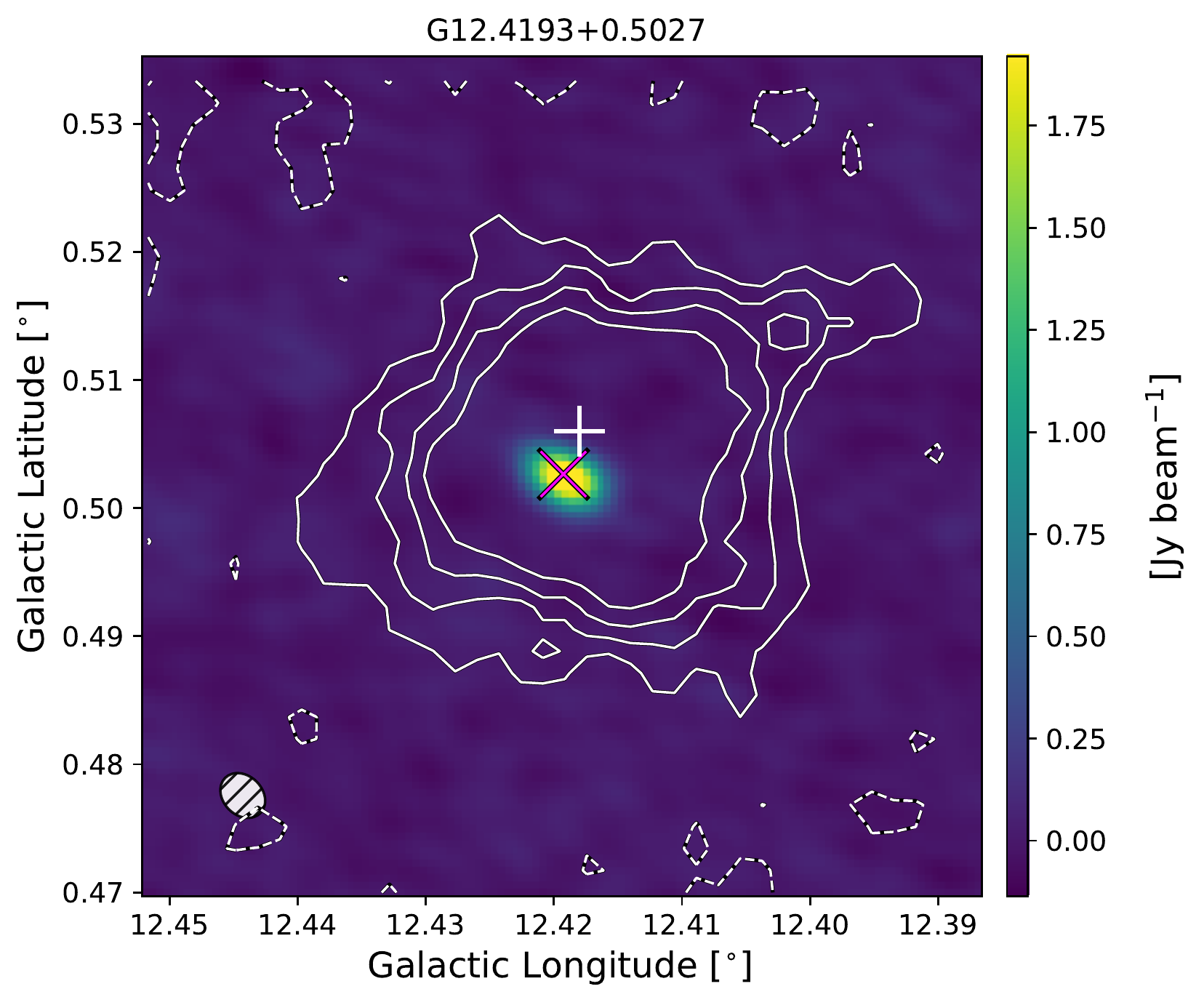} 
\includegraphics[width = 0.50\textwidth]{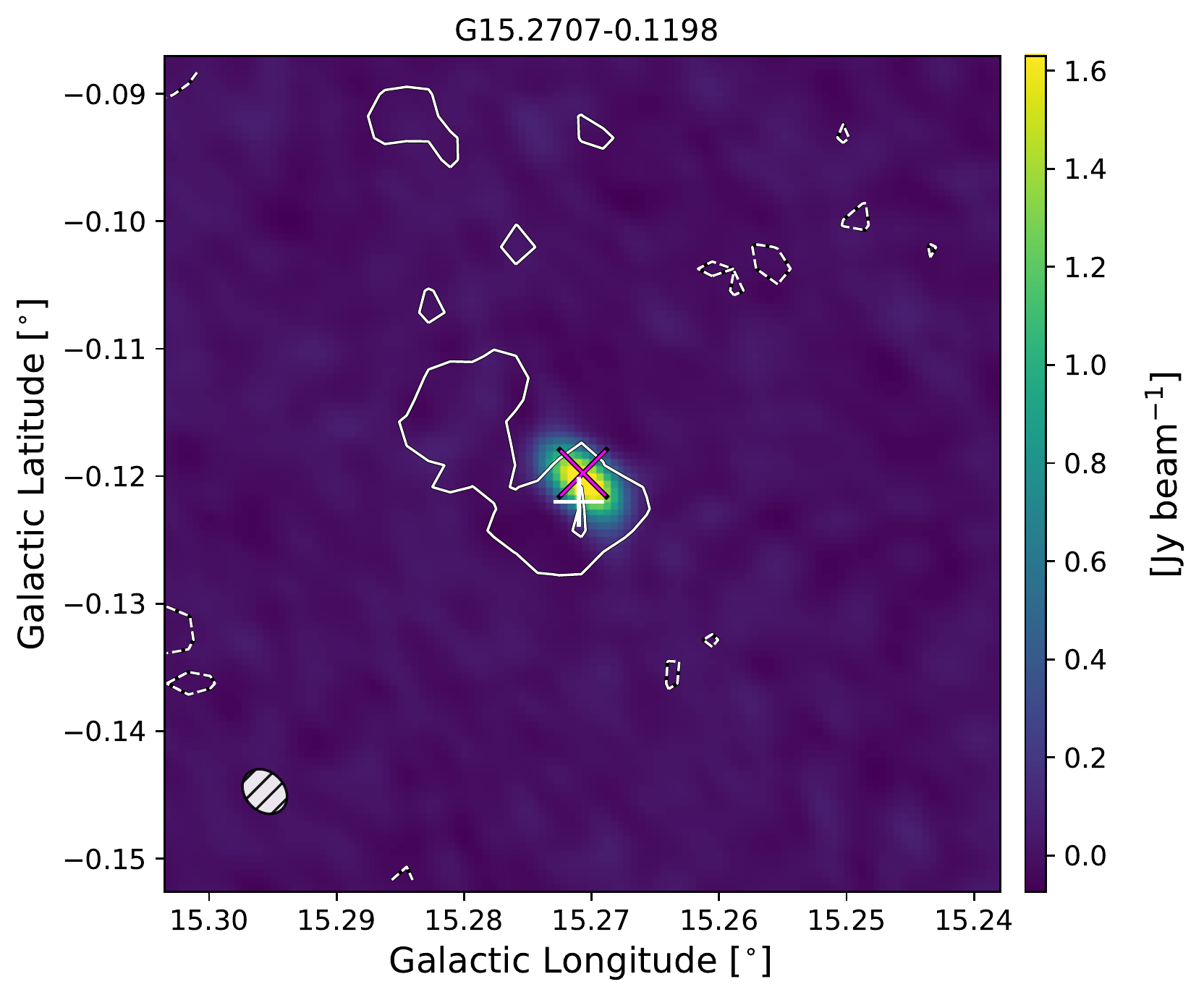} \\
\includegraphics[width = 0.50\textwidth]{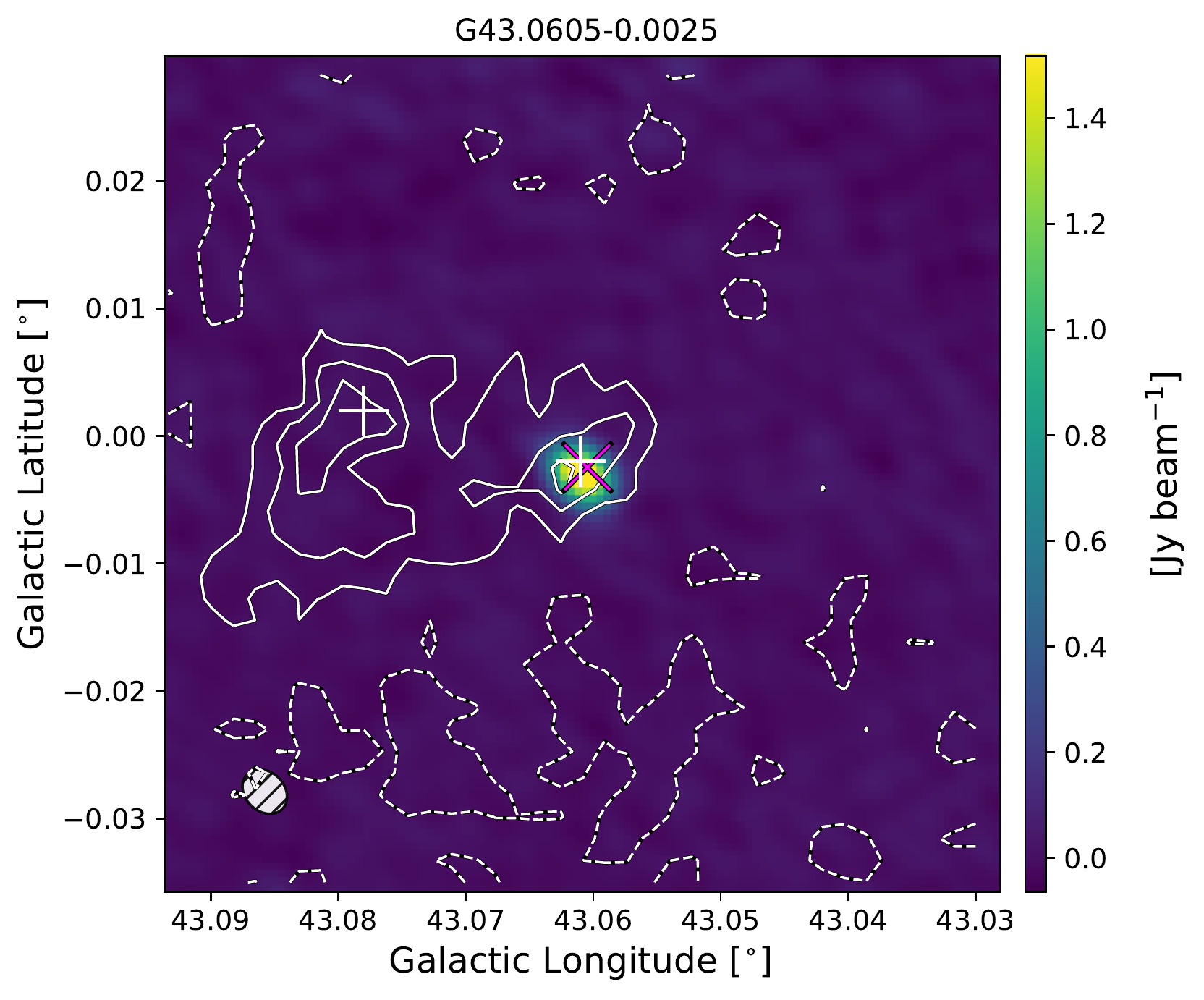} 
\includegraphics[width = 0.50\textwidth]{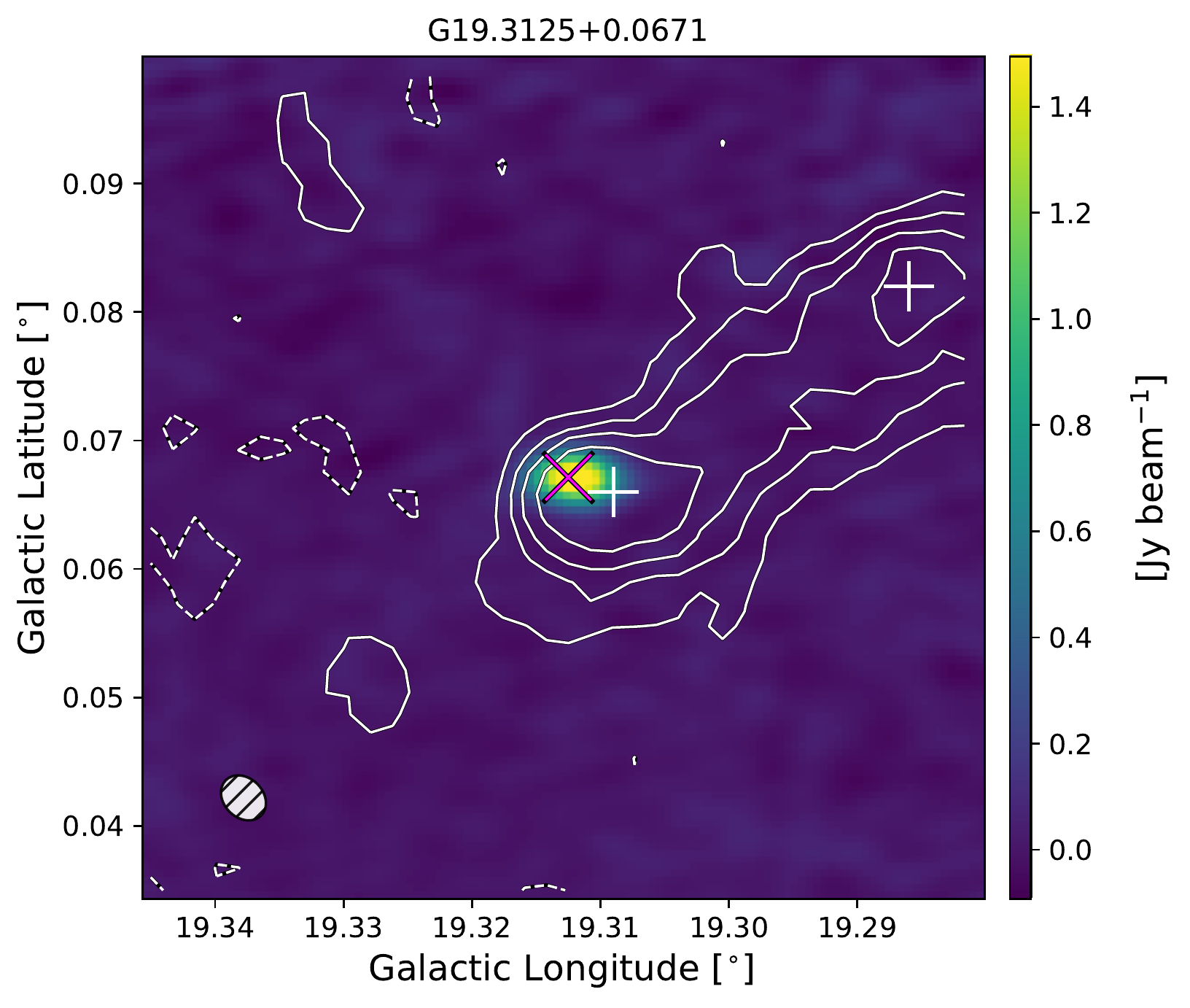} \\
\end{tabular}
\caption{Continued from Fig.~\ref{fig:new_maser_cutout}.}
\label{fig:new_maser_cutout2}
\end{figure*}

%%%%%%%%%%%%%%%%%%%%%%%%%%%%%%%%%%%%%%%%%%%%%%%%%%%%%%%%%%%%%%%%%%%%%%
\section{Association of masers with weaker ATLASGAL emission}
We visually inspect all maser positions that were not automatically matched with an ATLASGAL CSC counterpart for dust emission and find that most of them are still associated with dust.
\label{app:atlasgal_maser_examples}
\begin{figure*}
\centering
\begin{tabular}{cc}
\includegraphics[width = 0.50\textwidth]{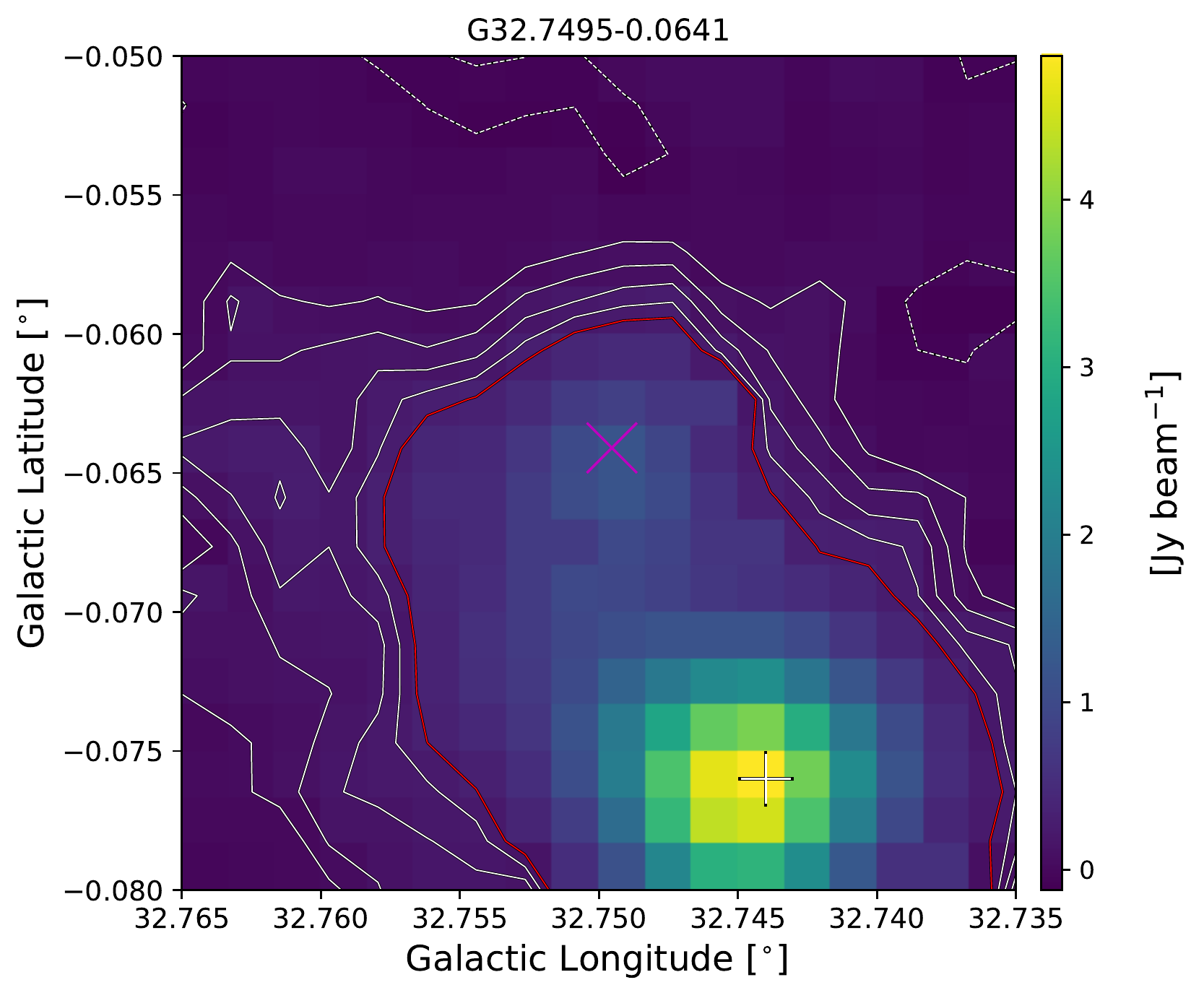} 
\includegraphics[width = 0.50\textwidth]{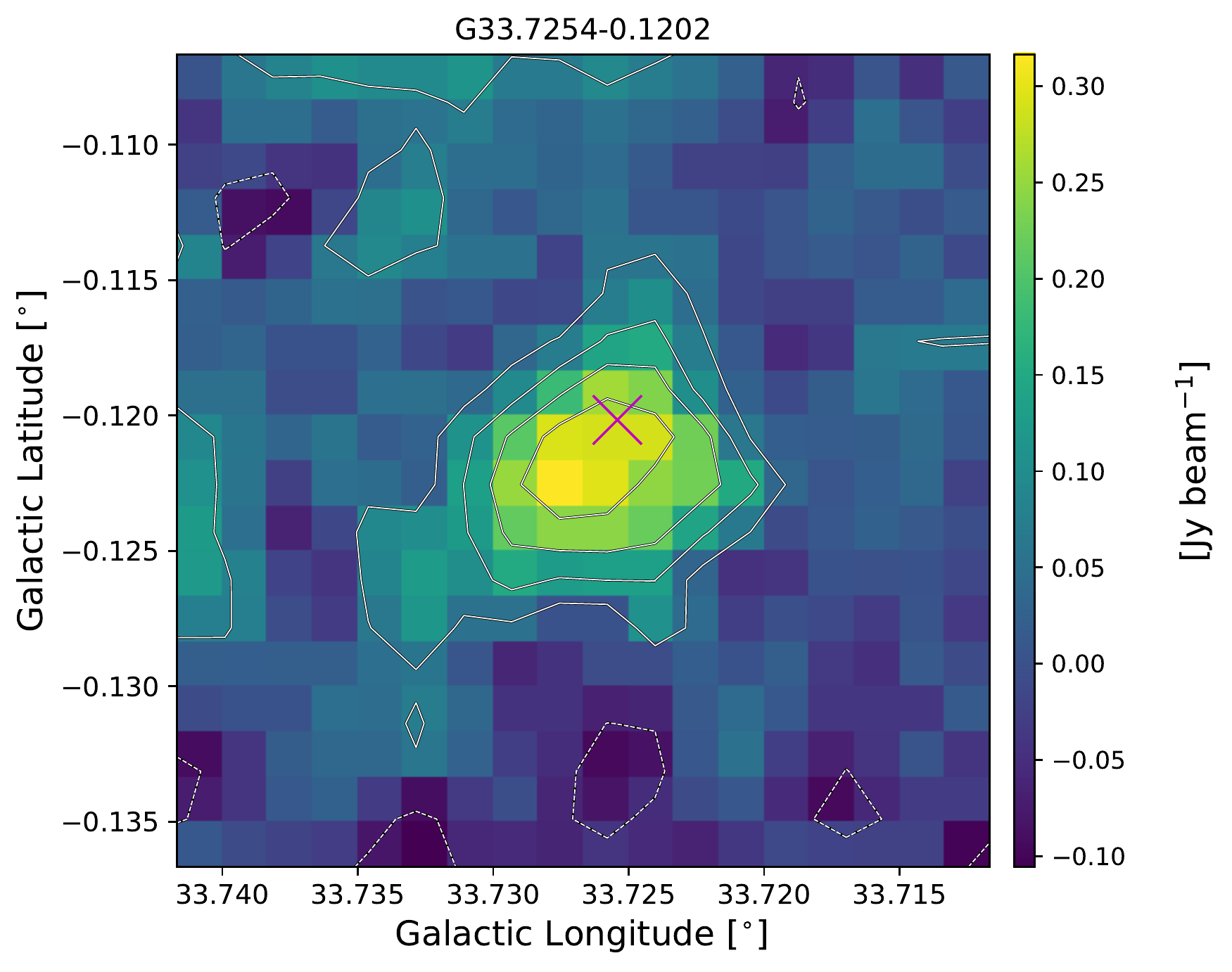}
\end{tabular}
\caption{ATLASGAL 870\,$\mu$m dust continuum cutouts centered on the position of a given maser, shown as a magenta `X'. The white `+' is the position of an ATLASGAL compact source from the \C{compact source catalogue} (CSC). The white contours are the ATLASGAL 1,2,3, and 4$\sigma$ levels, where the red contour is the 5$\sigma$ level. The left panel shows an example where it is clear that the methanol maser is associated with dust emission above 5$\sigma$, but farther than the 12\asec used for the association. The right panel shows an example of a weak compact source that shows a maser association, but was not considered for the ATLASGAL CSC. However, the association with the maser makes a strong argument for the veracity of the weak compact source.}
\label{fig:agal_cutoff}
\end{figure*}

%%%%%%%%%%%%%%%%%%%%%%%%%%%%%%%%%%%%%%%%%%%%%%%%%%%%%%%

%%%%%%%%%%%%%%%%%%%%%%%%%%%%%%%%%%%%%%%%%%%%%%%%%%%%%%%

\section{Spectra of \meth masers}\label{app:cutouts}
Contained in this section are spectra for each maser source.

%\include{spectra_latex_bk}
%Table 1
\begin{figure*}[!h]
\begin{tabular}{ccccc}
\colorbox{white}{\includegraphics[trim={2cm 0 0 0}, scale=0.195]{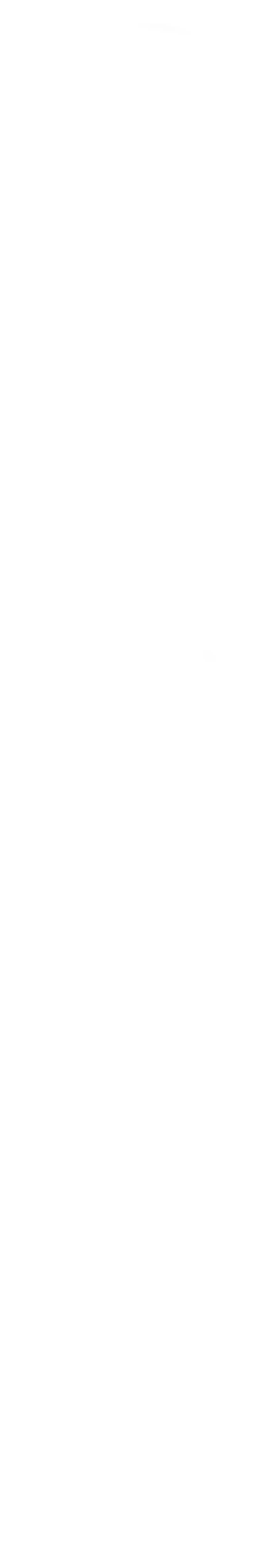}}
\includegraphics[trim={0.0cm 0.3cm 0.6cm 0.3cm}, clip, width = 0.235\textwidth]{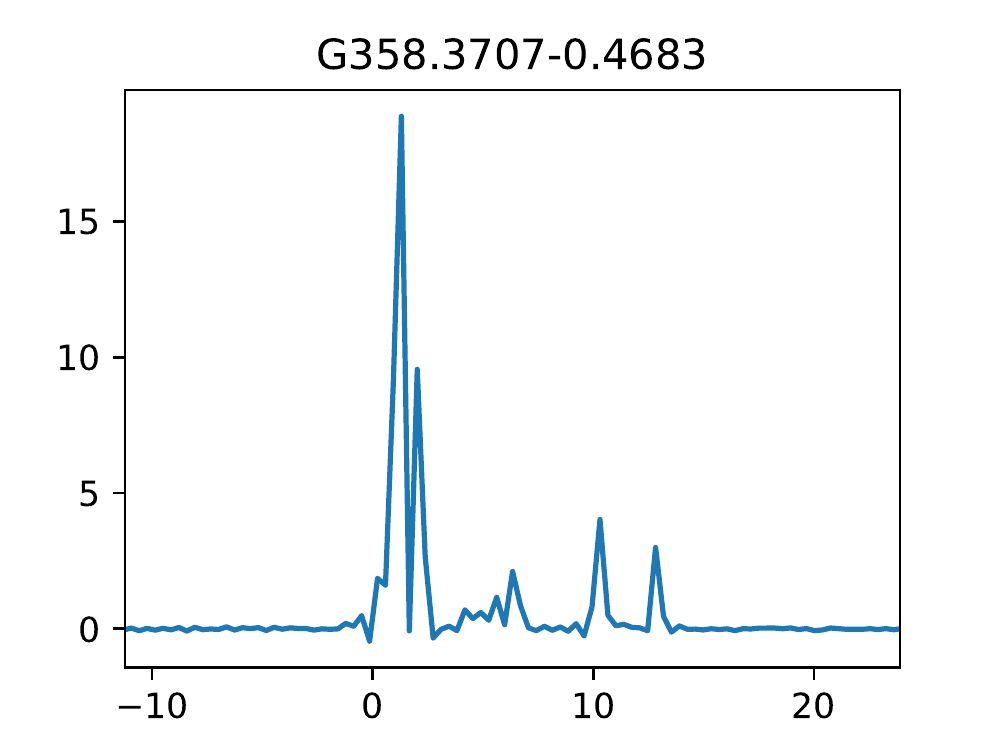}
\includegraphics[trim={0.0cm 0.3cm 0.6cm 0.3cm}, clip, width = 0.235\textwidth]{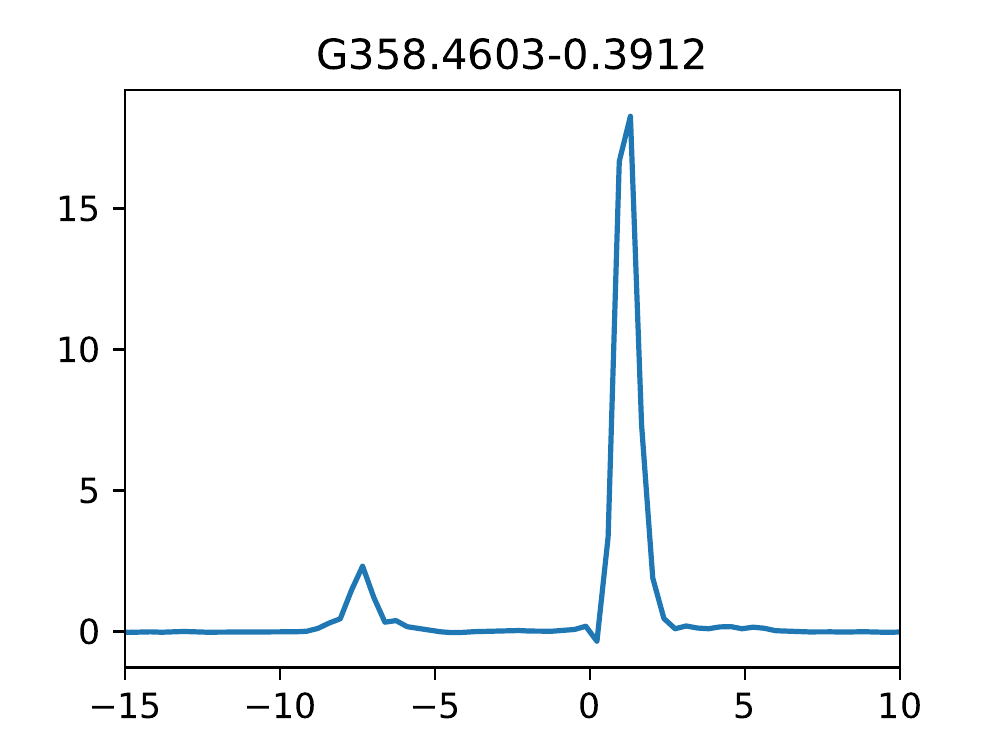}
\includegraphics[trim={0.0cm 0.3cm 0.6cm 0.3cm}, clip, width = 0.235\textwidth]{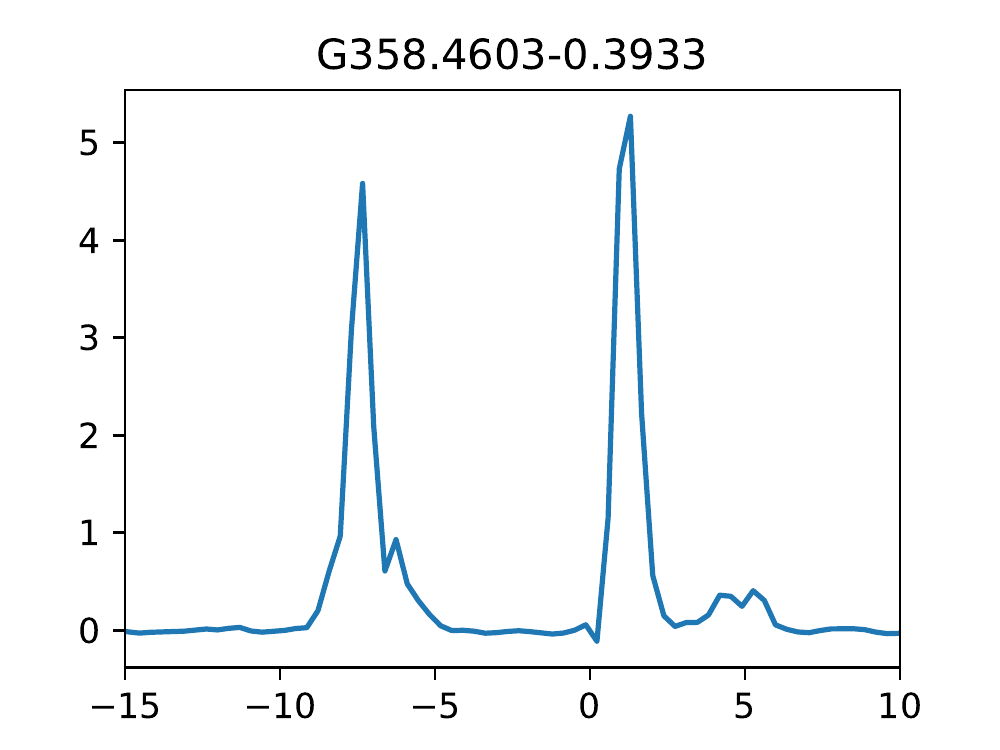}
\includegraphics[trim={0.0cm 0.3cm 0.6cm 0.3cm}, clip, width = 0.235\textwidth]{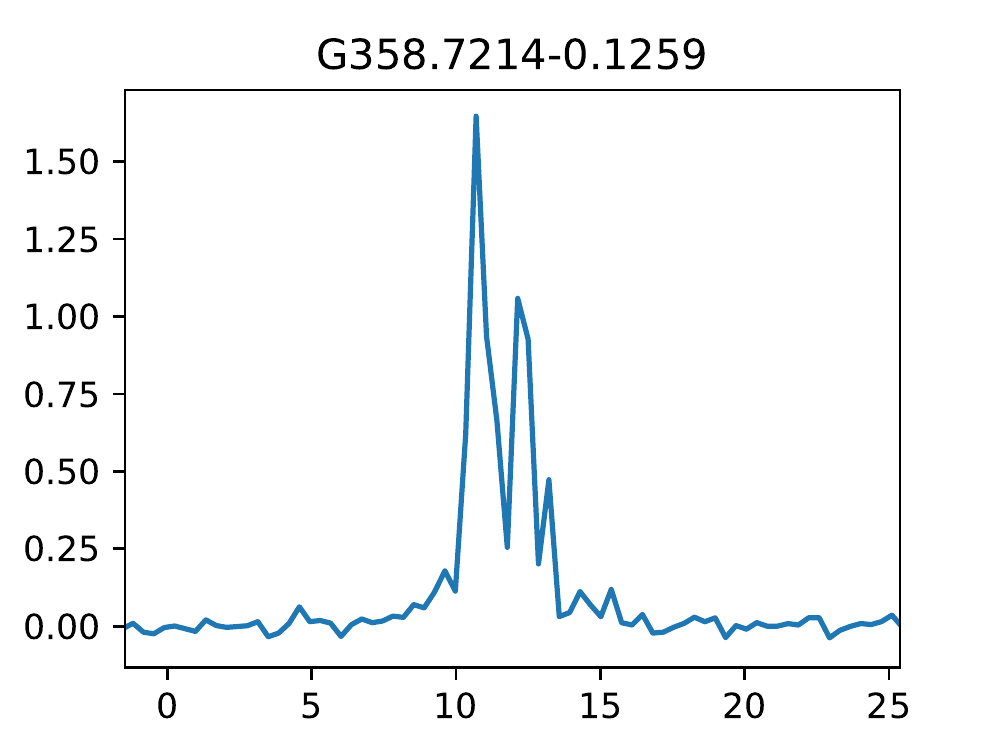}\\ 
\colorbox{white}{\includegraphics[trim={2cm 0 0 0}, scale=0.195]{figs/ylabel_filler.pdf}}
\includegraphics[trim={0.0cm 0.3cm 0.6cm 0.3cm}, clip, width = 0.235\textwidth]{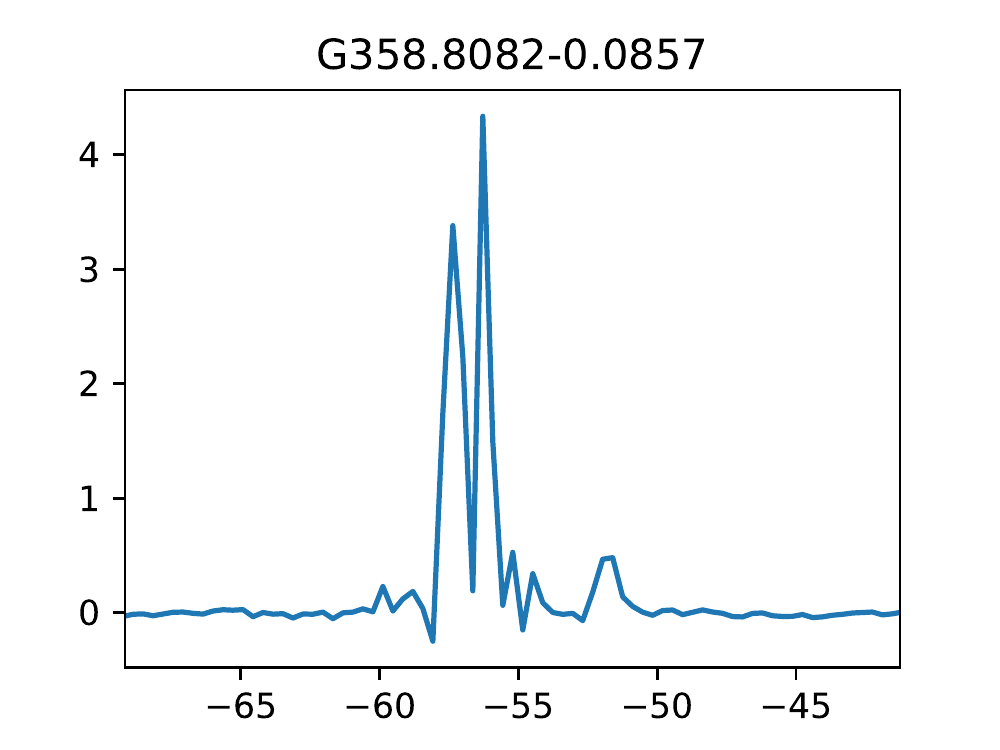}
\includegraphics[trim={0.0cm 0.3cm 0.6cm 0.3cm}, clip, width = 0.235\textwidth]{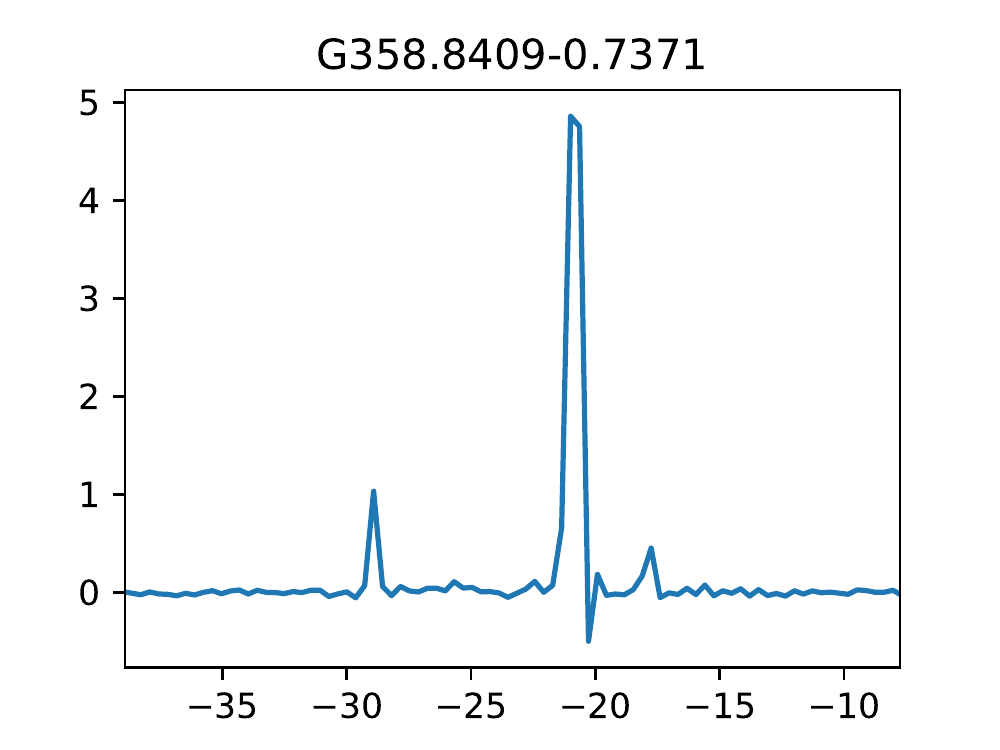}
\includegraphics[trim={0.0cm 0.3cm 0.6cm 0.3cm}, clip, width = 0.235\textwidth]{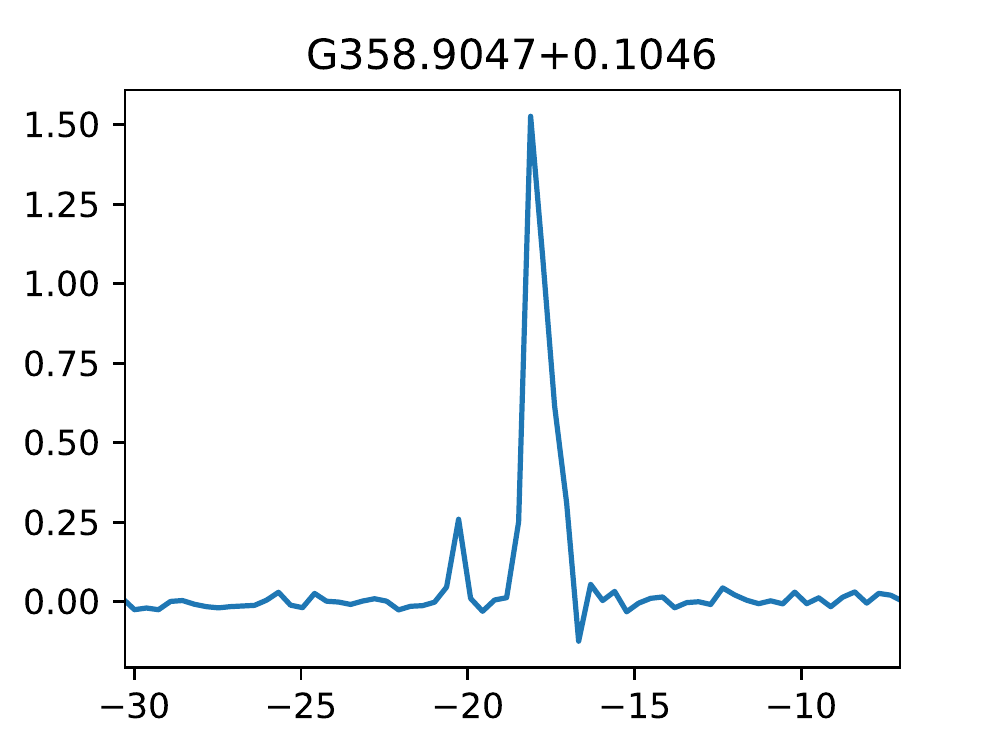}
\includegraphics[trim={0.0cm 0.3cm 0.6cm 0.3cm}, clip, width = 0.235\textwidth]{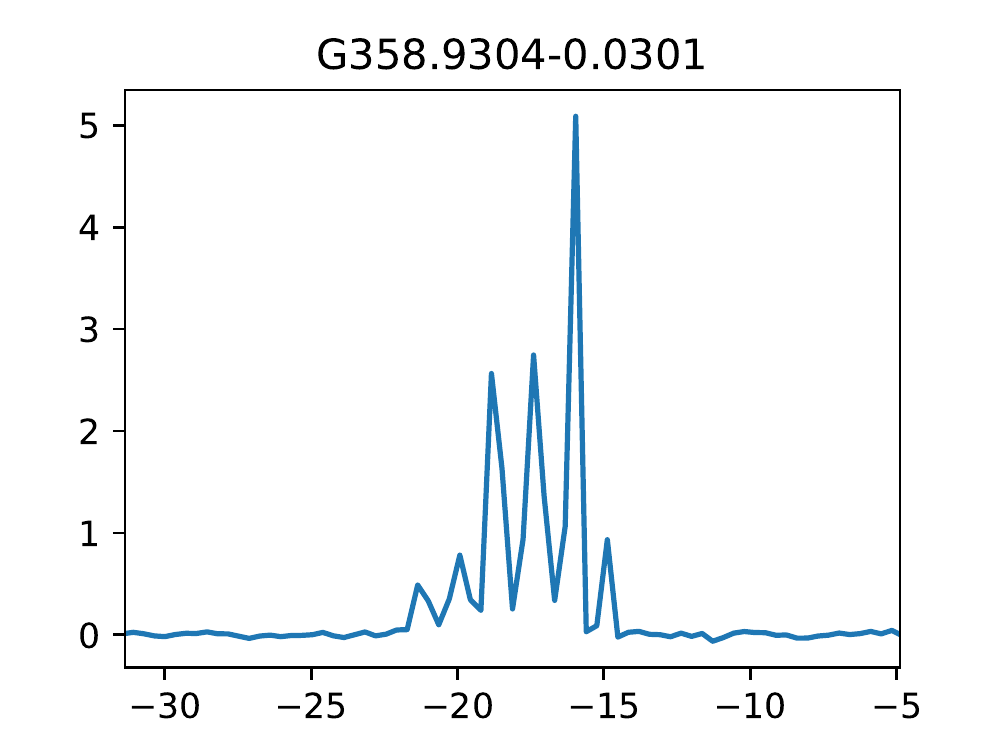}\\ 
\colorbox{white}{\includegraphics[trim={2cm 0 0 0}, scale=0.20]{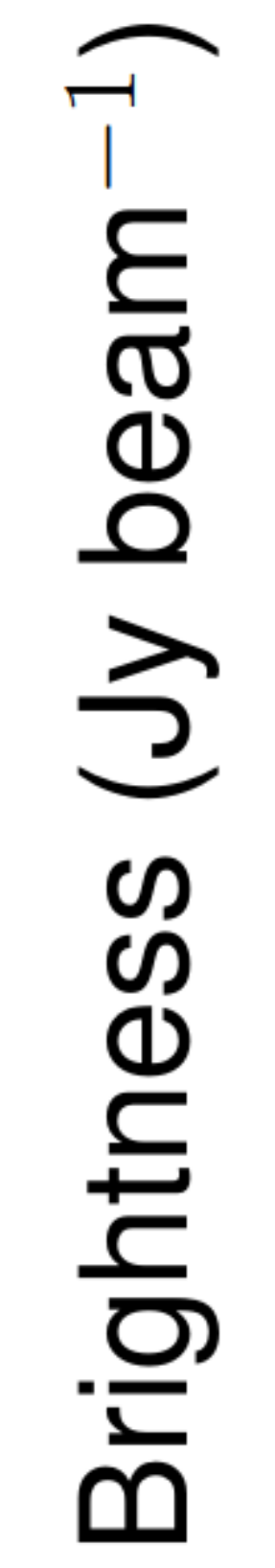}}
\includegraphics[trim={0.0cm 0.3cm 0.6cm 0.3cm}, clip, width = 0.235\textwidth]{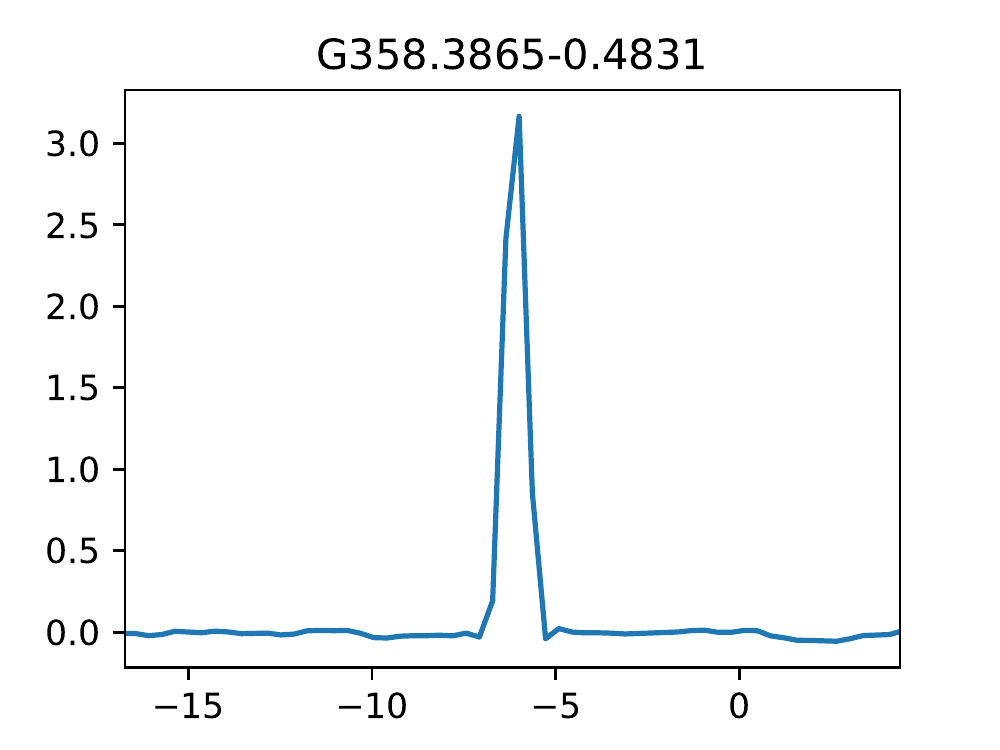}
\includegraphics[trim={0.0cm 0.3cm 0.6cm 0.3cm}, clip, width = 0.235\textwidth]{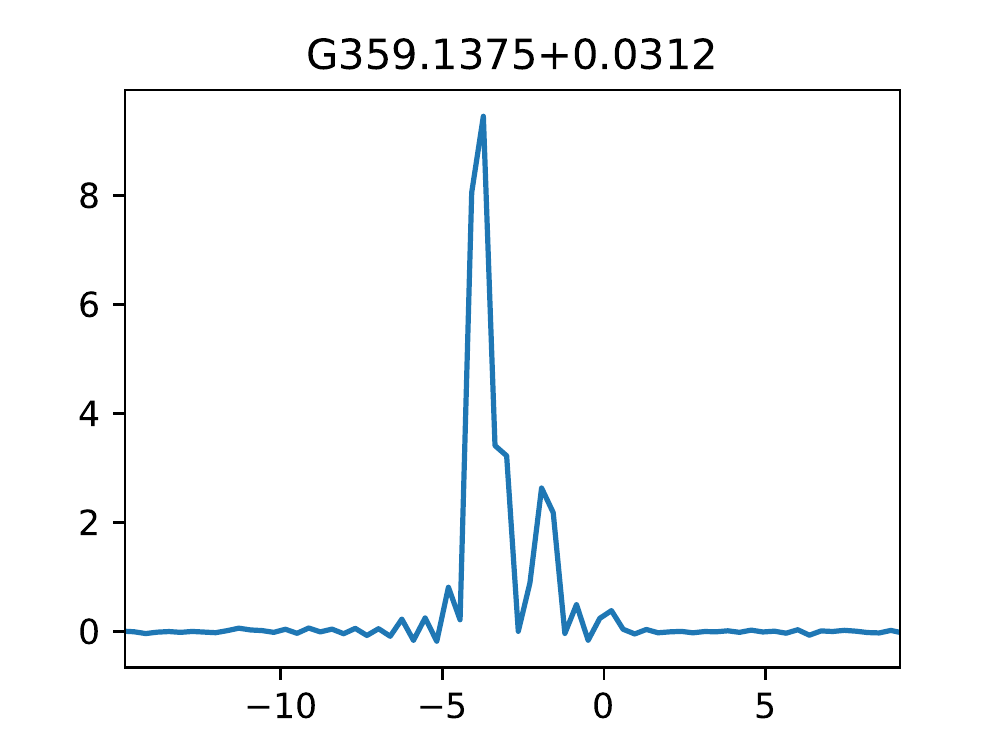}
\includegraphics[trim={0.0cm 0.3cm 0.6cm 0.3cm}, clip, width = 0.235\textwidth]{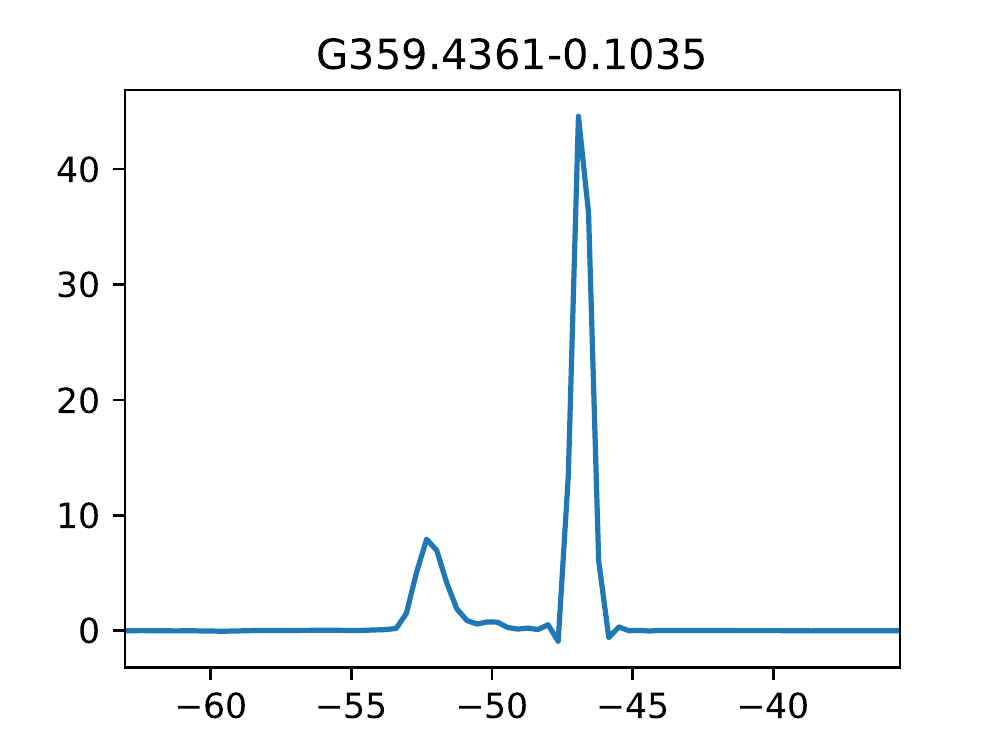}
\includegraphics[trim={0.0cm 0.3cm 0.6cm 0.3cm}, clip, width = 0.235\textwidth]{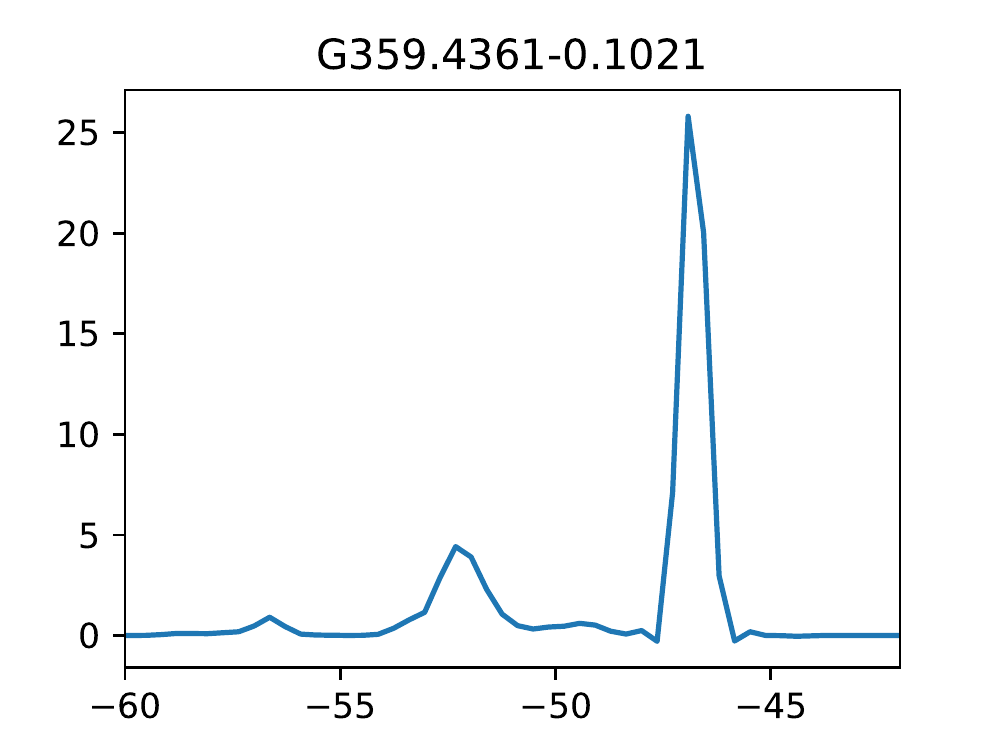}\\
\end{tabular}
\caption{Spectra of 6.7\,GHz methanol masers detected with GLOSTAR extracted at the peak pixel. The red dashed line indicates the ATLASGAL clump velocity \citep{urquhart2018,urquhart2022} in case of an associated 870\,$\mu$m compact source catalog (CSC) source.}
\label{fig:maser_many_spec}
\end{figure*}

%Table 2
\begin{figure*}[!h]
\begin{tabular}{ccccc}
\colorbox{white}{\includegraphics[trim={2cm 0 0 0}, scale=0.195]{figs/ylabel_filler.pdf}}
\includegraphics[trim={0.0cm 0.3cm 0.6cm 0.3cm}, clip, width = 0.235\textwidth]{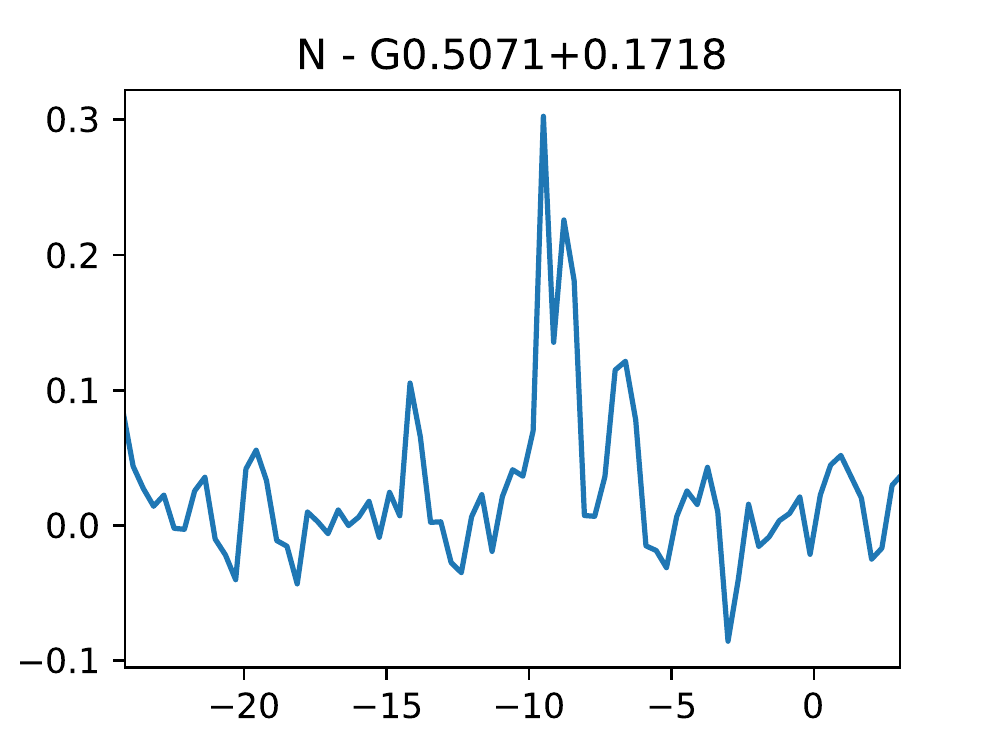}
\includegraphics[trim={0.0cm 0.3cm 0.6cm 0.3cm}, clip, width = 0.235\textwidth]{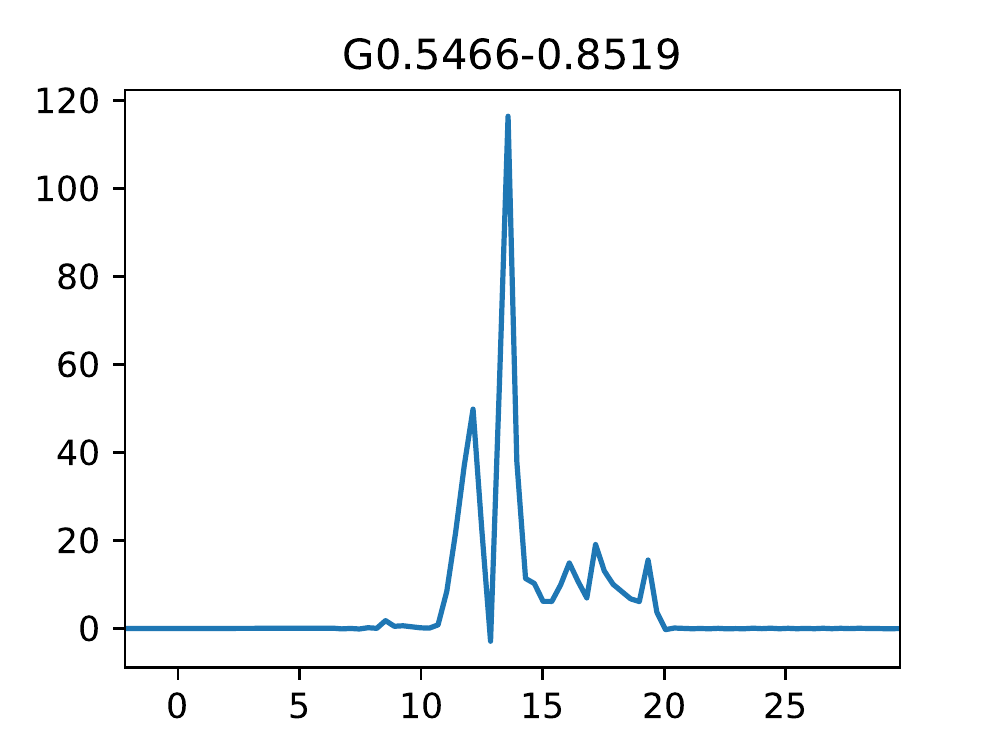}
\includegraphics[trim={0.0cm 0.3cm 0.6cm 0.3cm}, clip, width = 0.235\textwidth]{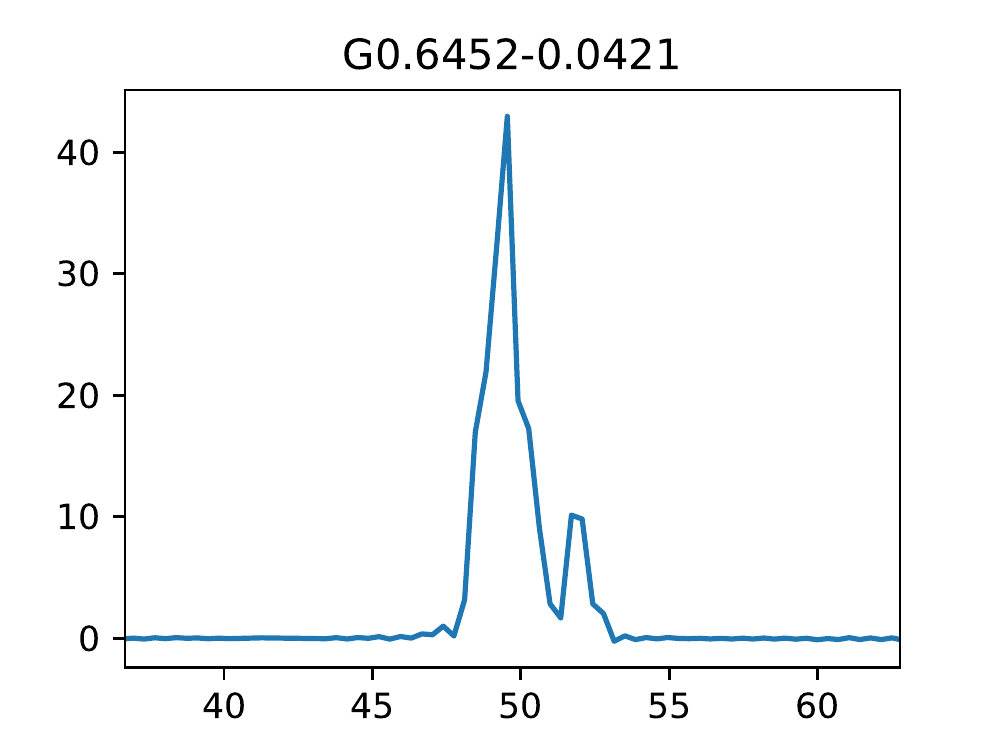}
\includegraphics[trim={0.0cm 0.3cm 0.6cm 0.3cm}, clip, width = 0.235\textwidth]{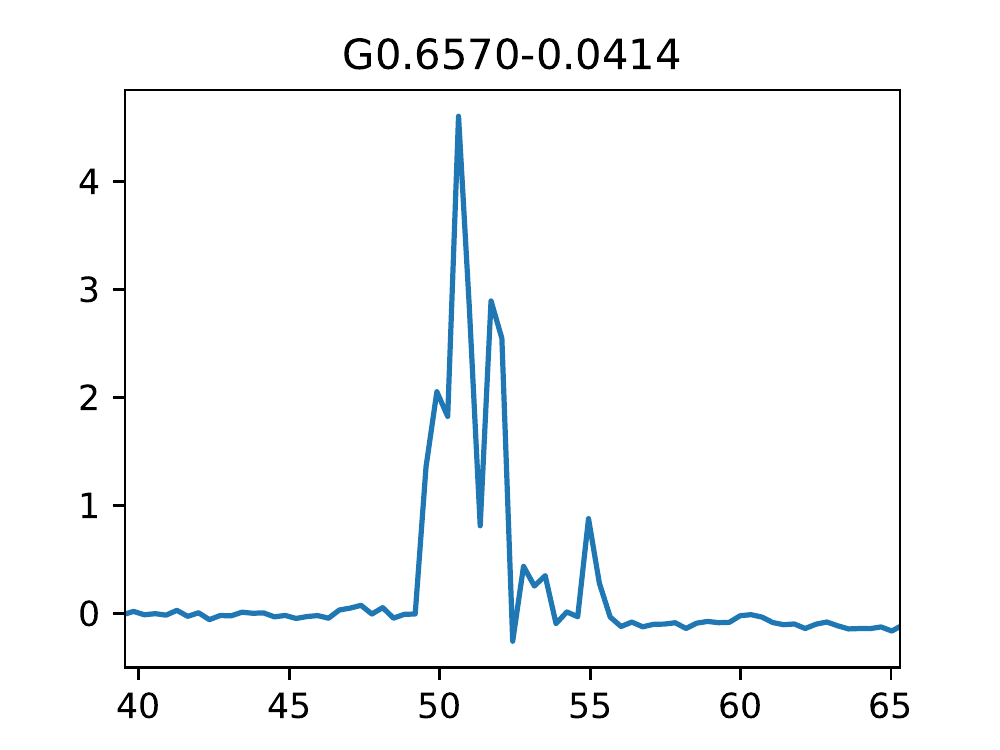}\\ 
\colorbox{white}{\includegraphics[trim={2cm 0 0 0}, scale=0.195]{figs/ylabel_filler.pdf}}
\includegraphics[trim={0.0cm 0.3cm 0.6cm 0.3cm}, clip, width = 0.235\textwidth]{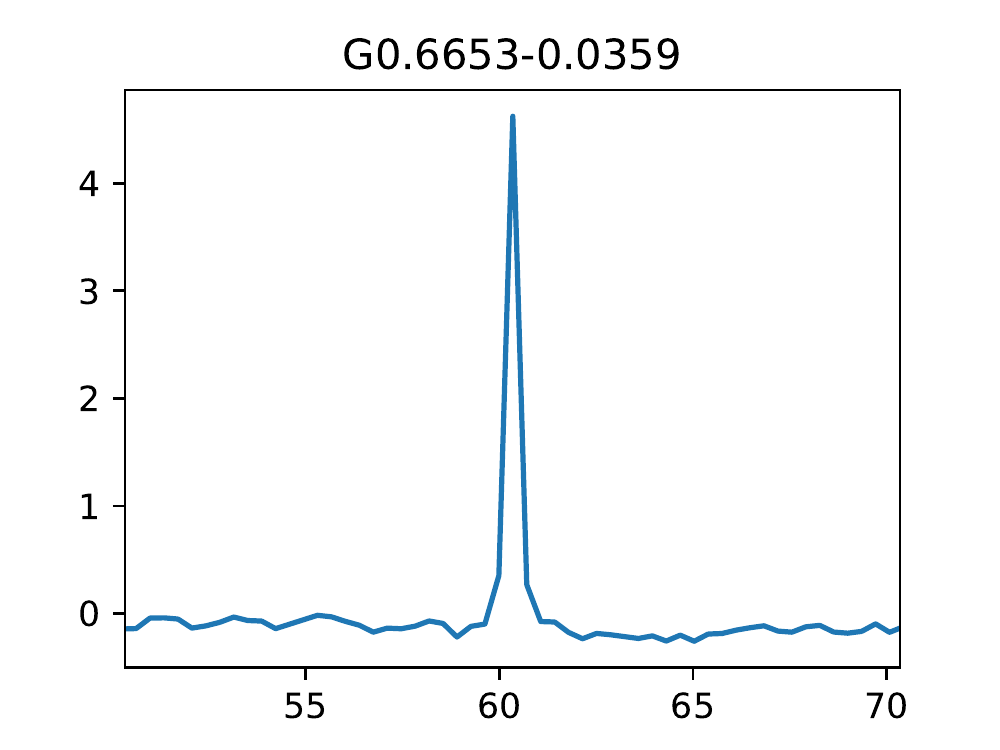}
\includegraphics[trim={0.0cm 0.3cm 0.6cm 0.3cm}, clip, width = 0.235\textwidth]{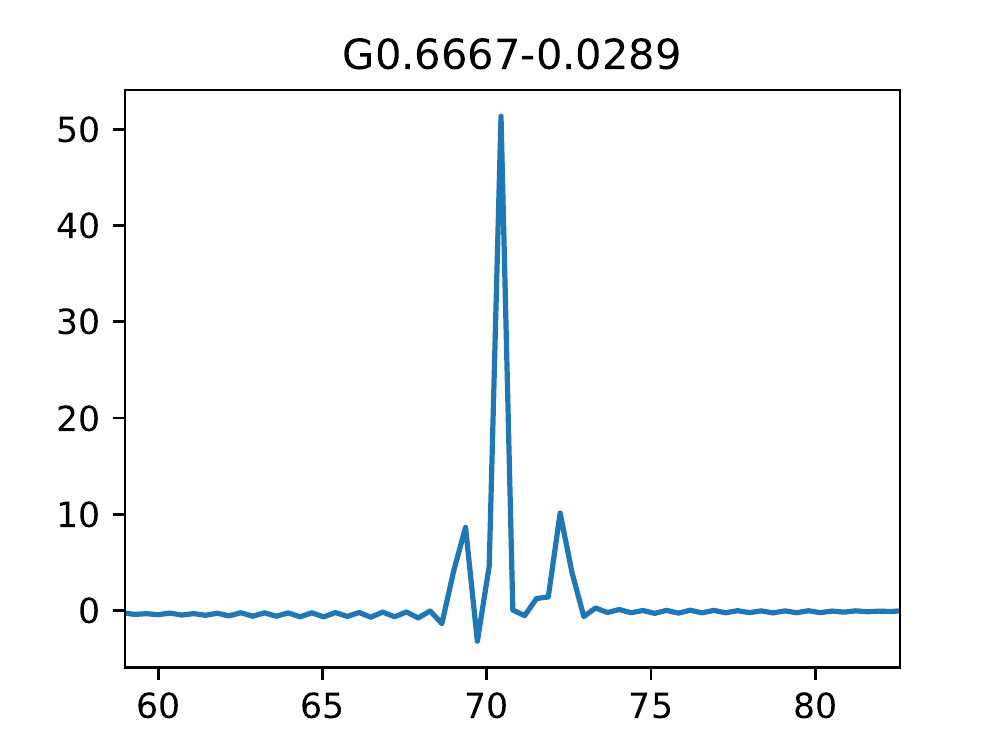}
\includegraphics[trim={0.0cm 0.3cm 0.6cm 0.3cm}, clip, width = 0.235\textwidth]{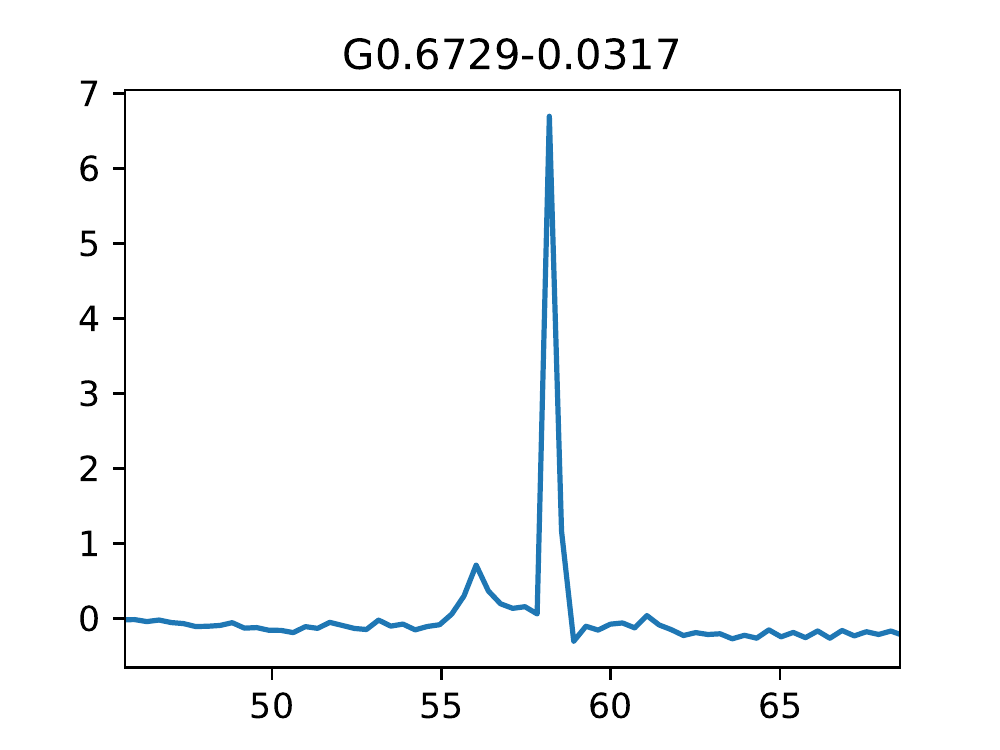}
\includegraphics[trim={0.0cm 0.3cm 0.6cm 0.3cm}, clip, width = 0.235\textwidth]{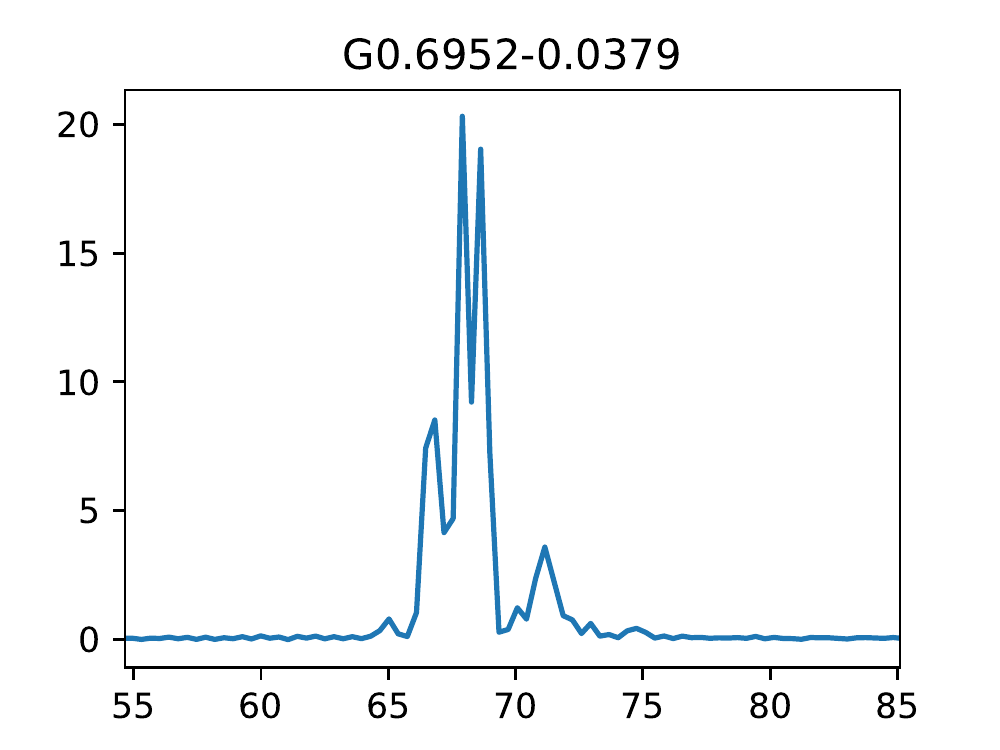}\\ 
\colorbox{white}{\includegraphics[trim={2cm 0 0 0}, scale=0.20]{figs/ylabel.pdf}}
\includegraphics[trim={0.0cm 0.3cm 0.6cm 0.3cm}, clip, width = 0.235\textwidth]{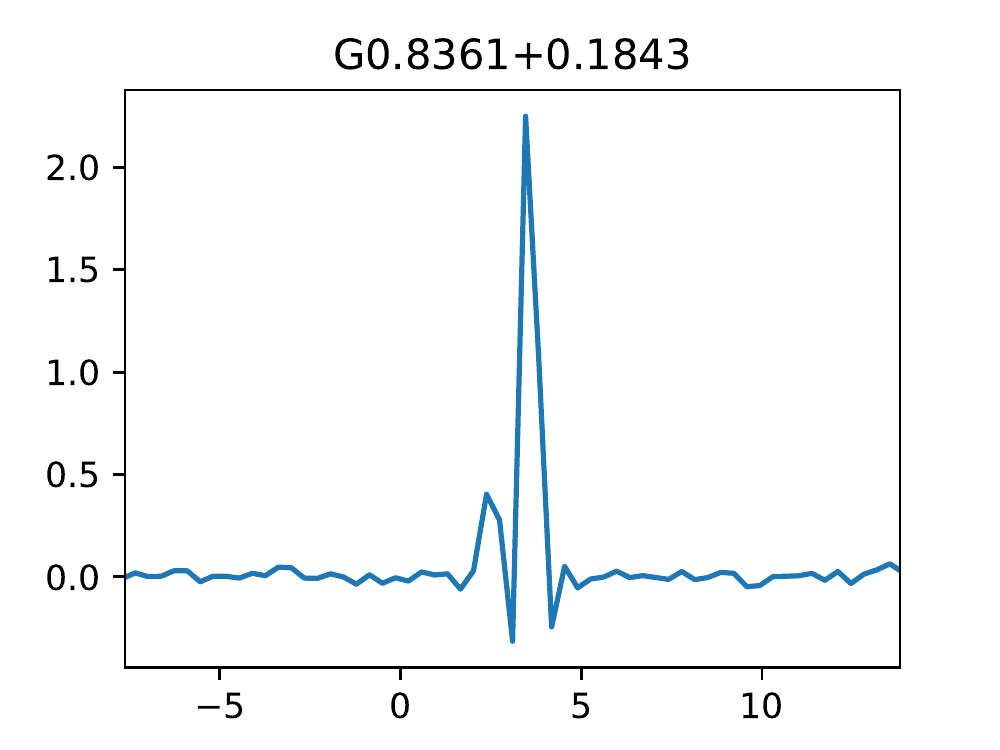}
\includegraphics[trim={0.0cm 0.3cm 0.6cm 0.3cm}, clip, width = 0.235\textwidth]{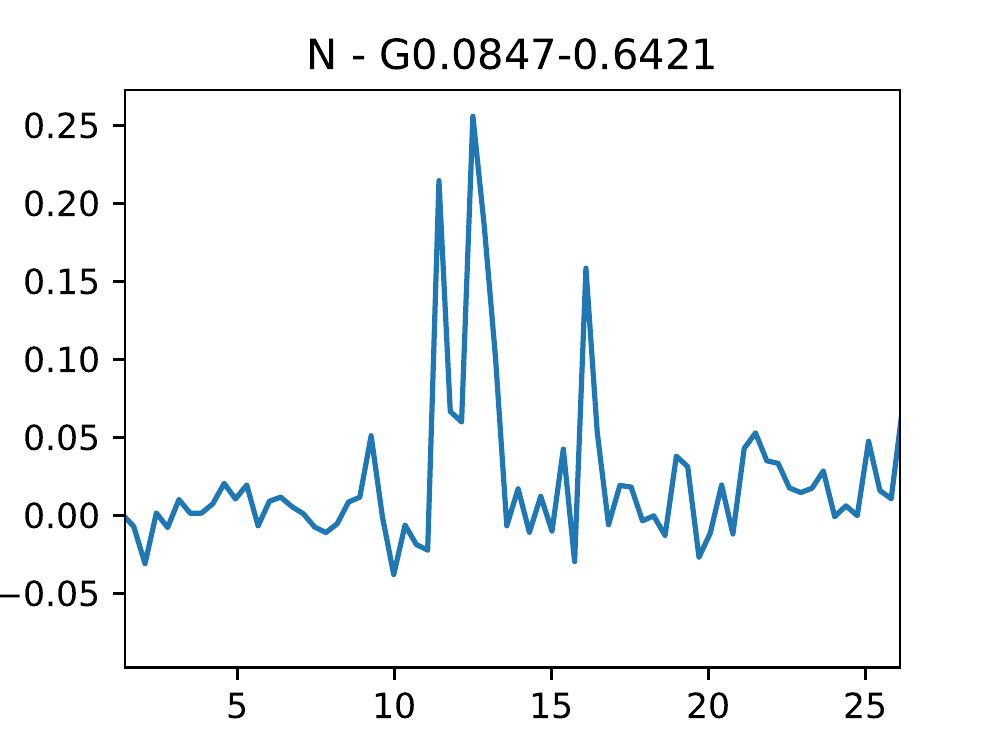}
\includegraphics[trim={0.0cm 0.3cm 0.6cm 0.3cm}, clip, width = 0.235\textwidth]{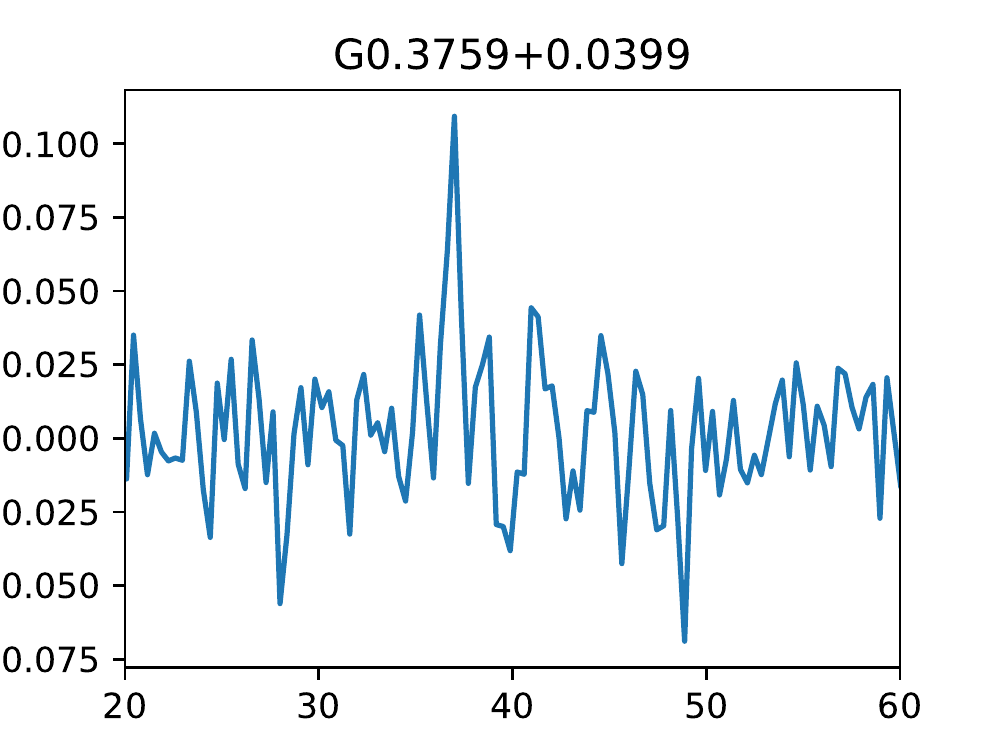}
\includegraphics[trim={0.0cm 0.3cm 0.6cm 0.3cm}, clip, width = 0.235\textwidth]{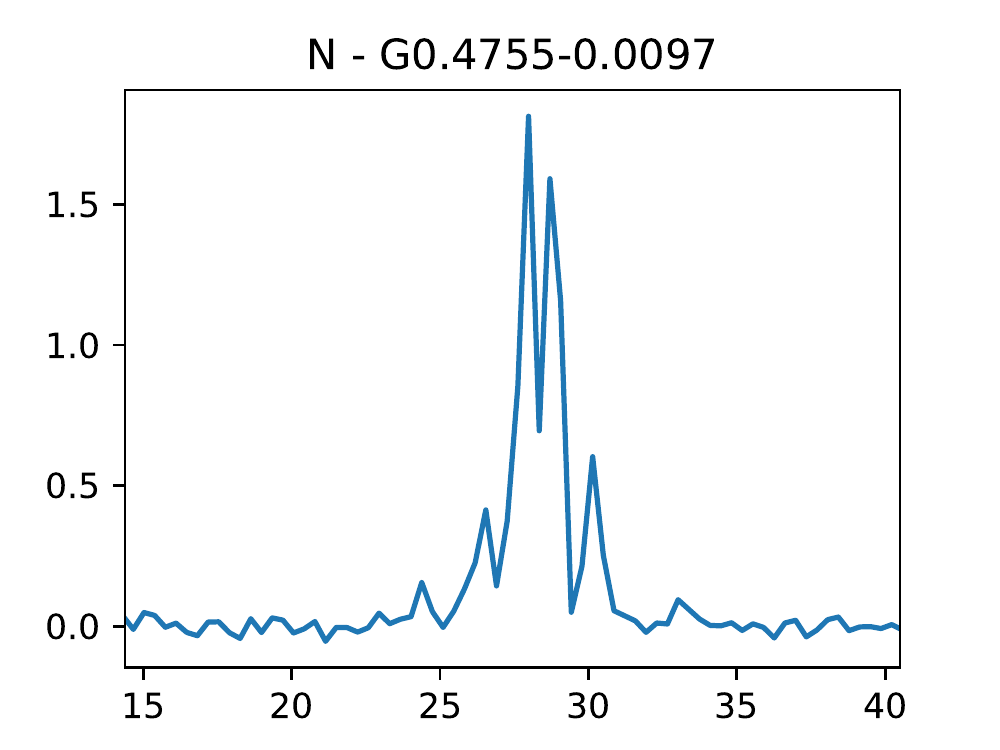}\\ 
\colorbox{white}{\includegraphics[trim={2cm 0 0 0}, scale=0.195]{figs/ylabel_filler.pdf}}
\includegraphics[trim={0.0cm 0.3cm 0.6cm 0.3cm}, clip, width = 0.235\textwidth]{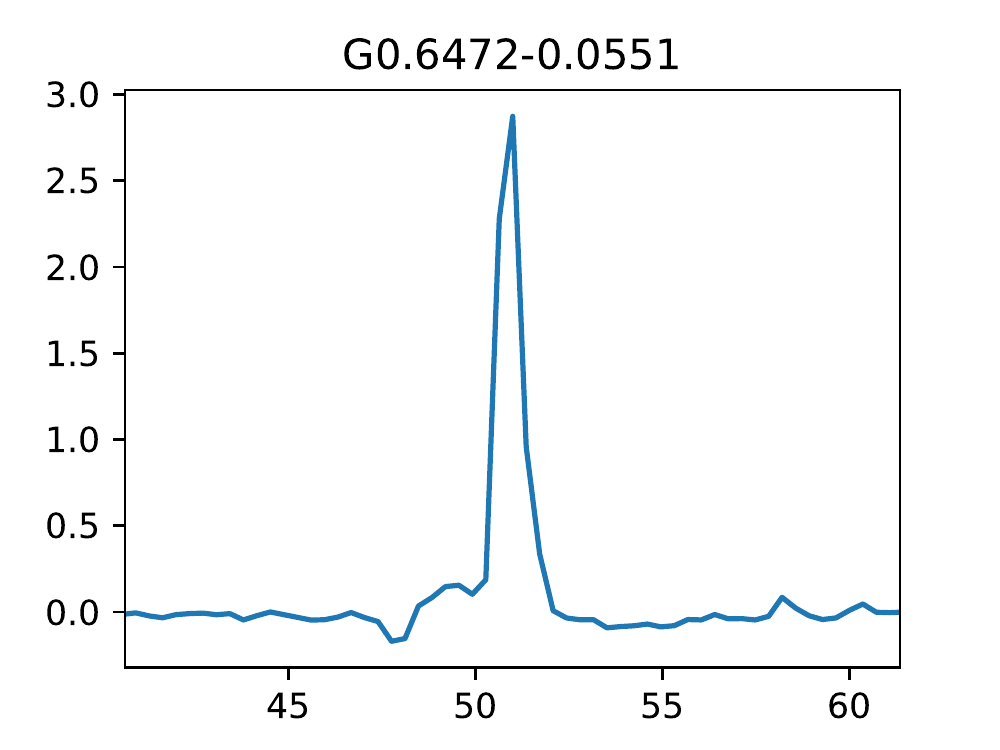}
\includegraphics[trim={0.0cm 0.3cm 0.6cm 0.3cm}, clip, width = 0.235\textwidth]{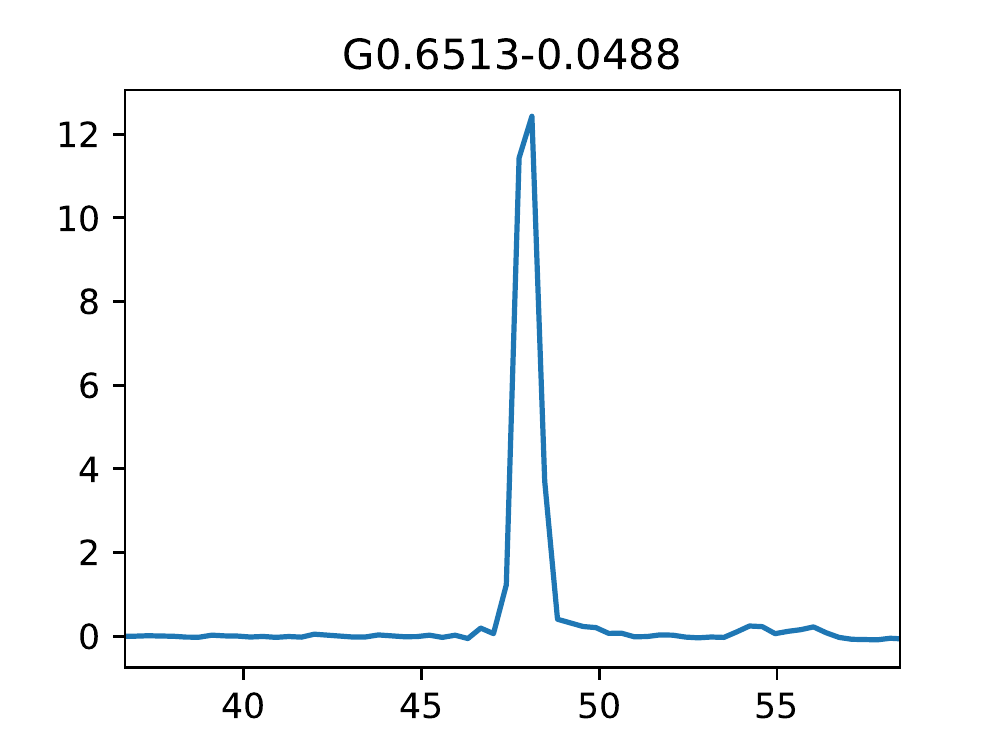}
\includegraphics[trim={0.0cm 0.3cm 0.6cm 0.3cm}, clip, width = 0.235\textwidth]{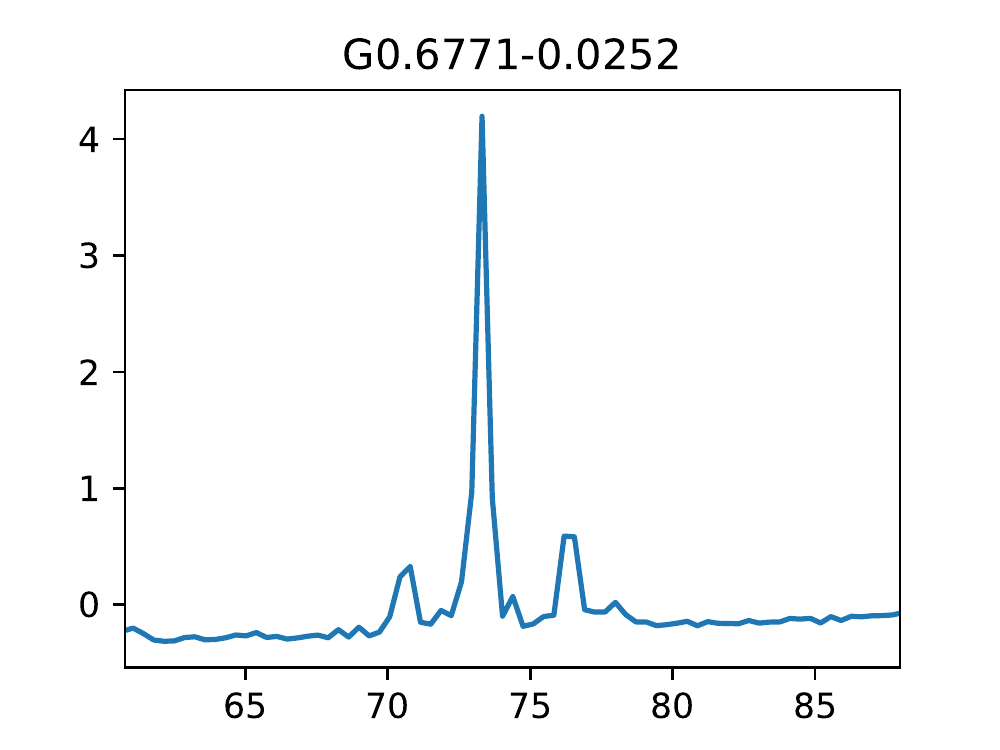}
\includegraphics[trim={0.0cm 0.3cm 0.6cm 0.3cm}, clip, width = 0.235\textwidth]{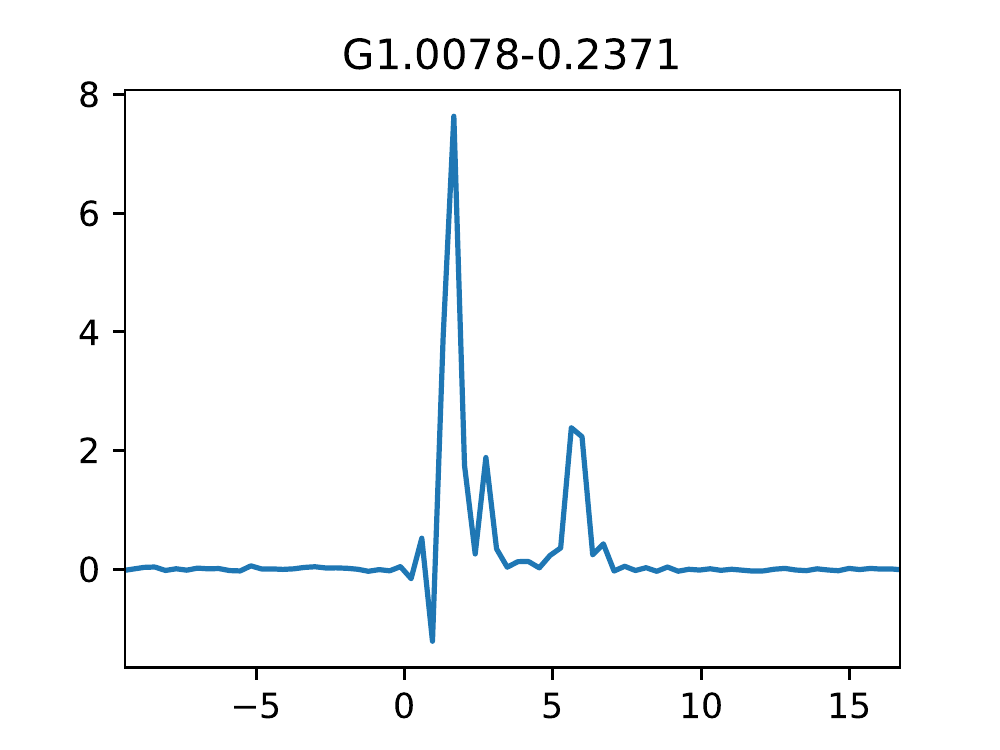}\\
\end{tabular}
\caption{Continued from Fig.~\ref{fig:maser_many_spec}.}
\label{fig:masers_48}
\end{figure*}

%Table 3
\begin{figure*}[!h]
\begin{tabular}{ccccc}
\colorbox{white}{\includegraphics[trim={2cm 0 0 0}, scale=0.195]{figs/ylabel_filler.pdf}}
\includegraphics[trim={0.0cm 0.3cm 0.6cm 0.3cm}, clip, width = 0.235\textwidth]{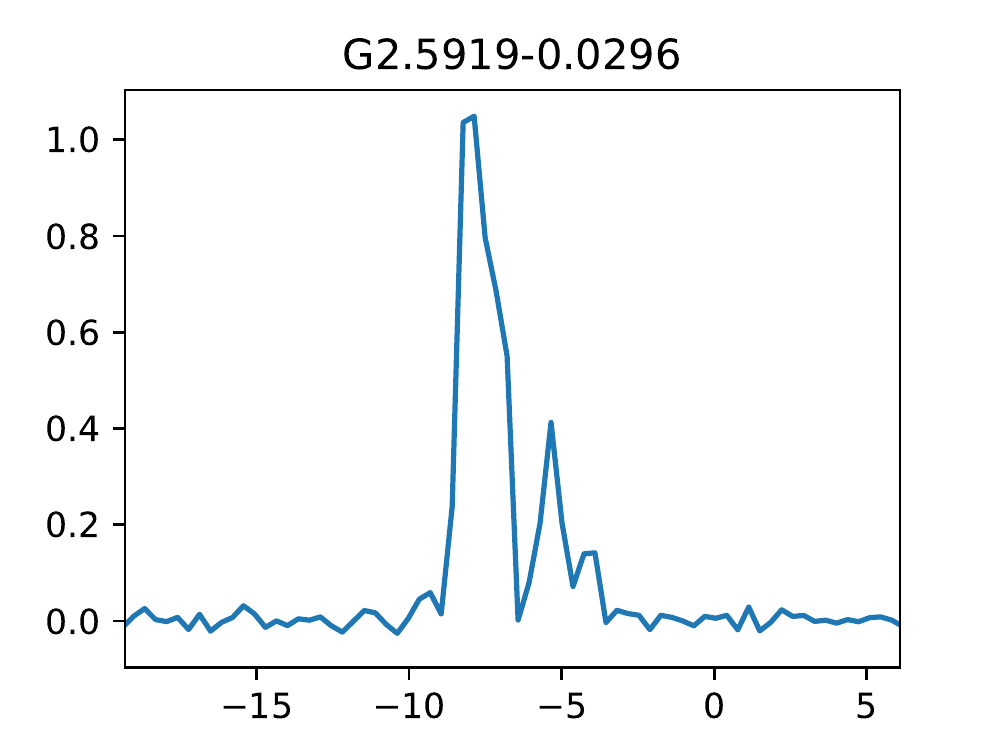}
\includegraphics[trim={0.0cm 0.3cm 0.6cm 0.3cm}, clip, width = 0.235\textwidth]{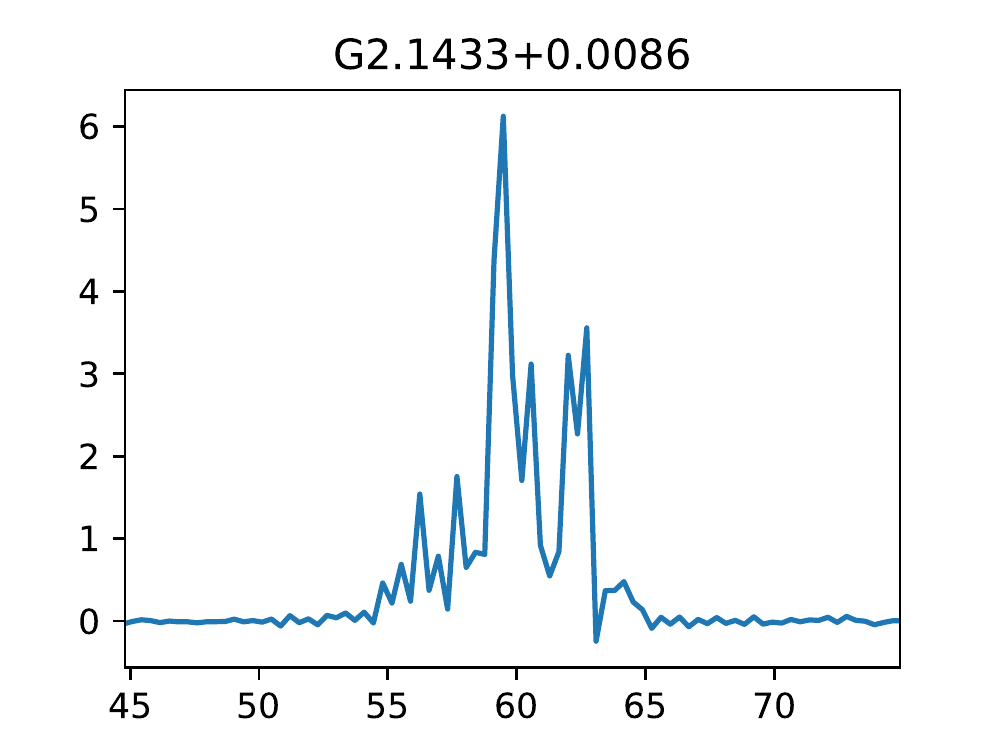}
\includegraphics[trim={0.0cm 0.3cm 0.6cm 0.3cm}, clip, width = 0.235\textwidth]{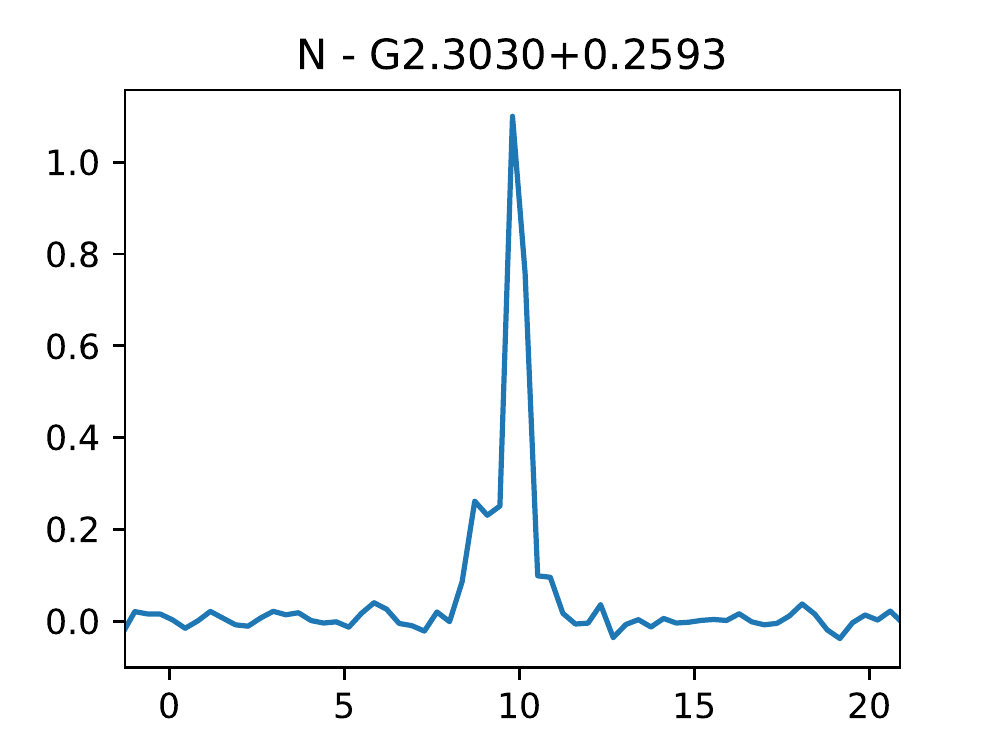}
\includegraphics[trim={0.0cm 0.3cm 0.6cm 0.3cm}, clip, width = 0.235\textwidth]{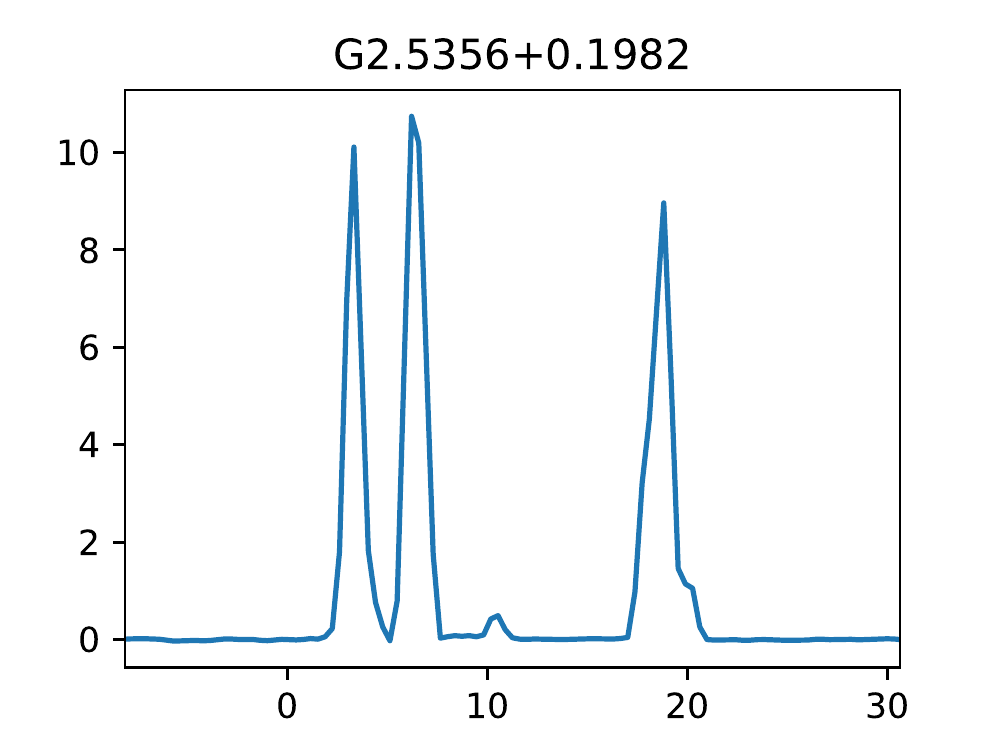}\\ 
\colorbox{white}{\includegraphics[trim={2cm 0 0 0}, scale=0.195]{figs/ylabel_filler.pdf}}
\includegraphics[trim={0.0cm 0.3cm 0.6cm 0.3cm}, clip, width = 0.235\textwidth]{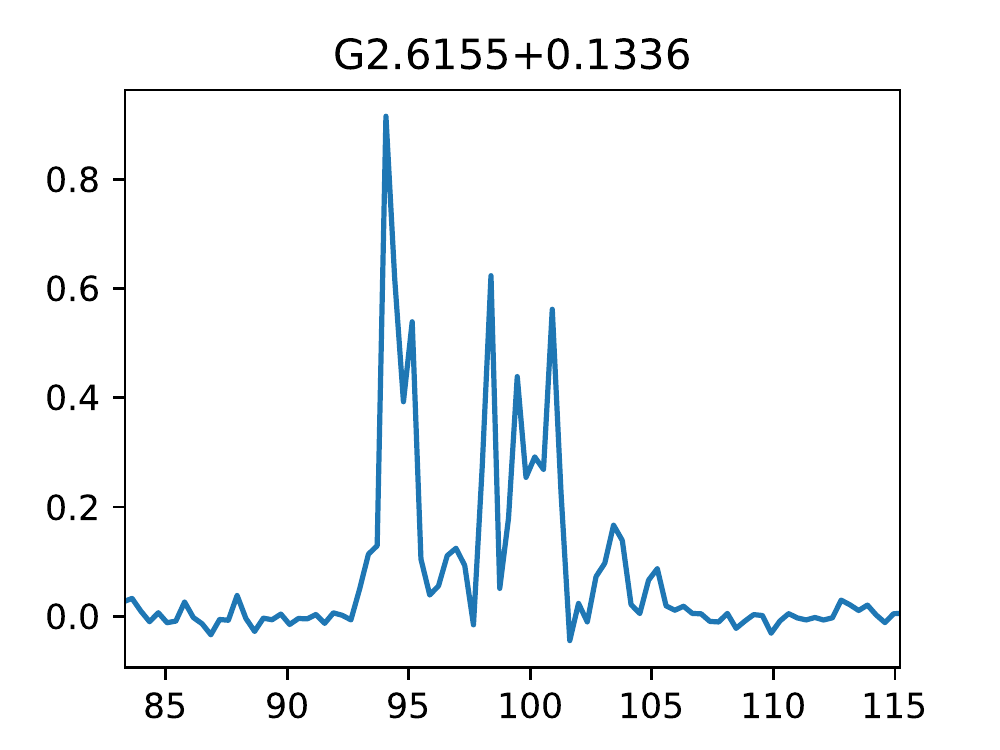}
\includegraphics[trim={0.0cm 0.3cm 0.6cm 0.3cm}, clip, width = 0.235\textwidth]{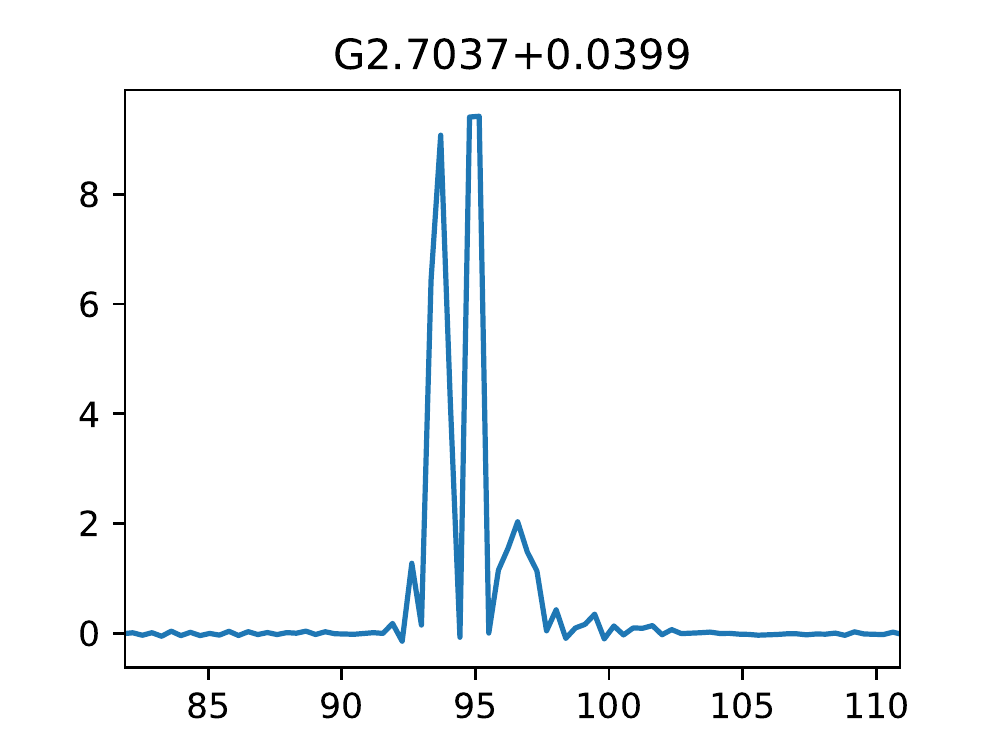}
\includegraphics[trim={0.0cm 0.3cm 0.6cm 0.3cm}, clip, width = 0.235\textwidth]{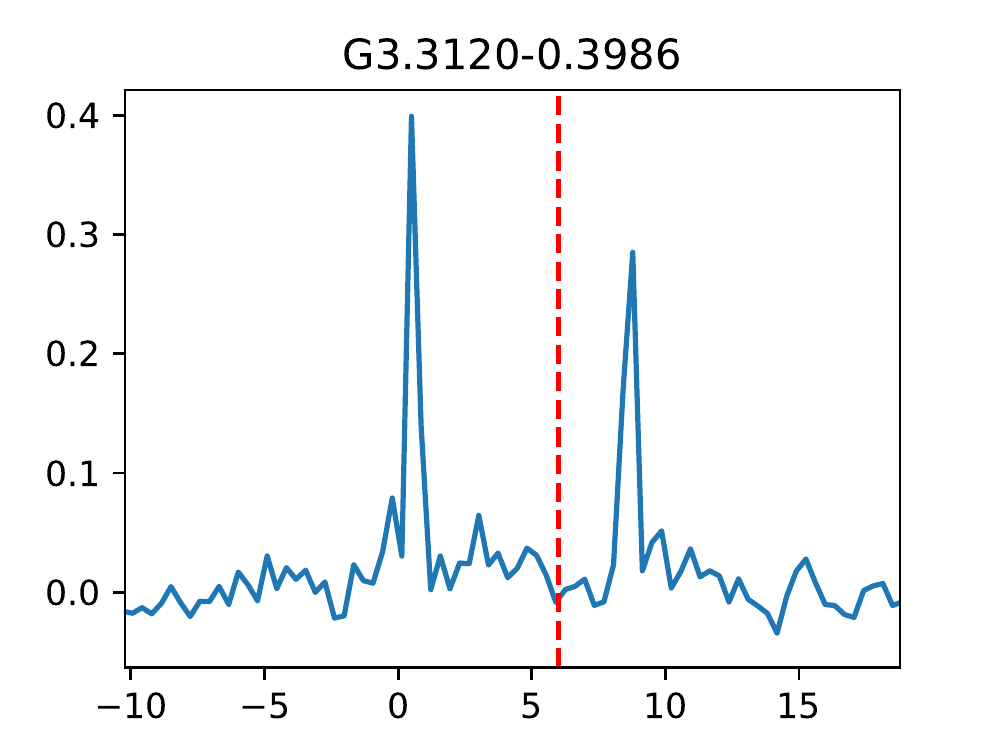}
\includegraphics[trim={0.0cm 0.3cm 0.6cm 0.3cm}, clip, width = 0.235\textwidth]{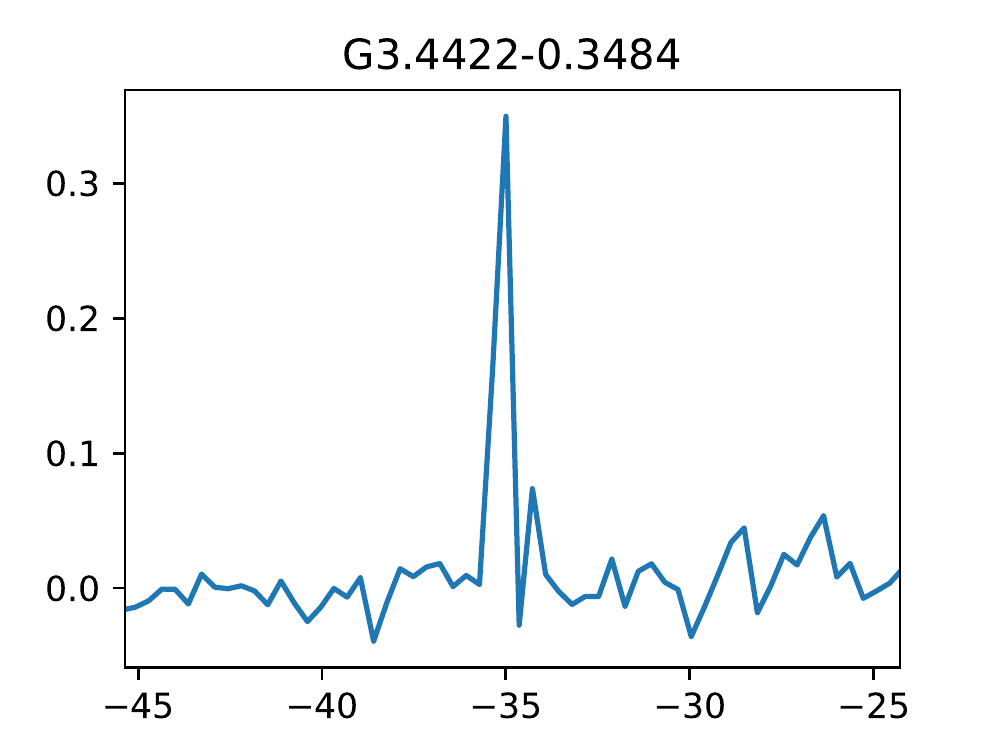}\\ 
\colorbox{white}{\includegraphics[trim={2cm 0 0 0}, scale=0.20]{figs/ylabel.pdf}}
\includegraphics[trim={0.0cm 0.3cm 0.6cm 0.3cm}, clip, width = 0.235\textwidth]{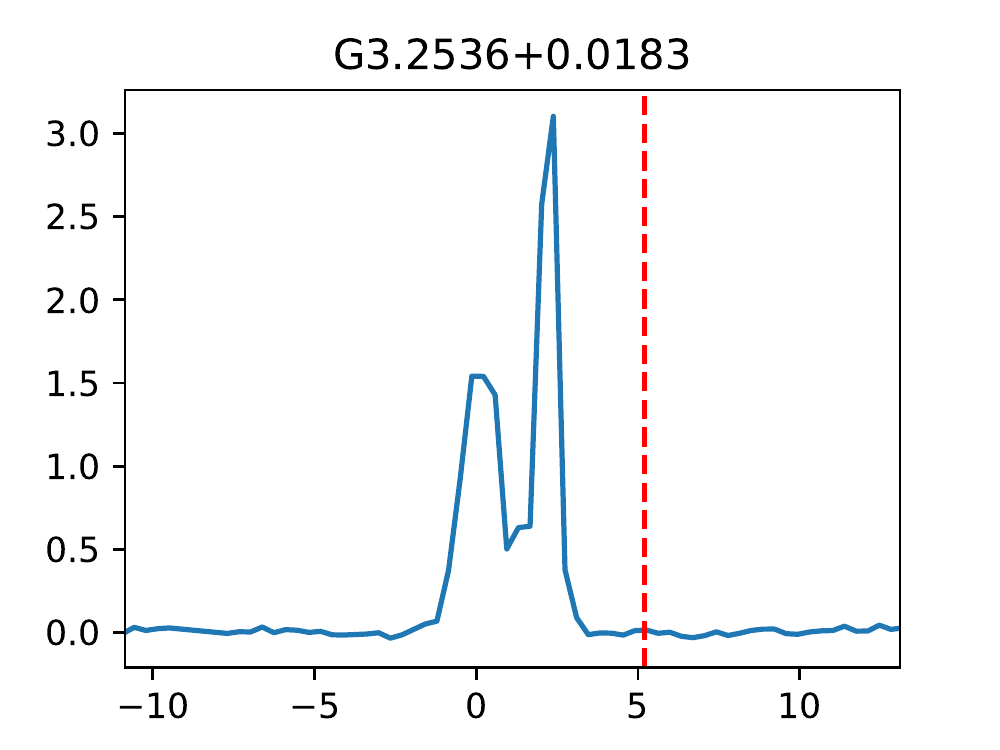}
\includegraphics[trim={0.0cm 0.3cm 0.6cm 0.3cm}, clip, width = 0.235\textwidth]{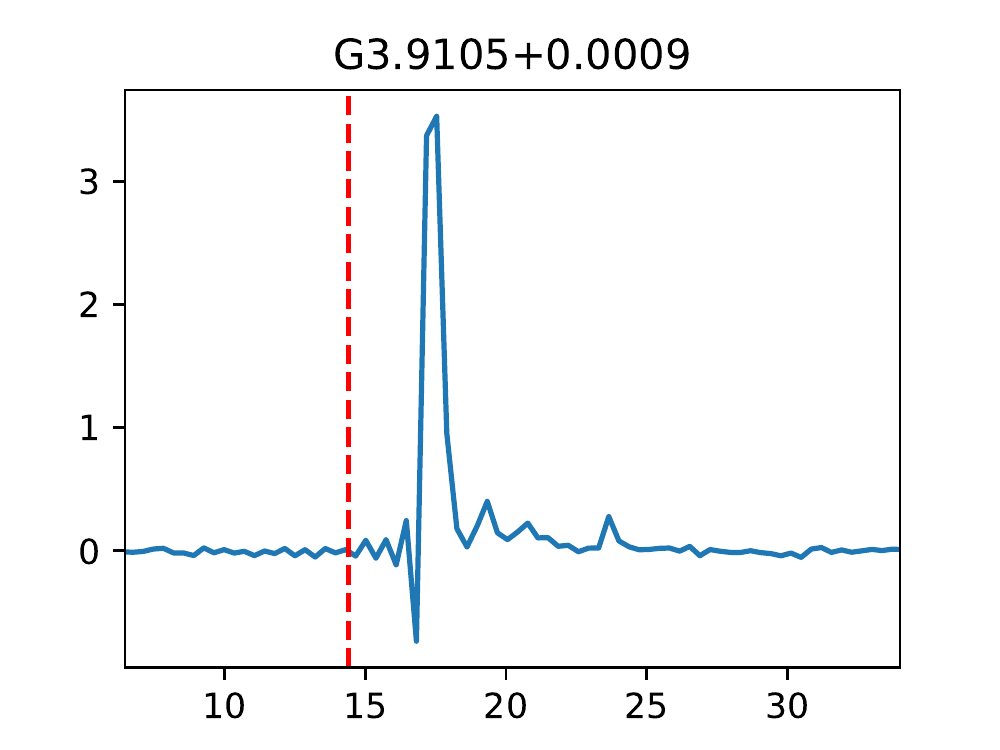}
\includegraphics[trim={0.0cm 0.3cm 0.6cm 0.3cm}, clip, width = 0.235\textwidth]{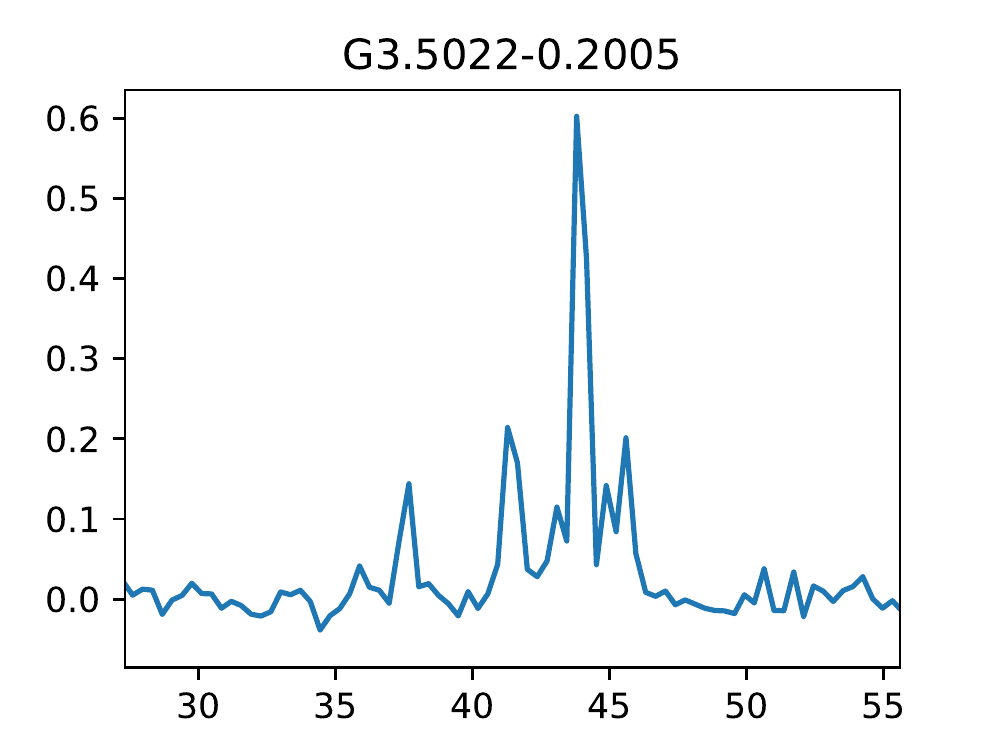}
\includegraphics[trim={0.0cm 0.3cm 0.6cm 0.3cm}, clip, width = 0.235\textwidth]{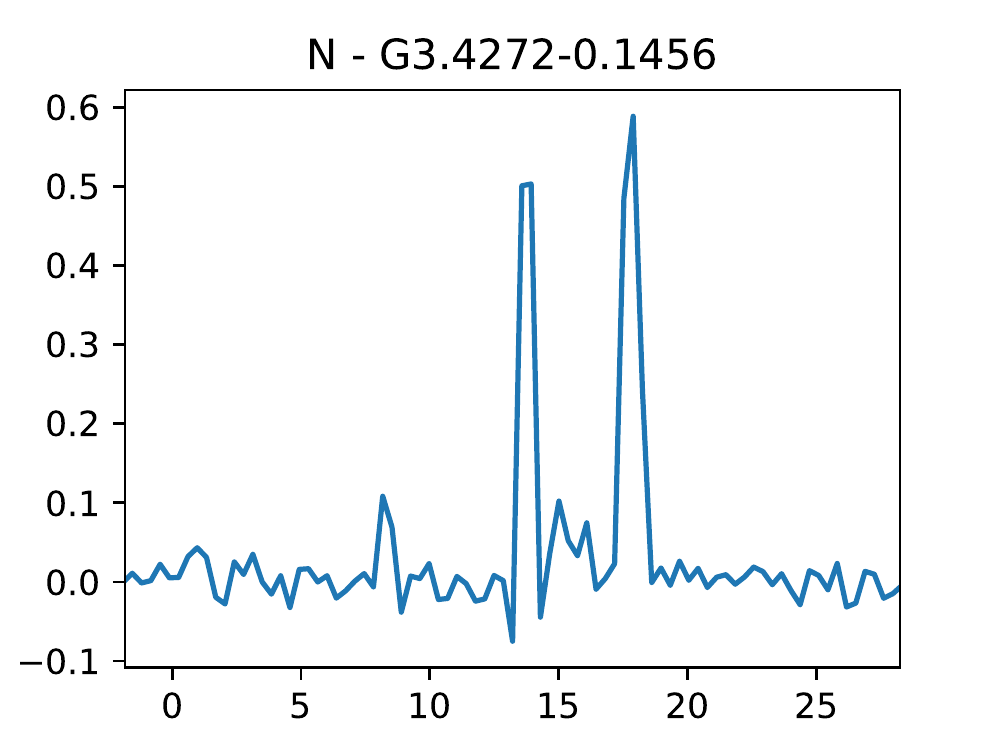}\\ 
\colorbox{white}{\includegraphics[trim={2cm 0 0 0}, scale=0.195]{figs/ylabel_filler.pdf}}
\includegraphics[trim={0.0cm 0.3cm 0.6cm 0.3cm}, clip, width = 0.235\textwidth]{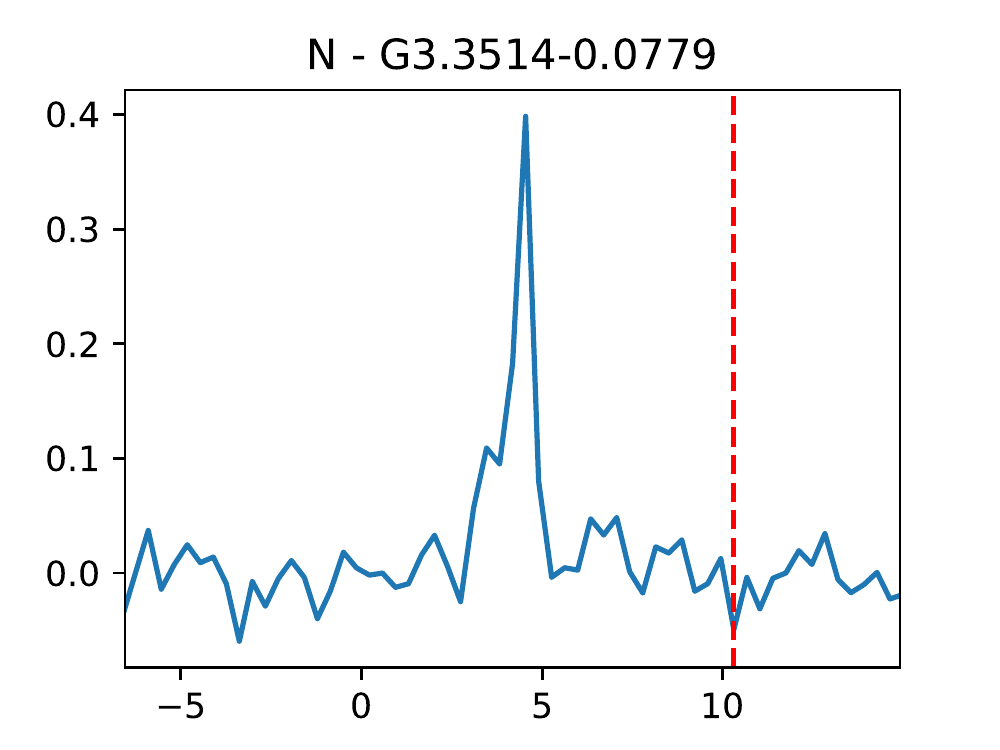}
\includegraphics[trim={0.0cm 0.3cm 0.6cm 0.3cm}, clip, width = 0.235\textwidth]{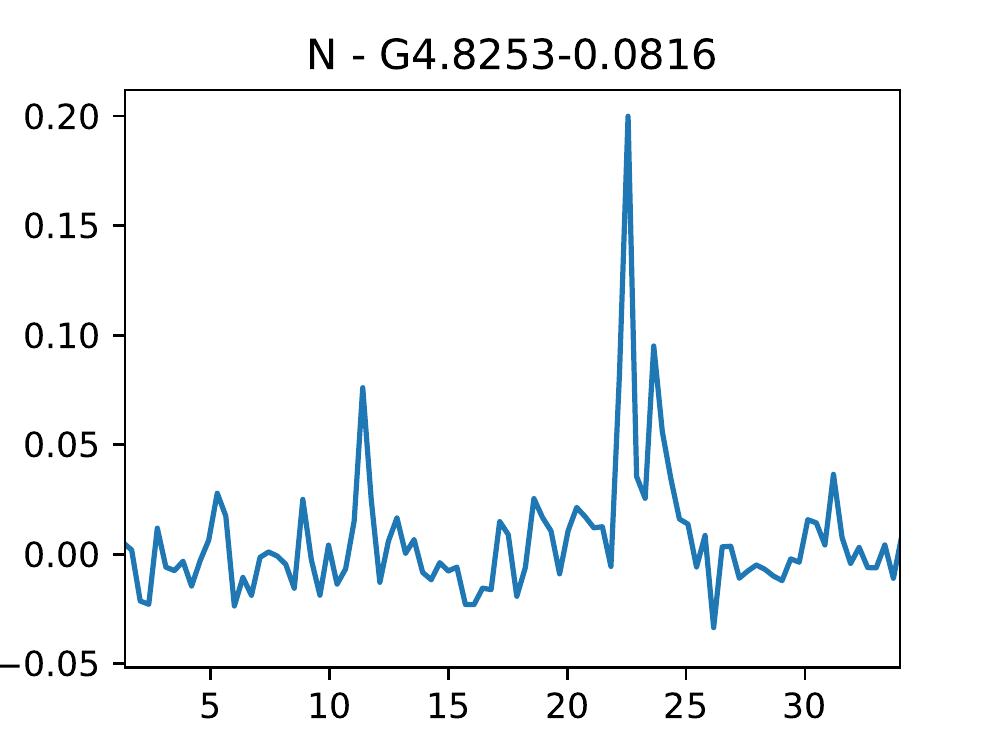}
\includegraphics[trim={0.0cm 0.3cm 0.6cm 0.3cm}, clip, width = 0.235\textwidth]{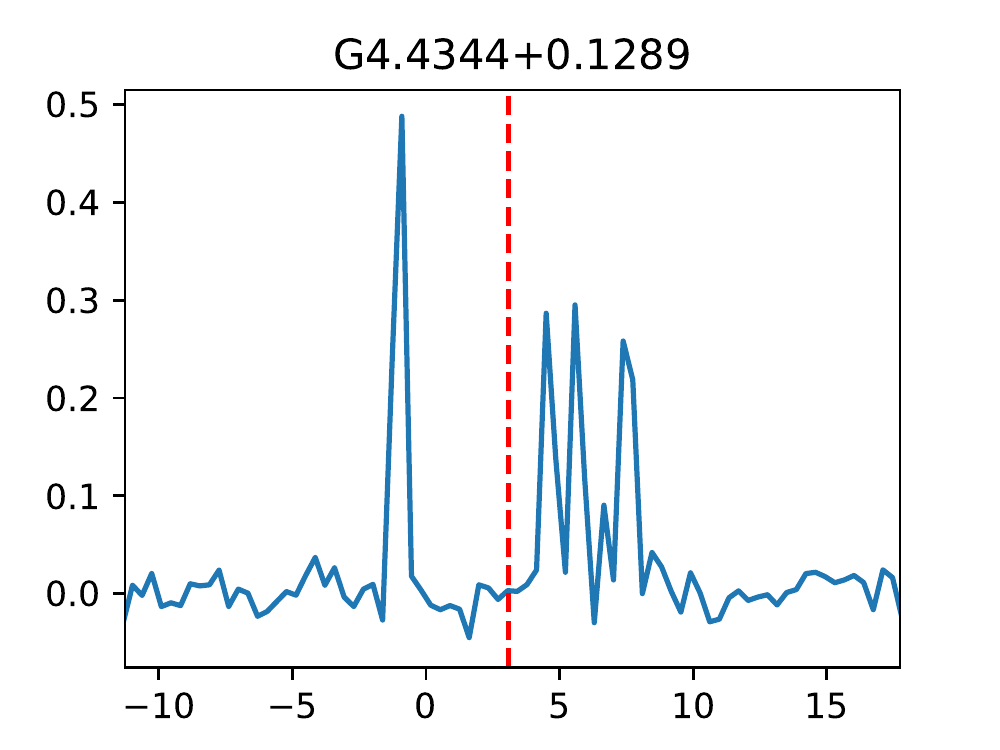}
\includegraphics[trim={0.0cm 0.3cm 0.6cm 0.3cm}, clip, width = 0.235\textwidth]{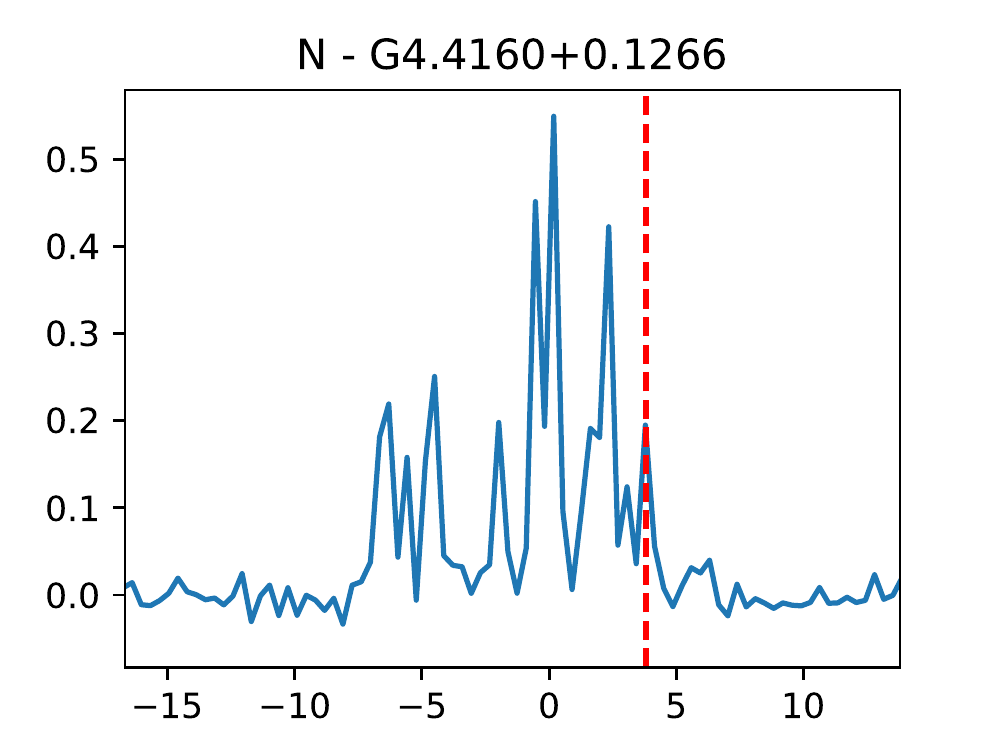}\\
\end{tabular}
\caption{Continued from Fig.~\ref{fig:maser_many_spec}.}
\label{fig:masers_72}
\end{figure*}

%Table 4
\begin{figure*}[!h]
\begin{tabular}{ccccc}
\colorbox{white}{\includegraphics[trim={2cm 0 0 0}, scale=0.195]{figs/ylabel_filler.pdf}}
\includegraphics[trim={0.0cm 0.3cm 0.6cm 0.3cm}, clip, width = 0.235\textwidth]{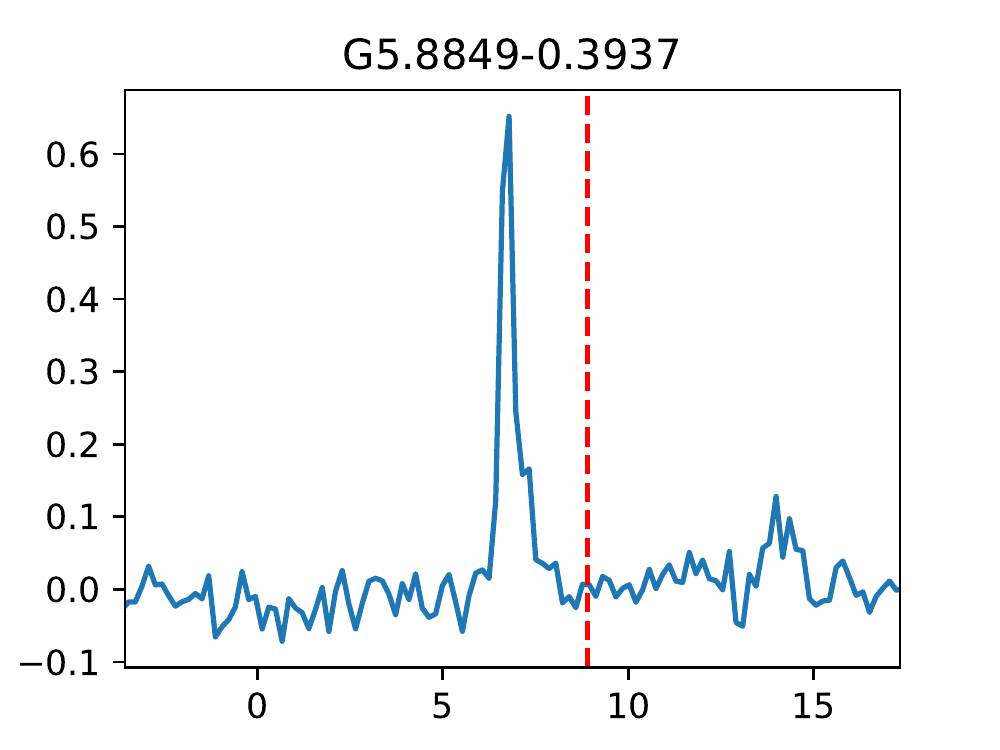}
\includegraphics[trim={0.0cm 0.3cm 0.6cm 0.3cm}, clip, width = 0.235\textwidth]{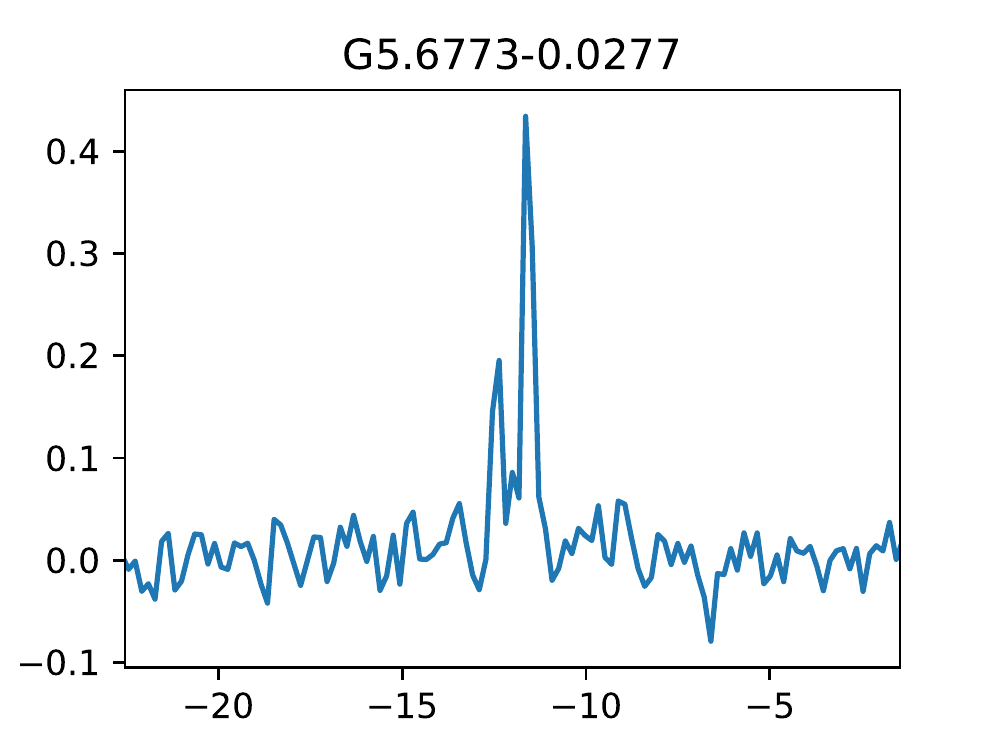}
\includegraphics[trim={0.0cm 0.3cm 0.6cm 0.3cm}, clip, width = 0.235\textwidth]{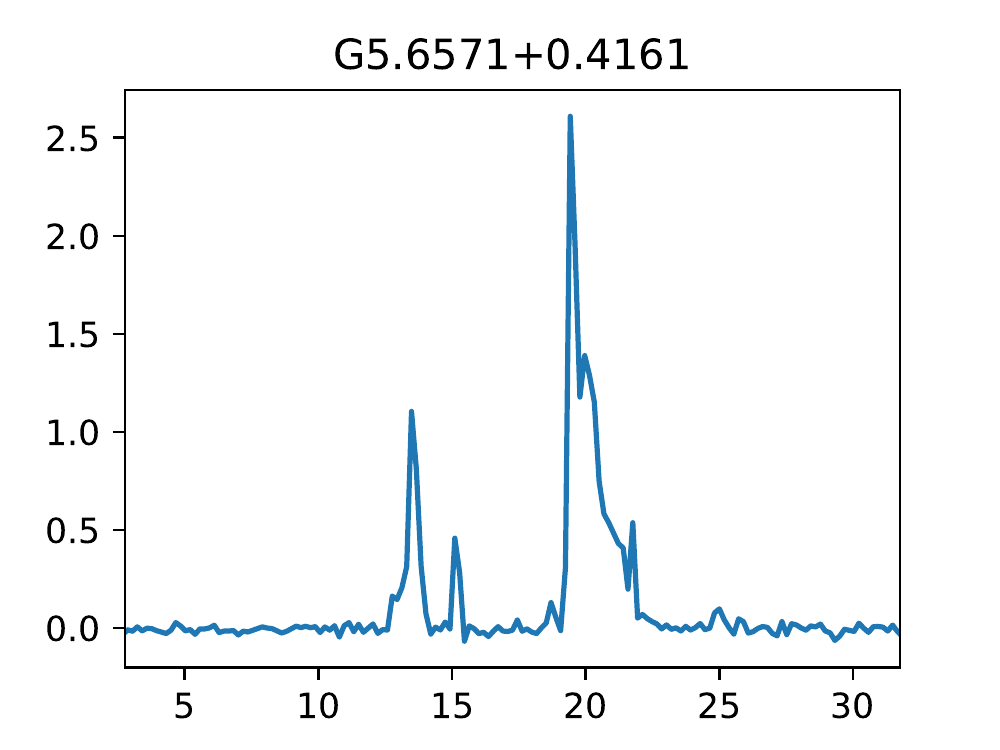}
\includegraphics[trim={0.0cm 0.3cm 0.6cm 0.3cm}, clip, width = 0.235\textwidth]{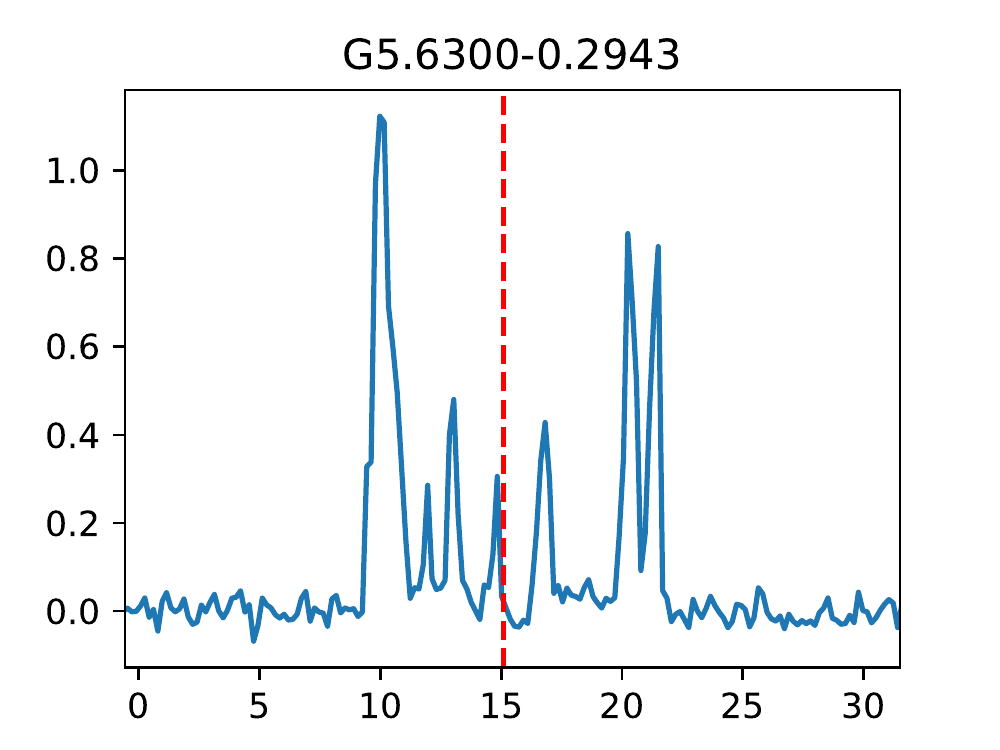}\\ 
\colorbox{white}{\includegraphics[trim={2cm 0 0 0}, scale=0.195]{figs/ylabel_filler.pdf}}
\includegraphics[trim={0.0cm 0.3cm 0.6cm 0.3cm}, clip, width = 0.235\textwidth]{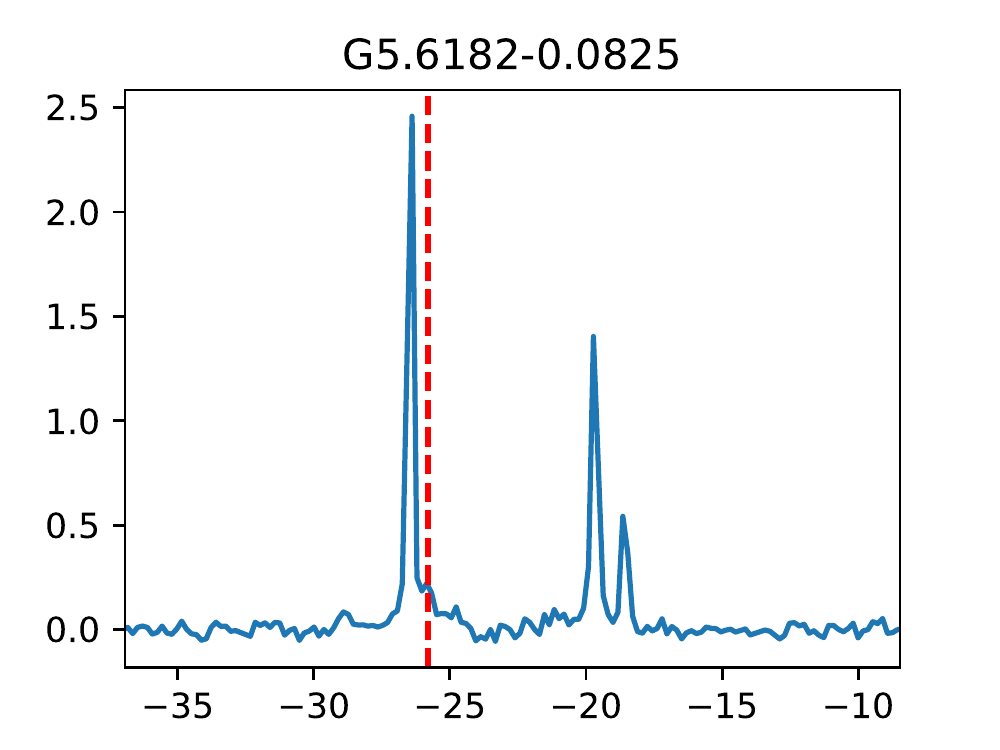}
\includegraphics[trim={0.0cm 0.3cm 0.6cm 0.3cm}, clip, width = 0.235\textwidth]{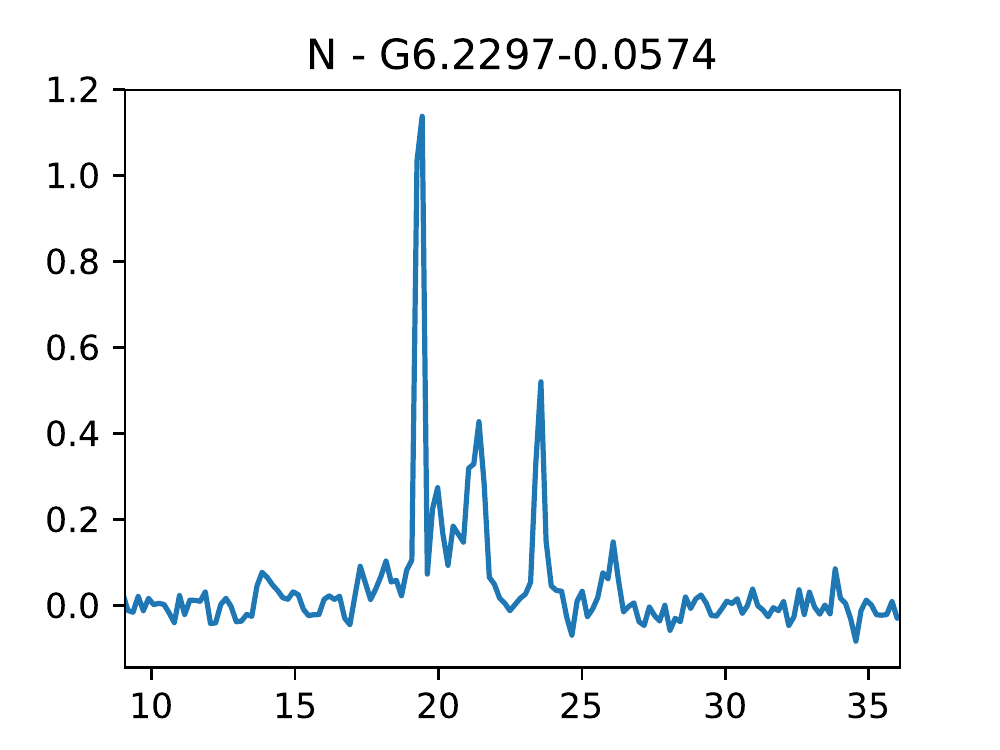}
\includegraphics[trim={0.0cm 0.3cm 0.6cm 0.3cm}, clip, width = 0.235\textwidth]{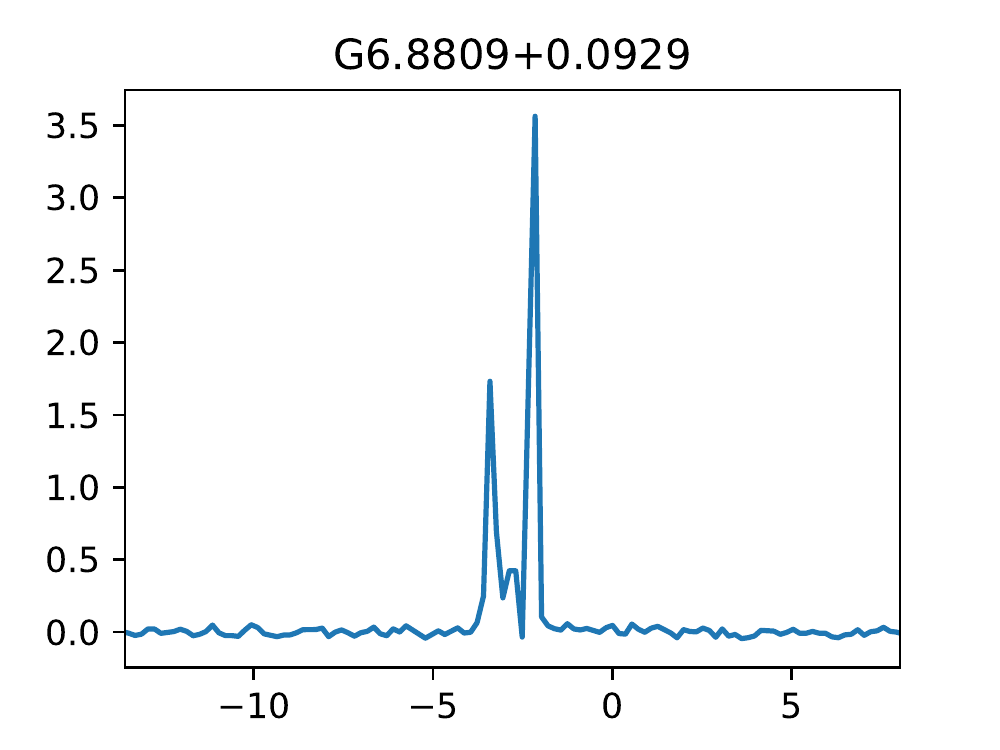}
\includegraphics[trim={0.0cm 0.3cm 0.6cm 0.3cm}, clip, width = 0.235\textwidth]{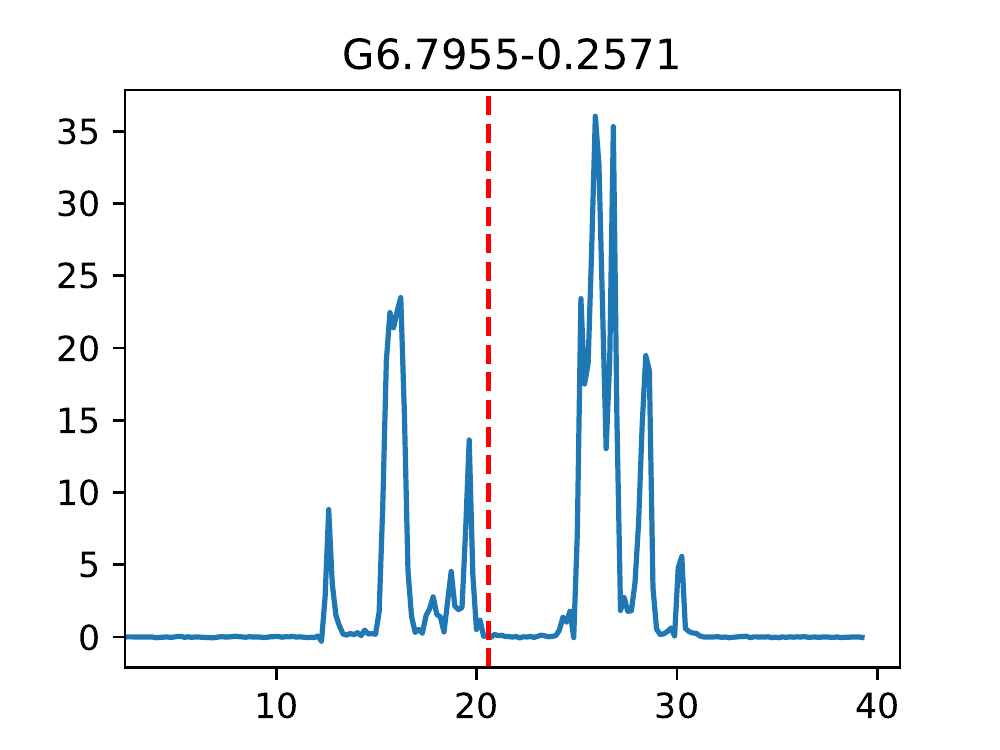}\\ 
\colorbox{white}{\includegraphics[trim={2cm 0 0 0}, scale=0.20]{figs/ylabel.pdf}}
\includegraphics[trim={0.0cm 0.3cm 0.6cm 0.3cm}, clip, width = 0.235\textwidth]{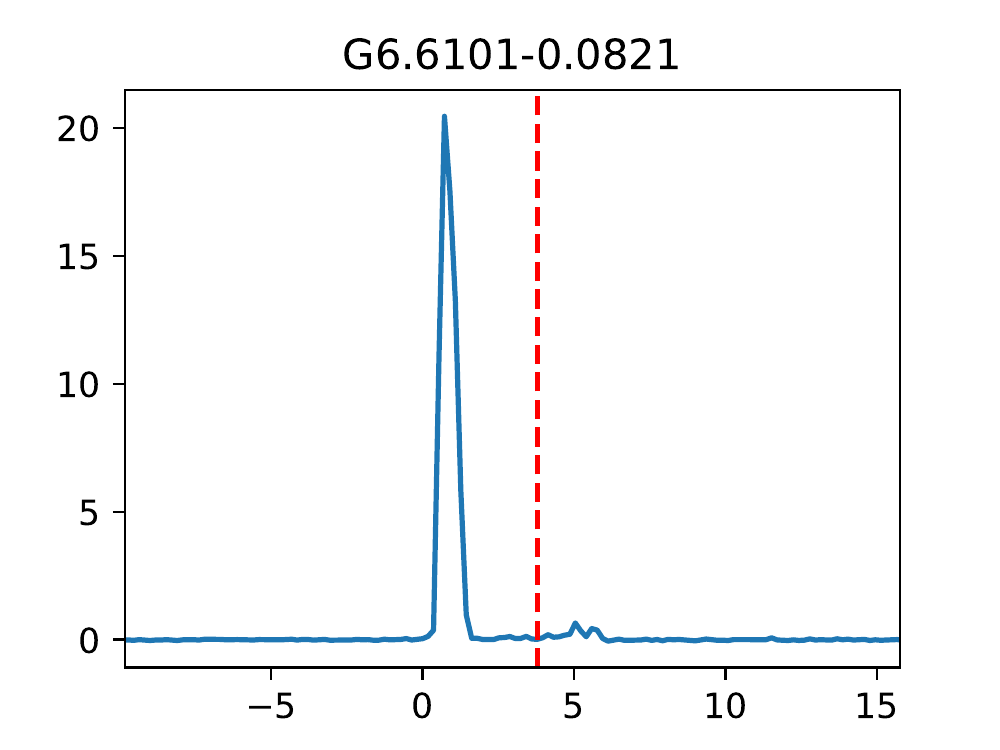}
\includegraphics[trim={0.0cm 0.3cm 0.6cm 0.3cm}, clip, width = 0.235\textwidth]{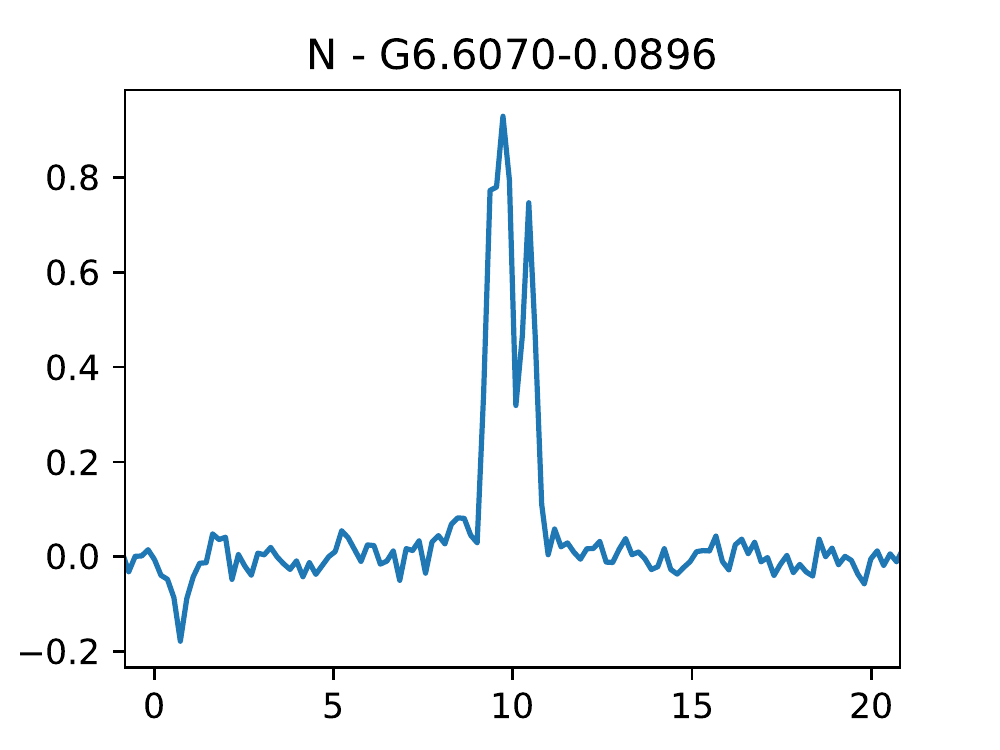}
\includegraphics[trim={0.0cm 0.3cm 0.6cm 0.3cm}, clip, width = 0.235\textwidth]{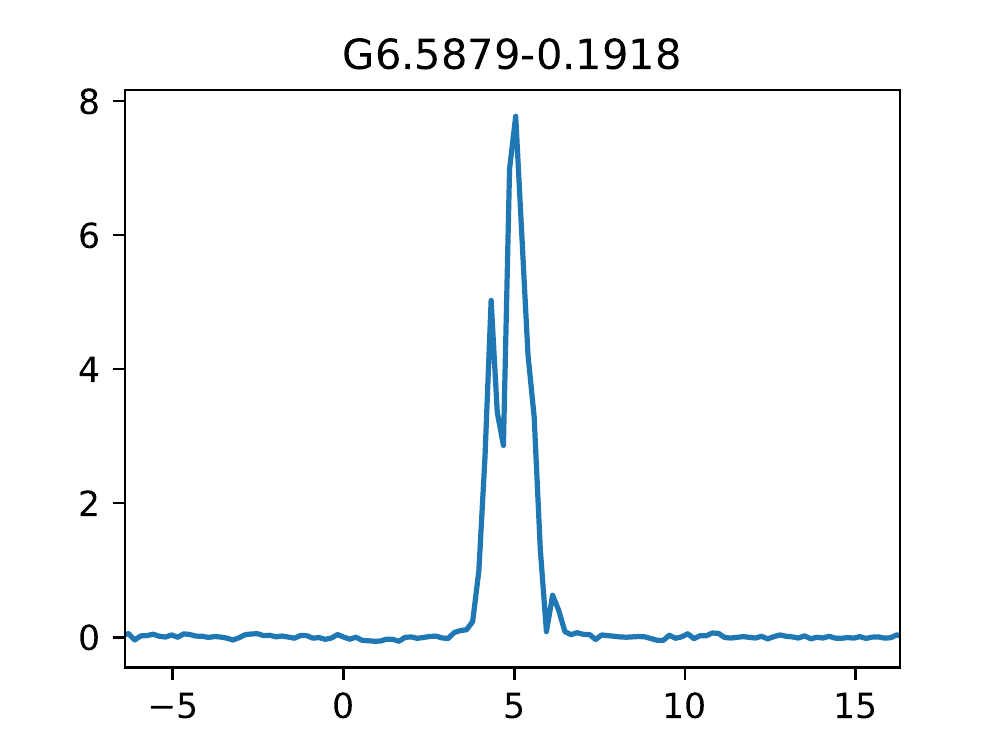}
\includegraphics[trim={0.0cm 0.3cm 0.6cm 0.3cm}, clip, width = 0.235\textwidth]{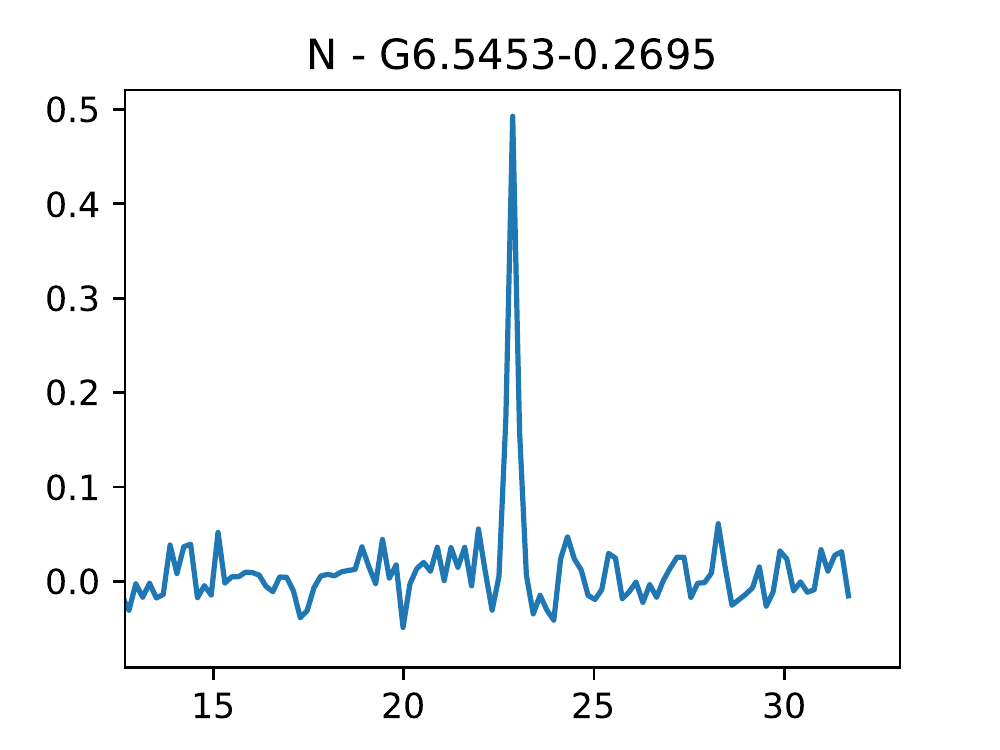}\\ 
\colorbox{white}{\includegraphics[trim={2cm 0 0 0}, scale=0.195]{figs/ylabel_filler.pdf}}
\includegraphics[trim={0.0cm 0.3cm 0.6cm 0.3cm}, clip, width = 0.235\textwidth]{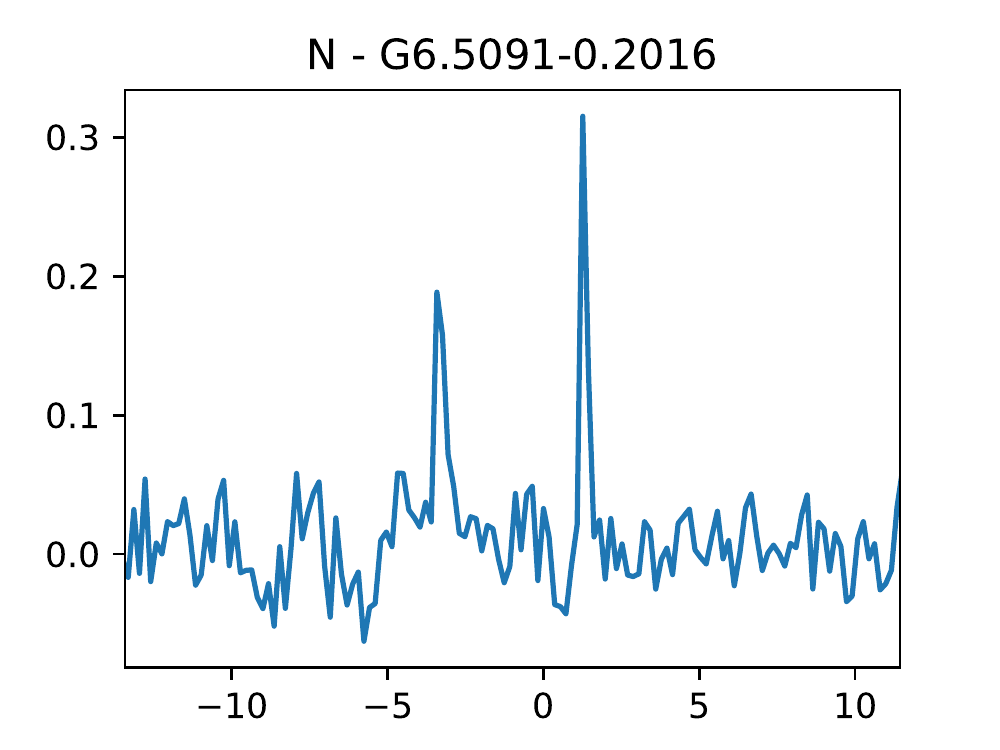}
\includegraphics[trim={0.0cm 0.3cm 0.6cm 0.3cm}, clip, width = 0.235\textwidth]{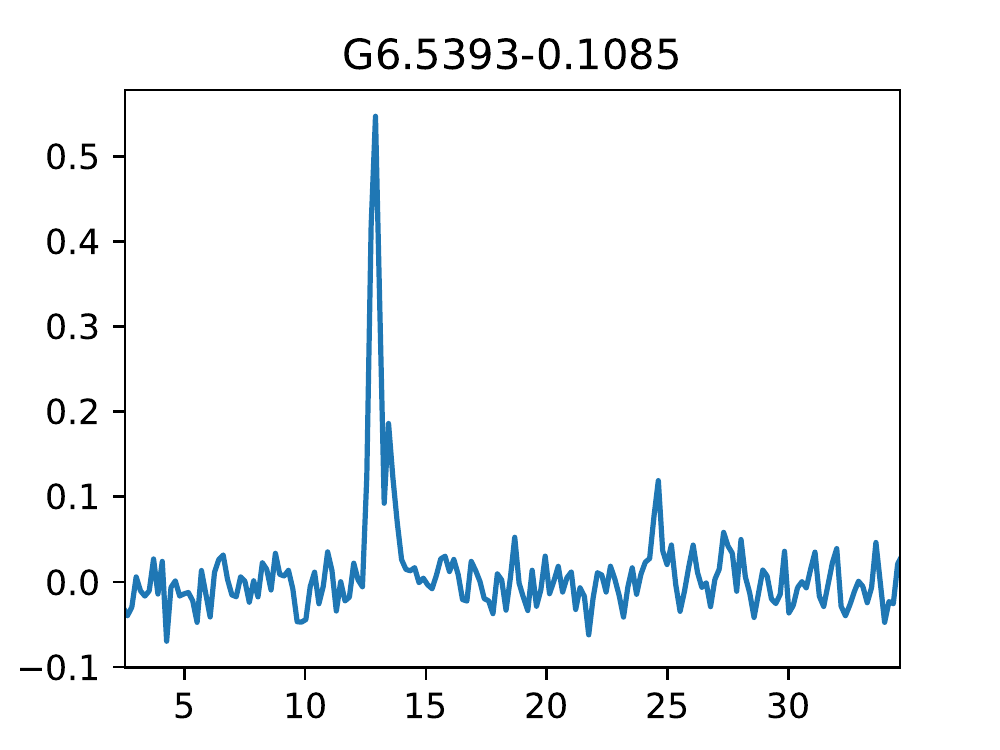}
\includegraphics[trim={0.0cm 0.3cm 0.6cm 0.3cm}, clip, width = 0.235\textwidth]{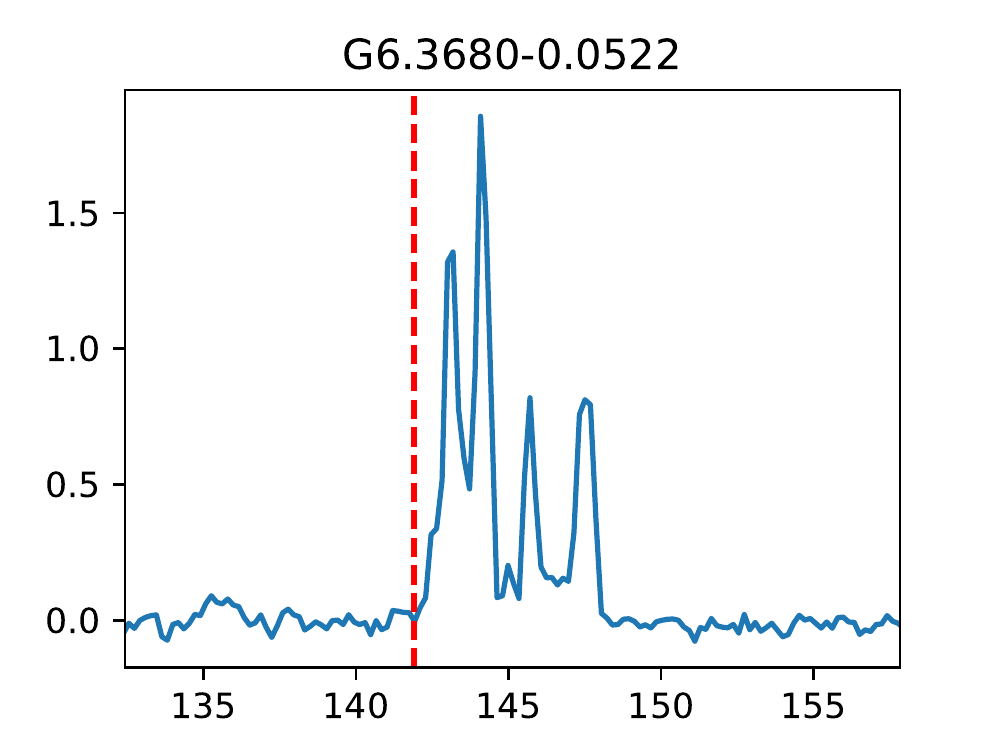}
\includegraphics[trim={0.0cm 0.3cm 0.6cm 0.3cm}, clip, width = 0.235\textwidth]{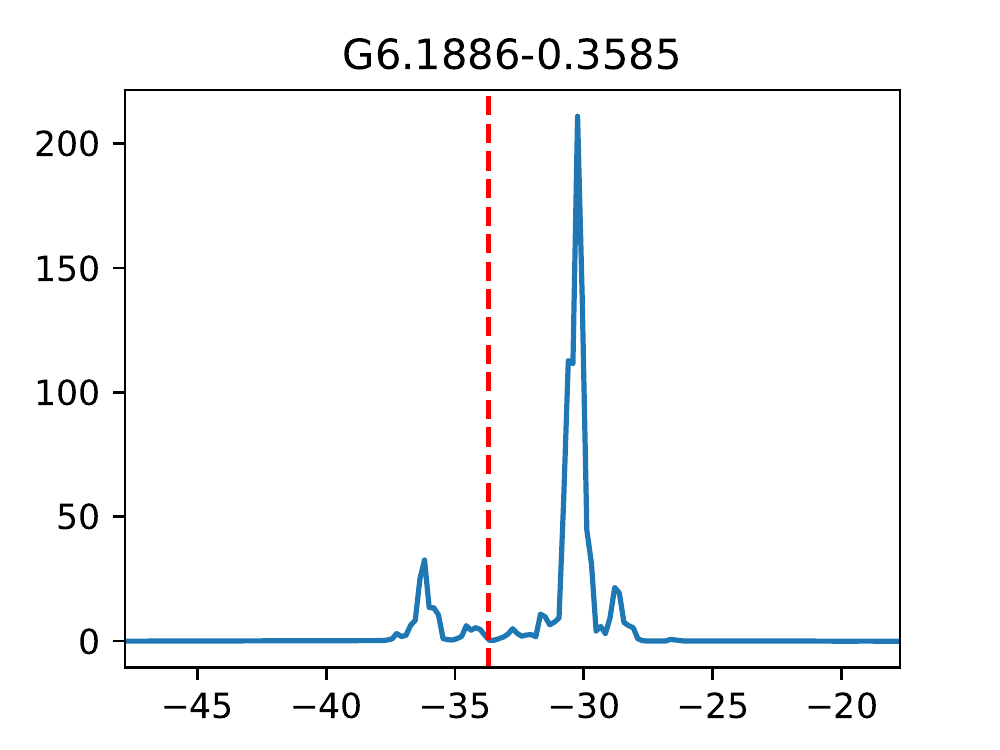}\\
\end{tabular}
\caption{Continued from Fig.~\ref{fig:maser_many_spec}.}
\label{fig:masers_96}
\end{figure*}

%Table 5
\begin{figure*}[!h]
\begin{tabular}{ccccc}
\colorbox{white}{\includegraphics[trim={2cm 0 0 0}, scale=0.195]{figs/ylabel_filler.pdf}}
\includegraphics[trim={0.0cm 0.3cm 0.6cm 0.3cm}, clip, width = 0.235\textwidth]{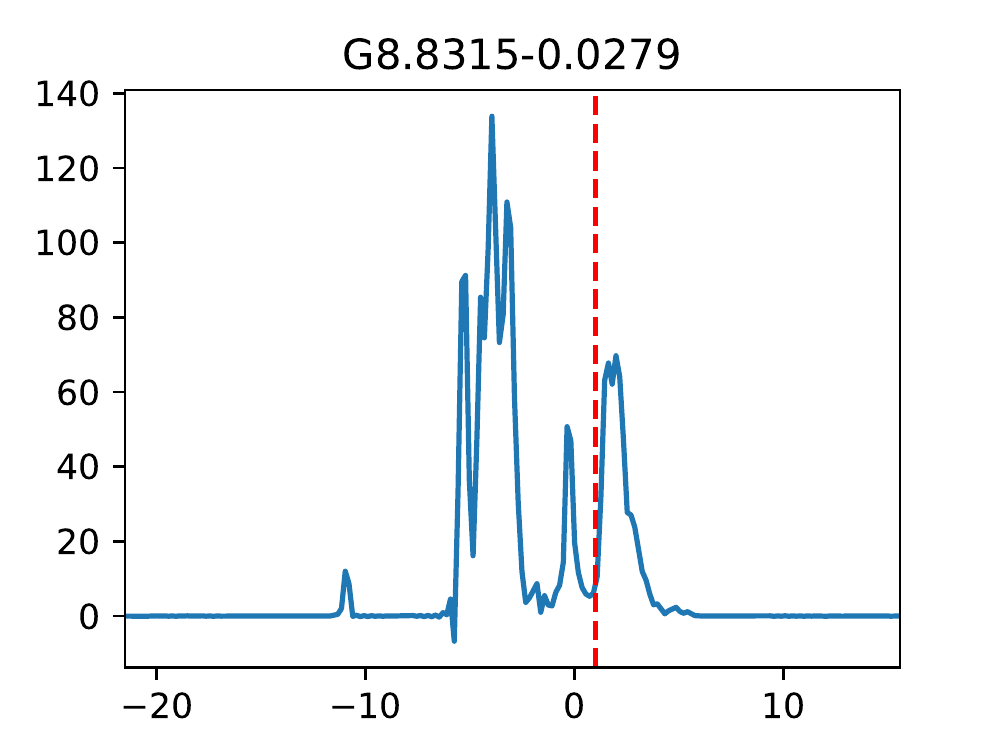}
\includegraphics[trim={0.0cm 0.3cm 0.6cm 0.3cm}, clip, width = 0.235\textwidth]{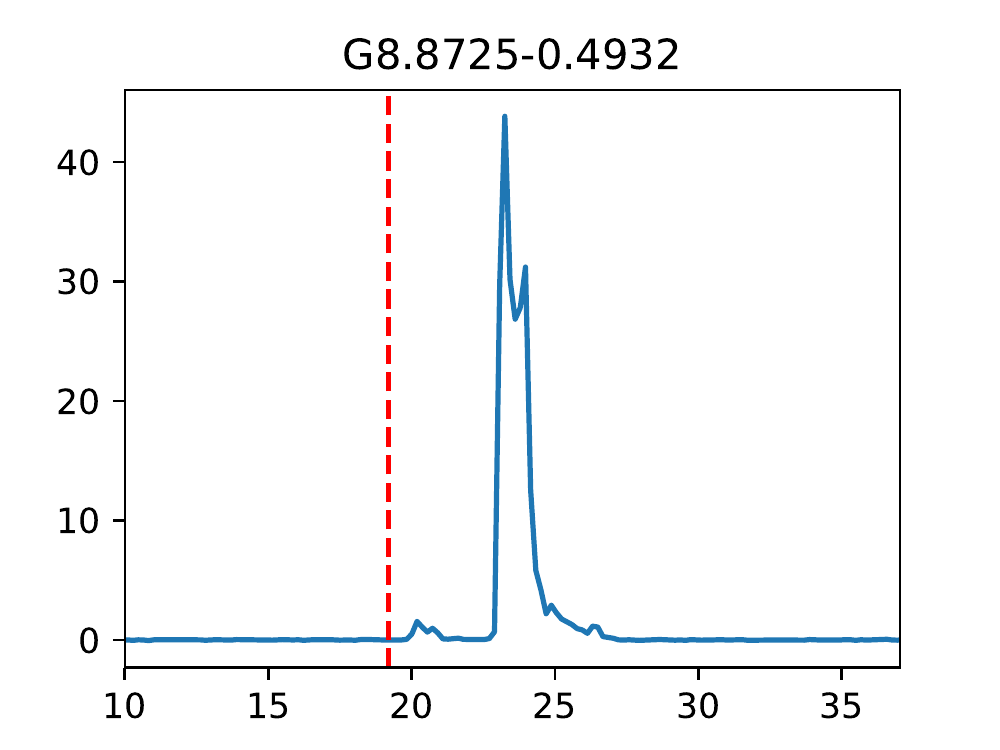}
\includegraphics[trim={0.0cm 0.3cm 0.6cm 0.3cm}, clip, width = 0.235\textwidth]{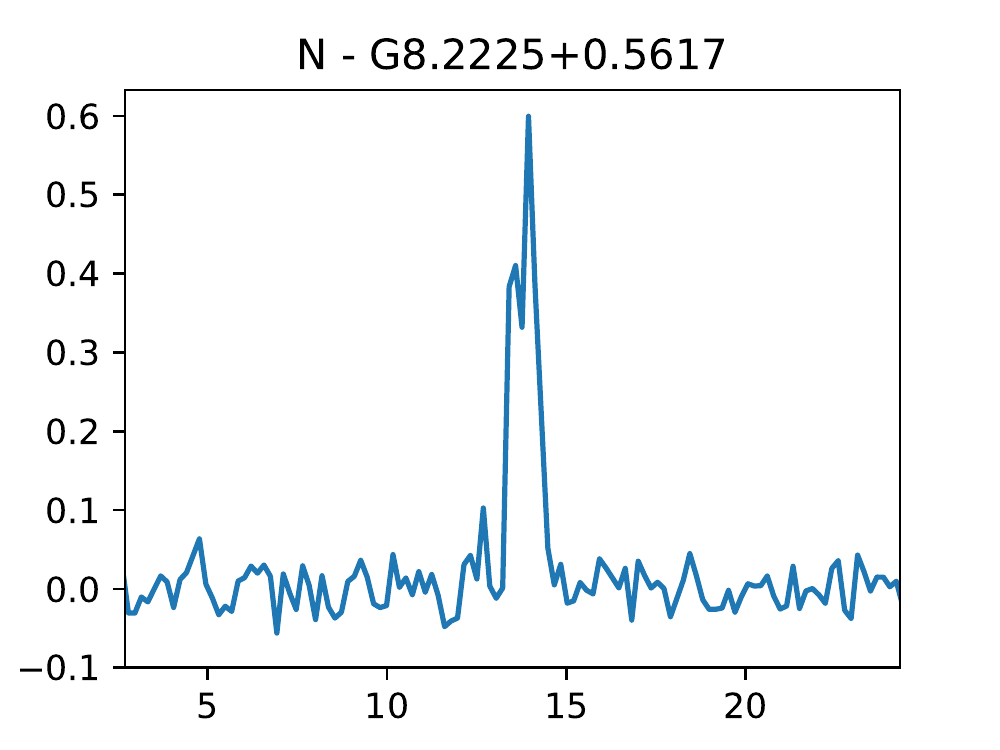}
\includegraphics[trim={0.0cm 0.3cm 0.6cm 0.3cm}, clip, width = 0.235\textwidth]{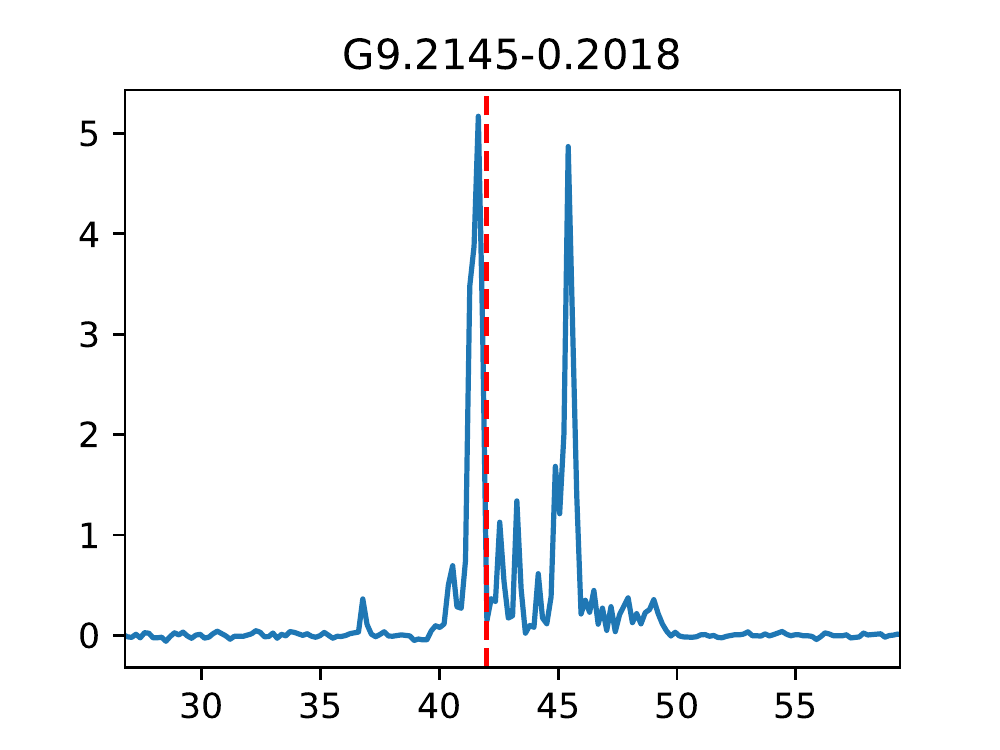}\\ 
\colorbox{white}{\includegraphics[trim={2cm 0 0 0}, scale=0.195]{figs/ylabel_filler.pdf}}
\includegraphics[trim={0.0cm 0.3cm 0.6cm 0.3cm}, clip, width = 0.235\textwidth]{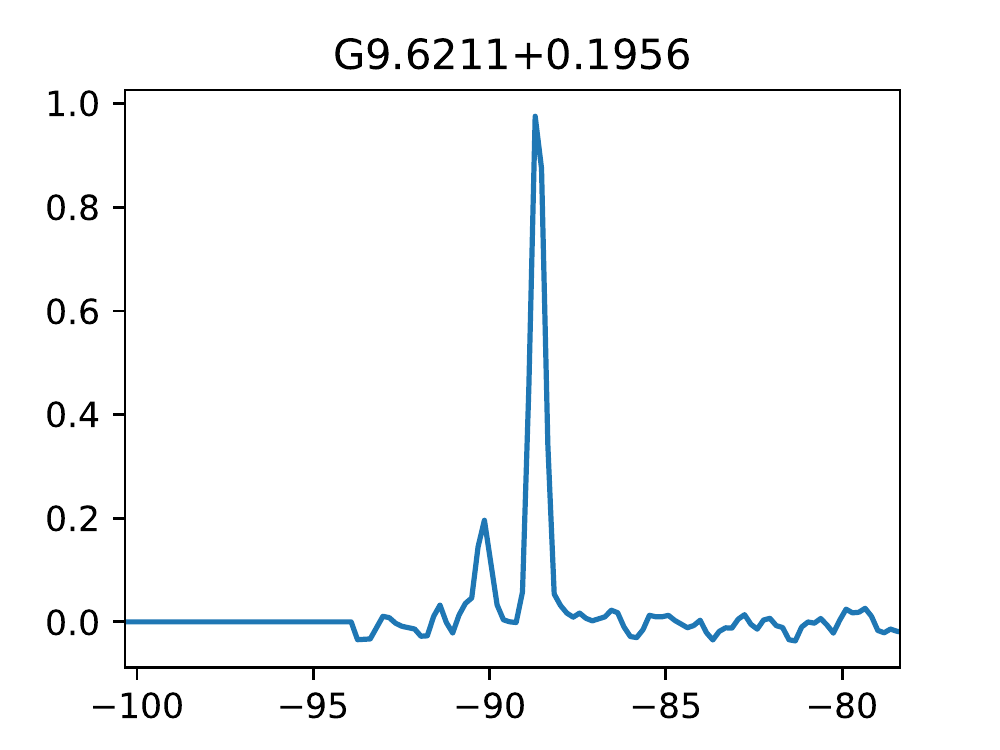}
\includegraphics[trim={0.0cm 0.3cm 0.6cm 0.3cm}, clip, width = 0.235\textwidth]{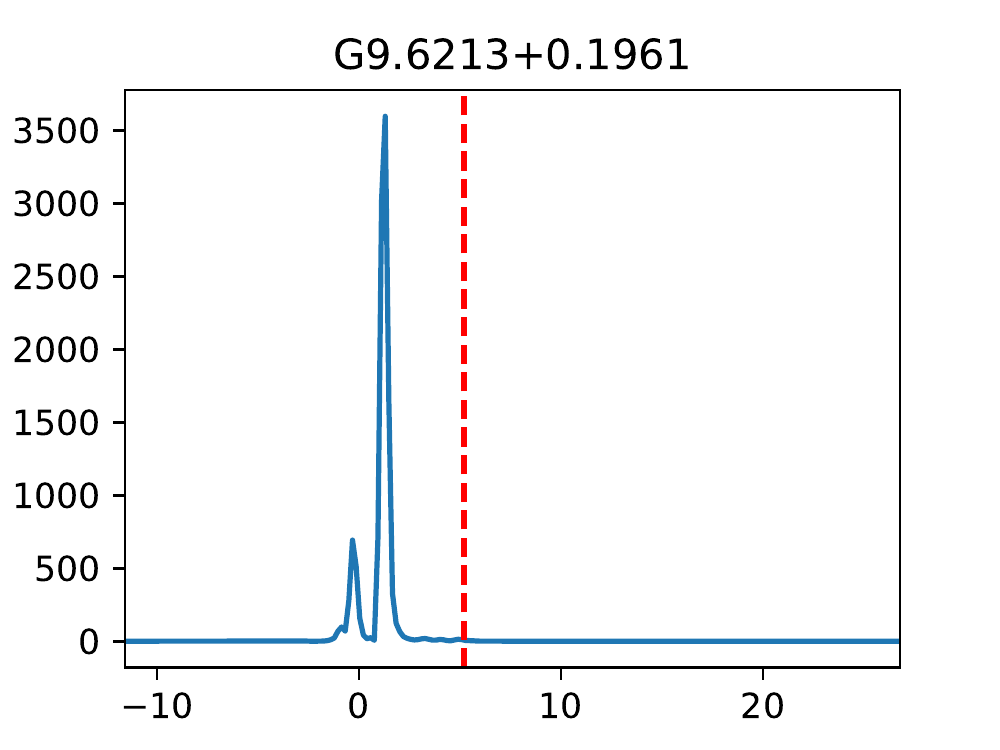}
\includegraphics[trim={0.0cm 0.3cm 0.6cm 0.3cm}, clip, width = 0.235\textwidth]{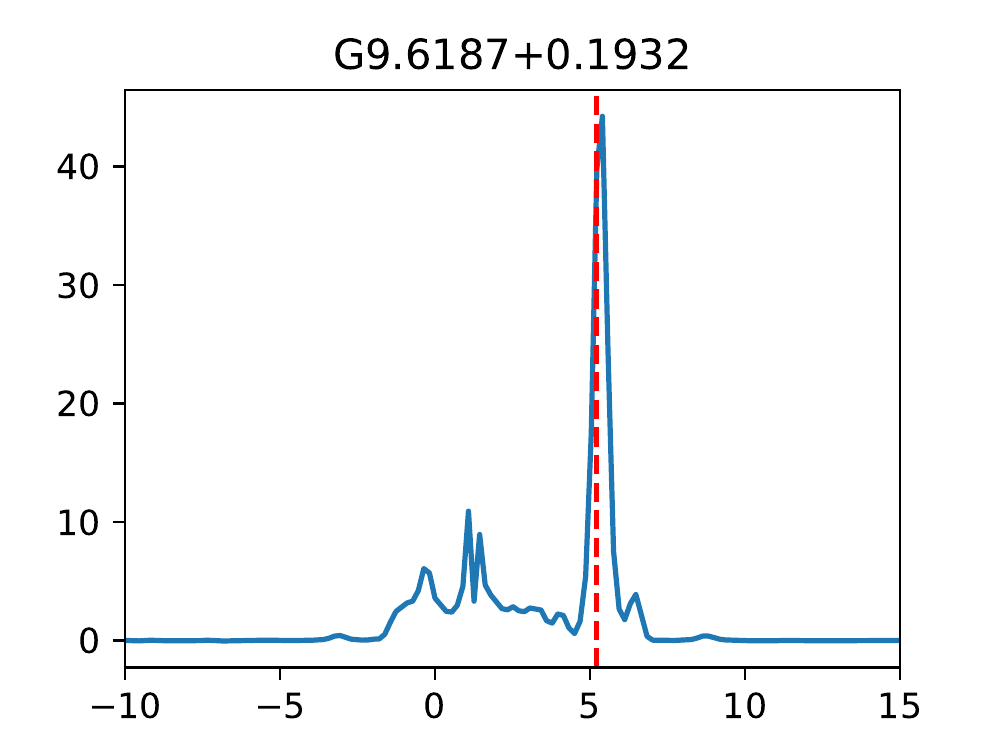}
\includegraphics[trim={0.0cm 0.3cm 0.6cm 0.3cm}, clip, width = 0.235\textwidth]{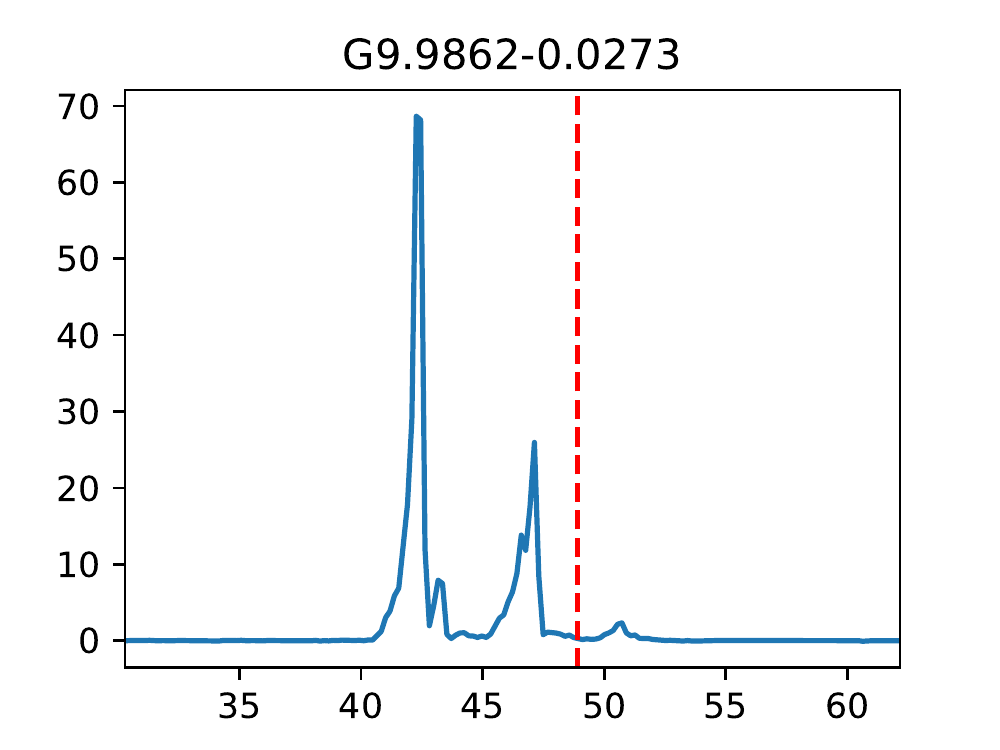}\\ 
\colorbox{white}{\includegraphics[trim={2cm 0 0 0}, scale=0.20]{figs/ylabel.pdf}}
\includegraphics[trim={0.0cm 0.3cm 0.6cm 0.3cm}, clip, width = 0.235\textwidth]{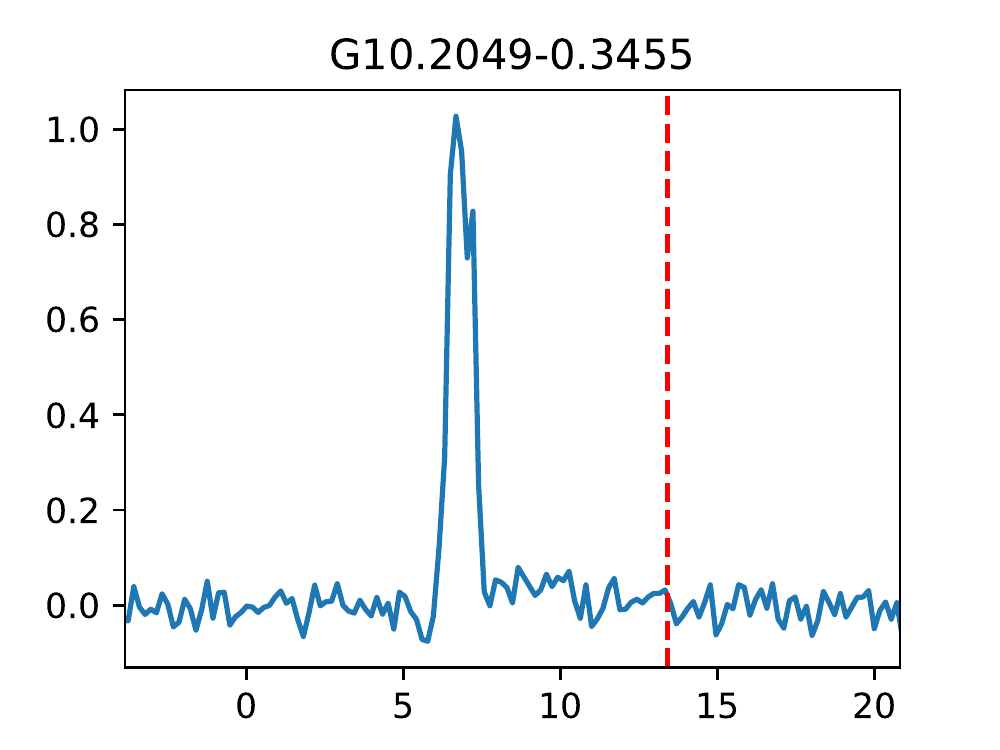}
\includegraphics[trim={0.0cm 0.3cm 0.6cm 0.3cm}, clip, width = 0.235\textwidth]{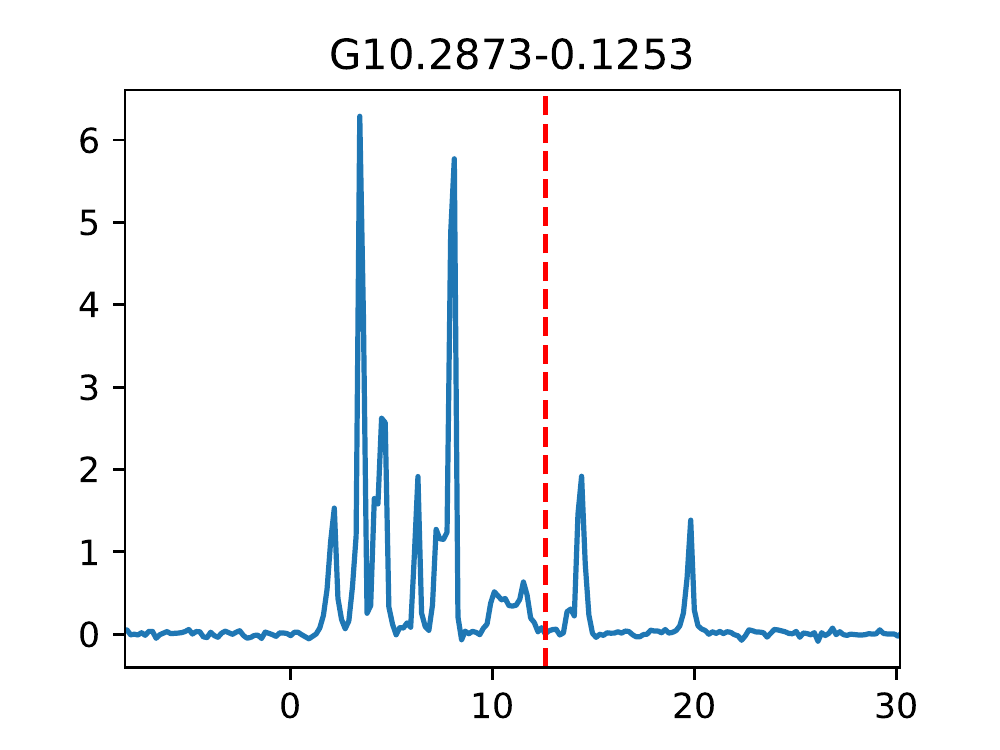}
\includegraphics[trim={0.0cm 0.3cm 0.6cm 0.3cm}, clip, width = 0.235\textwidth]{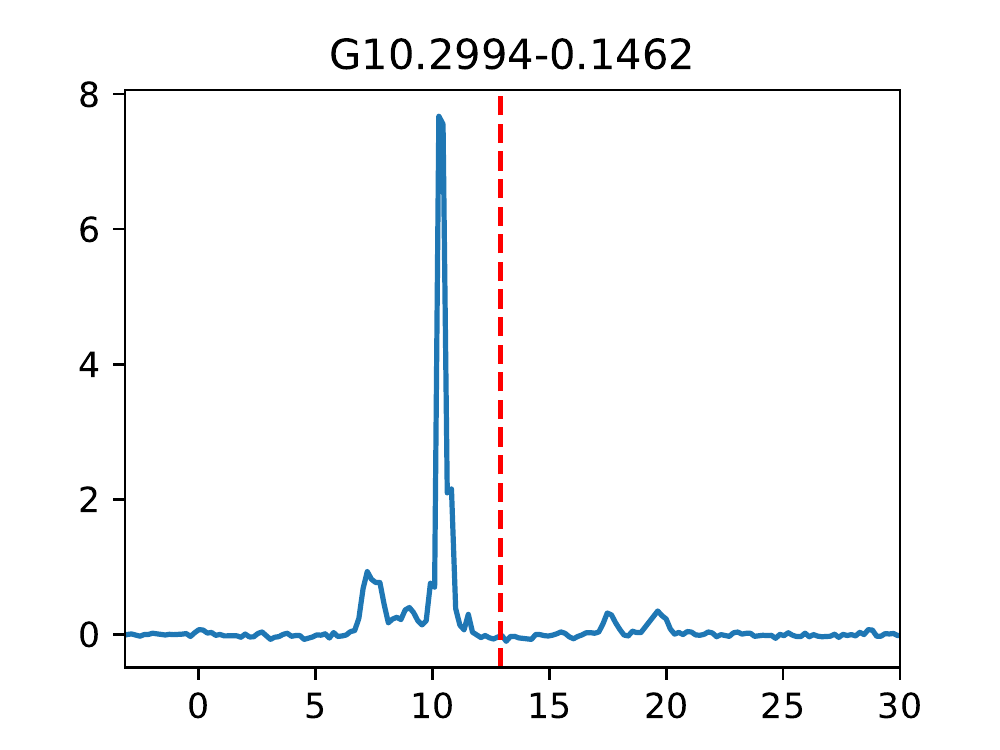}
\includegraphics[trim={0.0cm 0.3cm 0.6cm 0.3cm}, clip, width = 0.235\textwidth]{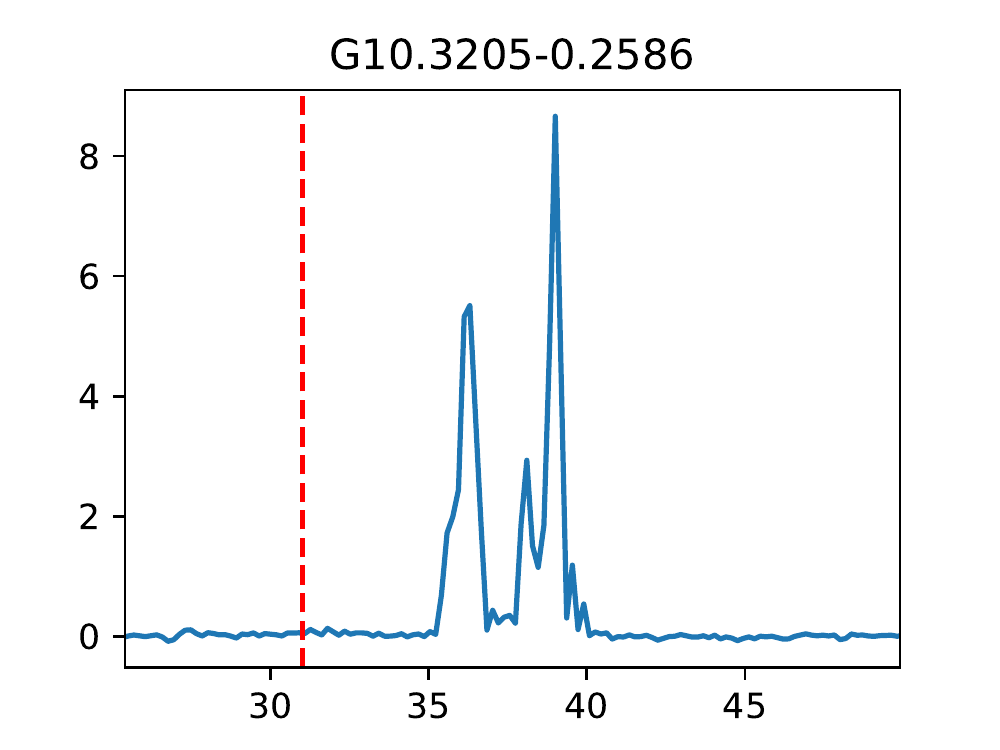}\\ 
\colorbox{white}{\includegraphics[trim={2cm 0 0 0}, scale=0.195]{figs/ylabel_filler.pdf}}
\includegraphics[trim={0.0cm 0.3cm 0.6cm 0.3cm}, clip, width = 0.235\textwidth]{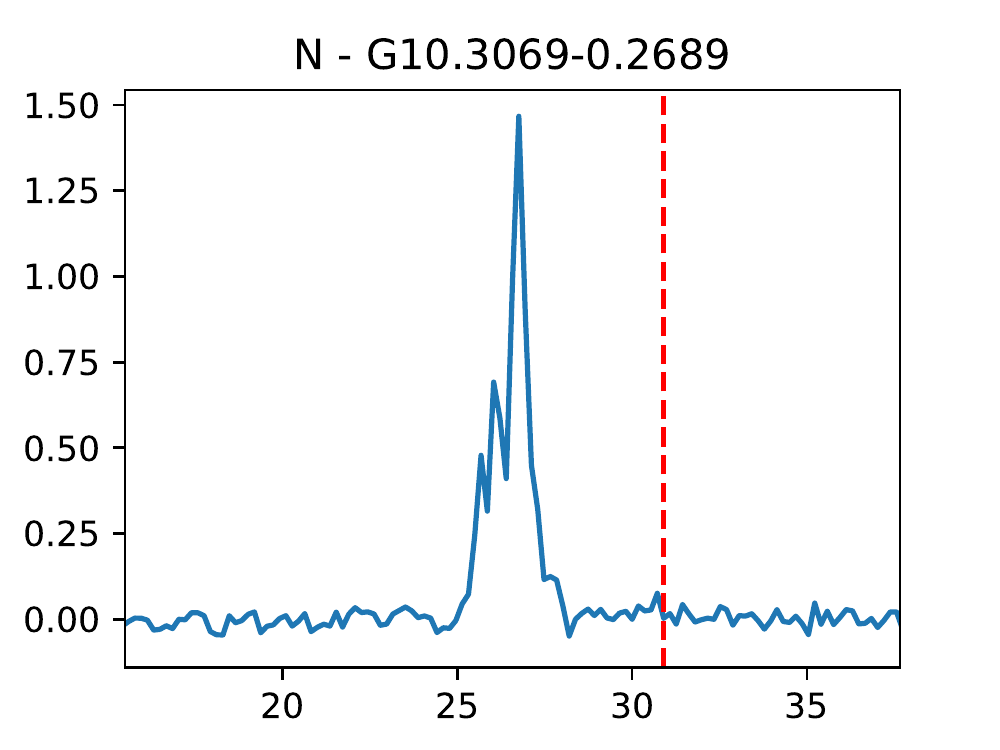}
\includegraphics[trim={0.0cm 0.3cm 0.6cm 0.3cm}, clip, width = 0.235\textwidth]{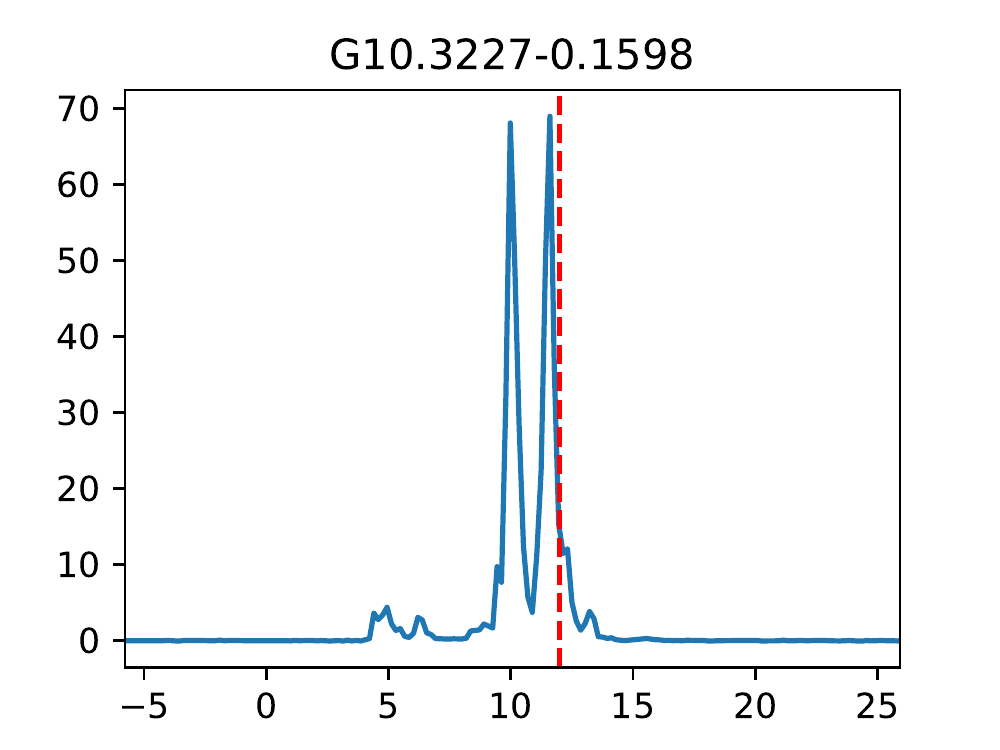}
\includegraphics[trim={0.0cm 0.3cm 0.6cm 0.3cm}, clip, width = 0.235\textwidth]{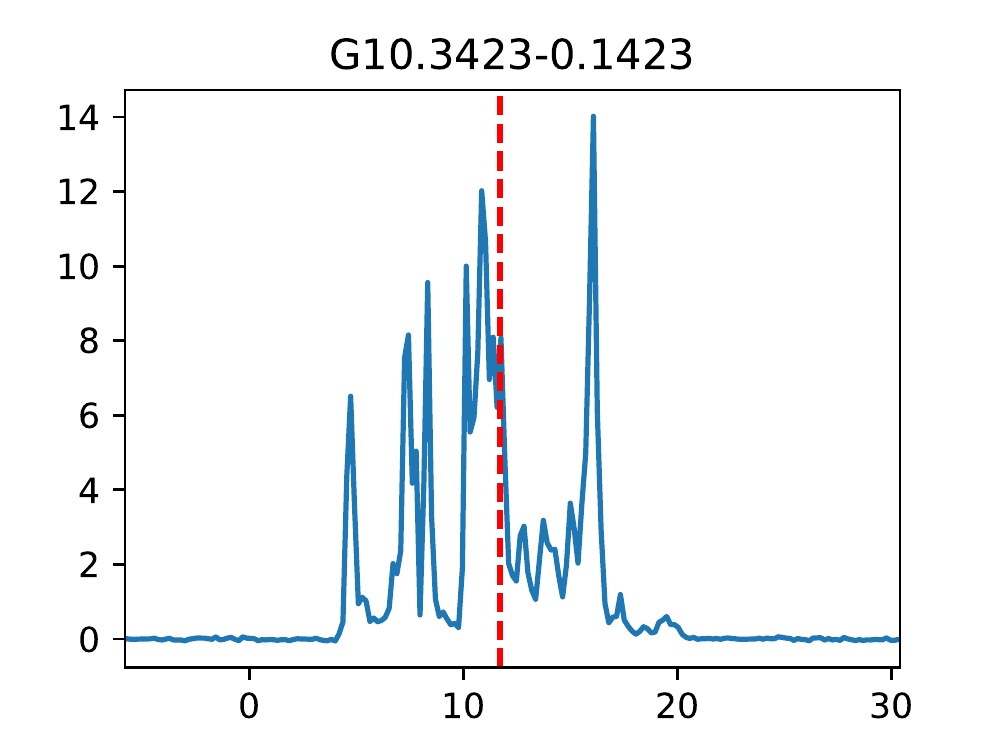}
\includegraphics[trim={0.0cm 0.3cm 0.6cm 0.3cm}, clip, width = 0.235\textwidth]{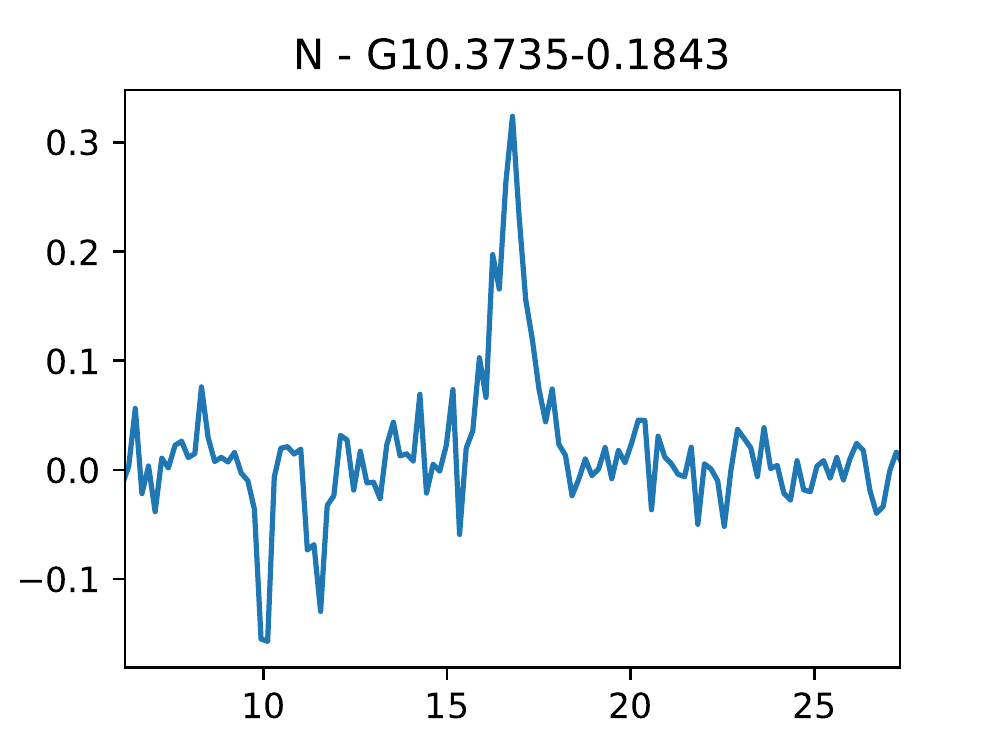}\\
\end{tabular}
\caption{Continued from Fig.~\ref{fig:maser_many_spec}.}
\label{fig:masers_120}
\end{figure*}

%Table 6
\begin{figure*}[!h]
\begin{tabular}{ccccc}
\colorbox{white}{\includegraphics[trim={2cm 0 0 0}, scale=0.195]{figs/ylabel_filler.pdf}}
\includegraphics[trim={0.0cm 0.3cm 0.6cm 0.3cm}, clip, width = 0.235\textwidth]{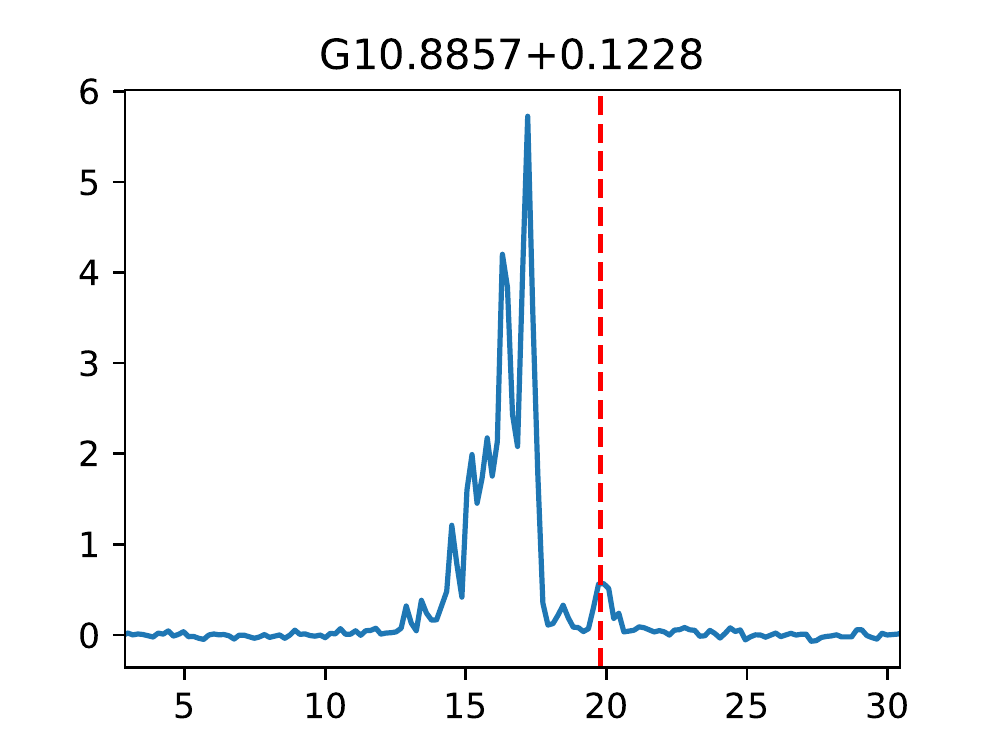}
\includegraphics[trim={0.0cm 0.3cm 0.6cm 0.3cm}, clip, width = 0.235\textwidth]{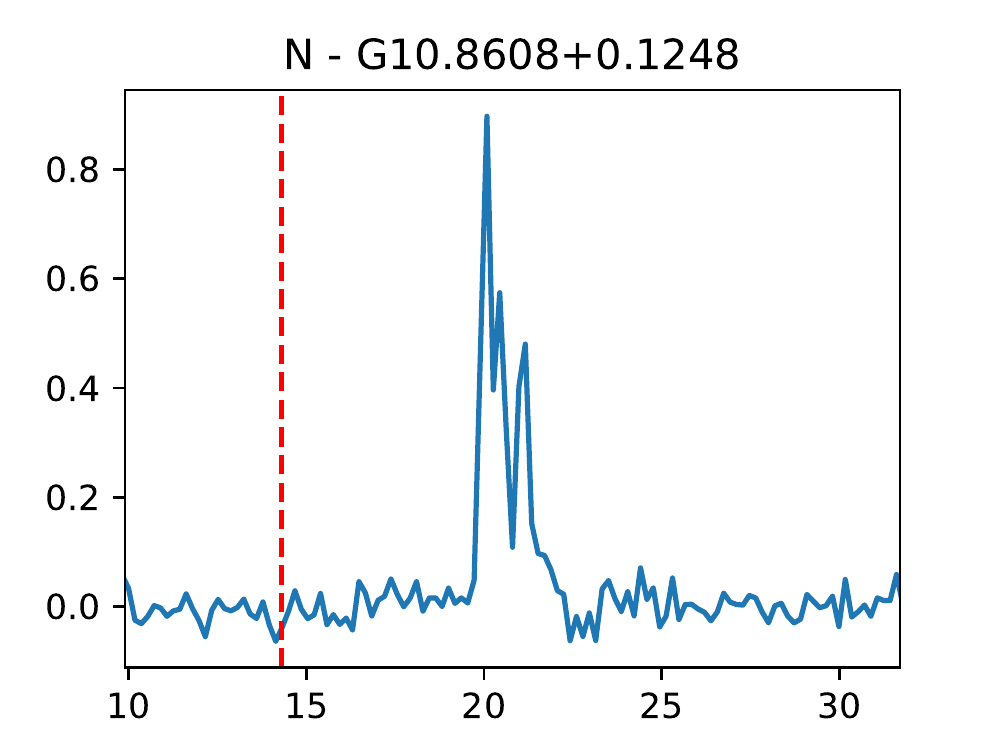}
\includegraphics[trim={0.0cm 0.3cm 0.6cm 0.3cm}, clip, width = 0.235\textwidth]{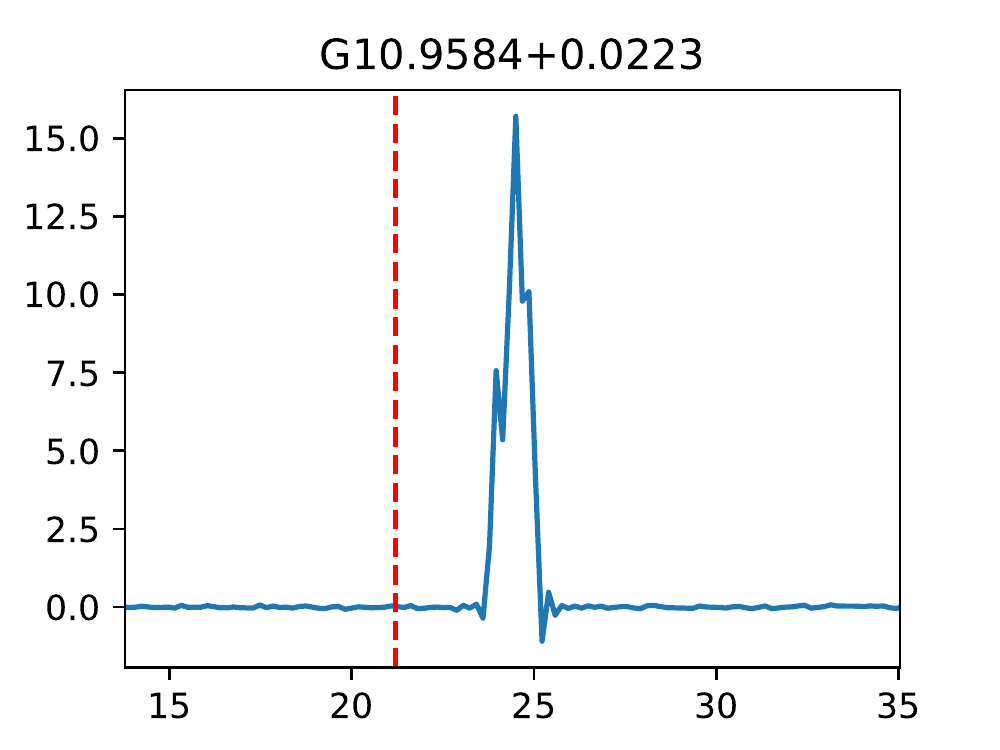}
\includegraphics[trim={0.0cm 0.3cm 0.6cm 0.3cm}, clip, width = 0.235\textwidth]{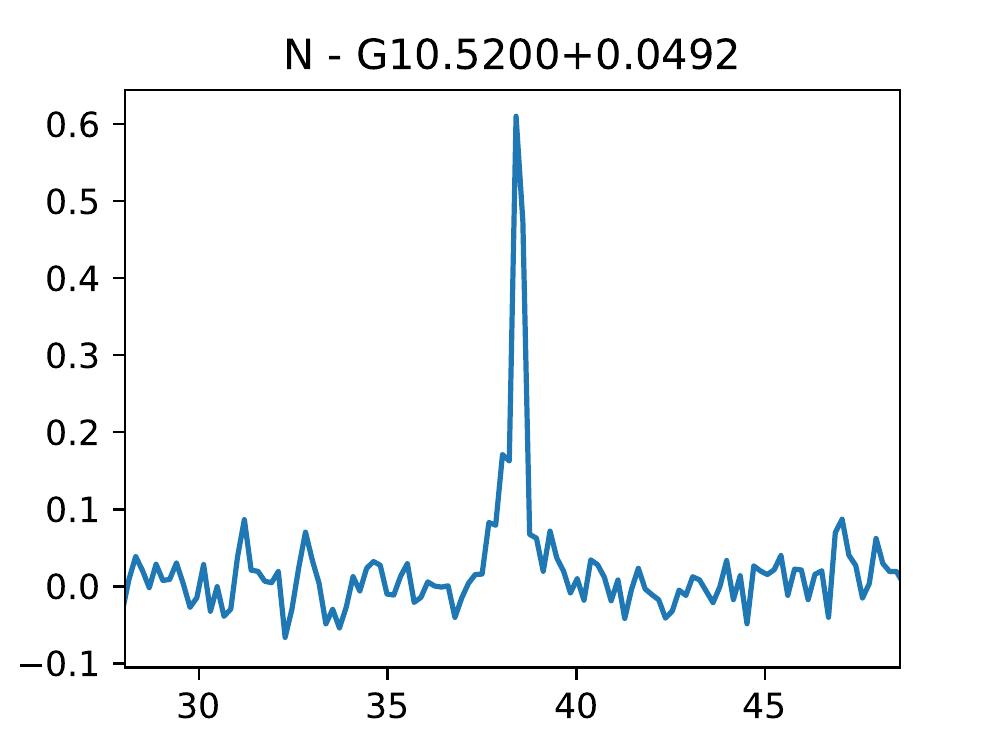}\\ 
\colorbox{white}{\includegraphics[trim={2cm 0 0 0}, scale=0.195]{figs/ylabel_filler.pdf}}
\includegraphics[trim={0.0cm 0.3cm 0.6cm 0.3cm}, clip, width = 0.235\textwidth]{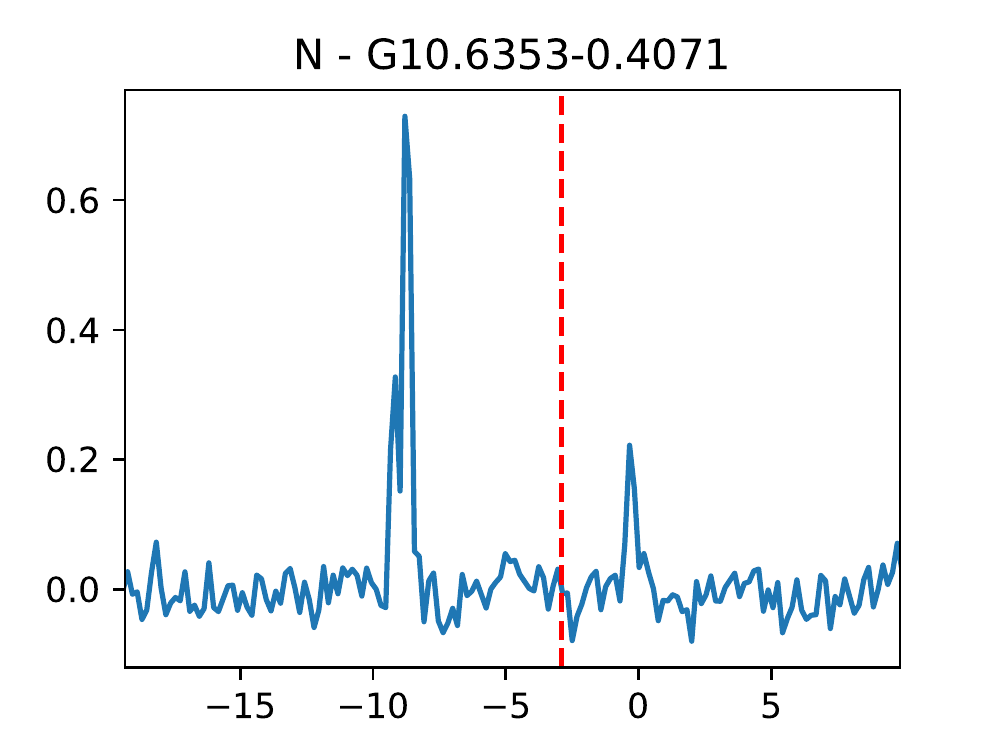}
\includegraphics[trim={0.0cm 0.3cm 0.6cm 0.3cm}, clip, width = 0.235\textwidth]{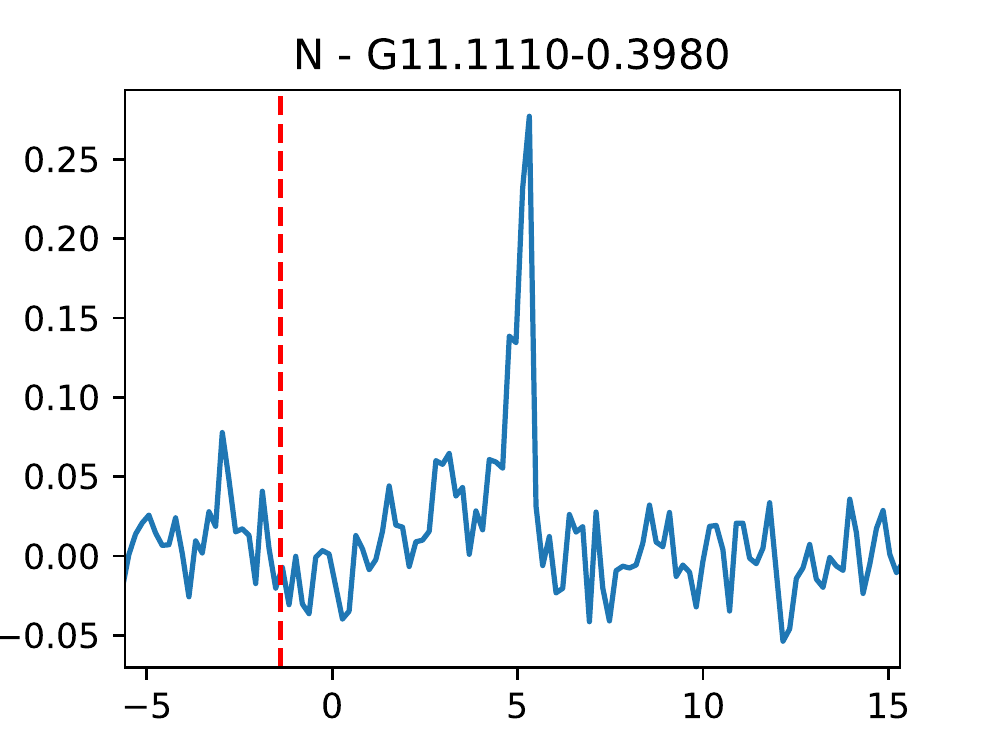}
\includegraphics[trim={0.0cm 0.3cm 0.6cm 0.3cm}, clip, width = 0.235\textwidth]{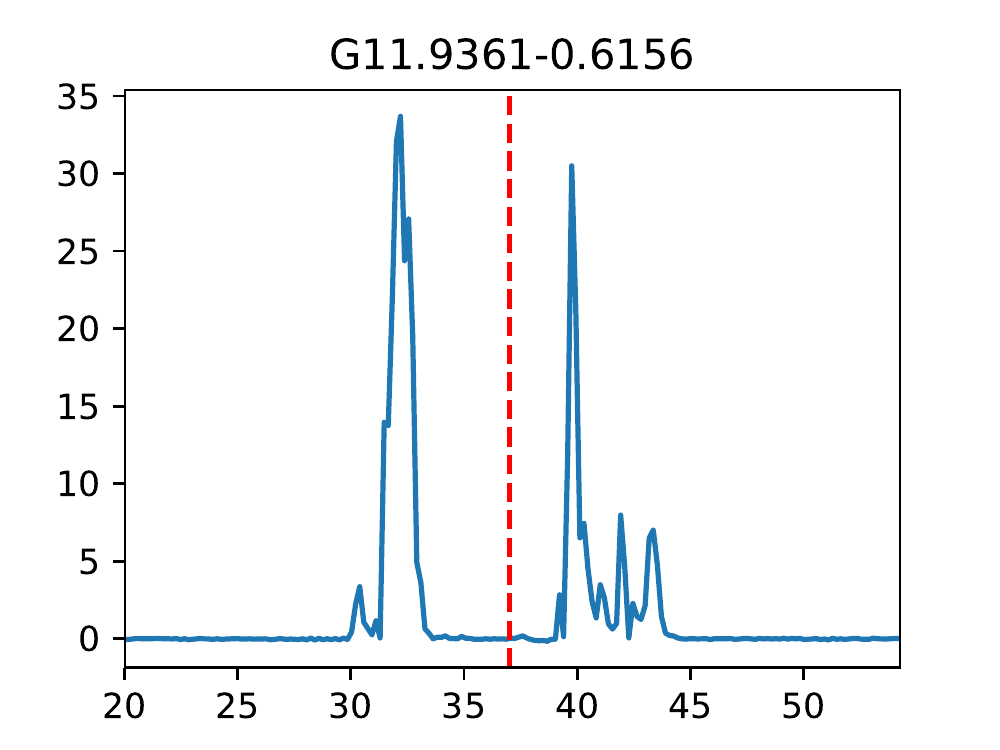}
\includegraphics[trim={0.0cm 0.3cm 0.6cm 0.3cm}, clip, width = 0.235\textwidth]{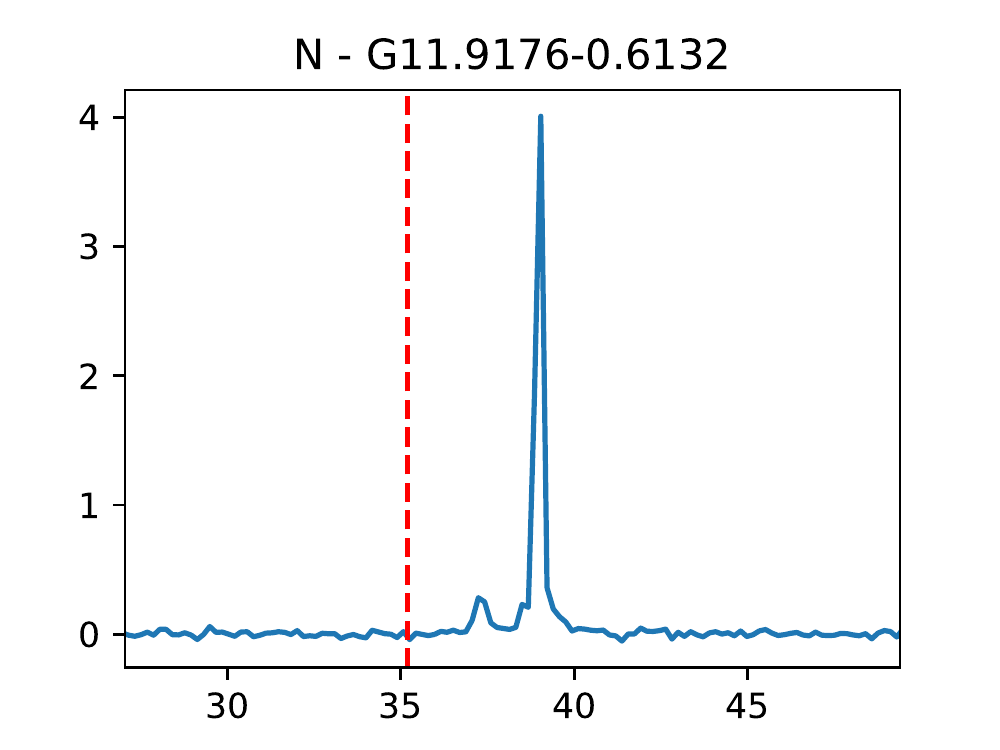}\\ 
\colorbox{white}{\includegraphics[trim={2cm 0 0 0}, scale=0.20]{figs/ylabel.pdf}}
\includegraphics[trim={0.0cm 0.3cm 0.6cm 0.3cm}, clip, width = 0.235\textwidth]{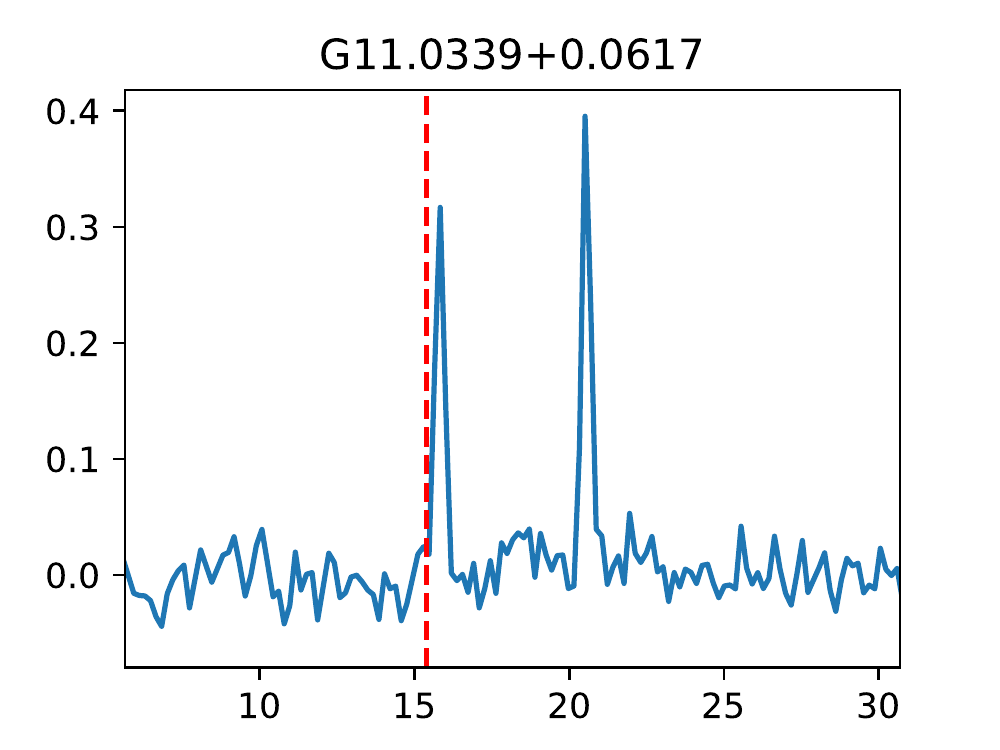}
\includegraphics[trim={0.0cm 0.3cm 0.6cm 0.3cm}, clip, width = 0.235\textwidth]{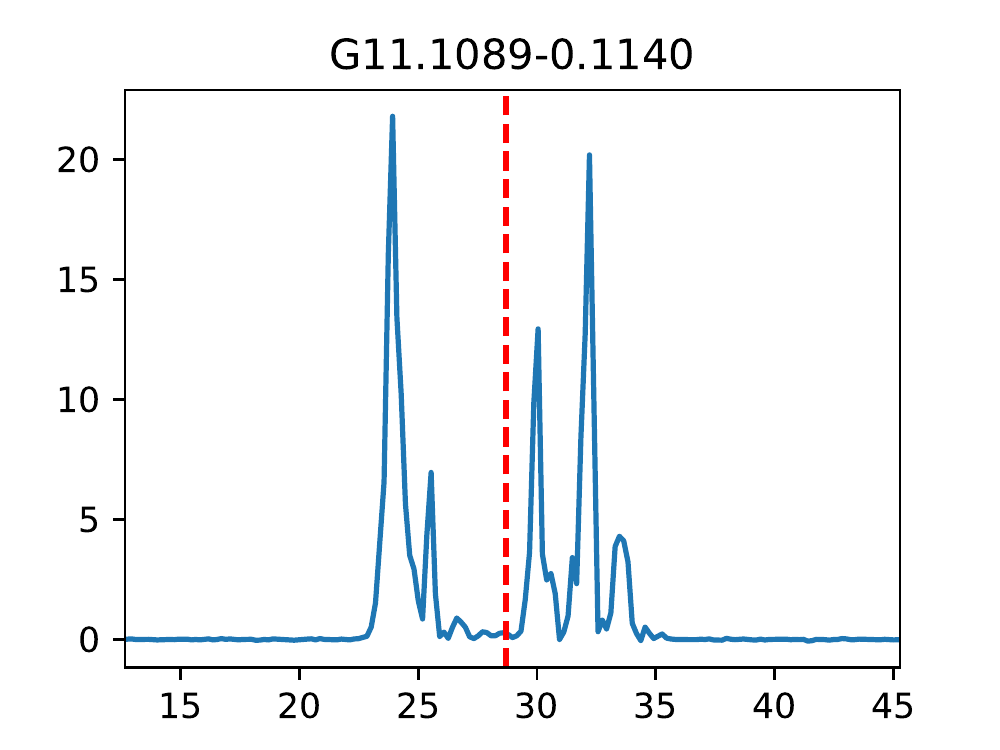}
\includegraphics[trim={0.0cm 0.3cm 0.6cm 0.3cm}, clip, width = 0.235\textwidth]{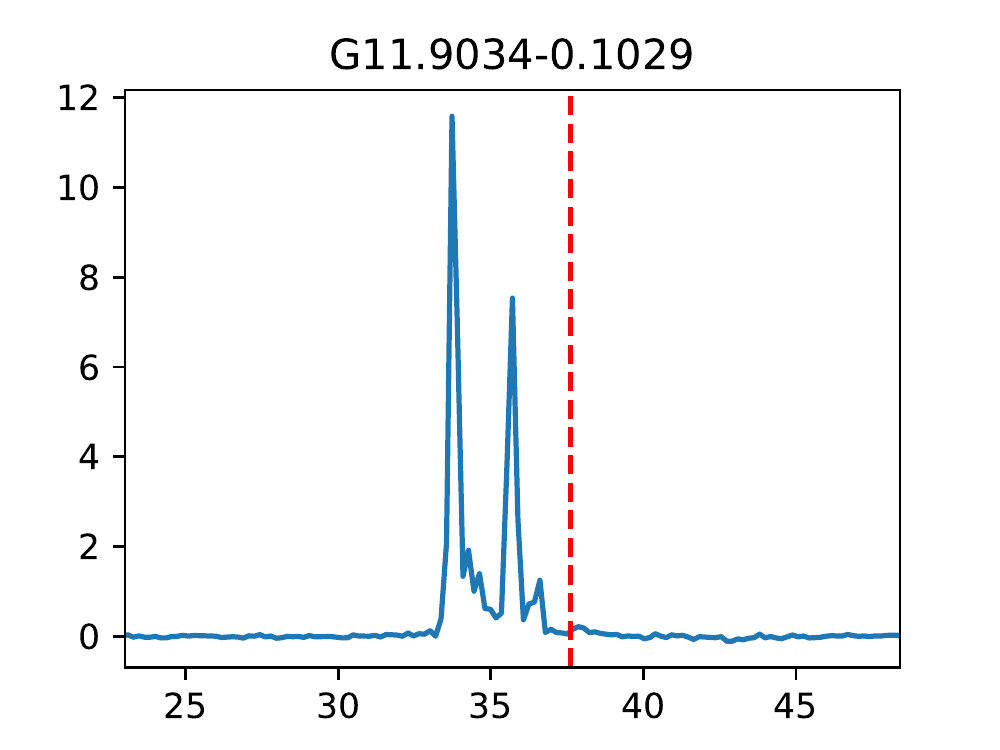}
\includegraphics[trim={0.0cm 0.3cm 0.6cm 0.3cm}, clip, width = 0.235\textwidth]{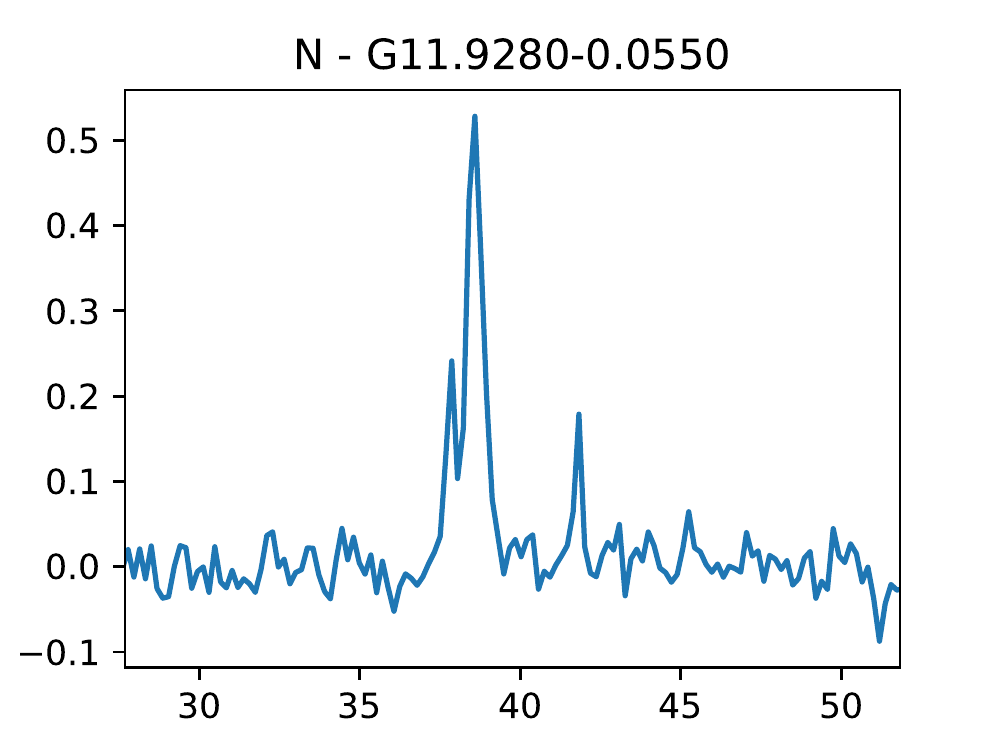}\\
\end{tabular}
\caption{Continued from Fig.~\ref{fig:maser_many_spec}.}
\label{fig:masers_144}
\end{figure*}

%Table 7
\begin{figure*}[!h]
\begin{tabular}{ccccc}
\colorbox{white}{\includegraphics[trim={2cm 0 0 0}, scale=0.195]{figs/ylabel_filler.pdf}}
\includegraphics[trim={0.0cm 0.3cm 0.6cm 0.3cm}, clip, width = 0.235\textwidth]{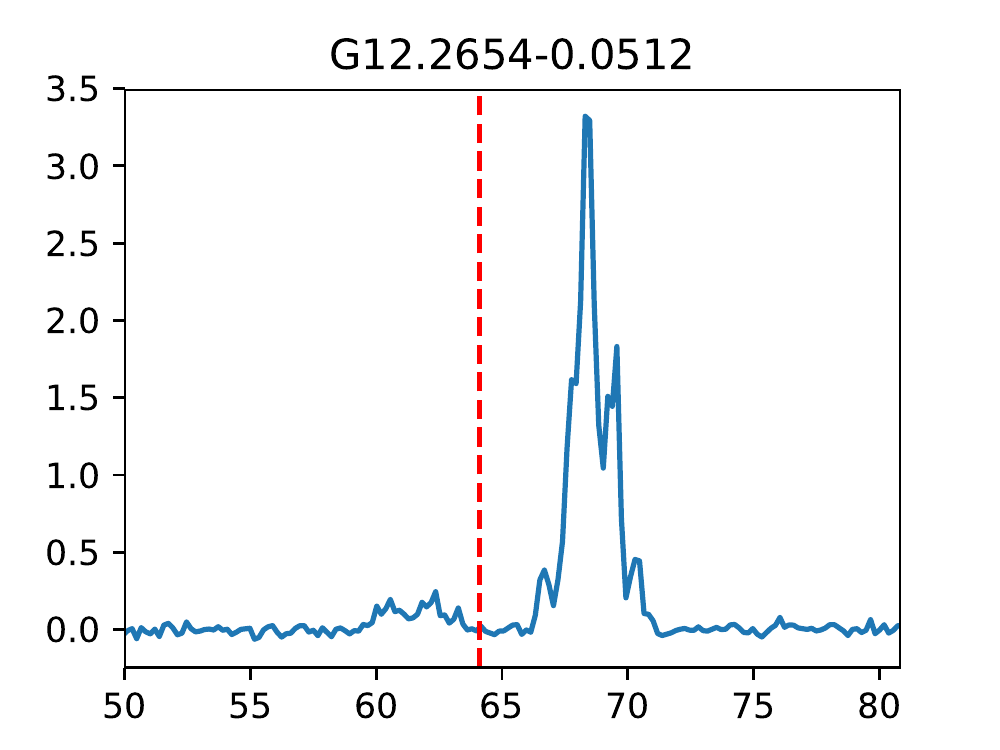}
\includegraphics[trim={0.0cm 0.3cm 0.6cm 0.3cm}, clip, width = 0.235\textwidth]{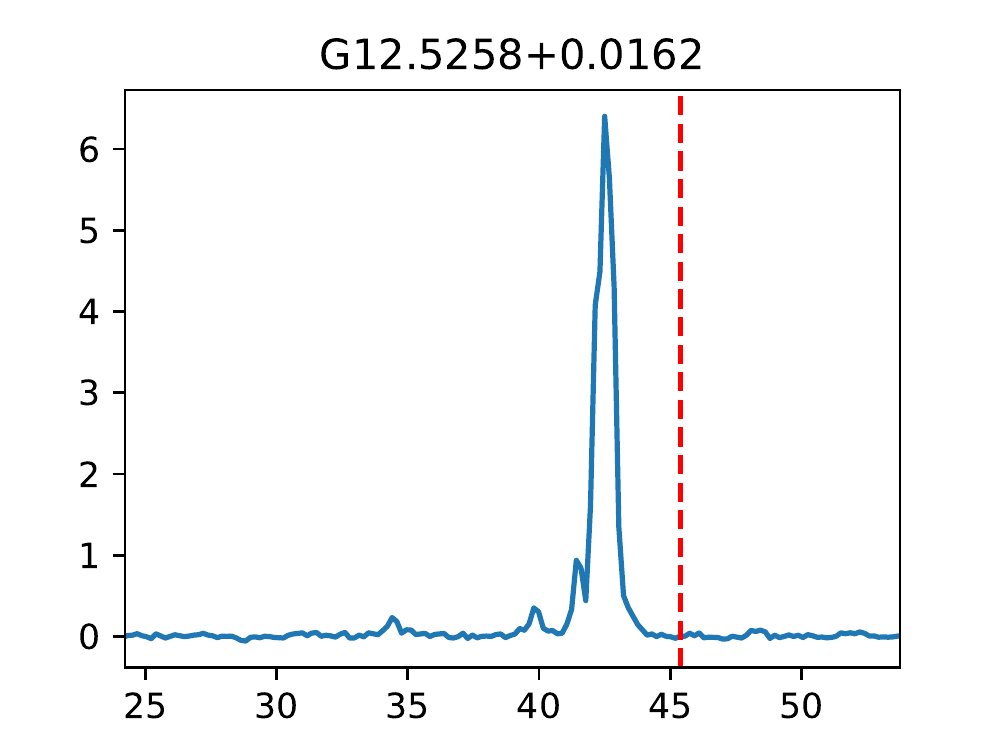}
\includegraphics[trim={0.0cm 0.3cm 0.6cm 0.3cm}, clip, width = 0.235\textwidth]{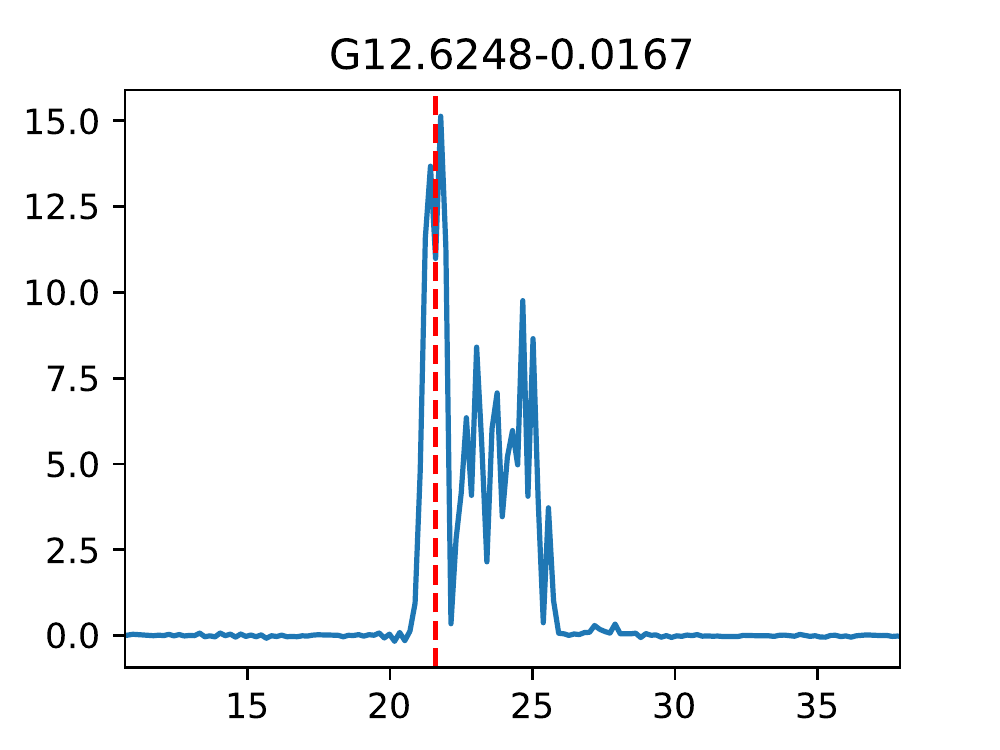}
\includegraphics[trim={0.0cm 0.3cm 0.6cm 0.3cm}, clip, width = 0.235\textwidth]{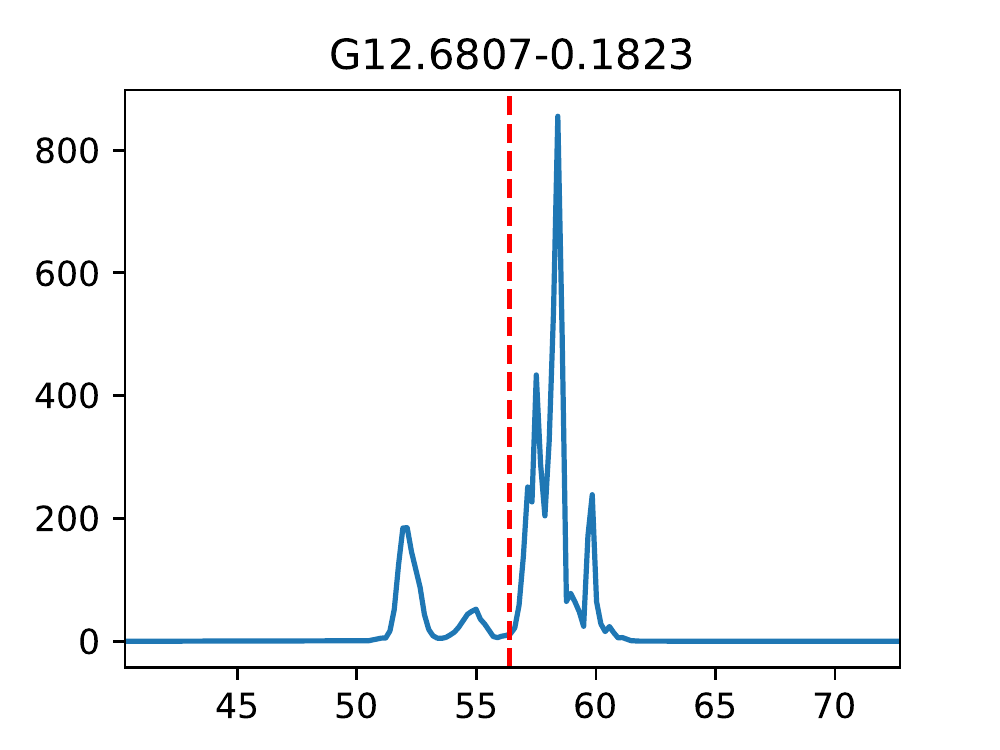}\\ 
\colorbox{white}{\includegraphics[trim={2cm 0 0 0}, scale=0.195]{figs/ylabel_filler.pdf}}
\includegraphics[trim={0.0cm 0.3cm 0.6cm 0.3cm}, clip, width = 0.235\textwidth]{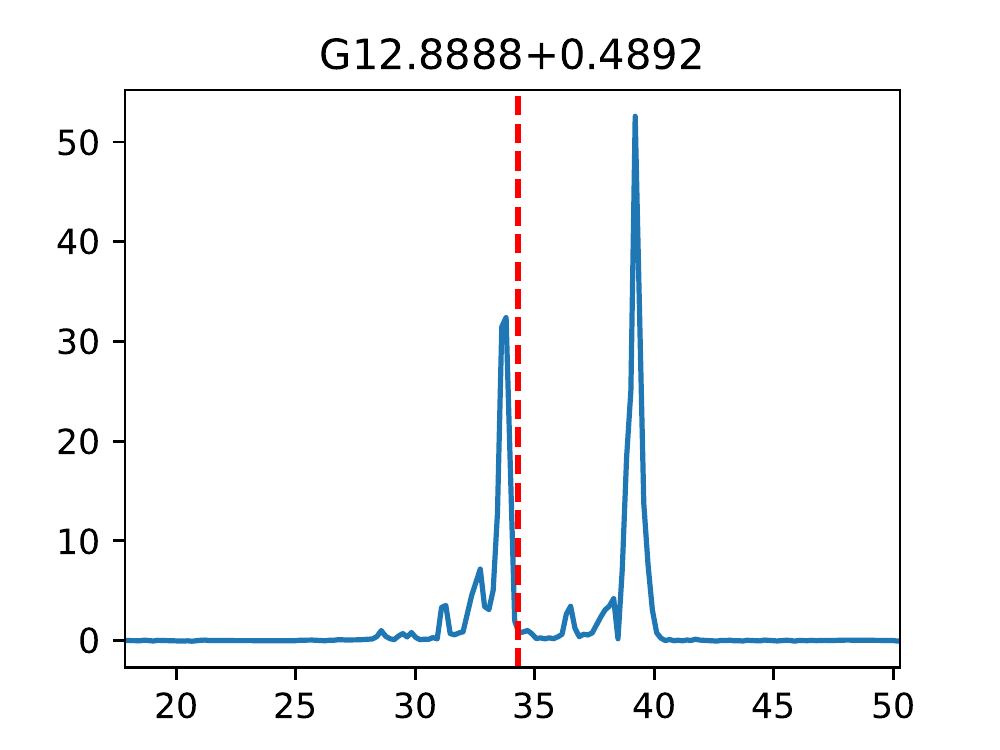}
\includegraphics[trim={0.0cm 0.3cm 0.6cm 0.3cm}, clip, width = 0.235\textwidth]{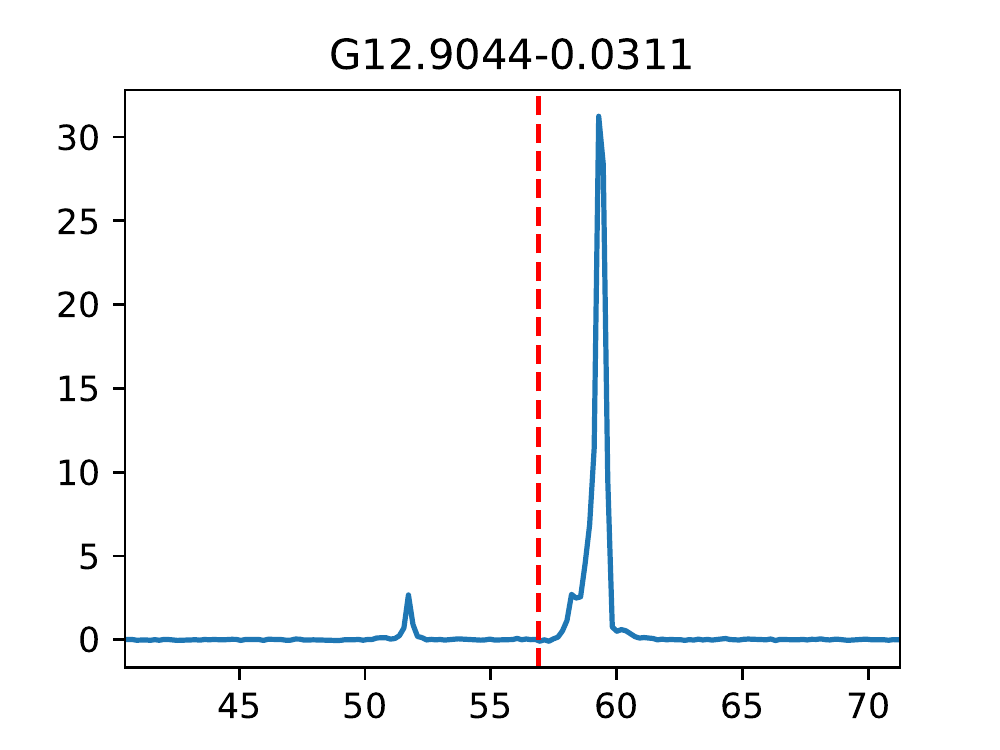}
\includegraphics[trim={0.0cm 0.3cm 0.6cm 0.3cm}, clip, width = 0.235\textwidth]{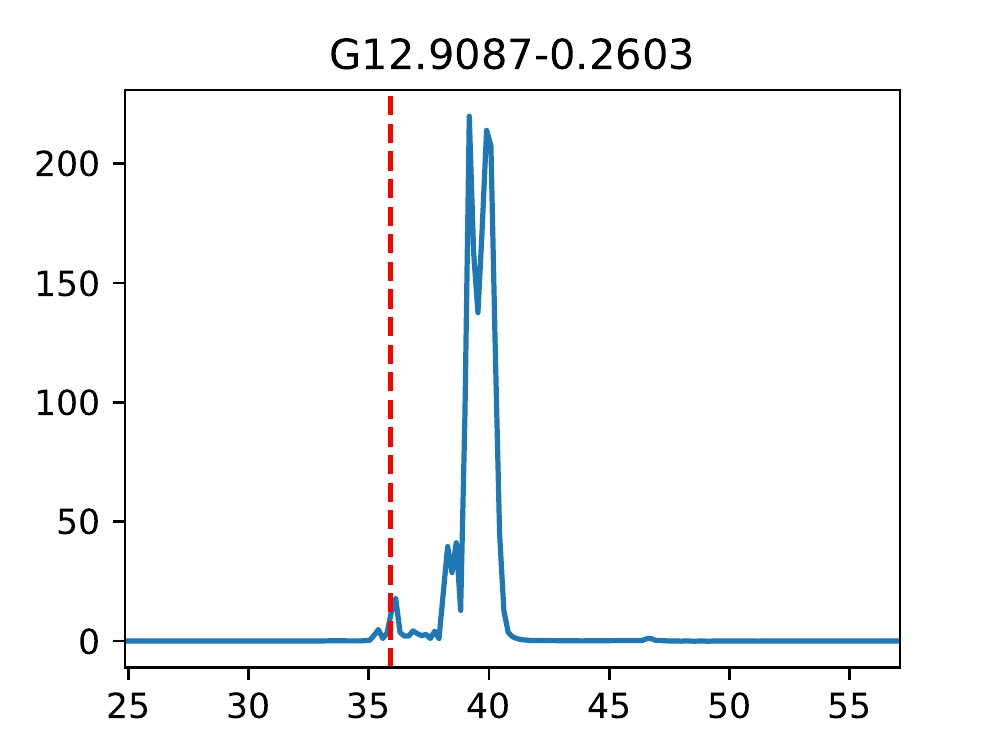}
\includegraphics[trim={0.0cm 0.3cm 0.6cm 0.3cm}, clip, width = 0.235\textwidth]{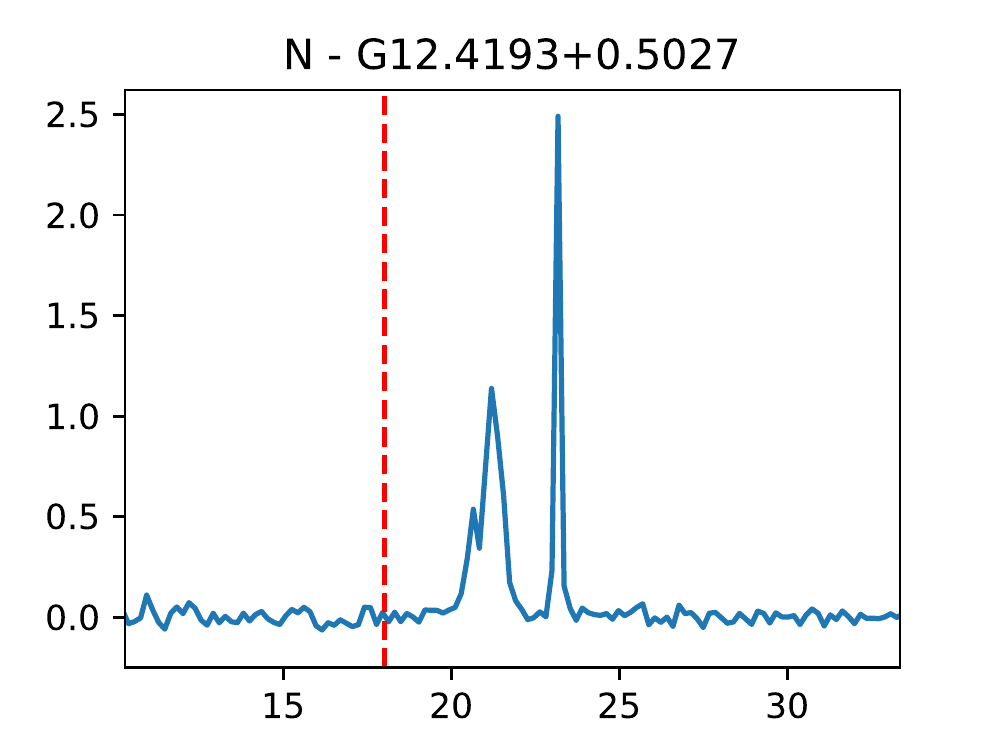}\\ 
\colorbox{white}{\includegraphics[trim={2cm 0 0 0}, scale=0.20]{figs/ylabel.pdf}}
\includegraphics[trim={0.0cm 0.3cm 0.6cm 0.3cm}, clip, width = 0.235\textwidth]{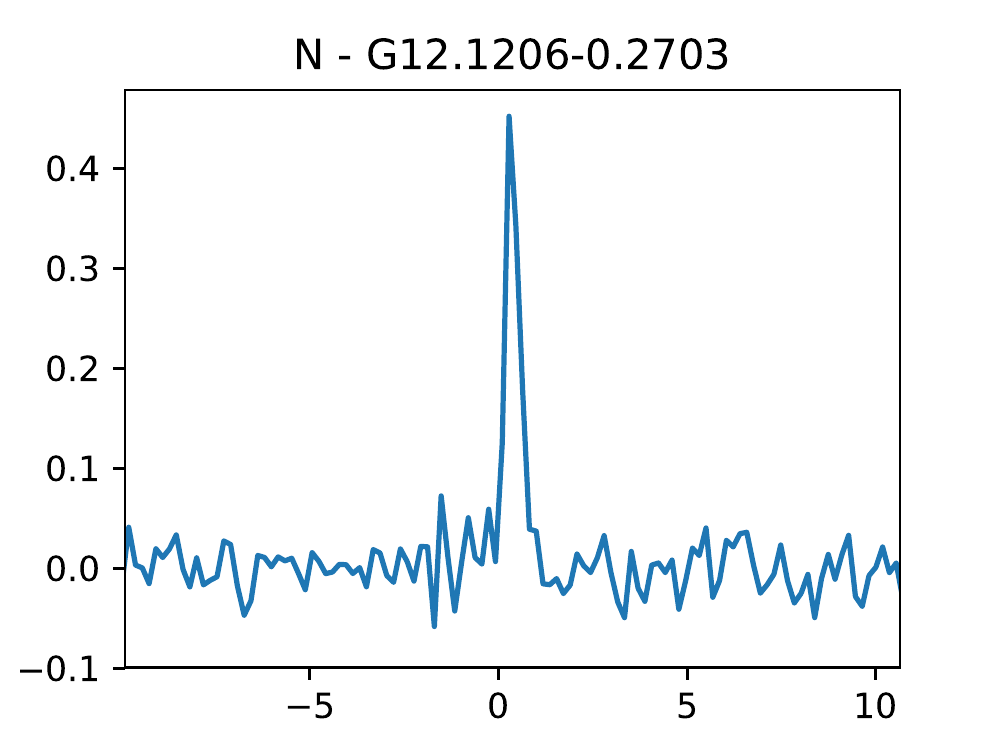}
\includegraphics[trim={0.0cm 0.3cm 0.6cm 0.3cm}, clip, width = 0.235\textwidth]{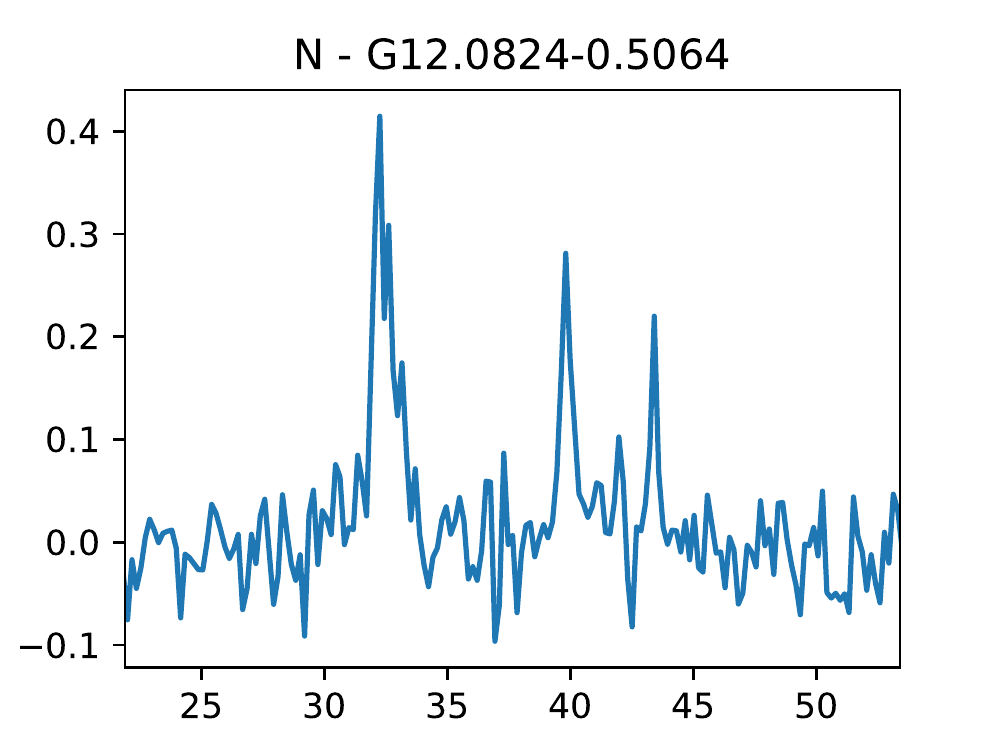}
\includegraphics[trim={0.0cm 0.3cm 0.6cm 0.3cm}, clip, width = 0.235\textwidth]{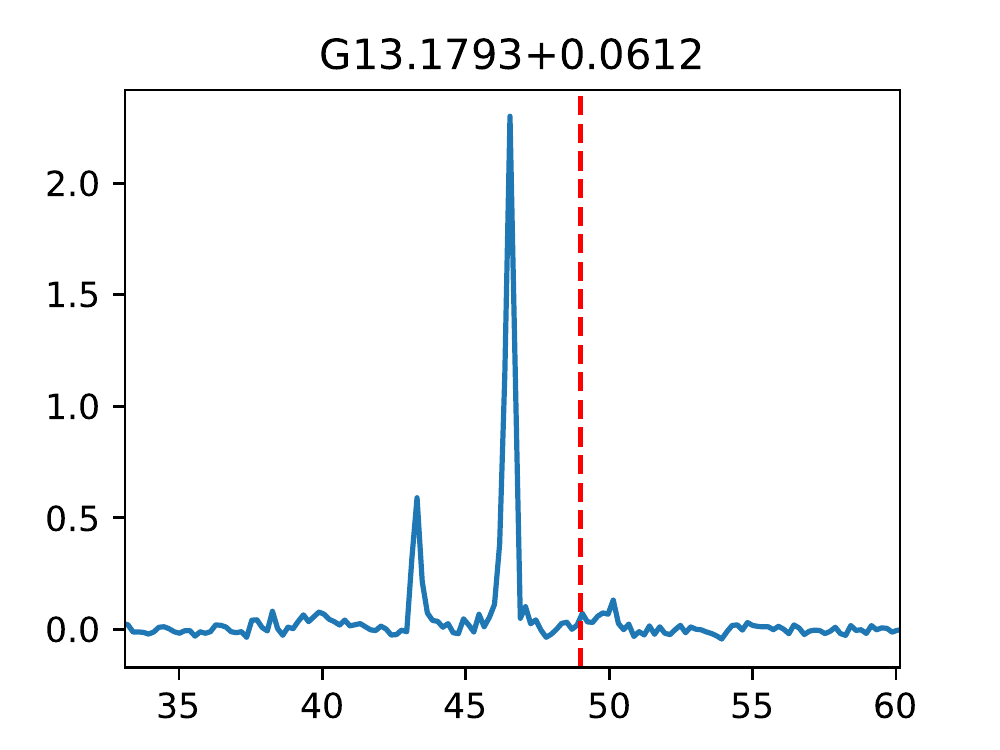}
\includegraphics[trim={0.0cm 0.3cm 0.6cm 0.3cm}, clip, width = 0.235\textwidth]{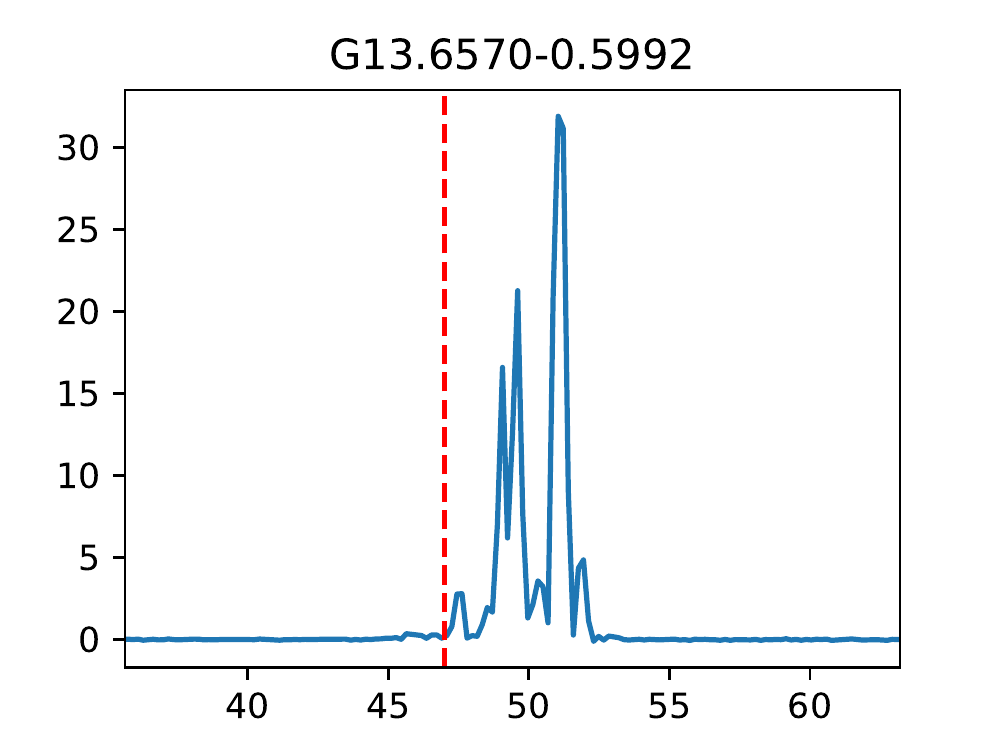}\\
\end{tabular}
\caption{Continued from Fig.~\ref{fig:maser_many_spec}.}
\label{fig:masers_168}
\end{figure*}

%Table 8
\begin{figure*}[!h]
\begin{tabular}{ccccc}
\colorbox{white}{\includegraphics[trim={2cm 0 0 0}, scale=0.195]{figs/ylabel_filler.pdf}}
\includegraphics[trim={0.0cm 0.3cm 0.6cm 0.3cm}, clip, width = 0.235\textwidth]{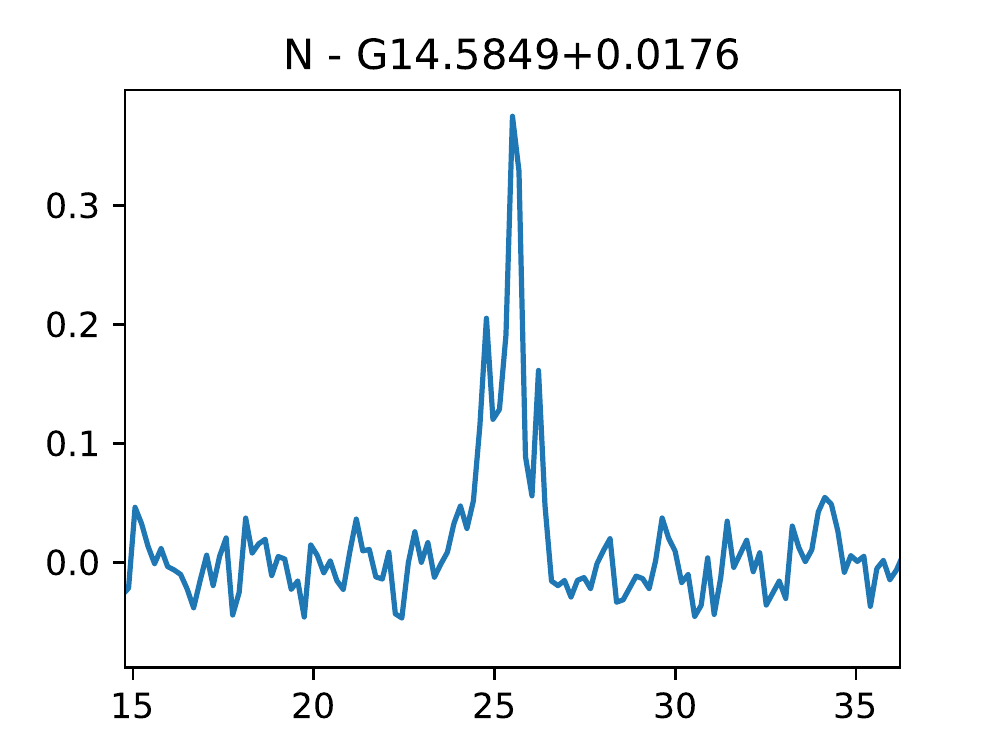}
\includegraphics[trim={0.0cm 0.3cm 0.6cm 0.3cm}, clip, width = 0.235\textwidth]{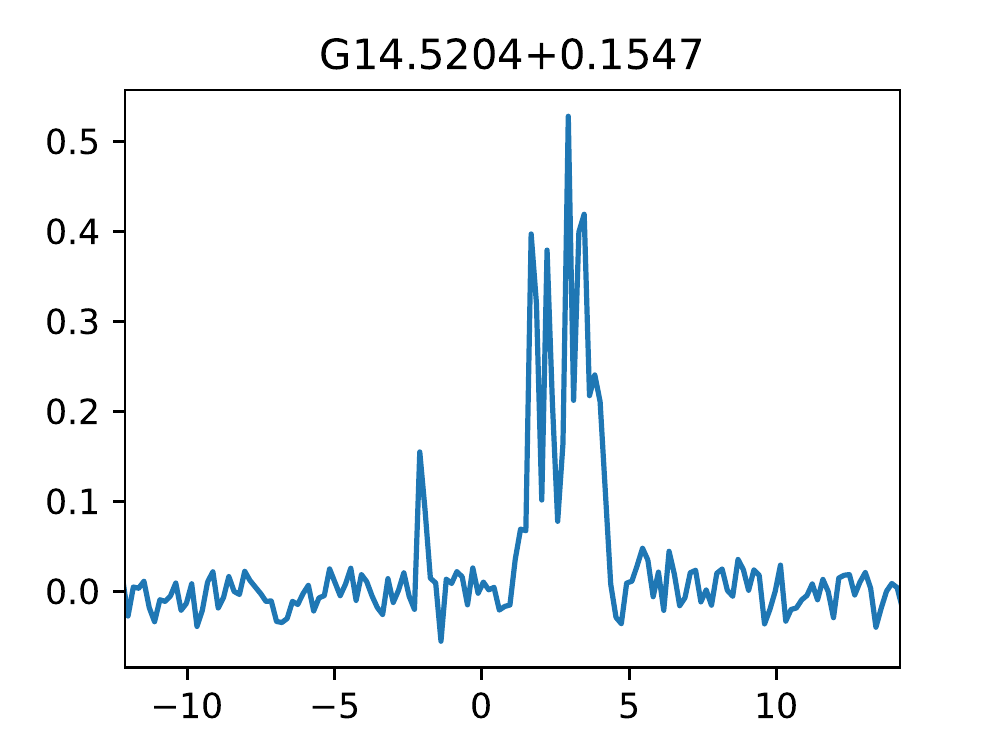}
\includegraphics[trim={0.0cm 0.3cm 0.6cm 0.3cm}, clip, width = 0.235\textwidth]{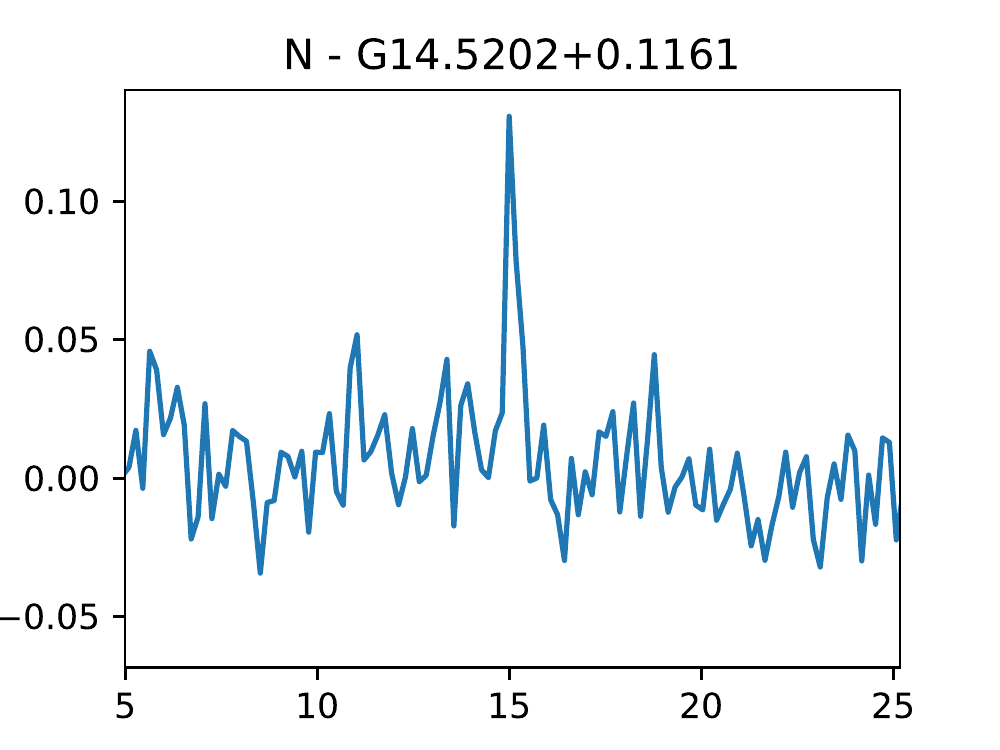}
\includegraphics[trim={0.0cm 0.3cm 0.6cm 0.3cm}, clip, width = 0.235\textwidth]{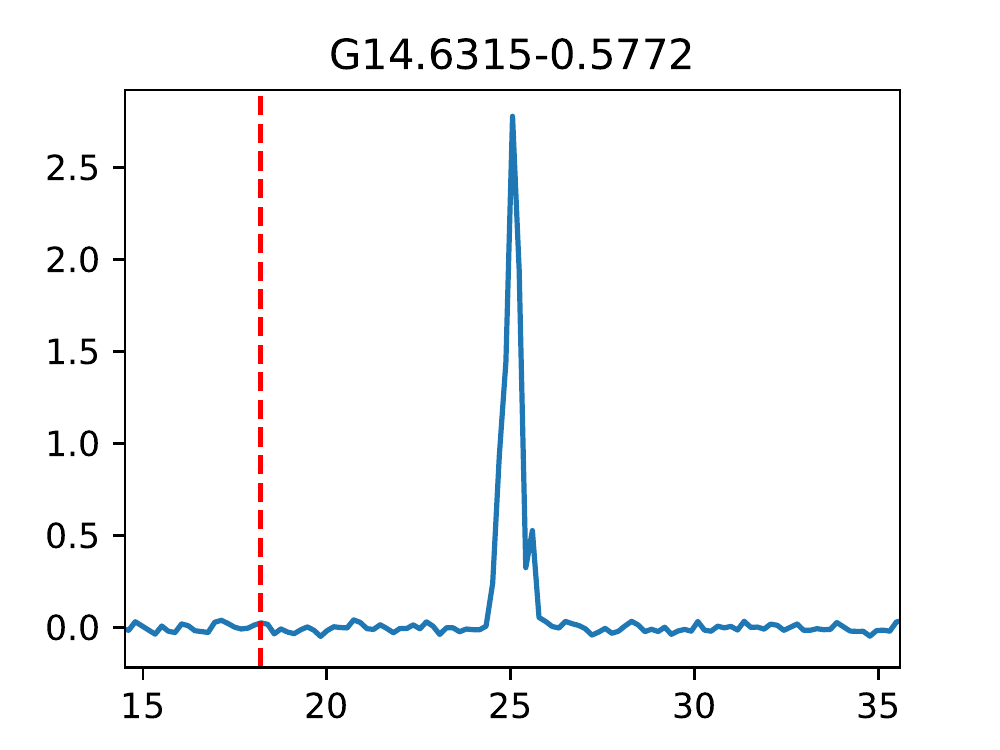}\\ 
\colorbox{white}{\includegraphics[trim={2cm 0 0 0}, scale=0.195]{figs/ylabel_filler.pdf}}
\includegraphics[trim={0.0cm 0.3cm 0.6cm 0.3cm}, clip, width = 0.235\textwidth]{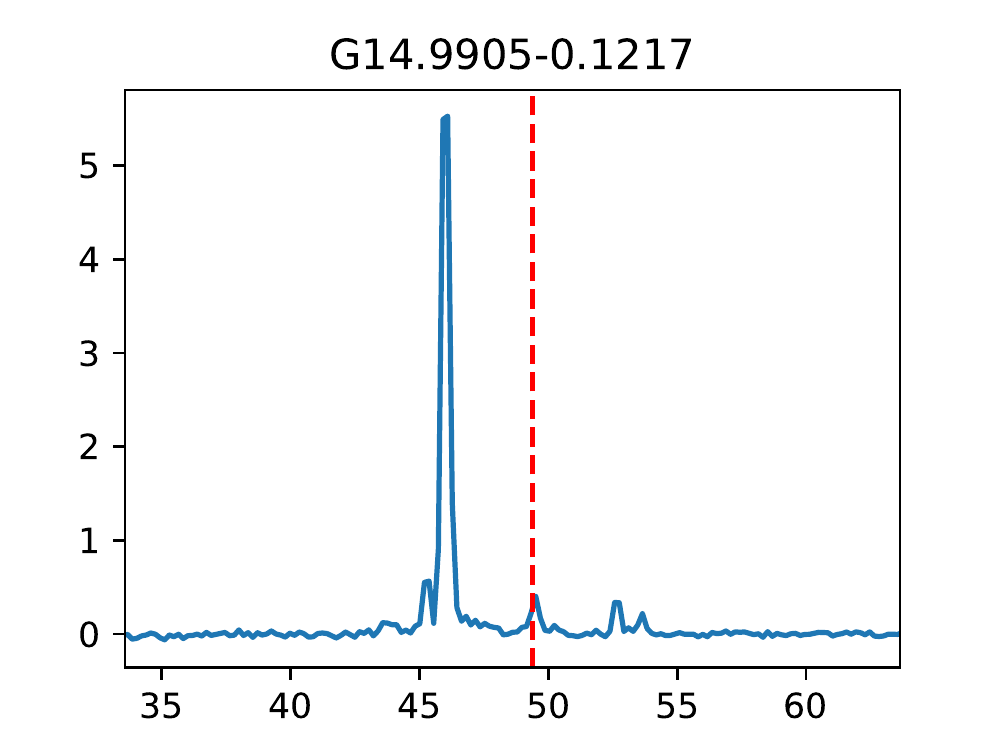}
\includegraphics[trim={0.0cm 0.3cm 0.6cm 0.3cm}, clip, width = 0.235\textwidth]{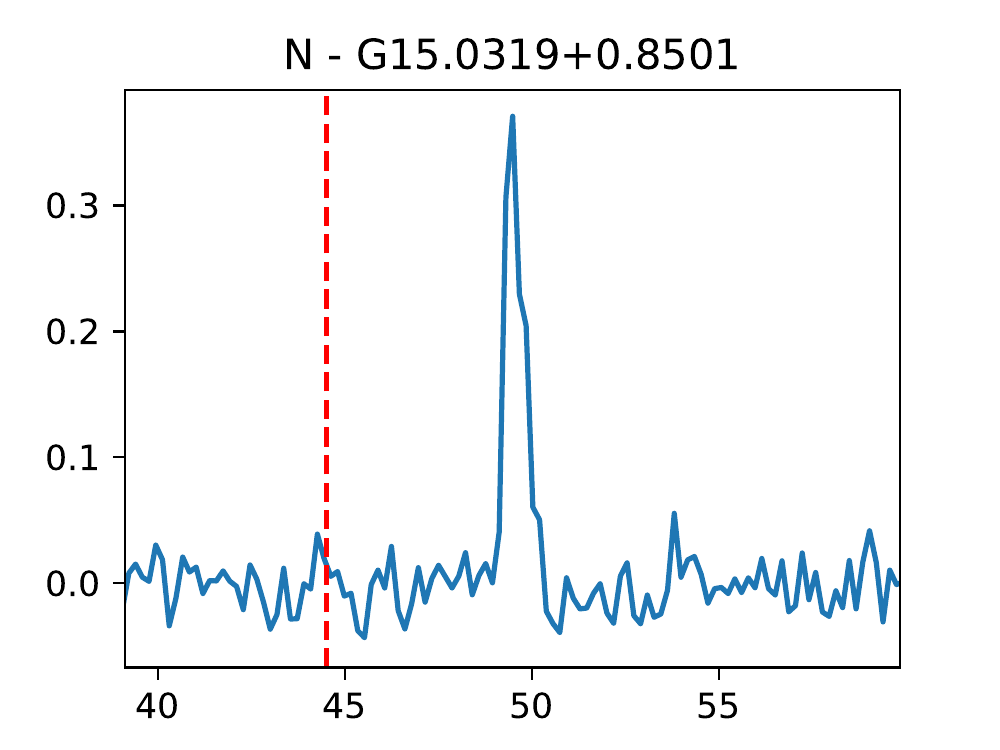}
\includegraphics[trim={0.0cm 0.3cm 0.6cm 0.3cm}, clip, width = 0.235\textwidth]{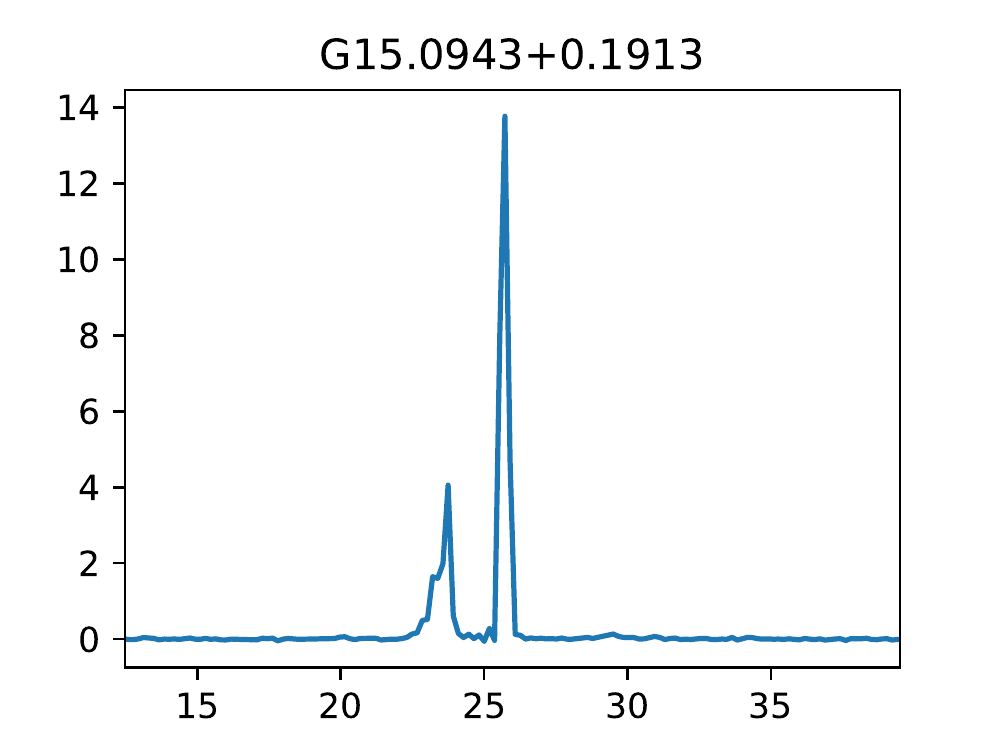}
\includegraphics[trim={0.0cm 0.3cm 0.6cm 0.3cm}, clip, width = 0.235\textwidth]{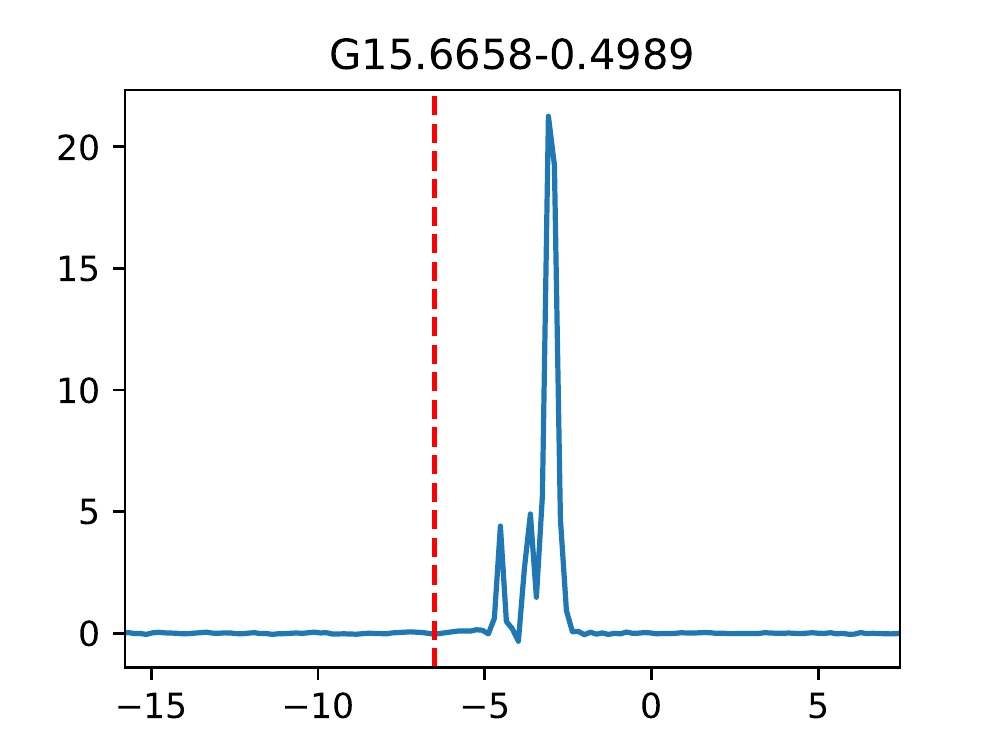}\\ 
\colorbox{white}{\includegraphics[trim={2cm 0 0 0}, scale=0.20]{figs/ylabel.pdf}}
\includegraphics[trim={0.0cm 0.3cm 0.6cm 0.3cm}, clip, width = 0.235\textwidth]{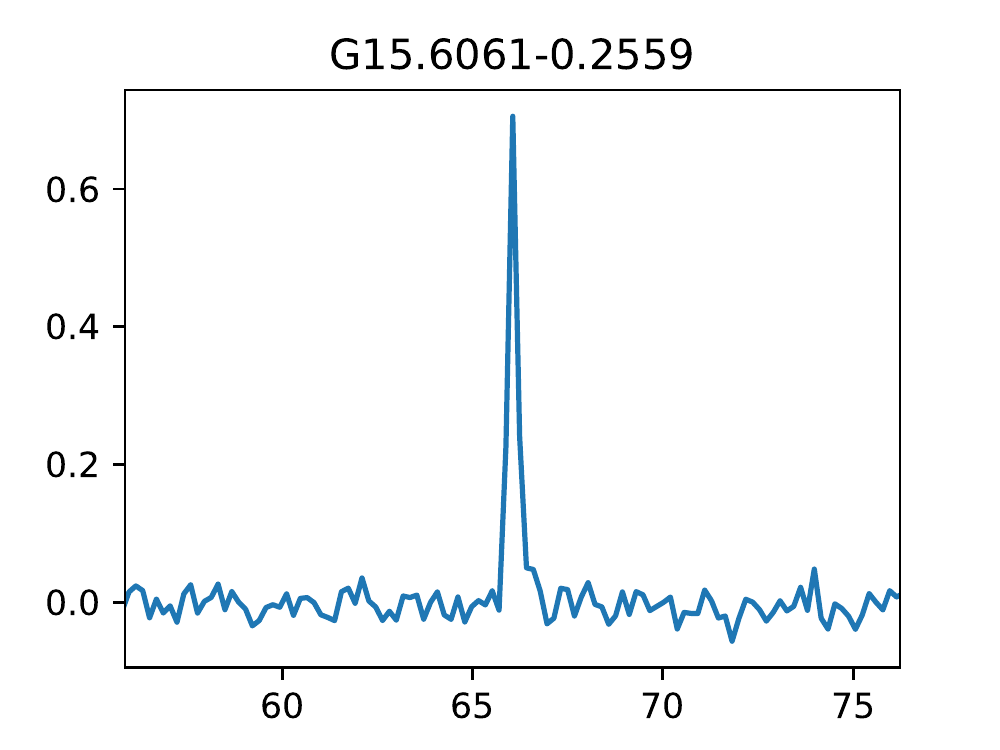}
\includegraphics[trim={0.0cm 0.3cm 0.6cm 0.3cm}, clip, width = 0.235\textwidth]{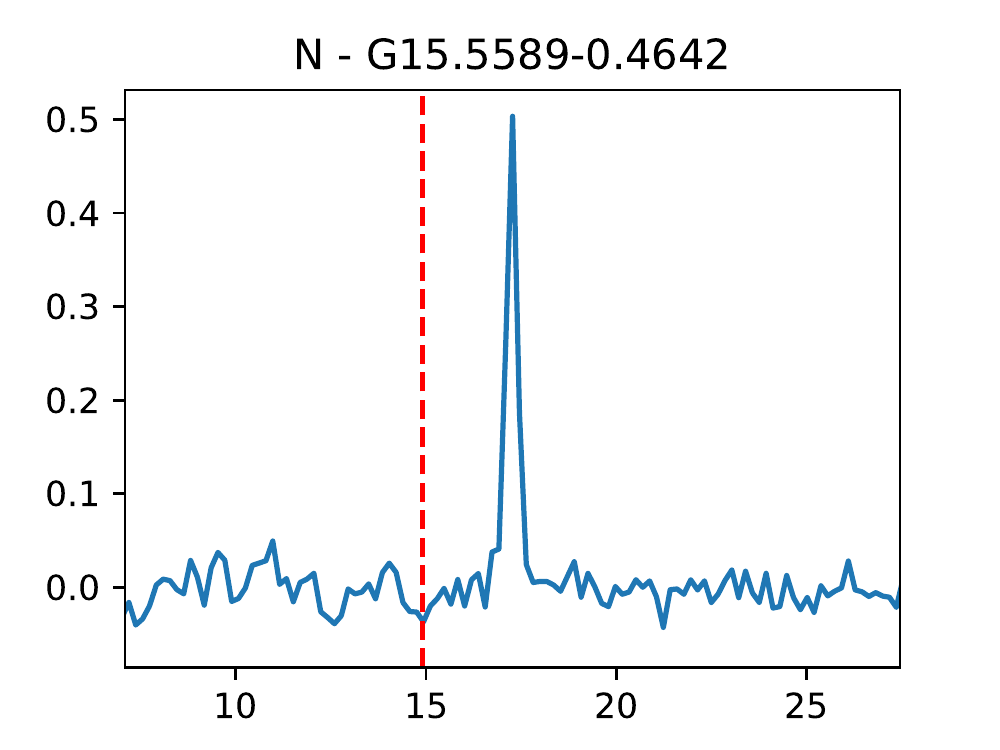}
\includegraphics[trim={0.0cm 0.3cm 0.6cm 0.3cm}, clip, width = 0.235\textwidth]{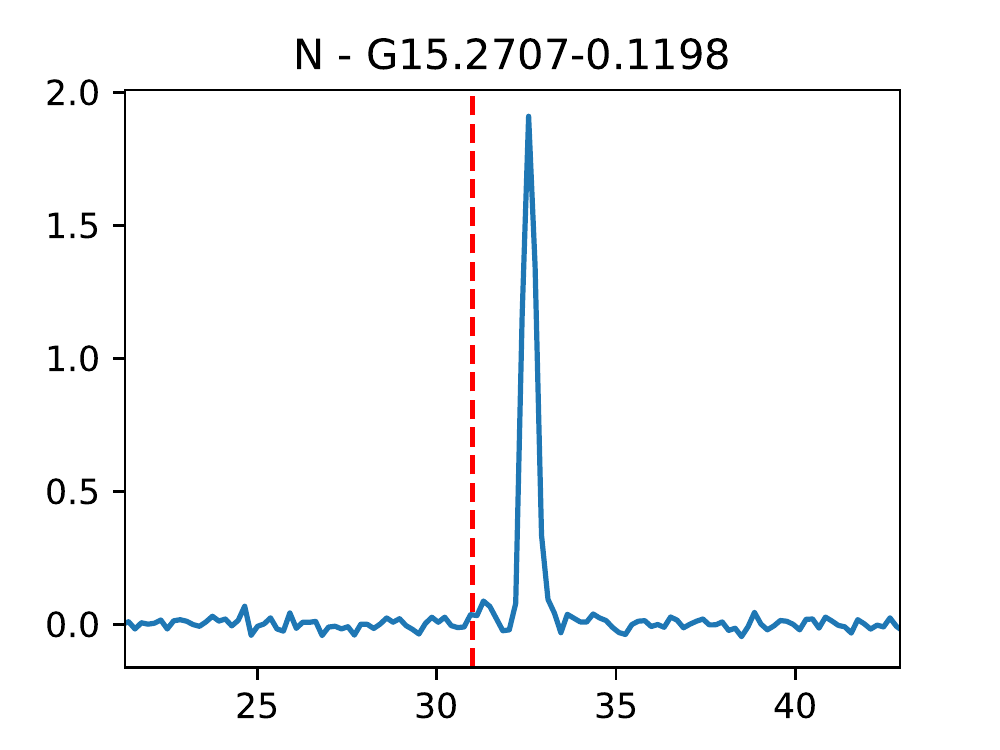}
\includegraphics[trim={0.0cm 0.3cm 0.6cm 0.3cm}, clip, width = 0.235\textwidth]{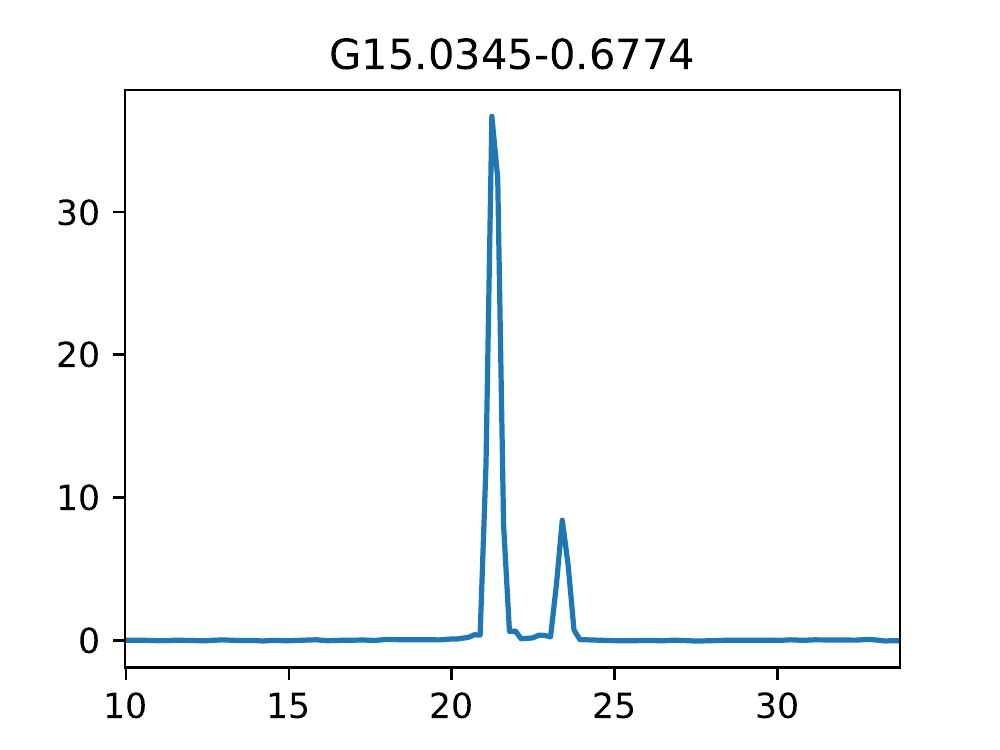}\\
\end{tabular}
\caption{Continued from Fig.~\ref{fig:maser_many_spec}.}
\label{fig:masers_192}
\end{figure*}

%Table 9
\begin{figure*}[!h]
\begin{tabular}{ccccc}
\colorbox{white}{\includegraphics[trim={2cm 0 0 0}, scale=0.195]{figs/ylabel_filler.pdf}}
\includegraphics[trim={0.0cm 0.3cm 0.6cm 0.3cm}, clip, width = 0.235\textwidth]{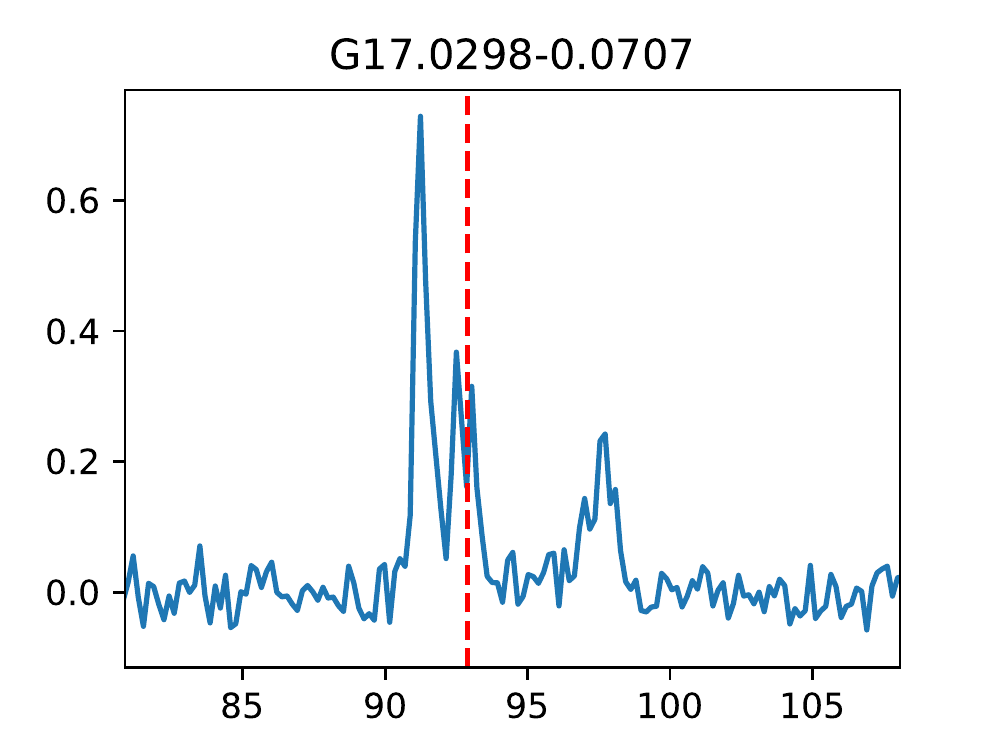}
\includegraphics[trim={0.0cm 0.3cm 0.6cm 0.3cm}, clip, width = 0.235\textwidth]{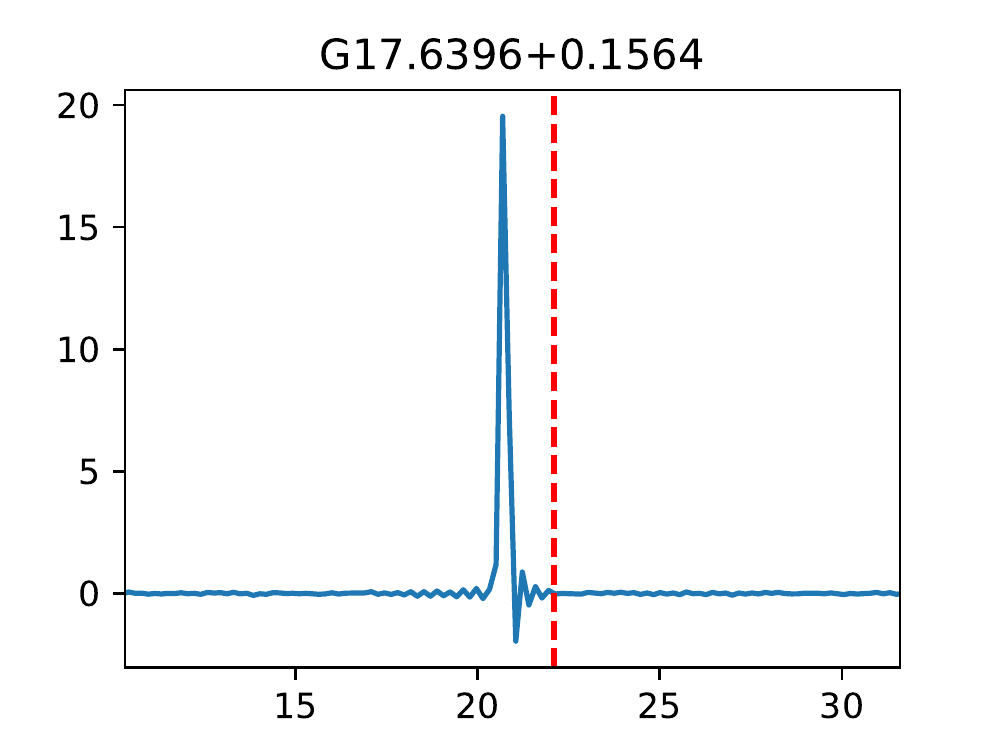}
\includegraphics[trim={0.0cm 0.3cm 0.6cm 0.3cm}, clip, width = 0.235\textwidth]{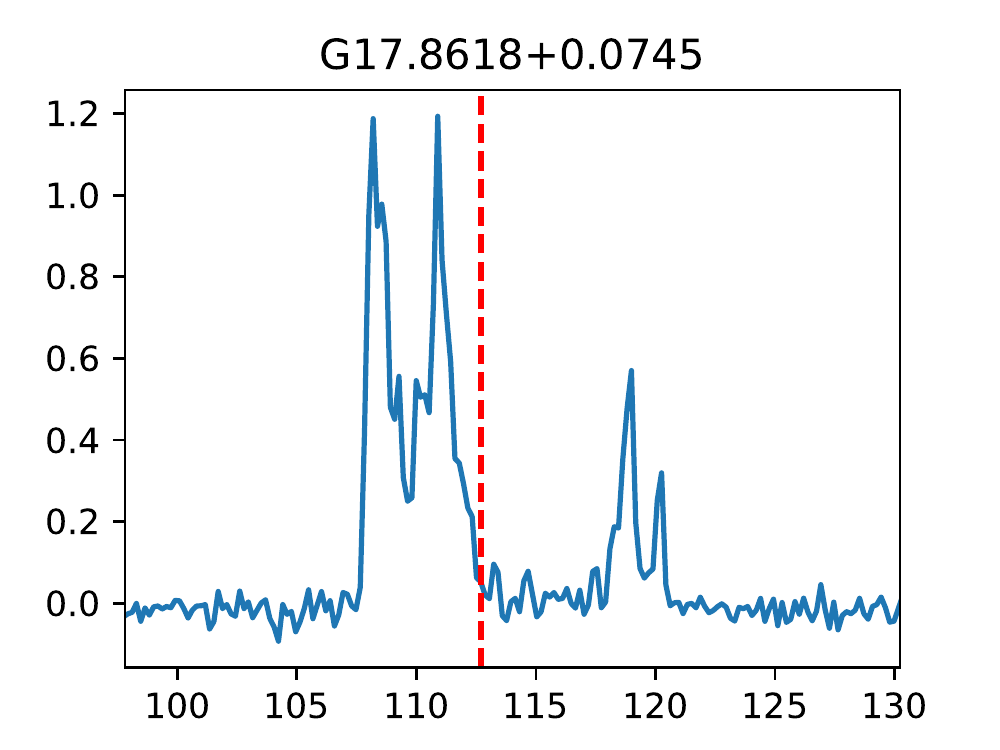}
\includegraphics[trim={0.0cm 0.3cm 0.6cm 0.3cm}, clip, width = 0.235\textwidth]{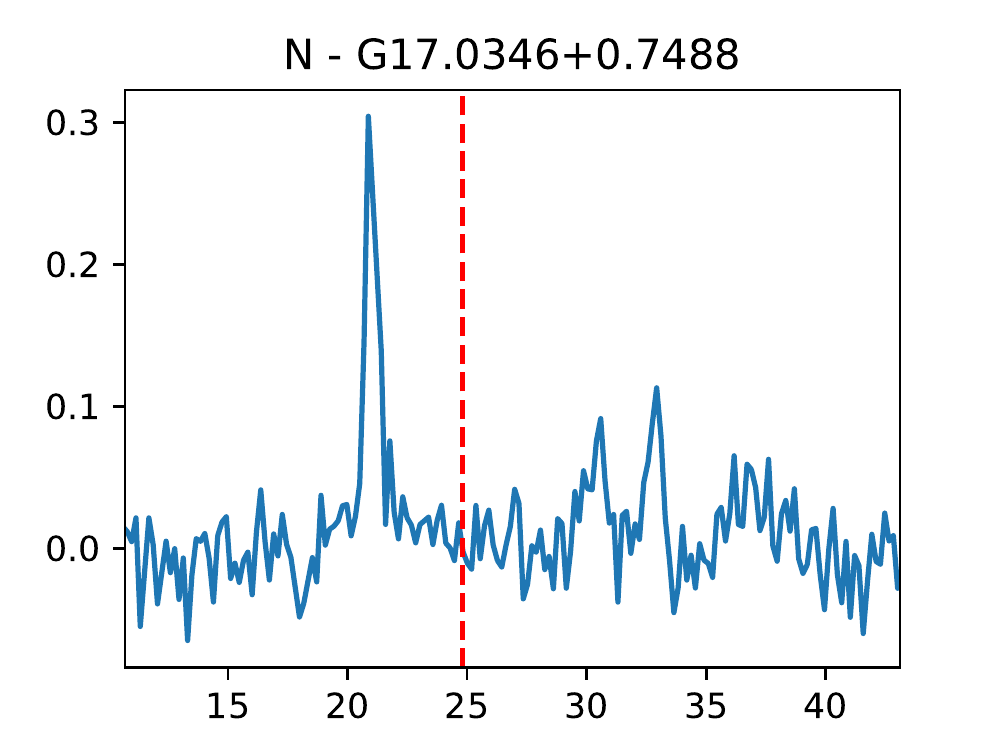}\\ 
\colorbox{white}{\includegraphics[trim={2cm 0 0 0}, scale=0.195]{figs/ylabel_filler.pdf}}
\includegraphics[trim={0.0cm 0.3cm 0.6cm 0.3cm}, clip, width = 0.235\textwidth]{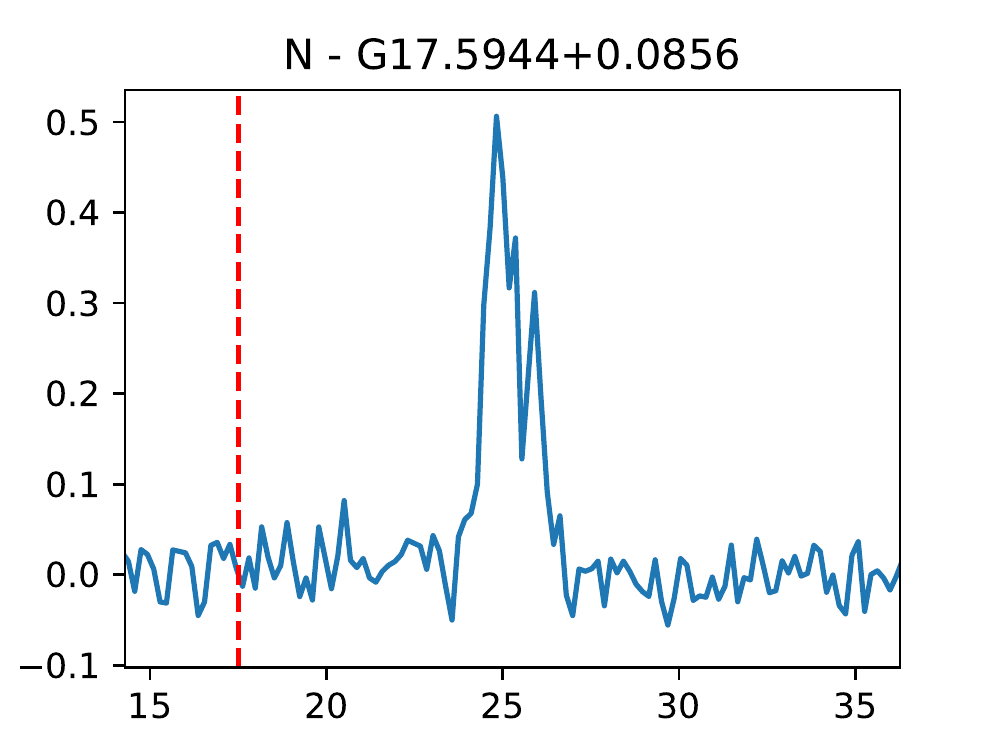}
\includegraphics[trim={0.0cm 0.3cm 0.6cm 0.3cm}, clip, width = 0.235\textwidth]{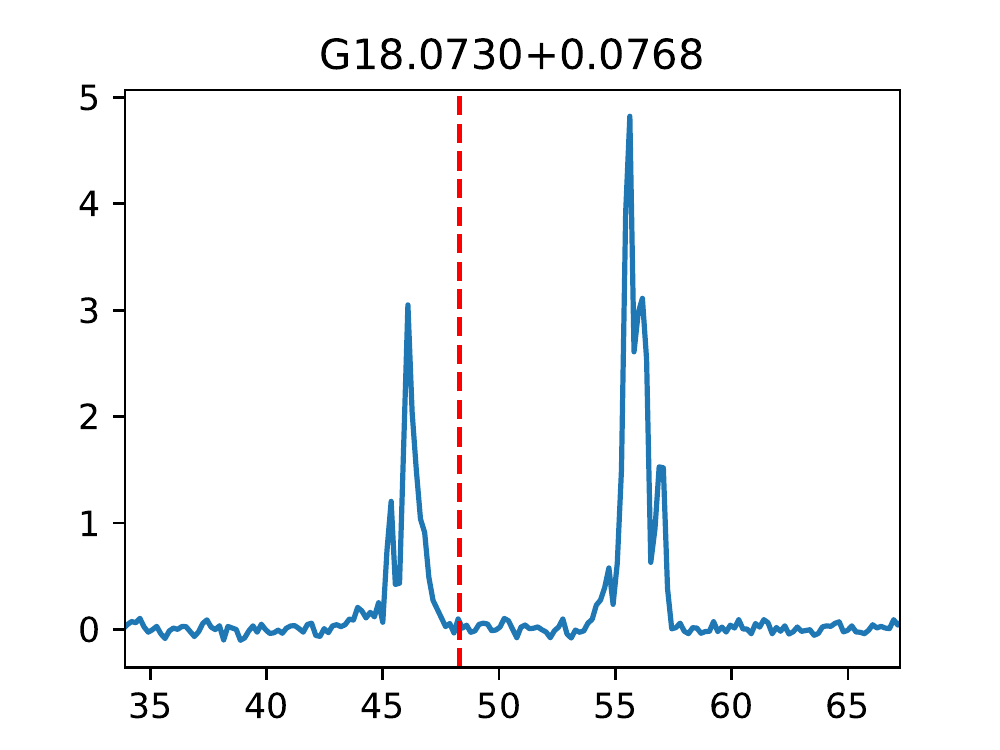}
\includegraphics[trim={0.0cm 0.3cm 0.6cm 0.3cm}, clip, width = 0.235\textwidth]{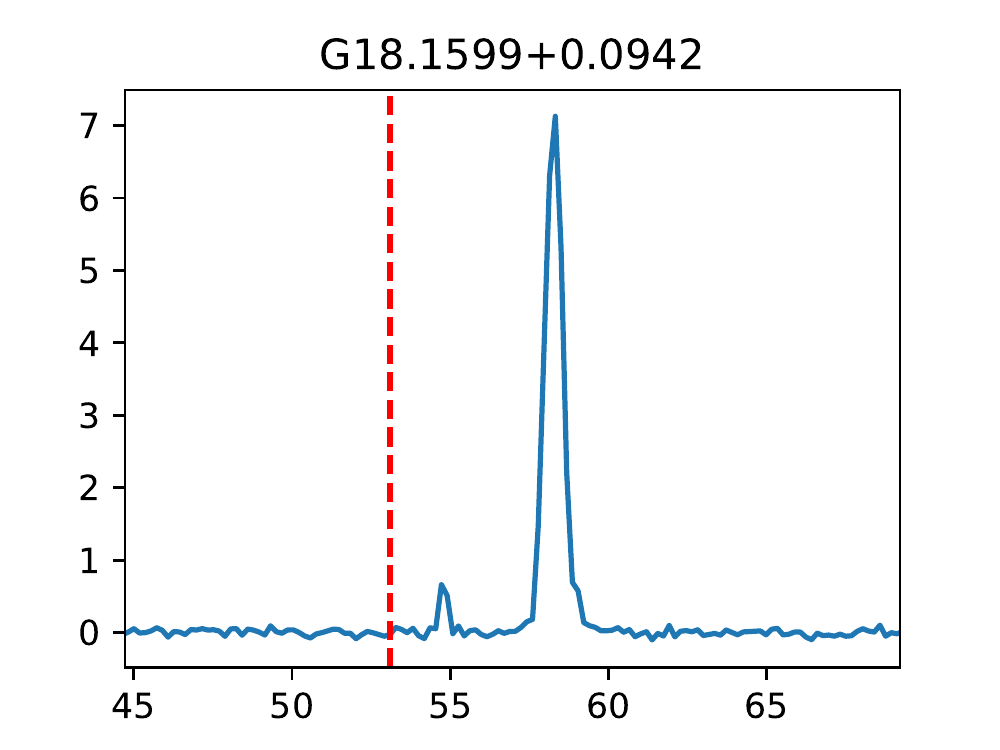}
\includegraphics[trim={0.0cm 0.3cm 0.6cm 0.3cm}, clip, width = 0.235\textwidth]{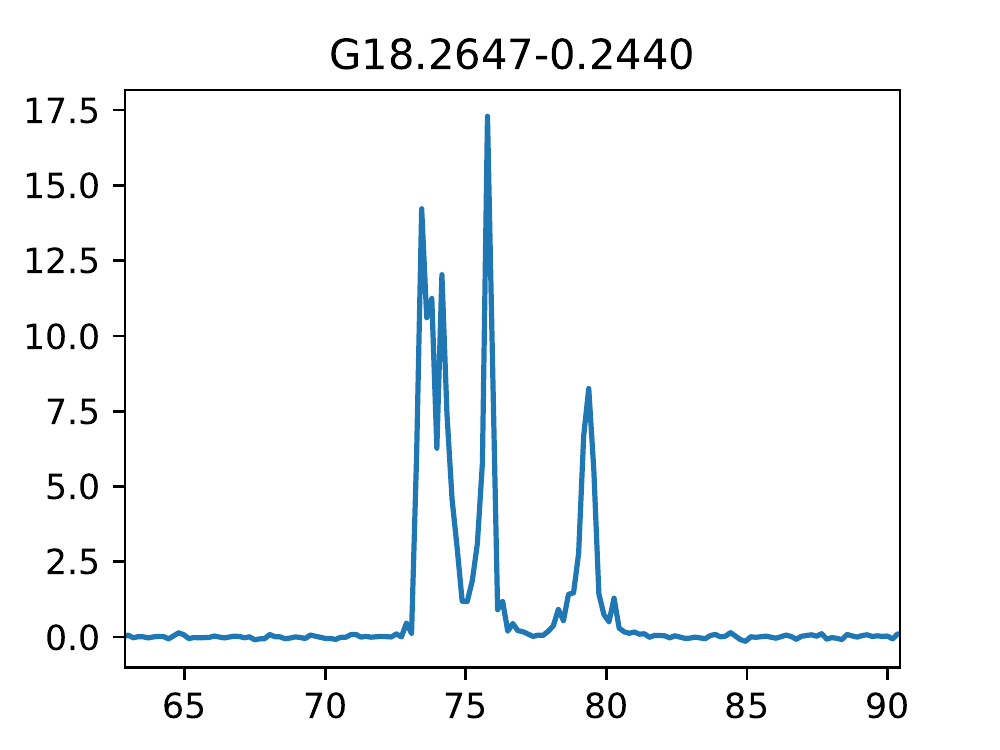}\\ 
\colorbox{white}{\includegraphics[trim={2cm 0 0 0}, scale=0.20]{figs/ylabel.pdf}}
\includegraphics[trim={0.0cm 0.3cm 0.6cm 0.3cm}, clip, width = 0.235\textwidth]{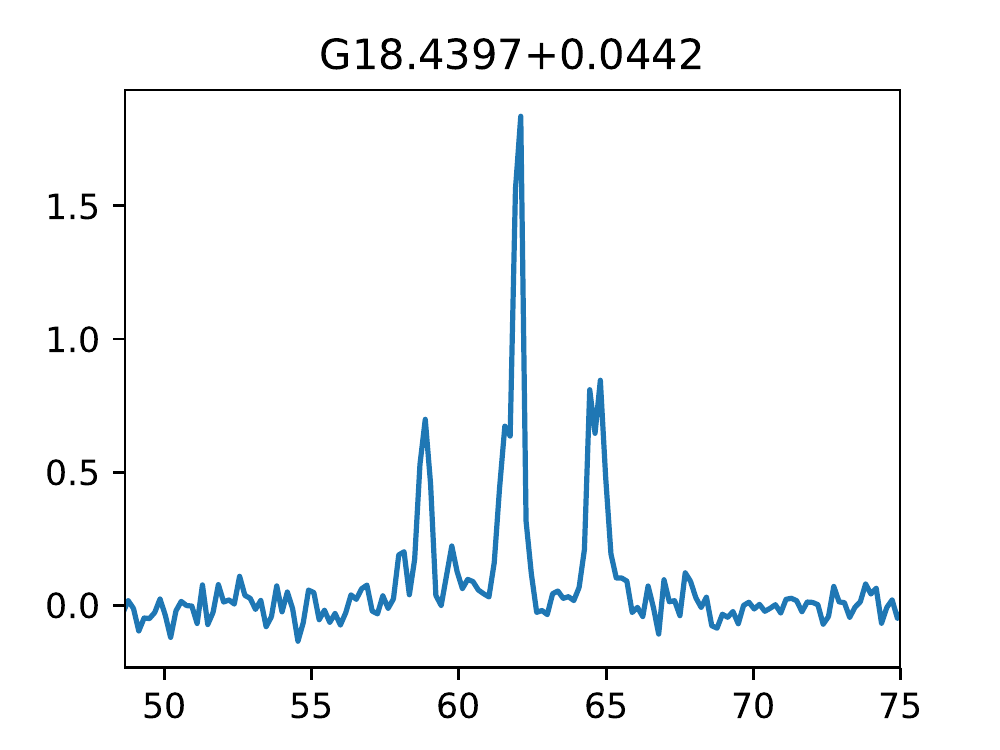}
\includegraphics[trim={0.0cm 0.3cm 0.6cm 0.3cm}, clip, width = 0.235\textwidth]{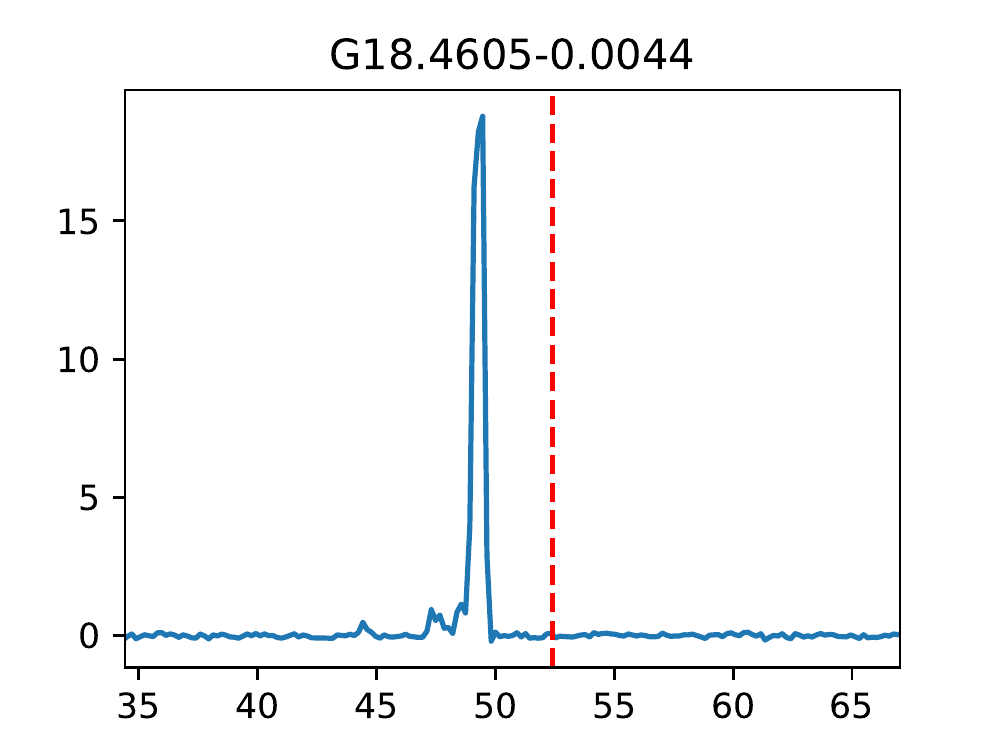}
\includegraphics[trim={0.0cm 0.3cm 0.6cm 0.3cm}, clip, width = 0.235\textwidth]{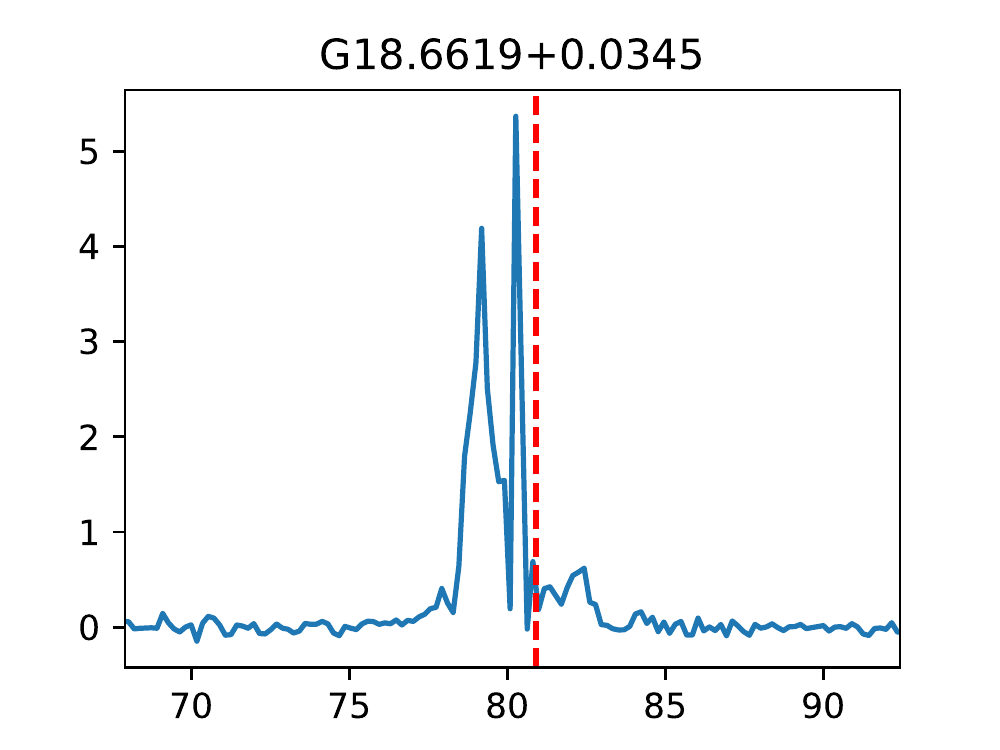}
\includegraphics[trim={0.0cm 0.3cm 0.6cm 0.3cm}, clip, width = 0.235\textwidth]{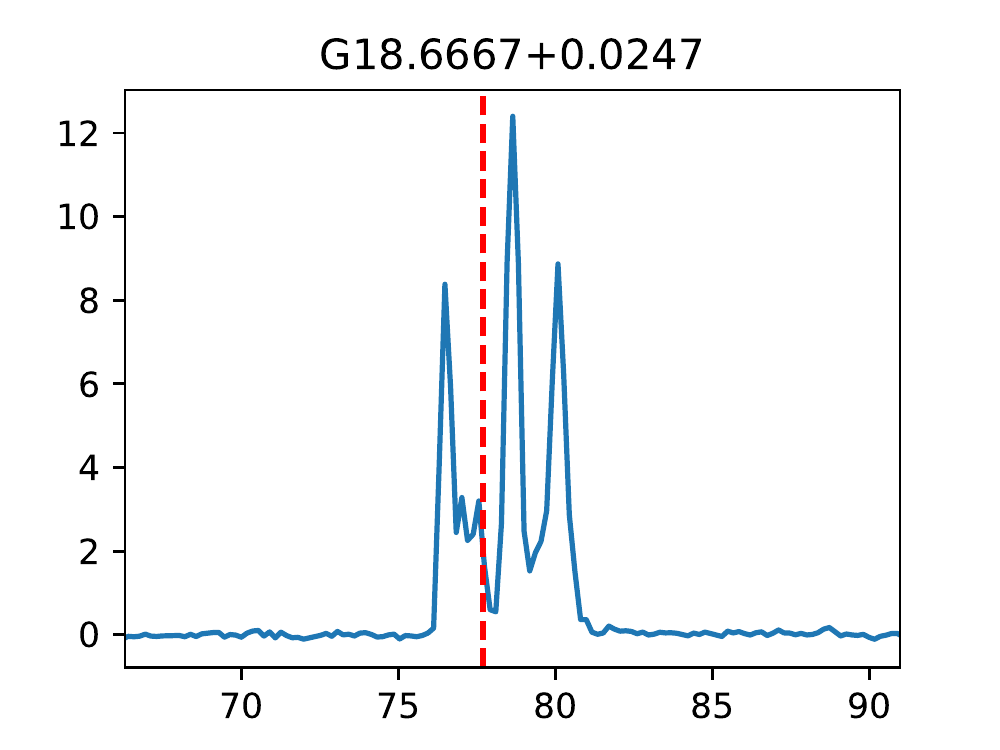}\\
\end{tabular}
\caption{Continued from Fig.~\ref{fig:maser_many_spec}.}
\label{fig:masers_216}
\end{figure*}

%Table 10
\begin{figure*}[!h]
\begin{tabular}{ccccc}
\colorbox{white}{\includegraphics[trim={2cm 0 0 0}, scale=0.195]{figs/ylabel_filler.pdf}}
\includegraphics[trim={0.0cm 0.3cm 0.6cm 0.3cm}, clip, width = 0.235\textwidth]{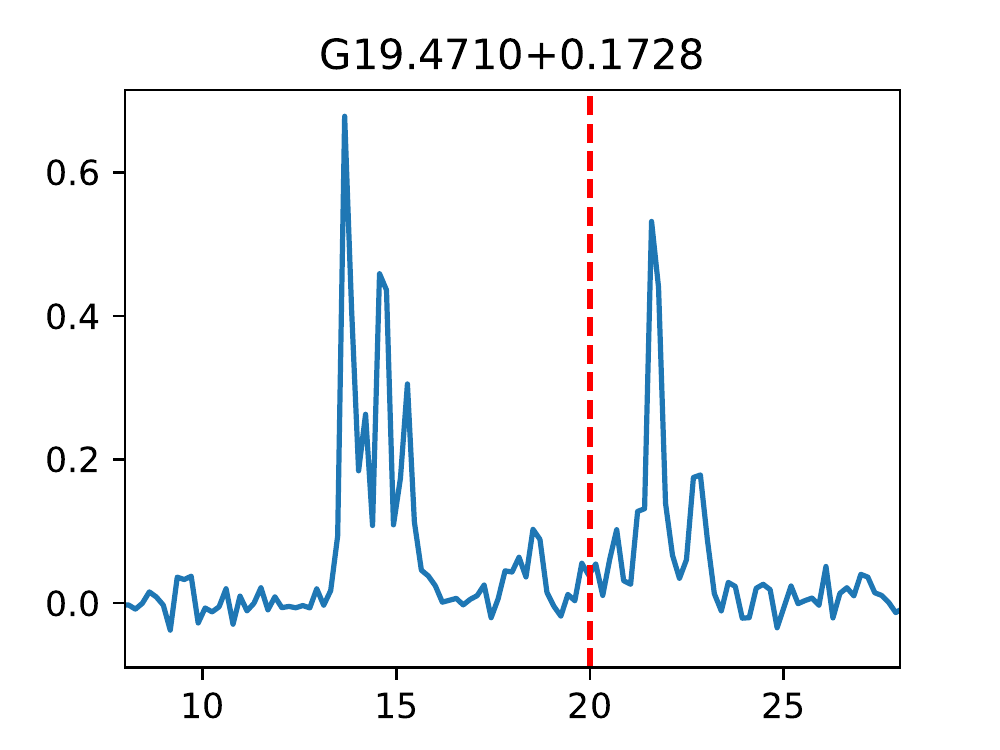}
\includegraphics[trim={0.0cm 0.3cm 0.6cm 0.3cm}, clip, width = 0.235\textwidth]{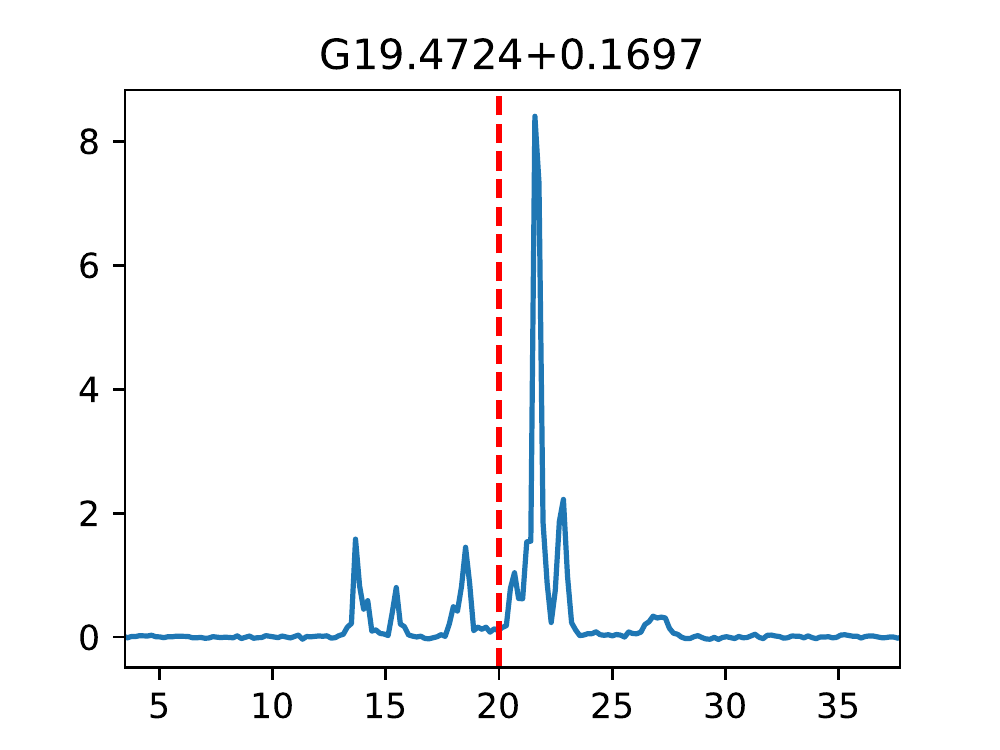}
\includegraphics[trim={0.0cm 0.3cm 0.6cm 0.3cm}, clip, width = 0.235\textwidth]{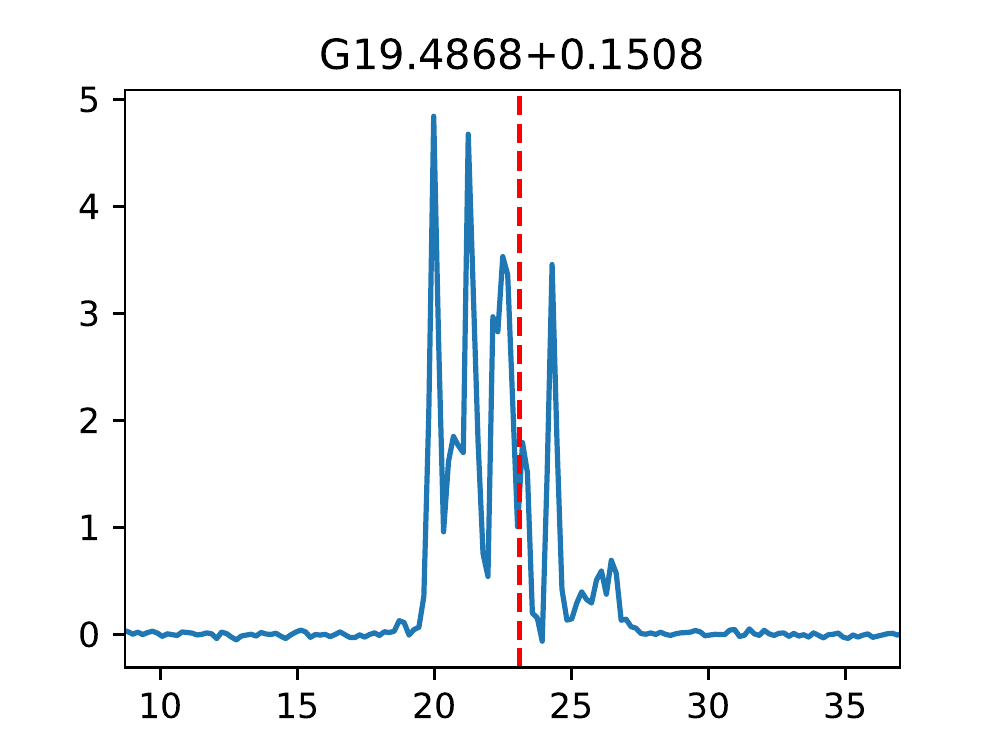}
\includegraphics[trim={0.0cm 0.3cm 0.6cm 0.3cm}, clip, width = 0.235\textwidth]{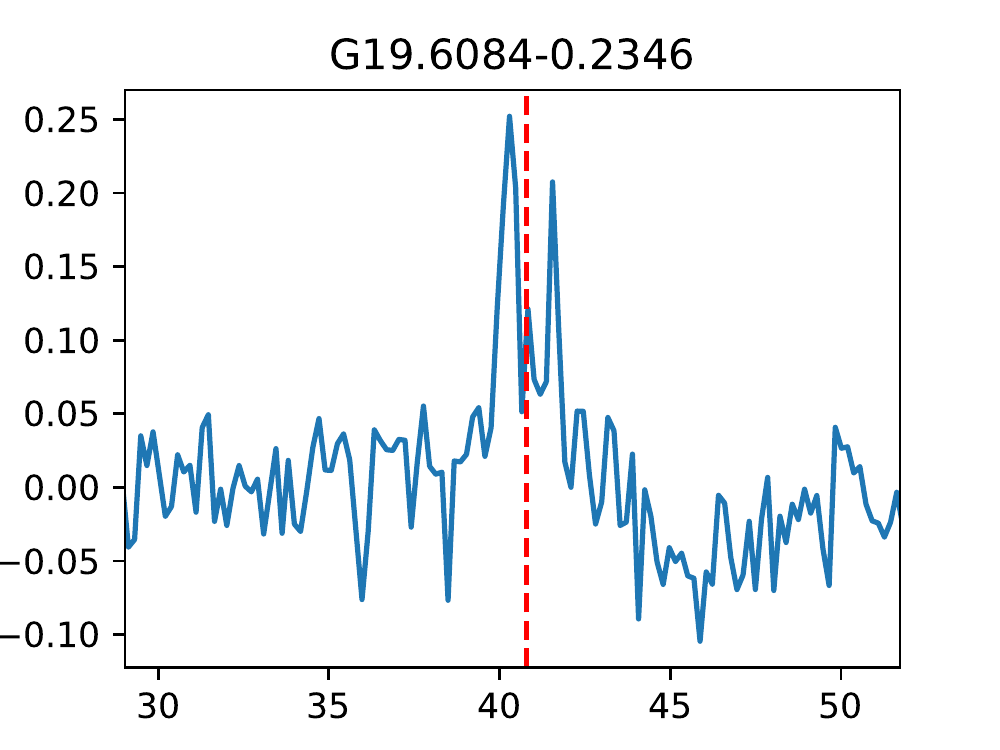}\\ 
\colorbox{white}{\includegraphics[trim={2cm 0 0 0}, scale=0.195]{figs/ylabel_filler.pdf}}
\includegraphics[trim={0.0cm 0.3cm 0.6cm 0.3cm}, clip, width = 0.235\textwidth]{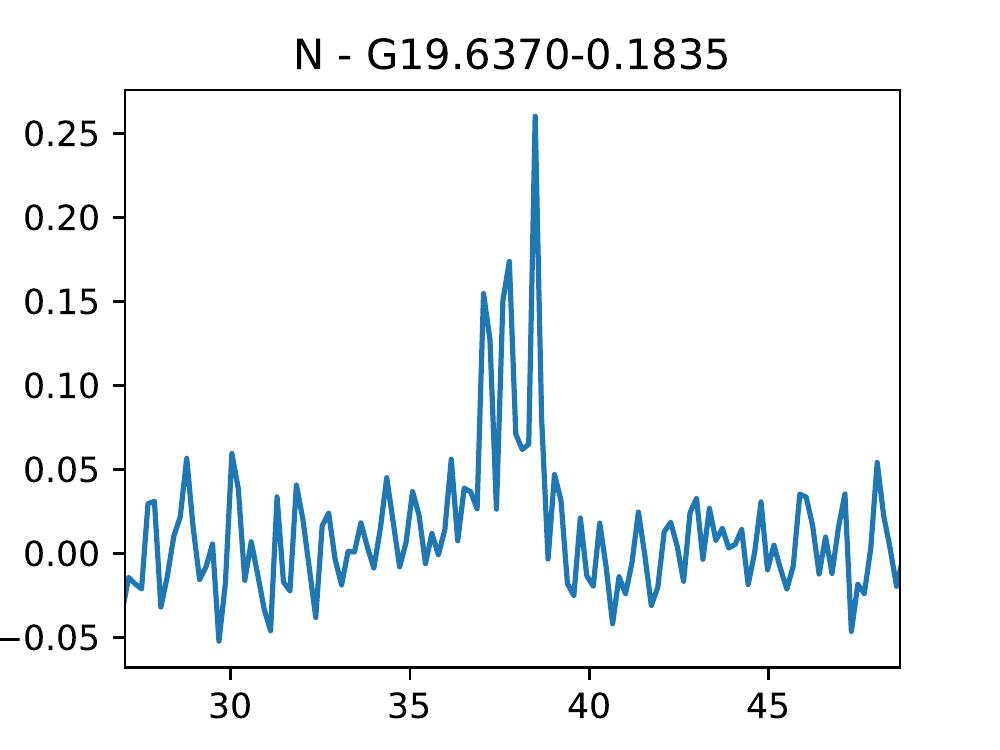}
\includegraphics[trim={0.0cm 0.3cm 0.6cm 0.3cm}, clip, width = 0.235\textwidth]{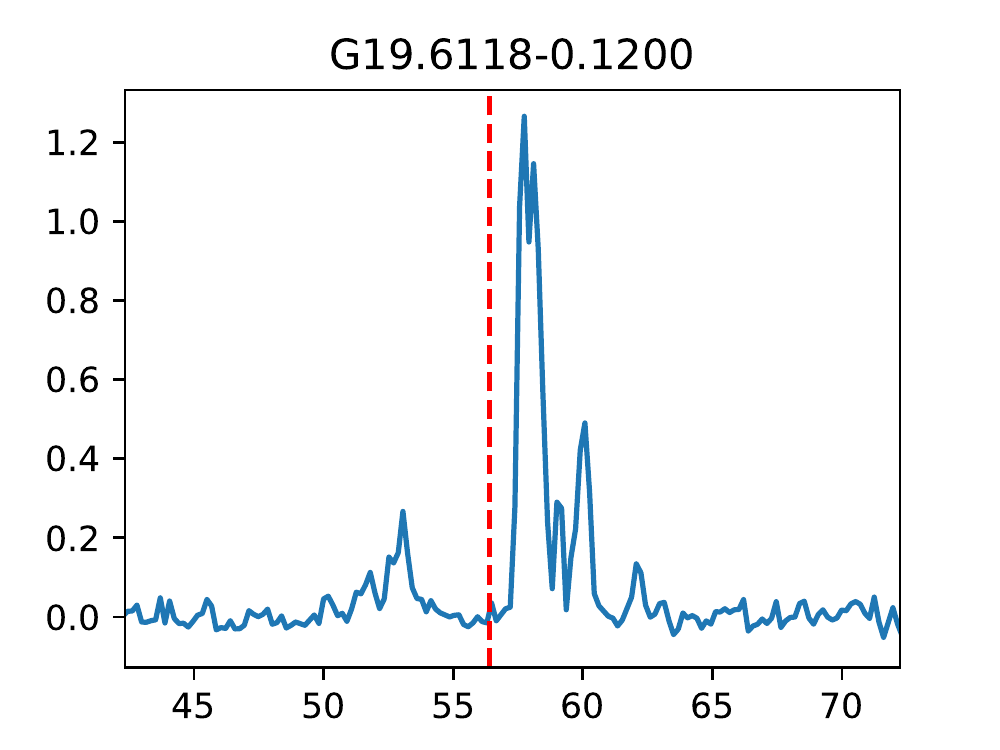}
\includegraphics[trim={0.0cm 0.3cm 0.6cm 0.3cm}, clip, width = 0.235\textwidth]{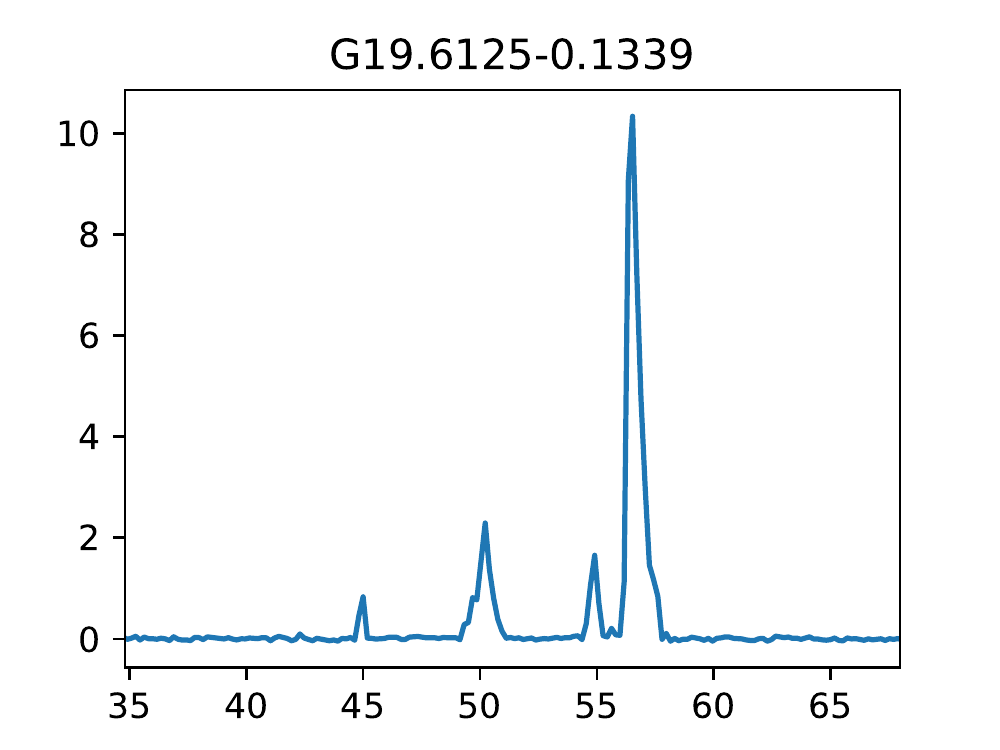}
\includegraphics[trim={0.0cm 0.3cm 0.6cm 0.3cm}, clip, width = 0.235\textwidth]{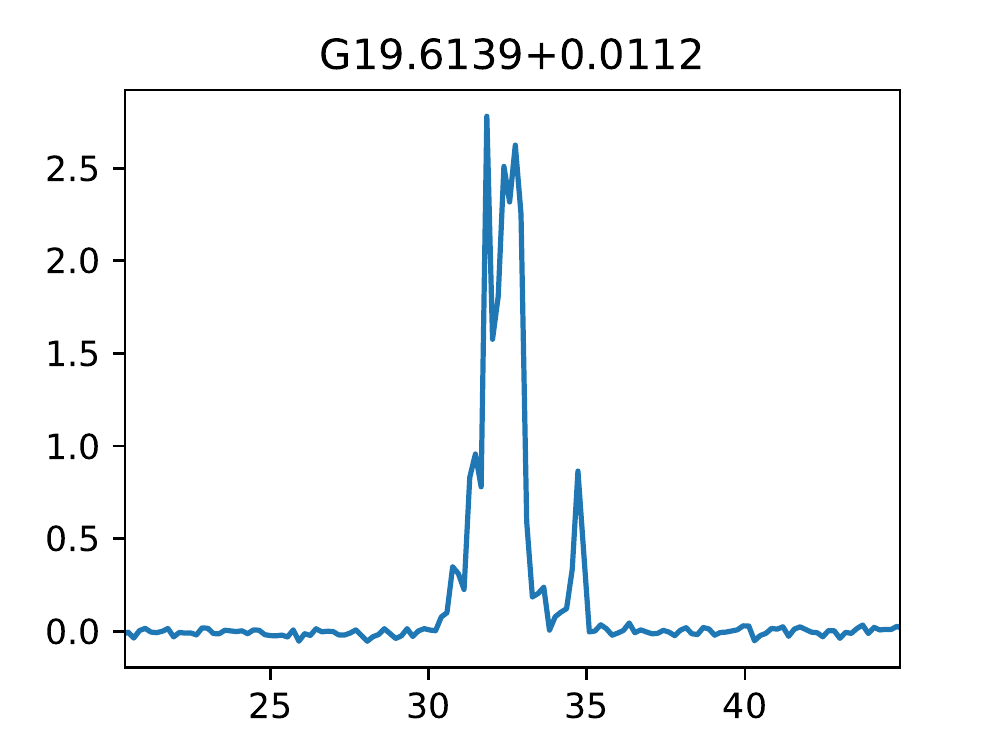}\\ 
\colorbox{white}{\includegraphics[trim={2cm 0 0 0}, scale=0.20]{figs/ylabel.pdf}}
\includegraphics[trim={0.0cm 0.3cm 0.6cm 0.3cm}, clip, width = 0.235\textwidth]{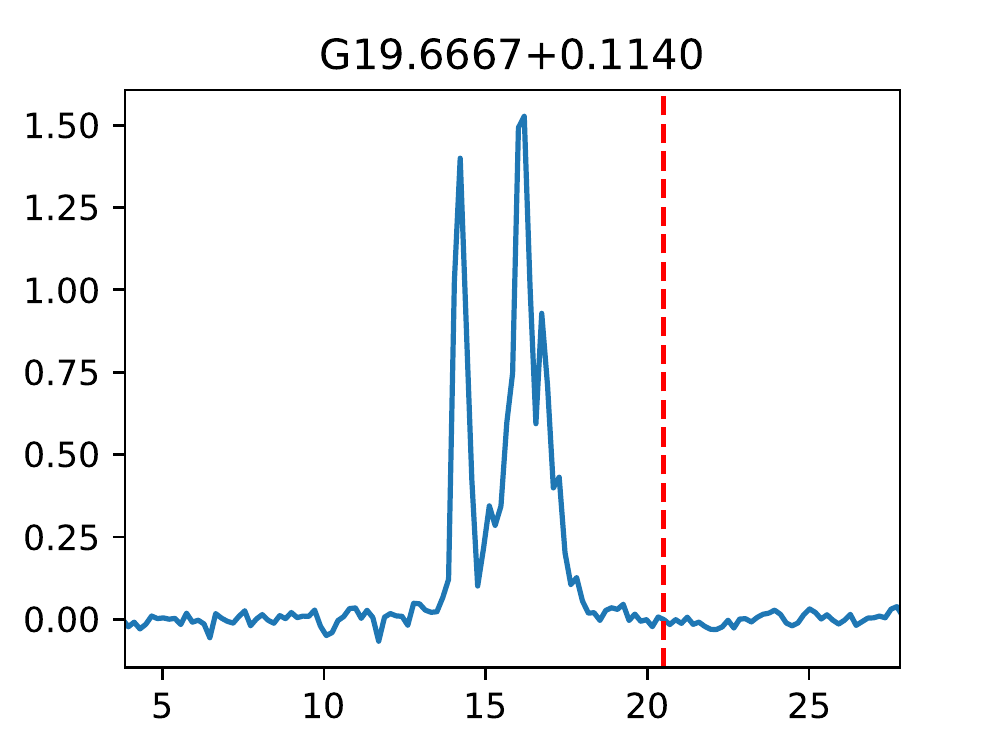}
\includegraphics[trim={0.0cm 0.3cm 0.6cm 0.3cm}, clip, width = 0.235\textwidth]{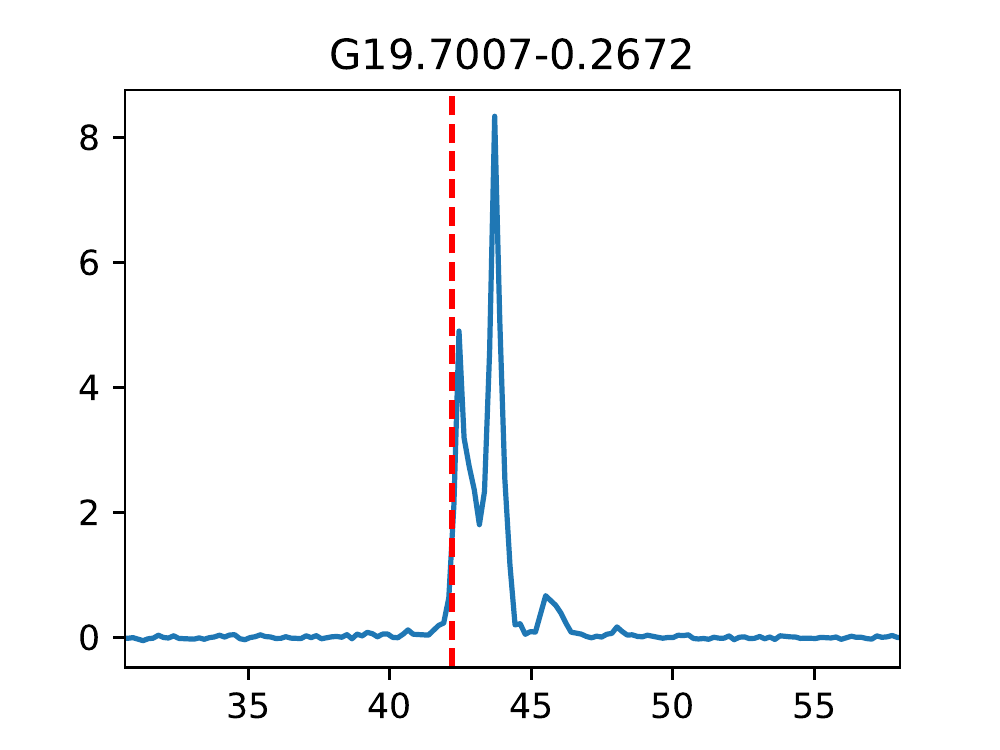}
\includegraphics[trim={0.0cm 0.3cm 0.6cm 0.3cm}, clip, width = 0.235\textwidth]{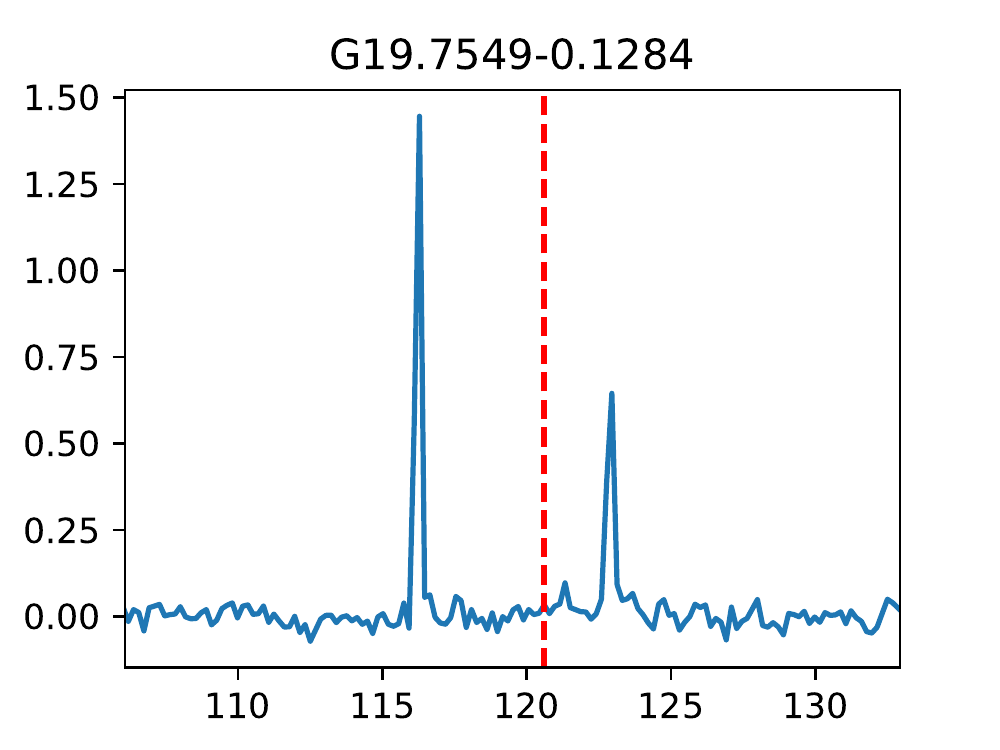}
\includegraphics[trim={0.0cm 0.3cm 0.6cm 0.3cm}, clip, width = 0.235\textwidth]{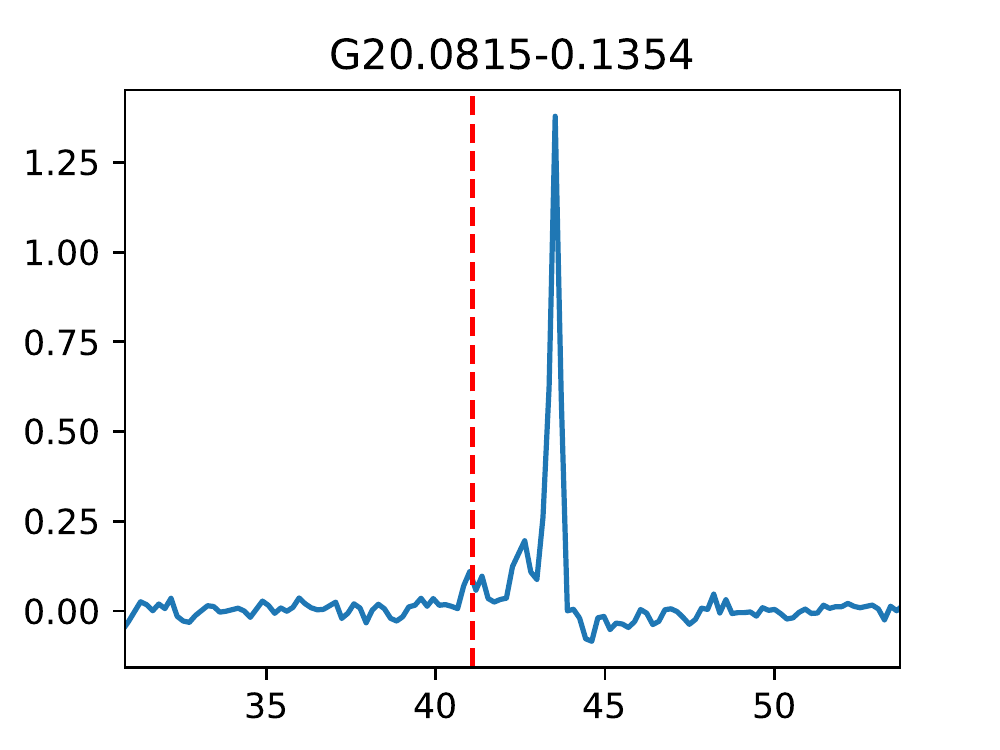}\\
\end{tabular}
\caption{Continued from Fig.~\ref{fig:maser_many_spec}.}
\label{fig:masers_240}
\end{figure*}

%Table 11
\begin{figure*}[!h]
\begin{tabular}{ccccc}
\colorbox{white}{\includegraphics[trim={2cm 0 0 0}, scale=0.195]{figs/ylabel_filler.pdf}}
\includegraphics[trim={0.0cm 0.3cm 0.6cm 0.3cm}, clip, width = 0.235\textwidth]{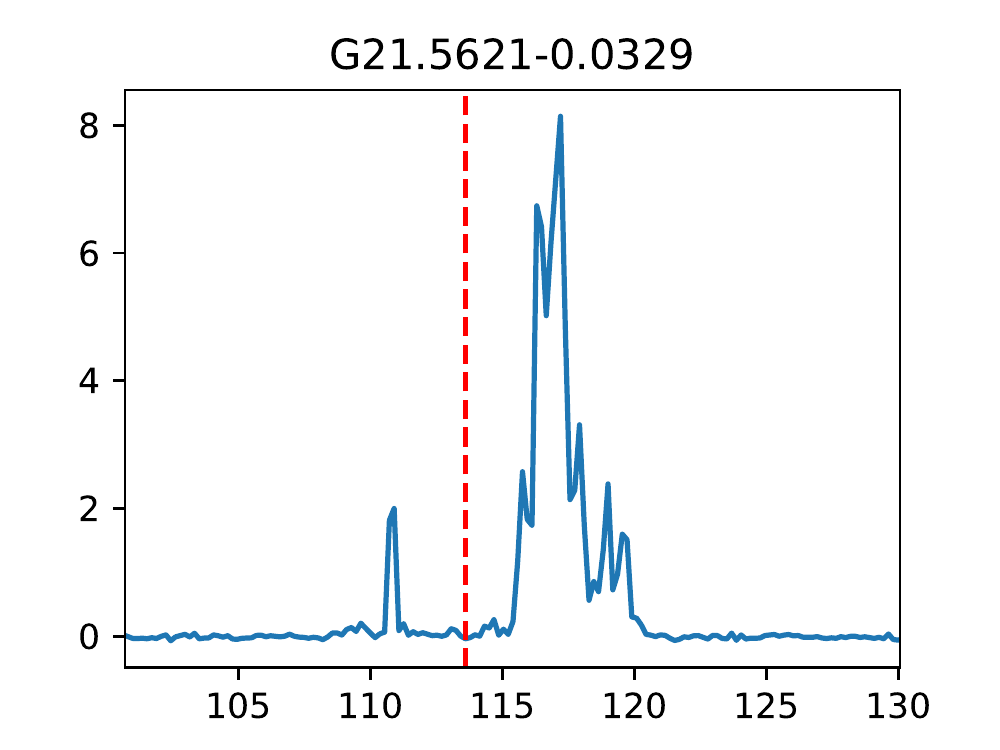}
\includegraphics[trim={0.0cm 0.3cm 0.6cm 0.3cm}, clip, width = 0.235\textwidth]{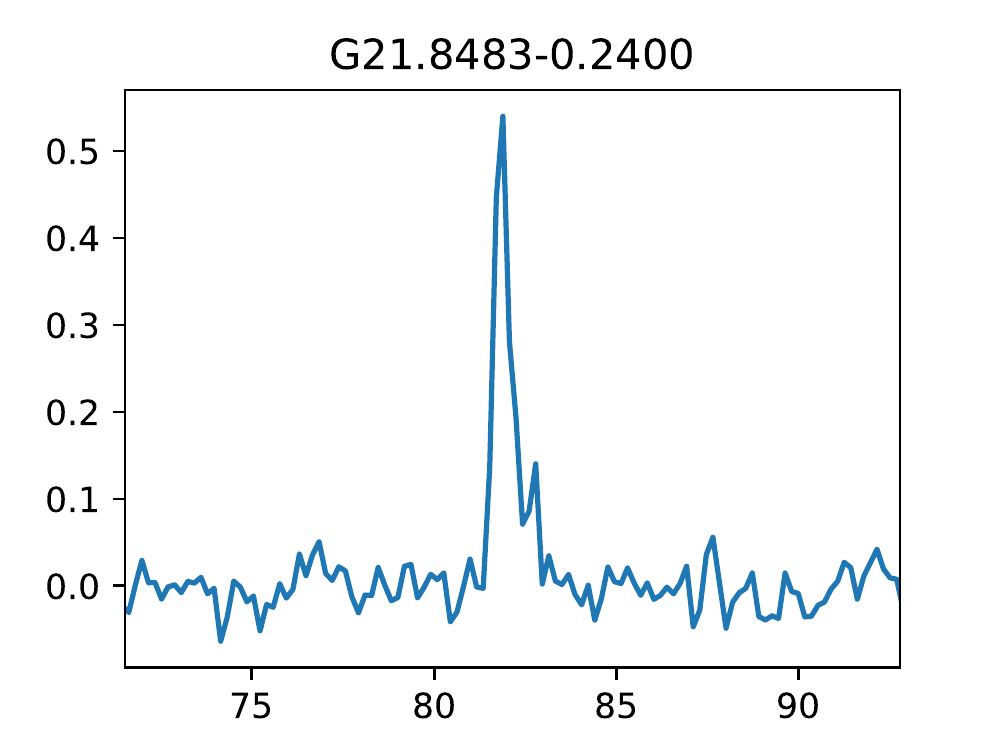}
\includegraphics[trim={0.0cm 0.3cm 0.6cm 0.3cm}, clip, width = 0.235\textwidth]{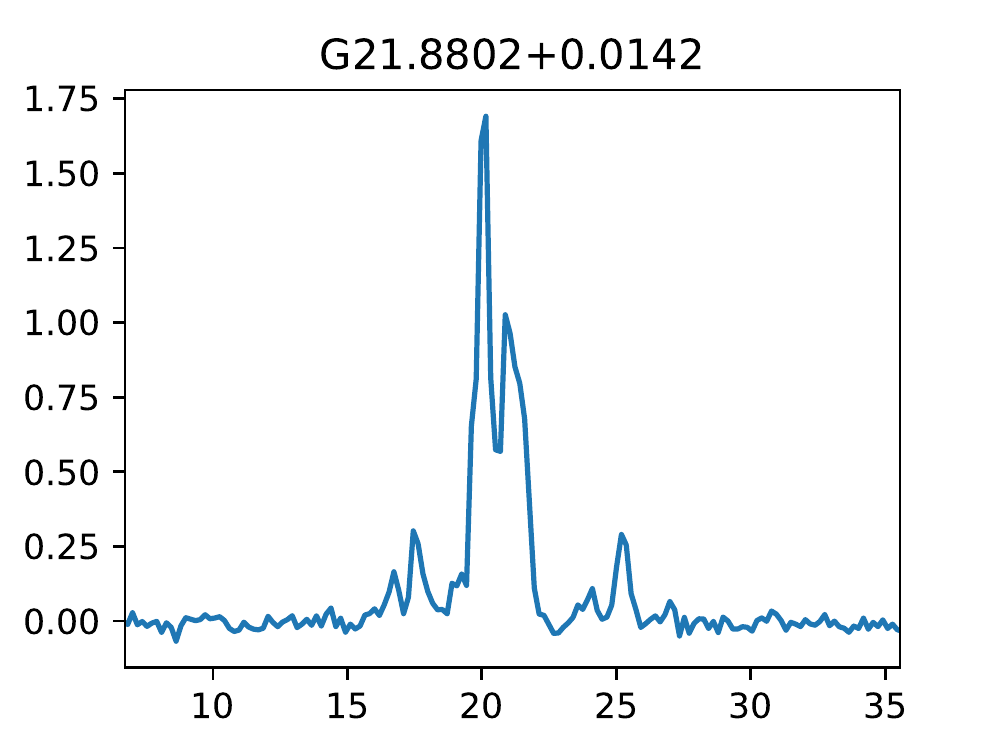}
\includegraphics[trim={0.0cm 0.3cm 0.6cm 0.3cm}, clip, width = 0.235\textwidth]{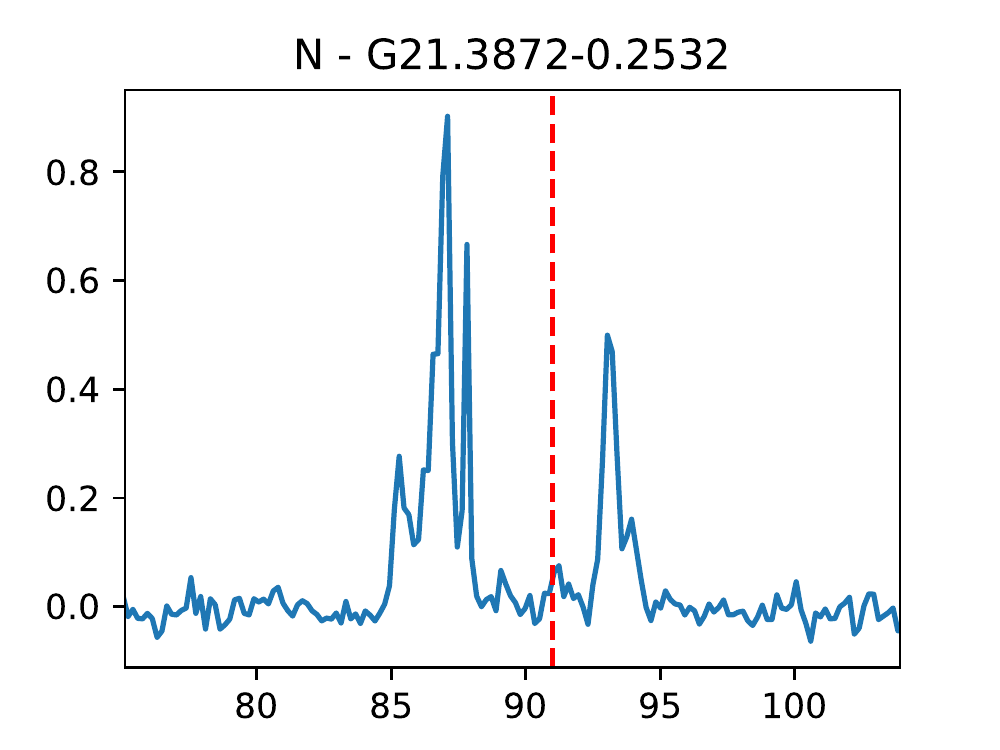}\\ 
\colorbox{white}{\includegraphics[trim={2cm 0 0 0}, scale=0.195]{figs/ylabel_filler.pdf}}
\includegraphics[trim={0.0cm 0.3cm 0.6cm 0.3cm}, clip, width = 0.235\textwidth]{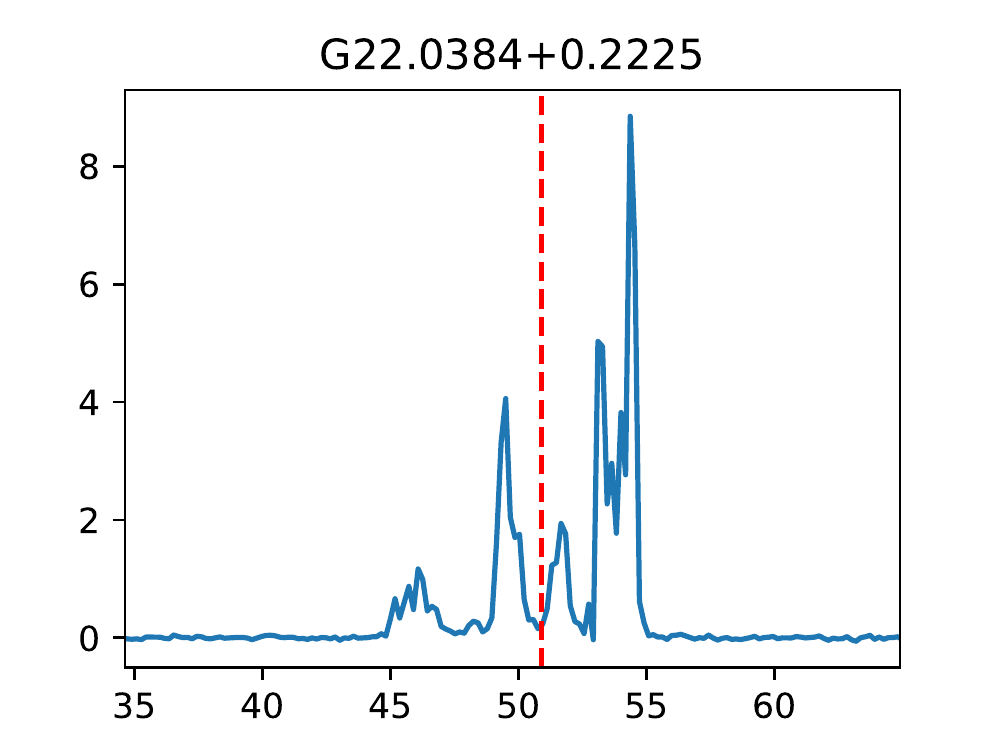}
\includegraphics[trim={0.0cm 0.3cm 0.6cm 0.3cm}, clip, width = 0.235\textwidth]{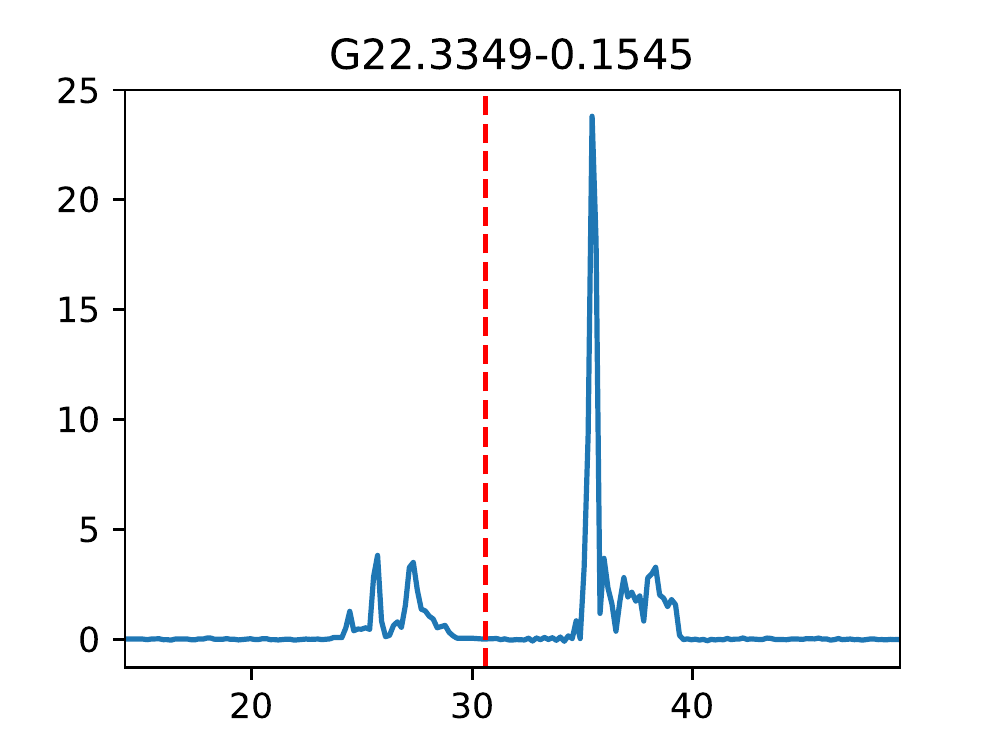}
\includegraphics[trim={0.0cm 0.3cm 0.6cm 0.3cm}, clip, width = 0.235\textwidth]{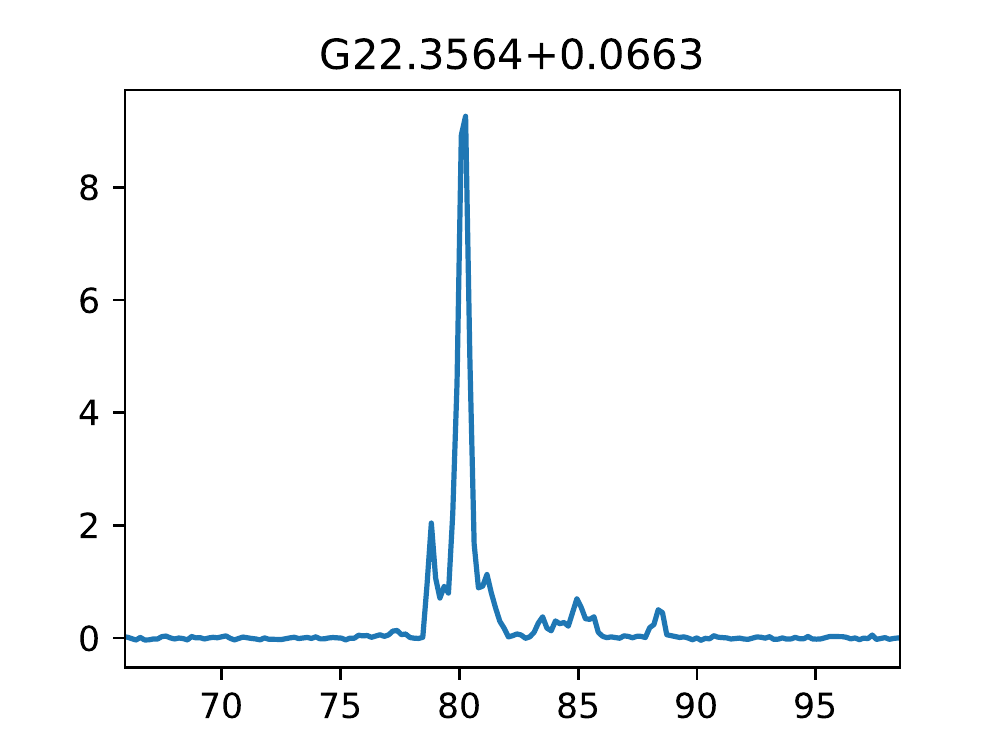}
\includegraphics[trim={0.0cm 0.3cm 0.6cm 0.3cm}, clip, width = 0.235\textwidth]{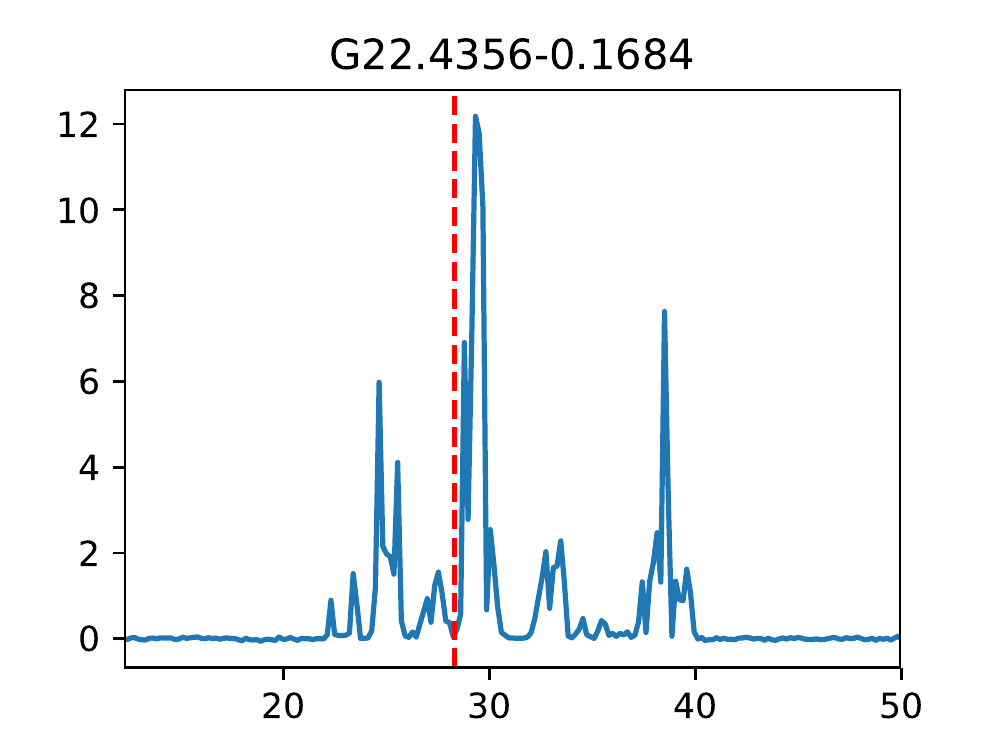}\\ 
\colorbox{white}{\includegraphics[trim={2cm 0 0 0}, scale=0.20]{figs/ylabel.pdf}}
\includegraphics[trim={0.0cm 0.3cm 0.6cm 0.3cm}, clip, width = 0.235\textwidth]{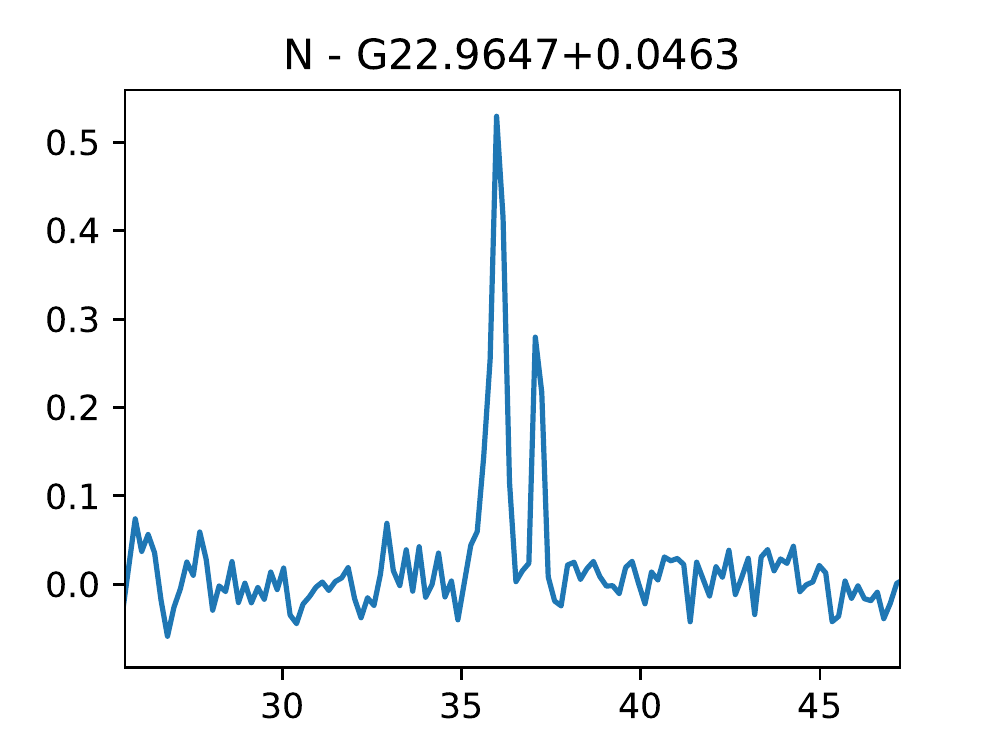}
\includegraphics[trim={0.0cm 0.3cm 0.6cm 0.3cm}, clip, width = 0.235\textwidth]{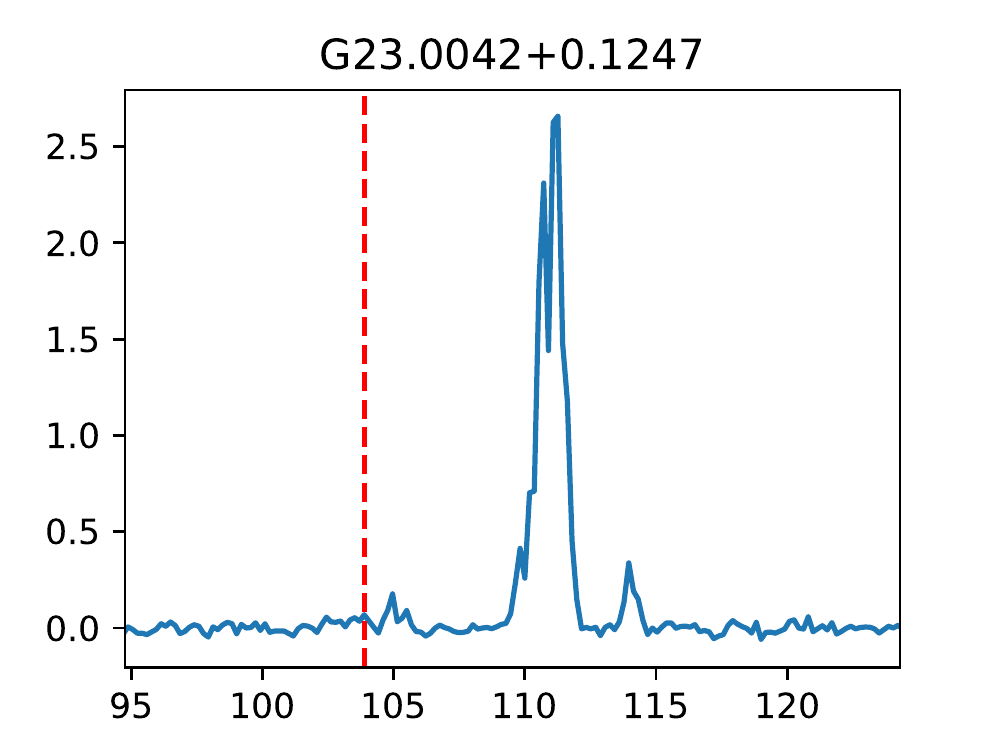}
\includegraphics[trim={0.0cm 0.3cm 0.6cm 0.3cm}, clip, width = 0.235\textwidth]{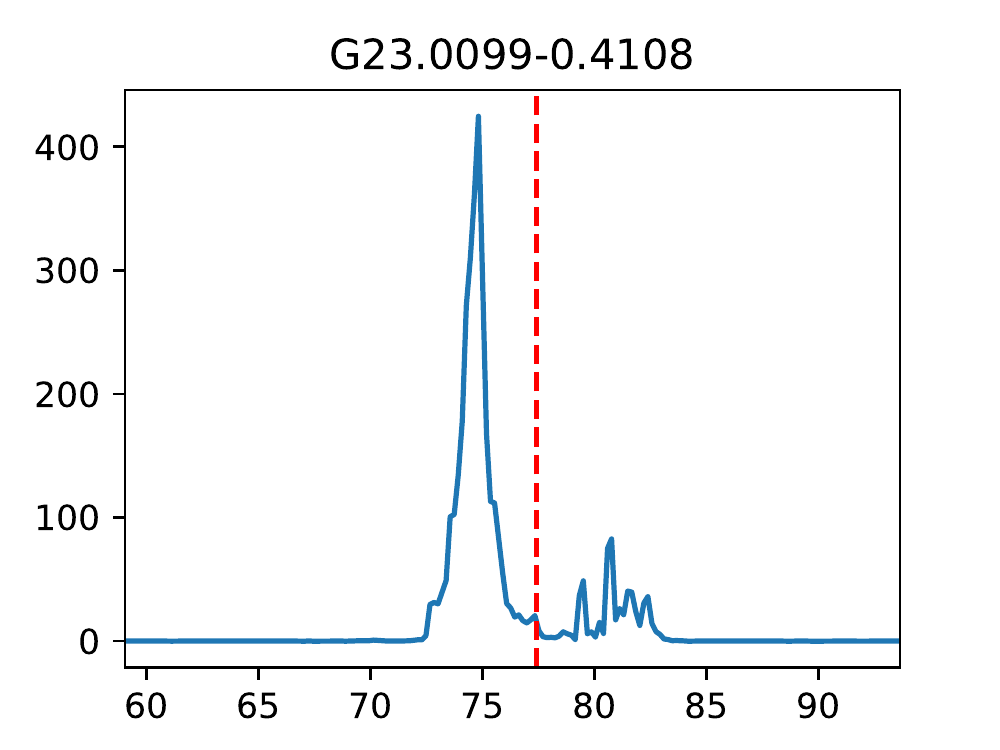}
\includegraphics[trim={0.0cm 0.3cm 0.6cm 0.3cm}, clip, width = 0.235\textwidth]{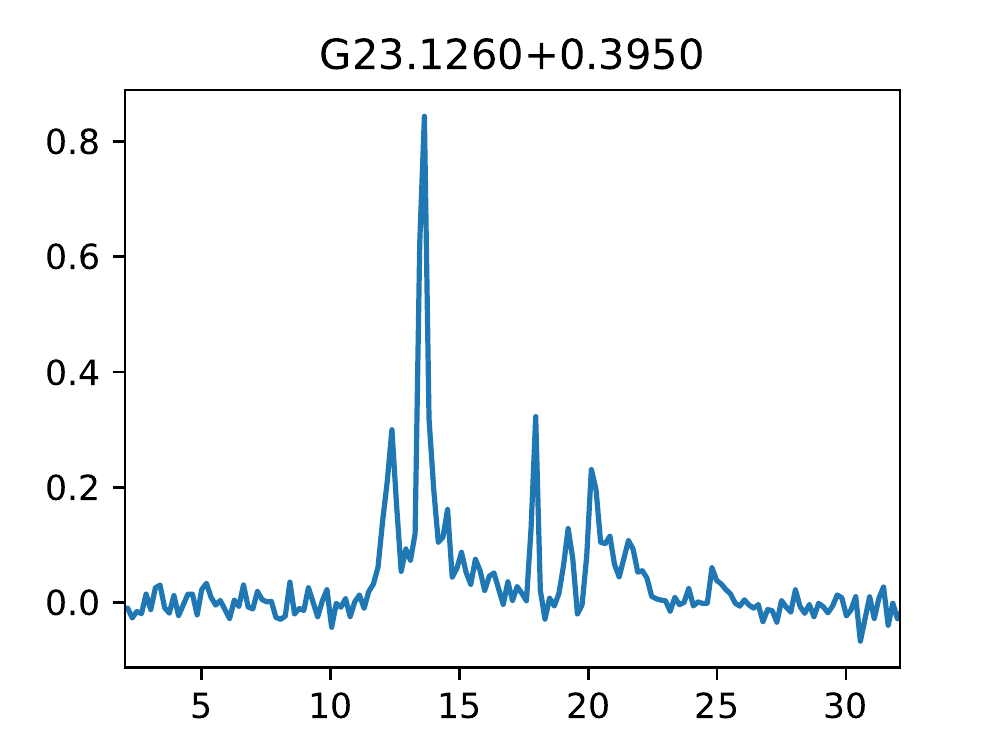}\\
\end{tabular}
\caption{Continued from Fig.~\ref{fig:maser_many_spec}.}
\label{fig:masers_264}
\end{figure*}

%Table 12
\begin{figure*}[!h]
\begin{tabular}{ccccc}
\colorbox{white}{\includegraphics[trim={2cm 0 0 0}, scale=0.195]{figs/ylabel_filler.pdf}}
\includegraphics[trim={0.0cm 0.3cm 0.6cm 0.3cm}, clip, width = 0.235\textwidth]{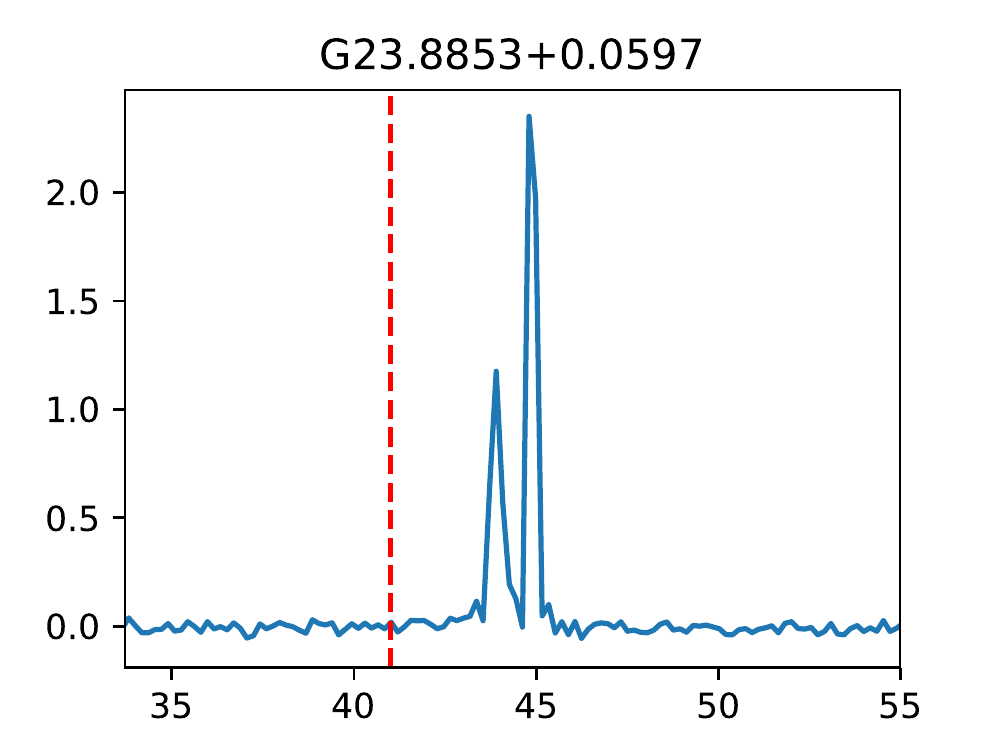}
\includegraphics[trim={0.0cm 0.3cm 0.6cm 0.3cm}, clip, width = 0.235\textwidth]{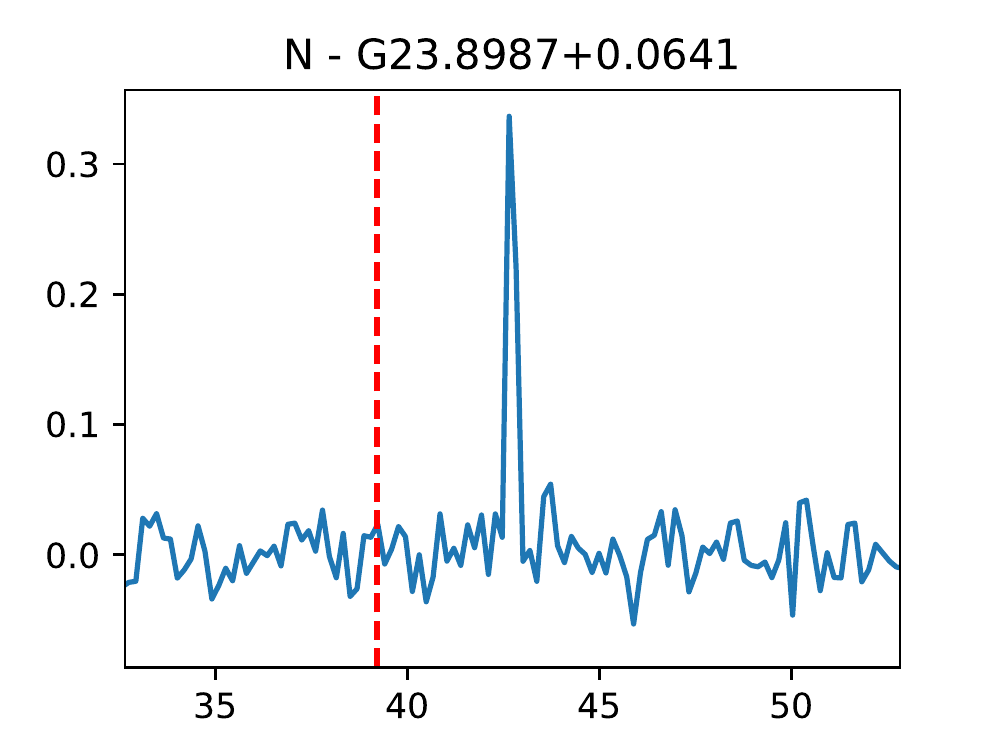}
\includegraphics[trim={0.0cm 0.3cm 0.6cm 0.3cm}, clip, width = 0.235\textwidth]{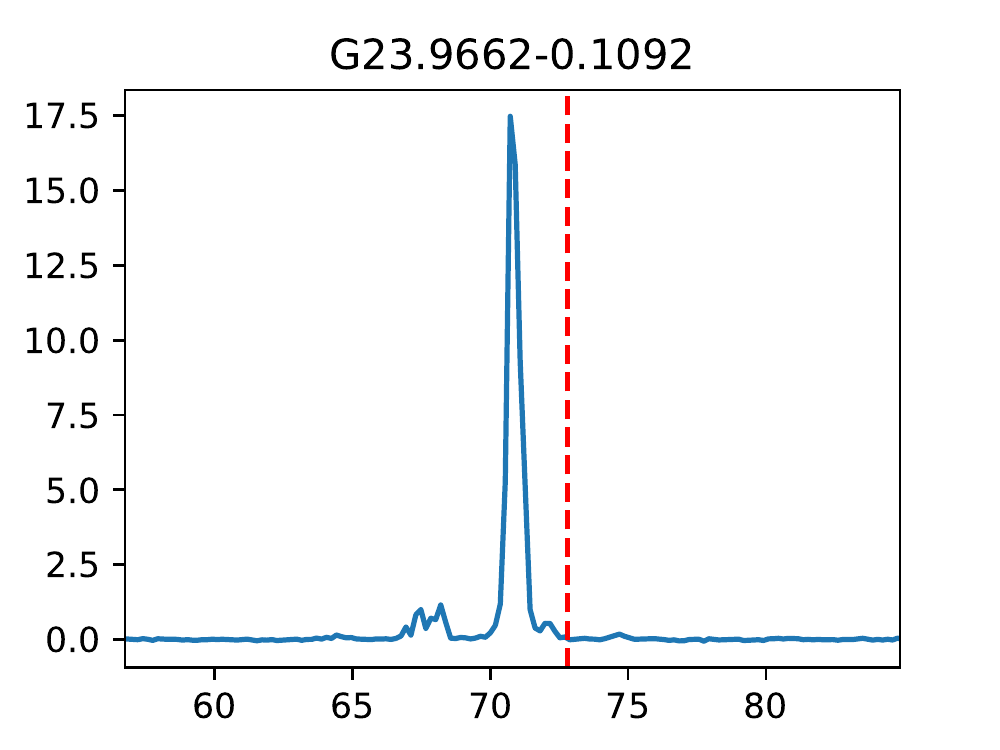}
\includegraphics[trim={0.0cm 0.3cm 0.6cm 0.3cm}, clip, width = 0.235\textwidth]{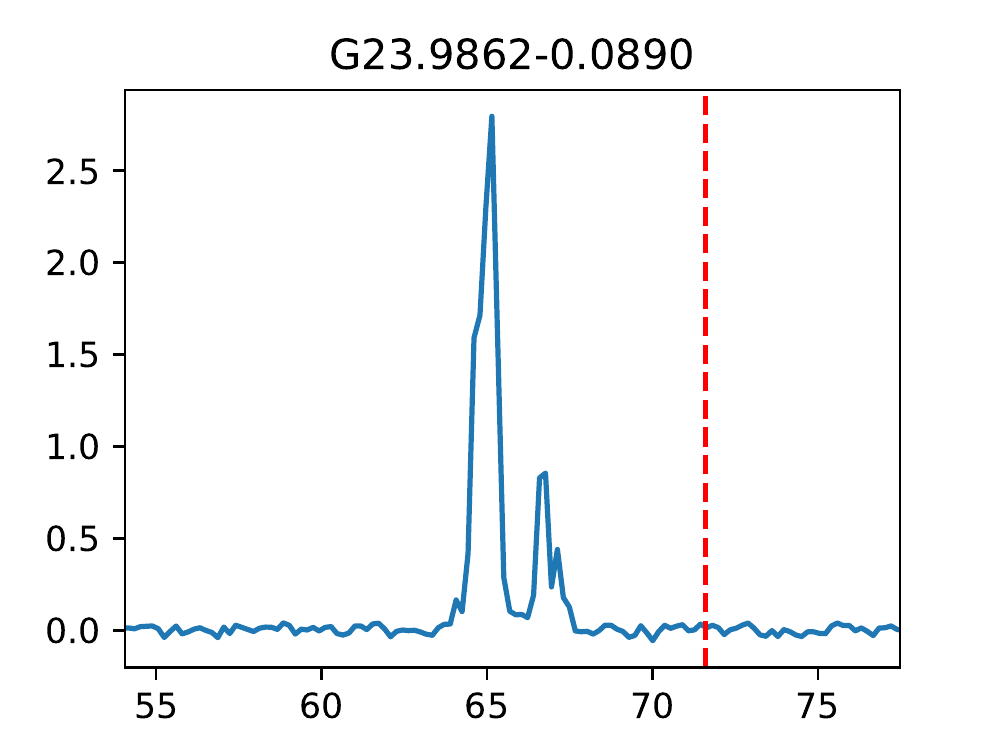}\\ 
\colorbox{white}{\includegraphics[trim={2cm 0 0 0}, scale=0.195]{figs/ylabel_filler.pdf}}
\includegraphics[trim={0.0cm 0.3cm 0.6cm 0.3cm}, clip, width = 0.235\textwidth]{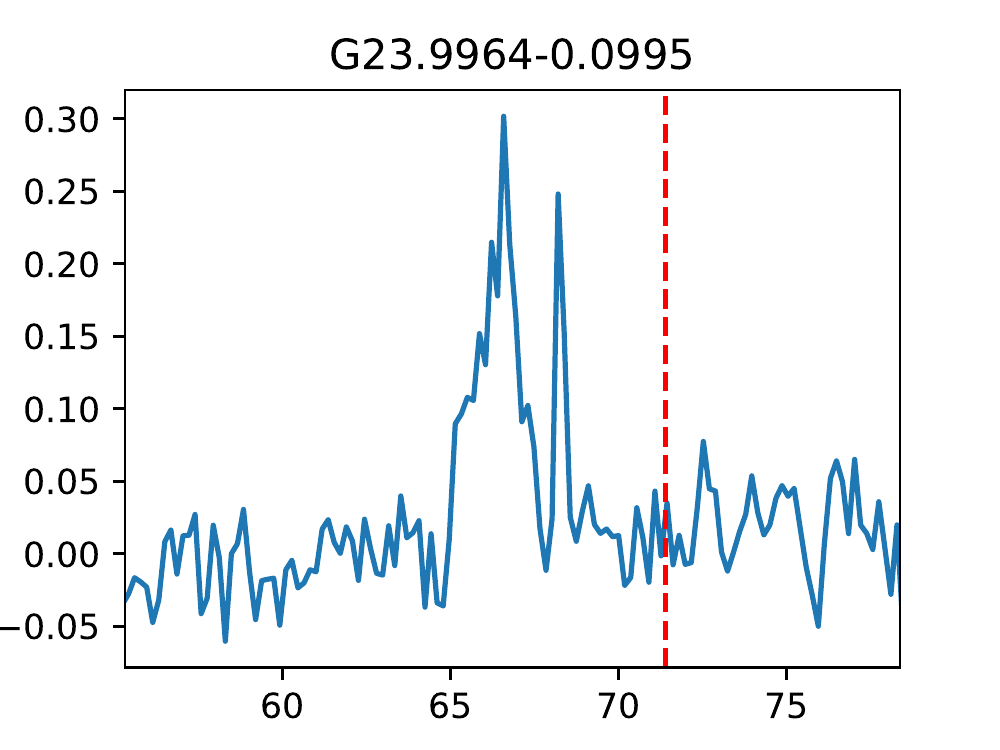}
\includegraphics[trim={0.0cm 0.3cm 0.6cm 0.3cm}, clip, width = 0.235\textwidth]{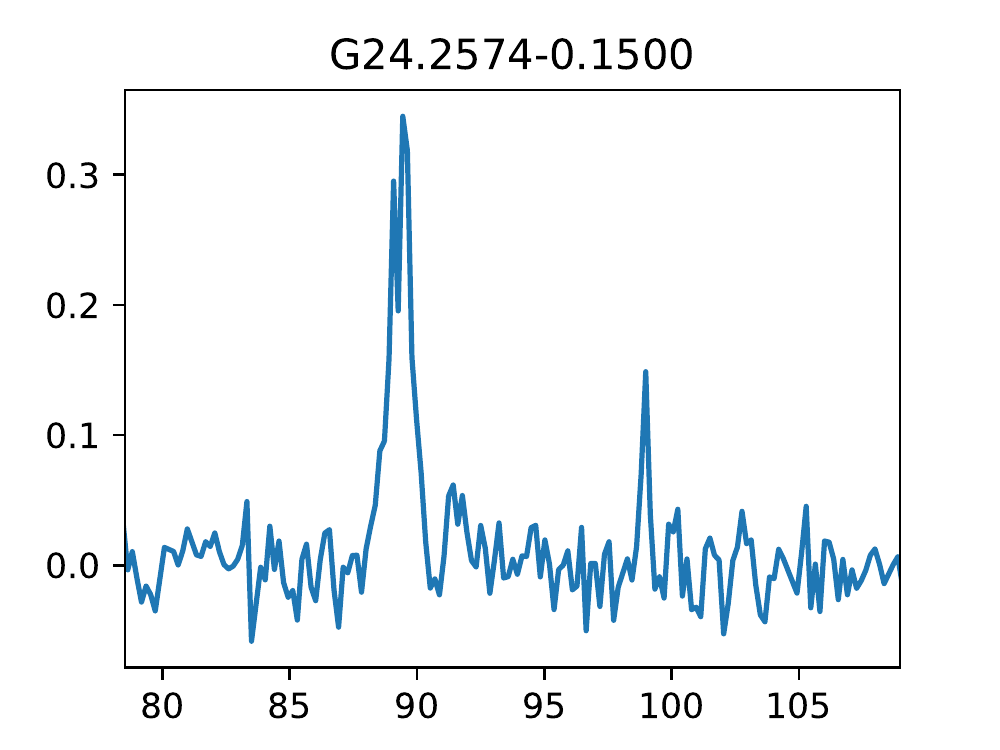}
\includegraphics[trim={0.0cm 0.3cm 0.6cm 0.3cm}, clip, width = 0.235\textwidth]{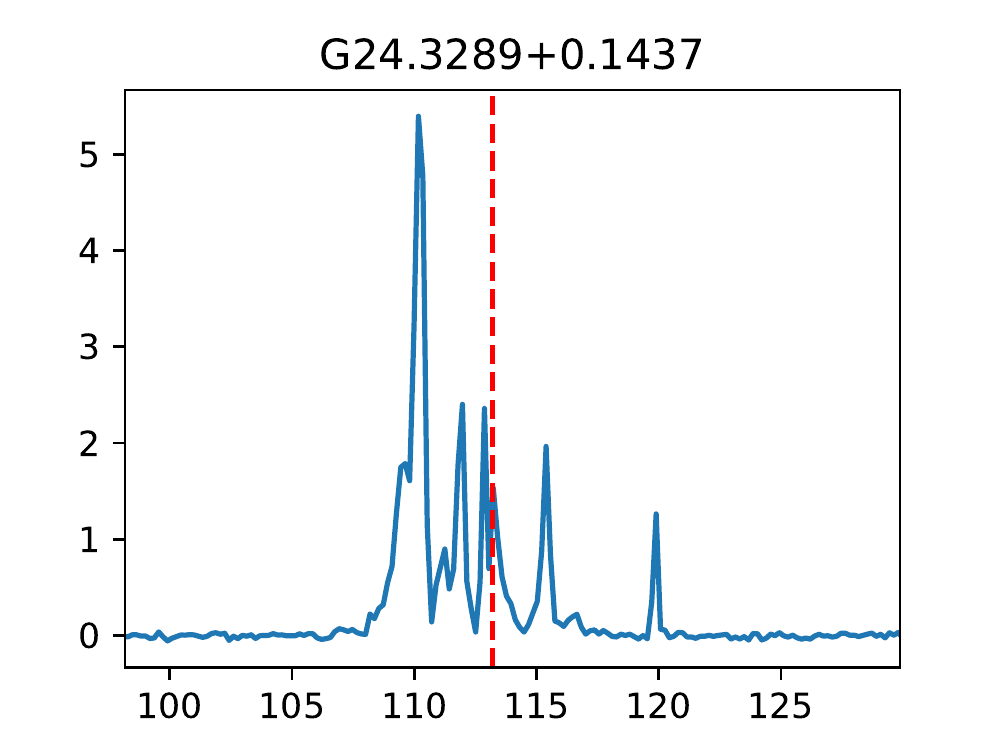}
\includegraphics[trim={0.0cm 0.3cm 0.6cm 0.3cm}, clip, width = 0.235\textwidth]{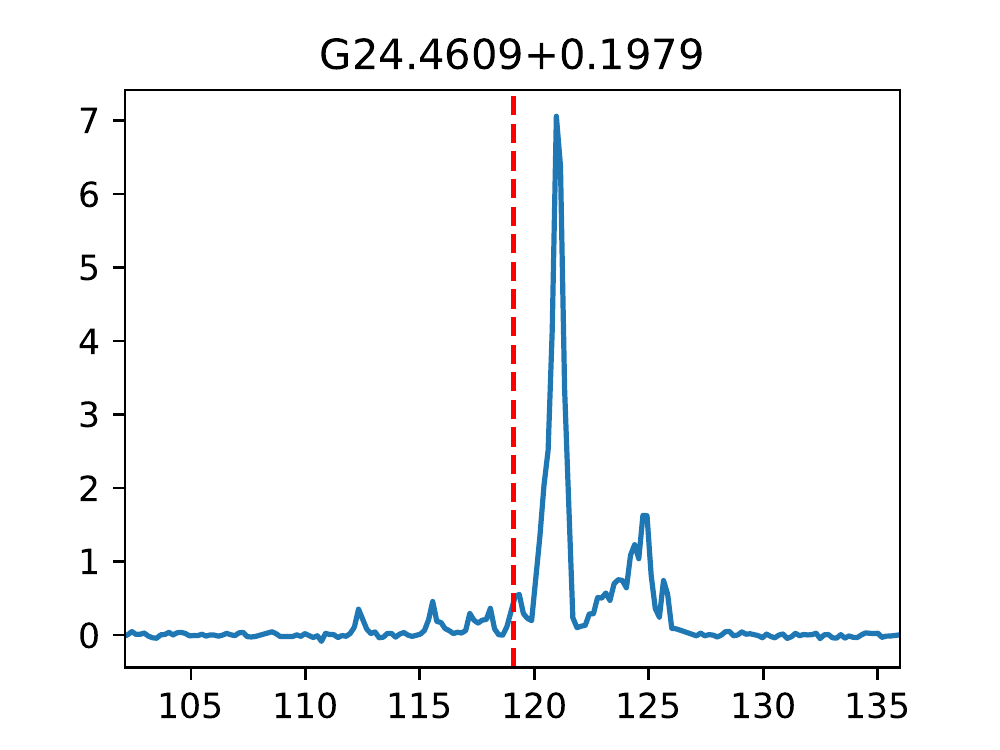}\\ 
\colorbox{white}{\includegraphics[trim={2cm 0 0 0}, scale=0.20]{figs/ylabel.pdf}}
\includegraphics[trim={0.0cm 0.3cm 0.6cm 0.3cm}, clip, width = 0.235\textwidth]{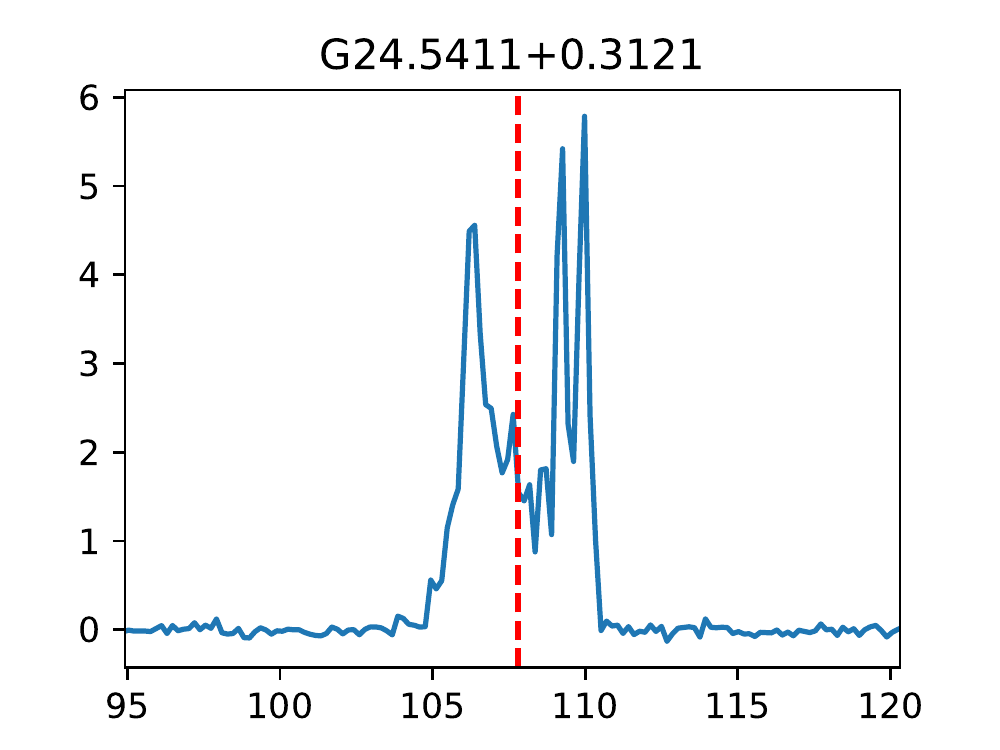}
\includegraphics[trim={0.0cm 0.3cm 0.6cm 0.3cm}, clip, width = 0.235\textwidth]{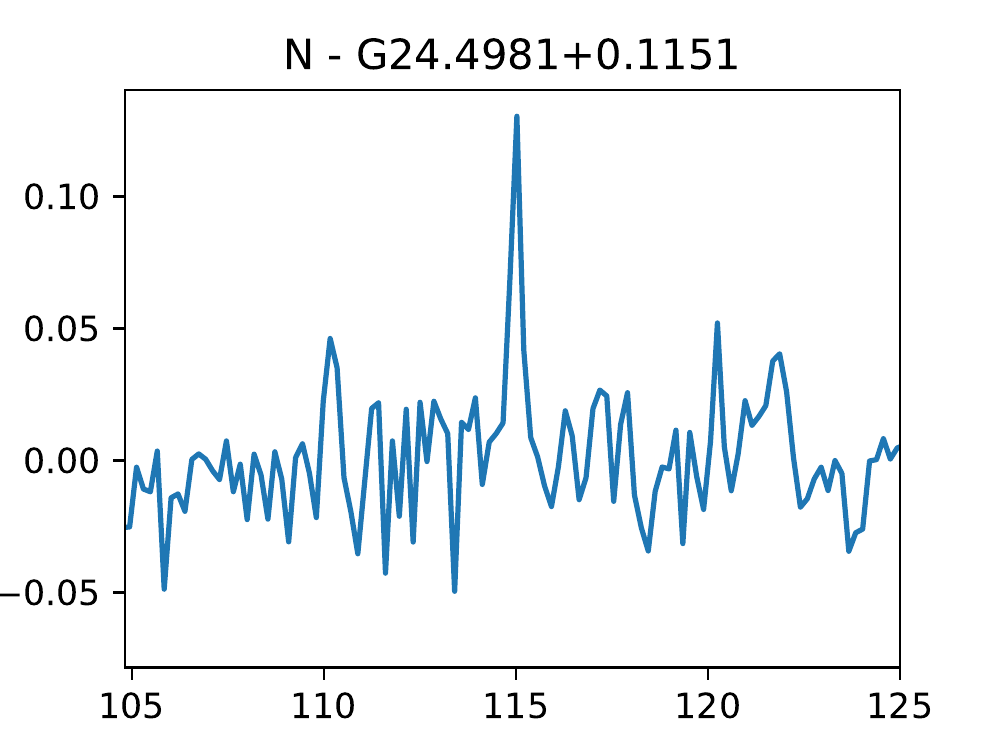}
\includegraphics[trim={0.0cm 0.3cm 0.6cm 0.3cm}, clip, width = 0.235\textwidth]{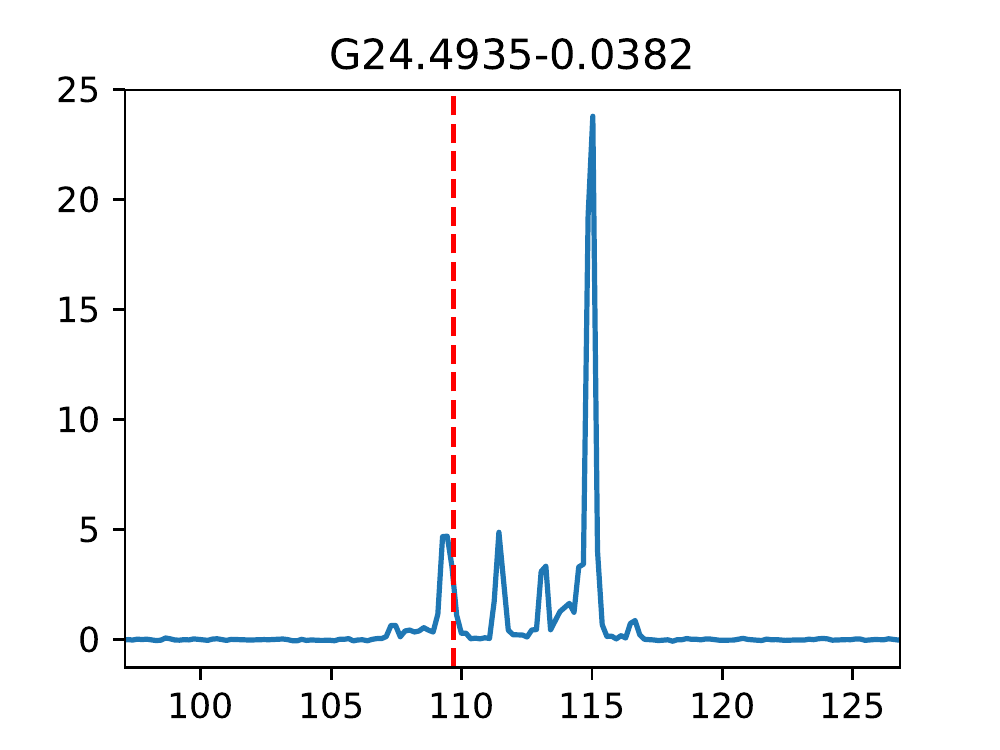}
\includegraphics[trim={0.0cm 0.3cm 0.6cm 0.3cm}, clip, width = 0.235\textwidth]{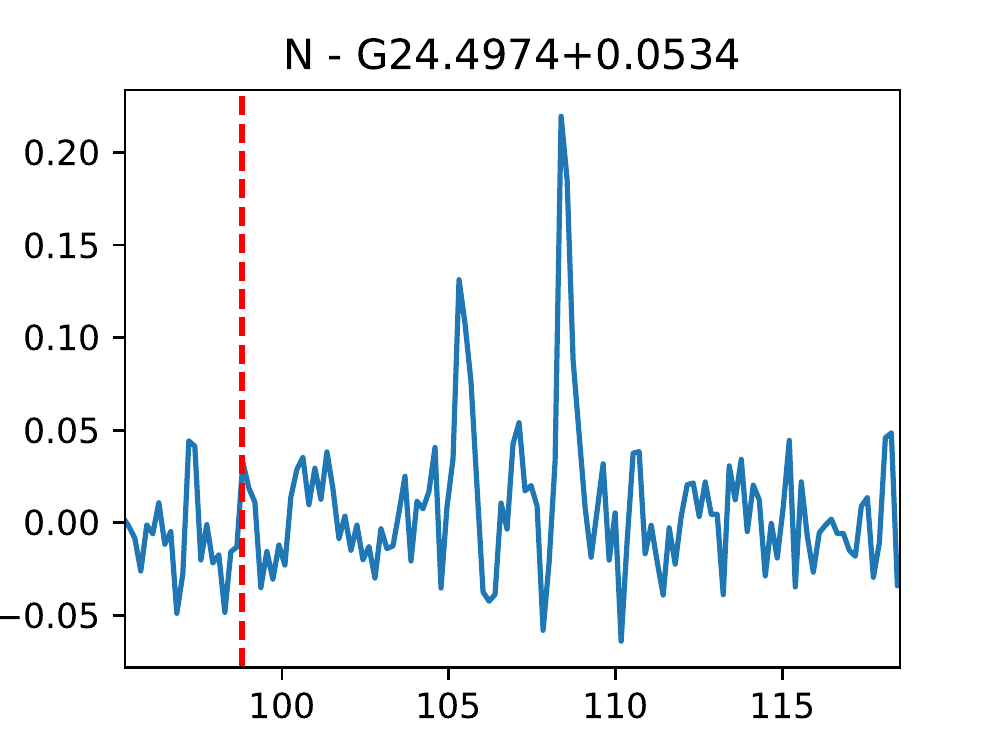}\\
\end{tabular}
\caption{Continued from Fig.~\ref{fig:maser_many_spec}.}
\label{fig:masers_288}
\end{figure*}

%Table 13
\begin{figure*}[!h]
\begin{tabular}{ccccc}
\colorbox{white}{\includegraphics[trim={2cm 0 0 0}, scale=0.195]{figs/ylabel_filler.pdf}}
\includegraphics[trim={0.0cm 0.3cm 0.6cm 0.3cm}, clip, width = 0.235\textwidth]{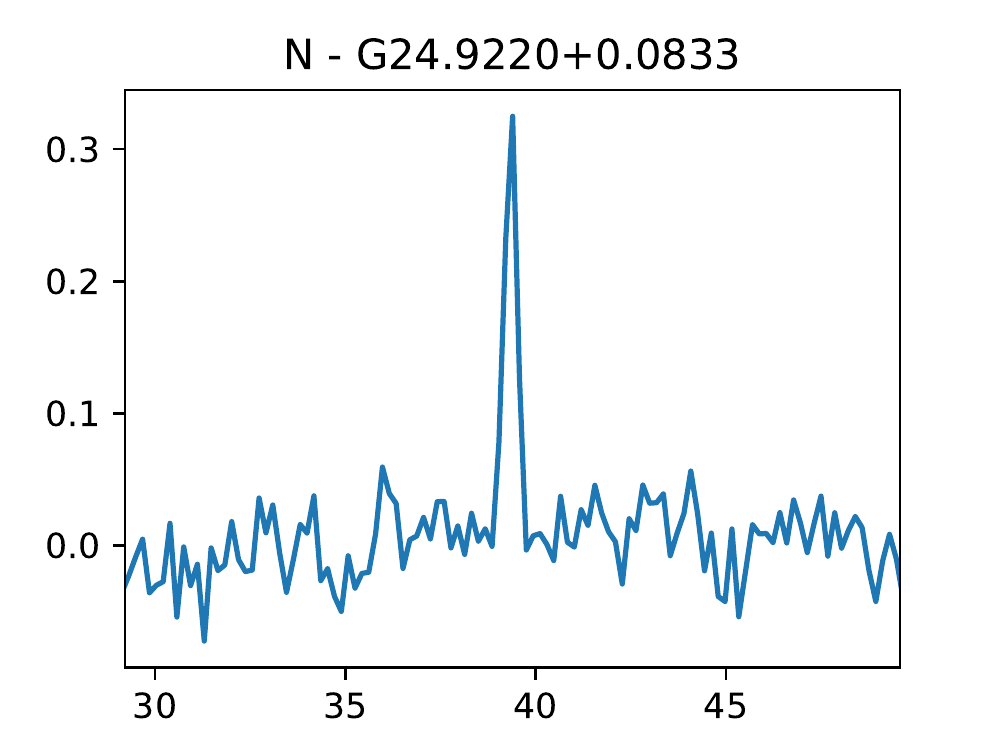}
\includegraphics[trim={0.0cm 0.3cm 0.6cm 0.3cm}, clip, width = 0.235\textwidth]{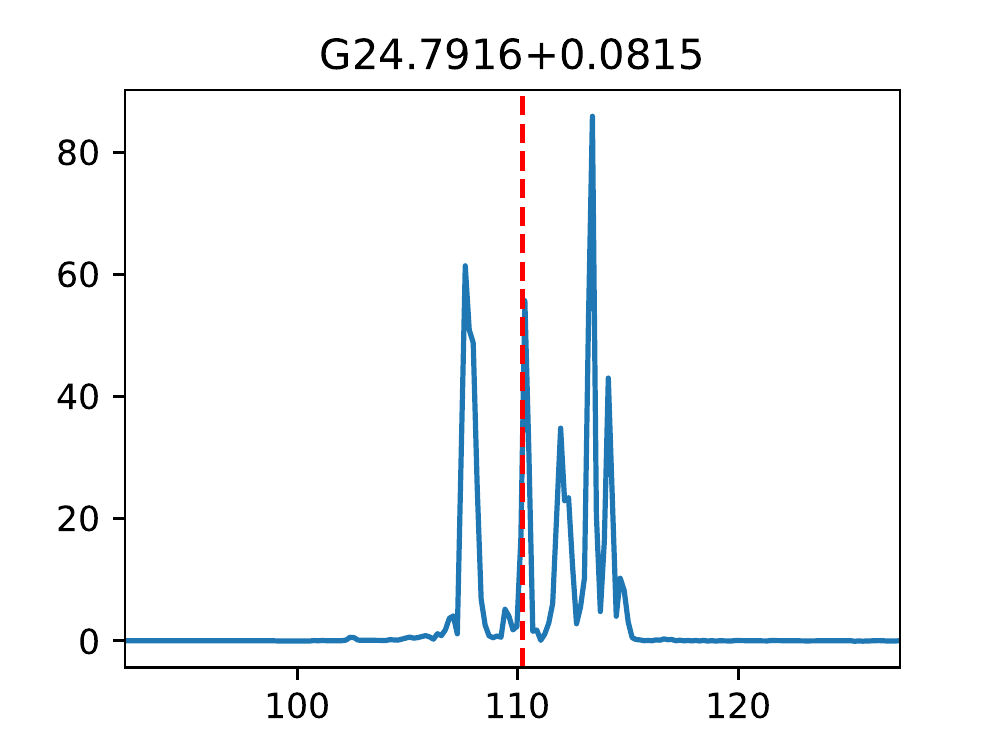}
\includegraphics[trim={0.0cm 0.3cm 0.6cm 0.3cm}, clip, width = 0.235\textwidth]{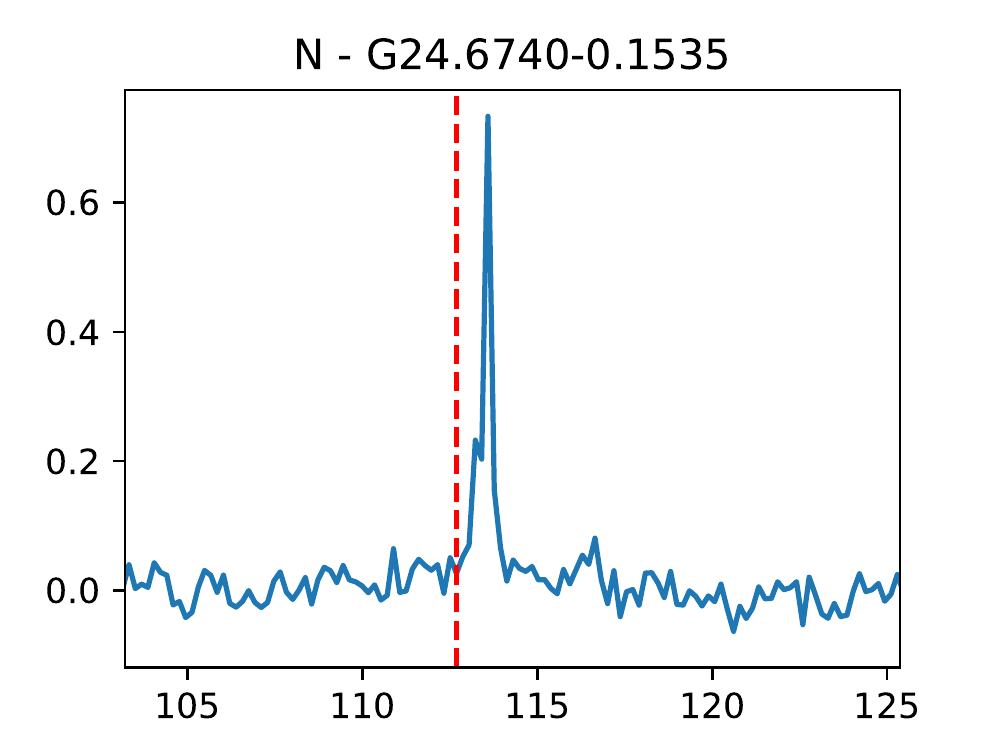}
\includegraphics[trim={0.0cm 0.3cm 0.6cm 0.3cm}, clip, width = 0.235\textwidth]{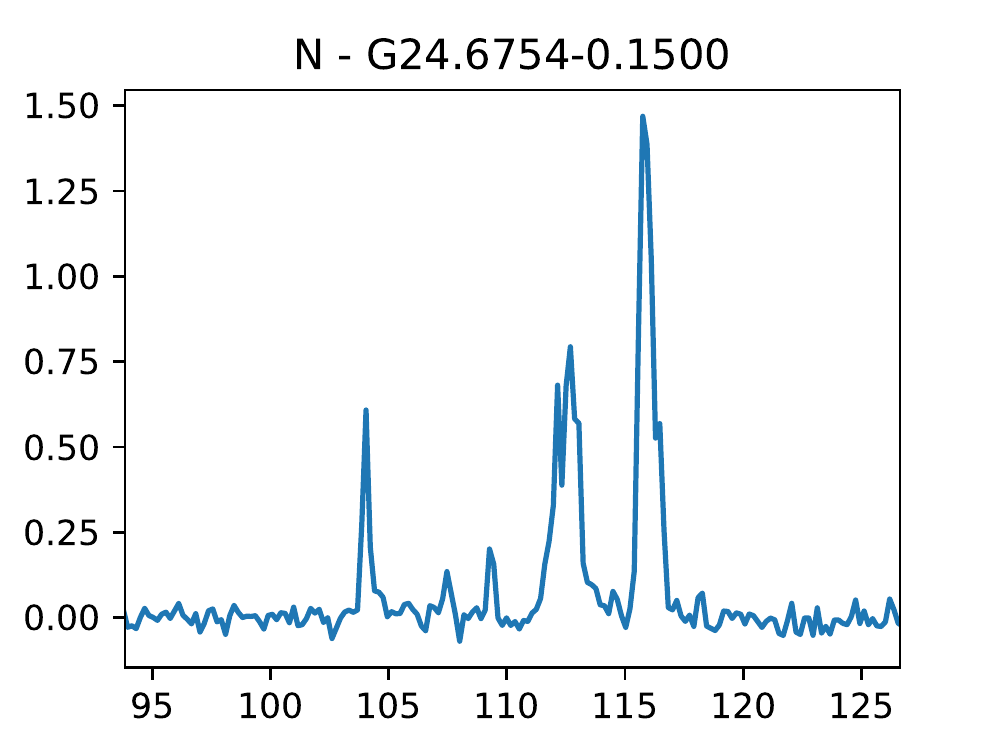}\\ 
\colorbox{white}{\includegraphics[trim={2cm 0 0 0}, scale=0.195]{figs/ylabel_filler.pdf}}
\includegraphics[trim={0.0cm 0.3cm 0.6cm 0.3cm}, clip, width = 0.235\textwidth]{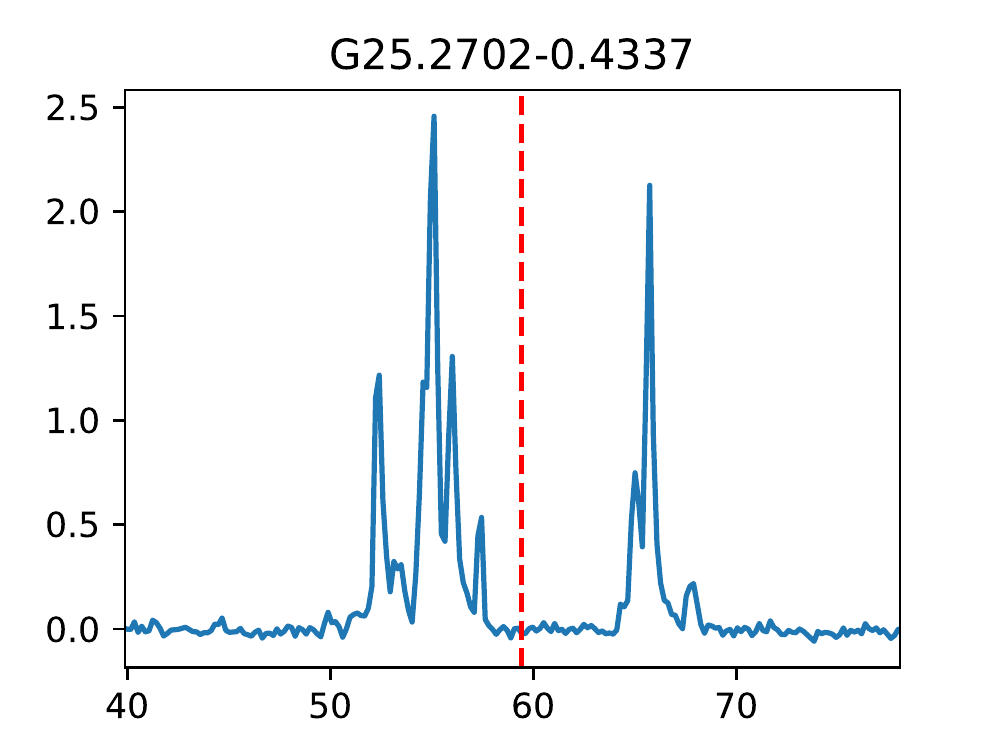}
\includegraphics[trim={0.0cm 0.3cm 0.6cm 0.3cm}, clip, width = 0.235\textwidth]{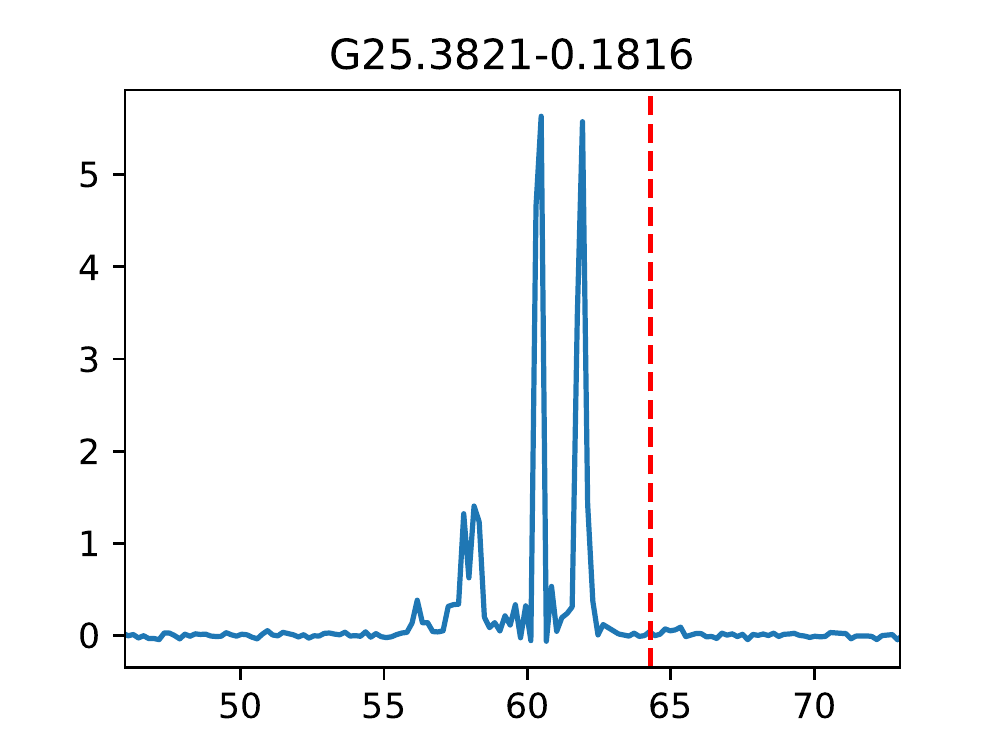}
\includegraphics[trim={0.0cm 0.3cm 0.6cm 0.3cm}, clip, width = 0.235\textwidth]{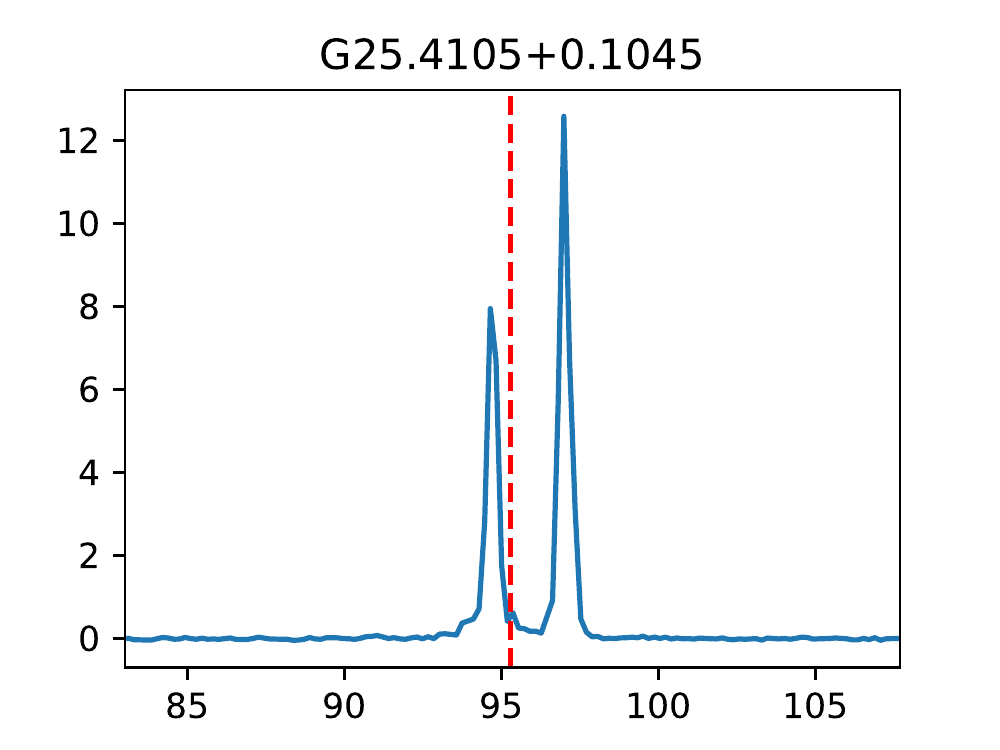}
\includegraphics[trim={0.0cm 0.3cm 0.6cm 0.3cm}, clip, width = 0.235\textwidth]{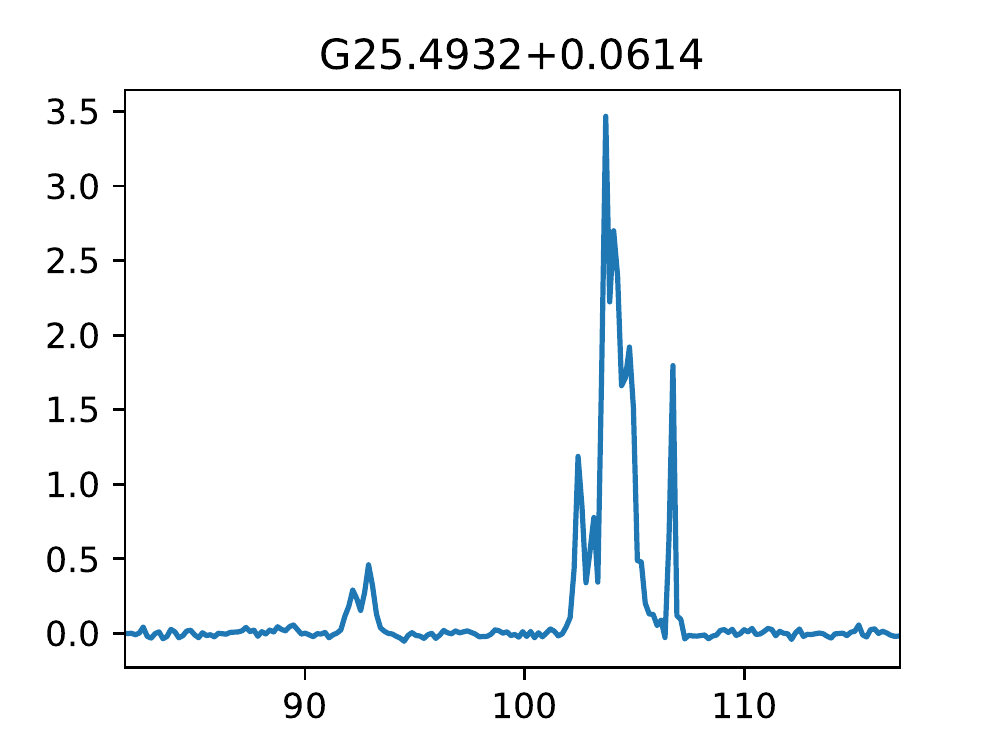}\\ 
\colorbox{white}{\includegraphics[trim={2cm 0 0 0}, scale=0.20]{figs/ylabel.pdf}}
\includegraphics[trim={0.0cm 0.3cm 0.6cm 0.3cm}, clip, width = 0.235\textwidth]{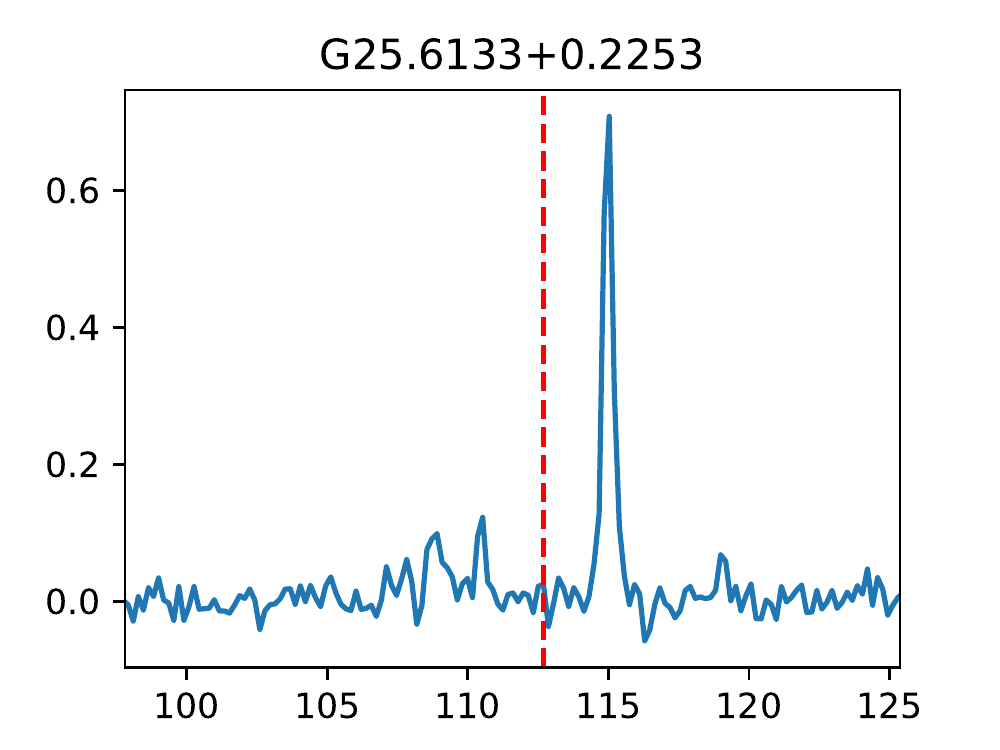}
\includegraphics[trim={0.0cm 0.3cm 0.6cm 0.3cm}, clip, width = 0.235\textwidth]{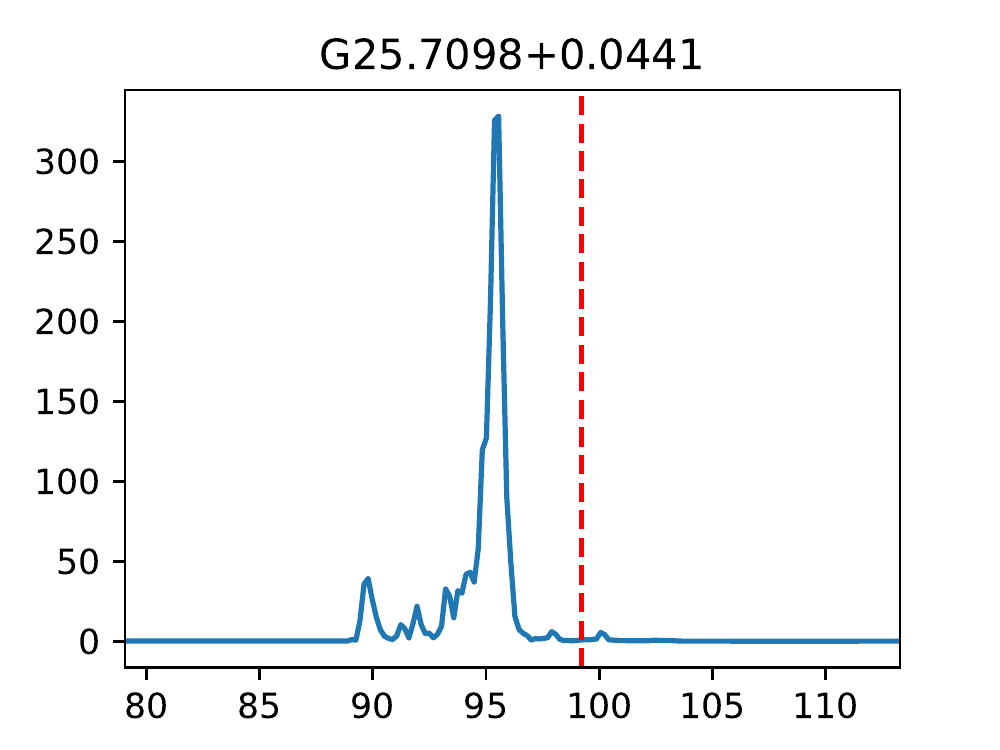}
\includegraphics[trim={0.0cm 0.3cm 0.6cm 0.3cm}, clip, width = 0.235\textwidth]{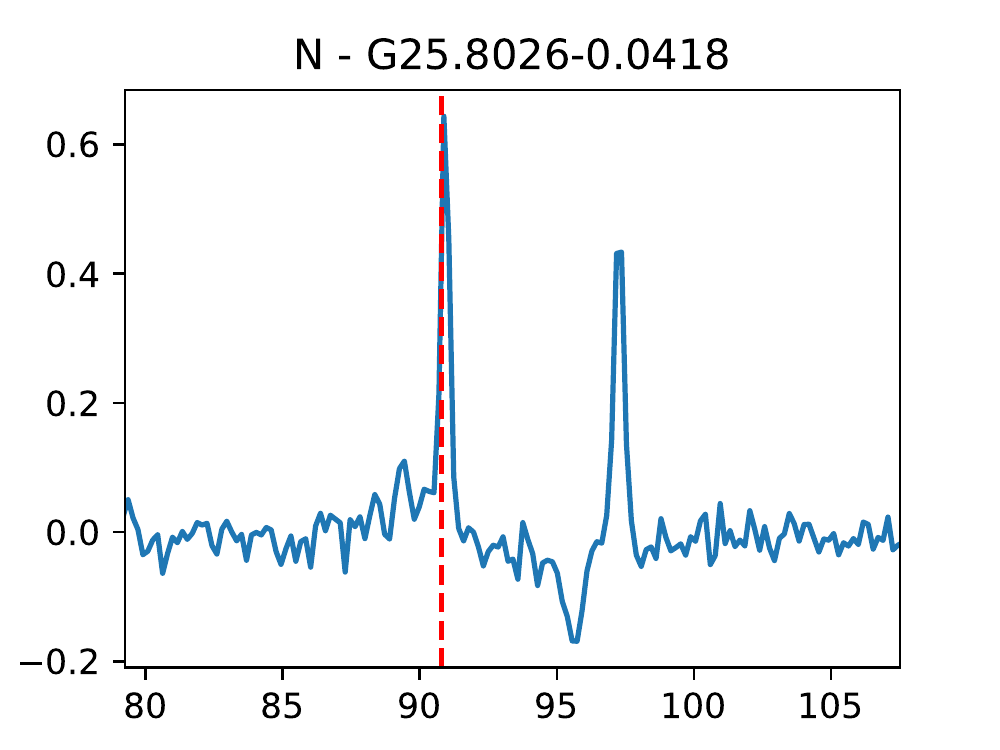}
\includegraphics[trim={0.0cm 0.3cm 0.6cm 0.3cm}, clip, width = 0.235\textwidth]{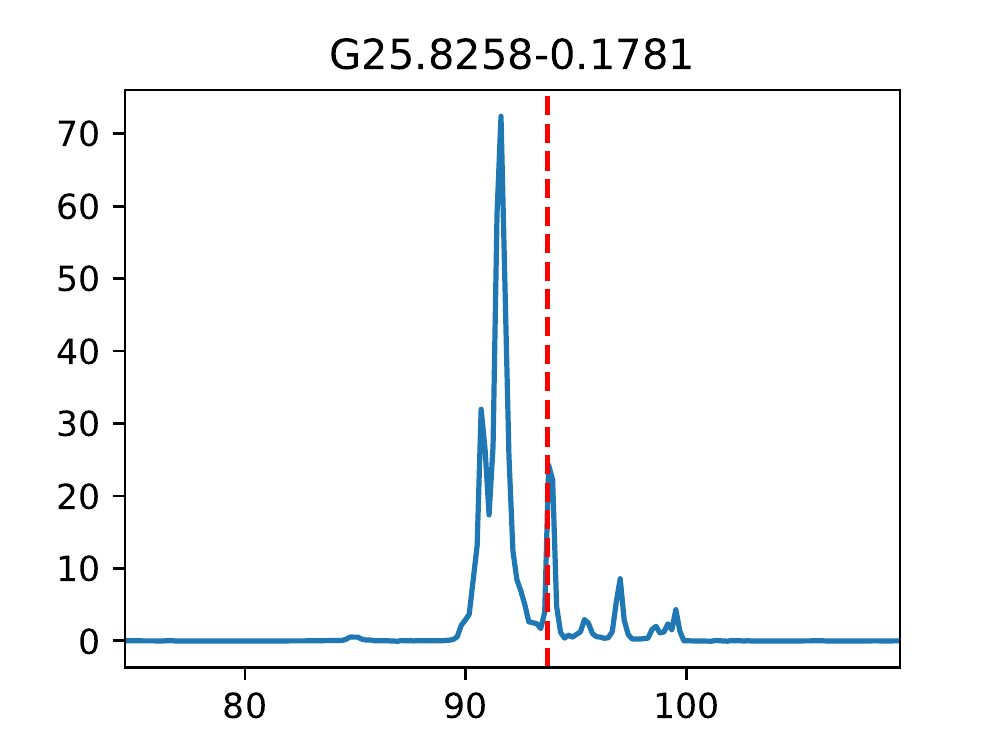}\\
\end{tabular}
\caption{Continued from Fig.~\ref{fig:maser_many_spec}.}
\label{fig:masers_312}
\end{figure*}

%Table 14
\begin{figure*}[!h]
\begin{tabular}{ccccc}
\colorbox{white}{\includegraphics[trim={2cm 0 0 0}, scale=0.195]{figs/ylabel_filler.pdf}}
\includegraphics[trim={0.0cm 0.3cm 0.6cm 0.3cm}, clip, width = 0.235\textwidth]{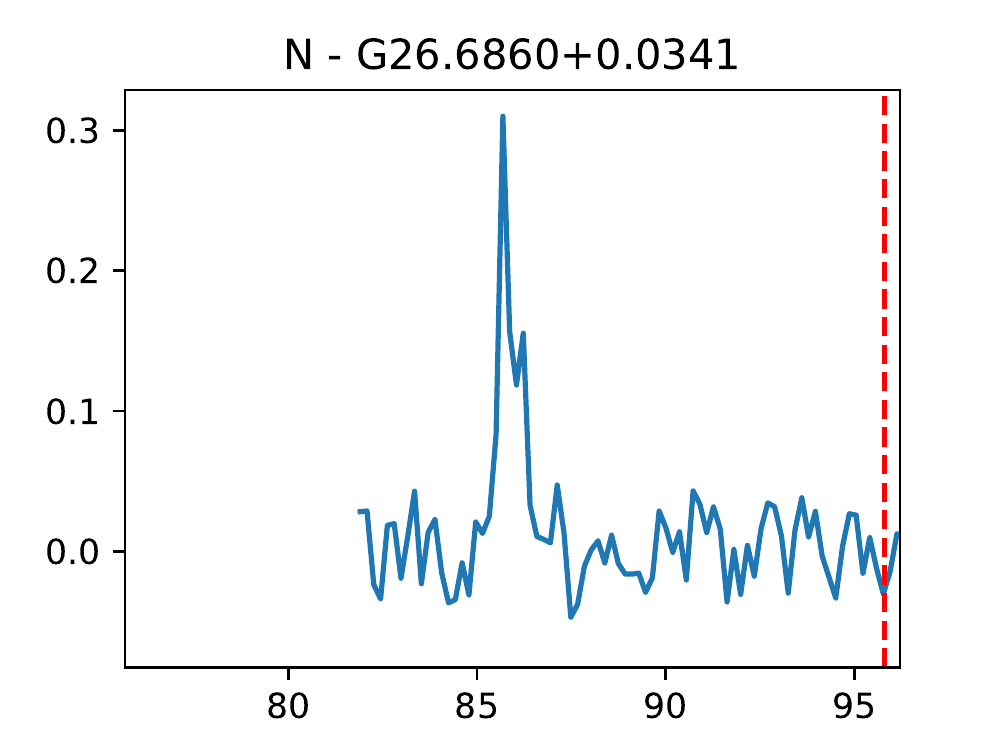}
\includegraphics[trim={0.0cm 0.3cm 0.6cm 0.3cm}, clip, width = 0.235\textwidth]{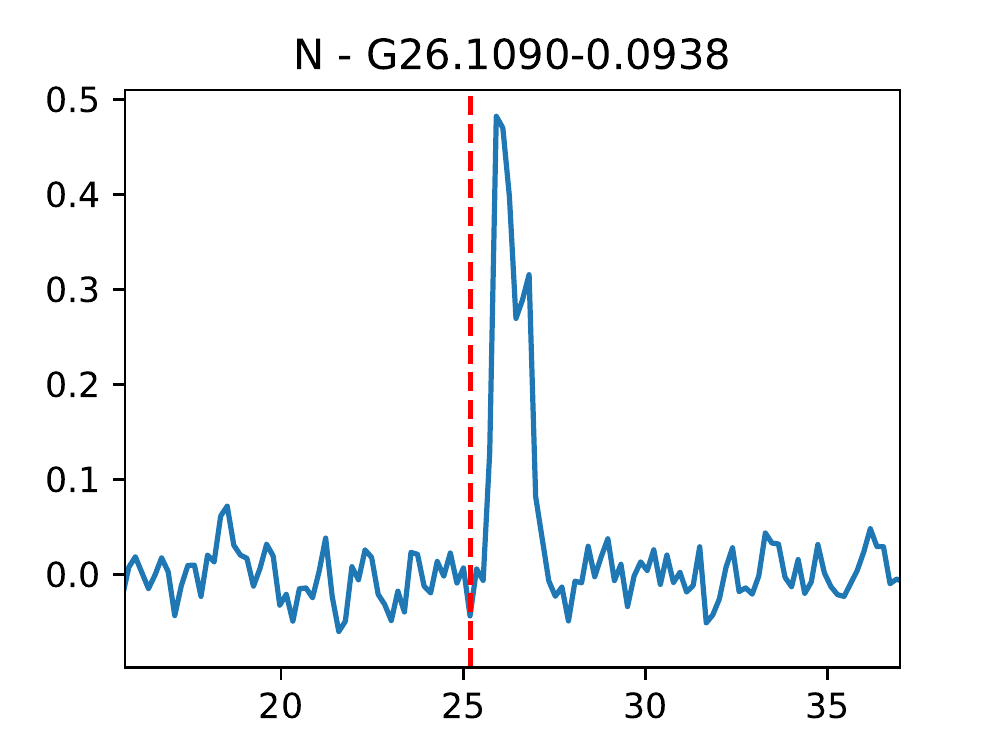}
\includegraphics[trim={0.0cm 0.3cm 0.6cm 0.3cm}, clip, width = 0.235\textwidth]{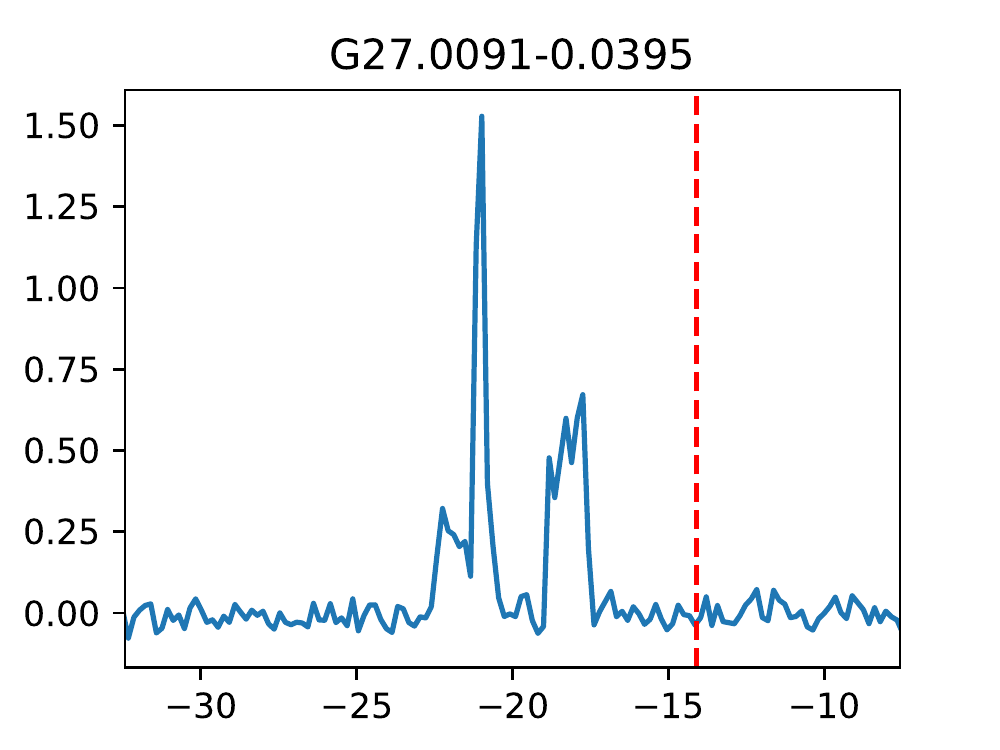}
\includegraphics[trim={0.0cm 0.3cm 0.6cm 0.3cm}, clip, width = 0.235\textwidth]{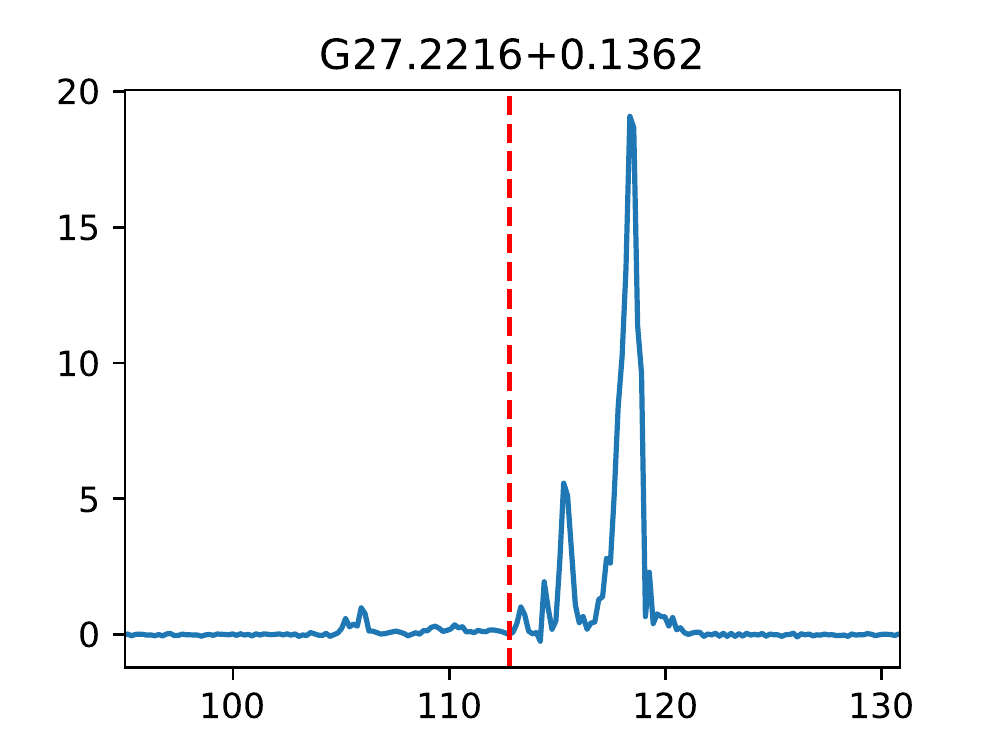}\\ 
\colorbox{white}{\includegraphics[trim={2cm 0 0 0}, scale=0.195]{figs/ylabel_filler.pdf}}
\includegraphics[trim={0.0cm 0.3cm 0.6cm 0.3cm}, clip, width = 0.235\textwidth]{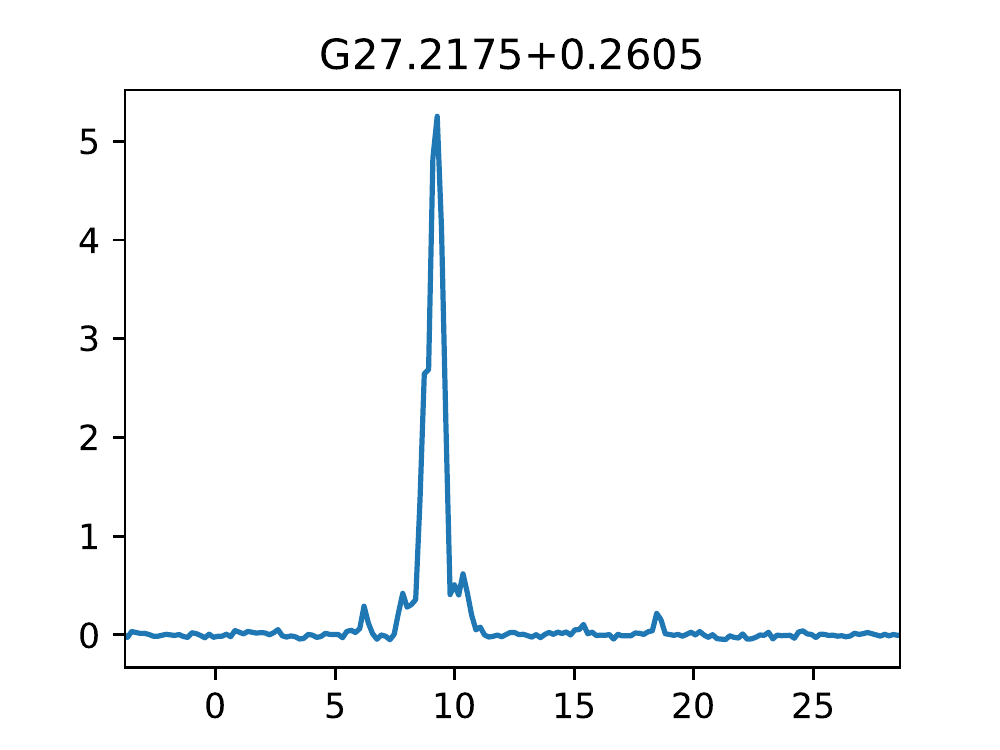}
\includegraphics[trim={0.0cm 0.3cm 0.6cm 0.3cm}, clip, width = 0.235\textwidth]{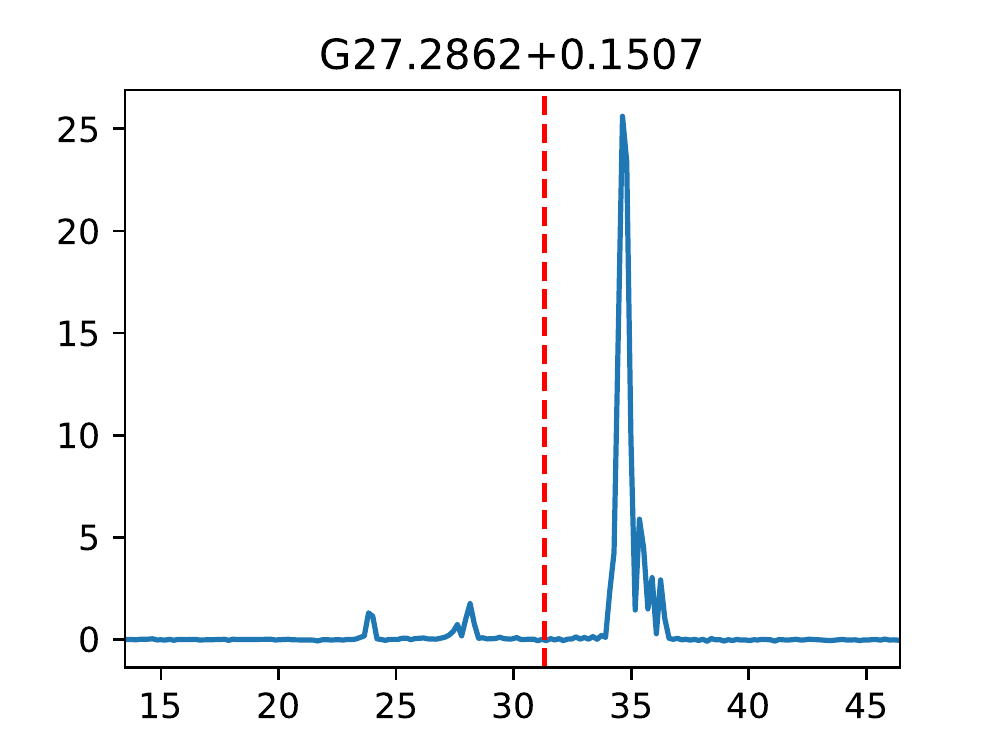}
\includegraphics[trim={0.0cm 0.3cm 0.6cm 0.3cm}, clip, width = 0.235\textwidth]{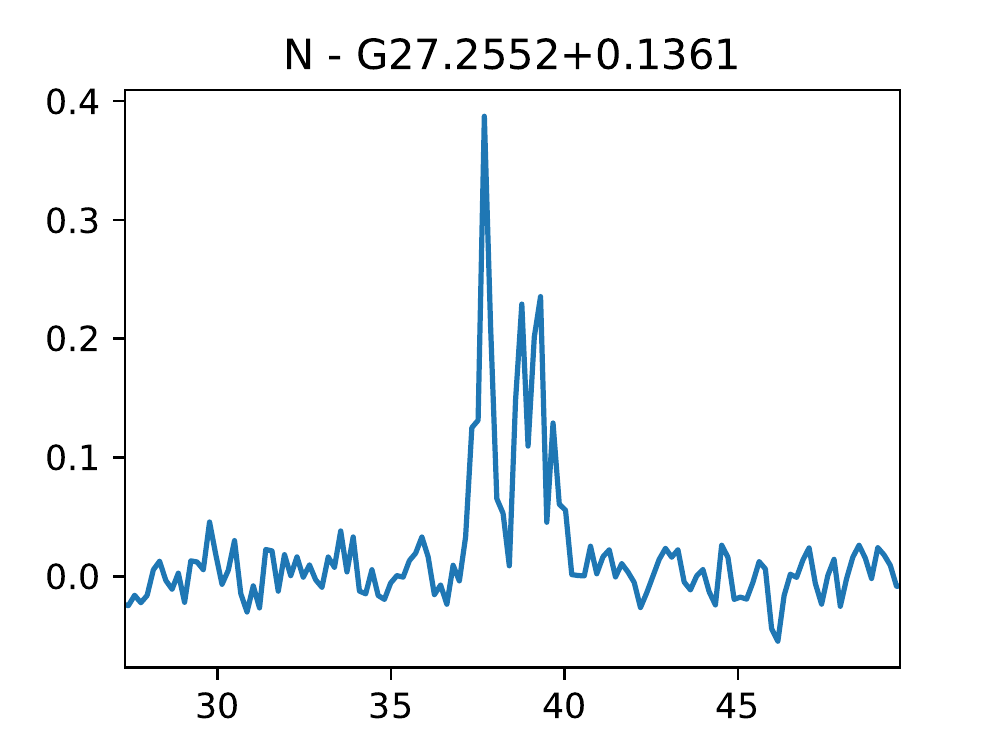}
\includegraphics[trim={0.0cm 0.3cm 0.6cm 0.3cm}, clip, width = 0.235\textwidth]{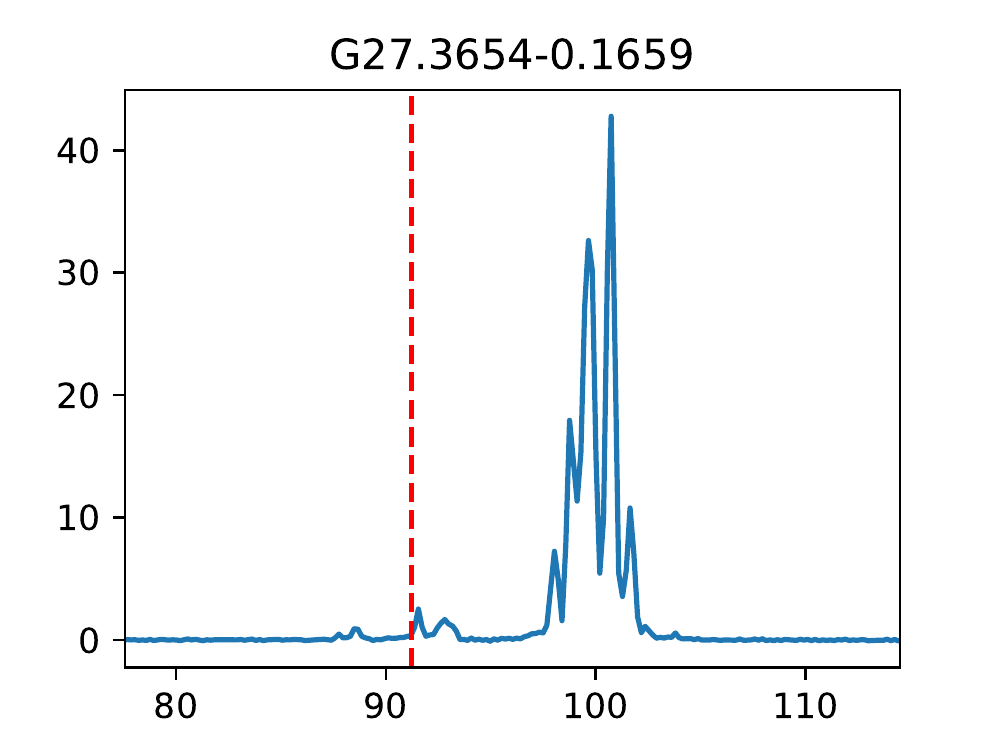}\\ 
\colorbox{white}{\includegraphics[trim={2cm 0 0 0}, scale=0.20]{figs/ylabel.pdf}}
\includegraphics[trim={0.0cm 0.3cm 0.6cm 0.3cm}, clip, width = 0.235\textwidth]{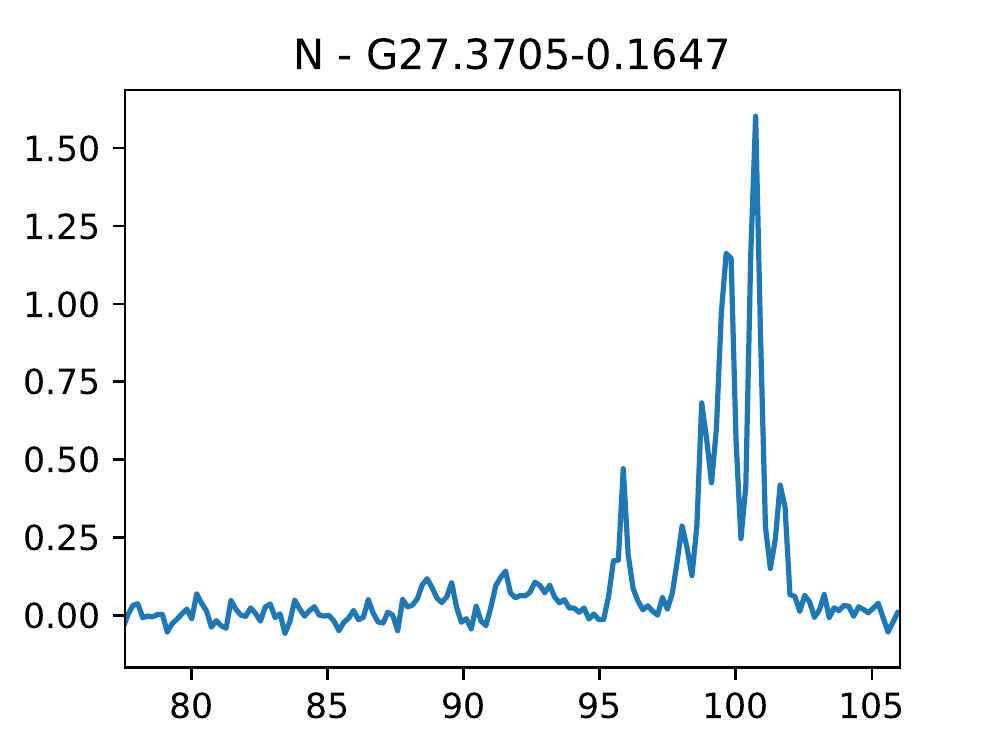}
\includegraphics[trim={0.0cm 0.3cm 0.6cm 0.3cm}, clip, width = 0.235\textwidth]{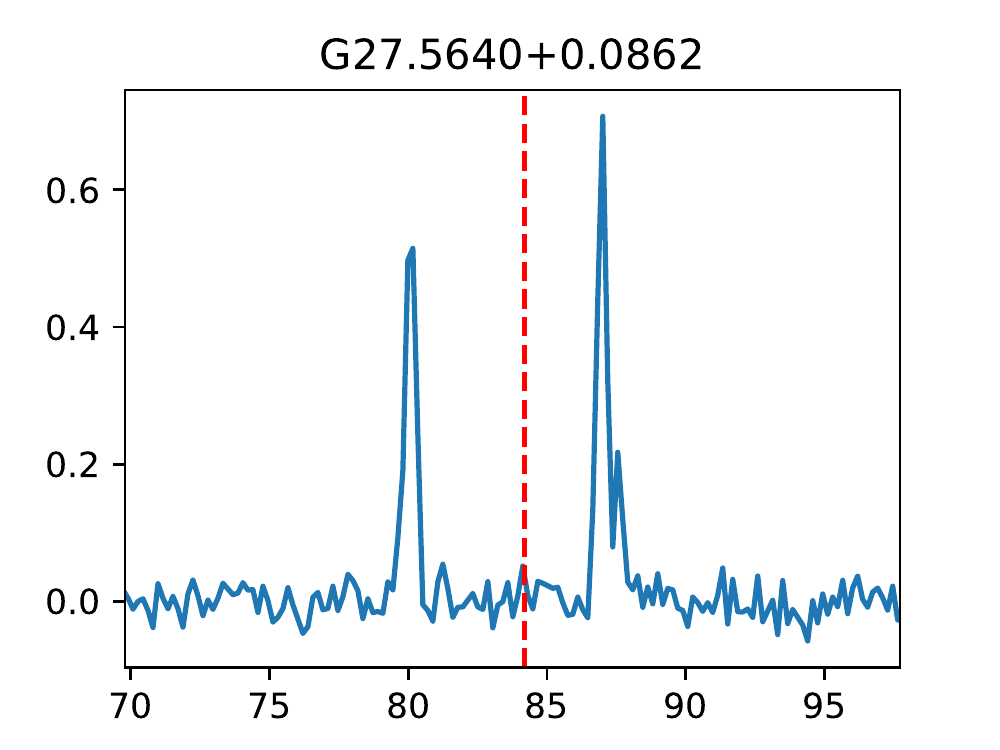}
\includegraphics[trim={0.0cm 0.3cm 0.6cm 0.3cm}, clip, width = 0.235\textwidth]{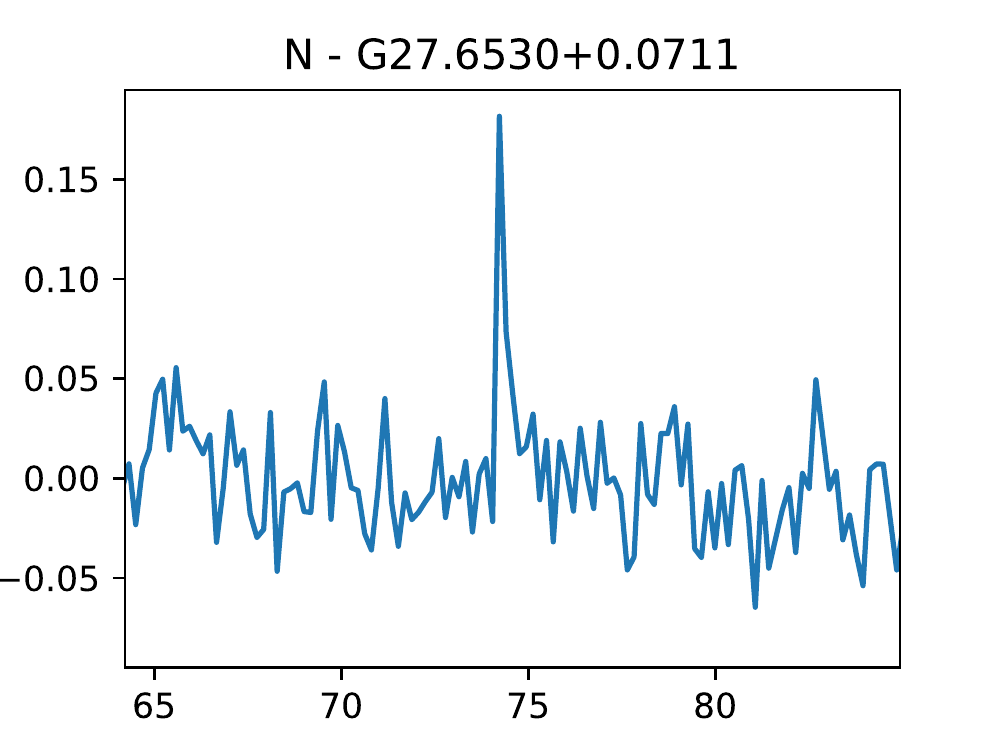}
\includegraphics[trim={0.0cm 0.3cm 0.6cm 0.3cm}, clip, width = 0.235\textwidth]{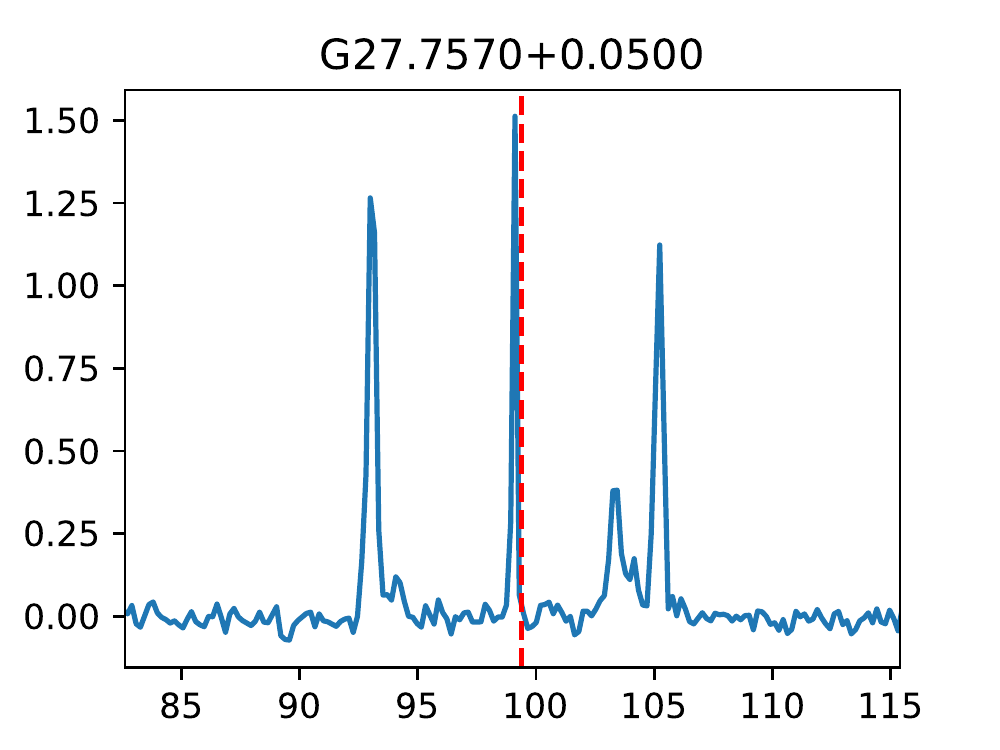}\\
\end{tabular}
\caption{Continued from Fig.~\ref{fig:maser_many_spec}.}
\label{fig:masers_336}
\end{figure*}

%Table 15
\begin{figure*}[!h]
\begin{tabular}{ccccc}
\colorbox{white}{\includegraphics[trim={2cm 0 0 0}, scale=0.195]{figs/ylabel_filler.pdf}}
\includegraphics[trim={0.0cm 0.3cm 0.6cm 0.3cm}, clip, width = 0.235\textwidth]{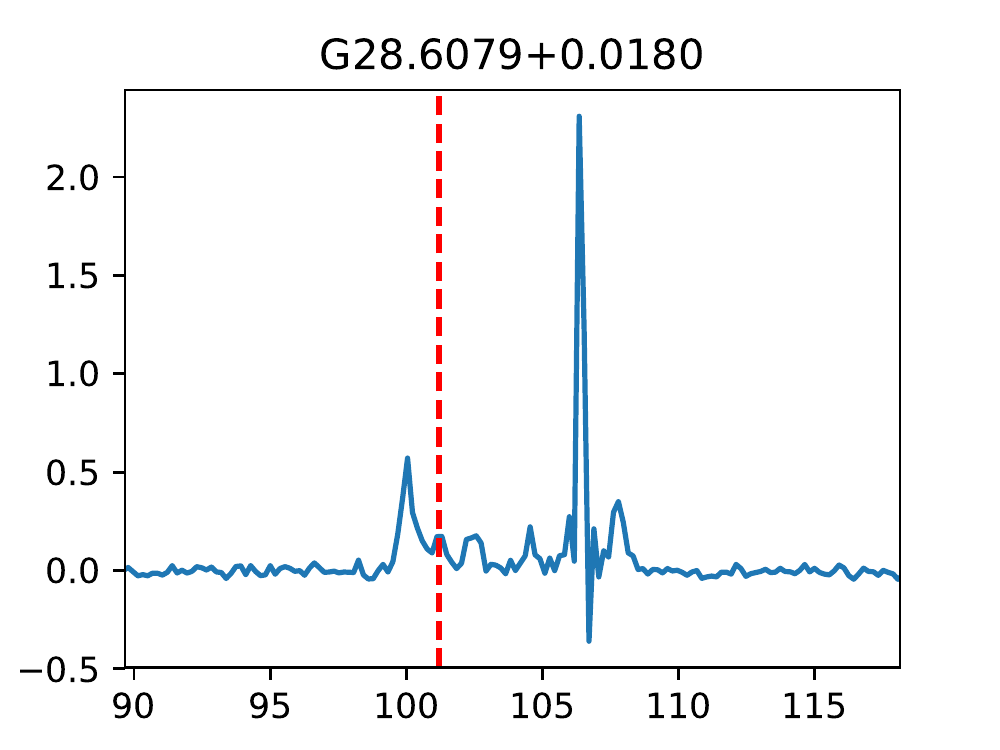}
\includegraphics[trim={0.0cm 0.3cm 0.6cm 0.3cm}, clip, width = 0.235\textwidth]{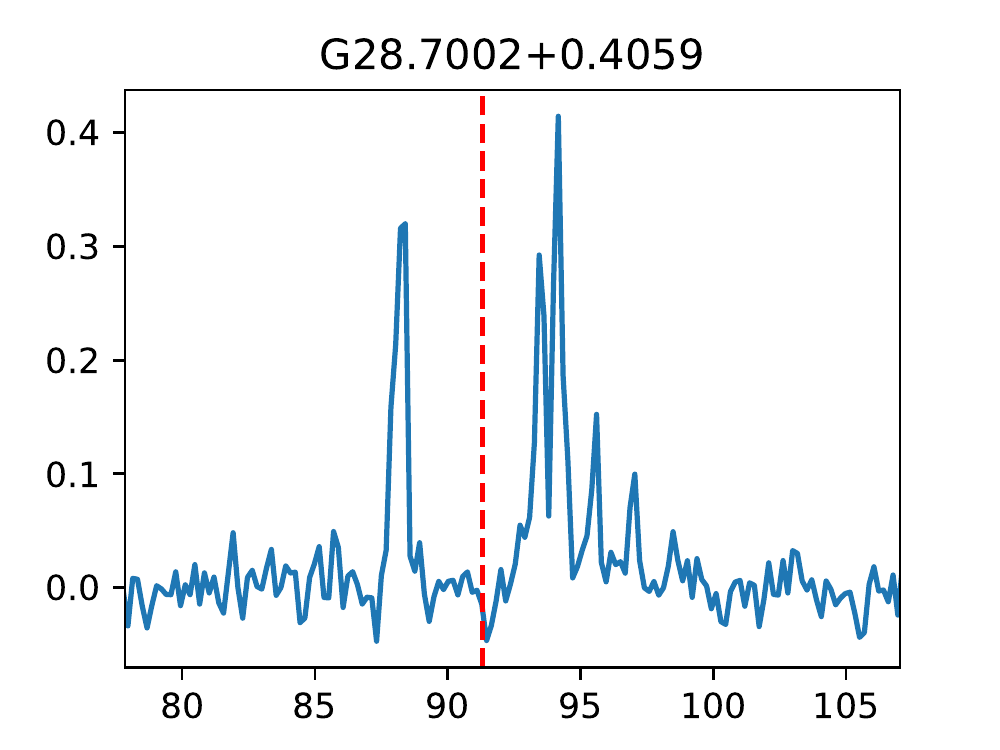}
\includegraphics[trim={0.0cm 0.3cm 0.6cm 0.3cm}, clip, width = 0.235\textwidth]{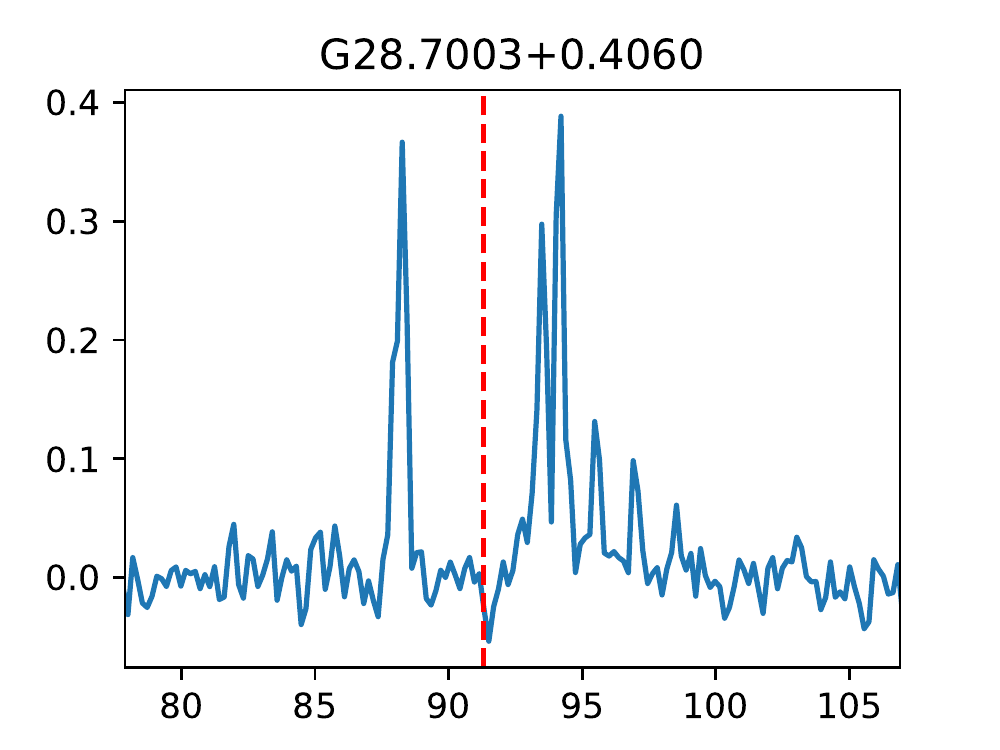}
\includegraphics[trim={0.0cm 0.3cm 0.6cm 0.3cm}, clip, width = 0.235\textwidth]{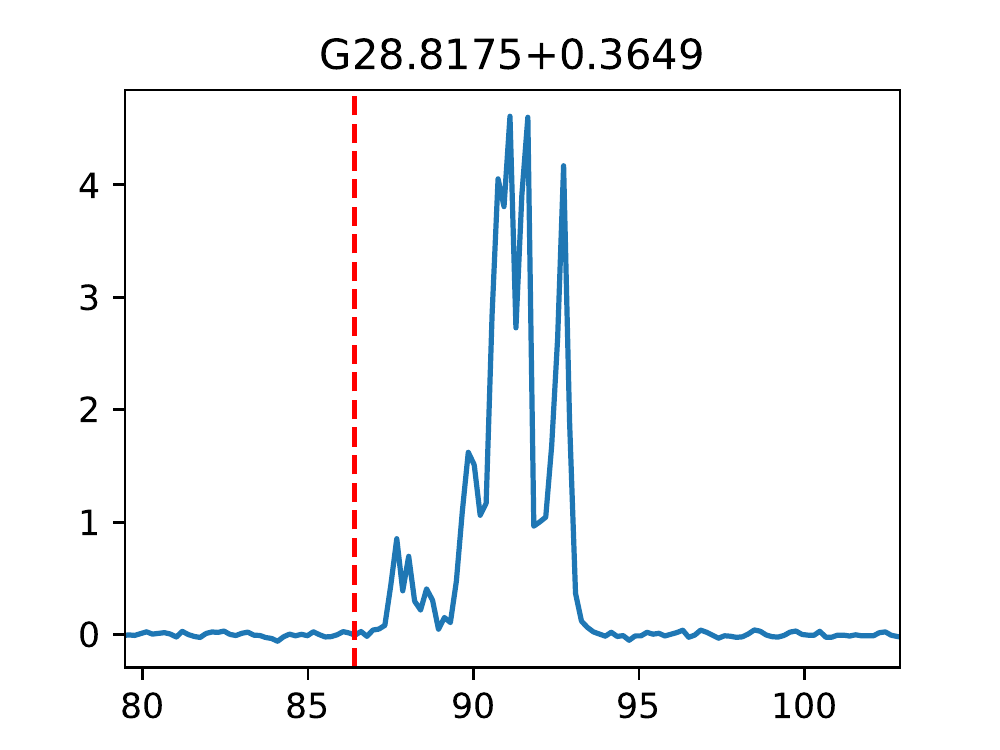}\\ 
\colorbox{white}{\includegraphics[trim={2cm 0 0 0}, scale=0.195]{figs/ylabel_filler.pdf}}
\includegraphics[trim={0.0cm 0.3cm 0.6cm 0.3cm}, clip, width = 0.235\textwidth]{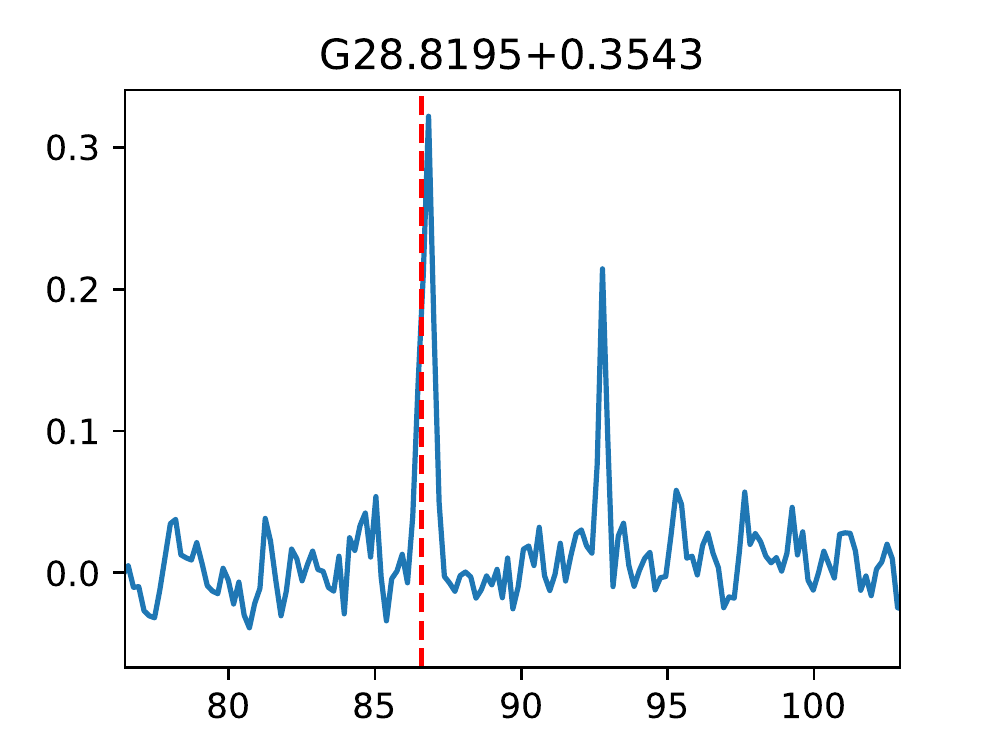}
\includegraphics[trim={0.0cm 0.3cm 0.6cm 0.3cm}, clip, width = 0.235\textwidth]{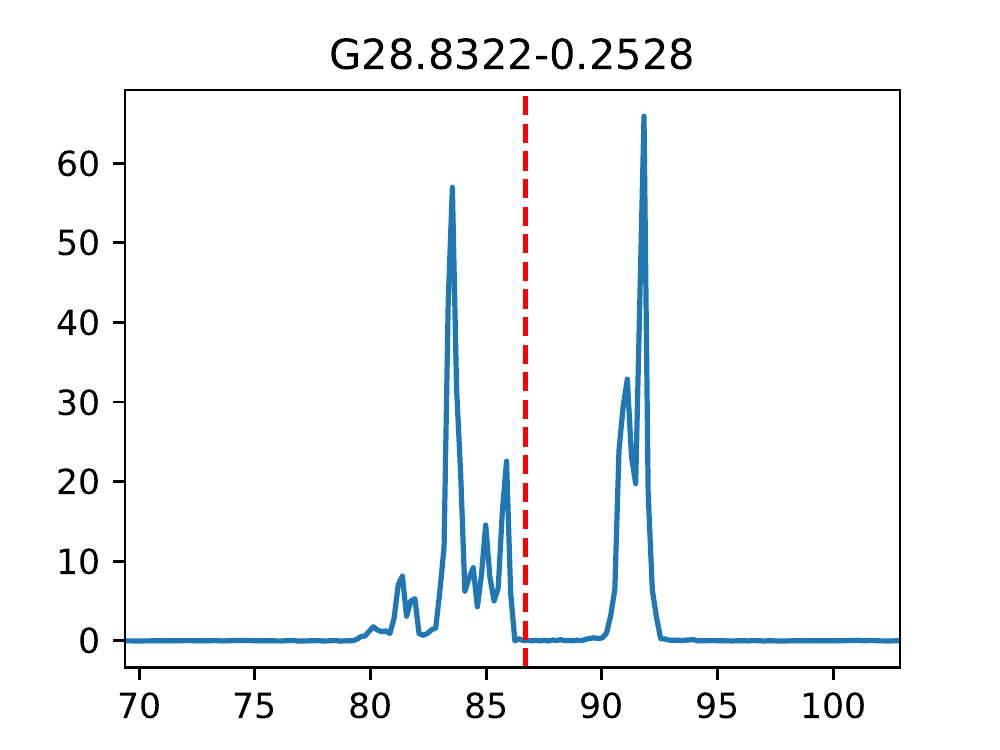}
\includegraphics[trim={0.0cm 0.3cm 0.6cm 0.3cm}, clip, width = 0.235\textwidth]{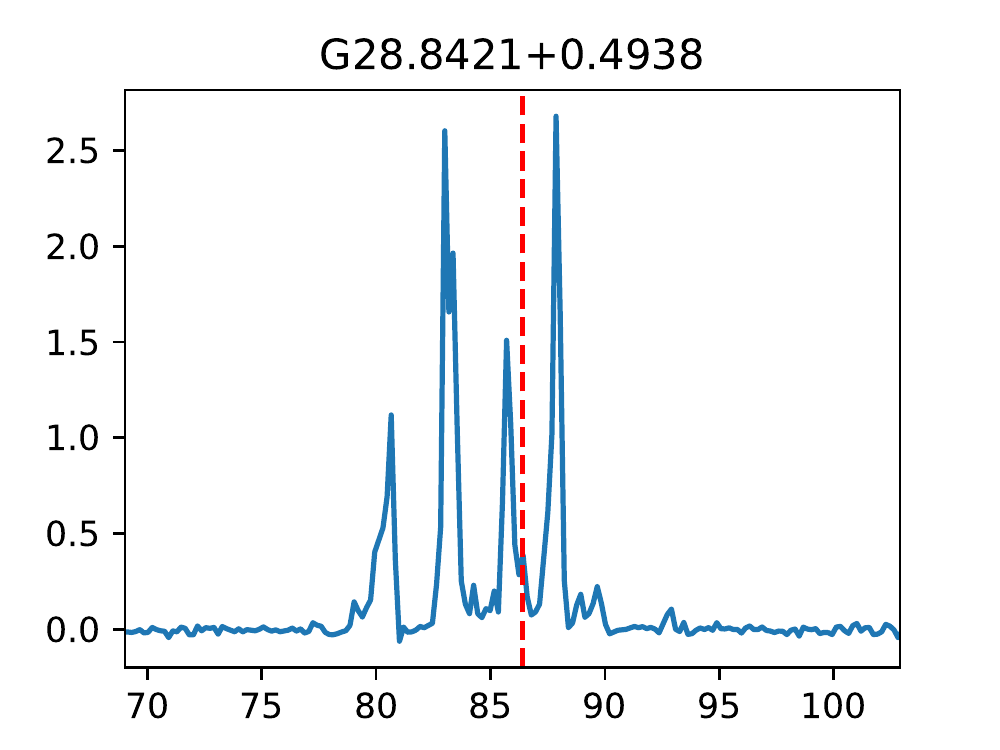}
\includegraphics[trim={0.0cm 0.3cm 0.6cm 0.3cm}, clip, width = 0.235\textwidth]{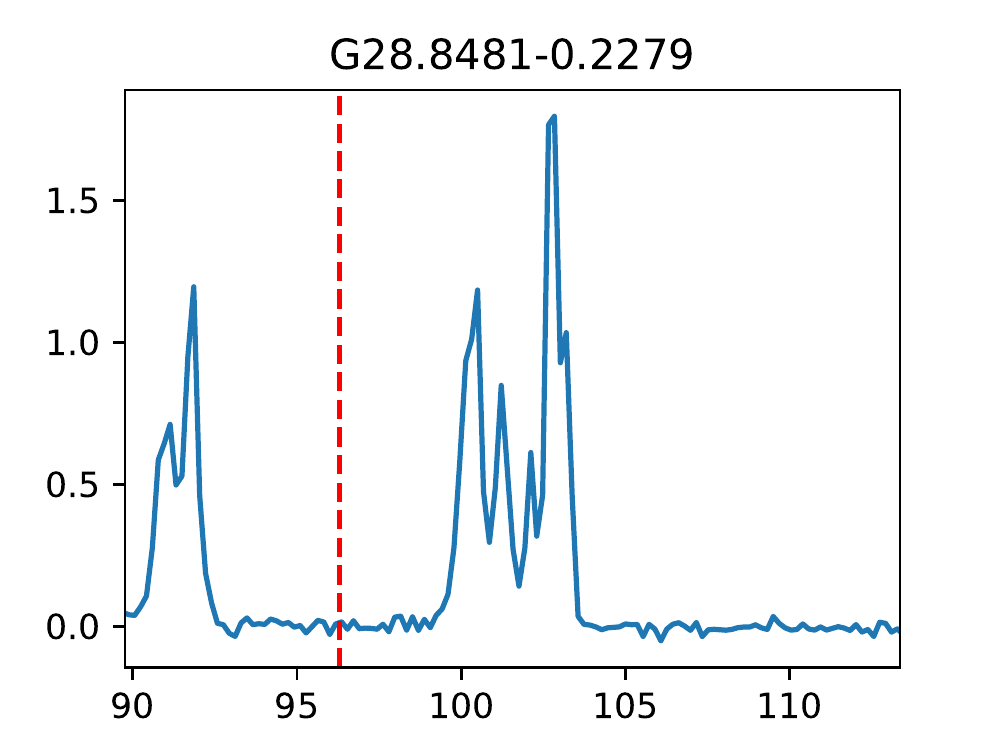}\\ 
\colorbox{white}{\includegraphics[trim={2cm 0 0 0}, scale=0.20]{figs/ylabel.pdf}}
\includegraphics[trim={0.0cm 0.3cm 0.6cm 0.3cm}, clip, width = 0.235\textwidth]{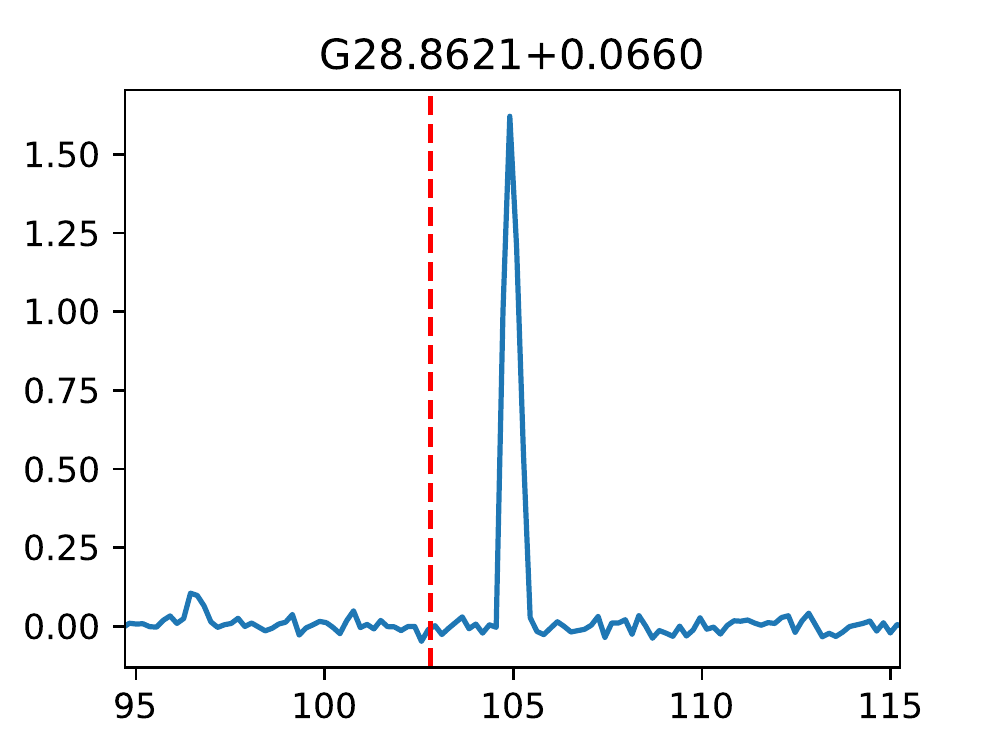}
\includegraphics[trim={0.0cm 0.3cm 0.6cm 0.3cm}, clip, width = 0.235\textwidth]{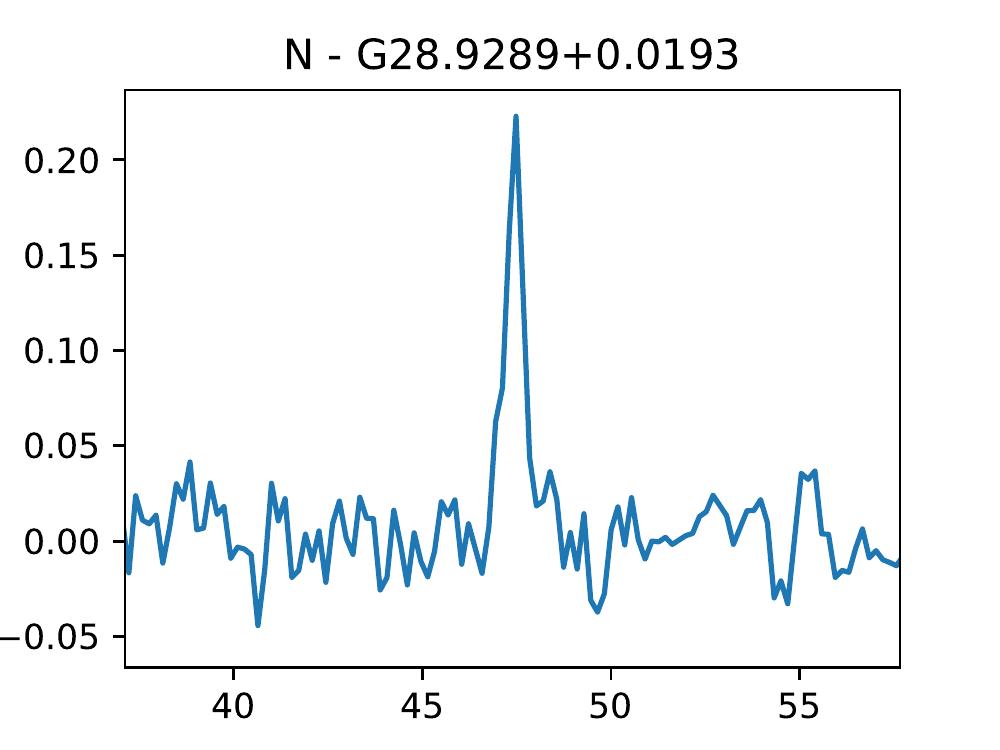}
\includegraphics[trim={0.0cm 0.3cm 0.6cm 0.3cm}, clip, width = 0.235\textwidth]{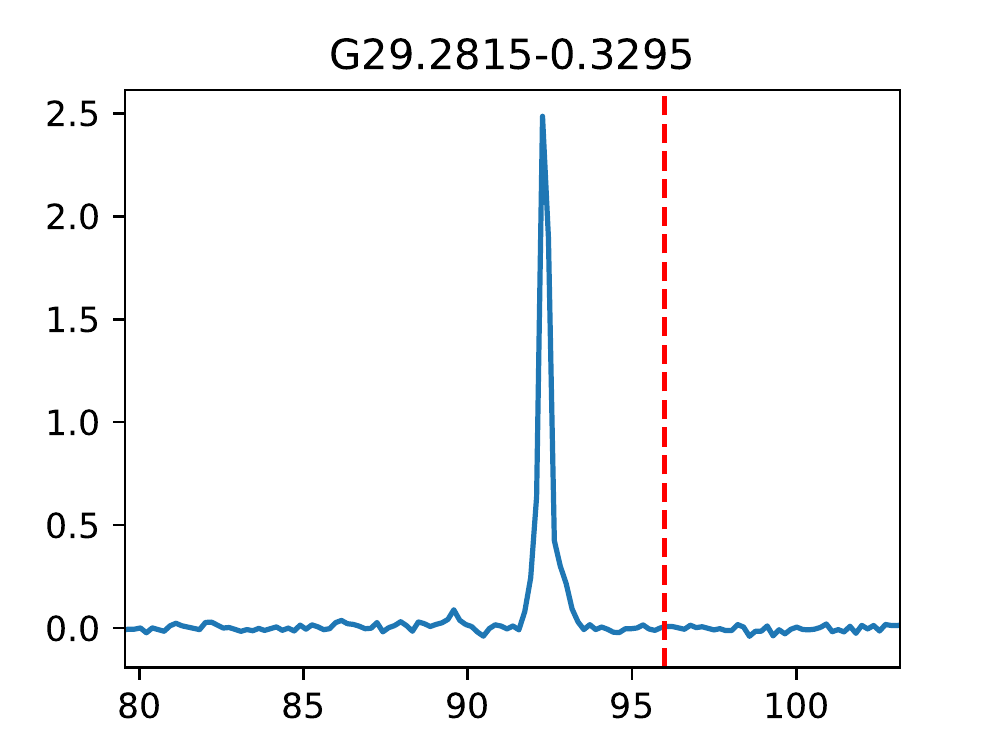}
\includegraphics[trim={0.0cm 0.3cm 0.6cm 0.3cm}, clip, width = 0.235\textwidth]{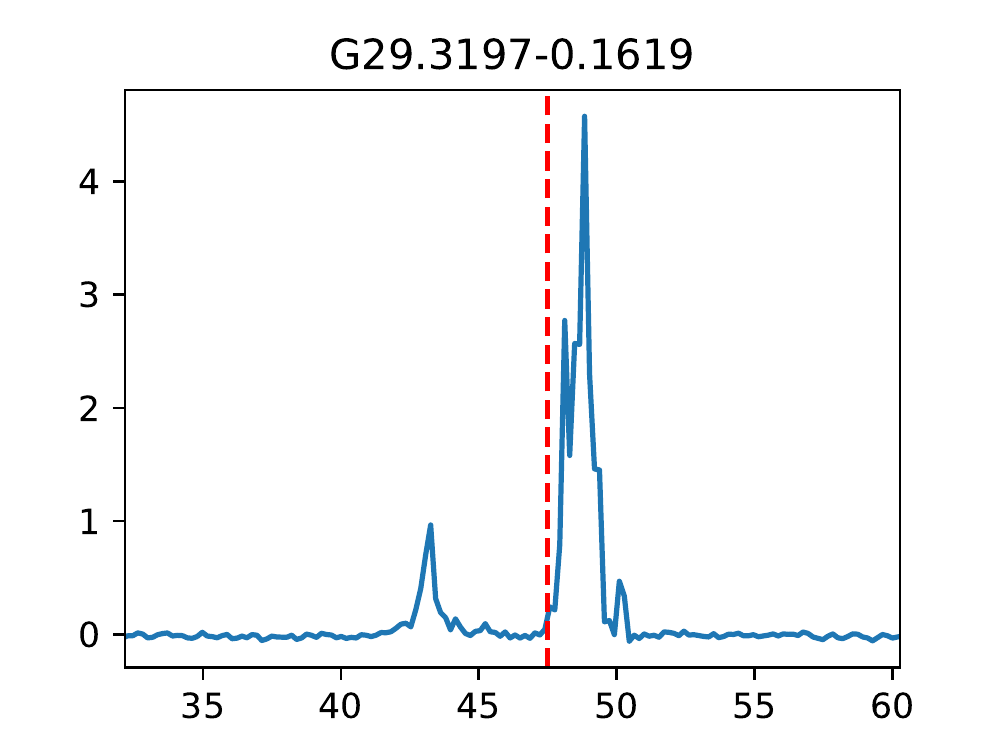}\\
\end{tabular}
\caption{Continued from Fig.~\ref{fig:maser_many_spec}.}
\label{fig:masers_360}
\end{figure*}

%Table 16
\begin{figure*}[!h]
\begin{tabular}{ccccc}
\colorbox{white}{\includegraphics[trim={2cm 0 0 0}, scale=0.195]{figs/ylabel_filler.pdf}}
\includegraphics[trim={0.0cm 0.3cm 0.6cm 0.3cm}, clip, width = 0.235\textwidth]{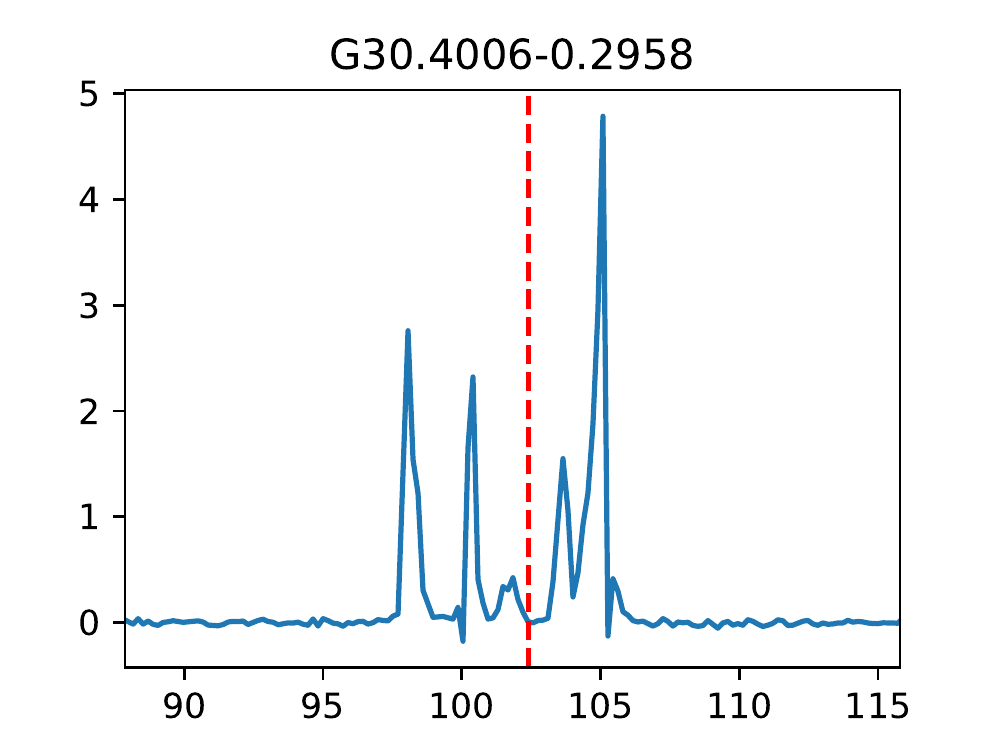}
\includegraphics[trim={0.0cm 0.3cm 0.6cm 0.3cm}, clip, width = 0.235\textwidth]{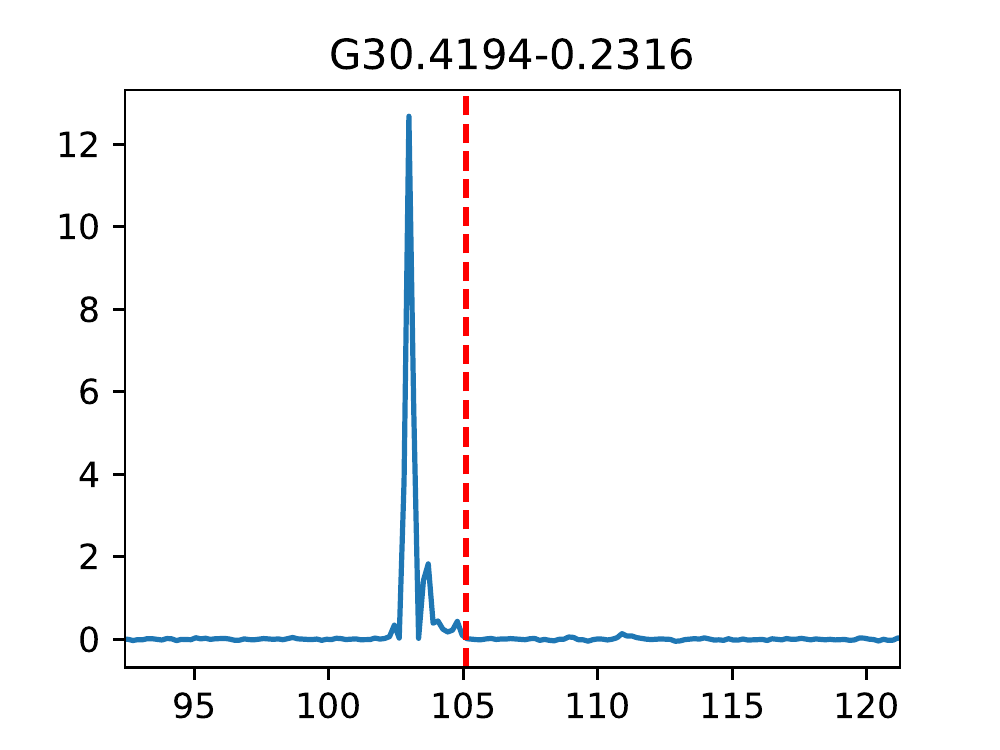}
\includegraphics[trim={0.0cm 0.3cm 0.6cm 0.3cm}, clip, width = 0.235\textwidth]{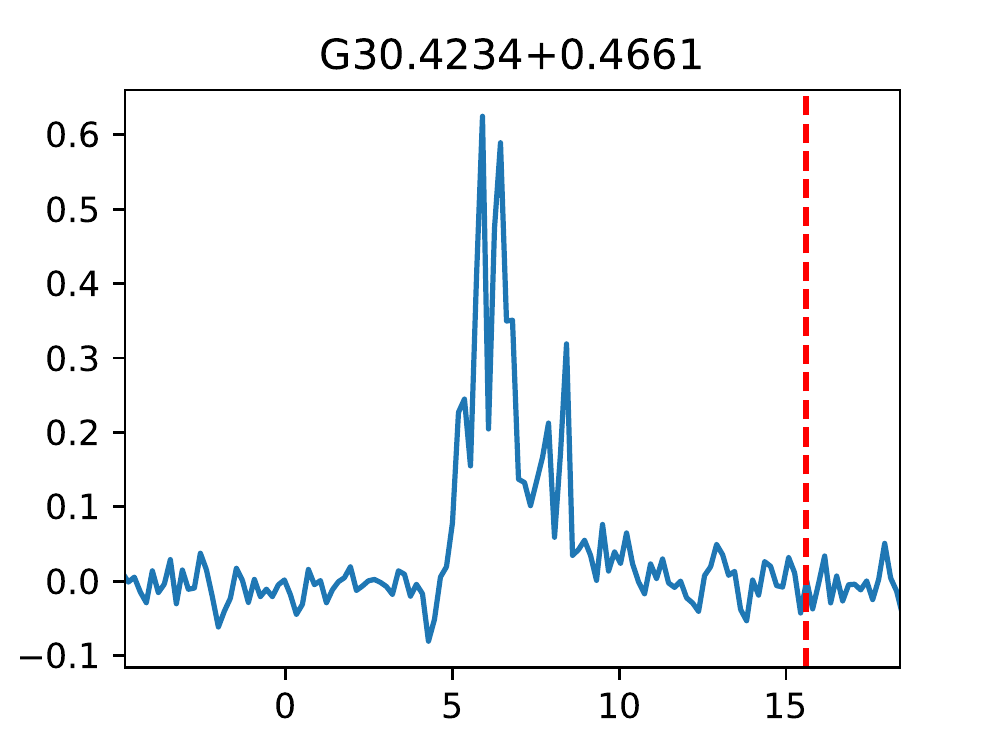}
\includegraphics[trim={0.0cm 0.3cm 0.6cm 0.3cm}, clip, width = 0.235\textwidth]{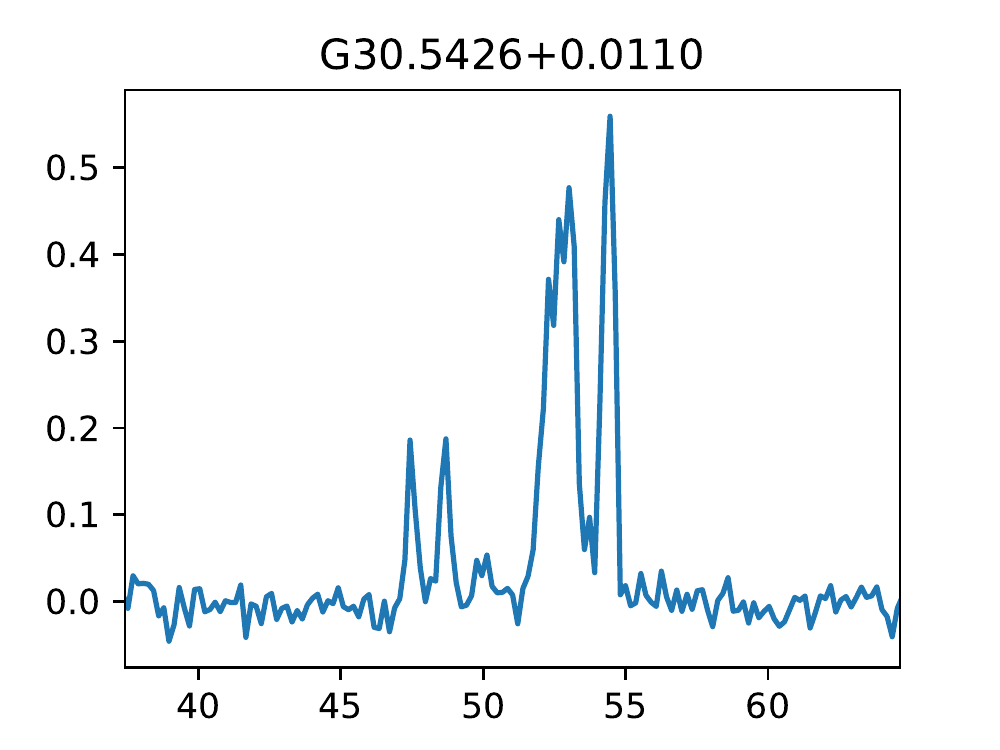}\\ 
\colorbox{white}{\includegraphics[trim={2cm 0 0 0}, scale=0.195]{figs/ylabel_filler.pdf}}
\includegraphics[trim={0.0cm 0.3cm 0.6cm 0.3cm}, clip, width = 0.235\textwidth]{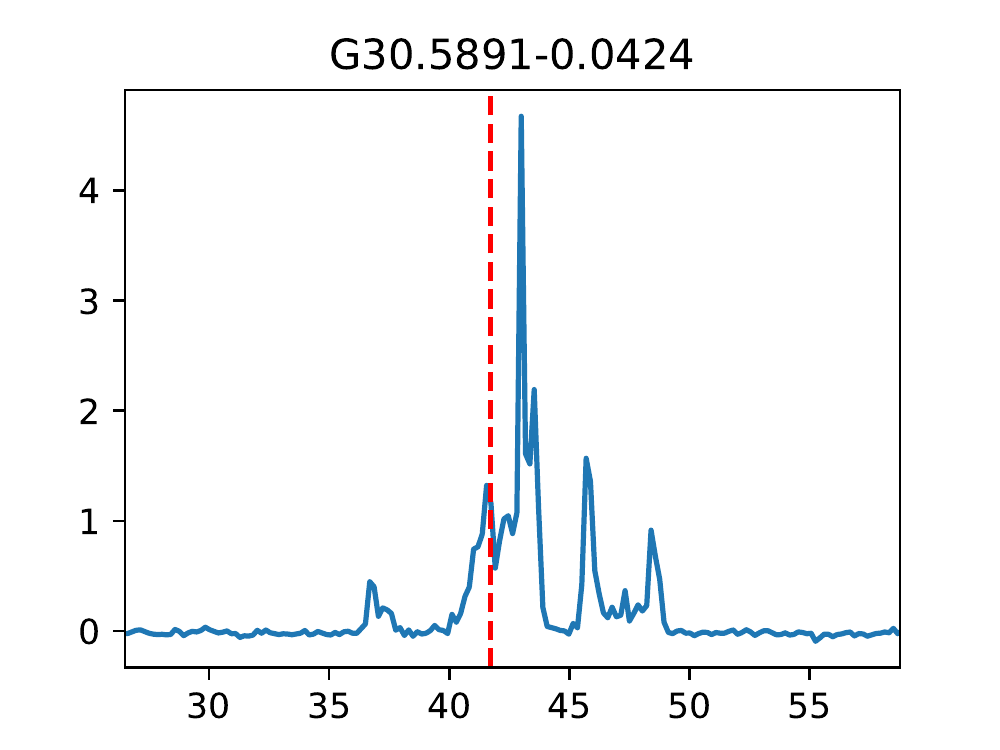}
\includegraphics[trim={0.0cm 0.3cm 0.6cm 0.3cm}, clip, width = 0.235\textwidth]{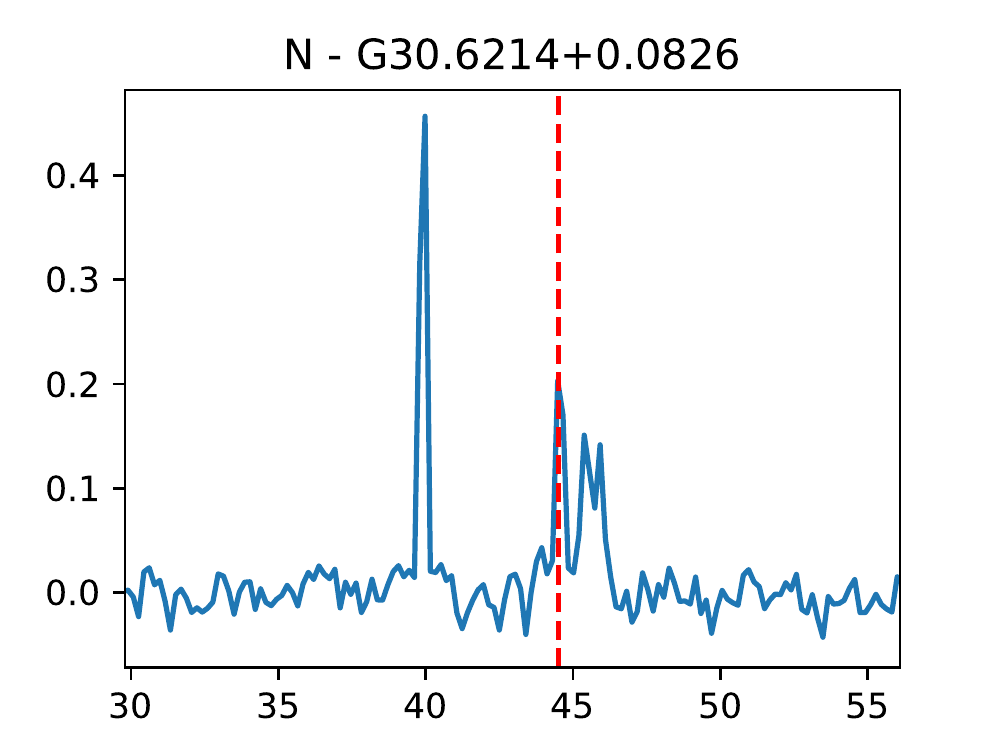}
\includegraphics[trim={0.0cm 0.3cm 0.6cm 0.3cm}, clip, width = 0.235\textwidth]{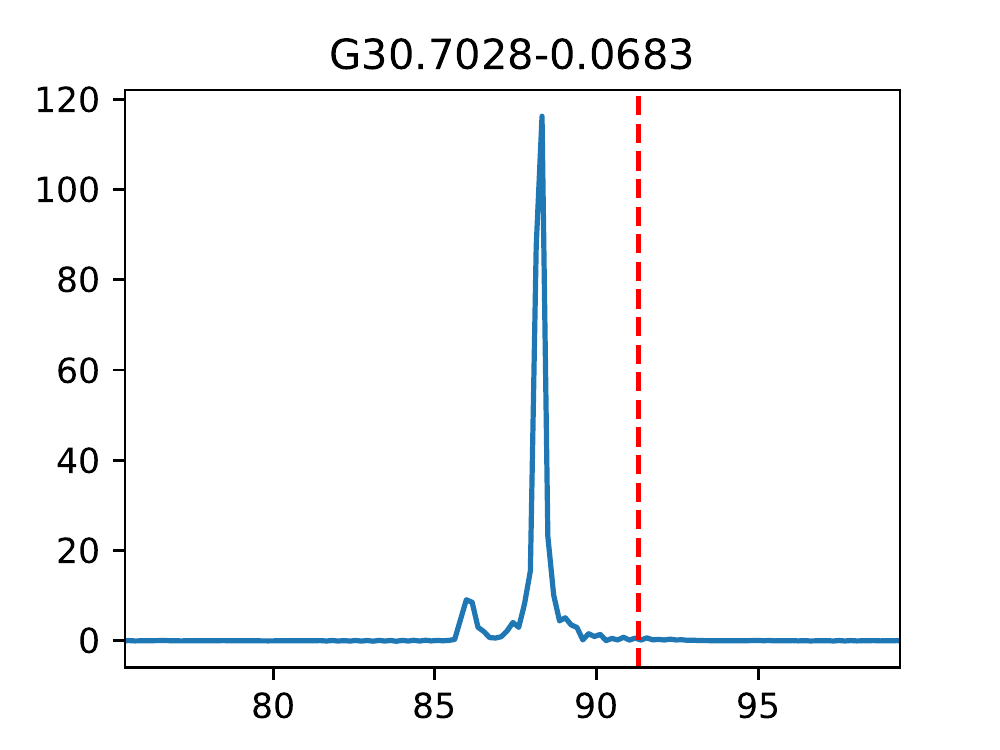}
\includegraphics[trim={0.0cm 0.3cm 0.6cm 0.3cm}, clip, width = 0.235\textwidth]{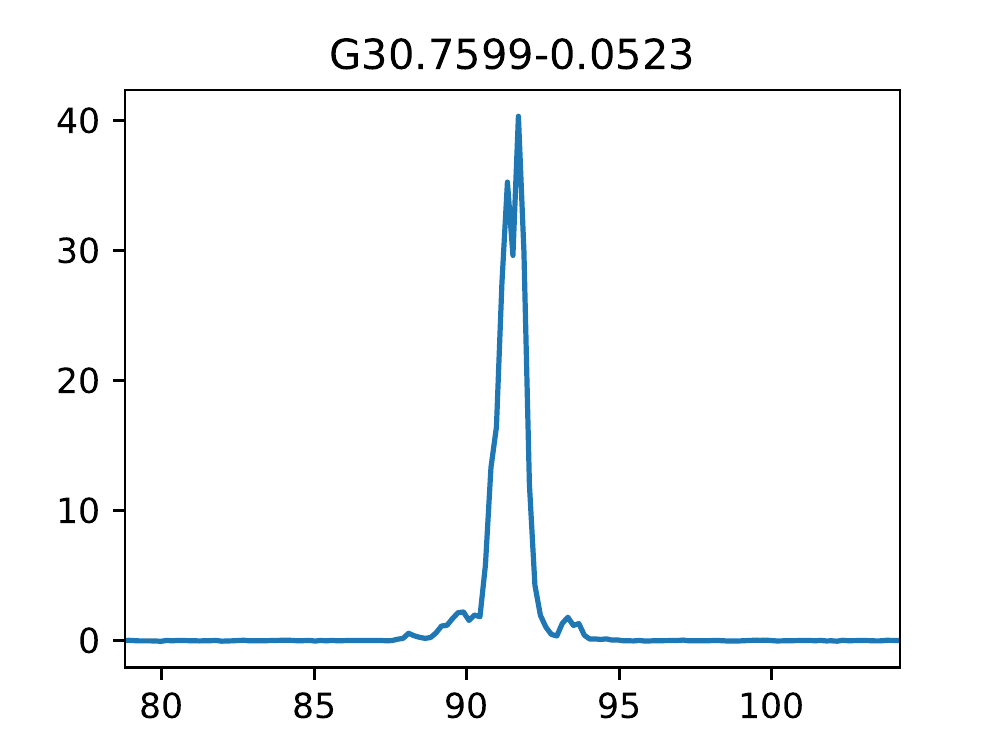}\\ 
\colorbox{white}{\includegraphics[trim={2cm 0 0 0}, scale=0.20]{figs/ylabel.pdf}}
\includegraphics[trim={0.0cm 0.3cm 0.6cm 0.3cm}, clip, width = 0.235\textwidth]{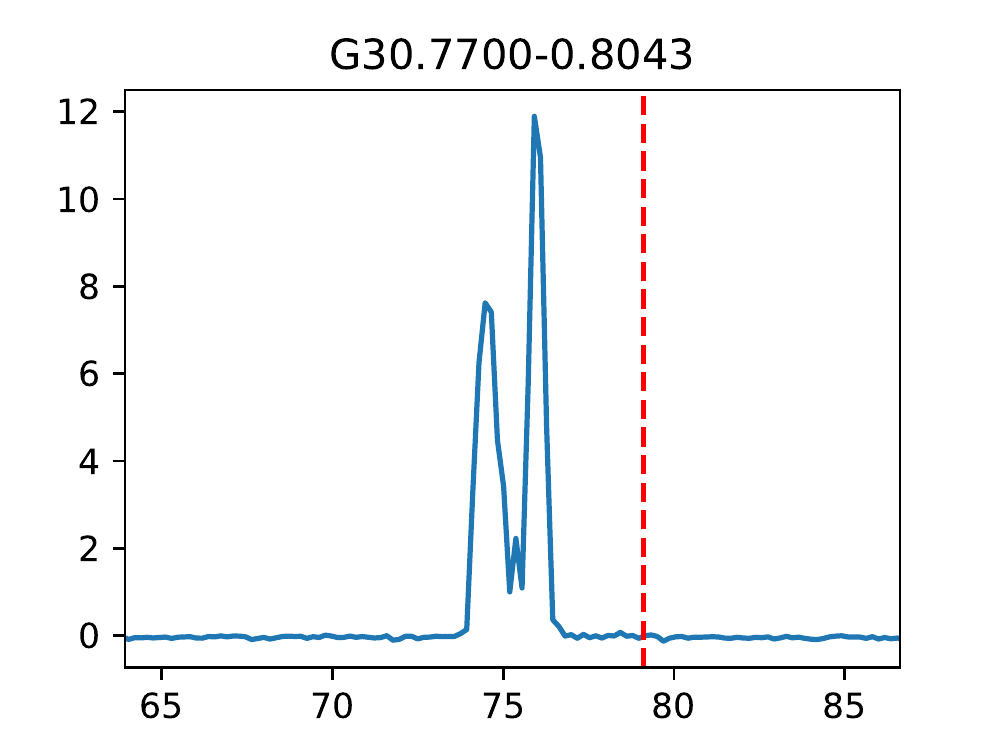}
\includegraphics[trim={0.0cm 0.3cm 0.6cm 0.3cm}, clip, width = 0.235\textwidth]{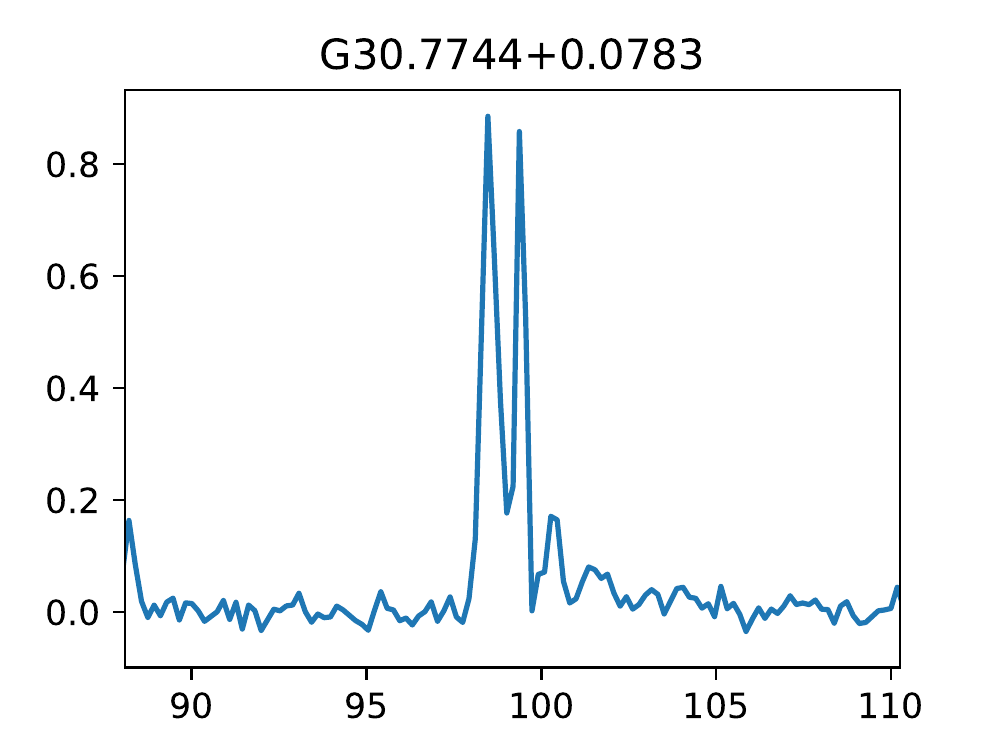}
\includegraphics[trim={0.0cm 0.3cm 0.6cm 0.3cm}, clip, width = 0.235\textwidth]{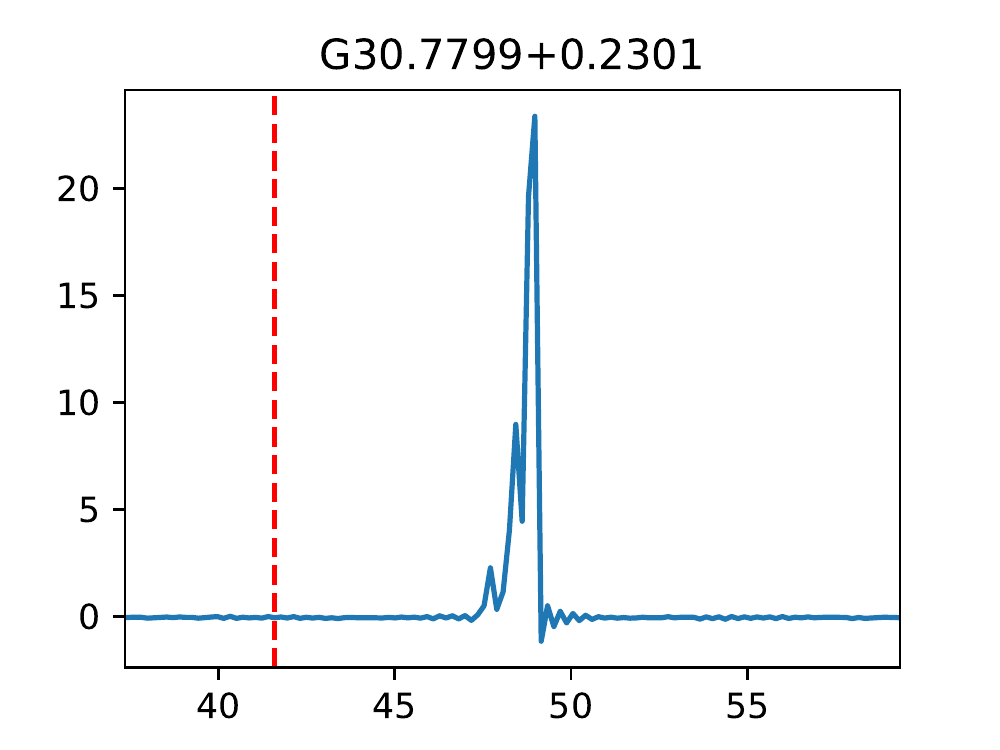}
\includegraphics[trim={0.0cm 0.3cm 0.6cm 0.3cm}, clip, width = 0.235\textwidth]{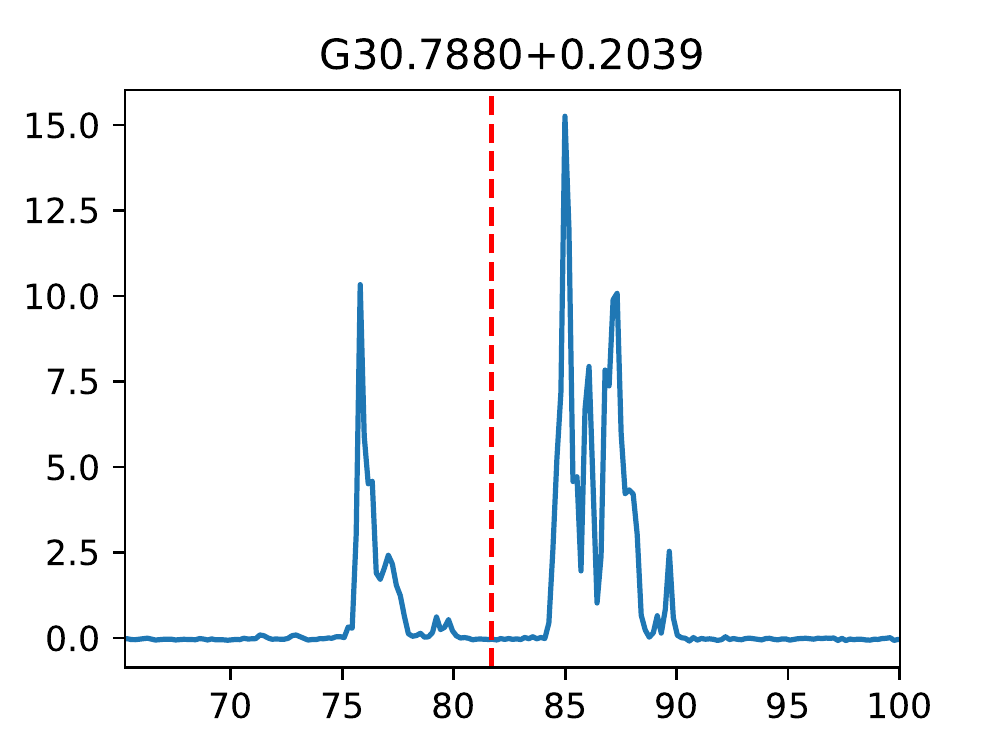}\\
\end{tabular}
\caption{Continued from Fig.~\ref{fig:maser_many_spec}.}
\label{fig:masers_384}
\end{figure*}

%Table 17
\begin{figure*}[!h]
\begin{tabular}{ccccc}
\colorbox{white}{\includegraphics[trim={2cm 0 0 0}, scale=0.195]{figs/ylabel_filler.pdf}}
\includegraphics[trim={0.0cm 0.3cm 0.6cm 0.3cm}, clip, width = 0.235\textwidth]{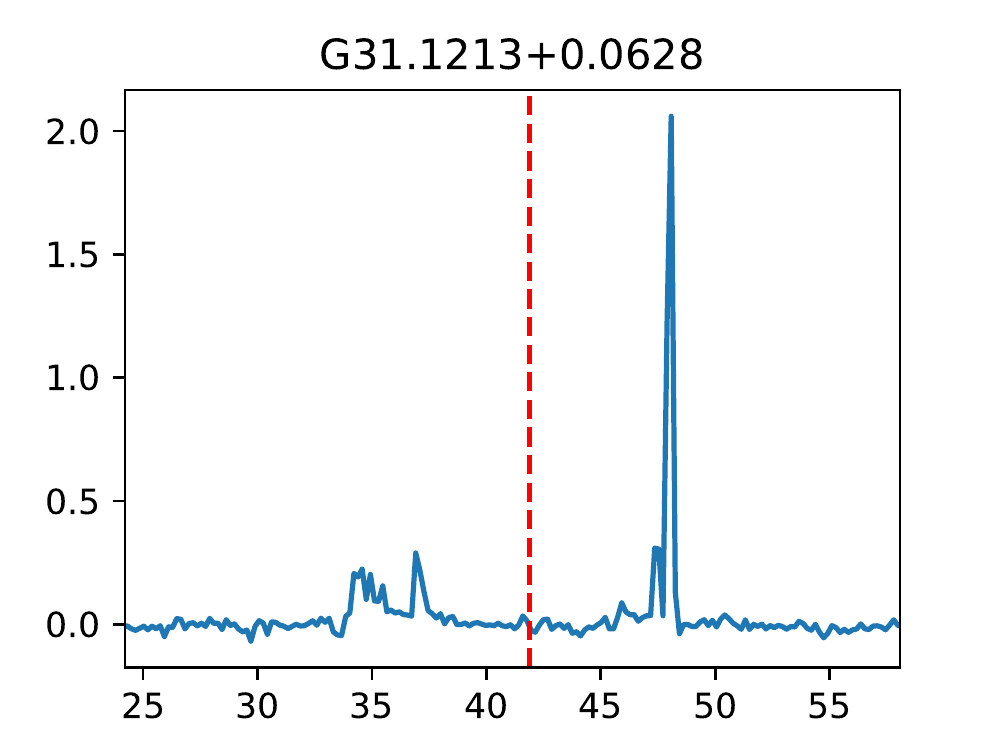}
\includegraphics[trim={0.0cm 0.3cm 0.6cm 0.3cm}, clip, width = 0.235\textwidth]{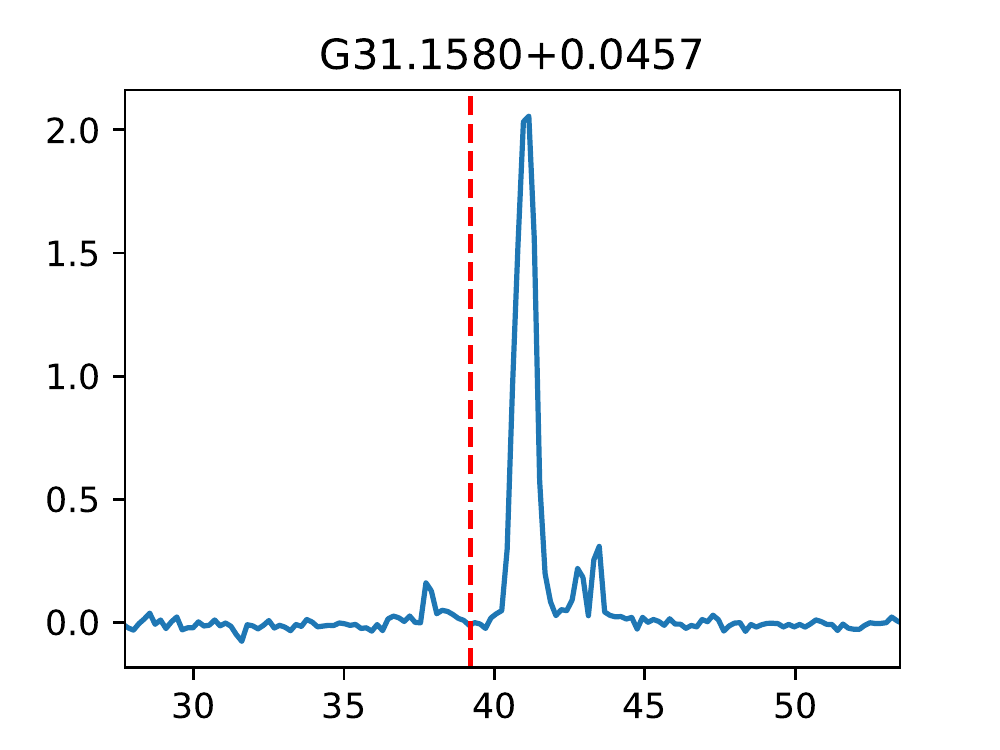}
\includegraphics[trim={0.0cm 0.3cm 0.6cm 0.3cm}, clip, width = 0.235\textwidth]{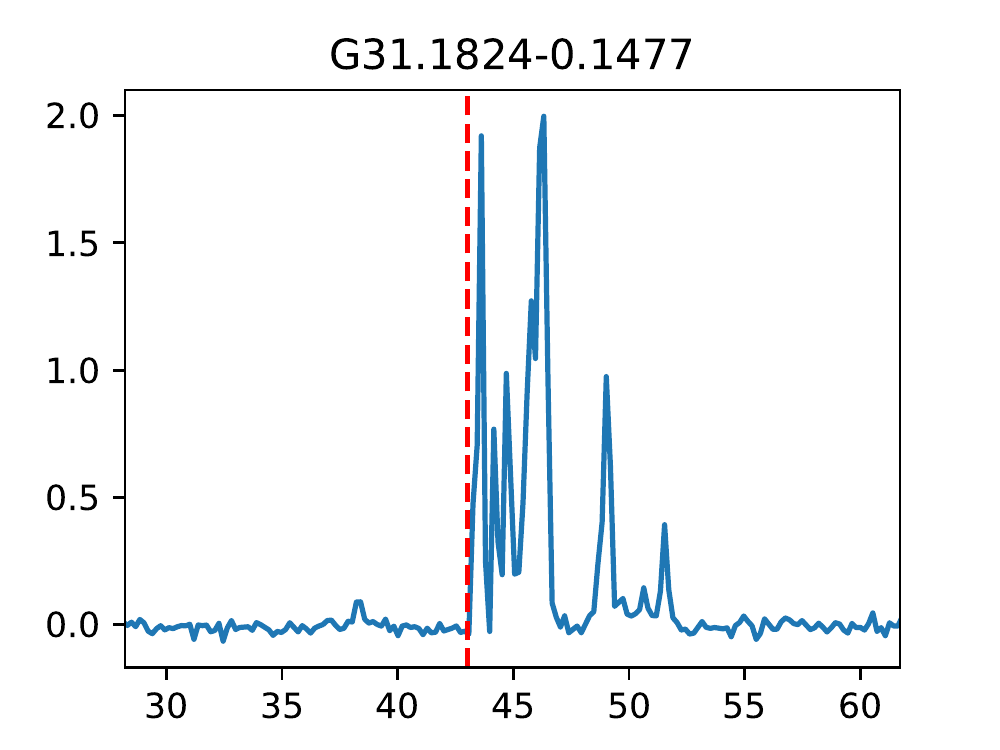}
\includegraphics[trim={0.0cm 0.3cm 0.6cm 0.3cm}, clip, width = 0.235\textwidth]{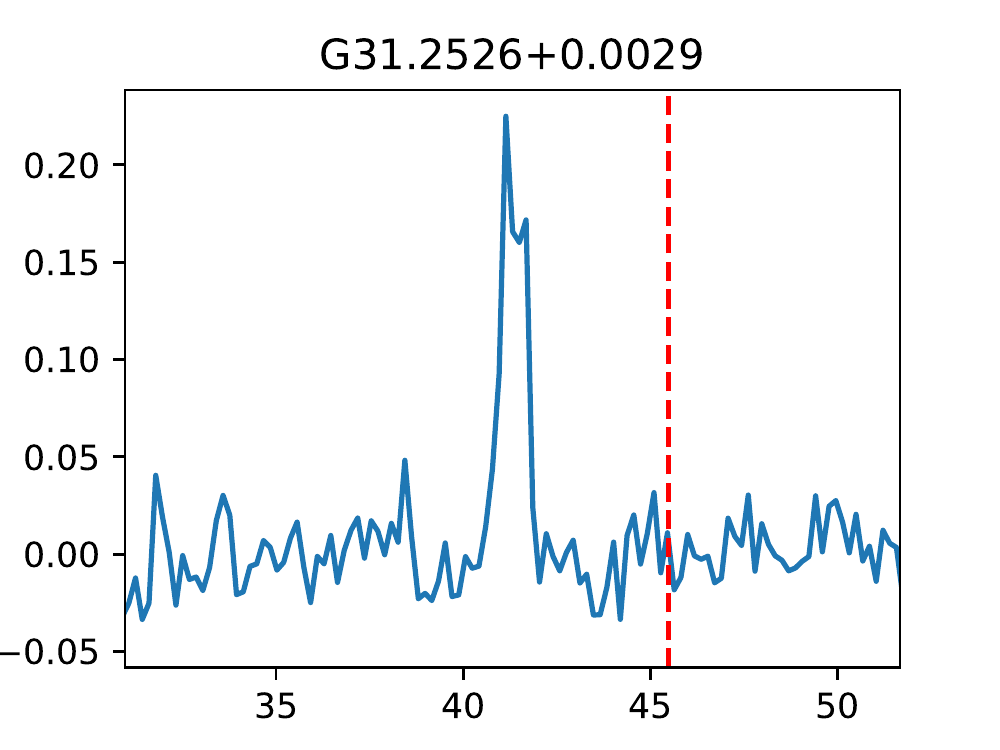}\\ 
\colorbox{white}{\includegraphics[trim={2cm 0 0 0}, scale=0.195]{figs/ylabel_filler.pdf}}
\includegraphics[trim={0.0cm 0.3cm 0.6cm 0.3cm}, clip, width = 0.235\textwidth]{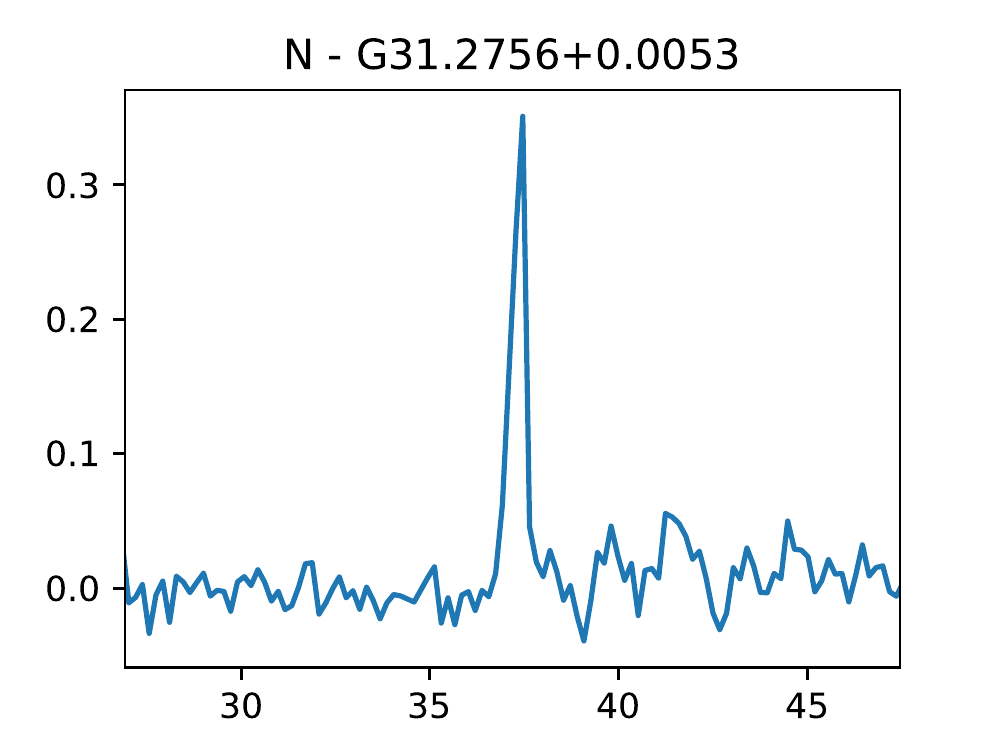}
\includegraphics[trim={0.0cm 0.3cm 0.6cm 0.3cm}, clip, width = 0.235\textwidth]{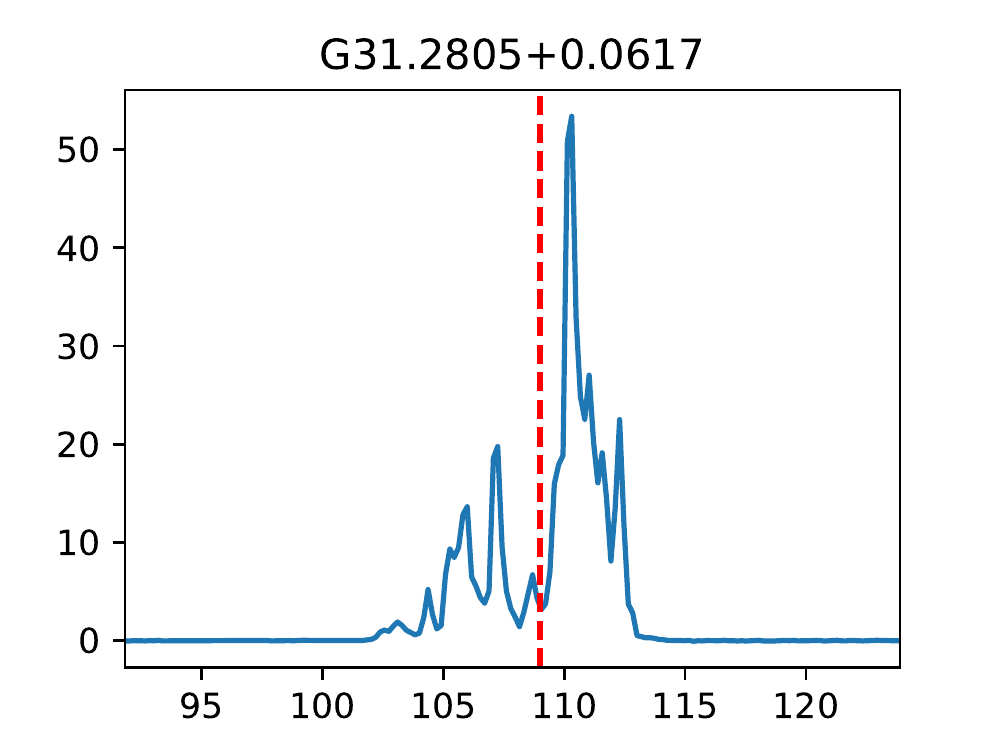}
\includegraphics[trim={0.0cm 0.3cm 0.6cm 0.3cm}, clip, width = 0.235\textwidth]{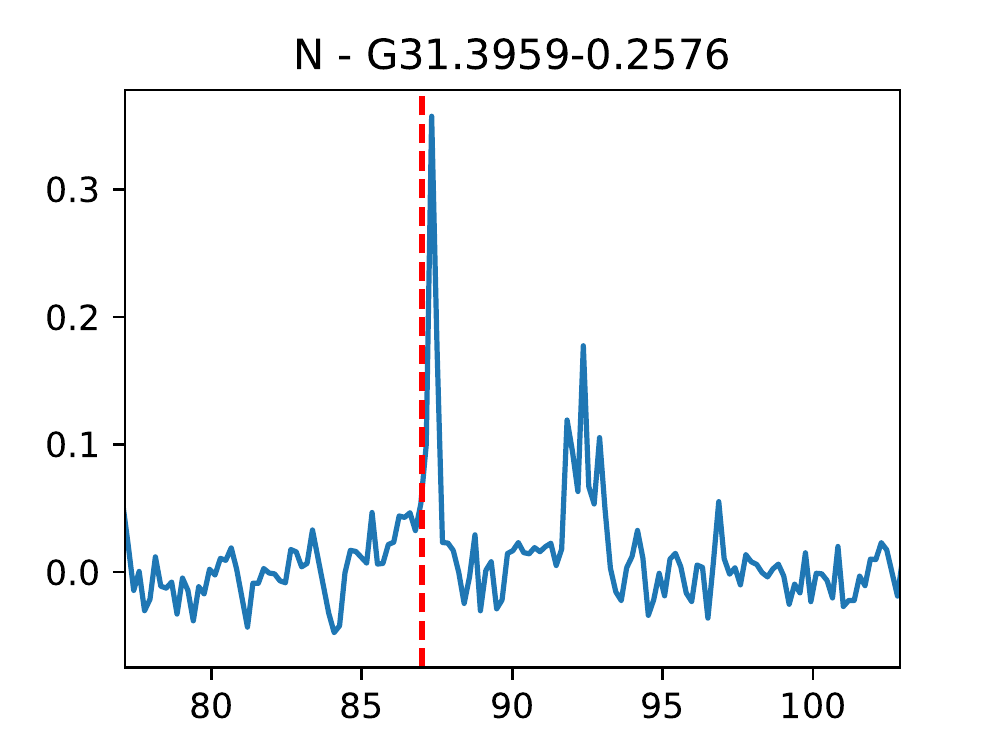}
\includegraphics[trim={0.0cm 0.3cm 0.6cm 0.3cm}, clip, width = 0.235\textwidth]{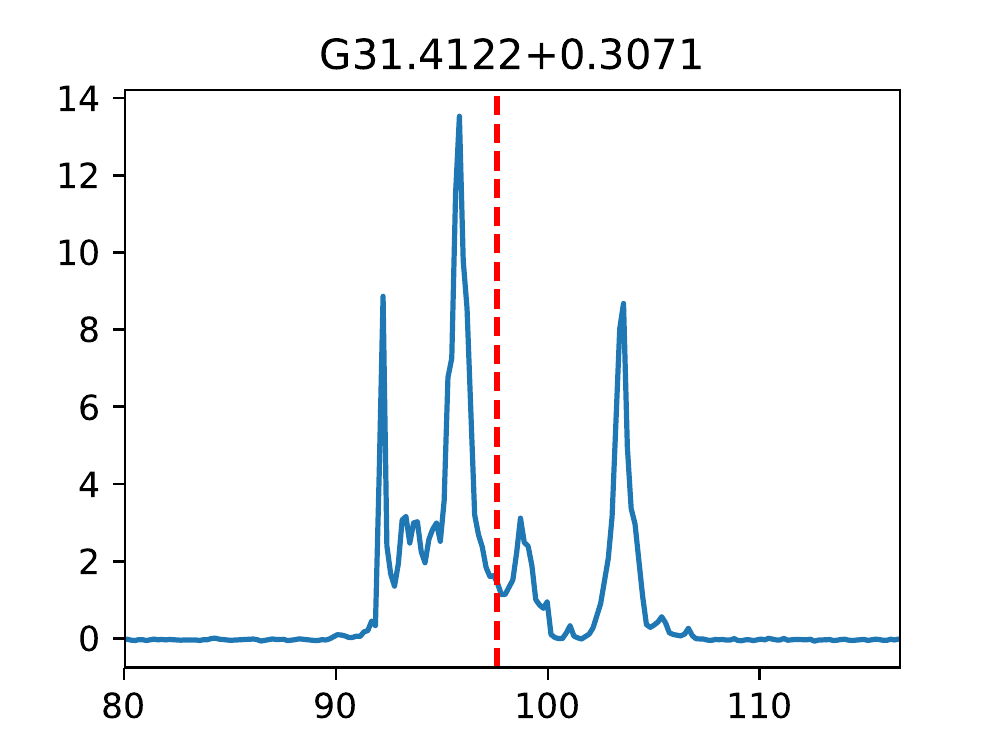}\\ 
\colorbox{white}{\includegraphics[trim={2cm 0 0 0}, scale=0.20]{figs/ylabel.pdf}}
\includegraphics[trim={0.0cm 0.3cm 0.6cm 0.3cm}, clip, width = 0.235\textwidth]{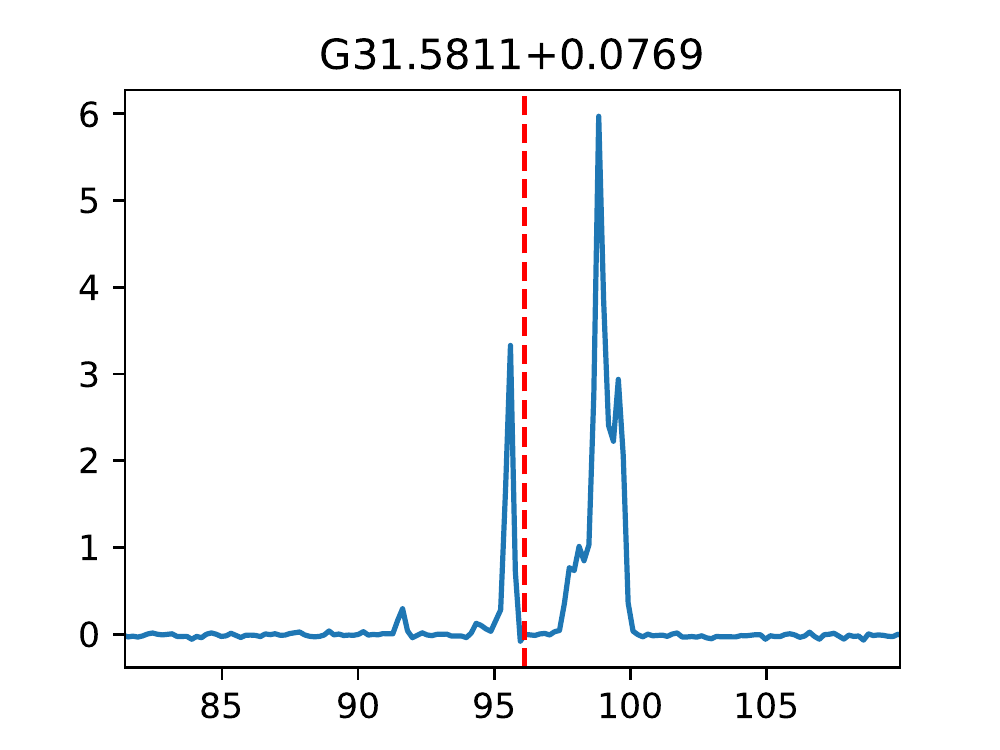}
\includegraphics[trim={0.0cm 0.3cm 0.6cm 0.3cm}, clip, width = 0.235\textwidth]{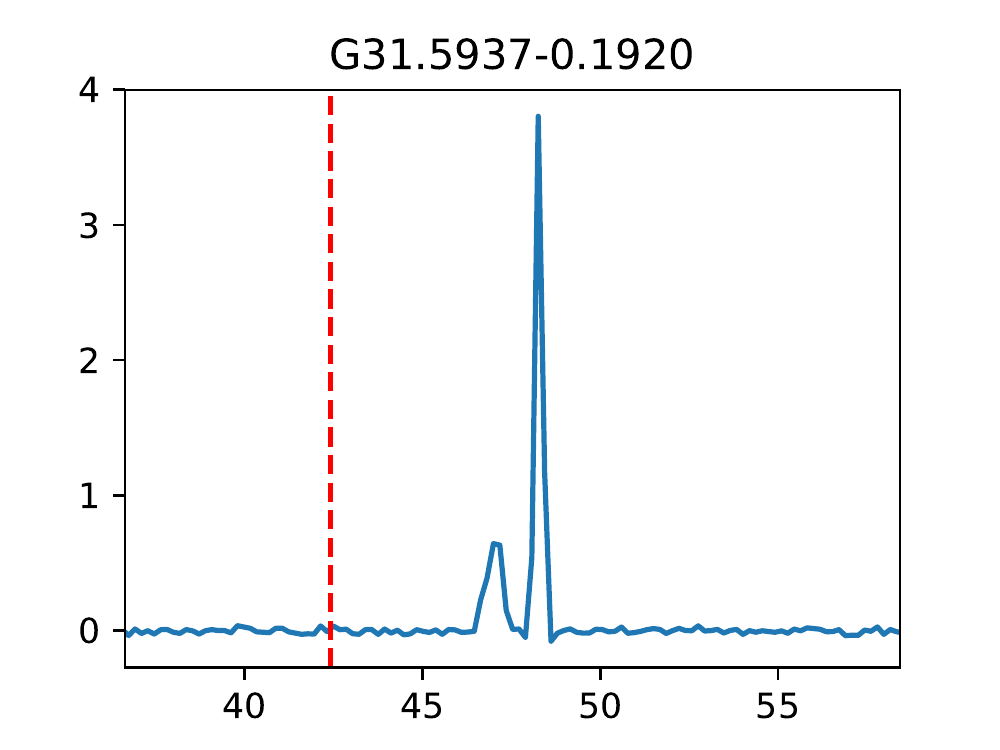}
\includegraphics[trim={0.0cm 0.3cm 0.6cm 0.3cm}, clip, width = 0.235\textwidth]{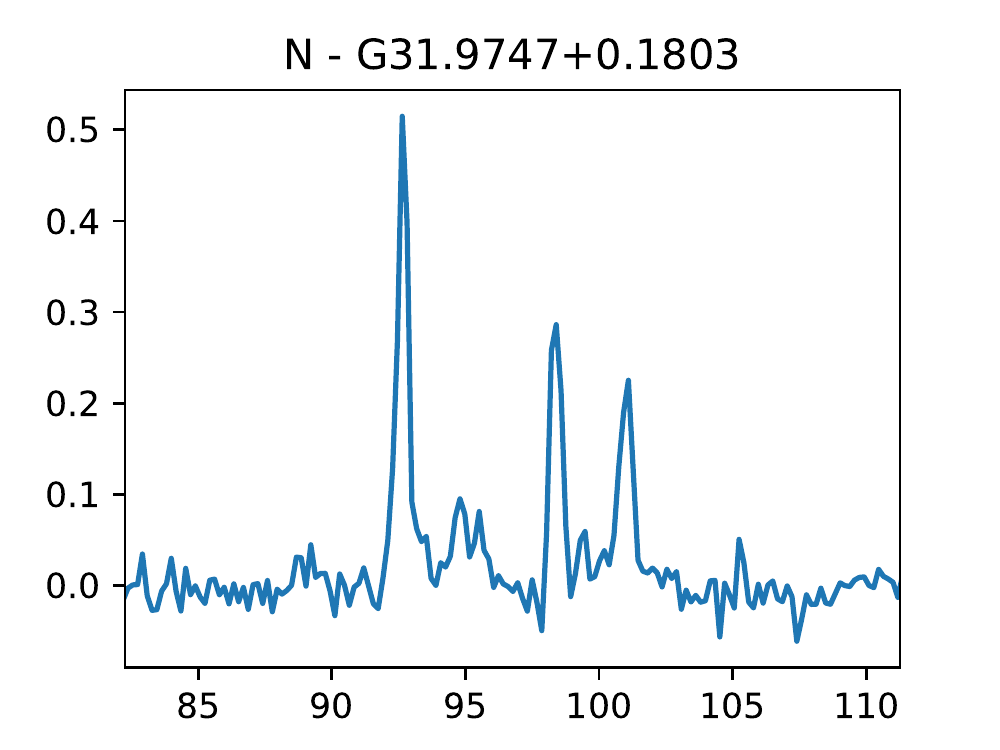}
\includegraphics[trim={0.0cm 0.3cm 0.6cm 0.3cm}, clip, width = 0.235\textwidth]{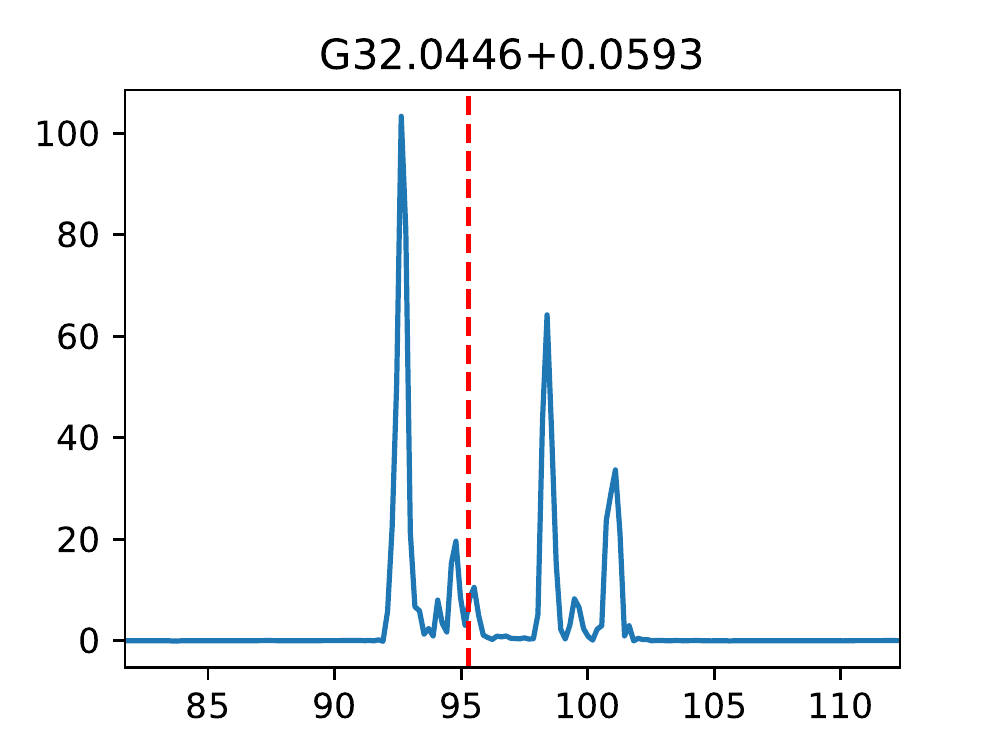}\\
\end{tabular}
\caption{Continued from Fig.~\ref{fig:maser_many_spec}.}
\label{fig:masers_408}
\end{figure*}

%Table 18
\begin{figure*}[!h]
\begin{tabular}{ccccc}
\colorbox{white}{\includegraphics[trim={2cm 0 0 0}, scale=0.195]{figs/ylabel_filler.pdf}}
\includegraphics[trim={0.0cm 0.3cm 0.6cm 0.3cm}, clip, width = 0.235\textwidth]{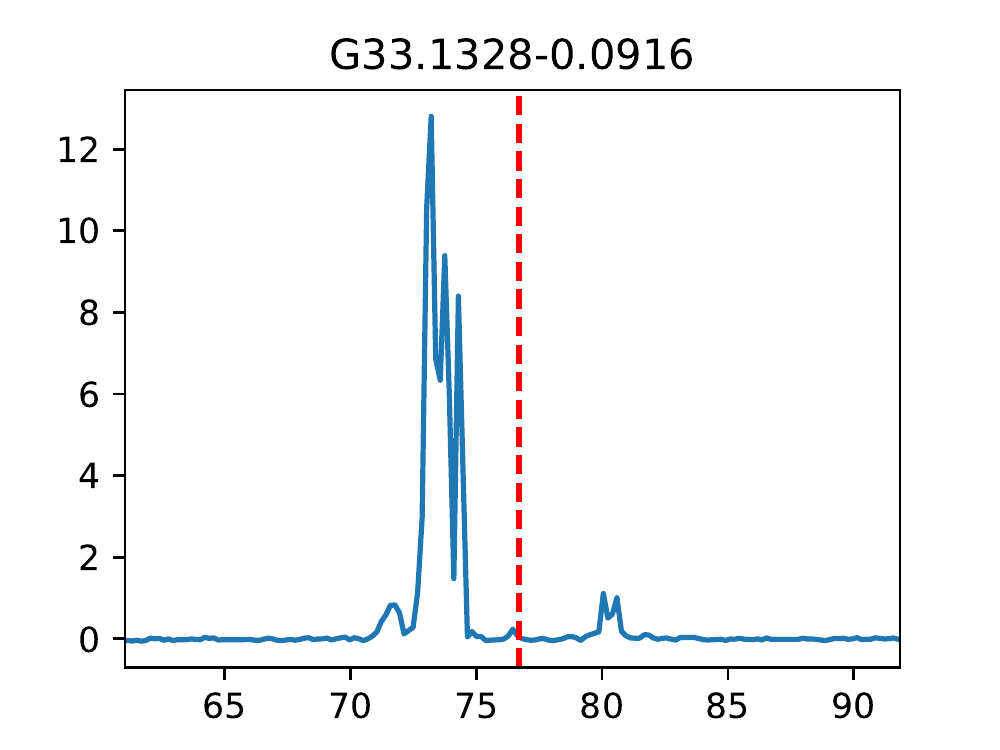}
\includegraphics[trim={0.0cm 0.3cm 0.6cm 0.3cm}, clip, width = 0.235\textwidth]{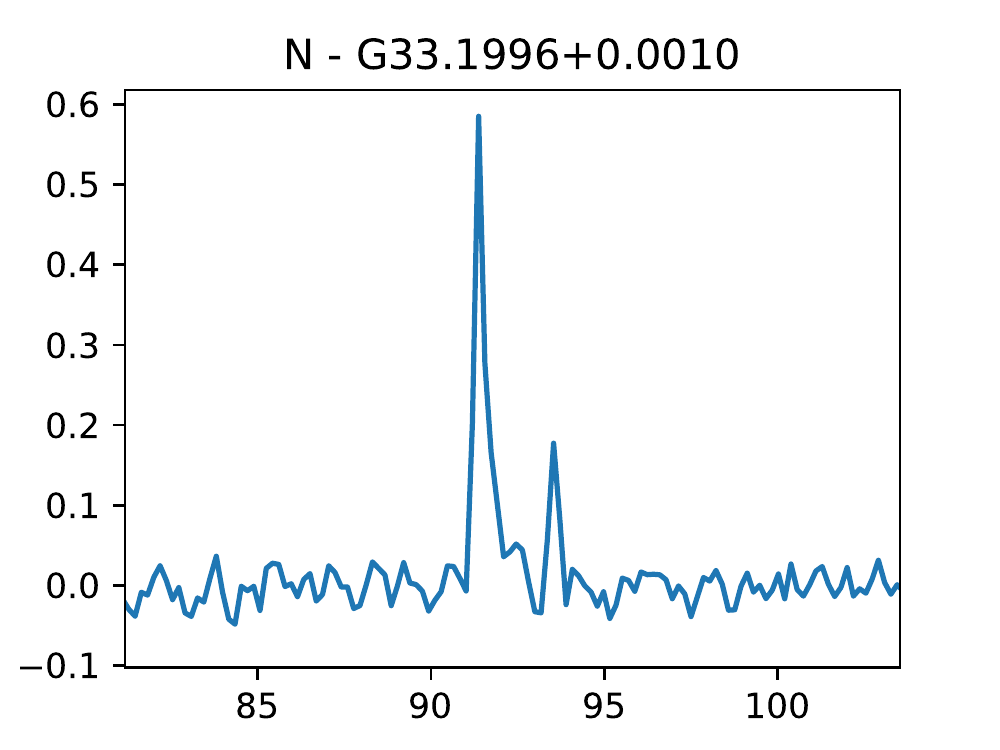}
\includegraphics[trim={0.0cm 0.3cm 0.6cm 0.3cm}, clip, width = 0.235\textwidth]{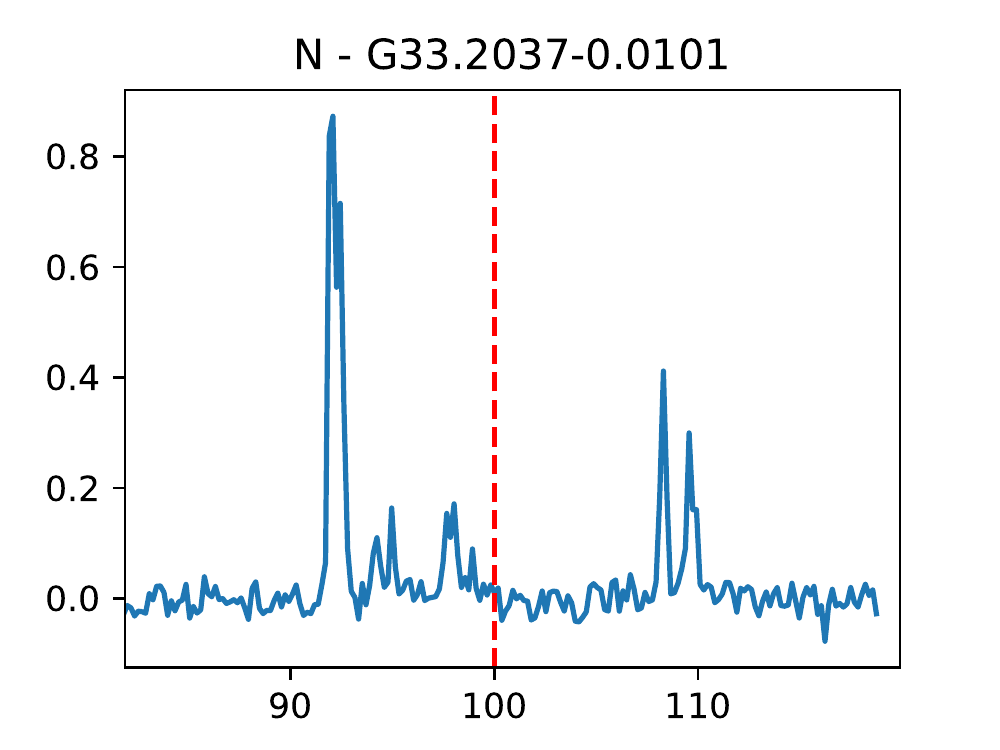}
\includegraphics[trim={0.0cm 0.3cm 0.6cm 0.3cm}, clip, width = 0.235\textwidth]{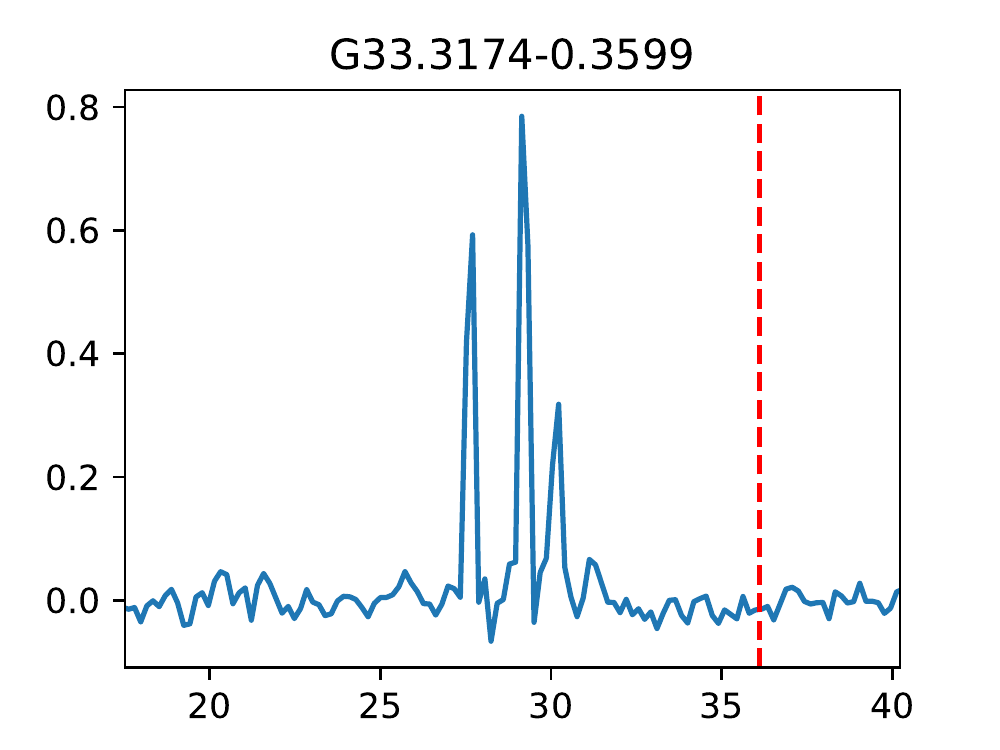}\\ 
\colorbox{white}{\includegraphics[trim={2cm 0 0 0}, scale=0.195]{figs/ylabel_filler.pdf}}
\includegraphics[trim={0.0cm 0.3cm 0.6cm 0.3cm}, clip, width = 0.235\textwidth]{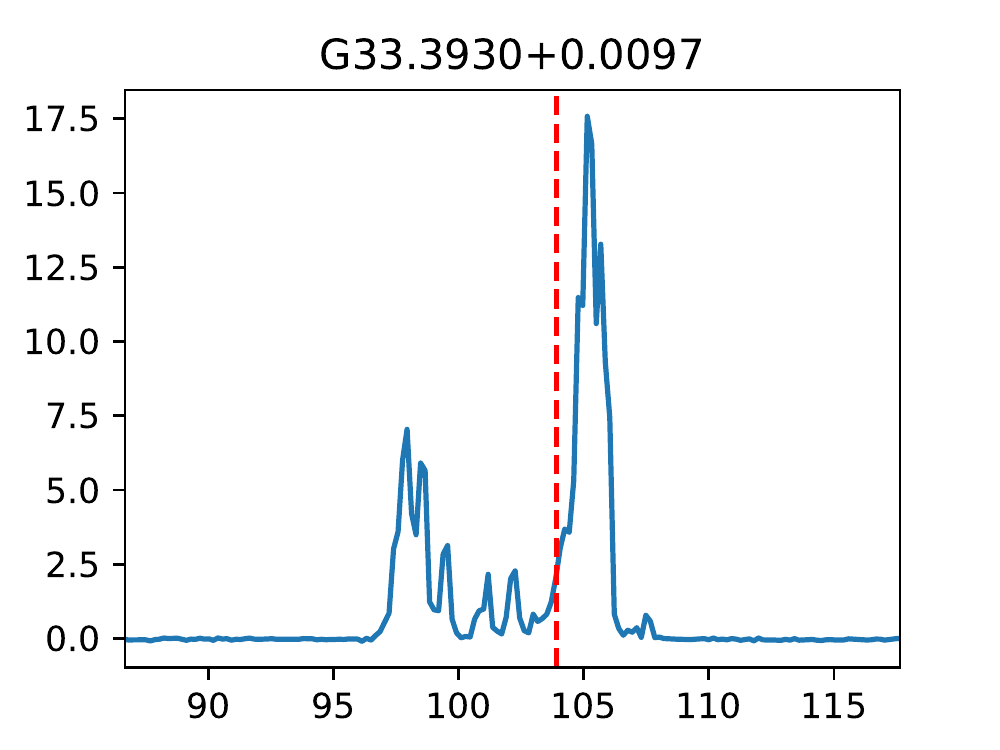}
\includegraphics[trim={0.0cm 0.3cm 0.6cm 0.3cm}, clip, width = 0.235\textwidth]{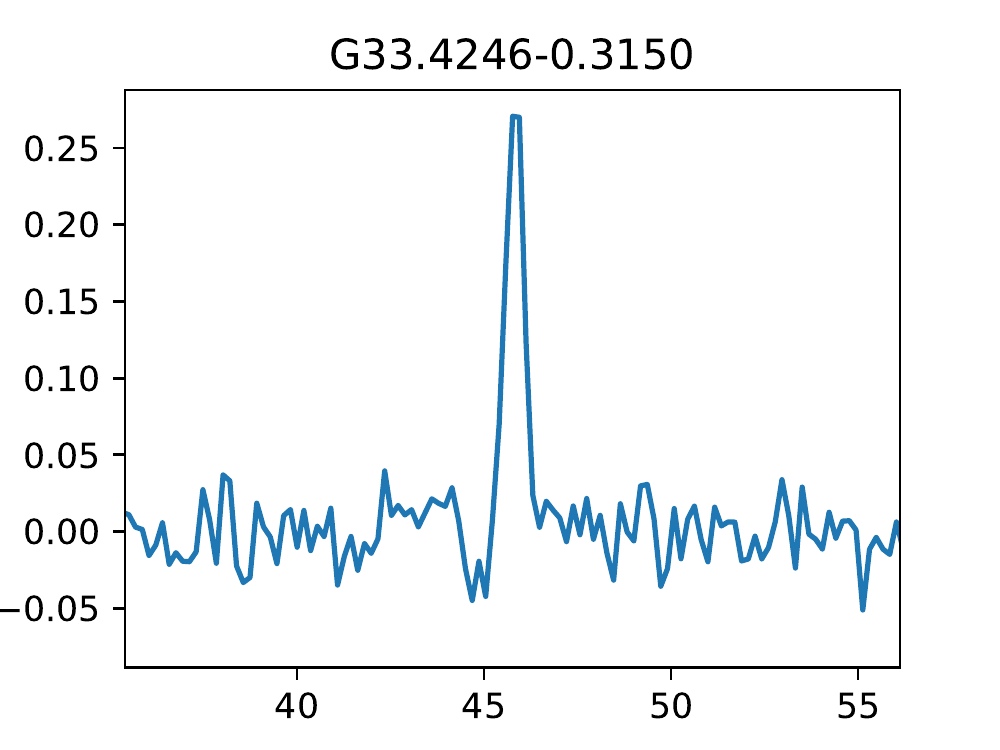}
\includegraphics[trim={0.0cm 0.3cm 0.6cm 0.3cm}, clip, width = 0.235\textwidth]{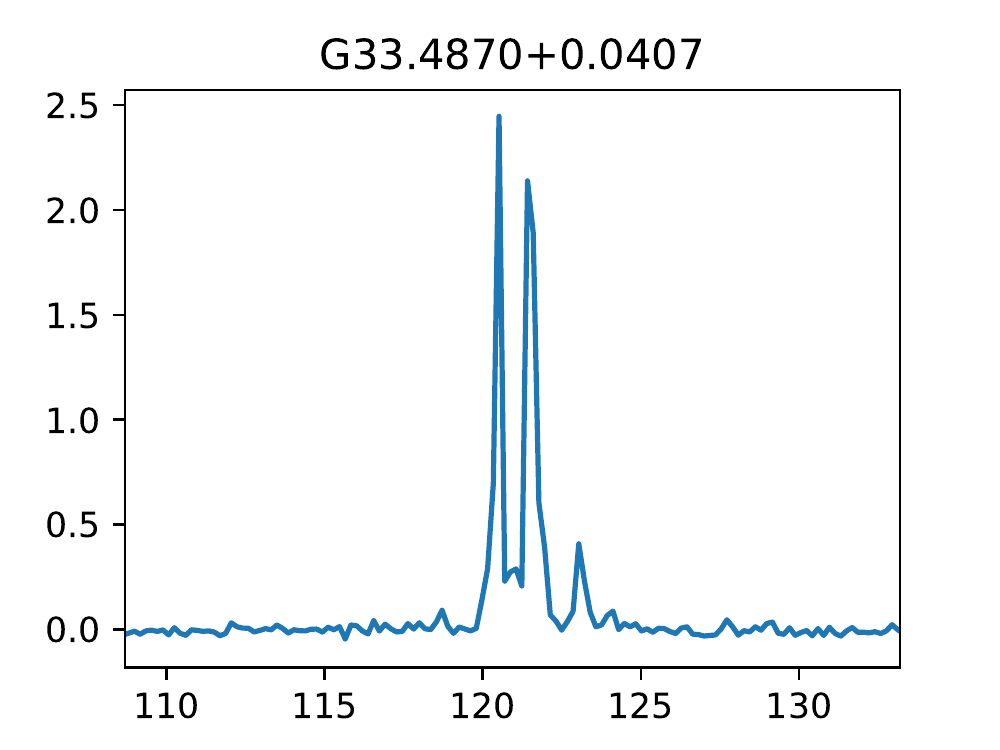}
\includegraphics[trim={0.0cm 0.3cm 0.6cm 0.3cm}, clip, width = 0.235\textwidth]{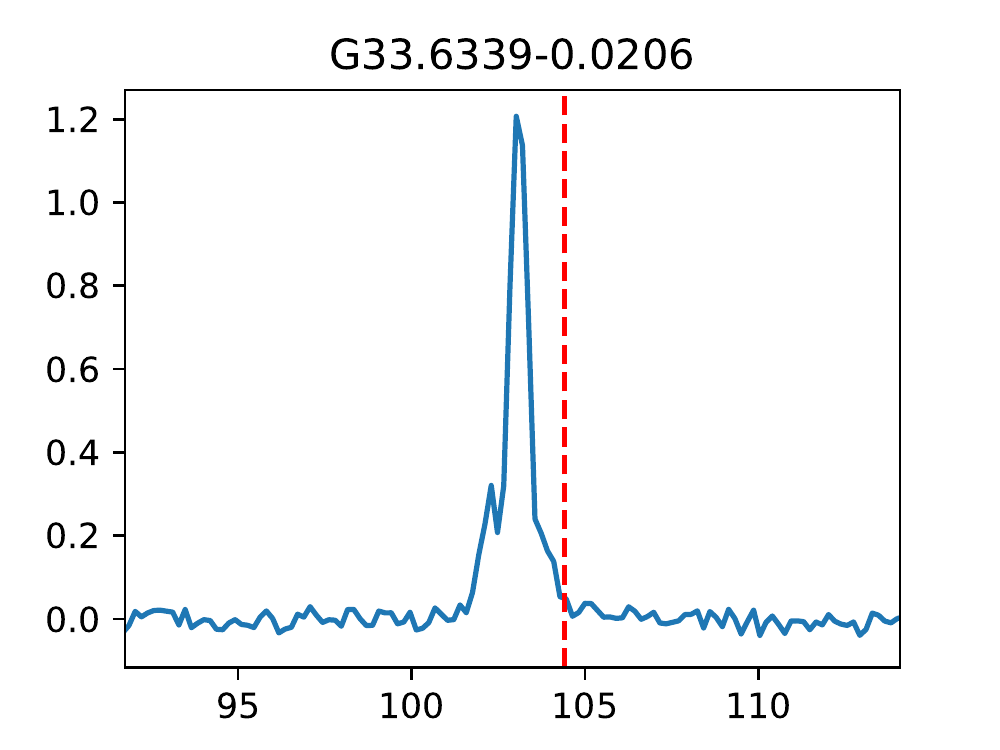}\\ 
\colorbox{white}{\includegraphics[trim={2cm 0 0 0}, scale=0.20]{figs/ylabel.pdf}}
\includegraphics[trim={0.0cm 0.3cm 0.6cm 0.3cm}, clip, width = 0.235\textwidth]{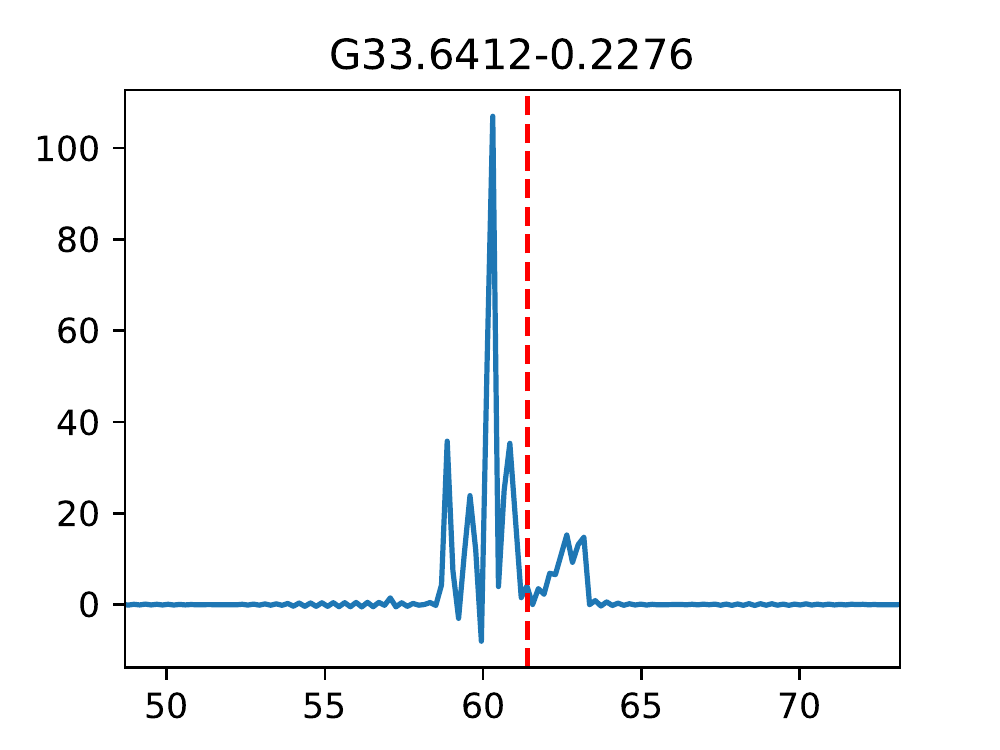}
\includegraphics[trim={0.0cm 0.3cm 0.6cm 0.3cm}, clip, width = 0.235\textwidth]{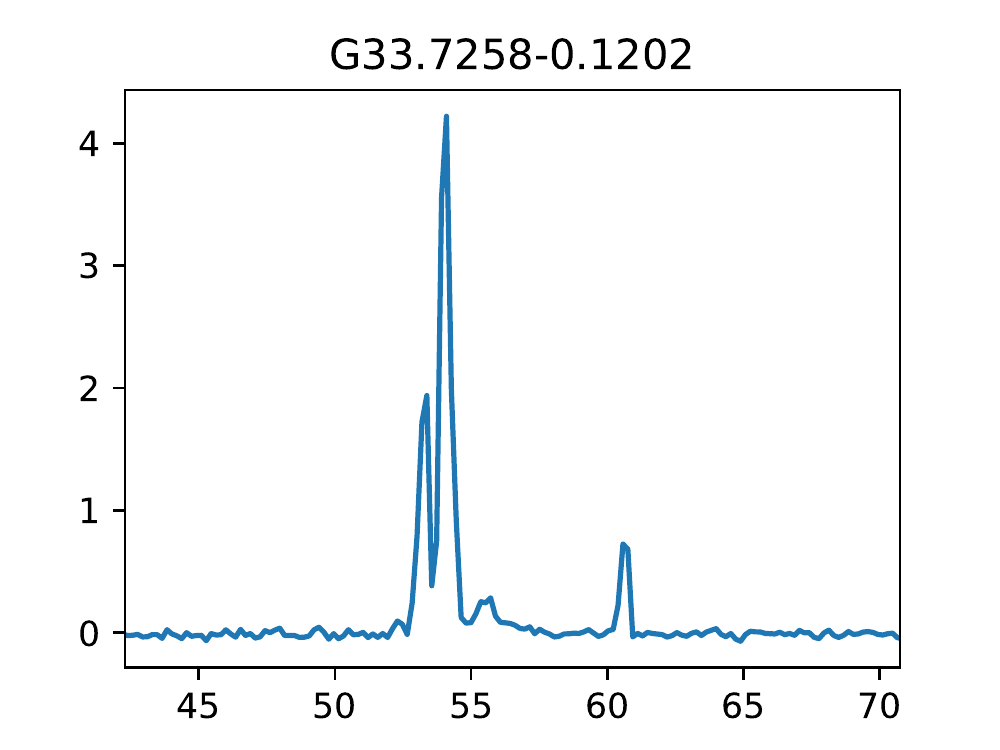}
\includegraphics[trim={0.0cm 0.3cm 0.6cm 0.3cm}, clip, width = 0.235\textwidth]{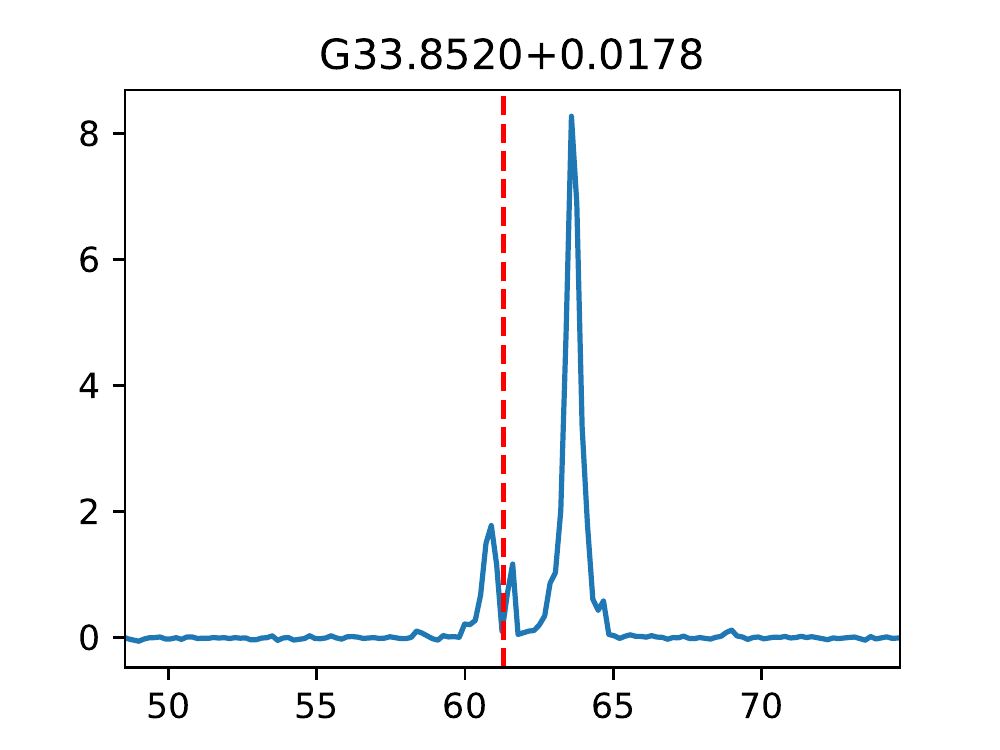}
\includegraphics[trim={0.0cm 0.3cm 0.6cm 0.3cm}, clip, width = 0.235\textwidth]{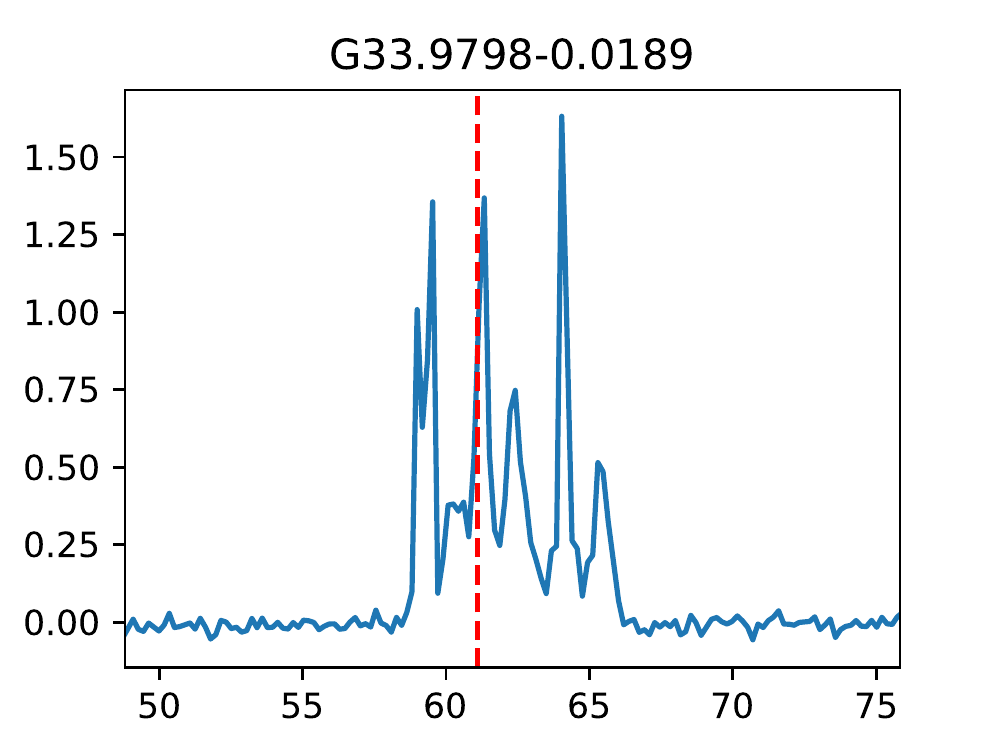}\\
\end{tabular}
\caption{Continued from Fig.~\ref{fig:maser_many_spec}.}
\label{fig:masers_432}
\end{figure*}

%Table 19
\begin{figure*}[!h]
\begin{tabular}{ccccc}
\colorbox{white}{\includegraphics[trim={2cm 0 0 0}, scale=0.195]{figs/ylabel_filler.pdf}}
\includegraphics[trim={0.0cm 0.3cm 0.6cm 0.3cm}, clip, width = 0.235\textwidth]{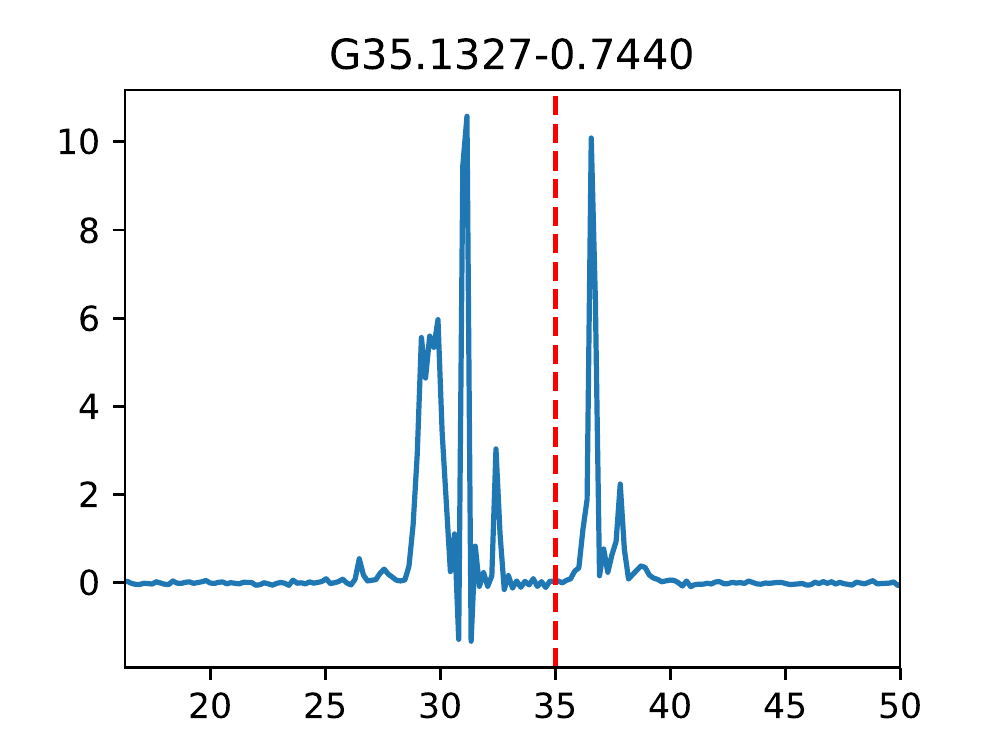}
\includegraphics[trim={0.0cm 0.3cm 0.6cm 0.3cm}, clip, width = 0.235\textwidth]{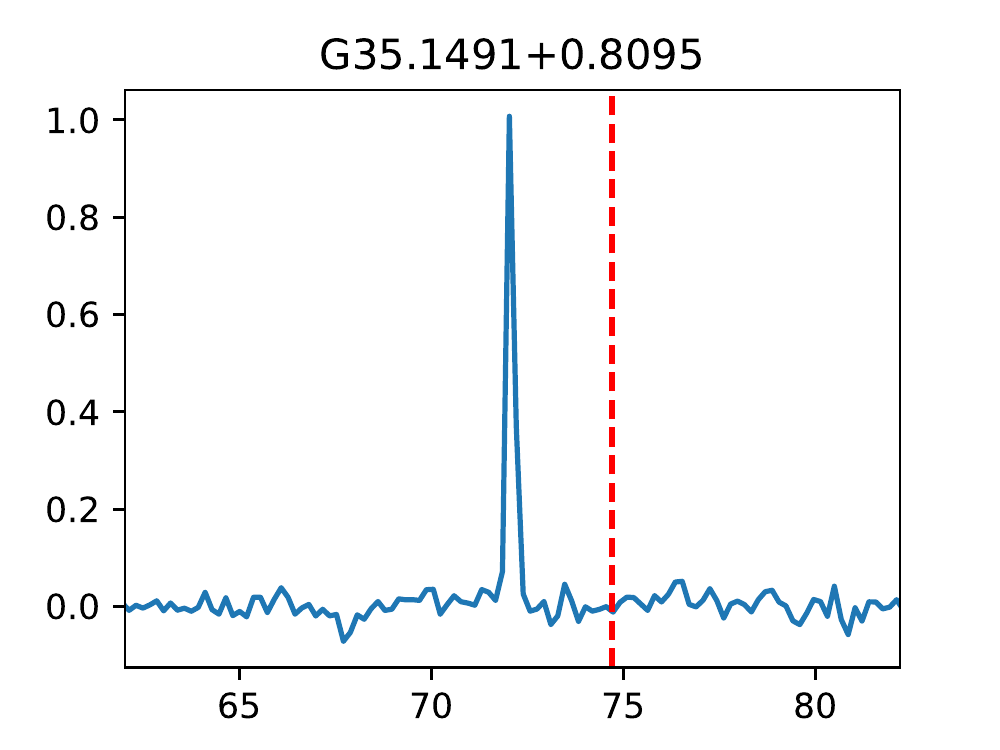}
\includegraphics[trim={0.0cm 0.3cm 0.6cm 0.3cm}, clip, width = 0.235\textwidth]{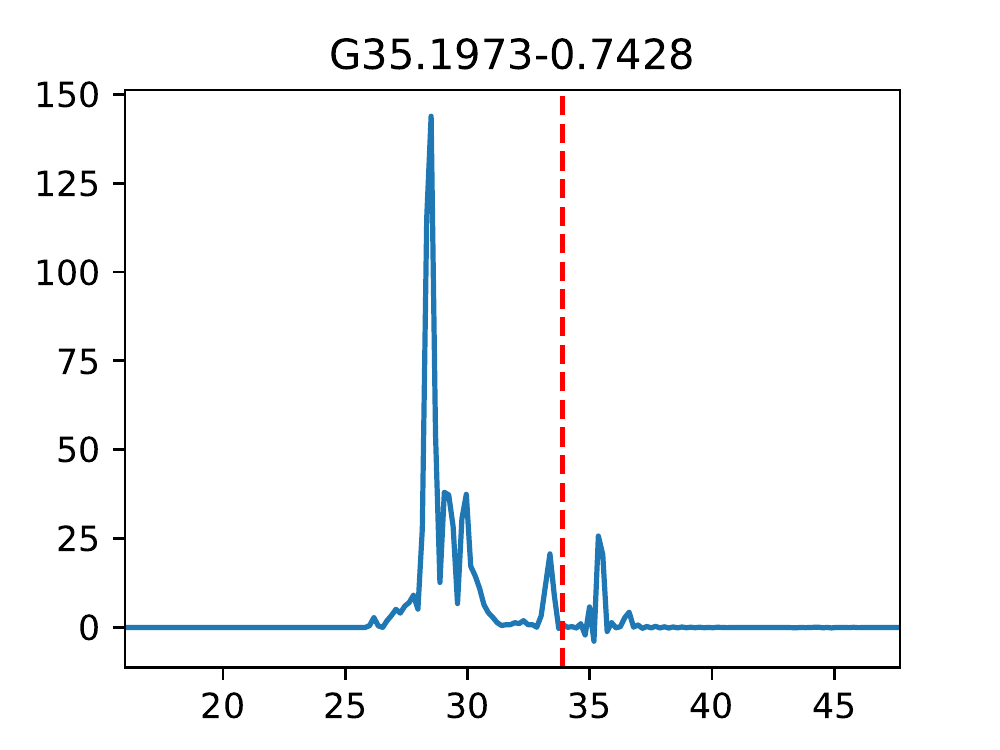}
\includegraphics[trim={0.0cm 0.3cm 0.6cm 0.3cm}, clip, width = 0.235\textwidth]{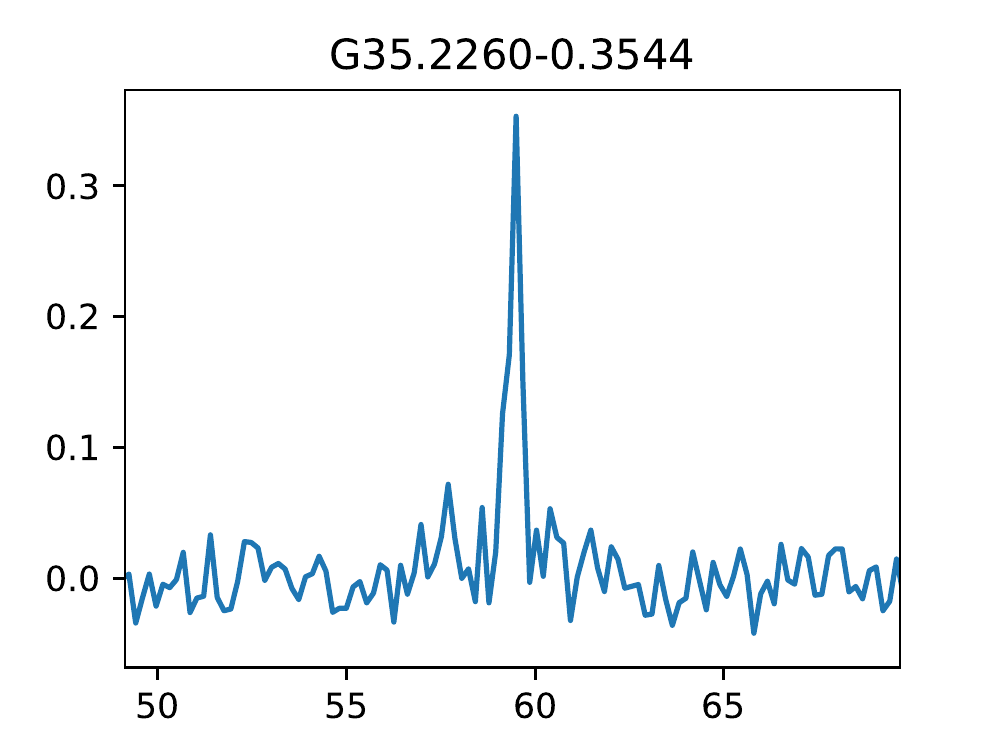}\\ 
\colorbox{white}{\includegraphics[trim={2cm 0 0 0}, scale=0.195]{figs/ylabel_filler.pdf}}
\includegraphics[trim={0.0cm 0.3cm 0.6cm 0.3cm}, clip, width = 0.235\textwidth]{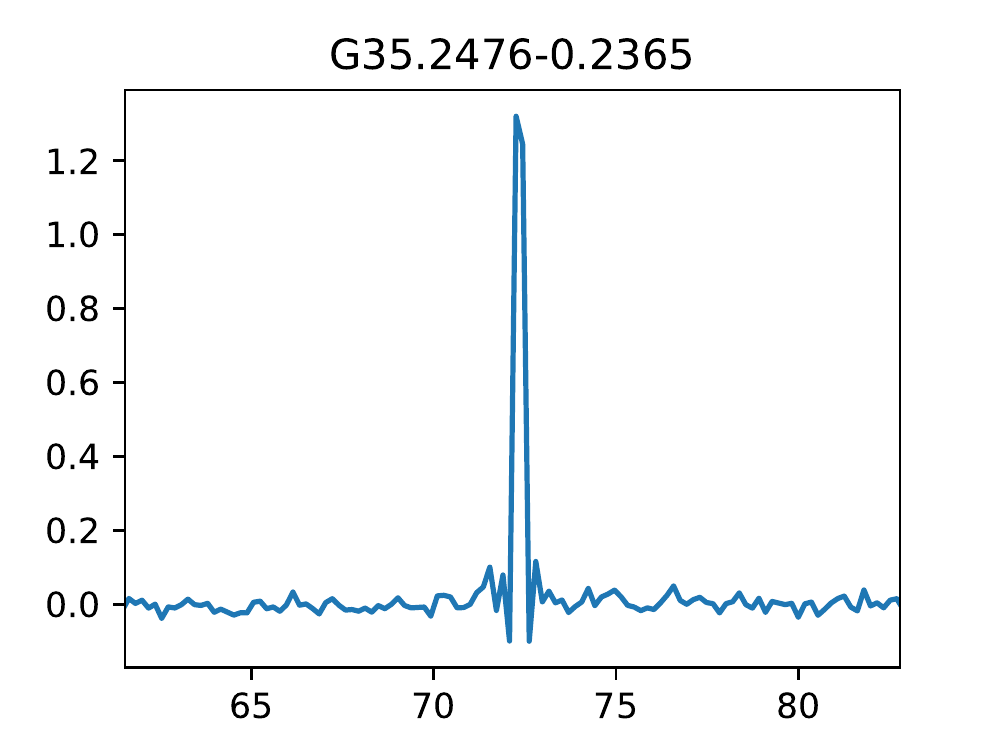}
\includegraphics[trim={0.0cm 0.3cm 0.6cm 0.3cm}, clip, width = 0.235\textwidth]{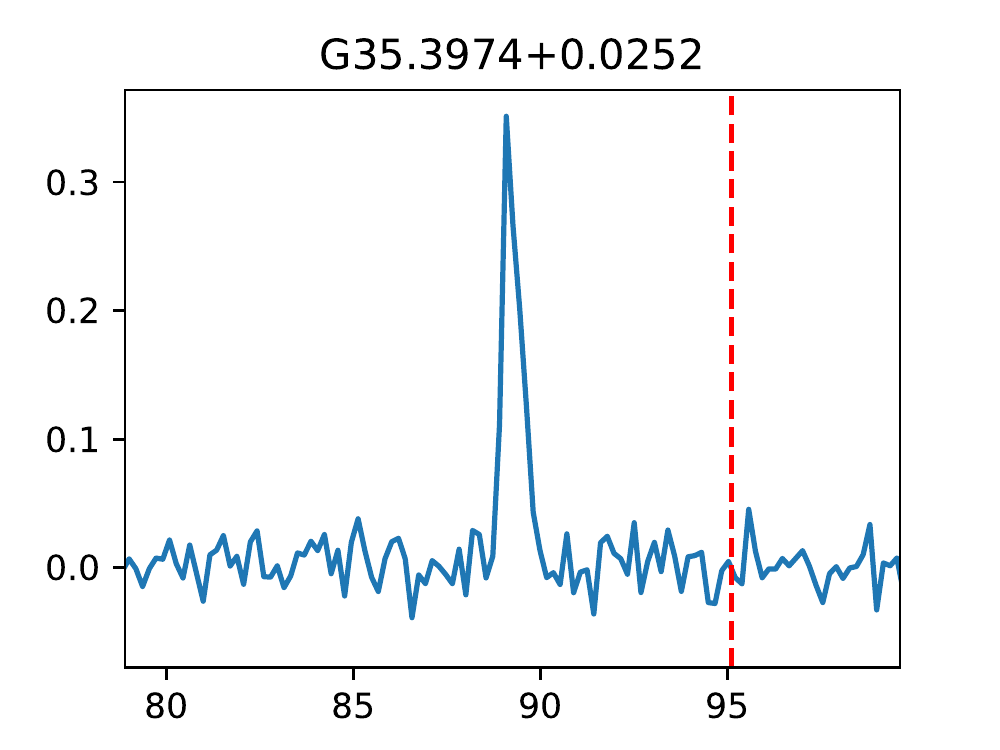}
\includegraphics[trim={0.0cm 0.3cm 0.6cm 0.3cm}, clip, width = 0.235\textwidth]{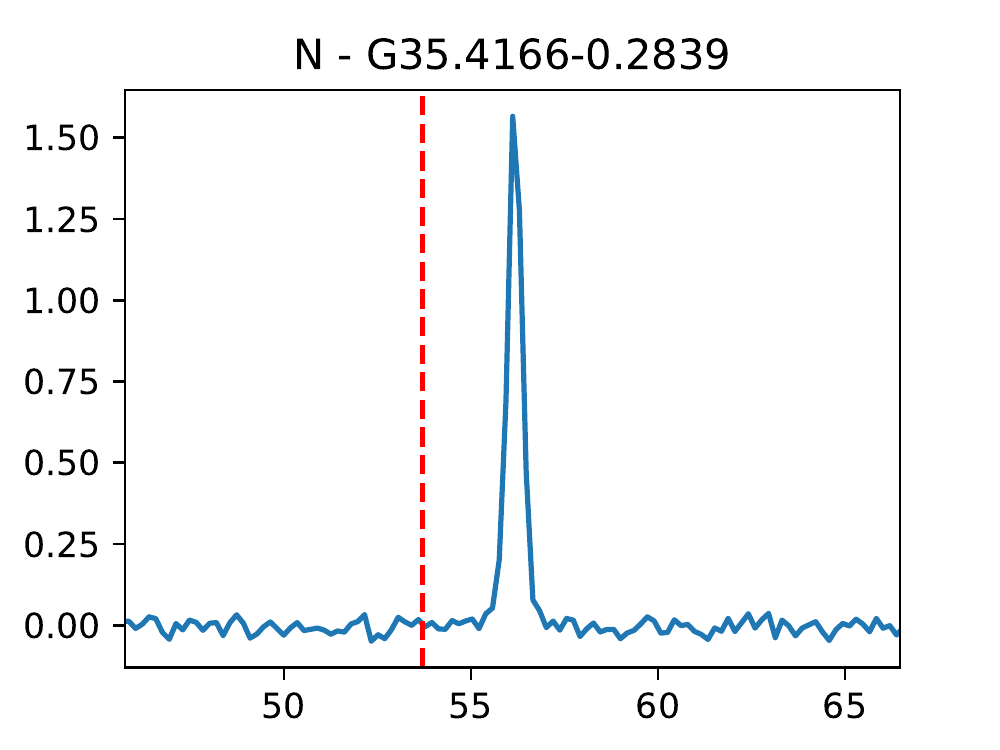}
\includegraphics[trim={0.0cm 0.3cm 0.6cm 0.3cm}, clip, width = 0.235\textwidth]{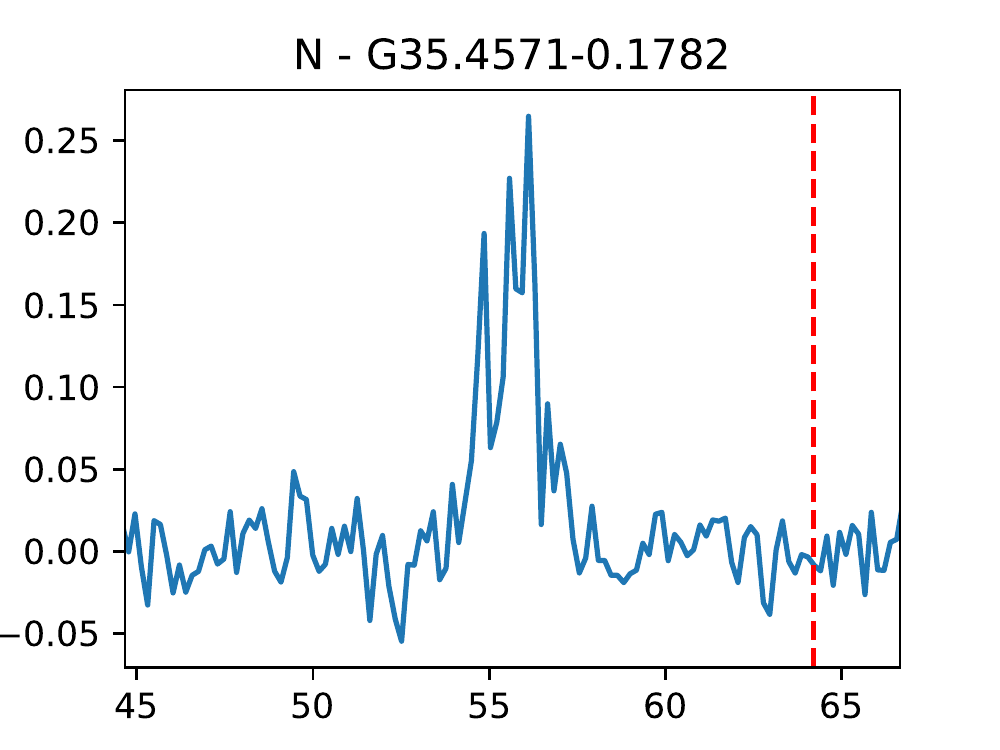}\\ 
\colorbox{white}{\includegraphics[trim={2cm 0 0 0}, scale=0.20]{figs/ylabel.pdf}}
\includegraphics[trim={0.0cm 0.3cm 0.6cm 0.3cm}, clip, width = 0.235\textwidth]{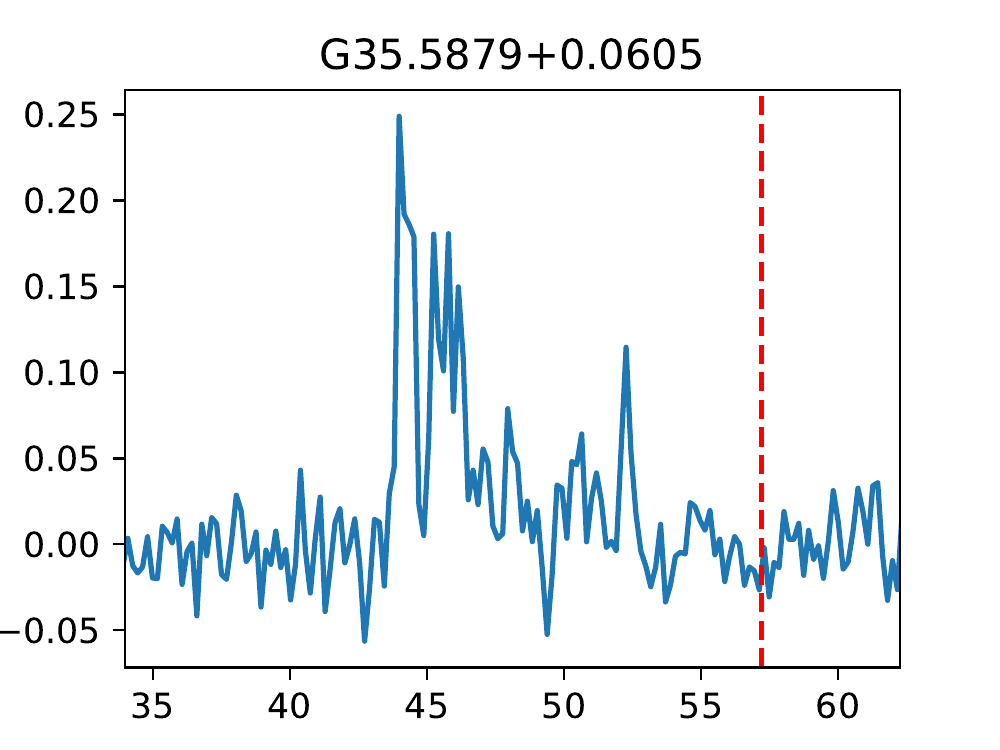}
\includegraphics[trim={0.0cm 0.3cm 0.6cm 0.3cm}, clip, width = 0.235\textwidth]{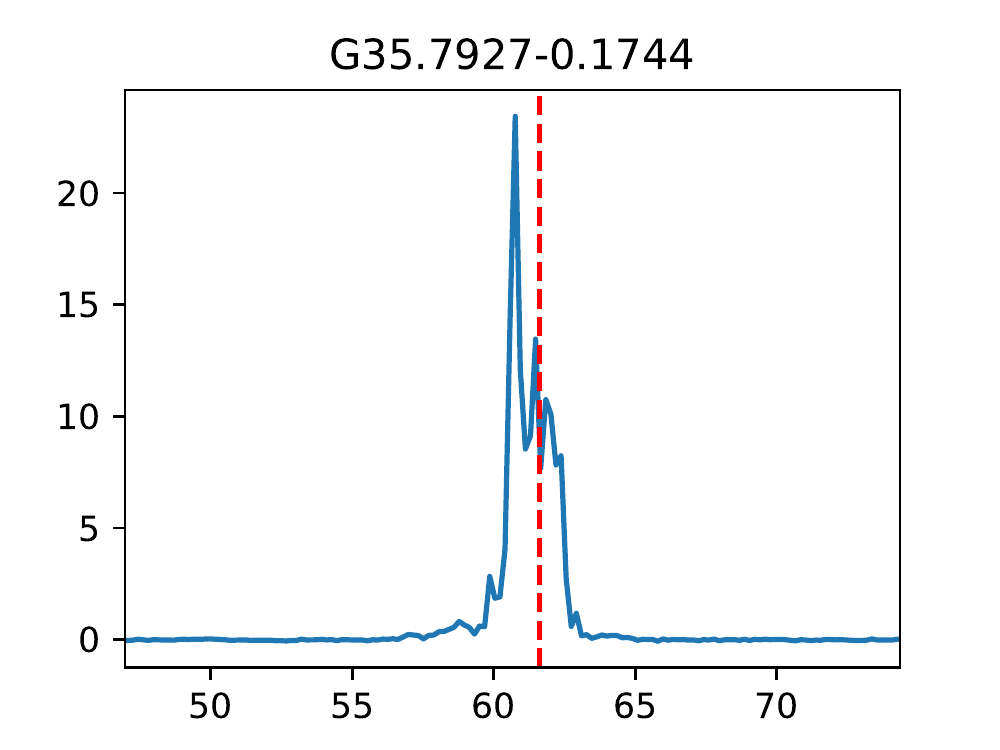}
\includegraphics[trim={0.0cm 0.3cm 0.6cm 0.3cm}, clip, width = 0.235\textwidth]{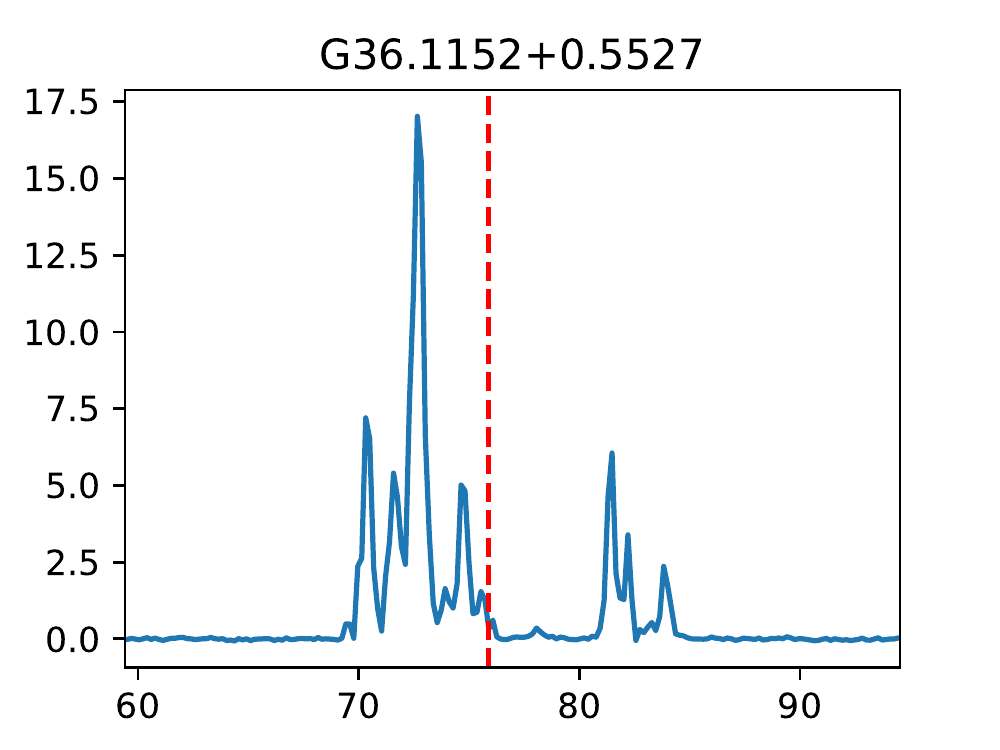}
\includegraphics[trim={0.0cm 0.3cm 0.6cm 0.3cm}, clip, width = 0.235\textwidth]{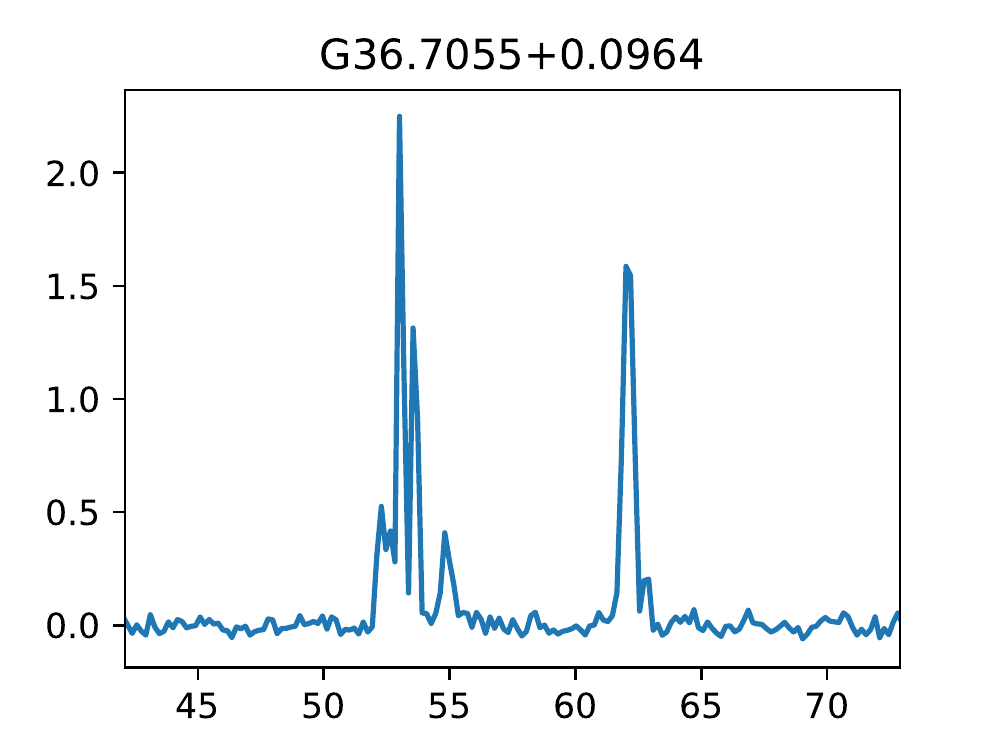}\\
\end{tabular}
\caption{Continued from Fig.~\ref{fig:maser_many_spec}.}
\label{fig:masers_456}
\end{figure*}

%Table 20
\begin{figure*}[!h]
\begin{tabular}{ccccc}
\colorbox{white}{\includegraphics[trim={2cm 0 0 0}, scale=0.195]{figs/ylabel_filler.pdf}}
\includegraphics[trim={0.0cm 0.3cm 0.6cm 0.3cm}, clip, width = 0.235\textwidth]{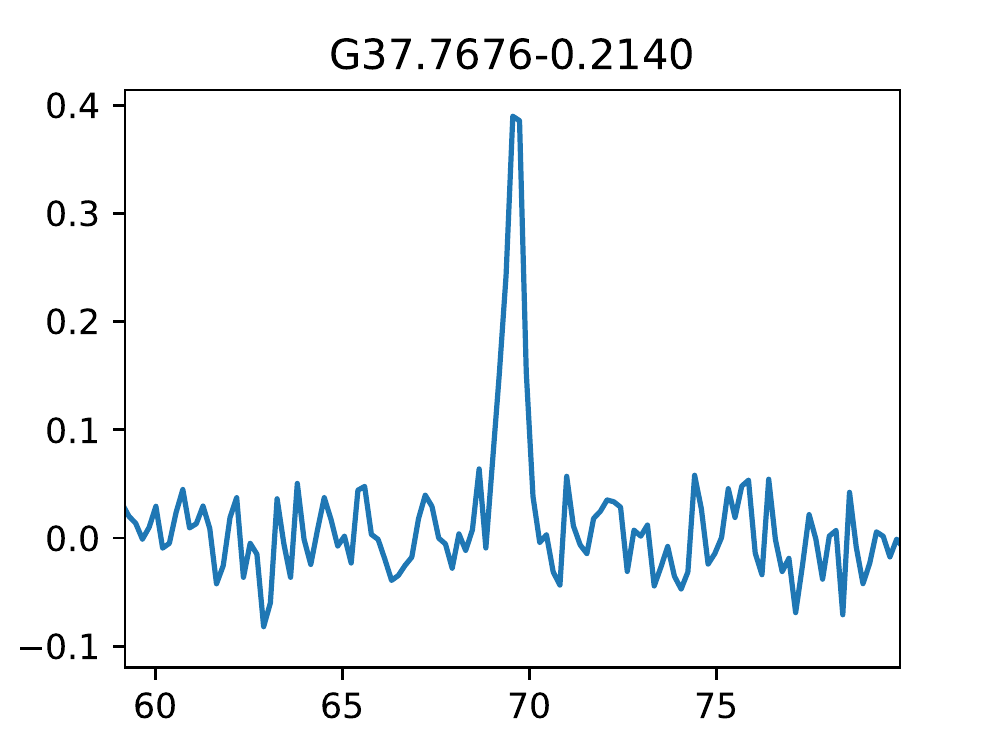}
\includegraphics[trim={0.0cm 0.3cm 0.6cm 0.3cm}, clip, width = 0.235\textwidth]{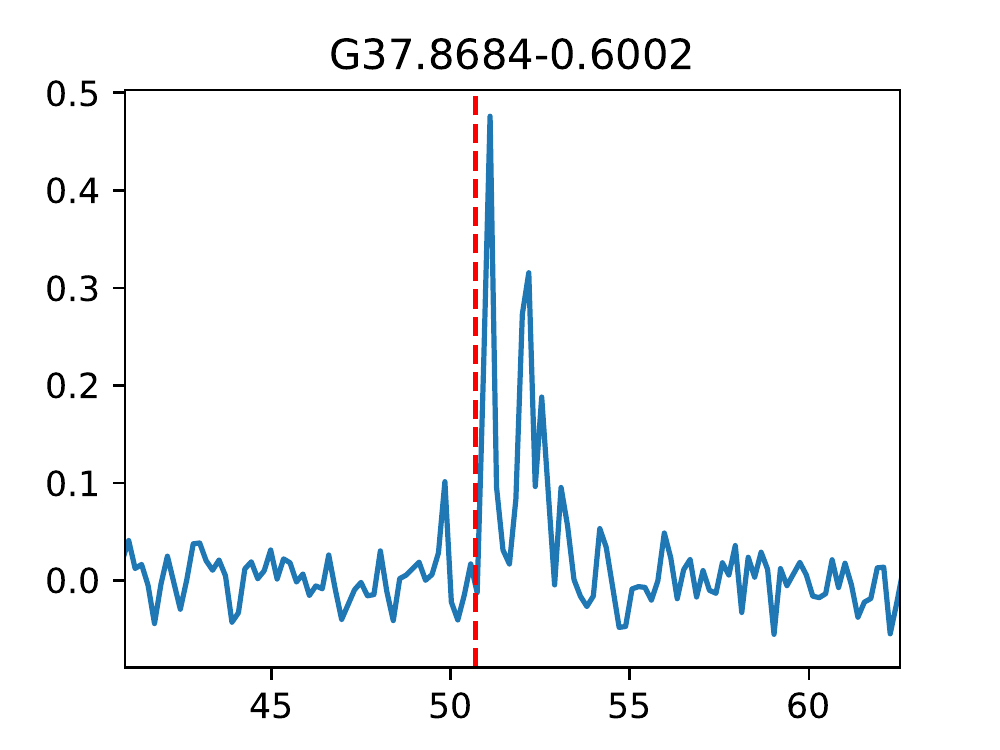}
\includegraphics[trim={0.0cm 0.3cm 0.6cm 0.3cm}, clip, width = 0.235\textwidth]{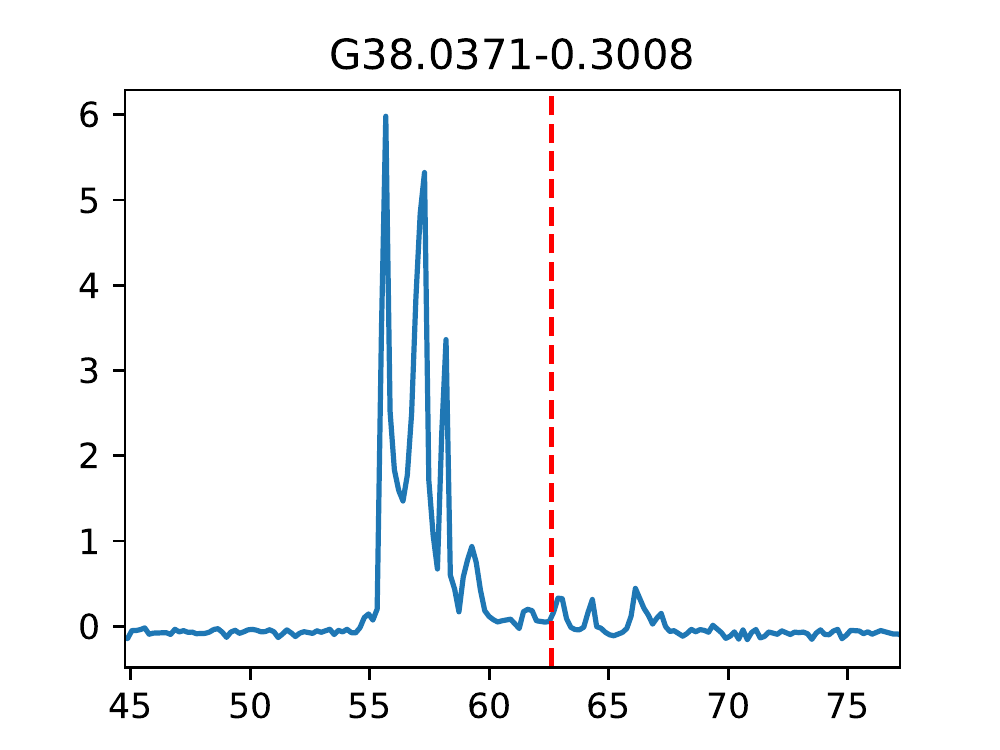}
\includegraphics[trim={0.0cm 0.3cm 0.6cm 0.3cm}, clip, width = 0.235\textwidth]{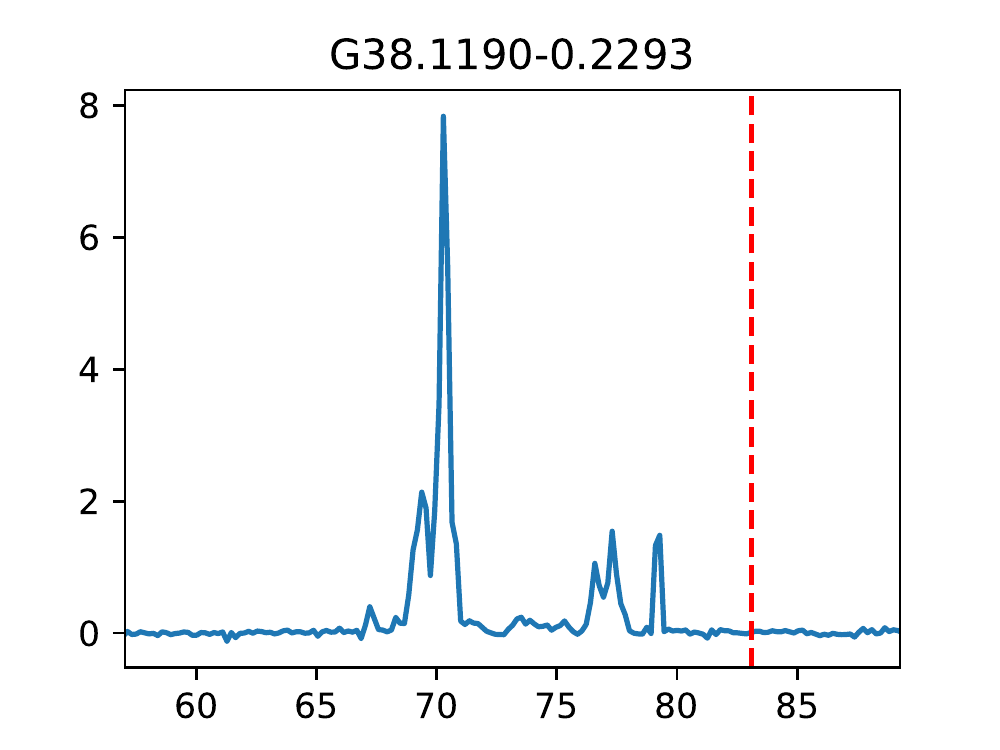}\\ 
\colorbox{white}{\includegraphics[trim={2cm 0 0 0}, scale=0.195]{figs/ylabel_filler.pdf}}
\includegraphics[trim={0.0cm 0.3cm 0.6cm 0.3cm}, clip, width = 0.235\textwidth]{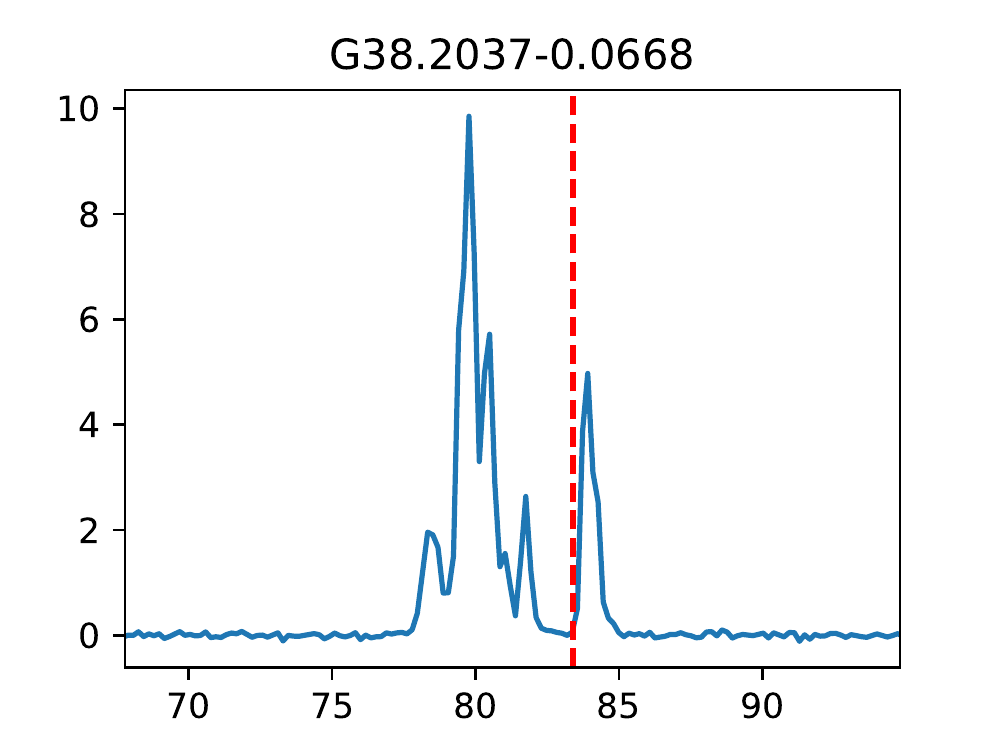}
\includegraphics[trim={0.0cm 0.3cm 0.6cm 0.3cm}, clip, width = 0.235\textwidth]{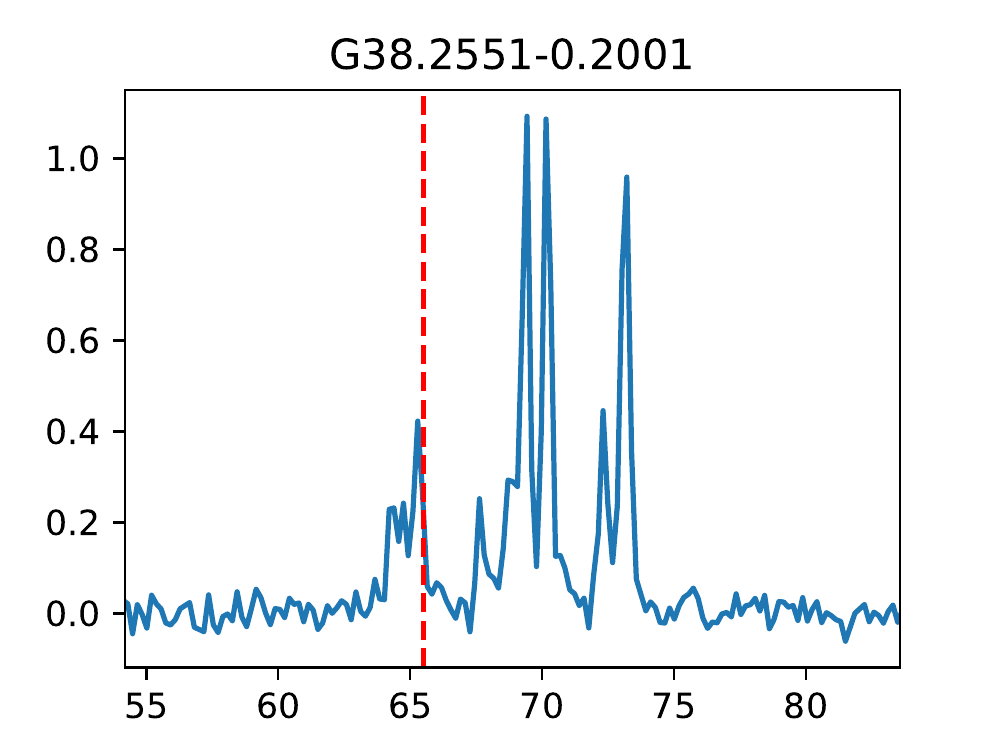}
\includegraphics[trim={0.0cm 0.3cm 0.6cm 0.3cm}, clip, width = 0.235\textwidth]{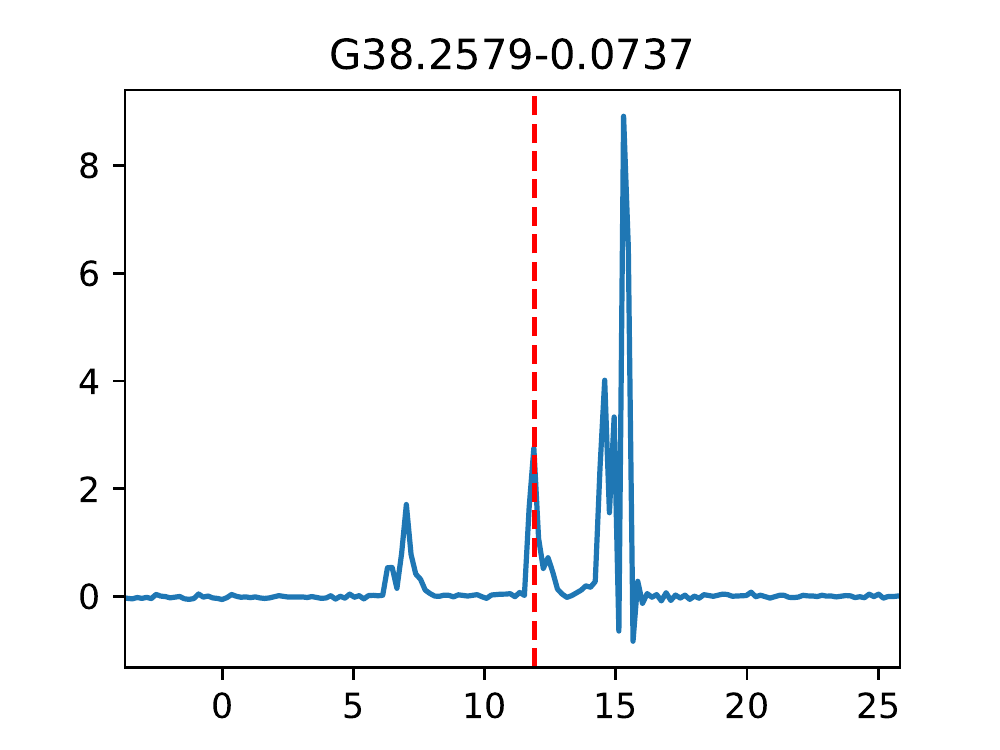}
\includegraphics[trim={0.0cm 0.3cm 0.6cm 0.3cm}, clip, width = 0.235\textwidth]{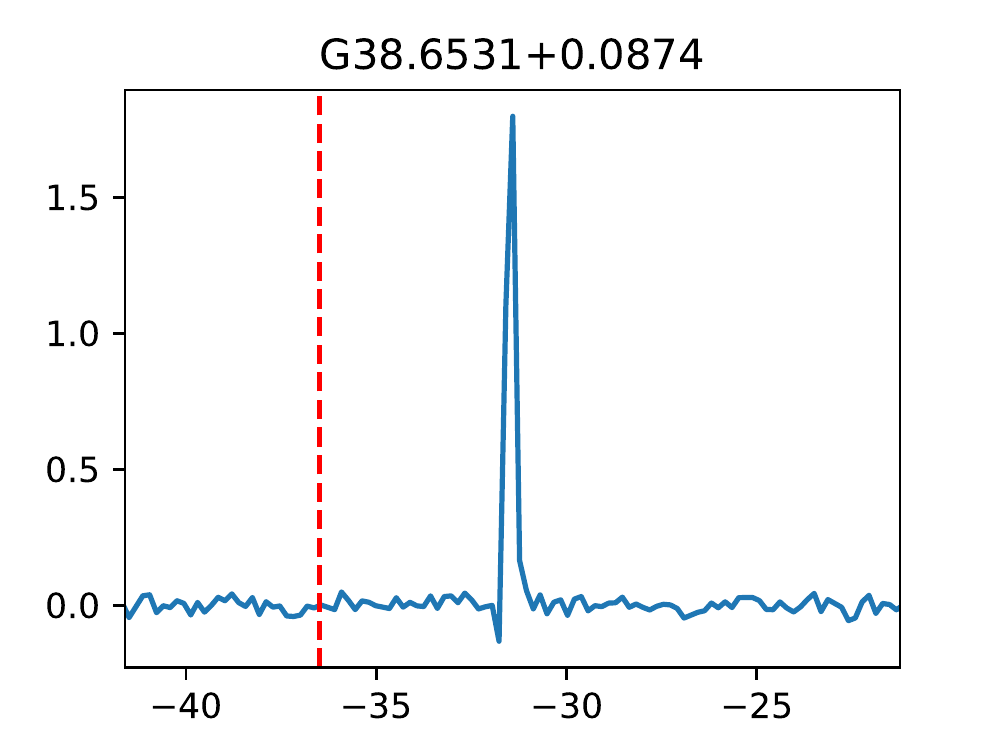}\\ 
\colorbox{white}{\includegraphics[trim={2cm 0 0 0}, scale=0.20]{figs/ylabel.pdf}}
\includegraphics[trim={0.0cm 0.3cm 0.6cm 0.3cm}, clip, width = 0.235\textwidth]{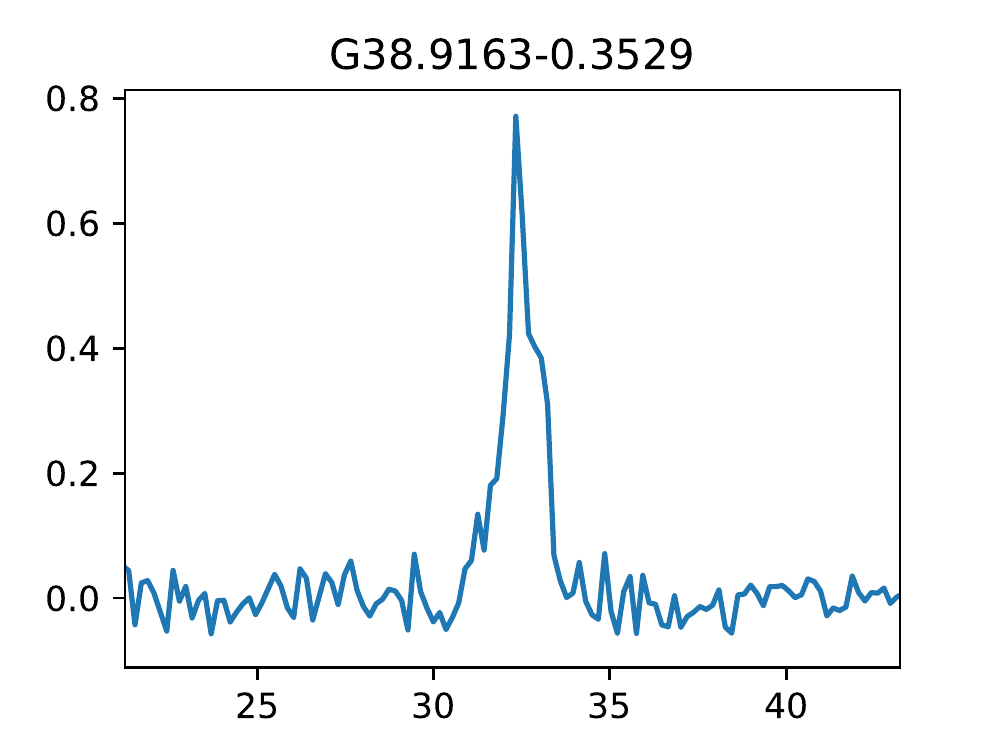}
\includegraphics[trim={0.0cm 0.3cm 0.6cm 0.3cm}, clip, width = 0.235\textwidth]{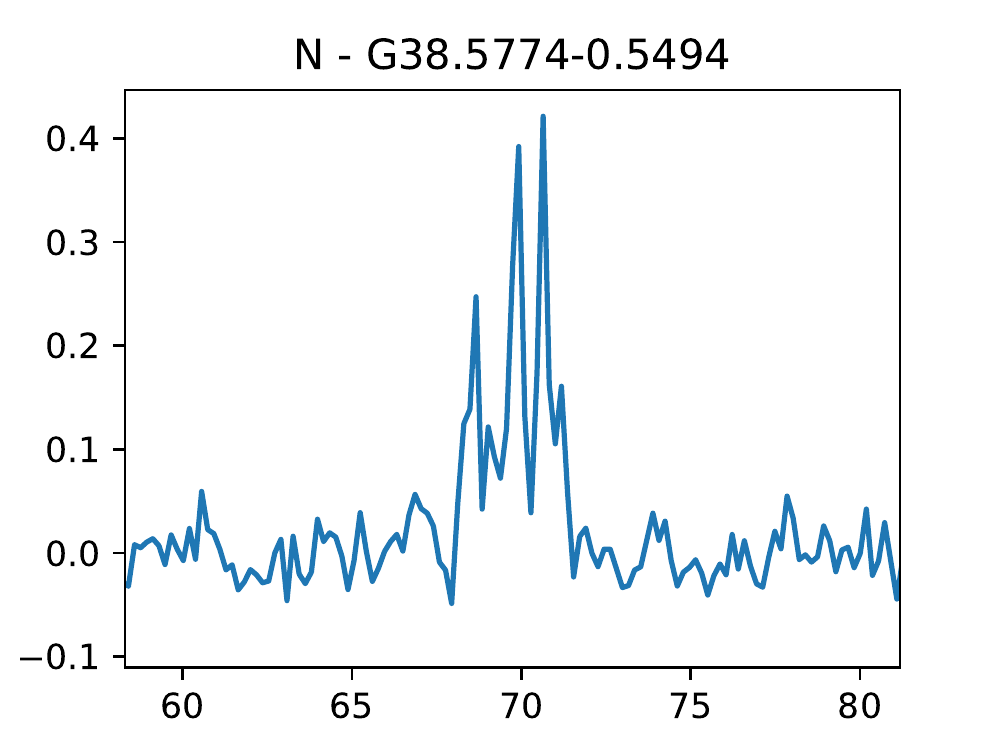}
\includegraphics[trim={0.0cm 0.3cm 0.6cm 0.3cm}, clip, width = 0.235\textwidth]{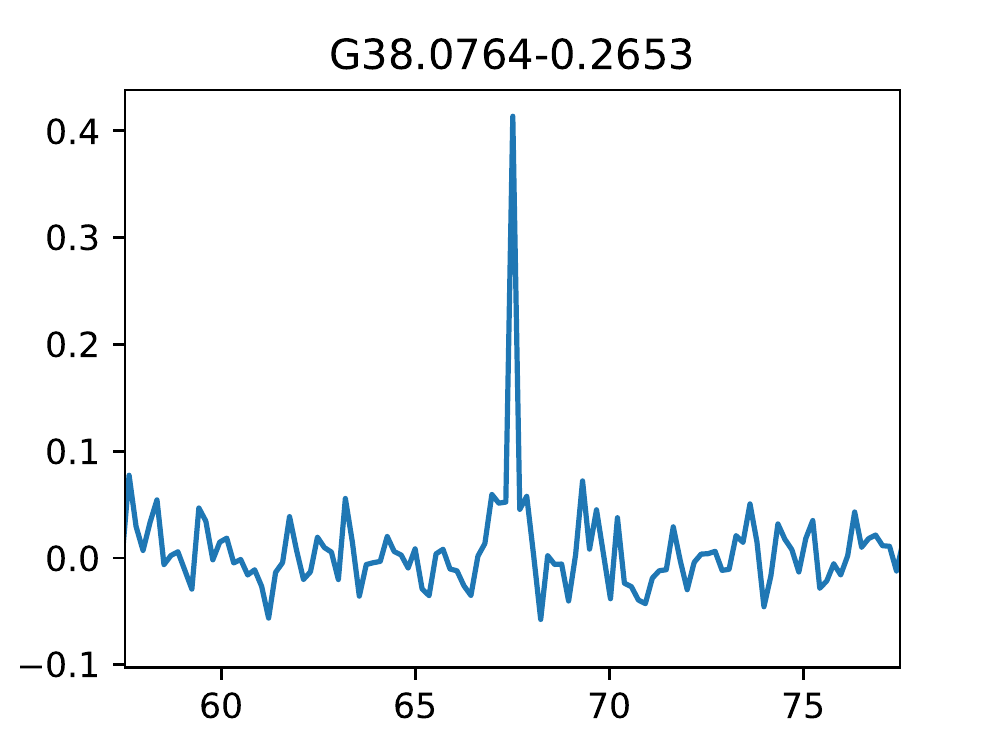}
\includegraphics[trim={0.0cm 0.3cm 0.6cm 0.3cm}, clip, width = 0.235\textwidth]{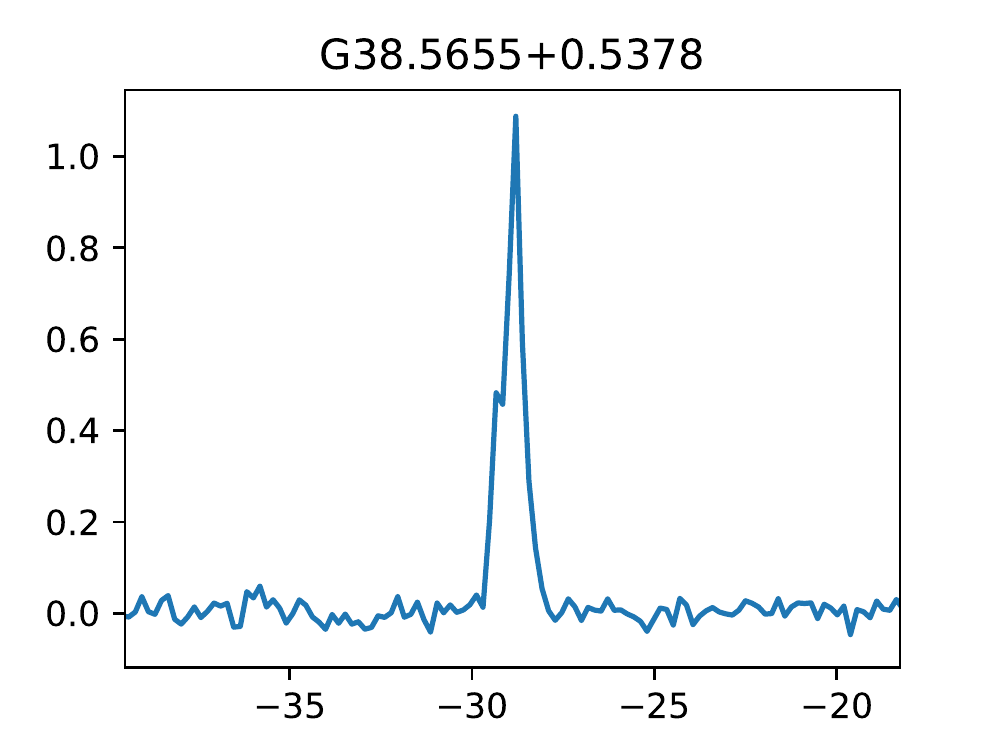}\\
\end{tabular}
\caption{Continued from Fig.~\ref{fig:maser_many_spec}.}
\label{fig:masers_480}
\end{figure*}

%Table 21
\begin{figure*}[!h]
\begin{tabular}{ccccc}
\colorbox{white}{\includegraphics[trim={2cm 0 0 0}, scale=0.195]{figs/ylabel_filler.pdf}}
\includegraphics[trim={0.0cm 0.3cm 0.6cm 0.3cm}, clip, width = 0.235\textwidth]{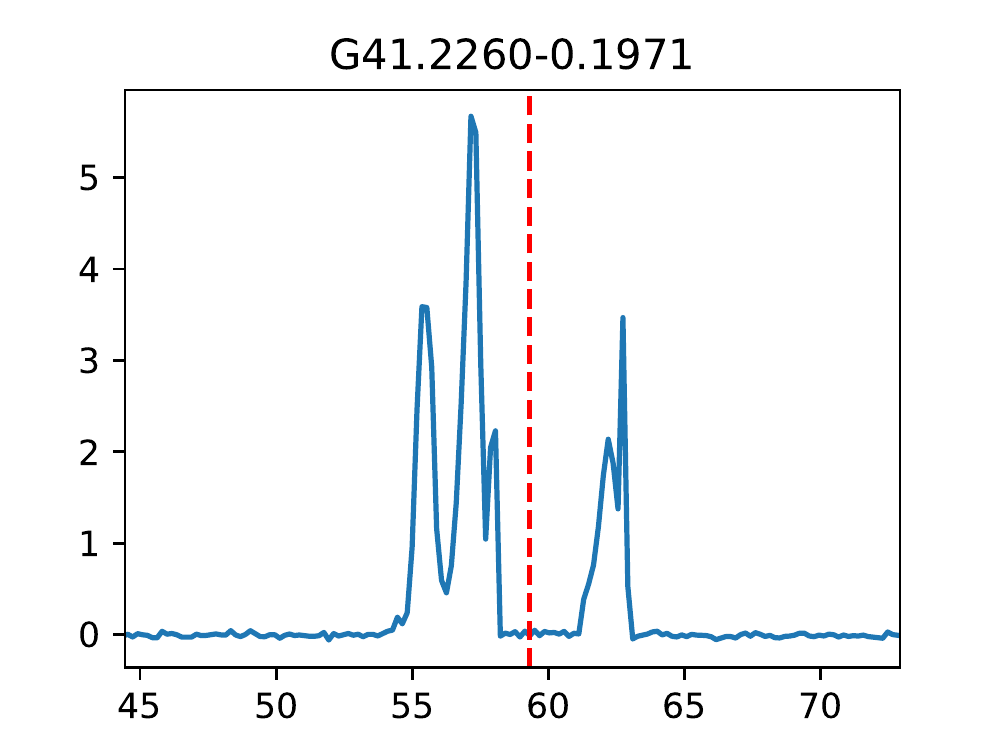}
\includegraphics[trim={0.0cm 0.3cm 0.6cm 0.3cm}, clip, width = 0.235\textwidth]{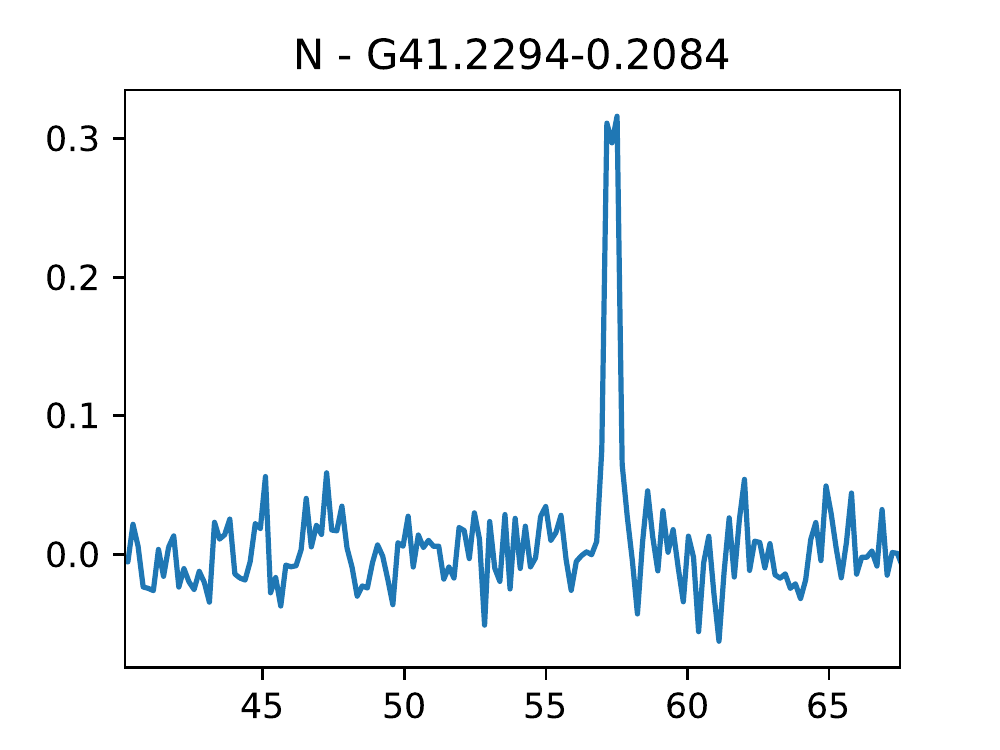}
\includegraphics[trim={0.0cm 0.3cm 0.6cm 0.3cm}, clip, width = 0.235\textwidth]{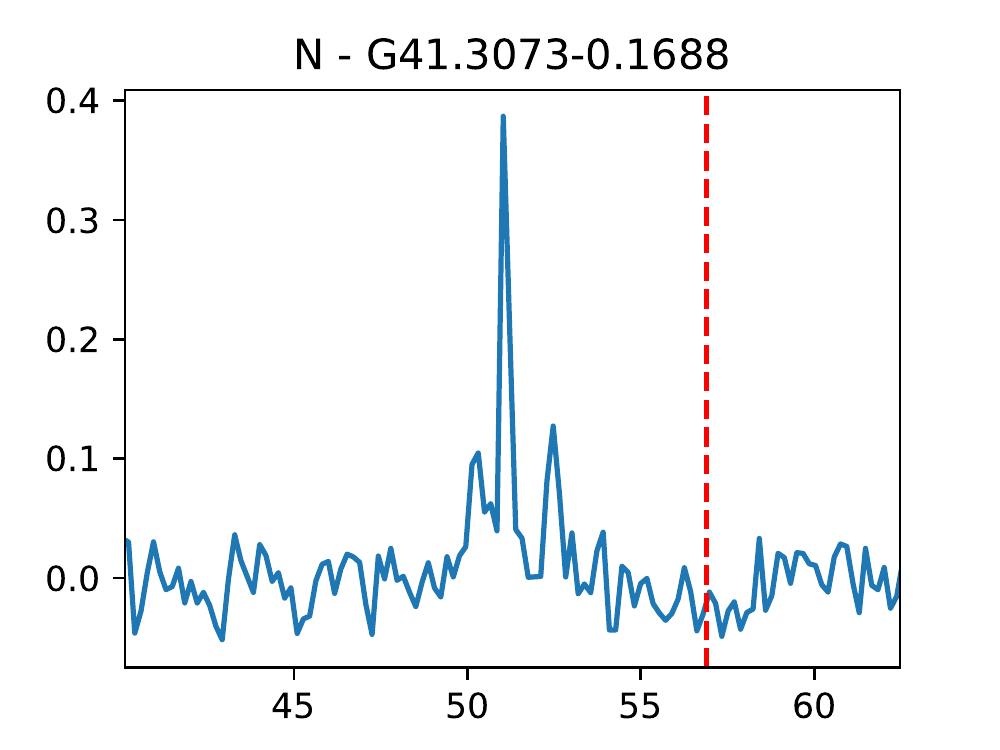}
\includegraphics[trim={0.0cm 0.3cm 0.6cm 0.3cm}, clip, width = 0.235\textwidth]{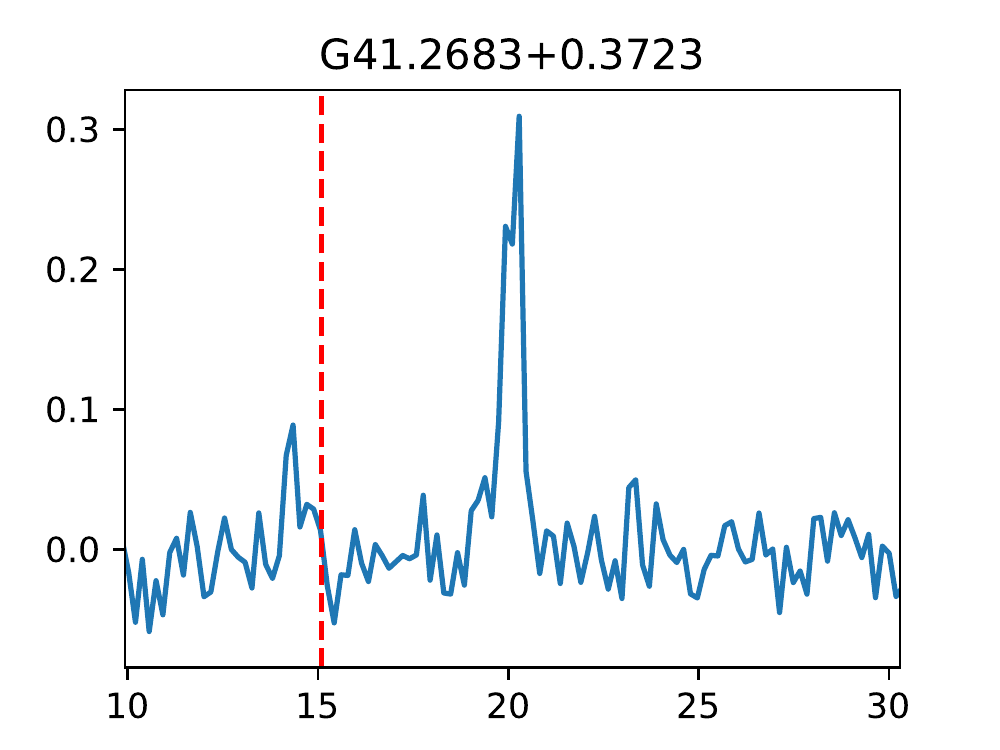}\\ 
\colorbox{white}{\includegraphics[trim={2cm 0 0 0}, scale=0.195]{figs/ylabel_filler.pdf}}
\includegraphics[trim={0.0cm 0.3cm 0.6cm 0.3cm}, clip, width = 0.235\textwidth]{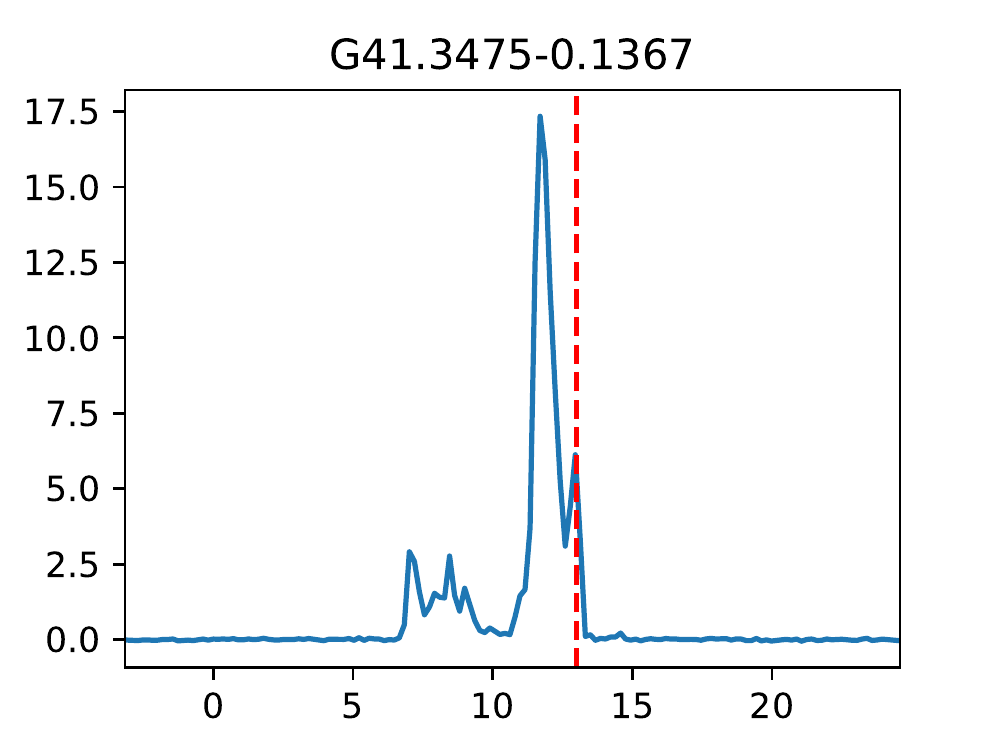}
\includegraphics[trim={0.0cm 0.3cm 0.6cm 0.3cm}, clip, width = 0.235\textwidth]{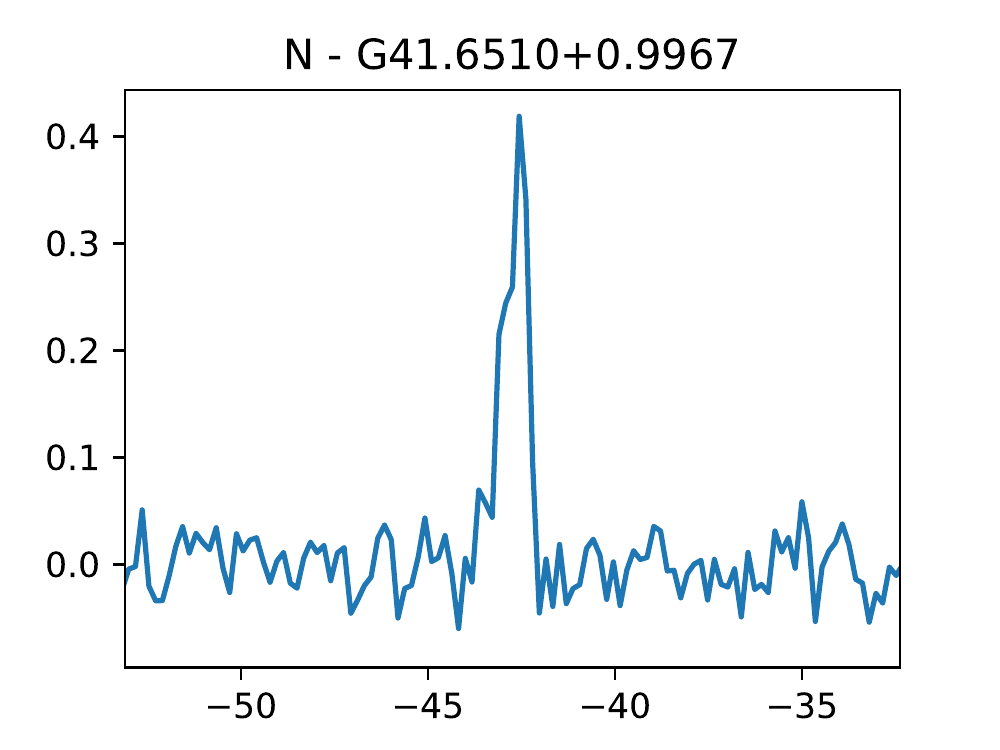}
\includegraphics[trim={0.0cm 0.3cm 0.6cm 0.3cm}, clip, width = 0.235\textwidth]{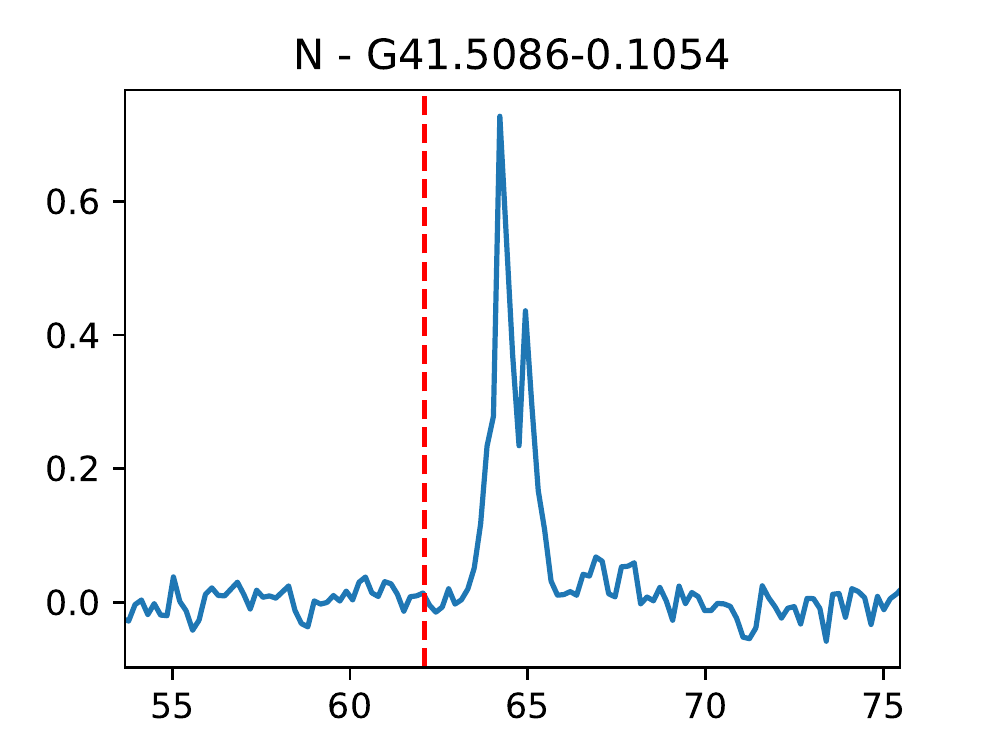}
\includegraphics[trim={0.0cm 0.3cm 0.6cm 0.3cm}, clip, width = 0.235\textwidth]{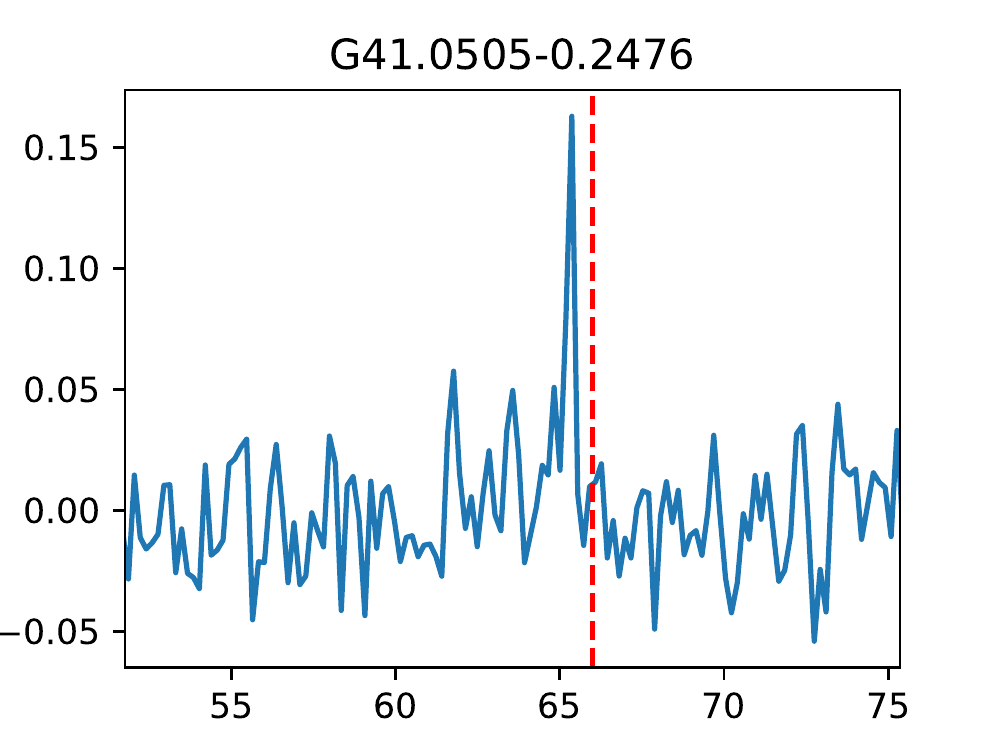}\\ 
\colorbox{white}{\includegraphics[trim={2cm 0 0 0}, scale=0.20]{figs/ylabel.pdf}}
\includegraphics[trim={0.0cm 0.3cm 0.6cm 0.3cm}, clip, width = 0.235\textwidth]{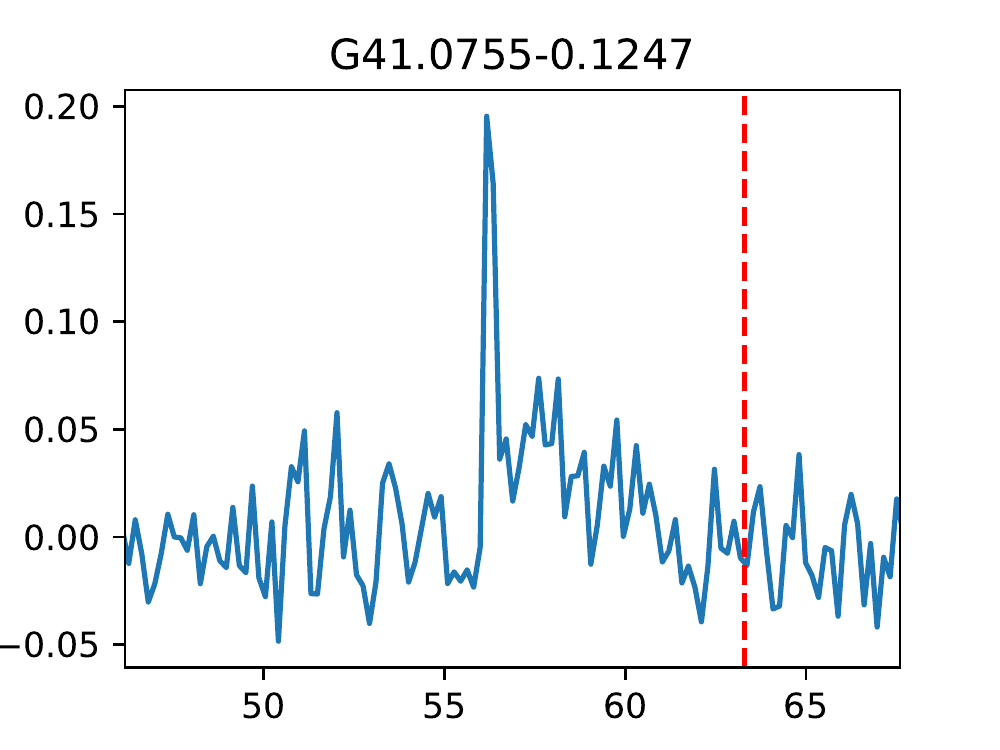}
\includegraphics[trim={0.0cm 0.3cm 0.6cm 0.3cm}, clip, width = 0.235\textwidth]{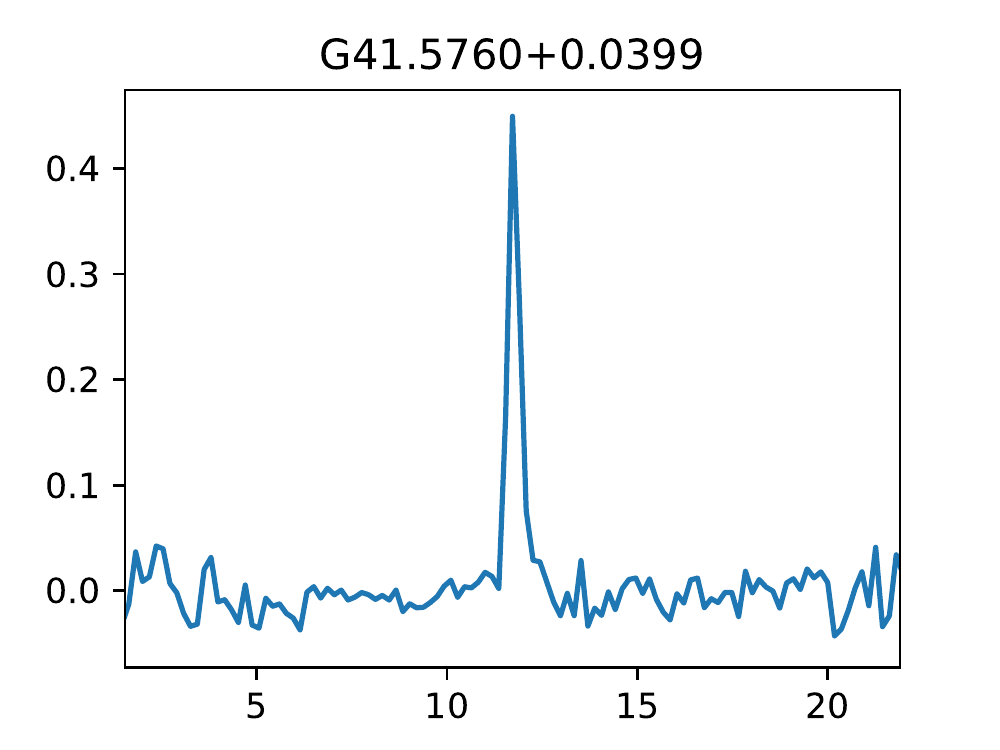}
\includegraphics[trim={0.0cm 0.3cm 0.6cm 0.3cm}, clip, width = 0.235\textwidth]{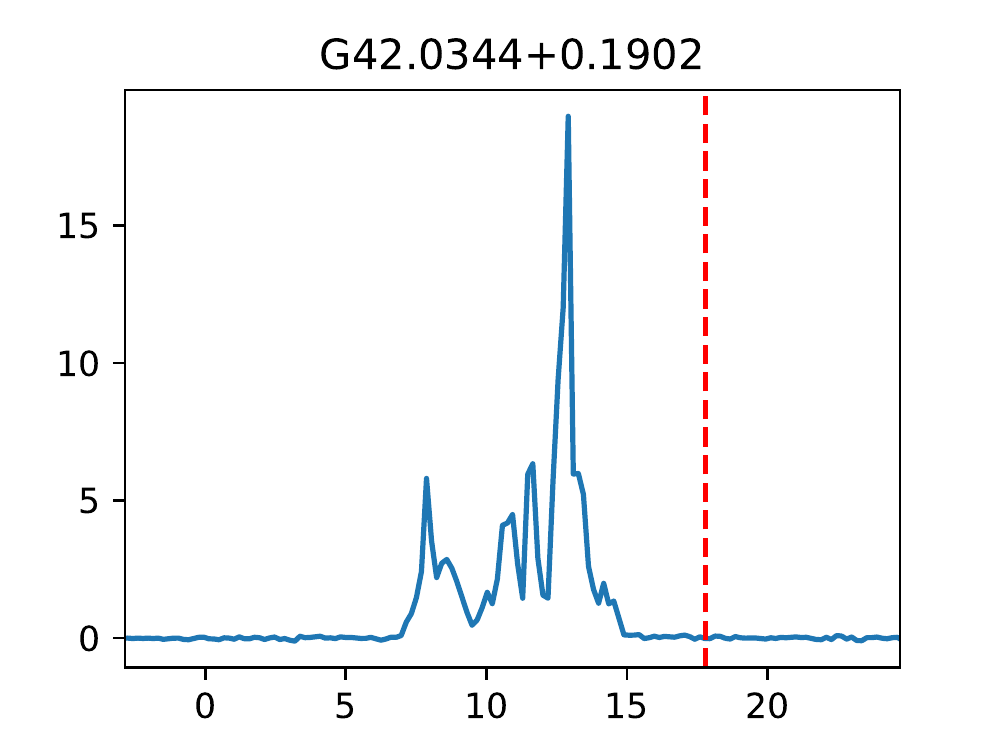}
\includegraphics[trim={0.0cm 0.3cm 0.6cm 0.3cm}, clip, width = 0.235\textwidth]{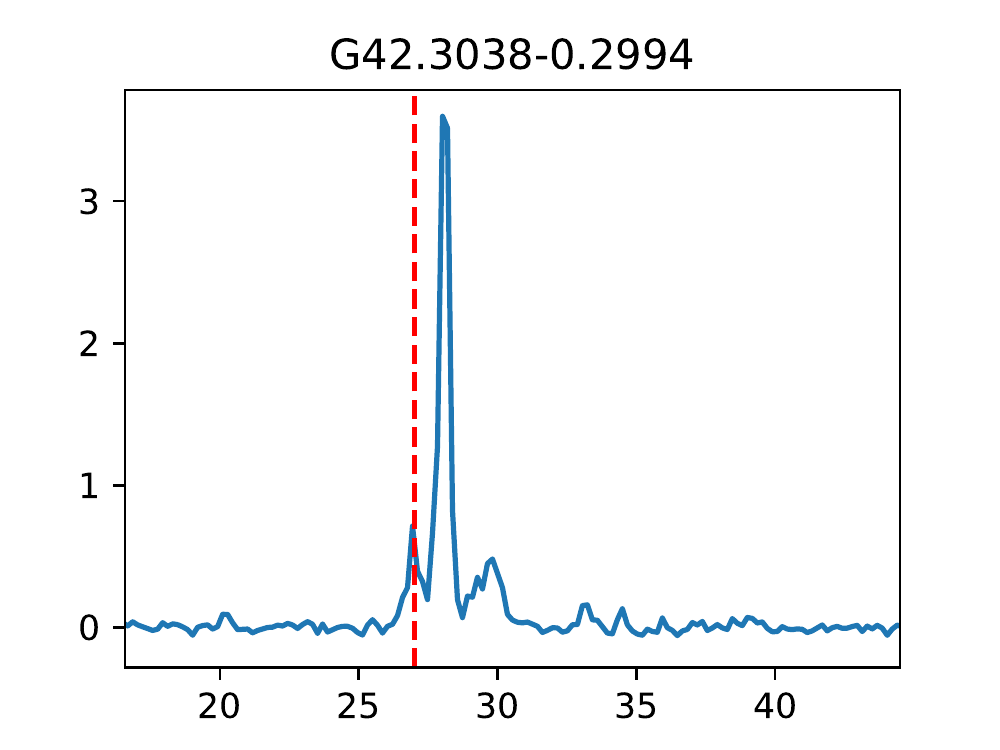}\\
\end{tabular}
\caption{Continued from Fig.~\ref{fig:maser_many_spec}.}
\label{fig:masers_504}
\end{figure*}

%Table 22
\begin{figure*}[!h]
\begin{tabular}{ccccc}
\colorbox{white}{\includegraphics[trim={2cm 0 0 0}, scale=0.195]{figs/ylabel_filler.pdf}}
\includegraphics[trim={0.0cm 0.3cm 0.6cm 0.3cm}, clip, width = 0.235\textwidth]{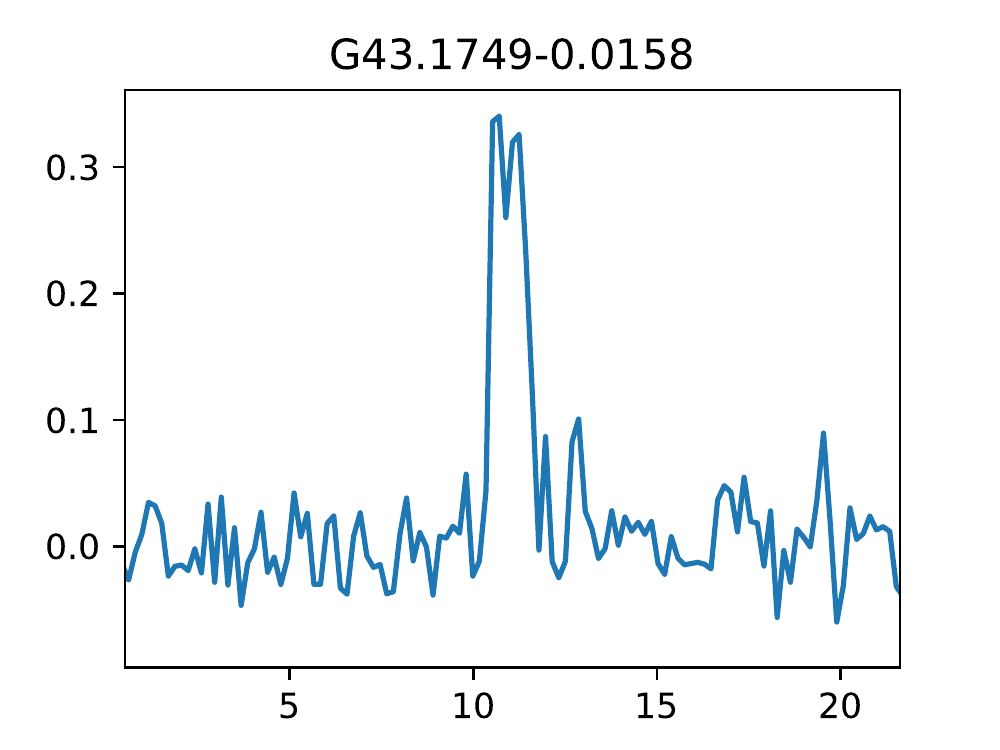}
\includegraphics[trim={0.0cm 0.3cm 0.6cm 0.3cm}, clip, width = 0.235\textwidth]{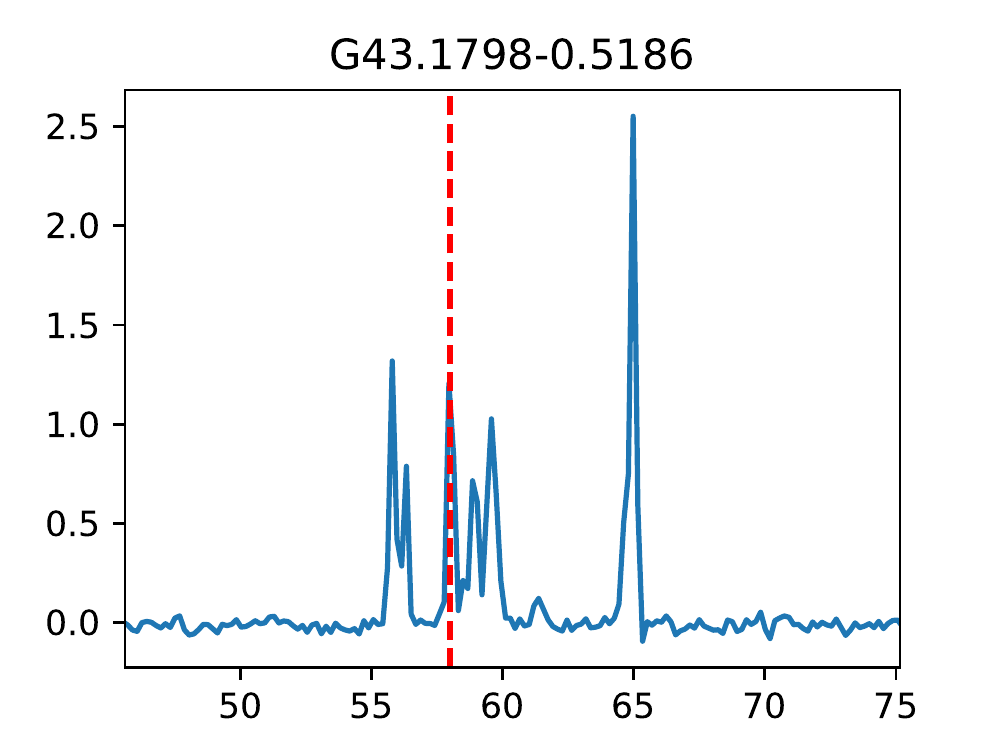}
\includegraphics[trim={0.0cm 0.3cm 0.6cm 0.3cm}, clip, width = 0.235\textwidth]{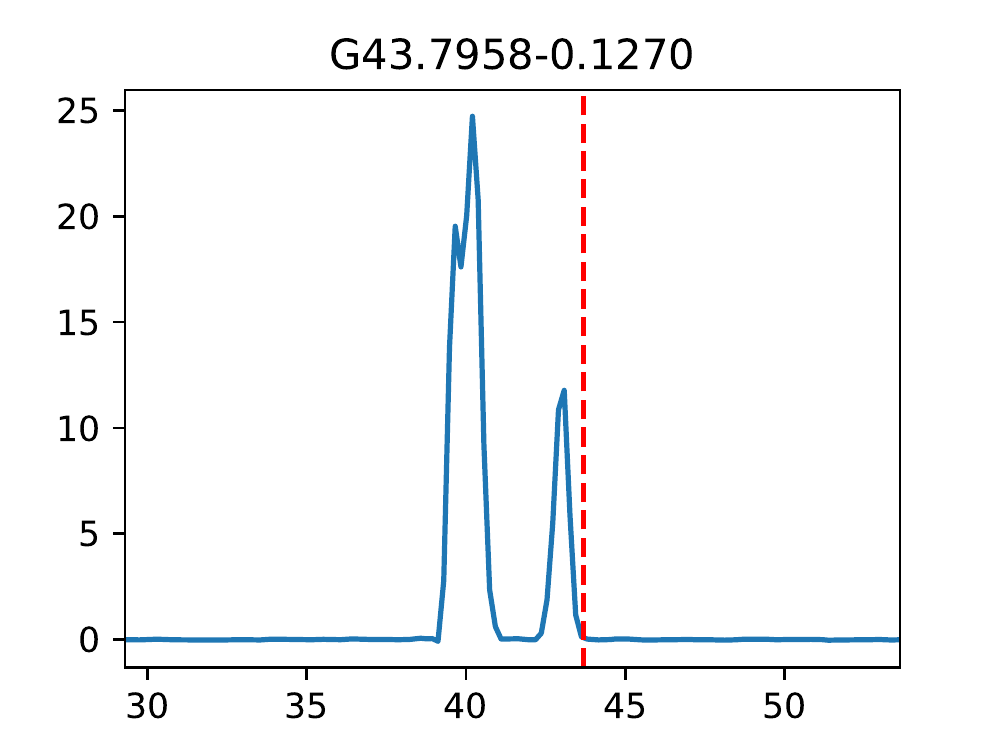}
\includegraphics[trim={0.0cm 0.3cm 0.6cm 0.3cm}, clip, width = 0.235\textwidth]{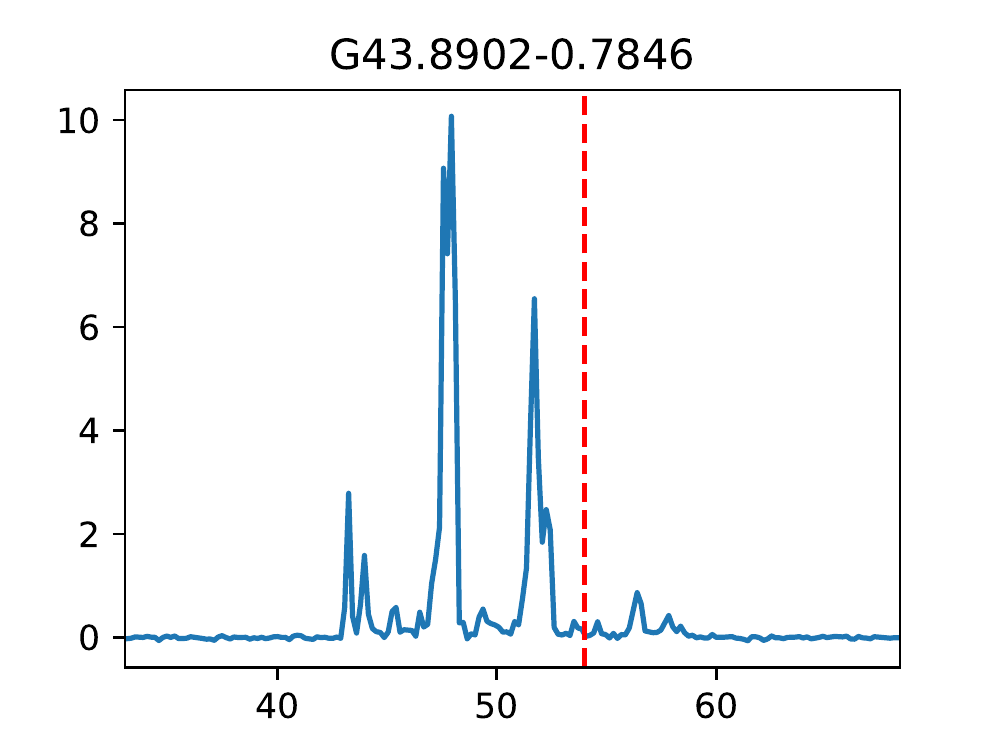}\\ 
\colorbox{white}{\includegraphics[trim={2cm 0 0 0}, scale=0.195]{figs/ylabel_filler.pdf}}
\includegraphics[trim={0.0cm 0.3cm 0.6cm 0.3cm}, clip, width = 0.235\textwidth]{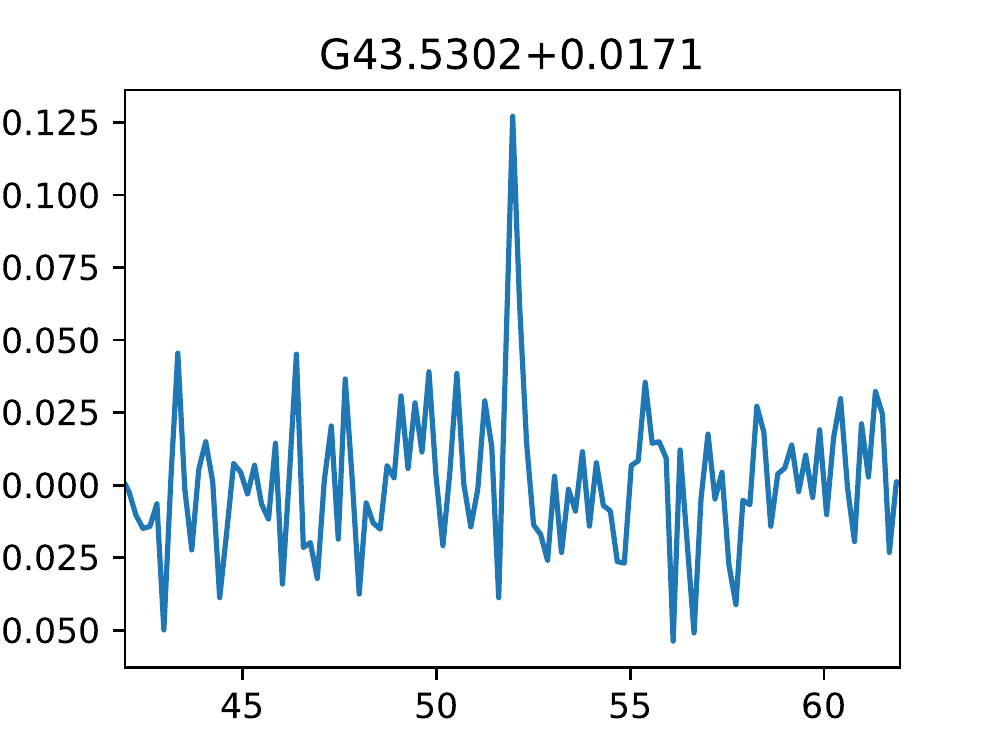}
\includegraphics[trim={0.0cm 0.3cm 0.6cm 0.3cm}, clip, width = 0.235\textwidth]{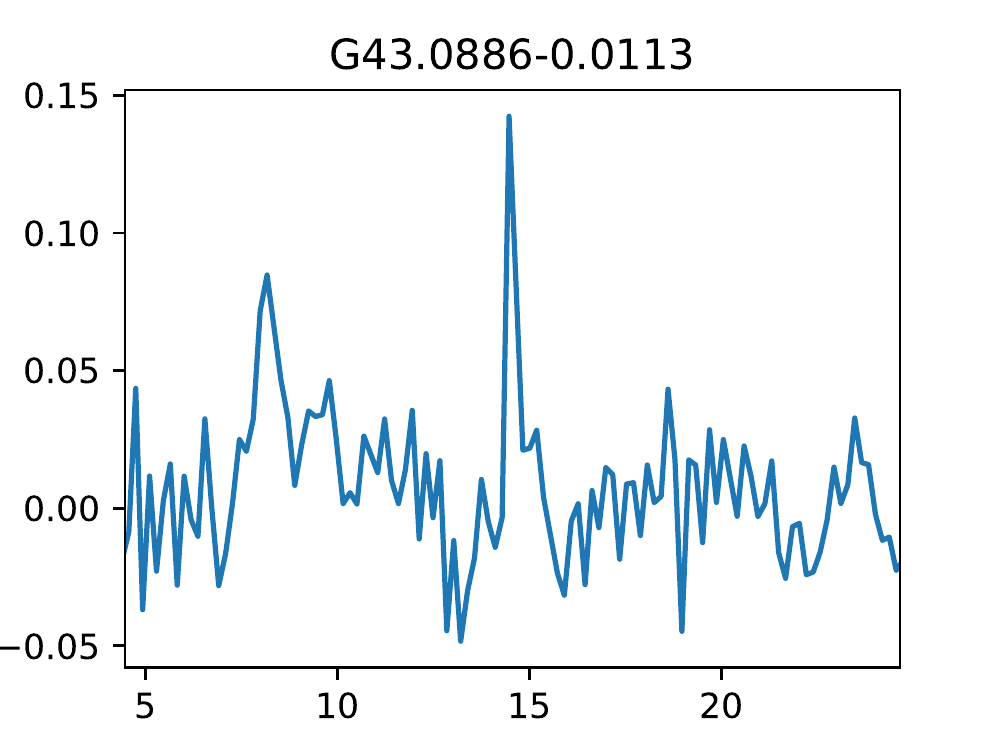}
\includegraphics[trim={0.0cm 0.3cm 0.6cm 0.3cm}, clip, width = 0.235\textwidth]{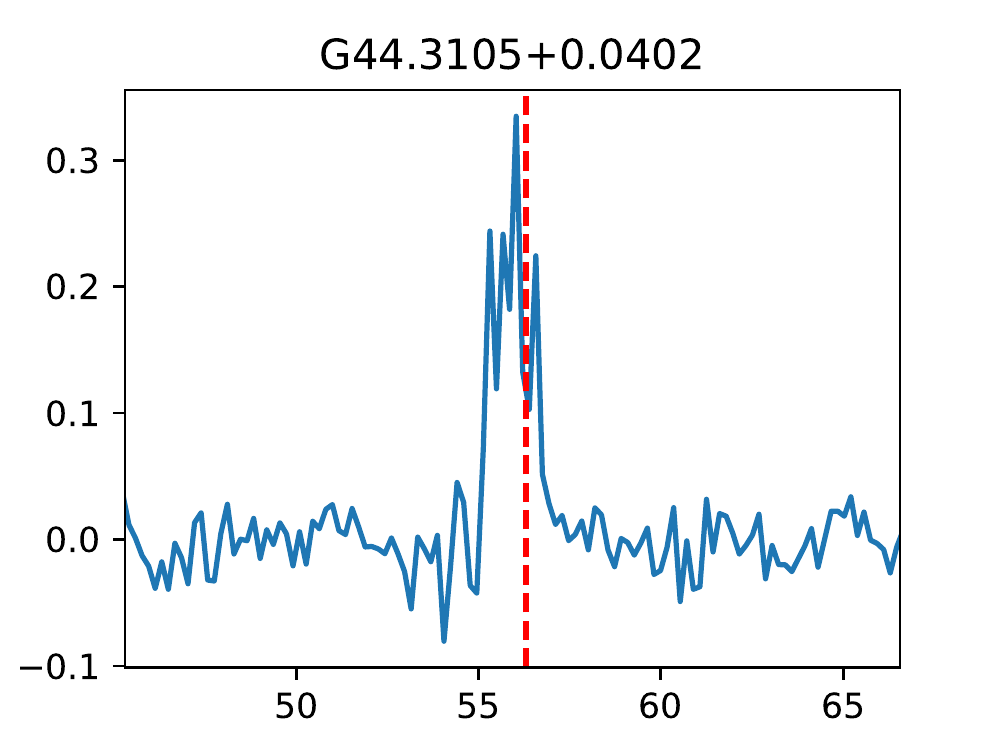}
\includegraphics[trim={0.0cm 0.3cm 0.6cm 0.3cm}, clip, width = 0.235\textwidth]{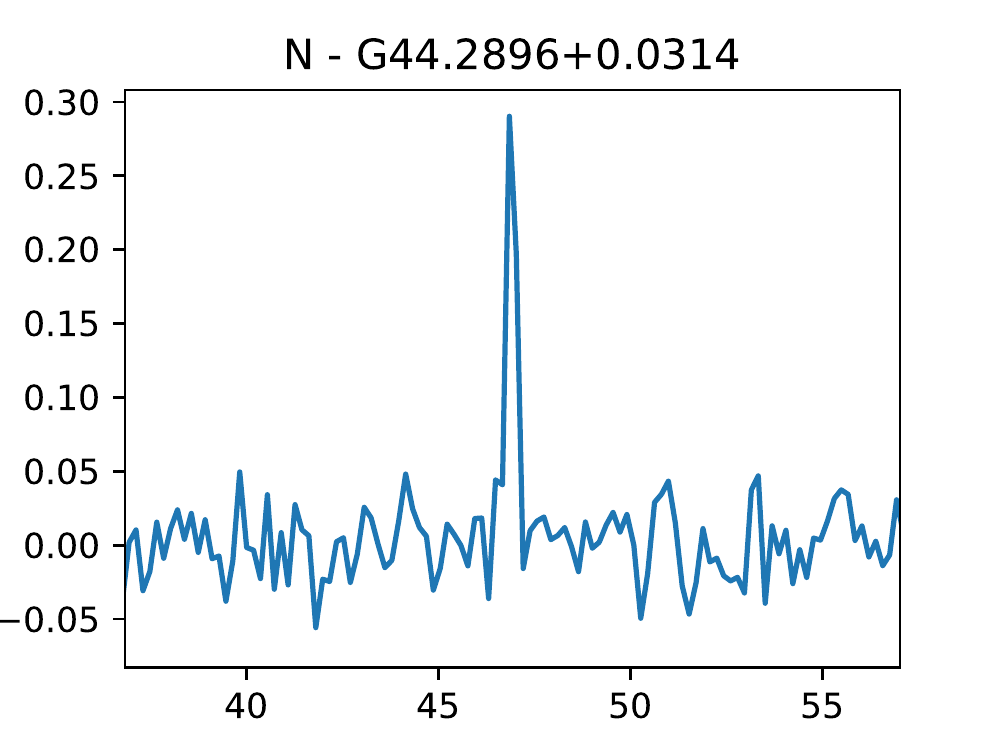}\\ 
\colorbox{white}{\includegraphics[trim={2cm 0 0 0}, scale=0.20]{figs/ylabel.pdf}}
\includegraphics[trim={0.0cm 0.3cm 0.6cm 0.3cm}, clip, width = 0.235\textwidth]{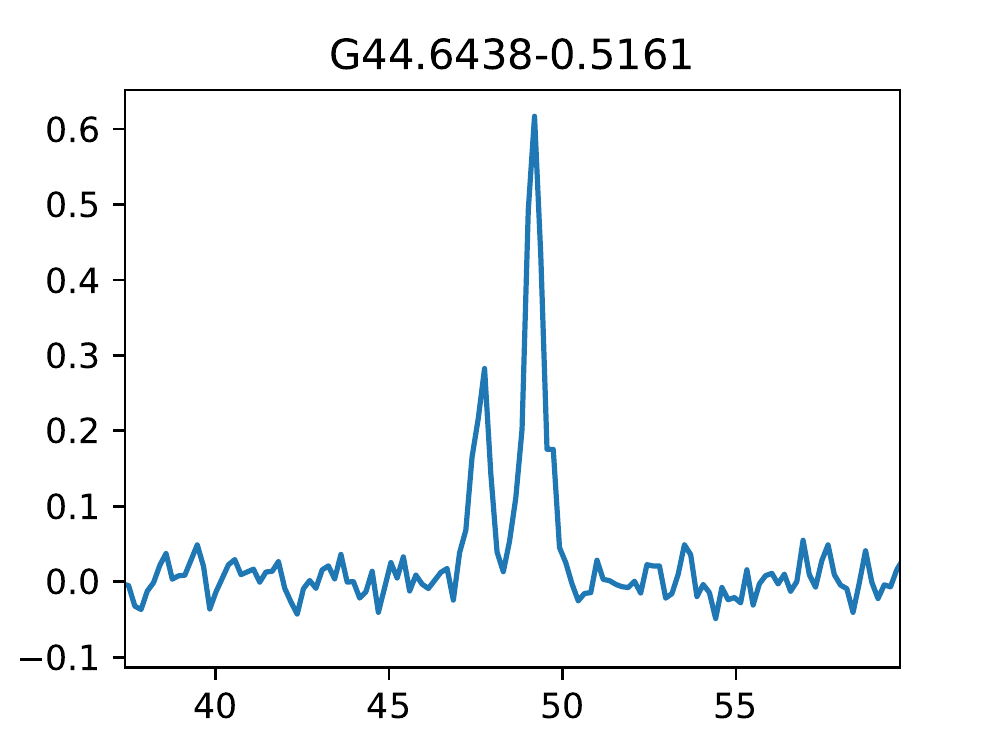}
\includegraphics[trim={0.0cm 0.3cm 0.6cm 0.3cm}, clip, width = 0.235\textwidth]{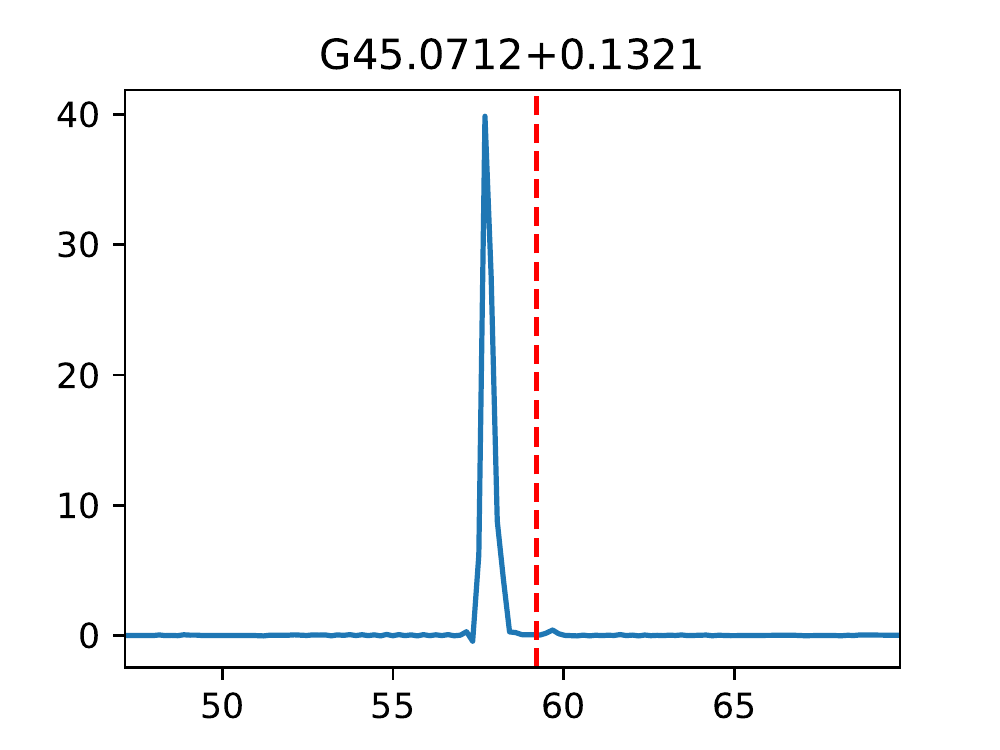}
\includegraphics[trim={0.0cm 0.3cm 0.6cm 0.3cm}, clip, width = 0.235\textwidth]{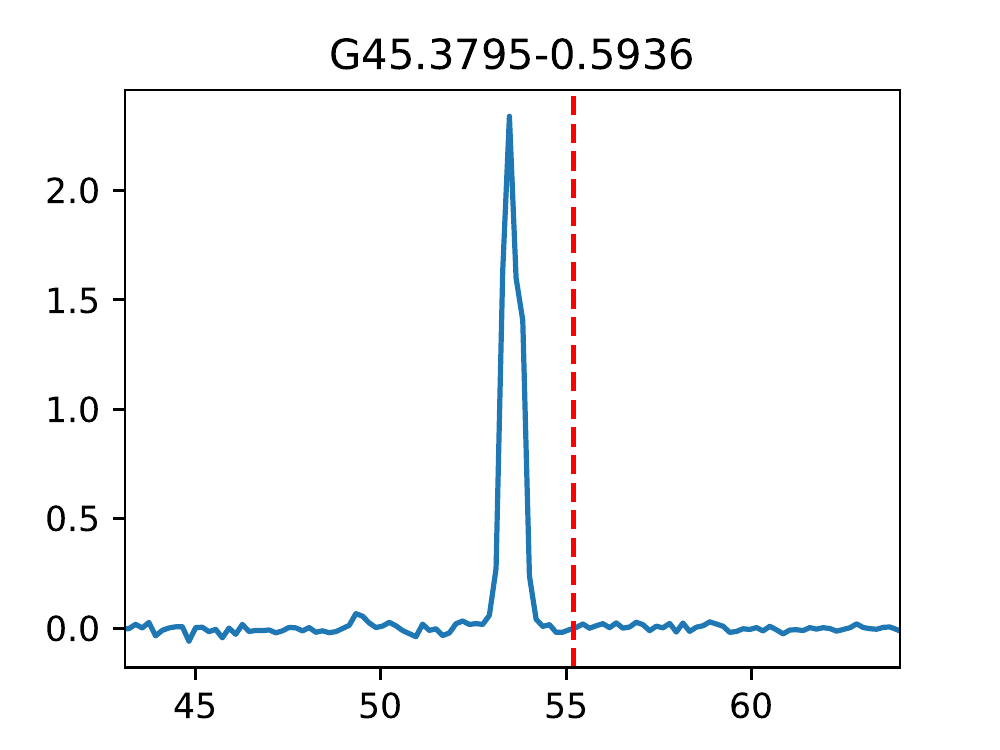}
\includegraphics[trim={0.0cm 0.3cm 0.6cm 0.3cm}, clip, width = 0.235\textwidth]{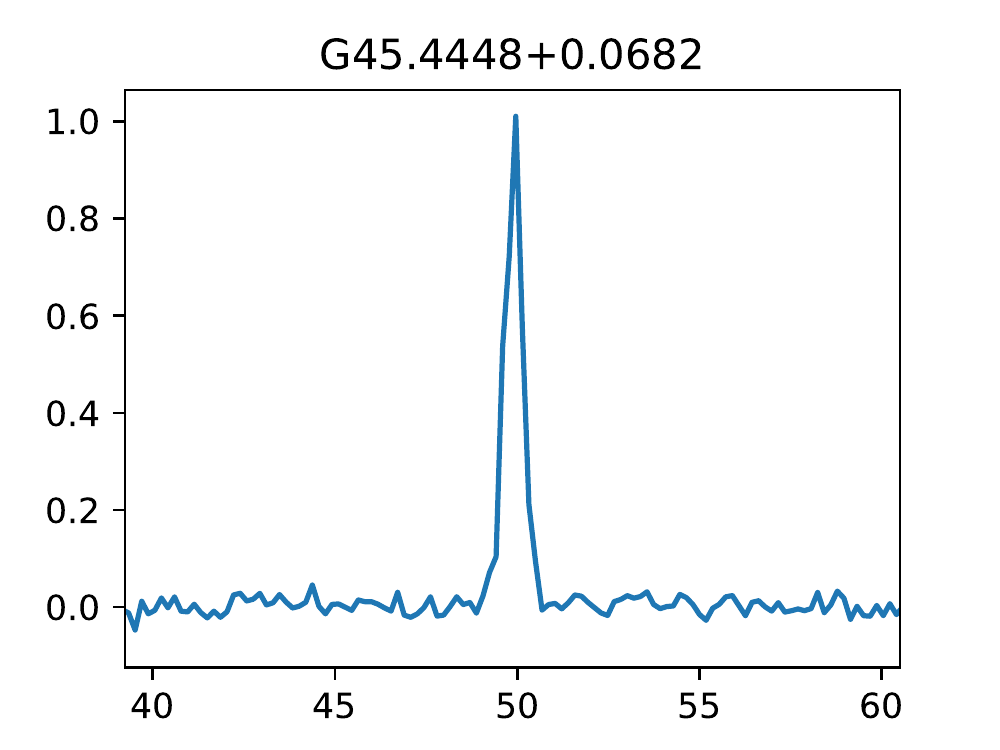}\\
\end{tabular}
\caption{Continued from Fig.~\ref{fig:maser_many_spec}.}
\label{fig:masers_528}
\end{figure*}

%Table 23
\begin{figure*}[!h]
\begin{tabular}{ccccc}
\colorbox{white}{\includegraphics[trim={2cm 0 0 0}, scale=0.195]{figs/ylabel_filler.pdf}}
\includegraphics[trim={0.0cm 0.3cm 0.6cm 0.3cm}, clip, width = 0.235\textwidth]{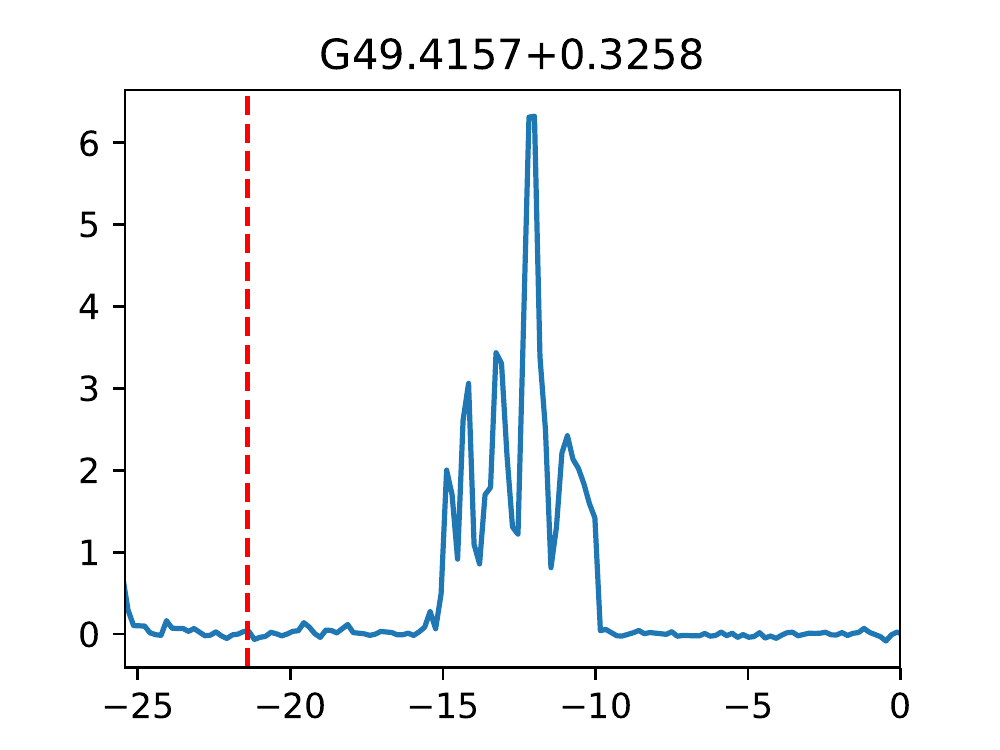}
\includegraphics[trim={0.0cm 0.3cm 0.6cm 0.3cm}, clip, width = 0.235\textwidth]{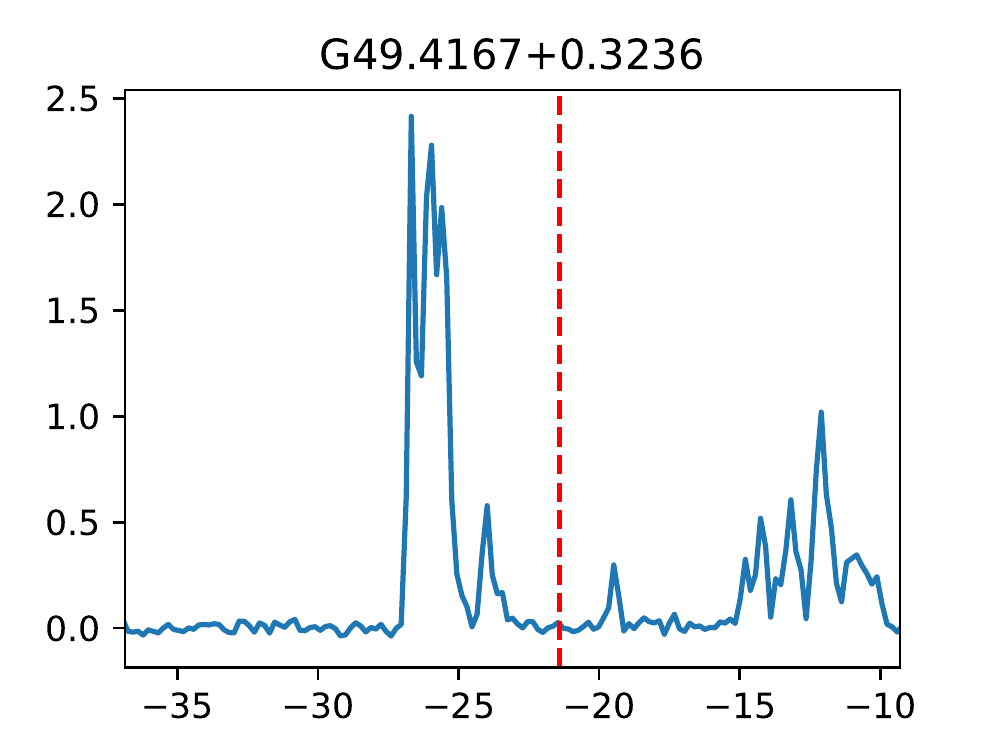}
\includegraphics[trim={0.0cm 0.3cm 0.6cm 0.3cm}, clip, width = 0.235\textwidth]{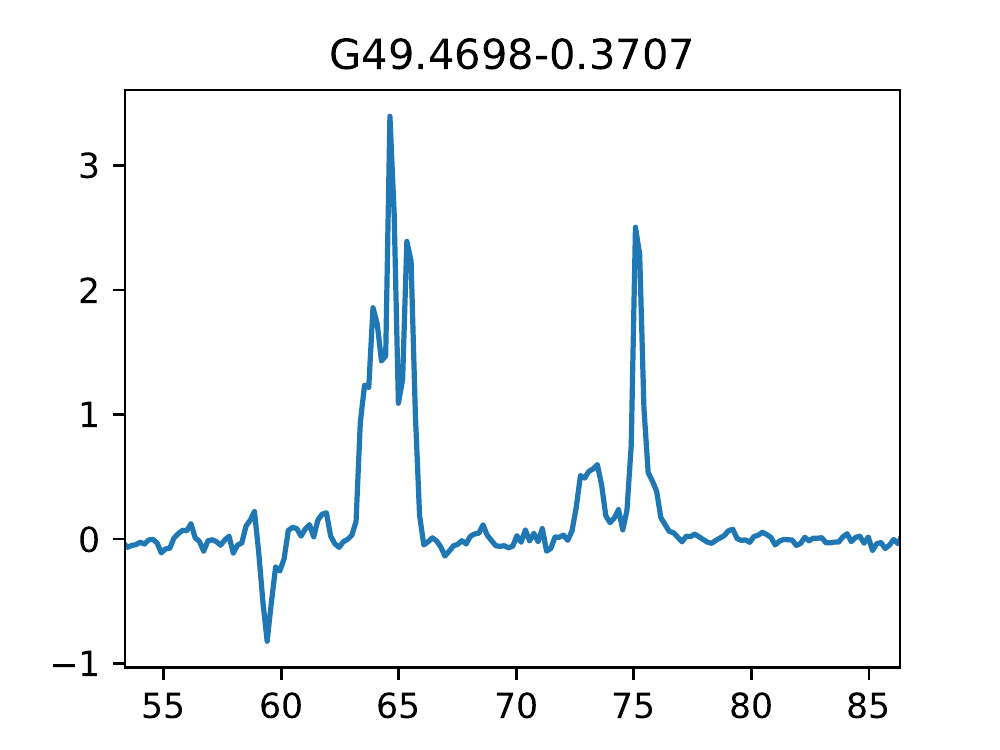}
\includegraphics[trim={0.0cm 0.3cm 0.6cm 0.3cm}, clip, width = 0.235\textwidth]{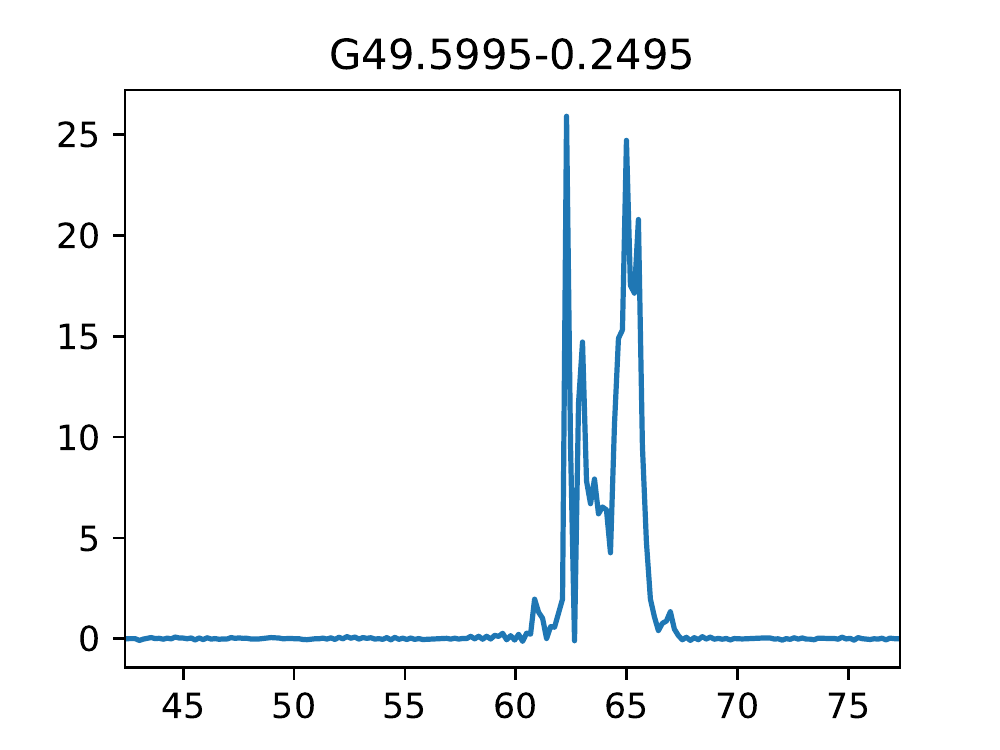}\\ 
\colorbox{white}{\includegraphics[trim={2cm 0 0 0}, scale=0.195]{figs/ylabel_filler.pdf}}
\includegraphics[trim={0.0cm 0.3cm 0.6cm 0.3cm}, clip, width = 0.235\textwidth]{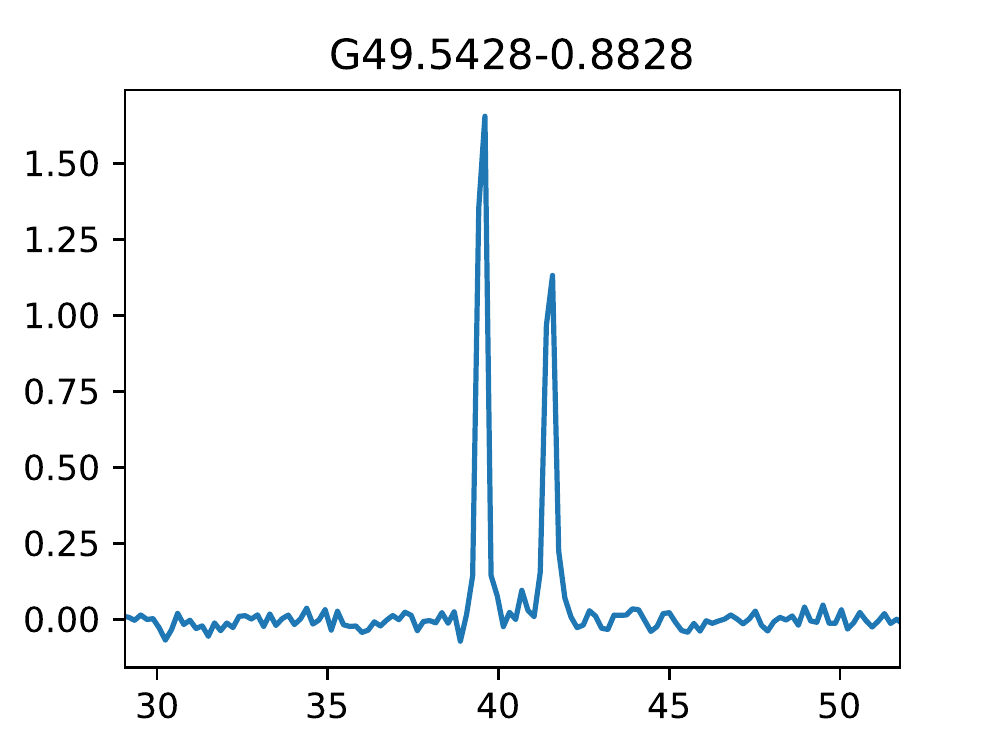}
\includegraphics[trim={0.0cm 0.3cm 0.6cm 0.3cm}, clip, width = 0.235\textwidth]{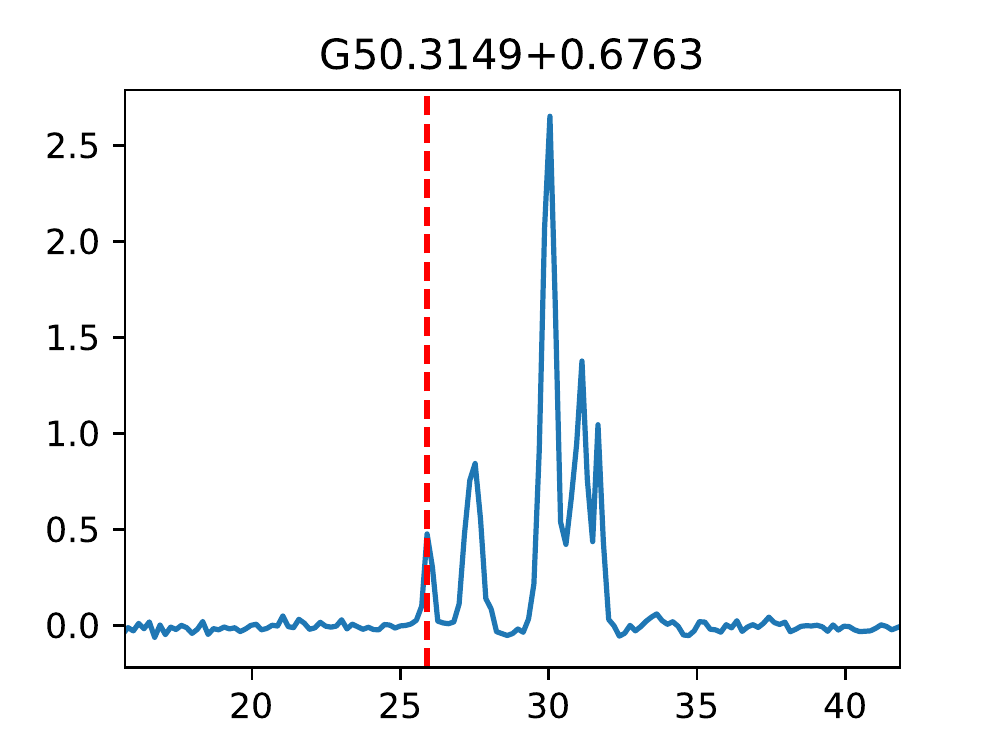}
\includegraphics[trim={0.0cm 0.3cm 0.6cm 0.3cm}, clip, width = 0.235\textwidth]{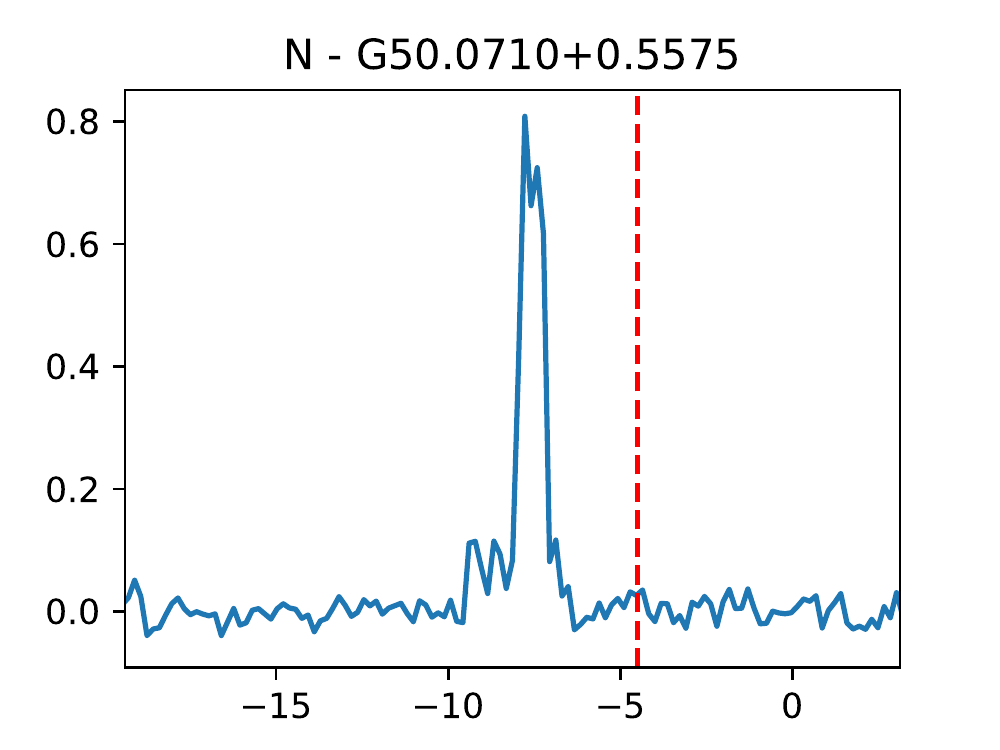}
\includegraphics[trim={0.0cm 0.3cm 0.6cm 0.3cm}, clip, width = 0.235\textwidth]{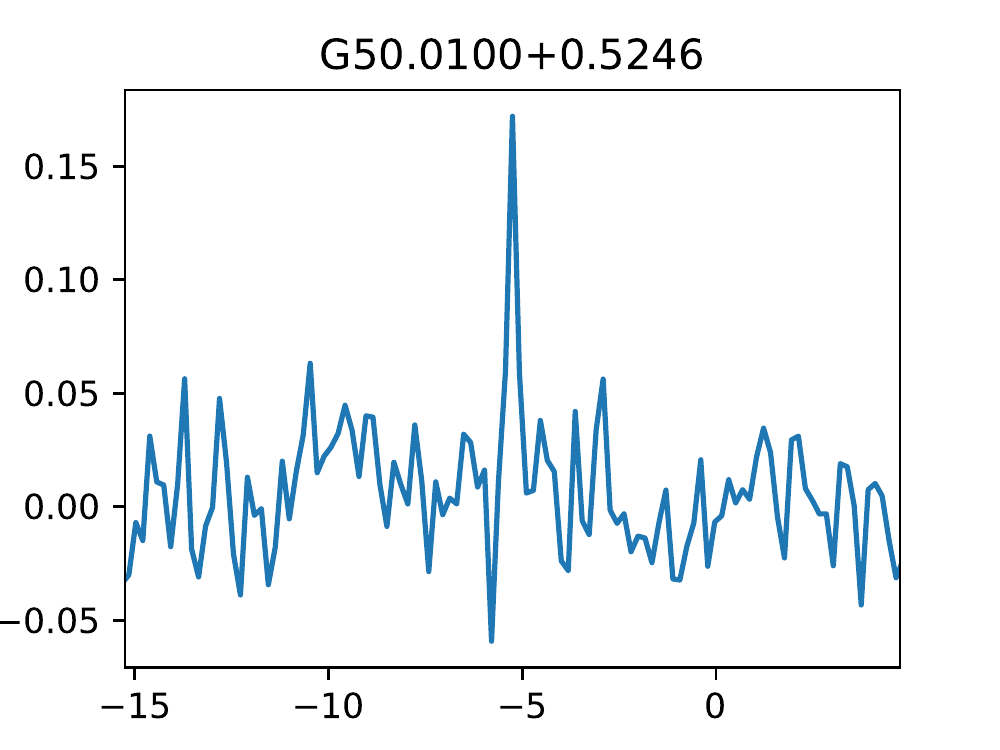}\\ 
\colorbox{white}{\includegraphics[trim={2cm 0 0 0}, scale=0.20]{figs/ylabel.pdf}}
\includegraphics[trim={0.0cm 0.3cm 0.6cm 0.3cm}, clip, width = 0.235\textwidth]{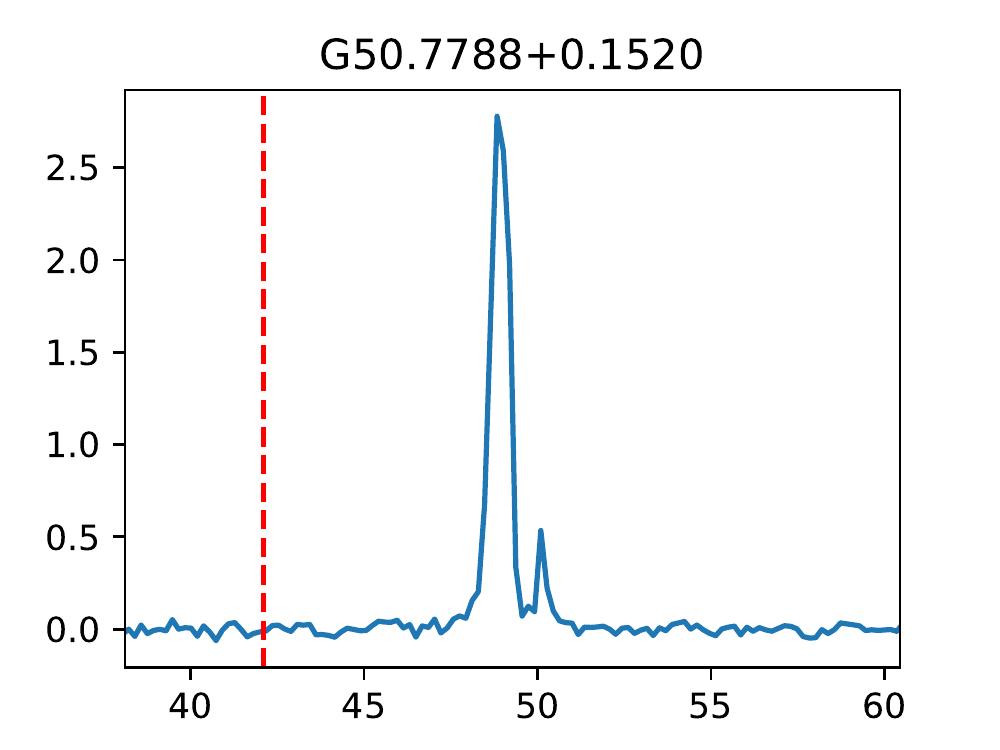}
\includegraphics[trim={0.0cm 0.3cm 0.6cm 0.3cm}, clip, width = 0.235\textwidth]{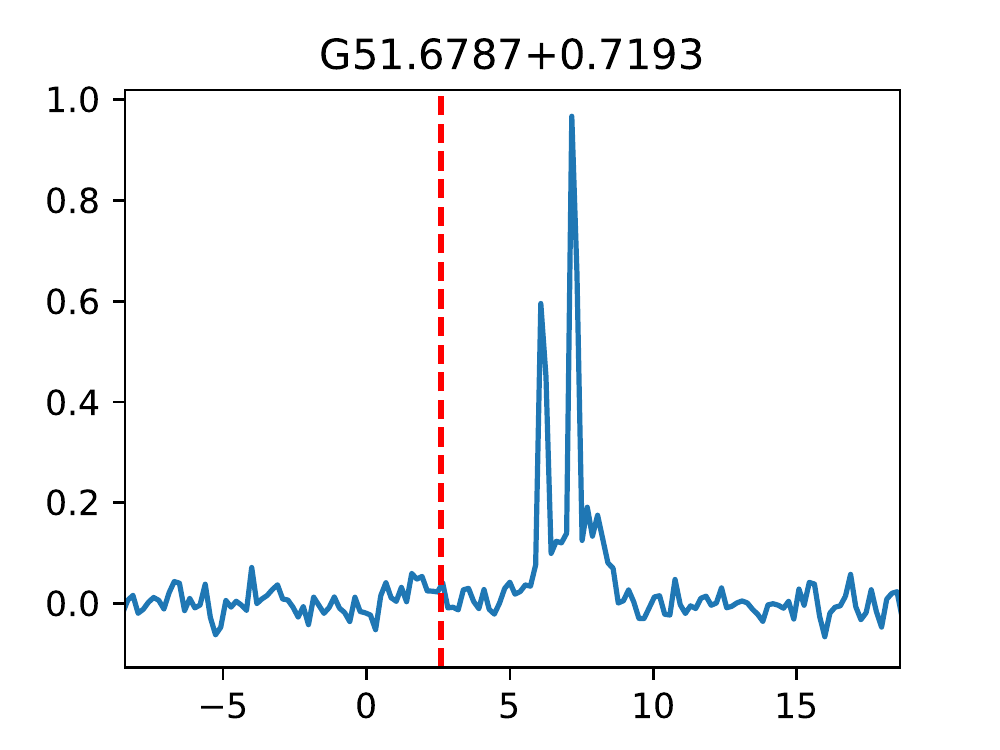}
\includegraphics[trim={0.0cm 0.3cm 0.6cm 0.3cm}, clip, width = 0.235\textwidth]{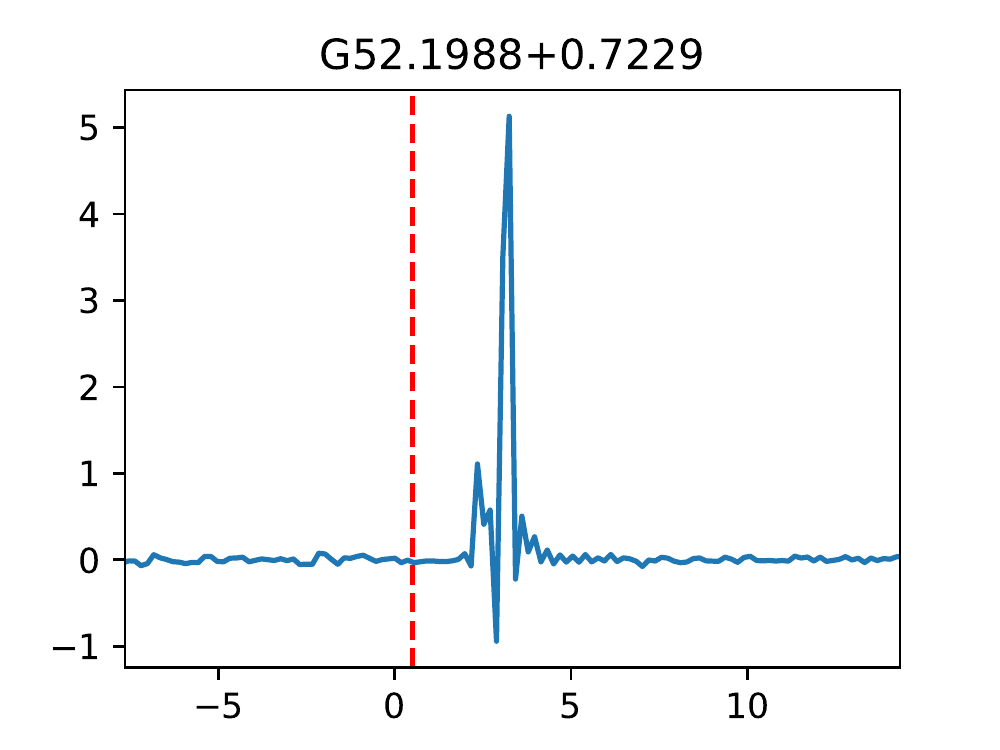}
\includegraphics[trim={0.0cm 0.3cm 0.6cm 0.3cm}, clip, width = 0.235\textwidth]{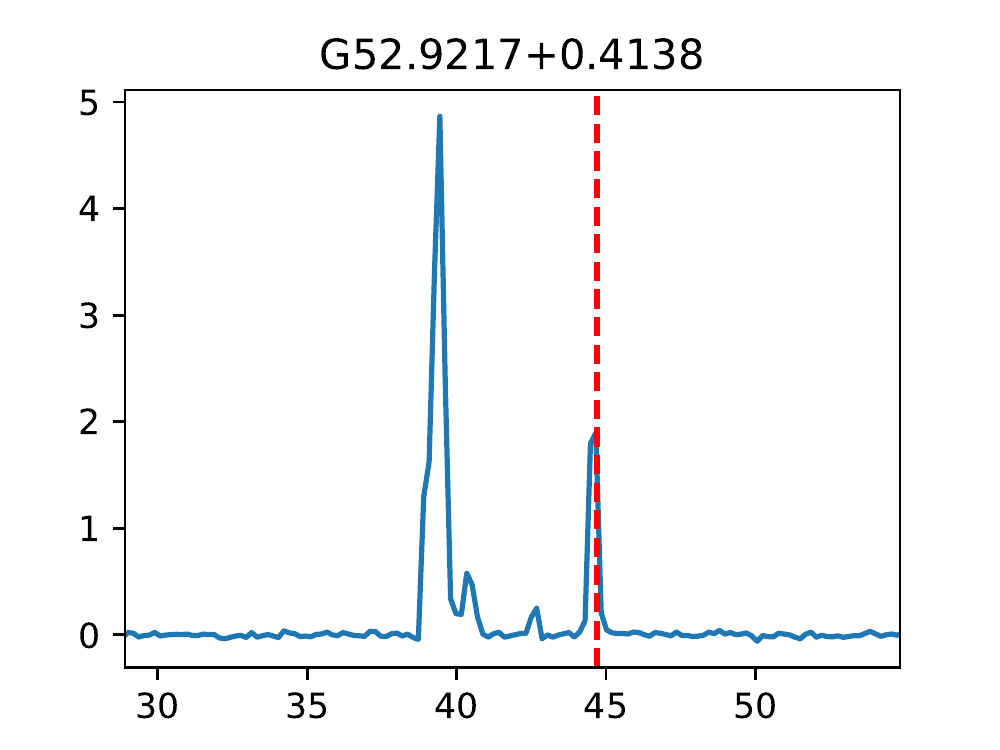}\\
\end{tabular}
\caption{Continued from Fig.~\ref{fig:maser_many_spec}.}
\label{fig:masers_552}
\end{figure*}

%Table 24
\begin{figure*}[!h]
\begin{tabular}{ccccc}
\colorbox{white}{\includegraphics[trim={2cm 0 0 0}, scale=0.195]{figs/ylabel.pdf}}
\includegraphics[trim={0.0cm 0.3cm 0.6cm 0.3cm}, clip, width = 0.235\textwidth]{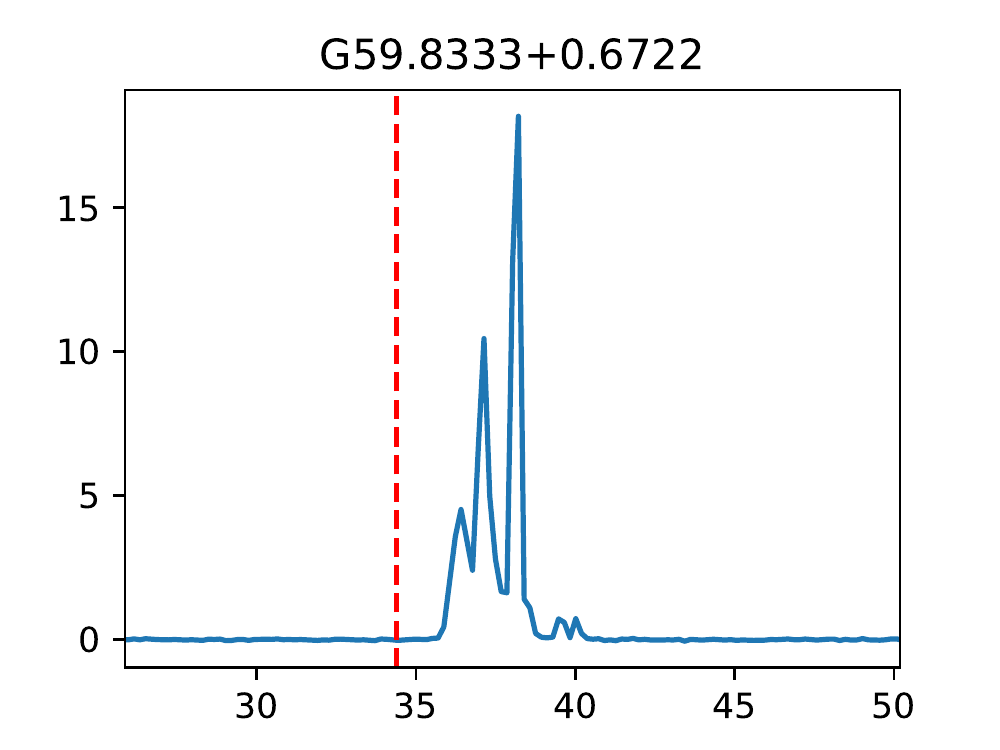}
\includegraphics[trim={0.0cm 0.3cm 0.6cm 0.3cm}, clip, width = 0.235\textwidth]{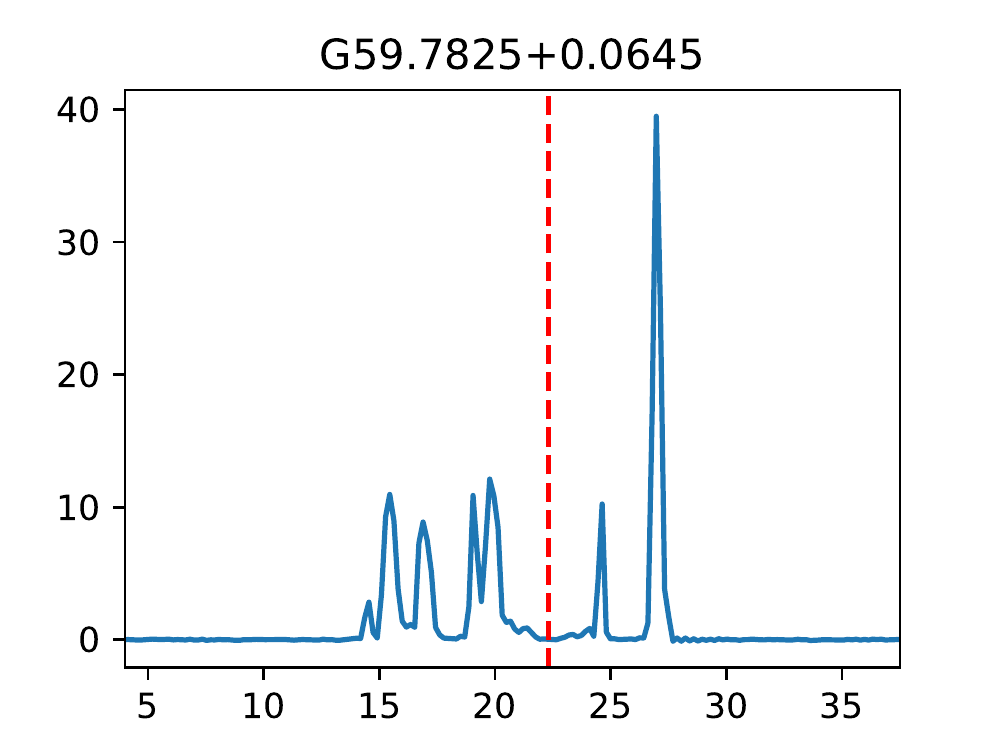}
\includegraphics[trim={0.0cm 0.3cm 0.6cm 0.3cm}, clip, width = 0.235\textwidth]{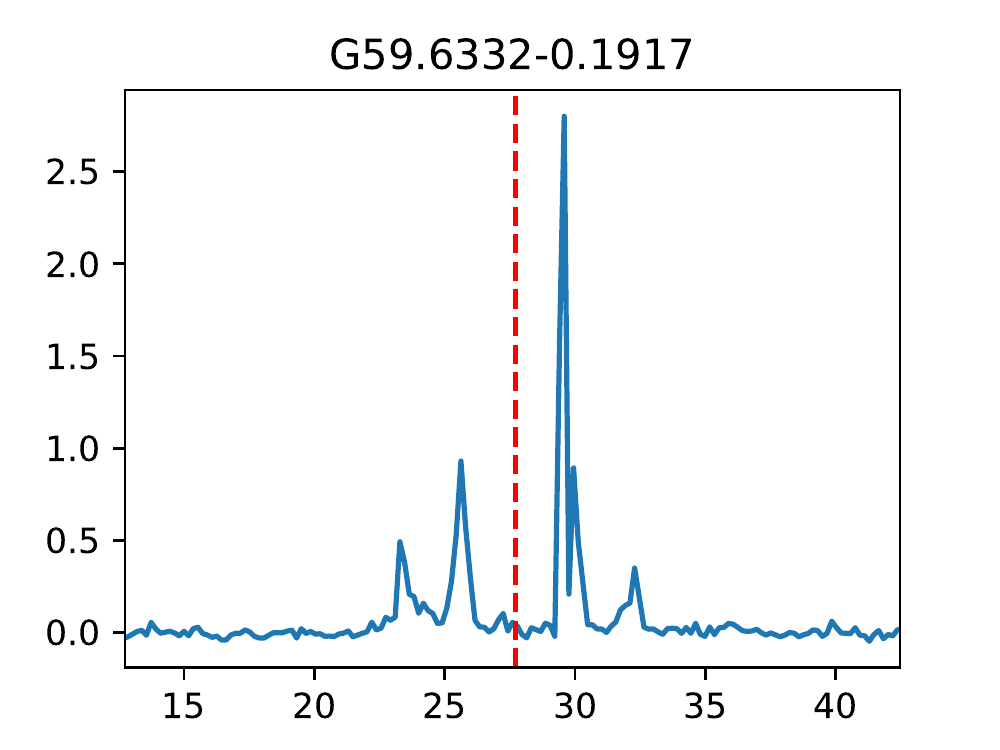}
\includegraphics[trim={0.0cm 0.3cm 0.6cm 0.3cm}, clip, width = 0.235\textwidth]{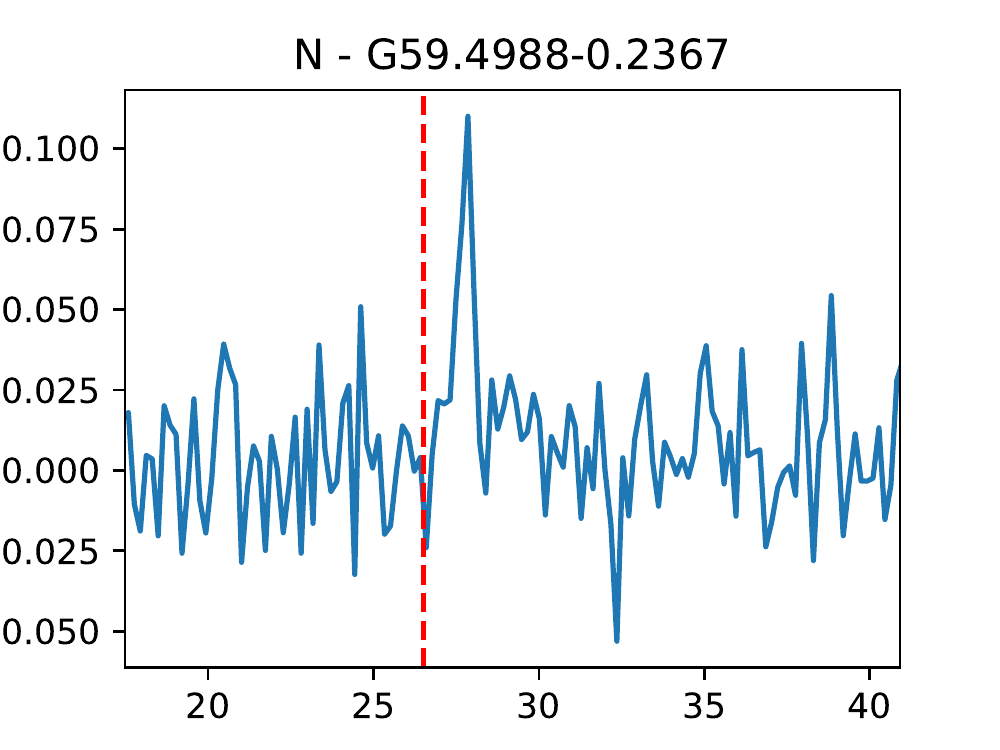}\\
\includegraphics[width = 0.175\textwidth]{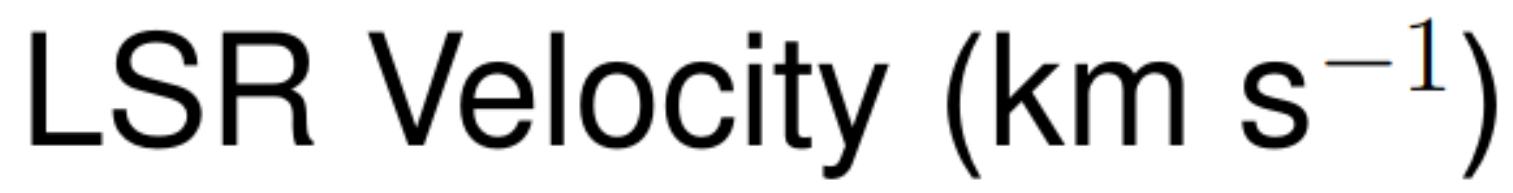}
\end{tabular}
\caption{Continued from Fig.~\ref{fig:maser_many_spec}.}
\label{fig:masers_576}
\end{figure*}

\end{appendix}

\end{document}